\documentclass[11pt,fleqn]{article}
\usepackage{amssymb, amsmath}
\usepackage{etoc}
\usepackage{amsfonts}
\usepackage{amsthm}
\usepackage{a4wide}
\newtheorem{assumption}{Assumption}
\usepackage{authblk}
\DeclareMathOperator*{\plim}{plim}
\usepackage{lscape}
\usepackage{longtable}
\usepackage{booktabs}
\usepackage{multirow}
\usepackage{adjustbox}
\usepackage[shortlabels]{enumitem}
\usepackage{xcolor}
\usepackage{mathpazo}
\usepackage[hidelinks]{hyperref}
\usepackage{chngcntr}
\usepackage{enumitem}
\counterwithin{table}{section}
\usepackage[round]{natbib}
\usepackage{rotating}
\usepackage{comment}
\numberwithin{equation}{section}
\newcommand{\y}{\mathbf{y}}
\newcommand{\X}{\mathbf{X}}
\newcommand{\x}{\mathbf{x}}
\newcommand{\F}{\mathbf{F}}
\newcommand{\f}{\mathbf{f}}
\newcommand{\C}{\mathbf{C}}

\newcommand{\W}{\mathbf{W}}
\newcommand{\w}{\mathbf{w}}

\newcommand{\z}{\mathbf{z}}
\newcommand{\U}{\mathbf{U}}

\newcommand{\A}{\mathbf{A}}

\newcommand{\V}{\mathbf{V}}

\newcommand{\g}{\mathbf{g}}
\newcommand{\I}{\mathbf{I}}

\newcommand{\M}{\mathbf{M}}

\newcommand{\e}{\mathbf{e}}

\def\*#1{\mathbf{#1}}
\def\+#1{\boldsymbol{#1}}

\newtheorem{algorithm}{Algorithm}

\newtheorem{remark}{REMARK}
\newtheorem{theorem}{Theorem}

\newtheorem{corollary}{Corollary}
\usepackage[margin=0.75in,paper=letterpaper]{geometry}
\begin{document}
{
		\title{\bf  Bootstrap Inference for Dynamic Panel Data Models with Common Correlated Effects}

        \author[1]{Artūras Juodis }
        \author[2]{Ovidijus Stauskas\thanks{Corresponding author. Financial support from the Dutch Research Council (NWO) under research grant VI.Vidi.231E.030 (Juodis) is gratefully acknowledged by all authors.}}
        \author[1]{Sander Tromp}
	
		\affil[1]{Amsterdam School of Economics, University of Amsterdam and Tinbergen Institute}
		\affil[2]{BI Norwegian Business School, Department of Economics}
		\maketitle
	}

    \begin{abstract}
We study recursive-design wild bootstrap inference for dynamic panel data models with unobserved common factors estimated by Common Correlated Effects. In the large $N,T$ setting, the bootstrap reproduces the biased limiting distribution in pure autoregressive models, but fails to capture all bias and factor-estimation variance components in models with additional regressors, particularly under weak exogeneity. We trace this failure to holding regressors fixed across bootstrap replications. We propose to combine bootstrap procedure with available bias-correction methods to conduct adjusted inference. Monte Carlo evidence shows substantial improvements over conventional strategies of using bias-correction paired with cross-sectional bootstrap methods.
    \end{abstract}



    \section{Introduction}
Dynamic panel data models with unobserved common shocks are now standard in empirical macroeconomics and finance, where $N$ cross-sectional units are exposed to global factors and exhibit substantial levels of persistence. In such environments, the common correlated effects (CCE) estimator of \citet{Pesaran2006} has become a central tool for dealing with cross-sectional dependence by augmenting the model with cross-sectional averages of the observed variables. Subsequent work has extended and refined this approach in multiple directions; see, e.g., \citet{JUODIS2026106120} for a recent review. 

Despite these advances, valid inference in dynamic CCE panels remains challenging, as emphasized in Lesson 2 of \citet{JUODIS2026106120}. When lagged dependent variables and individual effects are present, the CCE pooled (CCEP) estimator inherits the well-known ``Nickell bias'', whose magnitude can be amplified by the presence of unobserved factors; see \citet{de2021bias} and \citet{juodis2021robustness}. In addition to the ``Nickell bias'', which is inversely proportional to the time-series dimension $T$, the CCEP estimator also suffers from a cross-sectional-averages-induced bias, that is inversely related to the cross-sectional dimension $N$; see e.g., \citet{Westerlund2015}, \citet{juodis2021robustness}, and \citet{de2024cross}. 

Due to the presence of the ``Nickell bias'', the CCE method becomes inconsistent for a small number of time-series observations, in the so-called fixed-$T$ regime. In many situations, analytical bias correction, similar to that of \citet{moon2017dynamic}, or the half-panel jackknife (HPJ) of \citet{dhaene2015split}, paired with the cross-sectional (pairs) bootstrap, is used. These methods are based on an asymptotic approximation in which both $N$ and $T$ are large, subject to some rate restrictions. 

In response to the unsatisfactory finite-sample properties of such large-$N$, large-$T$ bias-correction approaches, \citet{de2021bias} proposed a modified CCEP-based methodology in which only $N$ is large, while $T$ can be fixed. This approach builds upon the insights of, among others, \citet{Bun2005} and \citet{Dhaene}, that, for the panel AR(1) model with strictly exogenous regressors, a fixed-$T$ bias-corrected estimator can be constructed using the formula for the corresponding ``Nickell bias''. While the approach works well within the class of models studied by \citet{de2021bias}, it is generally cumbersome to extend to settings with higher-order dynamics, unrestricted forms of heteroskedasticity, and unrestricted weakly exogenous regressors. 

In this paper, we use the finite-sample evidence in \citet{de2021bias} as motivation, but instead leverage the recent panel-data literature that highlights the benefits of using bootstrap-based methods; see, e.g., \citet{Higgins2024Bootstrap,HigginsJochmans2025}. Given the dynamic nature of the autoregressive model, this paper focuses on the properties of the recursive-design wild bootstrap (RD-WB) of \citet{gonccalves2015bootstrap} in the context of dynamic panel data models estimated using the CCEP estimator. We consider both a pure panel AR(1) specification and a more general ARX(1) model with additional regressors. To the best of our knowledge, this is the first paper in the literature that focuses on these settings. 

Within a double-asymptotic framework with large $N$ and $T$, we investigate conditions under which the procedure of combining dynamic CCEP with RD-WB yields valid inference for CCEP estimators. Our first main result shows that, in the AR(1) case, the RD-WB fully replicates the asymptotic distribution of the CCEP estimator: the bootstrap distribution centers at the same biased value as the original estimator, mirroring the behavior documented by \citet{gonccalves2015bootstrap} for the fixed effects estimator. In contrast, for ARX(1) models, the RD-WB neither fully replicates the bias nor the variances originating from the factor estimates; see \citet{Juodis2022CCER}. These problems are generally amplified when some of the regressors are only weakly exogenous, as the ``Nickell bias'' is then also not fully replicated. 

The failure of the bootstrap in this general design can be fully attributed to the fact that additional regressors are kept fixed throughout the bootstrap replications, similar to the fixed-design bootstrap studied by \citet{gonccalves2015bootstrap}. Hence, the bootstrap implementation shuts down any channels through which these regressors can affect the first-order asymptotic distribution of the CCEP estimator. In the Supplementary Online Appendix, we show how discrepancies between the asymptotic and bootstrap distributions diminish if some structure is imposed on the additional regressors, opening up channels for bootstrap variation originating from these regressors. 

Given the shortcomings of the RD-WB method in the general setting, we show how the bootstrap method can be combined with the available statistical CCEP toolkit. The proposed solution involves two main ingredients: (i) appropriate studentization that accounts for variance discrepancies between the true and bootstrap distributions; and (ii) the use of bias-corrected estimators, either analytical or jackknife-based, in the bootstrap recursions. Monte Carlo results, presented in the Supplementary Online Appendix, show that this combined approach substantially improves upon the naive approach based on bias correction with the cross-sectional (pairs) bootstrap. Improvements are especially stark for designs with small to moderate values of $T$.

The remainder of the paper is organized as follows. Section \ref{section::model} introduces the model and the pooled CCE estimator. Section \ref{section::bootstrap} summarizes the main theoretical results for both the AR(1) and ARX(1) models. Section \ref{section::takeaways} translates these theoretical results into takeaways for empirical researchers. Section \ref{section::empirical} provides an empirical illustration. Section \ref{section::conclusions} concludes.

\section{Dynamic linear regression with common factors}
\setcounter{equation}{0}
   \label{section::model}
   \subsection{The model}
   We consider the following first-order dynamic panel data model with regressors and unobserved interactive effects (common factors)
   \begin{equation}
   \label{eq::model_main}
       y_{i,t}=\alpha y_{i,t-1}+\+\beta'\*\x_{i,t}+\+\gamma_{y,i}'\+\f_{t}+\varepsilon_{i,t},
   \end{equation}
   for $i=1,\ldots,N$ cross-sectional units and $t=1,\ldots,T$ time periods. Here $y_{i,t}$ is the observed target variable, $\+\x_{i,t}$ is a $[k\times 1]$ vector of observed regressors (covariates), $\+\gamma_{y,i}'\+\f_{t}$ is the unobserved common factor, while $\varepsilon_{i,t}$ is the idiosyncratic error term. For simplicity, we assume that initial observations $(y_{i,0},i=1,\ldots,N)$ are observed. As standard in the literature, $\+\gamma_{y,i}$ is an $[R\times 1]$ vector of factor loadings, while $\+\f_{t}$ is the corresponding vector of common factors. Neither the factor loadings, nor the corresponding factors are observed, constituting the interactive fixed effects model.\footnote{For the sake of notational simplicity, we omit the individual effects $\eta_{i}$ from the model in Eq. \eqref{eq::model_main}. All theoretical results in this paper continue to hold in the presence of individual-specific intercepts and/or individual-specific deterministic trends.} 

   In what follows, the main object of interest is the full vector of common coefficient $\+\delta:=[\alpha,\+\beta']'$. Not accounting for unobserved factors results in inconsistent estimates for $\+\delta$ when the omitted factors are correlated with the included regressors, irrespective of the length of the time-series.
   
   In the literature multiple estimation methods for $\+\delta$ have been proposed. The suggested approaches mainly differ in whether $T$ is assumed to be small (the so-called fixed-$T$ regime), or that $T$ is allowed to be large and of similar magnitude to $N$ (the cross-sectional dimension). Contributions of \citet{Holtz-Eakin1988}, \citet{Ahn2013}, \citet{Robertson2015}, and \citet{JuodisSarafidis2019} (among others) fall into the fixed-$T$ category. The methods use the Generalized Method of Moments (GMM) methodology similar to the seminal work of \citet{Arellano1991}; see, e.g., \citet{JuodisSarafidis2016ER} for a detailed review.

   The large $T$ methods, on the other hand, use some least-squares objective function to estimate $\+\delta$ jointly with $\+\gamma_{y,i}$ and $\+\f_{t}$. The Interactive Fixed Effects (or the Principal components) estimator of \citet{bai2009panel}, and \citet{moon2015linear}, and the Common Correlated Effects (CCE) estimator of \citet{Pesaran2006} are the two prominent estimators of this class. Due to the dynamic nature of the model, these estimators generally suffer from the weak-exogeneity (or the \citealp{Nickell1981}) bias. As a result, some form of bias-correction is required to guarantee asymptotic validity of the corresponding inference procedures; see, e.g., \citet{moon2017dynamic} and \citet{juodis2021robustness}.

    In this paper, we contribute to the large $T$ literature and focus on bootstrap-based inference for the CCE estimator of \citet{Pesaran2006}. 
    
   \subsection{The Common Correlated Effects (CCE) estimator}
   In this section, we introduce the pooled CCE (CCEP) estimator of \citet{Pesaran2006} and discuss the theoretical results available to this estimator in the context of the model in Eq. \eqref{eq::model_main}. 
    
   Let $\+\w_{i,t}:=[y_{i,t-1},\+\x_{i,t}']'$, such that the original model can be expressed as
   \begin{equation}
        y_{i,t}=\+\delta'\*\w_{i,t}+\+\gamma_{y,i}'\+\f_{t}+\varepsilon_{i,t},
   \end{equation}
    or using the stacked matrix notation as
    \begin{equation}
        \+\y_{i}=\*\W_{i}\+\delta+\+\F \+\gamma_{y,i}+\+\varepsilon_{i}.
   \end{equation}   
    Here $\+\y_{i}$ is a $[T\times 1]$ vector of time-series stacked observations for the unit $i$, and similarly for all other quantities. 
    
    If all elements of $\+\F$ were known, the natural estimator for $\+\delta$ was the fixed effects (within group) type estimator that is defined as the joint minimizer of the least squares objective function together with the factor loadings $(\+\gamma_{y,1},\ldots,\+\gamma_{y,N})$. By the Frisch-Waugh-Lowell theorem the estimator $\widehat{\+\delta}_{LS}$ is given by
    \begin{equation}
    \label{eq::LS_infeasbile}
    \widehat{\+\delta}_{LS}=\left(\frac{1}{NT}\sum_{i=1}^{N}\+\W_{i}'\+\M_{\+\F}\+\W_{i}\right)^{-1}\left(\frac{1}{NT}\sum_{i=1}^{N}\+\W_{i}'\+\M_{\+\F}\+\y_{i}\right),
    \end{equation}
    where $\+\M_{\+\F}:= \+\I_{T} - \+\F \left(\+\F'\+\F\right)^{-1} \+\F'$. Unfortunately, not all columns of $\+\F$ are known; therefore, they need to be estimated from the data.
    
    For this, \citet{Pesaran2006} suggested estimating $\+\F$ by assuming that
    \begin{equation}
    \label{eq::model_w}
    \+\W_{i}=\+\F \+\Gamma_{w,i}+\+\U_{i},
    \end{equation}
    where $\+\U_{i}$ is the idiosyncratic error vector independent (or at least uncorrelated) with the factor component $\+\F \+\gamma_{y,i}$. The corresponding estimator $\widehat{\+\F}$ is formed by taking cross-sectional averages of all observed quantities (that are linear in common factors), i.e.,
    \begin{equation*}
    \widehat{\+\F}=[\overline{\+\y},\overline{\+\W}].
    \end{equation*}
    The resulting Common Correlated Effects Pooled (CCEP) estimator is just a feasible (plug-in) version of Eq. \eqref{eq::LS_infeasbile}
    \begin{equation}
    \label{eq::LS_feasbile}
    \widehat{\+\delta}_{CCEP}=\left(\frac{1}{NT}\sum_{i=1}^{N}\+\W_{i}'\+\M_{\widehat{\+\F}}\+\W_{i}\right)^{-1}\left(\frac{1}{NT}\sum_{i=1}^{N}\+\W_{i}'\+\M_{\widehat{\+\F}}\+\y_{i}\right).
    \end{equation}
    As long as $\widehat{\+\F}$ is a consistent estimator of $\+\F$ (the requirement usually formalized in the form of the rank condition on cross-sectional averages of all factor loadings), the CCEP estimator can be shown to be consistent as $N,T\to\infty$; see, e.g., \citet{Pesaran2006} and \citet{juodis2021robustness}.
\subsection{CCE framework for dynamic models}
While appealing, the dynamic nature of the model in Eq. \eqref{eq::model_main}, is generally incompatible for all regressors $\+\W_{i}$ in Eq. \eqref{eq::model_w}. This was first recognized in \citet{Chudik2015}, \citet{de2021bias}, and later formalized in \citet{juodis2021robustness} and \citet{Juodis2022CCER}. In particular, in dynamic models the following three features might appear: i) only rotation of $\widehat{\+\F}$ identifies $\+\F$; ii) factors different to $\+\F$ (nuisance/distinct factors) can enter $\+\W_{i}$; iii) some of these \emph{distinct} factors are not consistently estimable by cross-sectional averages.

Below, we illustrate these three features using two stylized Data Generating Processes (DGPs). First, consider the panel AR(1) model
   \begin{equation}
   \label{eq::model_main_ar}
       y_{i,t}=\alpha y_{i,t-1}+\+\gamma_{y,i}'\+\f_{t}+\varepsilon_{i,t},
   \end{equation}
   where $\varepsilon_{i,t} \sim i.i.d.(0,\sigma_{\varepsilon}^{2})$. In this model, the unobserved factor component $\+\gamma_{y,i}'\+\f_{t}$ can have at most one unobserved factor (thus, it collapses to a scalar $\gamma_{y,i} f_{t}$) for factors to be estimable by cross-sectional averages. For the true DGP we have
\begin{equation*}
    (\overline{y}_{t}-\alpha_{0}\overline{y}_{t-1})\overline{\gamma}_{y}^{-1}=f_{t}+O_P(N^{-1/2}).
\end{equation*}
Hence, the linear combination of the two cross-sectional averages consistently estimates the true factor $f_{t}$ as long as $\overline{\gamma}_{y}$ is non-zero in the limit. This is the property $i)$ mentioned above. 

Next, consider the cross-sectional average of the only regressor itself - $y_{i,t-1}$. It can be expanded
\begin{equation}
   \label{eq::average_lagy_expansion_AR1}
\overline{y}_{t-1}=\overline{\gamma}_{y}g_{t-1} + O_P(N^{-1/2}).
\end{equation}
Here, assuming infinite initialization of the process $y_{i,t}$, $g_{t-1}=\sum_{j=0}^\infty (\alpha_0)^jf_{t-j-1}$, and so we can expand $g_{t}=\alpha_{0}g_{t-1}+f_{t}$. Evidently, $\overline{y}_{t-1}$ will not identify $f_{t}$ (the factor that enters the model in Eq. \eqref{eq::model_main_ar}). Instead, the new factor - $g_{t-1}$ - is identified. This factor is a nuisance factor, as long as identification of $f_{t}$ is considered. This is the property $ii)$ mentioned above. 

The third property is best illustrated by adding an additional regressor $x_{i,t}$ into the above model
   \begin{align}
   \label{example:2eq1}
       y_{i,t}&=\alpha y_{i,t-1}+\beta x_{i,t}+\+\gamma_{y,i}'\+\f_{t}+\varepsilon_{i,t},\\
       x_{i,t}&=\+\gamma_{x,i}'\+\f_{t}+u_{i,t}.
          \label{example:2eq2}
   \end{align}
    Here we assume that $\+\f_{t}$ now has two unobserved factors, while $u_{i,t}$ is possibly serially correlated sequence independent of $\varepsilon_{i,t}$ for $t=1,\ldots,T$. Using similar algebraical manipulations as above, it can be easily seen that both factors $\+\f_{t}$ are identified from cross-sectional averages as long as the rank of $[\overline{\+\gamma}_{y},\overline{\+\gamma}_{x}]$ is full in the limit as $N\to\infty$. Next, consider a cross-sectional of $y_{i,t-1}$. It admits expansion of the form
\begin{equation}
   \label{eq::average_lagy_expansion_ARX}
\overline{y}_{t-1}=\overline{\+\gamma}_{y^{+}}'\+\g_{t-1} + O_P(N^{-1/2}),
\end{equation}
where (vector-valued) $\+\g_{t-1}$ is defined recursively as above and $\overline{\+\gamma}_{y^{+}}:=\overline{\+\gamma}_{y}+\beta_{0}\overline{\+\gamma}_{x}$. While Eq. \eqref{eq::average_lagy_expansion_ARX} looks similar to Eq. \eqref{eq::average_lagy_expansion_AR1}, the implications for the properties of the CCEP estimator are different: in the former case $\overline{y}_{t-1}$ is a scalar while $\+\g_{t-1}$ is a vector. As a result, $\overline{y}_{t-1}$ will not consistently estimate both factors, but only their linear combination - $\widetilde{g}_{t-1}$. This is the property $iii)$ mentioned above. 

In particular, upon adding and subtracting $\widetilde{g}_{t-1}$
\begin{equation*}
\overline{y}_{t-1}=\underbrace{\widetilde{g}_{t-1}}_{I}+\underbrace{(\overline{\+\gamma}_{y^{+}}'\+\g_{t-1}-\tilde{g}_{t-1})}_{II} + O_P(N^{-1/2}).
\end{equation*}
Here $II$ is a mean zero term (at least asymptotically) and is generally of the same order as the remainder. Unlike the remainder term, $II$ is not cross-sectionally independent and can be correlated with both $\+\gamma_{y,i}'\+\f_{t}$ and $\+\gamma_{x,i}'\+\f_{t}$. As shown by \citet{de2021bias}, \citet{juodis2021robustness}, and \citet{Juodis2022CCER}, the presence of such additional (distinct/nuisance) factors has non-negligible effect on the asymptotic distribution of the CCEP estimator. 

In particular, for the class of models considered in this paper, the CCEP estimator admits the asymptotic expansion of the from
\begin{equation}
\label{eq::asymptotic_distribution}
\sqrt{NT}(\widehat{\+\delta}_{CCEP}-\+\delta_{0})=\+\xi_{0}+\+\xi_{\perp}+\sqrt{\frac{N}{T}} \* b_{1}+\sqrt{\frac{T}{N}}\* b_{2}+\*r_{N,T},
\end{equation}
see also \citet{Juodis2022CCER}. 
Here $\+\xi_{0}$ is asymptotically normal term driven by innovations $(\varepsilon_{i,t}, t=1,\ldots,T)$; $\+\xi_{\perp}$ is asymptotically normal term (uncorrelated with $\+\xi_{0}$), present when $II$ is non-negligible; $\+ b_{1}$ is the  ``Nickell bias''; $\+ b_{2}$ is the factor-approximation bias (first derived in \citealp{Westerlund2015}); $\*r_{N,T}$ is an asymptotically negligible remainder term. Unless stated otherwise, all terms in Eq. \eqref{eq::asymptotic_distribution} are $O_P(1)$.

Hence, any asymptotically valid inference procedure should account for the two variance terms and the two bias terms present in Eq. \eqref{eq::asymptotic_distribution}. In the next section, we show how the Recursive Design Wild Bootstrap (RD-WB) of \citet{gonccalves2015bootstrap} can be adapted to our setting, and under which conditions it replicates the decomposition in Eq. \eqref{eq::asymptotic_distribution}. 

\section{Recursive design wild bootstrap}
\setcounter{equation}{0}
    \label{section::bootstrap}
\subsection{Implementation}

In the following, we summarize the main ingredients of the RD-WB procedure from \citet{gonccalves2015bootstrap} in the context of the CCEP estimator. Note that while innovations $\varepsilon_{i,t}$ are usually assumed to be serially uncorrelated, i.e., the model is dynamically complete, the dynamic structure of $\+\x_{i,t}$ is normally kept unrestricted. As a result, in what follows, we keep $\+\x_{i,t}$ fixed throughout all bootstrap replications.

Motivated by our empirical application, we assume that the factor estimates are given by
\begin{equation*}
\widehat{\+\F}=[\overline{\+\y},\overline{\+\y}_{-1},\overline{\+\X},\overline{\+\X}_{-1}].
\end{equation*}
The inclusion of $\overline{\+\X}_{-1}$ symmetrizes treatment of $y_{i,t}$ and $\+\x_{i,t}$, and aligns our suggested implementation to those in \citet{Chudik2015} and \citet{de2021bias}.

Let $\widehat{\+\delta}_{CCEP}$ be given as in Eq. \eqref{eq::LS_feasbile} then we set
\begin{align}
\label{eq::cce_fitted1}
\widehat{\+\gamma}_{y,i}&:=   \left(\sum_{t=1}^{T}\widehat{\+\f}_{t}\widehat{\+\f}_{t}'\right)^{-1}\sum_{t=1}^{T}\widehat{\+\f}_{t}(y_{i,t}-\+\w_{i,t}'\widehat{\+\delta}_{CCEP}),\\
\widehat{\varepsilon}_{i,t}&:= y_{i,t}-\+\w_{i,t}'\widehat{\+\delta}_{CCEP}-\widehat{\+\gamma}_{y,i}'\widehat{\+\f}_{t},
\label{eq::cce_fitted2}
\end{align}
where $\widehat{\+\f}_{t}$ is the $t$th column of $\widehat{\+\F}$.

\begin{algorithm}
\label{algo::naive}
\newcounter{bean}
\setcounter{bean}{0}
\begin{center}
\textnormal{
\begin{list}
{\textsc{Step} \arabic{bean}.}{\usecounter{bean}}
\item Obtain CCEP loadings and residuals as in Eqs. \eqref{eq::cce_fitted1}-\eqref{eq::cce_fitted2}.
\item Set $y_{i,0}^{*}=y_{i,0}$.
\item For $b=1,\ldots,B$ generate bootstrap dataset $y_{i,t}^{*}$ recursively for $t=1,\ldots,T$,
\begin{equation}
\label{eq::bootstrap_DGP}
    y_{i,t}^{*}:=\widehat{\alpha}_{CCEP} y_{i,t-1}^{*}+\widehat{\+\beta}_{CCEP}'\+\x_{i,t}+\widehat{\+\gamma}_{y,i}'\widehat{\+\f}_{t}+\omega_{i,t}^{*}\widehat{\varepsilon}_{i,t},
\end{equation}
for all $i=1,\ldots,N$, where $\omega_{i,t}^{*} \sim  i.i.d.(0,1)$, $E^*[(\omega_{i,t}^{*})^4]$ finite.
\item Given $((y_{i,t}^{*},\+\x_{i,t}')',i=1,\ldots,N,t=1,\ldots, T)$ obtain $\widehat{\+\delta}_{CCEP}^{*}$ from Eq. \eqref{eq::LS_feasbile} using $\widehat{\+\F}^{*}=[\overline{\+\y}^{*},\overline{\+\y}^{*}_{-1},\overline{\+\X},\overline{\+\X}_{-1}]$
\item Repeat for $b=1,\ldots,B$.
\end{list}
}
\end{center}
\end{algorithm}
The choice in Step 2 can be different in practice due to large $T$. However, it simplifies the demonstration on how the bootstrap averages consistently estimate latent factors in the bootstrap realm (see Section \ref{ssection::discussion} for a precise argument). In practice, we implement Algorithm \ref{algo::naive} using $\omega_{i,t}^{*} \sim  Rademacher(-1;1)$.
In what follows, we will refer to the bootstrap procedure described in Algorithm \ref{algo::naive} as ``naive'' bootstrap. This procedure does not impose any DGP-induced restrictions on cross-sectional averages of the data, see Section \ref{section:extensions} for further discussion.
\begin{remark}
\textnormal{As an alternative to re-estimating $\widehat{\+\f}_{t}^{*}$, we can alternatively consider the bootstrap scheme where $\widehat{\+\f}_{t}^{*}:=\widehat{\+\f}_{t}$; see e.g.,\citet{Westerlund2019}. Unfortunately, this implementation fails to replicate the variance term $\+\xi_{\perp}$, as well as the bias term $\* b_{2}$.}
\end{remark}
\subsection{Assumptions}
We assume that $\+\x_{i,t}$ are generated as follows
\begin{equation}
\label{eq::x}
    \+\x_{i,t}=\+\theta_0 y_{i,t-1}+\+\Gamma_{x,i}'\*f_t+\+\nu_{i,t}.
\end{equation}
Such that the reduced form for $\+\z_{i,t}:=[y_{i,t},\+\x_{i,t}']'$ is of the form
\begin{align}
\label{z_VAR}
    \+\z_{i,t}=\*A_0^{\prime}\+\z_{i,t-1}+\+\Gamma_i'\+\f_t+\+\e_{i,t}, 
\end{align}
where
\begin{align*}
    &\*A_0:=\begin{bmatrix} \alpha_0+\+\beta_0^{\prime}\+\theta_0 & \+\theta_0^{\prime}\\
   \*0_{k\times 1}  & \*0_{k\times k}\end{bmatrix},\\
    &\+\Gamma_i:=[\+\Gamma_{x,i}\+\beta_0+\+\gamma_{y,i}, \+\Gamma_{x,i}],\\
    &\*e_{i,t}:=[\+\nu_{i,t}'\+\beta_0+\varepsilon_{i,t}, \+\nu_{i,t}']'.
\end{align*}
 Below we summarize the set of assumptions used in the remainder of the paper. To fix the notation we set $K:=k+1$, and $0<\Delta<\infty$ is some arbitrary finite constant.
\begin{assumption}
\label{ass::1}(a) The error term $\varepsilon_{i,t}$ is i.i.d. over $i,t$ with $E[\varepsilon_{i,t}]=0$, $E[\varepsilon_{i,t}]=\sigma^2$ and $E[|\varepsilon_{i,t}|^{8}]<\Delta$; (b) The error terms $(\+\nu_t,t=1,\ldots,T)$ are covariance stationary for all $i=1,\ldots,N$; (c) $E[\+\nu_{i,t}]
=\*0_{k\times 1}$, $E[\+\nu_{i,t}\+\nu_{i,t}']=\+\Sigma_{\+\nu}(0)$, $E\left[\left\|\+\nu_{i,t} \right\|^{8}\right]<\Delta$; (d) The sequence $E\left[\+\nu_{i,h}\+\nu'_{i,0}\right]=\+\Sigma_{\+\nu}(h)$ is absolutely summable.
\end{assumption}

\begin{assumption}
\label{ass::2}(a) The factors $(\+\f_t,t=1,\ldots,T)$ are covariance stationary; (b) $E[\+\f_t\+\f_t']=\+\Sigma_{\+\f}(0)\in \mathbb{R}^{R\times R}$ is positive definite, and $E[\left\|\+\f_t \right\|^{8} ]<\Delta$; (c) The sequence $E[\+\f_{h}\+\f_{0}']=\+\Sigma_{\+\f}(h)$ is absolutely summable.
\end{assumption}

\begin{assumption}
\label{ass::3}(a) The loadings $\+\Gamma_{i}\in \mathbb{R}^{R\times K}$; (b) $\mathrm{rk}(\overline{\+\Gamma})=R=K$ as $N\to \infty$; (c) $\frac{1}{N}\sum_{i=1}^N\left\|\+\Gamma_i \right\|^{8}<\Delta$ for all $N$, including $N\to\infty$.
\end{assumption}

\begin{assumption}
\label{ass::4}Sequences $\+\f_t$, $\+\nu_{i,s}$ and $\varepsilon_{j,r}$ are independent for all $i,j,r,t$ and $s$.
\end{assumption}

\begin{assumption}
\label{ass::5}(a) The process $( \+\z_{i,t},t=1,\ldots,T)$ is initialized at an infinite past;  (b) $|\alpha_0|<1$, and the spectral radius of $\*A_0$ is bounded by 1; (c) $\+\z_{i,0}$ is available for all $i=1,\ldots, N$.
\end{assumption}

\begin{assumption}
\label{ass::6}$NT^{-1}\to \kappa\in (0,\infty)$ as $N,T\to\infty$ jointly.
\end{assumption}
Assumptions \ref{ass::1}-\ref{ass::6} are fairly standard for the CCE literature; see e.g., \citet{Pesaran2006}, \citet{Westerlund2015}, and \citet{juodis2021robustness}. Assumption \ref{ass::3} is the rank condition of \citet{Pesaran2006}. Here by treating factor loadings as fixed non-stochastic quantities, we deviate from \citet{de2021bias} and \citet{Juodis2022CCER}, but expect that quantitatively similar results can be derived under the assumption of stochastic loadings (as long as the joint distribution of $\+\Gamma_i$ is left unrestricted). Assumption \ref{ass::6} is standard in the interactive fixed effects literature with weakly exogenous regressors; see e.g., \citet{moon2017dynamic}.

Finally, we assume that all stochastic quantities are covariance stationary and homoscedastic. This is mostly a technical assumption that substantially simplifies the derivations. Given that we consider wild bootstrap-based inference, all results are expected to extend beyond this restricted setting.
\begin{remark}
\textnormal{When analyzing the results for the AR(1) model, we will continue to specify the results assuming that all Assumptions \ref{ass::1}-\ref{ass::6} are satisfied. Then it is implicitly assumed that the corresponding parts associated with $\+\x_{i,t}$ are omitted.}
\end{remark}
\begin{remark}
    \textnormal{The DGP in Eq. \eqref{eq::x} provides a convenient parameterization of $\+\x_{i,t}$ that accommodates three empirically relevant features: (i) a factor structure in the regressors; (ii) idiosyncratic serial correlation through $\+\nu_{i,t}$; and (iii) weak exogeneity with respect to $y_{i,t-1}$. We acknowledge that these features can be accommodated in other ways; see, for example, \citet{Juodis2022CCER} for an alternative DGP. However, our main conclusions do not depend critically on this particular specification.}
\end{remark}
\begin{remark}
\textnormal{In this paper, we assume that the rank condition is satisfied exactly, i.e. $\mathrm{rk}(\overline{\+\Gamma})=R=K$. \citet{Pesaran2006} allows for a less restricted version where $\mathrm{rk}(\overline{\+\Gamma})=R\leq K$. Allowing for such setup leads to additional non-trivial complications for asymptotic analysis under Assumption \ref{ass::6}; see e.g., \citet{Karabiyik2017} and \citet{de2024cross}. For this reason, we leave the analysis of this setup for future research.\footnote{Similarly to \citet{de2021bias}, the setting with $\mathrm{rk}(\overline{\+\Gamma})=R\leq K$ is expected to deliver quantitatively similar results as long as $\+\theta_{0}=\*0_{k}$ and $T/N\to 0$ and $N/T^{3}\to 0$. Assumption \ref{ass::6} needs to be modified appropriately to account for this possibility.} Alternatively, the number of factors can be selected using the tools suggested in the literature; see e.g., \citet{Juodis2022CCER}, \citet{10.1093/ectj/utad009}, and \citet{ditzen2026selection}.}
\end{remark}
\subsection{Results}
This section presents the paper’s two main results. We first discuss the properties of the recursive-design bootstrap in Algorithm \ref{algo::naive} for the special case of the panel AR(1) model; the results are stated in Theorem \ref{theorem::ar1}. The corresponding results for the more general ARX(1) model are presented in Theorem \ref{theorem::arx1}.

For what follows, let $\mathbb{P}^{*}(\cdot)$ be the bootstrap distribution function conditional on the realization of $\{(z_{i,t},i=1,\ldots,N; t=0,\ldots,T)\}$. In the AR(1) model the corresponding estimator is $\widehat{\alpha}_{CCEP}$ and $\alpha_{0}$ is the corresponding true value. For the general ARX(1) model, the corresponding quantities are given by $\widehat{\+\delta}_{CCEP}$ and $\+\delta_{0}$, respectively.

\begin{theorem}
\label{theorem::ar1}
If Assumptions \ref{ass::1}-\ref{ass::6} are satisfied, then in the AR(1) model
\begin{align*}
    \sup_{x\in\mathbb{R}} \Big|\mathbb{P}^{*}\Big(\sqrt{NT}(\widehat{\alpha}_{CCEP}^*-\widehat{\alpha}_{CCEP}) \leq x \Big) - \mathbb{P}\Big(\sqrt{NT}(\widehat{\alpha}_{CCEP}-\alpha_0) &\leq x\Big)\Big| =o_P(1).
\end{align*}
\end{theorem}
The result in Theorem \ref{theorem::ar1} is positive: under Assumptions \ref{ass::1}-\ref{ass::6}, the RD-WB procedure fully replicates the first-order asymptotic distribution of the CCEP estimator. In particular, the bootstrap distribution has the same asymptotic variance and bias as the sampling distribution under the true DGP. This result extends Theorem 3.1 of \citet{gonccalves2015bootstrap} from the standard fixed effects setting to the CCEP setting.

The results for the CCEP estimator in the AR(1) model resemble those for the FE estimator because the CCEP estimator has a particularly simple asymptotic distribution in this case. In particular, following \citet{juodis2021robustness}, we have $\+\xi_{\perp}=\* b_{2}=0$ for decomposition in Eq. \eqref{eq::asymptotic_distribution}. Hence, factor-estimation error has no impact on the first-order asymptotic properties of the CCEP estimator. Moreover, the model is dynamically complete, as it contains no additional regressors whose dynamic properties are left unspecified.

Our next result summarizes the implications of departing from the ideal AR(1) model and considering the more general setting with regressors satisfying Eq. \eqref{eq::x}.
\begin{theorem}
\label{theorem::arx1}
If Assumptions \ref{ass::1}-\ref{ass::6} are satisfied, then in the ARX(1) model
\begin{align*}
\sup_{\*x\in\mathbb{R}^{K}}\Big|\mathbb{P}^{*}\Big(\+\Xi\sqrt{NT}(\widehat{\+\delta}_{CCEP}^*-\widehat{\+\delta}_{CCEP}-\Delta\*b^*)\leq \*x \Big) - \mathbb{P}\Big(\sqrt{NT}(\widehat{\+\delta}_{CCEP}-\+\delta_{0}) &\leq \*x\Big)\Big|=o_P(1),
\end{align*}
where
\begin{equation*}
  \Delta\*b^{*}:= \frac{1}{T}\left(\*b_1^*-\*b_1\right) + \frac{1}{N}\left(\*b_2^*-\*b_2\right),  
\end{equation*}
and $\+\Xi$ is some positive-definite matrix. All inequalities are interpreted coordinatewise.
\end{theorem}
The exact expressions of all terms provided in Theorem \ref{theorem::arx1} are provided in the corresponding proof in the Supplementary Online Appendix.

Overall, unlike in the AR(1) case studied in Theorem \ref{theorem::ar1}, the RD-WB procedure does not fully replicate the first-order asymptotic distribution of the CCEP estimator in the more general ARX(1) setting. In particular, the bootstrap and sampling distributions exhibit asymptotic discrepancies in both bias and variance components.

This negative conclusion is primarily driven by the fact that the regressors $\+\x_{i,t}$ are held fixed in the proposed bootstrap scheme. The next corollary summarizes the results for the case in which all regressors $\+\x_{i,t}$ are assumed to be strictly exogenous as in \citet{de2021bias}.

\begin{corollary}
If Assumptions \ref{ass::1}-\ref{ass::6} are satisfied, then in the ARX(1) model with $\+\theta_{0}=\*0_{k}$ then
\begin{align*}
 \*b_1^*-\*b_1=\*0_{K}.   
\end{align*}
\end{corollary}
Hence, when the regressors are assumed to be strictly exogenous, the RD-WB procedure fully replicates the corresponding ``Nickell bias'' of the CCEP estimator.

In Section \ref{section::takeaways}, we draw practical implications from Theorem \ref{theorem::arx1} for empirical researchers using CCEP estimators in dynamic panel data models. Among other things, we suggest that bias-correction methods proposed in the literature can be combined with the RD-WB procedure, similarly to the recommendation in \citet{gonccalves2015bootstrap}. 

\begin{remark}
    \textnormal{It is easy to see that if we were to extend the FE results in \citet{gonccalves2015bootstrap} to the setting with potentially weakly exogenous regressors, the results would be qualitatively similar to Theorem \ref{theorem::arx1} (except for $\*b_{2}^{*}-\*b_{2}=\*0_{K}$ and $\+\Xi=\I_{K}$ in that case).}
\end{remark}
In the next section, we intuitively explain the mechanisms behind the main conclusions of Theorem \ref{theorem::arx1}.
\subsection{Discussion}
\label{ssection::discussion}
The remaining negative aspects of Theorem \ref{theorem::arx1}
are not affected by exogeneity properties of regressors, and are solely driven by the fact that factor proxies $(\overline{\+\x}_{t},t=1,\ldots,T)$ remain fixed for all bootstrap replications. Hence, in the bootstrap world, these factors are no longer \emph{latent}, but, rather, observed. 

We illustrate these features using the example presented in Eqs. \eqref{example:2eq1}-\eqref{example:2eq2}. For that model the bootstrap counterpart takes form
   \begin{align*}
       y_{i,t}^{*}&=\widehat{\alpha}_{CCEP} y_{i,t-1}^{*}+\widehat{\beta}_{CCEP} x_{i,t}+\widehat{\+\gamma}_{y,i}'\widehat{\+\f}_{t}+\omega_{i,t}\widehat{\varepsilon}_{i,t},\\
       x_{i,t}&=\+\gamma_{x,i}'\+\f_{t}+u_{i,t}.
   \end{align*}
Note that, although the original model contains only $R=2$ latent factors, the bootstrap DGP contains four factors. Two of these-namely, 
$\overline{\*x}$ and $\overline{\*x}_{-1}$ - are observed in the bootstrap world. Hence, only two factors remain latent.

As an intermediate step in the proof of Theorem \ref{theorem::arx1} we show that the loadings of the ($4-2=2$) ``excessive'' factors are asymptotically negligible, such that the bootstrap DGP is asymptotically equivalent to the DGP
   \begin{align*}
       y_{i,t}^{*}&=\widehat{\alpha}_{CCEP} y_{i,t-1}^{*}+\widehat{\beta}_{CCEP} x_{i,t}+\widetilde{\+\gamma}_{y,i}'\widetilde{\+\f}_{t}+\omega_{i,t}\widehat{\varepsilon}_{i,t},
   \end{align*}
where $\widetilde{\+\f}_{t}$ is a $[\widetilde{R}^{*}\times 1]$ vector of rotated cross-sectional averages with $\widetilde{R}^{*}=2$. Here, the first element is given by $\widetilde{f}_{t}^{(1)}:=\overline{y}_{t}-\alpha_{0}\overline{y}_{t-1}-\beta_{0}\overline{x}_{t}$, while $\widetilde{f}_{t}^{(2)}:=\overline{x}_{t}$. Given that the factor proxies in the bootstrap world are given by $\widehat{\+\f}_{t}^{*}=[\overline{y}_{t}^{*},\overline{y}_{t-1}^{*},\overline{x}_{t},\overline{x}_{t-1}]$, $\widetilde{f}_{t}^{(1)}$ is the only latent factor that drives $y_{i,t}^{*}$.

This has implications for the bootstrap asymptotic distribution of the CCEP estimator. In particular, in the asymptotic distribution in Eq. \eqref{eq::asymptotic_distribution}, the two CCEP-specific components, $\+\xi_{\perp}$ and $\*b_{2}$, are determined by the factor loadings associated with the latent factors driving $y_{i,t}$. In the true DGP, there are generally $R=2$ such factors, whereas in the bootstrap DGP only $R^{*}=1$ factor remains latent. Consequently, the corresponding bootstrap terms $\+\xi_{\perp}^{*}$ and $\*b_{2}^{*}$ cannot be asymptotically equivalent to $\+\xi_{\perp}$ and $\*b_{2}$, respectively, because one factor is missing from the bootstrap DGP. The resulting bias discrepancy, $\*b_{2}^{*}-\*b_{2}$, is shown explicitly in the definition of $\Delta\*b^{*}$, whereas the variance discrepancy, $\+\xi_{\perp}^{*}-\+\xi_{\perp}$, determines the scaling factors $\+\Xi$.

The above discussion extends directly to the more general setting with an arbitrary number of regressors, $k$, and, subsequently, to settings with weakly exogenous regressors. These cases are fully considered in the Supplementary Online Appendix.

Finally, why in Algorithm \ref{algo::naive} we set $y_{i,0}^{*}=y_{i,0}$. This choice is not innocuous and significantly simplifies the asymptotic analysis. Using the example above, note that
\begin{equation*}
          y_{i,t}^{*}=\widehat{\alpha}_{CCEP} y_{i,t-1}^{*}+\widehat{\beta}_{CCEP} x_{i,t}+\widehat{\+\gamma}_{y,i}'\widehat{\+\f}_{t}+\omega_{i,t}\widehat{\varepsilon}_{i,t}.
\end{equation*}
Consider now what happens with the corresponding cross-sectional average $\overline{y}_{t}^{*}$:
\begin{equation}
\label{eq::boostrap_cs_average1}
          \overline{y}_{t}^{*}=\widehat{\alpha}_{CCEP} \overline{y}_{t-1}^{*}+\widehat{\beta}_{CCEP} \overline{x}_{t}+\overline{\widehat{\+\gamma}}_{y}'\widehat{\+\f}_{t}+\overline{\omega\widehat{\varepsilon}}_{t}.
\end{equation}
Using the definition of $\widehat{\+\gamma}_{y,i}$ it is easy to see that $\overline{\widehat{\+\gamma}}_{y}'\widehat{\+\f}_{t}=\overline{y}_{t}-\widehat{\alpha}_{CCEP} \overline{y}_{t-1}-\widehat{\beta}_{CCEP} \overline{x}_{t}$.
Inserting this into (\ref{eq::boostrap_cs_average1}) together with $\overline{y}_{0}^{*}=\overline{y}_{0}$ gives
\begin{equation}
\label{eq::boostrap_cs_average2}
 \overline{y}_{t}^{*}-\overline{y}_{t}=\widehat{\alpha}_{CCEP}( \overline{y}_{t-1}^{*}-\overline{y}_{t-1})+\overline{\omega\widehat{\varepsilon}}_{t}\quad \Rightarrow\quad  \overline{y}_{t}^{*}=\overline{y}_{t}+\sum_{j=0}^{t-1}\widehat{\alpha}_{CCEP}^{j}\overline{\omega\widehat{\varepsilon}}_{t-j},
\end{equation}
for $t\geq 1$. It is evident that $\overline{y}_{t}^{*}$ is generally consistent for $\overline{y}_{t}$ (and likewise $\overline{y}_{t-1}^{*}$ is consistent for $\overline{y}_{t-1}$). This implies that the only latent factors in the bootstrap world - $\widetilde{f}_{t}^{(1)}$ - can be consistently estimated by a linear combination of $[\overline{y}_{t}^{*},\overline{y}_{t-1}^{*},\overline{x}_{t}]'$.

\section{Practical implications}
\setcounter{equation}{0}
   \label{section::takeaways}
The negative result in Theorem \ref{theorem::arx1} raises the following question: \emph{Should the RD-WB procedure be used with the CCEP estimator at all?} We argue that the answer is yes, subject to appropriate modifications. We discuss these modifications below.
\subsection{Bias-correction}
\label{section:bias_correction}
First, we discuss what can be done in practice with the \emph{bias} wedge $\Delta \*b^{*}$ derived in Theorem \ref{theorem::arx1}. This term consists of two wedges - the ``Nickell bias'' wedge - $T^{-1}(\*b_{1}^{*}-\*b_{1})$, and the factor-approximation bias wedge -  $N^{-1}(\*b_{2}^{*}-\*b_{2})$. 

The factor-approximation bias of the CCEP estimator has generally received little attention of empirical researchers. As reviewed by \citet{JUODIS2026106120}, in many cases the bias itself is either assumed away (e.g. \citealp{de2021bias} assume $T/N\to 0$) or simply ignored. If one wishes to account for this bias, thus also account for the corresponding wedge between the distributions, it can be accounted for by using analytical bias-correction methods; see e.g., \citet{Westerlund2015} and \citet{Juodis2022CCER}. Suggested bias-correction methods generally do not fully remove this bias (hence, also the wedge), as part of the bias can be non-deterministic (see the corresponding discussion in \citealp{Juodis2022CCER}). On the other hand, the empirical consequence of the wedge $N^{-1}(\*b_{2}^{*}-\*b_{2})$ is expected to be limited, as the Monte Carlo results in the Supplementary Online Appendix indicate \footnote{All estimators considered in the Monte Carlo study do not explicitly account for the presence of the factor-estimation bias. This, however, has little impact both on reported biases as well as rejection rates.}

The wedge in the ``Nickell bias'' (when suspected to be present) generally should not be ignored in typical datasets where either $N>T$ or $N\approx T$ holds. As suggested in \citet{gonccalves2015bootstrap}, the RD-WB approach can be combined with any bias-corrected CCEP estimator that targets the ``Nickell bias'' of the CCEP estimator. The commonly used approaches are the Half Panel Jackknife (HPJ) bias-correction approach of \citet{dhaene2015split}, and the analytical bias-corrected estimator of \citet{HahnKursteiner2011BiasReduction} and \citet{moon2017dynamic}.\footnote{For the panel AR(1) model with additive fixed effects, \citet{gonccalves2015bootstrap} suggested using the analytical bias-corrected estimator of \citet{Hahn2002c}. Unfortunately, no similarly simple bias-corrected estimator is available for the factor-augmented setting considered here.} 

Such bias-corrected CCEP estimators $\widehat{\+\delta}_{CCEP-bc}$ can be then used to re-estimate the factor loadings and the corresponding residuals
\begin{align}
    \label{eq::cce_fitted1_bc}
    \widehat{\+\gamma}_{y,i}&:=   \left(\sum_{t=1}^{T}\widehat{\+\f}_{t}\widehat{\+\f}_{t}'\right)^{-1}\sum_{t=1}^{T}\widehat{\+\f}_{t}(y_{i,t}-\+\w_{i,t}'\widehat{\+\delta}_{CCEP-bc}),\\
    \widehat{\varepsilon}_{i,t}&:= y_{i,t}-\+\w_{i,t}'\widehat{\+\delta}_{CCEP-bc}-\widehat{\+\gamma}_{y,i}'\widehat{\+\f}_{t}.
    \label{eq::cce_fitted2_bc}
\end{align}
The modified RD-WB algorithm is summarized below.
\begin{algorithm}
\label{algo::naive_bc}
\newcounter{bean2}
\setcounter{bean2}{0}
\begin{center}
\textnormal{
\begin{list}
{\textsc{Step} \arabic{bean}.}{\usecounter{bean}}
\item Obtain CCEP-bc loadings and residuals as in Eqs. \eqref{eq::cce_fitted1_bc}-\eqref{eq::cce_fitted2_bc}.
\item Set $y_{i,0}^{*}=y_{i,0}$.
\item For $b=1,\ldots,B$ generate bootstrap dataset $y_{i,t}^{*}$ recursively for $t=1,\ldots,T$,
\begin{equation}
\label{eq::bootstrap_DGP}
    y_{i,t}^{*}:=\widehat{\alpha}_{CCEP-bc} y_{i,t-1}^{*}+\widehat{\+\beta}_{CCEP-bc}'\+\x_{i,t}+\widehat{\+\gamma}_{y,i}'\widehat{\+\f}_{t}+\omega_{i,t}^{*}\widehat{\varepsilon}_{i,t},
\end{equation}
for all $i=1,\ldots,N$, where $\omega_{i,t}^{*} \sim  Rademacher(-1;1)$.
\item Given $((y_{i,t}^{*},\+\x_{i,t}')',i=1,\ldots,N,t=1,\ldots, T)$ obtain $\widehat{\+\delta}_{CCEP-bc}^{*}$ from Eq. \eqref{eq::LS_feasbile} using $\widehat{\+\F}^{*}=[\overline{\+\y}^{*},\overline{\+\y}^{*}_{-1},\overline{\+\X},\overline{\+\X}_{-1}]$
\item Repeat for $b=1,\ldots,B$.
\end{list}
}
\end{center}
\end{algorithm}
\subsection{Studentization}
\label{section:Studentization}
In general, the proposed bootstrap procedure cannot replicate the asymptotic variance of the CCEP estimator, since $\+\Xi\neq \I_{K}$. Below, we propose an empirical strategy that accounts for this feature. First, however, we show why the most natural, or naive, approach is not appropriate in this setting.

The most natural starting point is to use the usual sandwich variance estimator, with the clustered covariance matrix (CCM) of \citet{doi:10.1111/j.1468-0084.1987.mp49004006.x} as its middle component. This approach is intuitively appealing, as \citet{Cui03072023} show that the CCM consistently estimates the asymptotic variance of the IFE estimator of \citet{bai2009panel}.

Unfortunately, for the CCEP estimator, the CCM approach accounts only for variation arising from the $\+\xi_{0}$ term, but not for the variation associated with factor approximation, $\+\xi_{\bot}$; see the decomposition in Eq. \eqref{eq::asymptotic_distribution}. As discussed in Section \ref{ssection::discussion}, it is precisely the $\+\xi_{\bot}$ term that causes $\+\Xi\neq \I_{K}$.

This issue has been discussed by \citet{JuodisSarafidis2019}, \citet{Juodis2022CCER}, and \citet{Brown28052026}. In response, \citet{Juodis2022CCER} advocate the use of the pairs bootstrap,\footnote{In the context of their bias-corrected estimator, \citet{de2021bias} provide both a CCM-type estimator of the asymptotic variance and an estimator based on the pairs bootstrap. It can be shown that their proposed CCM-type estimator is generally inconsistent because it does not replicate the variance of $\boldsymbol{\+\xi}_\perp$ from (\ref{eq::asymptotic_distribution}). In contrast, the pairs bootstrap, also primarily advocated by \citet{de2021bias}, is consistent because it fully accounts for sampling uncertainty induced by factor estimation.} whereas \citet{JuodisSarafidis2019} and \citet{Brown28052026} propose modified CCM-type variance-matrix estimators. Unfortunately, these modifications do not translate directly to our setting, because the corresponding estimators are not consistent under fixed-$T$.
\footnote{\citet{Brown28052026} consider the CCEP setting with strictly exogenous regressors.}

As a solution, we suggest a jackknife-based variance estimator $\widehat{\+\delta}_{CCEP}$ (see \citealp{Tukey1958jackknife} and \citealp{Mackinnon2023leverage})
\begin{align}
\label{eq::jackknife_conventional}
    \widehat{\text{Avar}}_{CCEP} = \frac{N-1}{N} \sum_{i=1}^N \left(\widehat{\+\delta}_{CCEP}^{(-i)} - \overline{\widehat{\+\delta}_{CCEP}}\right)\left(\widehat{\+\delta}_{CCEP}^{(-i)} - \overline{\widehat{\+\delta}_{CCEP}}\right)'.
\end{align}
Here $\widehat{\+\delta}_{CCEP}^{(-i)}$ is the CCEP estimate with $i-$th cross-sectional unit removed, while $\overline{\widehat{\+\delta}_{CCEP}}$ is the corresponding average of $N$ such estimates. For our purpose, it is critical, that within every $\widehat{\+\delta}_{CCEP}^{(-i)}$ also the factor estimates omit the $i-$th cross-sectional unit. The estimator in Eq. \eqref{eq::jackknife_conventional} is the ``conventional'' jackknife-based variance estimator using the terminology of \citet{Hansen2026Jackknife}. For the cross-sectional (clustered) setting \citet{Hansen2026Jackknife} suggested other variants of Eq. \eqref{eq::jackknife_conventional}. However, given relatively large $N$ available in a typical panel application, we do not expect any major differences between different jackknife estimators.

The jackknife approach in Eq. \eqref{eq::jackknife_conventional} while applicable for true data, cannot be directly used for bootstrap data $((y_{i,t}^{*},\+\x_{i,t}')',i=1,\ldots,N,t=1,\ldots, T)$. In particular, the delete-one analogue of Eq. \eqref{eq::boostrap_cs_average2} takes the form
\begin{equation}
\label{eq::boostrap_cs_average2_delete}
 \overline{y}_{t}^{*(-i)}-\frac{N}{N-1}\overline{y}_{t}=\widehat{\alpha}_{CCEP}\left( \overline{y}_{t-1}^{*(-i)}-\frac{N}{N-1}\overline{y}_{t-1}\right)+\overline{\omega\widehat{\varepsilon}}_{t}^{(-i)}-\frac{1}{N-1}\left(\widehat{\beta}_{CCEP}x_{i,t}+\widehat{\+\gamma}_{y,i}'\widehat{\+
\f}_{t}\right).
\end{equation}
This extra term in Eq. \eqref{eq::boostrap_cs_average2_delete} prevents straightforward use of the jackknife methodology in this case. Intuitively, in the bootstrap realm there is no error from factor uncertainty that stems from $\+\x_{i,t}$, because we keep $\+\x_{i,t}$ fixed. The jackknife procedure reintroduces this error by omitting the $i-th$ cross-sectional unit, such that another wedge is created between the bootstrap realm variance and the jackknife estimate. Instead, as a solution to this problem, we use the idea recently highlighted in \citet{heller_jochmans_2026_iterated_bootstrap} and run a second round of $d=1,\ldots,D$ RD-WB bootstrap procedure within each $b=1,\ldots,B$ replications (though their application and motivation is different).

Iterated bootstrap of this form is applicable (even if do not attempt to formally prove validity), as long as regressors remain to be kept fixed in the second round of RD-WB iterations. The corresponding variance estimator for $\widehat{\+\delta}_{CCEP}^{*}$ is given by
\begin{align}
\label{eq::bootstrap_conventional}
    \widehat{\text{Avar}}_{CCEP^{*}} = \frac{1}{D-1} \sum_{d=1}^D \left(\widehat{\+\delta}_{CCEP}^{**(d)} - \overline{\widehat{\+\delta}_{CCEP}^{**}}\right)\left(\widehat{\+\delta}_{CCEP}^{**(d)} - \overline{\widehat{\+\delta}_{CCEP}^{**}}\right)',
\end{align}
where $\widehat{\+\delta}_{CCEP}^{**(b)}$ is the CCEP estimate (given any bootstrap sample $b=1,\ldots,B$) in Algorithm \ref{algo::naive} for every second round iteration $d=1,\ldots,D$. In practice, we simply set $D=B$.\footnote{While theoretically Theorems \ref{theorem::ar1}-\ref{theorem::arx1} are not sufficient to prove consistency of the variance estimators of the form Eq. \eqref{eq::bootstrap_conventional}, and some form of trimming is needed to ensure consistency (see e.g., \citealp{Goncalves01092005}). We note that the simple proposal in Eq. \eqref{eq::bootstrap_conventional} works reasonably well in practice.}

\begin{remark}
    \textnormal{Note that the studentization approach discussed above is only necessary in the context of models covered in Theorem \ref{theorem::arx1}. For the simple AR(1) model, while not necessary in practice, studentization can be nevertheless implemented. For that case (as well as other special cases we cover in Section \ref{section:extensions}) the simple CCM-based variance estimator suffices.}
\end{remark}

\subsection{Extensions}
\label{section:extensions}
Although the conclusions of Theorem \ref{theorem::arx1} are largely negative, the theorem also identifies conditions under which the positive conclusions of Theorem \ref{theorem::ar1} extend to more general settings. The two most straightforward extensions are panel autoregressive and panel vector autoregressive models of a finite order $p\geq 1$.

In the Supplementary Online Appendix, we sketch the argument establishing the validity of the RD-WB in the AR(p) setting. Unsurprisingly, the resulting conclusions fully mirror those of Theorem \ref{theorem::ar1}. For VAR(p) models, the failure to replicate the asymptotic distribution in the presence of additional regressors, as documented in Theorem \ref{theorem::arx1}, is driven solely by the fact that the regressors $\+\x_{i,t}$ and their corresponding cross-sectional averages, $\overline{\+\x}_{t}$, are held fixed throughout the bootstrap replications.

As a result, if we are willing to impose some structure on regressors and \emph{exploit} that structure explicitly in the construction of the bootstrap DGP, i.e. to make the full model for $\+\z_{i,t}$ dynamically complete, the original problem simplifies dramatically. For example, take the VAR(1) model in Eq. \eqref{z_VAR}
\begin{align*}
    \+\z_{i,t}=\*A_0^{\prime}\+\z_{i,t-1}+\+\Gamma_i'\+\f_t+\+\e_{i,t}. 
\end{align*}
The bootstrap counterpart takes the form 
\begin{align}
\label{eq::var_z_star}
    \+\z_{i,t}^{*}=\widehat{\+\A}_{CCEP}\+\z_{i,t-1}^{*}+\widehat{\+\Gamma}_i'\widehat{\+\f}_t+\omega_{i,t}\widehat{\e}_{i,t},
\end{align}
with $\+\z_{i,0}^{*}=\+\z_{i,0}$ as previously. Hence, unlike in the ARX(1) setting, all elements of $\*A_0$ must be estimated using CCEP-not only the equation for $y_{i,t}$- and the corresponding vector of CCEP-based residuals, $(\widehat{\e}_{i,t},i=1,\ldots,N;t=1,\ldots,T)$, must be constructed.\footnote{ DGP-consistent restrictions can be imposed when constructing $\widehat{\+\A}_{CCEP}$, such as known zero restrictions.} To replicate the asymptotic distribution of the CCEP estimator, it is crucial to use scalar weights $\omega_{i,t}$.\footnote{We provide the theoretical evidence on this case in Section 4 of Supplementary Online Appendix.}

Finally, although the RD-WB procedure summarized in Algorithm \ref{algo::naive} is expected to be consistent for dynamically complete models beyond the AR(1) model, it is not necessarily the most efficient CCEP-based implementation of the RD-WB. In Eq. \eqref{eq::var_z_star}, we use
\begin{equation*}
   \widehat{\+\f}_t=[\overline{y}_{t},\overline{y}_{t-1}, \overline{\+\x}_{t}',\overline{\+\x}_{t-1}']',
\end{equation*}
so that $\widehat{\+\f}_t$ has $2K$ elements, whereas the original vector $\+\f_{t}$ has only $K=R$ elements under Assumptions \ref{ass::2}-\ref{ass::3}. Thus, after an appropriate rotation, the remaining $2K-K=K$ elements are asymptotically redundant. Although this rotation is generally unknown, as it depends on the full matrix $\*A_0$, it can be consistently estimated within the multi-equation VAR(1) model—and hence also in the single-equation AR(1) model.

Hence, the alternative version of the bootstrap algorithm uses
\begin{equation}
\label{eq::f_hat_restricted}
\ddot{\+\f}_t:=\overline{\+\z}_{t}-\widehat{\+\A}_{CCEP}'\overline{\+\z}_{t-1}.
\end{equation}
The factor loading matrix $\widehat{\+\Gamma}_i$ should be re-estimated accordingly using the factor estimates $\ddot{\+\F}$. For obvious reasons, the implementation of the bootstrap algorithm that uses the restricted factors Eq. \eqref{eq::f_hat_restricted} is referred to as the \emph{sophisticated}, while the original implementation as the \emph{naive} one.\footnote{Unlike \citet{Everaert2016}, we do not suggest a restricted (non-linear) version of the original CCEP estimator in the dynamically complete setting. The restrictions of the form Eq. \eqref{eq::f_hat_restricted} are only used in the construction of the bootstrap samples.} The theoretical properties of sophisticated schemes for AR(1) and AR(p) cases are explicitly addressed in Section 4 of Supplementary Online Appendix.

\begin{remark}
    \textnormal{The decomposition in Eq. \eqref{eq::x} can be utilized without the need to fully specify the correlation structure in $\+\nu_{i,t}$ using some forms of (panel) Autoregressive Wild Bootstrap; see e.g., \citet{Juodis2020}. However, such approaches require stronger assumptions on the true DGP than utilized by the Algorithm \ref{algo::naive}.}
\end{remark}
   \section{Empirical illustration}
   \setcounter{equation}{0}
      \label{section::empirical}
      \subsection{Setup}
We apply the proposed methodology to re-investigate the dynamic effects of temperature shocks on Gross Domestic Product (GDP) growth. The premise was initially investigated by \citet{Dell2012Temperature} to assess the role of temperature in economic development and the impact of global warming on the future. Further work by \citet{Dell2014Weather} extended the model to capture unobserved heterogeneity by using a factor-augmented approach. 

The climate panel data set contains $125$ countries observed over 1961-2003. Following \citet{Dell2012Temperature}, we account for the different effects of temperature of GDP growth for poor and rich countries. After applying this split and removing countries that have missing observations in either panel, we are left with a balanced panel of $93$ countries for the first panel, between 1962 and 1982.\footnote{See \citet{Dell2014Weather} who advocate for the splitting of the panel in two sub-periods, due to weather intensification or adaptation in recent years. The classification is based on an initial above or below median PPP-adjusted per capita GDP.} For the second panel over the period 1983-2003, the cross-section increases slightly to $118$ countries. 

Given that $T$ is small in the resulting panels, \citet{de2021bias} advocate the use of their fixed-$T$ bias-corrected estimator, as temperature variable - the main regressor of interest - is expected to be strictly exogenous within the given time span. As indicated by the Monte Carlo results in the Supplementary Online Appendix, our proposed bootstrap procedure should be competitive for such values of $N,T$ even if it relies on the large $N,T$ asymptotic approximation.

Similarly to \citet{de2021bias}, we consider the following  model
\begin{align}\label{eq:empirical}
    g_{i,t} &= \alpha g_{i,t-1} + c_i + \beta_1 T_{i,t} + \beta_2 T_{i,t-1}+ \gamma_if_{t} + \varepsilon_{i,t}.
\end{align}
Here $g_{i,t}$ denotes the real per capita GDP growth, and $T_{i,t}$ denotes temperature. We further interact both temperature variables with a dummy variable indicating whether a country is rich or poor. This allows for heterogeneous exposure to temperature between developed and undeveloped economies. To proxy for the factors, we take the cross-sectional averages of all the available regressors. 

The inclusion of $g_{i,t-1}$ in Equation \ref{eq:empirical} allows for persistent output growth, whereas the lagged temperature helps to distinguish the dynamic nature of the effect. The initial effect of a $1^{\circ}C$ increase in temperature on GDP growth is measured through $\beta_1$. A transitory shock has no permanent effect on output if $\beta_1 + \beta_2 = 0$. The implied cumulative growth effects (GE) in Equation \ref{eq:empirical} can be shown to be $\frac{\beta_1 + \beta_2}{1-\alpha}$. The vector $\widehat{\+\f}_{t}$ used in all our estimators  is $[7\times 1]$. It includes averages of all right-hand-side (5 in total and the fixed effect) as well as the left-hand-side variable.  

\subsection{Implementation}
Given the exogeneity of the temperature variable, the bias-correction from \citet{de2021bias} is the most obvious benchmark estimator in this setting (``DVS''). Moreover, we report the original CCE estimator (``CCEP''), as well as the Half-Panel-Jackknife procedure of \citet{dhaene2015split} (``HPJ''), and the analytical correction to the CCEP estimator from \citet{HahnKursteiner2011BiasReduction} (``AN''). Next to the label of the estimator considered, we also report the type of inference procedure used either ``(CS)'', that refers to cross-sectional bootstrap based inference methods or ``(RDn)'' that refers to the ``naive'' implementation of the recursive design bootstrap procedure.

For ``(CS)'' we follow \citet{de2021bias} and report the bootstrap-based standard errors (in the corresponding ``SE'' row), as well as bootstrap-based (equal-tailed) reverse percentile confidence intervals (the corresponding ``CI'' row). As advocated in Section \ref{section:Studentization}, for the recursive design based procedures ``(RDn)'', we report the leave-one-out jackknife based standard errors and double bootstrap-based confidence intervals (with studentization). Finally, following \citet{Higgins2024Bootstrap}, all point-estimates of the ``(RDn)'' estimators are bias-corrected using the median of the bootstrap distribution.
\subsection{Results}
In Tables \ref{tab:1962-1982}-\ref{tab:1983-2003}, we report the results for the first and the second sub-panels, respectively. 

For the first sub-panel, we find that (focusing on the temperature variables) the main conclusions derived from the DVS and AN as well as CCEP (bias corrected by RDn) are comparable. If a poor country experiences a temperature shock it is found that there exists a statistically significant negative effect on GDP growth. However, this loss in growth is mostly compensated in the following year. As such, roughly 90\% of the growth loss is temporary, and the remainder being permanent. HPJ-based estimators generally result in smaller (in absolute value) temperature effect that is not found to be statistically significant (irrespective of the implementation used). This might serve as an indication of further time-series instabilities in the data beyond the original split suggested by \citet{Dell2014Weather}, or the fact that the length of the time-series is too short for precise estimation in every half-panel.\footnote{Note that for $T=20$ every half-panel has approximately $10$ observations, resulting in only $10-7$ effective degrees of freedom.}

For the second sub-panel, it is now found that the temperature effect is no longer statistically significant. In this implementation, both the contemporaneous and the lagged temperature variable are not statistically significant for a given statistical significance level. As \citet{Dell2012Temperature} argues, this could be seen as evidence that countries are adapting to more frequent swings in temperature. It is possible that countries which were classified as poor at the beginning of the sample, have developed into industries that are less affected by temperature shocks. This would reduce the exposure of the GDP growth rate of a country with respect to temperature. However, we find again, that the HPJ-based results tend to deviate the most from other estimators. 

\begin{table}[h!]
\begin{center}
\caption{Estimation results for 1962-1982.}\label{tab:1962-1982}
\begin{adjustbox}{width=1\textwidth}
\begin{tabular}{l|cccc|ccc}
\hline
\hline
 & CCEP (CS) & DVS (CS) & HPJ (CS) & AN (CS) & CCEP (RDn) & HPJ (RDn) & AN (RDn) \\
\hline
$g_{i,t-1}$ & 0.15** & 0.24** & 0.21** & 0.23** & 0.21** & 0.24** & 0.26** \\
SE & (0.08) & (0.08) & (0.09) & (0.08) & (0.09) & (0.10) & (0.09) \\
CI & (0.03, 0.32) & (0.11, 0.38) & (0.04, 0.38) & (0.10, 0.41) & (0.04, 0.52) & (0.06, 0.47) & (0.06, 0.50) \\
$T_{i,t}^{rich}$ & 0.47 & 0.48 & 0.18 & 0.48 & 0.44 & 0.10 & 0.42 \\
SE & (0.60) & (0.52) & (1.05) & (0.60) & (0.61) & (0.96) & (0.60) \\
CI & (-0.74, 1.49) & (-0.53, 1.41) & (-1.62, 2.45) & (-0.63, 1.56) & (-1.08, 1.78) & (-2.16, 2.29) & (-1.14, 1.76) \\
$T_{i,t-1}^{rich}$ & -0.35 & -0.39 & -0.82 & -0.38 & -0.49 & -0.99 & -0.48 \\
SE & (0.56) & (0.53) & (0.85) & (0.53) & (0.64) & (0.97) & (0.62) \\
CI & (-1.77, 0.45) & (-1.70, 0.40) & (-2.63, 0.50) & (-1.79, 0.45) & (-1.96, 0.86) & (-3.09, 1.05) & (-1.94, 0.94) \\
$T_{i,t}^{poor}$ & -1.94** & -1.93** & -0.48 & -1.93** & -2.05** & -0.34 & -2.10 \\
SE & (0.90) & (0.91) & (1.51) & (0.84) & (0.92) & (1.53) & (0.93) \\
CI & (-4.09, -0.40) & (-4.00, -0.33) & (-2.20, 3.31) & (-3.75, -0.31) & (-4.38, -0.09) & (-4.29, 2.80) & (-4.77, 0.24) \\
$T_{i,t-1}^{poor}$ & 1.76** & 1.84** & 1.77 & 1.83** & 1.95** & 1.86 & 1.92 \\
SE & (0.95) & (0.92) & (1.39) & (0.89) & (0.99) & (1.70) & (1.01) \\
CI & (0.13, 3.84) & (0.07, 3.73) & (-0.31, 5.26) & (0.22, 3.62) & (0.06, 4.35) & (-1.58, 6.94) & (-0.40, 4.39) \\
\hline
GE Rich Countries & 0.14 & 0.12 & -0.80 & 0.12 & -0.07 & -1.17 & -0.08 \\
SE & (1.11) & (1.08) & (1.92) & (1.14) & (1.36) & (2.21) & (1.39) \\
CI & (-2.31, 1.88) & (-2.24, 1.95) & (-3.99, 3.51) & (-2.04, 2.15) & (-3.71, 2.07) & (-4.74, 4.69) & (-3.49, 2.65) \\
GE Poor Countries & -0.21 & -0.12 & 1.62 & -0.13 & -0.12 & 2.00 & -0.24 \\
SE & (1.18) & (1.35) & (2.40) & (1.36) & (1.40) & (2.96) & (1.44) \\
CI & (-2.33, 2.11) & (-3.16, 2.63) & (-0.80, 8.48) & (-2.60, 2.60) & (-2.59, 2.98) & (-6.46, 6.19) & (-3.22, 2.81) \\
\hline
\hline
\end{tabular}
\end{adjustbox}
\end{center}
   \footnotesize
   \renewcommand{\baselineskip}{11pt}
   \textbf{Note:} Bootstrapped standard deviations (SE) are shown in brackets. The 95\% confidence interval (CI) are shown as the tuple in brackets. ** denotes significance at level 5\%, using the confidence interval. \\ From left to right, the shown estimators are (CCEP) \citet{Pesaran2006}; (DVS) \citet{de2021bias}; (HPJ) \citet{dhaene2015split}; (AN) \citet{HahnKursteiner2011BiasReduction}. Estimators with the suffix (RDn) are supplied with the proposed recursive design bootstrap. In the remaining cases, (CS) is used to denote the cross-sectional bootstrap.
\end{table}

\begin{table}[h!]
\begin{center}
\caption{Estimation results for 1983-2003.}\label{tab:1983-2003}
\begin{adjustbox}{width=\textwidth}
\begin{tabular}{l|cccc|ccc}
\hline
\hline
 & CCEP (CS) & DVS (CS) & HPJ (CS) & AN (CS) & CCEP (RDn) & HPJ (RDn) & AN (RDn) \\
\hline
$g_{i,t-1}$ & 0.07 & 0.22** & 0.27** & 0.16** & 0.16** & 0.23 & 0.21** \\
SE & (0.06) & (0.07) & (0.13) & (0.06) & (0.09) & (0.22) & (0.09) \\
CI & (-0.08, 0.17) & (0.07, 0.35) & (0.03, 0.53) & (0.05, 0.27) & (0.06, 0.48) & (-0.44, 0.72) & (0.08, 0.45) \\
$T_{i,t}^{rich}$ & 0.47 & 0.44 & 0.81 & 0.45 & 0.47 & 0.82 & 0.53 \\
SE & (0.39) & (0.38) & (0.65) & (0.38) & (0.45) & (0.83) & (0.43) \\
CI & (-0.29, 1.27) & (-0.26, 1.23) & (-0.34, 2.12) & (-0.34, 1.15) & (-0.60, 1.36) & (-1.23, 2.77) & (-0.52, 1.55) \\
$T_{i,t-1}^{rich}$ & 0.09 & 0.08 & 0.41 & 0.08 & 0.14 & 0.44 & 0.08 \\
SE & (0.33) & (0.35) & (0.60) & (0.33) & (0.39) & (0.80) & (0.37) \\
CI & (-0.56, 0.75) & (-0.64, 0.73) & (-0.70, 1.72) & (-0.55, 0.75) & (-0.85, 0.92) & (-1.09, 2.30) & (-0.64, 0.81) \\
$T_{i,t}^{poor}$ & -1.11 & -1.24 & -0.56 & -1.19** & -1.09 & -0.49 & -1.07 \\
SE & (0.67) & (0.66) & (1.00) & (0.62) & (0.75) & (1.22) & (0.77) \\
CI & (-2.31, 0.47) & (-2.47, 0.13) & (-2.39, 1.60) & (-2.39, -0.03) & (-2.56, 0.80) & (-2.89, 2.30) & (-2.70, 0.71) \\
$T_{i,t-1}^{poor}$ & 0.30 & 0.57 & 0.46 & 0.47 & 0.42 & 0.48 & 0.50 \\
SE & (0.72) & (0.70) & (1.23) & (0.72) & (0.81) & (1.74) & (0.83) \\
CI & (-1.26, 1.43) & (-0.78, 1.86) & (-2.08, 2.49) & (-0.91, 1.92) & (-1.40, 2.41) & (-3.44, 4.86) & (-1.37, 2.64) \\
\hline
GE Rich Countries & 0.60 & 0.66 & 1.67 & 0.64 & 0.72 & 1.63 & 0.78 \\
SE & (0.59) & (0.69) & (1.43) & (0.63) & (0.79) & (1.71) & (0.80) \\
CI & (-0.58, 1.70) & (-0.67, 1.98) & (-1.07, 4.67) & (-0.56, 1.88) & (-2.00, 1.11) & (-6.53, 1.76) & (-1.94, 1.24) \\
GE Poor Countries & -0.87 & -0.87 & -0.13 & -0.87 & -0.79 & -0.02 & -0.73 \\
SE & (0.85) & (0.87) & (2.08) & (0.86) & (0.88) & (2.60) & (0.90) \\
CI & (-2.49, 0.81) & (-2.49, 0.76) & (-4.38, 3.61) & (-2.49, 0.89) & (-1.13, 2.31) & (-6.05, 8.80) & (-1.16, 2.20) \\
\hline
\hline
\end{tabular}
\end{adjustbox}
\end{center}
   \footnotesize
   \renewcommand{\baselineskip}{11pt}
   \textbf{Note:} Bootstrapped standard deviations (SE) are shown in brackets. The 95\% confidence interval (CI) are shown as the tuple in brackets. ** denotes significance at level 5\%, using the confidence interval. \\ From left to right, the shown estimators are (CCEP) \citet{Pesaran2006}, (DVS) \citet{de2021bias}, (HPJ) \citet{dhaene2015split}, and (AN) \citet{HahnKursteiner2011BiasReduction}. Estimators with the suffix (RDn) are supplied with the proposed recursive design bootstrap. In the remaining cases, (CS) is used to denote the cross-sectional bootstrap.
\end{table}
\section{Concluding remarks}
\setcounter{equation}{0}
   \label{section::conclusions}
This paper studies the validity of the recursive-design wild bootstrap (RD-WB) for inference in linear dynamic panel-data models with common factors (interactive fixed effects). Our analysis is based on the CCEP estimator of \citet{Pesaran2006}, which approximates the factor structure using cross-sectional averages of observed variables. We establish the asymptotic validity of the bootstrap in the panel AR(1) setting. The Monte Carlo results show that the proposed procedure yields coverage rates that are at least comparable to, and often improve upon, those of commonly used alternatives across a range of designs.

We then examine  this bootstrap algorithm in settings with additional weakly exogenous regressors. In the bootstrap world, we adopt an agnostic approach by holding the regressors fixed. This induces discrepancies in both the bias and variance between the sampling distribution and its bootstrap counterpart. For empirical applications, we propose a statistical toolkit that combines the RD-WB procedure with commonly used bias-correction methods and appropriate studentization. Although the asymptotic results for the ARX(1) setting are more nuanced, Monte Carlo evidence shows that RD-WB achieves coverage rates close to their nominal levels over a broad range of designs.

 We restrict attention to the CCEP estimator of \citet{Pesaran2006} motivated by the fixed-$T$ consistency of factor estimates based on cross-sectional averages. This choice is primarily responsible for the negative aspects identified in Theorem \ref{theorem::arx1}. Alternatively, we could use the principal-components (PC) approach of \citet{GreenawayMcGrevy201248}; see also \citet{Westerlund2015} and \citet{juodis2026factoraugmentedpanelregressionsvarianceweighted}. We conjecture that the RD-WB is valid for this class of estimators, at least when the regressors are strictly exogenous. A formal analysis of this extension is beyond the scope of this study.
    \section*{Acknowledgments}
    \thanks{We thank Otilia Boldea, S\'{i}lvia Gon\c{c}alves, Ayden Higgens, and the participants of the NESG 2026 (Tilburg), Conference in Honour of James Mackinnon (Aarhus), IPDC 2026 (Exeter). Financial support from the Dutch Research Council (NWO) under research grant VI.Vidi.231E.030 is gratefully acknowledged by all authors. This paper benefited from the use of generative AI tools to assist with language editing and \LaTeX   formatting. All output was carefully reviewed by the authors. All substantive content, results, and any remaining errors are the authors’ responsibility.}



    \bibliographystyle{chicago}
    \bibliography{biblio_ectj_new}

@article{Juodis2022CCER,
author = {Juodis, Artūras},
doi = {10.1002/jae.2899},
issn = {10991255},
journal = {Journal of Applied Econometrics},
	volume={37},
	number={4},
pages = {788--810},
title = {{A regularization approach to common correlated effects estimation}},
year = {2022}
}

@techreport{HigginsJochmans2025,
  author       = {Higgins, Ayden and Jochmans, Koen},
  title        = {Inference in dynamic models for panel data using the moving block bootstrap},
  year         = {2025},
  url          = {https://arxiv.org/abs/2502.08311},
  note         = {Working paper}
}

@Article{Bun2005,
  author    = {M. J. G. Bun and M. A. Carree},
  title     = {Bias-Corrected Estimation in Dynamic Panel Data Models},
  pages     = {200-210},
  volume    = {23(2)},
  journal   = {Journal of Business \& Economic Statistics},
  year      = {2005},
}

@Article{Dhaene,
  author    = {G. Dhaene and K. Jochmans},
  title     = {Likelihood Inference in an Autoregression with Fixed Effects},
  number    = {5},
  pages     = {1178-1215},
  volume    = {32},
  journal   = {Econometric Theory},
  year      = {2016},
}

@article{JUODIS2026106120,
title = {Five lessons for applied researchers from twenty years of common correlated effects estimation},
journal = {Journal of Econometrics},
volume = {253},
pages = {106120},
year = {2026},
issn = {0304-4076},
doi = {https://doi.org/10.1016/j.jeconom.2025.106120},
url = {https://www.sciencedirect.com/science/article/pii/S0304407625001745},
author = {Artūras Juodis and Simon Reese}
}

@ARTICLE{Ahn2013,
	author={Ahn, Seung C. and Lee, Young H. and Schmidt, Peter},
	title={{Panel data models with multiple time-varying individual effects}},
	journal={Journal of Econometrics},
	year={2013},
	volume={174},
	number={1},
	pages={1-14},
	month={},
	url={http://ideas.repec.org/a/eee/econom/v174y2013i1p1-14.html}
}

@Article{doi:10.1111/j.1468-0084.1987.mp49004006.x,
  author  = {Arellano, M.},
  title   = {Computing Robust Standard Errors for Within-groups Estimators},
  number  = {4},
  pages   = {431-434},
  volume  = {49},
  journal = {Oxford Bulletin of Economics and Statistics},
  year    = {1987},
}

@ARTICLE{Arellano1991,
  author = {Arellano, M. and S. Bond},
  title = {{Some Tests of Specification for Panel Data: Monte Carlo Evidence
	and an Application to Employment Equations}},
  journal = {Review of Economic Studies},
  year = {1991},
  volume = {58},
  pages = {277-297}
}

@article{ditzen2026selection,
  title={On Selection of Cross-Section Averages in Non-Stationary Environments},
  author={Ditzen, Jan and Stauskas, Ovidijus},
  journal={Journal of Time Series Analysis},
  year={2026},
  publisher={Wiley Online Library}
}

@article{Goncalves01092005,
author = {Silvia Gon{\c{c}}alves and Halbert White},
title = {Bootstrap Standard Error Estimates for Linear Regression},
journal = {Journal of the American Statistical Association},
volume = {100},
number = {471},
pages = {970--979},
year = {2005},
publisher = {Taylor \& Francis},
doi = {10.1198/016214504000002087},


URL = { 
     
        https://doi.org/10.1198/016214504000002087
    
    

},
eprint = { 
    
        https://doi.org/10.1198/016214504000002087
    
    

}

}

@Article{GreenawayMcGrevy201248,
  author   = {Ryan Greenaway-McGrevy and Chirok Han and Donggyu Sul},
  journal  = {Journal of Econometrics},
  title    = {Asymptotic Distribution of Factor Augmented Estimators for Panel Regression},
  year     = {2012},
  issn     = {0304-4076},
  number   = {1},
  pages    = {48 - 53},
  volume   = {169},
  doi      = {http://dx.doi.org/10.1016/j.jeconom.2012.01.003},
  url      = {http://www.sciencedirect.com/science/article/pii/S0304407612000048},
}

@ARTICLE{Chudik2015,
	title = "Common correlated effects estimation of heterogeneous dynamic panel data models with weakly exogenous regressors ",
	journal = "Journal of Econometrics ",
	volume = "188",
	number = "2",
	pages = "393 - 420",
	year = "2015",
	issn = "0304-4076",
	doi = "http://dx.doi.org/10.1016/j.jeconom.2015.03.007",
	url = "http://www.sciencedirect.com/science/article/pii/S0304407615000767",
	author = "Alexander Chudik and M. Hashem Pesaran"
}

@ARTICLE{DeVos2019,
	title={Bias-corrected common correlated effects pooled estimation in dynamic panels},
  author={De Vos, Ignace and Everaert, Gerdie},
  journal={Journal of Business \& Economic Statistics},
  volume={39},
  number={1},
  pages={294--306},
  year={2021},
  publisher={Taylor \& Francis}
}

@ARTICLE{Everaert2016,
	author = {Gerdie Everaert and Tom {De Groote}},
	title = {Common Correlated Effects Estimation of Dynamic Panels with Cross-Sectional Dependence},
	journal = {Econometric Reviews},
	volume = {35},
	number = {3},
	pages = {428-463},
	year = {2016},
	doi = {10.1080/07474938.2014.966635},
	URL = { 
	http://dx.doi.org/10.1080/07474938.2014.966635	
	},
	eprint = { 
	http://dx.doi.org/10.1080/07474938.2014.966635	
	}
	
}

@ARTICLE{Goncalves2014,
	author = {Sílvia Gonçalves and Benoit Perron},
	title = {Bootstrapping factor-augmented regression models},
	journal = {Journal of Econometrics},
	volume = {182},
	number = {1},
	pages = {156 - 173},
	year = {2014},
	note = {Causality, Prediction, and Specification Analysis: Recent Advances and Future Directions},
	issn = {0304-4076},
	doi = {https://doi.org/10.1016/j.jeconom.2014.04.015},
	url = {http://www.sciencedirect.com/science/article/pii/S0304407614000748}
}

@ARTICLE{Hahn2002c,
  author = {Hahn, JY and Kuersteiner, G},
  title = {{Asymptotically unbiased inference for a dynamic panel model with
	fixed effects when both N and T are large}},
  journal = {{Econometrica}},
  year = {{2002}},
  volume = {{70}},
  pages = {{1639-1657}},
  number = {{4}},
  address = {{108 COWLEY RD, OXFORD OX4 1JF, OXON, ENGLAND}},
  doc-delivery-number = {{568EK}},
  doi = {{10.1111/1468-0262.00344}},
  issn = {{0012-9682}},
  journal-iso = {{Econometrica}},
  keywords-plus = {{EFFICIENT ESTIMATION; MOMENT RESTRICTIONS; GROWTH EMPIRICS}},
  language = {{English}},
  number-of-cited-references = {{21}},
  publisher = {{BLACKWELL PUBL LTD}},
  research-areas = {{Business \& Economics; Mathematics; Mathematical Methods In Social
	Sciences}},
  times-cited = {{60}},
  type = {{Article}},
  unique-id = {{ISI:000176529300011}},
  web-of-science-categories = {{Economics; Mathematics, Interdisciplinary Applications; Social Sciences,
	Mathematical Methods; Statistics \& Probability}}
}

@ARTICLE{Holtz-Eakin1988,
  author = {Holtz-Eakin, D and Newey, W and Rosen, HS},
  title = {Estimating Vector Autoregressions with Panel Data},
  journal = {Econometrica},
  year = {1988},
  volume = {56},
  pages = {1371-1395},
  number = {6},
  address = {{C/O BASIL BLACKWELL LTD, C/O MARSTON BOOK SERVICES, PO BOX 87, OXFORD,
	OXON, ENGLAND OX2 0DT}},
  affiliation = {{HOLTZEAKIN, D (Reprint Author), COLUMBIA UNIV,NEW YORK,NY 10027.
	PRINCETON UNIV,PRINCETON,NJ 08544.}},
  doc-delivery-number = {{R5835}},
  issn = {{0012-9682}},
  journal-iso = {{Econometrica}},
  language = {{English}},
  number-of-cited-references = {{26}},
  publisher = {{ECONOMETRIC SOCIETY}},
  subject-category = {{Economics; Mathematics, Interdisciplinary Applications; Social Sciences,
	Mathematical Methods; Statistics \& Probability}},
  times-cited = {{252}},
  type = {{Article}},
  unique-id = {{ISI:A1988R583500008}}
}

@ARTICLE{Kapetanios2008,
  author = {Kapetanios, G.},
  title = {{A bootstrap procedure for panel data sets with many cross-sectional
	units}},
  journal = {{Econometrics Journal}},
  year = {{2008}},
  volume = {{11}},
  pages = {{377-395}},
  number = {{2}},
  address = {{9600 GARSINGTON RD, OXFORD OX4 2DQ, OXON, ENGLAND}},
  affiliation = {{Kapetanios, G (Reprint Author), Univ London, Dept Econ, Mile End
	Rd, London E1 4NS, England. Univ London, Dept Econ, London E1 4NS,
	England.}},
  author-email = {{G.Kapetanios@qmul.ac.uk}},
  doc-delivery-number = {{325KF}},
  doi = {{10.1111/j.1368-423X.2008.00243.x}},
  issn = {{1368-4221}},
  journal-iso = {{Econom. J.}},
  keywords-plus = {{FACTOR MODELS}},
  language = {{English}},
  number-of-cited-references = {{14}},
  publisher = {{BLACKWELL PUBLISHING}},
  subject-category = {{Economics; Mathematics, Interdisciplinary Applications; Social Sciences,
	Mathematical Methods; Statistics \& Probability}},
  times-cited = {{1}},
  type = {{Article}},
  unique-id = {{ISI:000257587100008}}
}

@ARTICLE{Karabiyik2017,
	title = "On the role of the rank condition in {CCE} estimation of factor-augmented panel regressions ",
	journal = "Journal of Econometrics ",
	volume = "197",
	number = "1",
	pages = "60 - 64",
	year = "2017",
	note = "",
	issn = "0304-4076",
	doi = "http://dx.doi.org/10.1016/j.jeconom.2016.10.006",
	url = "http://www.sciencedirect.com/science/article/pii/S0304407616302007",
	author = "Hande Karabiyik and Simon Reese and Joakim Westerlund"
}

@ARTICLE{Nickell1981,
  author = {Nickell, S.},
  title = {{Biases in Dynamic Models with Fixed Effects}},
  journal = {Econometrica},
  year = {1981},
  volume = {49},
  pages = {1417-1426},
  number = {6}
}

@article{gonccalves2004bootstrapping,
  title={Bootstrapping autoregressions with conditional heteroskedasticity of unknown form},
  author={Gon{\c{c}}alves, S{\i}lvia and Kilian, Lutz},
  journal={Journal of econometrics},
  volume={123},
  number={1},
  pages={89--120},
  year={2004},
  publisher={Elsevier}
}

@ARTICLE{Pesaran2006,
  author = {Pesaran, M.H.},
  title = {{Estimation and Inference in Large Heterogeneous Panels with a Multifactor
	Error Structure}},
  journal = {Econometrica},
  year = {2006},
  volume = {74},
  pages = {967-1012},
  number = {4},
  address = {{9600 GARSINGTON RD, OXFORD OX4 2DQ, OXON, ENGLAND}},
  affiliation = {{Pesaran, MH (Reprint Author), Univ Cambridge, Fac Econ, Sidgwick
	Ave, Cambridge CB3 9DD, England. Univ Cambridge, Fac Econ, Cambridge
	CB3 9DD, England.}},
  author-email = {{hashem.pesaran@econ.cam.ac.uk}},
  doc-delivery-number = {{054CF}},
  issn = {{0012-9682}},
  journal-iso = {{Econometrica}},
  keywords-plus = {{VECTOR AUTOREGRESSIONS; FACTOR MODELS}},
  language = {{English}},
  number-of-cited-references = {{29}},
  publisher = {{BLACKWELL PUBLISHING}},
  subject-category = {{Economics; Mathematics, Interdisciplinary Applications; Social Sciences,
	Mathematical Methods; Statistics \& Probability}},
  times-cited = {{37}},
  type = {{Article}},
  unique-id = {{ISI:000238352900004}}
}

@ARTICLE{Robertson2015,
	author={Robertson, Donald and Sarafidis, Vasilis},
	title={{IV estimation of panels with factor residuals}},
	journal={Journal of Econometrics},
	year={2015},
	volume={185},
	number={2},
	pages={526-541},
	month={},
	url={http://ideas.repec.org/a/eee/econom/v185y2015i2p526-541.html}
}

@ARTICLE{Westerlund2015,
	title = "Cross-sectional averages versus principal components ",
	journal = "Journal of Econometrics ",
	volume = "185",
	number = "2",
	pages = "372 - 377",
	year = "2015",
	note = "",
	issn = "0304-4076",
	doi = "http://dx.doi.org/10.1016/j.jeconom.2014.09.014",
	url = "http://www.sciencedirect.com/science/article/pii/S0304407614002784",
	author = "Joakim Westerlund and Jean-Pierre Urbain"
}

@ARTICLE{Westerlund2019,
	author = {Westerlund, Joakim and Petrova, Yana and Norkut\.{e}, Milda},
	title = {{CCE in fixed-T panels}},
	journal = {Journal of Applied Econometrics},
	volume = {34},
	number = {5},
	pages = {746-761},
	doi = {10.1002/jae.2707},
	url = {https://onlinelibrary.wiley.com/doi/abs/10.1002/jae.2707},
	eprint = {https://onlinelibrary.wiley.com/doi/pdf/10.1002/jae.2707},
	year = {2019}
}

@article{Brown28052026,
author = {Nicholas L. Brown and Peter Schmidt and Jeffrey M. Wooldridge},
title = {A simple reformulation of the common correlated effects model},
journal = {Econometric Reviews},
volume = {45},
number = {5},
pages = {760--772},
year = {2026},
publisher = {Taylor \& Francis},
doi = {10.1080/07474938.2026.2615434},


URL = { 
    
        https://doi.org/10.1080/07474938.2026.2615434
    
    

},
eprint = { 
    
        https://doi.org/10.1080/07474938.2026.2615434
    
    

}

}

@Article{JuodisSarafidis2019,
  author    = {A. Juodis and V. Sarafidis},
  journal   = {Journal of Business and Economic Statistics},
  title     = {A Linear Estimator for Factor-Augmented Fixed-T Panels with Endogenous Regressors},
  year      = {2022},
  number    = {1},
  pages     = {1--15},
  volume    = {40},
}

@Article{JuodisSarafidis2016ER,
  author =    {A. Juodis and V. Sarafidis},
  title =     {Fixed T Dynamic Panel Data Estimators with Multi-Factor Errors},
  journal =   {Econometric Reviews},
  year =      {2018},
  volume =    {37},
  number =    {8},
  pages =     {893-929}
}

@article{Juodis2020,
  author = {A. Juodis},
  title  = {This Shock is Different: Estimation and Inference in Misspecified Two-Way Fixed Effects Regressions},
  journal={Econometric Theory},
    volume={42},
  number={4},
  pages={869--902},
  year   = {2026},
}

@article{de2021bias,
  title={Bias-corrected common correlated effects pooled estimation in dynamic panels},
  author={De Vos, Ignace and Everaert, Gerdie},
  journal={Journal of Business \& Economic Statistics},
  volume={39},
  number={1},
  pages={294--306},
  year={2021},
  publisher={Taylor \& Francis}
}

@article{juodis2021robustness,
  title={On the robustness of the pooled CCE estimator},
  author={Juodis, Art{\=u}ras and Karabiyik, Hande and Westerlund, Joakim},
  journal={Journal of Econometrics},
  volume={220},
  number={2},
  pages={325--348},
  year={2021},
  publisher={Elsevier}
}

@article{gonccalves2015bootstrap,
  title={Bootstrap inference for linear dynamic panel data models with individual fixed effects},
  author={Gon{\c{c}}alves, S{\'\i}lvia and Kaffo, Maximilien},
  journal={Journal of Econometrics},
  volume={186},
  number={2},
  pages={407--426},
  year={2015},
  publisher={Elsevier}
}

@article{dhaene2015split,
  title={Split-panel jackknife estimation of fixed-effect models},
  author={Dhaene, Geert and Jochmans, Koen},
  journal={The Review of Economic Studies},
  volume={82},
  number={3},
  pages={991--1030},
  year={2015},
  publisher={Oxford University Press}
}

@ARTICLE{Hall1988,
	ISSN = {00905364},
	URL = {http://www.jstor.org/stable/2241604},
	author = {Peter Hall},
	journal = {The Annals of Statistics},
	number = {3},
	pages = {927--953},
	publisher = {Institute of Mathematical Statistics},
	title = {Theoretical Comparison of Bootstrap Confidence Intervals},
	volume = {16},
	year = {1988}
}

@article{moon2015linear,
  title={Linear regression for panel with unknown number of factors as interactive fixed effects},
  author={Moon, Hyungsik Roger and Weidner, Martin},
  journal={Econometrica},
  volume={83},
  number={4},
  pages={1543--1579},
  year={2015},
  publisher={Wiley Online Library}
}

@article{bai2009panel,
  title={Panel data models with interactive fixed effects},
  author={Bai, Jushan},
  journal={Econometrica},
  volume={77},
  number={4},
  pages={1229--1279},
  year={2009},
  publisher={Wiley Online Library}
}

@article{de2024cross,
  title={Cross-section bootstrap for CCE regressions},
  author={De Vos, Ignace and Stauskas, Ovidijus},
  journal={Journal of Econometrics},
  volume={240},
  number={1},
  pages={105648},
  year={2024},
  publisher={Elsevier}
}

@article{moon2017dynamic,
  title={Dynamic linear panel regression models with interactive fixed effects},
  author={Moon, Hyungsik Roger and Weidner, Martin},
  journal={Econometric Theory},
  volume={33},
  number={1},
  pages={158--195},
  year={2017},
  publisher={Cambridge University Press}
}

@article{Dell2012Temperature,
Author = {Dell, Melissa and Jones, Benjamin F. and Olken, Benjamin A.},
Title = {Temperature Shocks and Economic Growth: Evidence from the Last Half Century},
Journal = {American Economic Journal: Macroeconomics},
Volume = {4},
Number = {3},
Year = {2012},
Pages = {66–95},
DOI = {10.1257/mac.4.3.66},
URL = {https://www.aeaweb.org/articles?id=10.1257/mac.4.3.66}}

@article{Dell2014Weather,
Author = {Dell, Melissa and Jones, Benjamin F. and Olken, Benjamin A.},
Title = {What Do We Learn from the Weather? The New Climate-Economy Literature},
Journal = {Journal of Economic Literature},
Volume = {52},
Number = {3},
Year = {2014},
Pages = {740–98},
DOI = {10.1257/jel.52.3.740},
URL = {https://www.aeaweb.org/articles?id=10.1257/jel.52.3.740}}

@article{Cui03072023,
author = {Guowei Cui and Kazuhiko Hayakawa and Shuichi Nagata and Takashi Yamagata},
title = {A Robust Approach to Heteroscedasticity, Error Serial Correlation and Slope Heterogeneity in Linear Models with Interactive Effects for Large Panel Data},
journal = {Journal of Business \& Economic Statistics},
volume = {41},
number = {3},
pages = {862--875},
year = {2023},
publisher = {Taylor \& Francis},
doi = {10.1080/07350015.2022.2077349},


URL = { 
    
        https://doi.org/10.1080/07350015.2022.2077349
    
    

},
eprint = { 
    
        https://doi.org/10.1080/07350015.2022.2077349
    
    

}

}

@Article{HahnKursteiner2011BiasReduction,
journal={Econometric Theory},
author={Hahn, Jinyong and Kuersteiner, Guido},
title={Bias Reduction For Dynamic Nonlinear Panel Models With Fixed Effects},
year={2011},
pages={1152-1191},
volume={27},
number={6},
doi={None},
url={https://ideas.repec.org/a/cup/etheor/v27y2011i06p1152-1191_00.html},
}

@article{Higgins2024Bootstrap,
author = {Higgins, Ayden and Jochmans, Koen},
title = {Bootstrap Inference for Fixed-Effect Models},
journal = {Econometrica},
volume = {92},
number = {2},
pages = {411-427},
doi = {https://doi-org.proxy.uba.uva.nl/10.3982/ECTA20712},
url = {https://onlinelibrary-wiley-com.proxy.uba.uva.nl/doi/abs/10.3982/ECTA20712},
eprint = {https://onlinelibrary-wiley-com.proxy.uba.uva.nl/doi/pdf/10.3982/ECTA20712},
year = {2024}
}

@article{10.1093/ectj/utad009,
    author = {Margaritella, Luca and Westerlund, Joakim},
    title = {Using information criteria to select averages in CCE},
    journal = {The Econometrics Journal},
    volume = {26},
    number = {3},
    pages = {405-421},
    year = {2023},
    issn = {1368-4221},
    doi = {10.1093/ectj/utad009},
    url = {https://doi.org/10.1093/ectj/utad009},
    eprint = {https://academic.oup.com/ectj/article-pdf/26/3/405/51707962/utad009.pdf},
}

@article{Tukey1958jackknife,
  author  = {Tukey, John W.},
  title   = {Bias and confidence in not-quite large sample},
  journal = {Annals of Mathematical Statistics},
  year    = {1958},
  volume  = {29},
  pages   = {614}
}

@Unpublished{heller_jochmans_2026_iterated_bootstrap,
  author      = {Heller, Val{\'e}rie and Jochmans, Koen},
  title       = {Iterated-Bootstrap Inference for Panel-Data Models},
  institution = {Toulouse School of Economics},
  type        = {TSE Working Paper},
  number      = {26-1754},
  year        = {2026},
  url         = {https://jochmans.github.io/preprints/doubleb/doubleb.pdf},
  note        = {Working paper}
}

@misc{juodis2026factoraugmentedpanelregressionsvarianceweighted,
      title={Factor-Augmented Panel Regressions and Variance-Weighted Treatment Effects}, 
      author={Artūras Juodis and Martin Weidner},
      year={2026},
      eprint={2604.18078},
      archivePrefix={arXiv},
      primaryClass={econ.EM},
    note         = {Working paper},
      url={https://arxiv.org/abs/2604.18078} 
}

@article{menzel2021bootstrap,
  title={Bootstrap with cluster-dependence in two or more dimensions},
  author={Menzel, Konrad},
  journal={Econometrica},
  volume={89},
  number={5},
  pages={2143--2188},
  year={2021},
  publisher={Wiley Online Library}
}

@article{Hansen2026Jackknife,
    author = {Hansen, Bruce E},
    title = {Jackknife Standard Errors for Clustered Regression},
    journal = {The Review of Economic Studies},
    pages = {(forthcoming)},
    year = {2026},
    issn = {0034-6527},
    doi = {10.1093/restud/rdag050},
    url = {https://doi.org/10.1093/restud/rdag050},
    eprint = {https://academic.oup.com/restud/advance-article-pdf/doi/10.1093/restud/rdag050/68332934/rdag050.pdf},
}

@article{Mackinnon2023leverage,
  author  = {James G. MacKinnon and Morten {\O}rregaard Nielsen and Matthew D. Webb},
  title   = {Leverage, Influence, and the Jackknife in Clustered Regression Models: Reliable Inference Using summclust},
  journal = {The Stata Journal},
  volume  = {23},
  number  = {4},
  pages   = {942--982},
  year    = {2023},
  doi     = {10.1177/1536867X231212433}
}

    \clearpage
\appendix
\section{Appendix}
\setcounter{assumption}{0}
\localtableofcontents 
 \subsection{Simulation study}\label{2}
   \subsubsection{General remarks: Computations}\label{2.1}
    Throughout the Monte Carlo exercise, we use $M=4000$ replications for each specification in the design. We use $B=399$ bootstrap draws when a single bootstrap iteration is used, giving $B+1=400$ bootstrap observations including the original estimate. When a double bootstrap design is applied, we employ $B=D=199$ draws in the first and second bootstrap round. 
    
    When using the recursive design bootstrap, we report the point-estimates that are bias-corrected using the median of the bootstrap distribution as in \citet{Higgins2024Bootstrap}. The bias results refer to the bias scaled by $\sqrt{NT}$. 
    
    \subsubsection{General remarks: Estimators}
    To facilitate the discussion, we introduce the estimators considered along with their abbreviations. As the baseline estimator, we employ the Common Correlated Effects estimator in pooled fashion from \citet{Pesaran2006} (CCEP). We consider three bias-correction procedures that are applied to CCEP: the analytical bias-correction from \citet{HahnKursteiner2011BiasReduction} (AN), the half-panel jackknife from \citet{dhaene2015split} (HPJ), and the fixed-$T$ bias-correction of \citet{DeVos2019} (DVS). For all estimators considered, the vector of ones $(1,\ldots,1)'$ is always included among the columns of $\widehat{ \*F}$.

    Coverage rates are obtained using either the cross-sectional bootstrap (CS) or the proposed recursive design bootstrap, with ``naive'' implementation denoted (RDn) and ``sophisticated'' implementation denoted (RDs).

    When using the CS bootstrap, coverage rates are computed by taking the reverse quantile 95\% confidence interval from the bootstrap sample, as is standard in the literature. For the RD bootstrap we make a distinction between the AR(1) and ARX(1) setting. 
    
    In AR(1), we apply studentization whenever the resulting t-statistic is asymptotically pivotal. This is the case when applying HPJ and AN. This does not apply to the CCEP estimator due to the remaining incidental parameter bias as shown in \citet{juodis2021robustness}. \citet{Hall1988} illustrates that an improved rate on the coverage error is obtained when studentizing a pivotal statistic. In turn, for the CCEP estimator we report the reverse quantile confidence intervals, while for HPJ and AN we report the studentized confidence intervals.

    For the ARX(1) setting, Section \ref{4.2} illustrates that the variance in the bootstrap realm is different compared to the variance in the real world. To resolve this mismatch, one is required to studentize the confidence intervals to obtain coverage rates that converge to the nominal level asymptotically.  
    
    The improvements on the order of the coverage rate error of studentization depend on a consistent estimator for the variance. We outline the methodology to estimate the respective variances in the studentization section of the paper.

    In general, a bootstrap procedure with $B$ draws produces a set of estimated parameters; taking $\widehat{\alpha}$ as an example, these are denoted by $\left\{ \widehat{\alpha}^*_i\right\}_{i=1}^B$. If one can consistently estimate the variance $\mathbb{V}\left(\widehat{\alpha}\right)$ and $\mathbb{V}\left(\widehat{\alpha}^*_i\right)$, one can compute the bootstrap t-statistic:
    \begin{align*}
        t^*_i = \frac{\widehat{\alpha}^*_i-\widehat{\alpha}}{\sqrt{\mathbb{V}\left(\widehat{\alpha}^*_i\right)}}.
    \end{align*}
    We define $Q_p\left( \{.\}\right)$ as the quantile function for the $p$-th percentile of the set $\{.\}$. With this notation, the reverse-quantile (RQ) confidence interval with nominal level $0.05$ can be computed as:
    \begin{align*}
        CI_{RQ}(\widehat{\alpha}) = &\left(2\widehat{\alpha} - Q_{0.975}\left(\left\{ \widehat{\alpha}^*_i\right\}_{i=1}^B\right), 2\widehat{\alpha} - Q_{0.025}\left(\left\{ \widehat{\alpha}^*_i\right\}^{B}_{i=1}\right) \right).
    \end{align*}
    Furthermore, we can compute the studentized (S) confidence interval with nominal level $0.05$ as:
    \begin{align*}
        CI_S(\widehat{\alpha}) = &\left(\widehat{\alpha} - Q_{0.975}\left(\left\{ t^*_i\right\}_{i=1}^B\right) \sqrt{\mathbb{V}(\widehat{\alpha})} , \widehat{\alpha} - Q_{0.025}\left(\left\{ t^*_i\right\}^B_{i=1}\right) \sqrt{\mathbb{V}(\widehat{\alpha})}\right).
    \end{align*}
    Similarly, one can construct confidence intervals for a bias-corrected estimate, $\widehat{\alpha}_{bc} = \widehat{\alpha} - \widehat{b}$, where $\widehat{b}$ is the estimated bias term. We proceed with the bootstrap, but now obtain a set of bias-corrected estimates, $\left\{ \widehat{\alpha}^*_{bc,i}\right\}_{i=1}^B$. For studentization we compute the variance in the real world and in the bootstrap realm, $\mathbb{V}\left(\widehat{\alpha}_{bc}\right)$ and $\mathbb{V}\left(\widehat{\alpha}^*_{bc,i}\right)$ respectively. Following this, one can compute the t-statistic of the bias-corrected estimates:
    \begin{align*}
        t^*_{bc,i} = \frac{\widehat{\alpha}^*_{bc,i}-\widehat{\alpha}_{bc}}{\sqrt{\mathbb{V}\left(\widehat{\alpha}^*_{bc,i}\right)}}.
    \end{align*}
    Note that, as a consequence of bias-correction, we have that $t^*_{bc,i} \to_{d^*}\mathcal{N}\left(0,1\right)$. We can see this by rewriting the difference as follows,
    \begin{align*}
        \widehat{\alpha}^*_{bc,i}-\widehat{\alpha}_{bc} &= (\widehat{\alpha}_i^*-\widehat{b}_i^*)-(\widehat{\alpha}-\widehat{b})\notag\\
        &=(\widehat{\alpha}_i^*-\widehat{\alpha})-(\widehat{b}_i^*-\widehat{b}),
    \end{align*}
   where $\widehat{b}^*_i$ and $\widehat{b}$ are bootstrap (for iteration $i$) and original sample analytical estimators of the ``Nickel bias''. In our theoretical results (Theorem 3.1 and Lemma 3), we show that 
   \begin{align*}
       \widehat{\alpha}_i^*-\widehat{\alpha}= T^{-1}D^{-1}b+O_{P^*}((NT)^{-1/2}),
   \end{align*}
   where $D^{-1}b$ is the true ```Nickell bias'', which is a function of the limit of the denominator of the CCEP estimator ($D$), and the bias in the numerator ($b$ from Lemma 3). This leads to
    \begin{align}
        \widehat{\alpha}_i^*-\widehat{\alpha}=( \widehat{\alpha}_i^*-\alpha_0)-(\widehat{\alpha}-\alpha_0)&=(\widehat{\alpha}_i^*-\alpha_0)-T^{-1}D^{-1}b+o_{P^*}(1).
    \end{align}
   This implies that 
    \begin{align*}
        \widehat{\alpha}_i^*-\alpha_0=2T^{-1}D^{-1}b+o_{P^*}(1)
    \end{align*}
    which means that $\widehat{\alpha}_i^*$ has twice the original bias, and so $\widehat{b}_i^*$ converges to the same quantity. Combining these results, 
    \begin{align*}
        \widehat{\alpha}^*_{bc,i}-\widehat{\alpha}_{bc}&=(\widehat{\alpha}_i^*-\widehat{\alpha})-(\widehat{b}_i^*-\widehat{b})\notag\\
        &=T^{-1}D^{-1}b-(2T^{-1}D^{-1}b-T^{-1}D^{-1}b)+o_{P^*}(1)\notag\\
        &=o_{P^*}(1)\notag
    \end{align*}
    and 
    \begin{align*}
        \sqrt{NT}(\widehat{\alpha}^*_{bc,i}-\widehat{\alpha}_{bc})\to_{d^*}\mathcal{N}(0,\phi^{-1}),
    \end{align*}
   where $\phi$ is defined in Theorem 3.1. This is in contrast to the uncorrected t-statistic $t^*_i \to_{p^*} \mathcal{N}\left(\text{b},1\right)$, which is centered at the ``Nickel bias''. It is precisely because of this bias that the uncorrected t-statistic is not asymptotically pivotal as required in \citet{Hall1988}, whereas the corrected t-statistic is asymptotically pivotal. On the other hand, as we know that the limiting distribution of the CCEP estimator $\widehat{\alpha}$ contains the ``Nickel bias'', it must be the case that this bias is replicated in the bootstrap realm if we want to have a test statistic that is size correct. 
    Finally, we can proceed with obtaining the reverse-quantile (RQ) confidence interval with nominal level $0.05$, but now for the bias-corrected estimate:
    \begin{align*}
        CI_{RQ}(\widehat{\alpha}_{bc}) = &\left(2\widehat{\alpha}_{bc} - Q_{0.975}\left(\left\{ \widehat{\alpha}^*_{bc,i}\right\}_{i=1}^B\right), 2\widehat{\alpha}_{bc} - Q_{0.025}\left(\left\{ \widehat{\alpha}^*_{bc,i}\right\}^{B}_{i=1}\right) \right).
    \end{align*}
    The studentized (S) confidence interval with nominal level $0.05$ for the bias-corrected estimate is computed in similar fashion:
    \begin{align*}
        CI_S(\widehat{\alpha}_{bc}) = &\left(\widehat{\alpha}_{bc} - Q_{0.975}\left(\left\{ t^*_{bc,i}\right\}_{i=1}^B\right) \sqrt{\mathbb{V}(\widehat{\alpha}_{bc})} , \widehat{\alpha}_{bc} - Q_{0.025}\left(\left\{ t^*_{bc,i}\right\}^B_{i=1}\right) \sqrt{\mathbb{V}(\widehat{\alpha}_{bc})}\right).
    \end{align*}

    \begin{table}[h]
        \begin{center}
        \caption{Parameter values for the simulation designs considered.}
        \begin{tabular}{c|c|cccc}
        & AR(1) & \multicolumn{4}{c}{ARX(1)} \\
        \toprule
             & (1) & (2) & (3) & (4) & (5) \\
        \midrule
            $\alpha_0$ & 0.8 & 0.8 & 0.8 & 0.6 & 0.6 \\
            $\beta_0$ & 0.0 & 0.2 & 0.2 & 0.4 & 0.4 \\
            $\theta_0$ & 0.0 & 0.0 & 0.0 & 0.5 & 0.3\\
            $\phi_0$ & 0.0 & 0.0 & 0.4 & 0.0 & 0.4 \\
        \bottomrule
        \end{tabular}
        \end{center}
        \label{tab:simulation_design_param}
    \end{table}
    \subsubsection{AR(1): Design}\label{2.2}
    First, we introduce the pure AR(1) DGP that forms the basis of our analysis.
    \begin{align*}
        y_{i,t}=\sqrt{1-\alpha_0^2}c_i + \alpha_0 y_{i,t-1}+\gamma_if_t+\sqrt{1-\alpha_0^2}\varepsilon_{i,t}, \quad \varepsilon_{i,t} \sim \mathcal{N}(0,1),
    \end{align*}
    where $f_t$ and $\gamma_i$ are the univariate factor and loading pair that produce the interactive fixed effects, and $c_i \sim \mathcal{N}(0,1)$ is the individual fixed-effect. As shown by \citet{Everaert2016} and \citet{juodis2021robustness}, the bias of the CCEP estimator increases as $\alpha_{0}\to 1$. To examine a setting with pronounced bias, we consider a strongly persistent setting where $\alpha_0=0.8$ shown in Table \ref{tab:simulation_design_param}.
    
    The factors are generated as 
    \begin{align*}
        f_{t} &= a_0 f_{t-1} + \sqrt{1-a_0^2}e_{t}, \quad e_t \sim \mathcal{N}(0,1),
    \end{align*}
    where we set $a_0=0.6$. The factor loadings are generated as $\gamma_i \sim \mathcal{N}(1,\sigma^2_\gamma)$ such that $E[\gamma_i]=1\neq 0$, i.e. the CCE rank condition is satisfied. We set $\sigma^2_\gamma = 1-\alpha_{0}^2$ such that the variability of the factor exposure is scaled relative to the persistence of the autoregressive component in $y_{i,t}$, so that the contribution of the factor remains comparable across different values of $\alpha_0$.
    
    We initialize the process with $y_{i,-T_b} = c_i$ and $f_{-T_b} = e\sim \mathcal{N}(0,1)$
    for $T_b=200$ burn-in periods.

    \subsubsection{AR(1): Results}\label{2.3}
    We begin our evaluation with the baseline AR(1) results from Table \ref{tab:MC_DGP1}. As predicted by the theoretical results of \citet{Everaert2016} and \citet{juodis2021robustness}, the asymptotic bias or ``Nickell bias'' of the CCEP estimator increases in $\alpha_0$. As a result, without bias-correction, the CCEP estimator is biased across all pairs $(N,T)$. 
    
    To address this, we consider three popular bias-correction methods from the literature. The HPJ method achieves lower bias than the AN method across all designs, though both are outperformed by DVS. The differences in bias across methods shrink as $(N/T)\to 0$. 

    One of the attractive features of the RDn bootstrap is that the ``Nickell bias'' is replicated in the bootstrap world. \citet{Higgins2024Bootstrap} illustrate that this allows for bias-correction of the point-estimate using the median of the bootstrap distribution. The RDn bootstrap method substantially reduces bias. When $N,T \geq 50$, RDn provides bias correction of the point-estimate that is more accurate compared to both the HPJ and AN. The RDs bootstrap provides bias-correction of the point-estimate that is comparable to the HPJ even when $T$ is small. Beyond the standalone bootstrap methods, adapting the approach of \citet{gonccalves2015bootstrap} yields additional finite sample improvements when combining bias-correction with the RDn bootstrap. This extension provides substantially better finite sample properties in terms of bias, compared to using one or the other. Most notably, the extension provides bias-correction that is comparable to the DVS method.

    In terms of coverage, the confidence intervals based on the uncorrected CCEP estimator are generally unreliable, as the cross-sectional bootstrap does not replicate the ``Nickell bias'', as illustrated in \citet{gonccalves2015bootstrap}. As evident from Table \ref{tab:MC_DGP1}, coverage rates are far below nominal level even when $N=25$ and $T=100$. In line with the general form of the ``Nickell bias'', coverage decreases when $(N / T)$ increases.
    
    After correcting for the bias using the HPJ or AN, inference can be conducted using CS bootstrap. Table \ref{tab:MC_DGP1} illustrates that non-negligible bias remains for both methods, resulting in coverage rates below the nominal levels in most designs. The coverage of DVS is close to nominal level for all designs considered. 
    
    As established in Section \ref{4.2.7}, the RDn bootstrap fully replicates the $(N/T)$-bias term, such that asymptotically bias-correction prior to inference is unnecessary. For the ``naive'' construction of the factor space, the coverage is close to the nominal level whenever $T = 50$ for moderate $N$, and approaches the nominal level when $T=100$. Coverage rates can be substantially improved in the smaller $T$ settings by adopting the ``sophisticated'' construction of factors. 
    
    More substantial improvements are achieved by combining RDs bootstrap with bias-correction. Table \ref{tab:MC_DGP1} indicates that the combination provides superior coverage compared to either bias-correction or the RDs bootstrap alone. The combination of HPJ and RDs bootstrap attains coverage rates that are close to nominal level in all simulation designs. These rates are comparable to those of DVS, at substantially lower computational costs.

\begin{landscape}
\begin{table}[h]
\begin{center}
\caption{AR(1): Monte Carlo results for $\widehat{\alpha}$ in DGP (1).}
\small{
\begin{tabular}{ll|cccc|ccc|ccc}
\toprule
\multicolumn{12}{c}{Bias}\\
\toprule
 &  & CCEP (CS) & DVS (CS) & HPJ (CS) & AN (CS) & CCEP (RDn) & HPJ (RDn) & AN (RDn) & CCEP (RDs) & HPJ (RDs) & AN (RDs) \\
$N$ & $T$ &  &  &  &  &  &  &  &  &  &  \\
\midrule
\multirow[t]{3}{*}{25} & 25 & -3.946 & -0.030 & 0.451 & -1.830 & -0.978 & -0.208 & -0.566 & -0.531 & -0.121 & -0.358 \\
 & 50 & -2.531 & -0.101 & 0.509 & -0.905 & -0.309 & -0.074 & -0.201 & -0.136 & -0.090 & -0.141 \\
 & 100 & -1.647 & -0.027 & 0.230 & -0.378 & -0.077 & -0.008 & -0.045 & -0.017 & -0.007 & -0.027 \\
\cline{1-12}
\multirow[t]{3}{*}{50} & 25 & -5.493 & 0.006 & 0.669 & -2.531 & -1.289 & -0.284 & -0.730 & -0.662 & -0.165 & -0.441 \\
 & 50 & -3.487 & -0.037 & 0.850 & -1.195 & -0.327 & -0.059 & -0.186 & -0.091 & -0.084 & -0.104 \\
 & 100 & -2.321 & -0.026 & 0.326 & -0.536 & -0.097 & 0.004 & -0.051 & -0.017 & 0.001 & -0.032 \\
\cline{1-12}
\multirow[t]{3}{*}{100} & 25 & -7.740 & 0.024 & 0.913 & -3.567 & -1.782 & -0.471 & -1.011 & -0.916 & -0.309 & -0.610 \\
 & 50 & -4.907 & -0.029 & 1.125 & -1.691 & -0.440 & -0.088 & -0.253 & -0.112 & -0.121 & -0.143 \\
 & 100 & -3.261 & -0.020 & 0.453 & -0.762 & -0.121 & 0.002 & -0.068 & -0.006 & -0.007 & -0.041 \\
\cline{1-12}
\bottomrule
\multicolumn{12}{c}{Coverage rates}\\
\toprule
 &  & CCEP (CS) & DVS (CS) & HPJ (CS) & AN (CS) & CCEP (RDn) & HPJ (RDn) & AN (RDn) & CCEP (RDs) & HPJ (RDs) & AN (RDs) \\
$N$ & $T$ &  &  &  &  &  &  &  &  &  &  \\
\midrule
\multirow[t]{3}{*}{25} & 25 & 0.070 & 0.887 & 0.767 & 0.637 & 0.792 & 0.913 & 0.873 & 0.888 & 0.918 & 0.861 \\
 & 50 & 0.172 & 0.925 & 0.865 & 0.811 & 0.933 & 0.947 & 0.899 & 0.948 & 0.944 & 0.890 \\
 & 100 & 0.393 & 0.937 & 0.924 & 0.908 & 0.956 & 0.950 & 0.926 & 0.958 & 0.943 & 0.923 \\
\cline{1-12}
\multirow[t]{3}{*}{50} & 25 & 0.002 & 0.944 & 0.664 & 0.388 & 0.703 & 0.901 & 0.879 & 0.868 & 0.913 & 0.869 \\
 & 50 & 0.019 & 0.933 & 0.784 & 0.706 & 0.924 & 0.952 & 0.905 & 0.950 & 0.949 & 0.896 \\
 & 100 & 0.112 & 0.938 & 0.906 & 0.877 & 0.952 & 0.948 & 0.928 & 0.958 & 0.943 & 0.923 \\
\cline{1-12}
\multirow[t]{3}{*}{100} & 25 & 0.000 & 0.955 & 0.517 & 0.144 & 0.543 & 0.838 & 0.841 & 0.791 & 0.883 & 0.841 \\
 & 50 & 0.001 & 0.940 & 0.638 & 0.469 & 0.904 & 0.946 & 0.913 & 0.946 & 0.943 & 0.901 \\
 & 100 & 0.003 & 0.942 & 0.871 & 0.810 & 0.945 & 0.951 & 0.927 & 0.957 & 0.944 & 0.926 \\
\cline{1-12}
\bottomrule
\end{tabular}
\label{tab:MC_DGP1}
}
\end{center}
\footnotesize
\renewcommand{\baselineskip}{11pt}
\textbf{Note:} $(\sqrt{NT})$ Scaled bias and coverage rate based on confidence intervals. From left to right, the shown estimators are conventional CCEP \citet{Pesaran2006}, and the bias corrected versions: DVS \citet{DeVos2019}, HPJ \citet{dhaene2015split}, and AN \citet{HahnKursteiner2011BiasReduction}. The label (CS) refers to the cross-sectional bootstrap as in \citet{Kapetanios2008}, while (RDn) refers to the ``naive'' implementation of the proposed recursive design bootstrap. (RDs) refers to the ``sophisticated'' implementation.      
\end{table}
\end{landscape}
    \clearpage
    \subsubsection{ARX(1): Design}\label{2.4}
    We extend our pure AR(1) DGP by including one explanatory variable $x_{i,t}$,
    \begin{align*}
        &y_{i,t}=\sqrt{1-\alpha_0^2}c_i+\alpha_0 y_{i,t-1}+\beta_{0} x_{i,t}+\+\gamma_{y,i}'\*f_t+\sqrt{1-\alpha_0^2}\varepsilon_{i,t},\\
        &x_{i,t}=\theta_0 y_{i,t-1}+ \phi_0 x_{i,t-1} +\+\Gamma_{x,i}'\*f_t+\sqrt{1-\phi_0^2}\nu_{i,t}.
    \end{align*}
    We allow for $R=2$ pairs of common factors and loadings, where the number of regressors equals $R$. The factor error terms, $e_{t,j}$, with $j=1, \dots,r$, are independent across $j$. Moreover, to control the contribution of the factors to the variance of $y$, we divide the factor error variance by $R$. The error terms $\varepsilon_{i,t}$ and $\nu_{i,t}$ are both i.i.d. standard normally distributed. To satisfy the rank condition of the CCEP estimator, we impose the following restrictions on the mean loadings:
    \begin{align*}
        E \left[ \begin{pmatrix}
            \gamma_{y,i,1} & \Gamma_{x,i,1} \\
            \gamma_{y,i,2} & \Gamma_{x,i,2}
        \end{pmatrix}\right] = \begin{pmatrix}
            1 & 1 \\ 0 & 1
        \end{pmatrix}.
    \end{align*}
     For the case with regressors, we construct a setting where the dynamic structure of the AR case is replicated as closely as possible. We replicate the dynamics of the unconditional expectation of $y$ by imposing $\alpha_{0,AR} = \alpha_{0,ARX} + \frac{\beta_0 \theta_0}{1-\phi_0} $. As is common in the literature, we normalize the long-run dynamics of the additional regressor on $y$ to one, such that $\beta_0=1-\alpha_0$. Throughout, this normalization is implemented irrespective of the exogeneity properties of the corresponding regressor.
     
     In Table \ref{tab:simulation_design_param}, DGPs (2) and (3) correspond to a setting with strict exogeneity of the regressor $x_{i,t}$, while in DGPs (4) and (5) we impose weak exogeneity.

     Similar to the AR(1)-setting, we initialize the process as
    \begin{align*}
        y_{i,-T_b} &= c_i, \\
        x_{i,-T_b} &= \nu_{i,0}, \\
        \*f_{-T_b} &= \*e.
    \end{align*}
    Next, we discard the first $T_b=200$ observations to ensure the results are invariant to the choice of initialization.

    \subsubsection{ARX(1): Results}\label{2.5}
    \paragraph{Bias in DGP (2) and (3): Strict exogeneity of $x_{i,t}$}
    Contrary to the pure AR$(1)$ case, we know from \citet{juodis2021robustness} that there are now two sources of bias in the CCEP estimator. The first $(N/T)$-term originates from the weakly exogenous regressors, and the second $(T/N)$-term from the estimation of the factors. In DGPs (2) and (3), Section \ref{4.2} illustrates that this second bias term is not fully replicated in the bootstrap world, as we keep $x_{i,t}$ fixed. In this setting, the cross-sectional averages of $x_{i,t}$ are not a proxy for the factor, but serve as the actual factors in the bootstrap world and are projected out completely. It is therefore natural to investigate to what extent the additional regressor affects bias and coverage relative to the AR(1) case. We compare the bias and coverage of $\widehat{\alpha}$ in Table \ref{tab:MC_DGP1} with the results from Tables \ref{tab:MC_DGP2_a} and \ref{tab:MC_DGP3_a}. 
    
    Compared to the AR(1)-setting, the bias of $\widehat{\alpha}$ in the baseline CCEP estimator decreases in the ARX(1)-model. Nevertheless, the bias on $\widehat{\alpha}$ remains substantial such that bias-correction is required. The bias of $\widehat{\beta}$ is considerably lower, as this estimate is only indirectly affected by the weakly exogenous $(N/T)$-bias. The AN method has lower bias of $\widehat{\alpha}$ in the ARX(1)-setting compared to the AR(1)-case. On the other hand, the bias of $\widehat{\alpha}$ increases marginally for HPJ. For $\widehat{\beta}$, the AN method achieves the lowest bias. Interestingly, for DVS, the bias of $\widehat{\alpha}$ under ARX(1)-specification exceeds that of the AR(1) specification. For HPJ and AN the bias decreases when $(N/T)$ shrinks. This is in contrast to DVS: Monte Carlo evidence indicates that the bias of DVS is increasing, rather than decreasing, when $(N/T)$ shrinks. We conjecture this arises because of the non-negligible $(T/N)$-bias term in the method. 
    
    When the RDn bootstrap is applied to the CCEP estimator, we observe that the bias of $\widehat{\alpha}$ decreases in the ARX(1)-setting, compared to the AR(1)-setting. The bias on $\widehat{\beta}$ is negligible for any of the designs considered. In contrast, when the RDn bootstrap is applied to the AN and HPJ methods, the bias of $\widehat{\alpha}$ is slightly larger when $T=25$ in the ARX(1)-case compared to the AR(1)-setting. In the remaining designs, the respective bias is of similar magnitude. The bias of $\widehat{\beta}$ with bias-correction is marginally larger compared to the CCEP estimator.

    \paragraph{Bias in DGP (4) and (5): Weak exogeneity of $x_{i,t}$}
    In DGPs (4) and (5), we extend the case to weakly exogenous regressors, as shown in Tables \ref{tab:MC_DGP4_a} and \ref{tab:MC_DGP5_a}. This implies that a bias from weak exogeneity is present in both regressors. As illustrated in Section \ref{4.2}, this has two consequences in the bootstrap realm. The $(N/T)$-term from the weakly exogenous regressors is now only partially replicated, and the variance of the bootstrap distribution itself is distorted. Again, this is a consequence of keeping $x_{i,t}$ fixed in the bootstrap world, and so $y_{i,t-1}$ (which sits in $x_{i,t}$ and acts as the source of weak exogeneity) is not stochastic in bootstrap realm. On the other hand, we know that the bias-correction formula from DVS is also invalid when $x_{i,t}$ is weakly exogenous.

    The bias of $\widehat{\alpha}$ for the CCEP estimator does not increase when $x_{i,t}$ is weakly exogenous instead of strictly exogenous. Conversely, the bias on $\widehat{\beta}$ is almost twice as large. Both of the considered bias-correction tools AN and HPJ can accommodate additional weakly exogenous regressors. As expected, the bias in $\widehat{\alpha}$ decreases slightly. Only for $T=25$ is the bias of $\widehat{\beta}$ increased compared to the strict exogeneity setting. The DVS method produces bias for $\widehat{\alpha}$ in the weakly exogenous setting that is comparable to the strict exogeneity setting. However, the bias for $\widehat{\beta}$ is now considerably increased. 

    The RDn bootstrap substantially reduces the bias of $\widehat{\alpha}$ relative to the standard CCEP estimator. Under the RDn bootstrap, the bias of $\widehat{\alpha}$ has increased marginally relative to the strict exogeneity setting. The bias of $\widehat{\beta}$, however, is substantially higher than in the strict exogeneity setting. However, combining the RDn bootstrap with the bias-correction tools AN and HPJ achieves considerable bias reduction. The HPJ method achieves the lowest bias across all estimators. Most notably, the bias is substantially reduced even for $\widehat{\beta}$.

     \paragraph{Coverage rates in DGP (2) and (3): Strict exogeneity of $x_{i,t}$}
    The coverage rates of $\widehat{\alpha}$ for the baseline CCEP estimator are improved in the ARX setting compared to the AR setting. This follows from the reduction in the bias of $\widehat{\alpha}$ documented above. As discussed previously, the bias is non-negligible, which drives coverage rates that are far from nominal level. The bias of $\widehat{\beta}$ is negligible, such that the coverage rates are close to nominal level.

    The bias-correction methods AN and HPJ paired with the CS bootstrap provide coverage rates of $\widehat{\alpha}$ that are near the nominal level only when $(N/T) \approx 0$. Unsurprisingly, coverage of $\widehat{\beta}$ is close to nominal level for all designs considered. The increase in bias of $\widehat{\alpha}$ for DVS is not reflected in the relevant coverage rates. We conjecture that this finding is driven by the CS bootstrap, though the precise mechanism remains unclear. We leave a theoretical explanation to future work. Across the designs considered, coverage rates are at the nominal level, both for $\widehat{\alpha}$ and $\widehat{\beta}$.

    The RDn bootstrap markedly improves coverage rates of $\widehat{\alpha}$ for the CCEP estimator over the CS bootstrap. The CCEP (RDn) estimator achieves coverage rates ranging from 0.875 to 0.930 regardless of $N$. The coverage rate of $\widehat{\beta}$ is also near the nominal level. In the designs considered, the bias-correction methods AN and HPJ achieve higher coverage of $\widehat{\alpha}$ under the RDn bootstrap than under the CS bootstrap. The coverage rates of $\widehat{\alpha}$ and $\widehat{\beta}$ are comparable across the two methods and close to nominal level. 

   \paragraph{Coverage rates in DGP (4) and (5): Weak exogeneity of $x_{i,t}$}
    The coverage rates of $\widehat{\alpha}$ for the baseline CCEP estimator are slightly improved, although still considerably below nominal level. Intuitively, the minor decrease in bias translates to a correspondingly small improvement in coverage. The bias-correction tools AN and HPJ have slightly higher coverage of $\widehat{\alpha}$ compared to the strict exogeneity case. Most notably, even the coverage rates of $\widehat{\beta}$ decrease only marginally. DVS maintains coverage rates close to nominal level for $\widehat{\alpha}$, while the coverage rates of $\widehat{\beta}$ are considerably below nominal level. Tables \ref{tab:MC_DGP4_b} and \ref{tab:MC_DGP5_b} highlight that DVS is strongly susceptible to misspecification of the strict exogeneity assumption. In the most adverse designs, DVS (CS) coverage for  $\widehat{\beta}$ falls as low as 0.076, driven by the severe bias documented in Table \ref{tab:MC_DGP4_b}.

    The RDn bootstrap achieves coverage close to nominal level for $\widehat{\alpha}$. However, as established in Section \ref{4.2}, the $(N/T)$-bias term is not replicated in the bootstrap realm. For larger values of $(N/T)$ this problem is more severe. Consequently, the coverage rates of $\widehat{\beta}$ are substantially below nominal level, presenting a clear limitation of the RDn bootstrap.
    
    One attempt to address the coverage distortions is to eliminate the bias that cannot be replicated prior to bootstrap inference. This further motivates combining a bias-correction tool with the RDn bootstrap to address the partial replication of the $(N/T)$-term, adapting the approach of \citet{gonccalves2015bootstrap}. Combining the RDn bootstrap with either the AN or HPJ method yields coverage substantially closer to the nominal level whenever $T \geq 50$ for both $\widehat{\alpha}$ and $\widehat{\beta}$. 

\begin{landscape}
\begin{table}[h]
\begin{center}
\caption{ARX(1): Monte Carlo results for $\widehat{\alpha}$ in DGP (2).}
\small{
\begin{tabular}{ll|cccc|ccc}
\toprule
\multicolumn{9}{c}{Bias}\\
\toprule
 &  & CCEP (CS) & DVS (CS) & HPJ (CS) & AN (CS) & CCEP (RDn) & HPJ (RDn) & AN (RDn) \\
$N$ & $T$ &  &  &  &  &  &  &  \\
\midrule
\multirow[t]{3}{*}{25} & 25 & -2.424 & 0.143 & 1.419 & -1.150 & -0.603 & 0.446 & -0.286 \\
 & 50 & -1.237 & 0.217 & 0.869 & -0.284 & -0.009 & 0.165 & 0.106 \\
 & 100 & -0.614 & 0.333 & 0.557 & 0.122 & 0.177 & 0.240 & 0.229 \\
\cline{1-9}
\multirow[t]{3}{*}{50} & 25 & -3.381 & 0.118 & 1.970 & -1.634 & -0.838 & 0.447 & -0.424 \\
 & 50 & -1.858 & 0.145 & 1.065 & -0.563 & -0.135 & 0.106 & 0.004 \\
 & 100 & -1.044 & 0.243 & 0.547 & -0.058 & 0.077 & 0.164 & 0.136 \\
\cline{1-9}
\multirow[t]{3}{*}{100} & 25 & -4.784 & 0.071 & 2.644 & -2.382 & -1.210 & 0.489 & -0.672 \\
 & 50 & -2.695 & 0.088 & 1.411 & -0.912 & -0.257 & 0.054 & -0.086 \\
 & 100 & -1.593 & 0.185 & 0.582 & -0.247 & 0.011 & 0.114 & 0.078 \\
\cline{1-9}
\bottomrule
\multicolumn{9}{c}{Coverage rates}\\
\toprule
 &  & CCEP (CS) & DVS (CS) & HPJ (CS) & AN (CS) & CCEP (RDn) & HPJ (RDn) & AN (RDn) \\
 $N$ & $T$ &  &  &  &  &  &  &  \\
\midrule
\multirow[t]{3}{*}{25} & 25 & 0.370 & 0.921 & 0.734 & 0.743 & 0.913 & 0.910 & 0.919 \\
 & 50 & 0.557 & 0.929 & 0.852 & 0.875 & 0.903 & 0.956 & 0.914 \\
 & 100 & 0.729 & 0.945 & 0.939 & 0.934 & 0.906 & 0.962 & 0.933 \\
\cline{1-9}
\multirow[t]{3}{*}{50} & 25 & 0.146 & 0.953 & 0.650 & 0.590 & 0.903 & 0.874 & 0.934 \\
 & 50 & 0.282 & 0.941 & 0.790 & 0.813 & 0.917 & 0.956 & 0.923 \\
 & 100 & 0.502 & 0.954 & 0.930 & 0.913 & 0.911 & 0.957 & 0.936 \\
\cline{1-9}
\multirow[t]{3}{*}{100} & 25 & 0.028 & 0.949 & 0.516 & 0.348 & 0.891 & 0.820 & 0.928 \\
 & 50 & 0.067 & 0.944 & 0.665 & 0.693 & 0.916 & 0.947 & 0.934 \\
 & 100 & 0.223 & 0.947 & 0.895 & 0.885 & 0.913 & 0.957 & 0.935 \\
\cline{1-9}
\bottomrule
\end{tabular}
\label{tab:MC_DGP2_a}
}
\end{center}
\footnotesize
\renewcommand{\baselineskip}{11pt}
\textbf{Note:} $(\sqrt{NT})$ Scaled bias and coverage rate based on confidence intervals. From left to right, the shown estimators are conventional CCEP \citet{Pesaran2006}, and the bias corrected versions: DVS \citet{DeVos2019}, HPJ \citet{dhaene2015split}, and AN \citet{HahnKursteiner2011BiasReduction}. The label (CS) refers to the cross-sectional bootstrap as in \citet{Kapetanios2008}, while (RDn) refers to the ``naive'' implementation of the proposed recursive design bootstrap.
\end{table}
\end{landscape}
\begin{landscape}
\begin{table}[h]
\begin{center}
\caption{ARX(1): Monte Carlo results for $\widehat{\beta}$ in DGP (2).}
\small{
\begin{tabular}{ll|cccc|ccc}
\toprule
\multicolumn{9}{c}{Bias}\\
\toprule
 &  & CCEP (CS) & DVS (CS) & HPJ (CS) & AN (CS) & CCEP (RDn) & HPJ (RDn) & AN (RDn) \\
$N$ & $T$ &  &  &  &  &  &  &  \\
\midrule
\multirow[t]{3}{*}{25} & 25 & -0.136 & 0.002 & 0.231 & -0.107 & -0.060 & 0.108 & -0.040 \\
 & 50 & -0.028 & 0.013 & 0.118 & -0.011 & -0.003 & 0.018 & 0.008 \\
 & 100 & -0.013 & -0.003 & 0.031 & -0.006 & -0.003 & 0.002 & 0.005 \\
\cline{1-9}
\multirow[t]{3}{*}{50} & 25 & -0.191 & 0.004 & 0.324 & -0.155 & -0.078 & 0.115 & -0.054 \\
 & 50 & -0.050 & 0.016 & 0.156 & -0.026 & -0.006 & 0.019 & 0.004 \\
 & 100 & -0.015 & 0.002 & 0.052 & -0.005 & -0.003 & 0.006 & 0.001 \\
\cline{1-9}
\multirow[t]{3}{*}{100} & 25 & -0.263 & 0.010 & 0.443 & -0.220 & -0.098 & 0.128 & -0.077 \\
 & 50 & -0.096 & 0.003 & 0.199 & -0.067 & -0.029 & -0.000 & -0.023 \\
 & 100 & -0.024 & 0.004 & 0.069 & -0.009 & -0.002 & 0.006 & 0.000 \\
\cline{1-9}
\bottomrule
\multicolumn{9}{c}{Coverage rates}\\
\toprule
 &  & CCEP (CS) & DVS (CS) & HPJ (CS) & AN (CS) & CCEP (RDn) & HPJ (RDn) & AN (RDn) \\
$N$ & $T$ &  &  &  &  &  &  &  \\
\midrule
\multirow[t]{3}{*}{25} & 25 & 0.941 & 0.947 & 0.940 & 0.943 & 0.964 & 0.959 & 0.967 \\
 & 50 & 0.970 & 0.969 & 0.966 & 0.970 & 0.964 & 0.966 & 0.965 \\
 & 100 & 0.987 & 0.986 & 0.986 & 0.986 & 0.963 & 0.965 & 0.964 \\
\cline{1-9}
\multirow[t]{3}{*}{50} & 25 & 0.939 & 0.952 & 0.929 & 0.939 & 0.971 & 0.951 & 0.973 \\
 & 50 & 0.956 & 0.958 & 0.953 & 0.956 & 0.960 & 0.960 & 0.958 \\
 & 100 & 0.975 & 0.975 & 0.974 & 0.975 & 0.959 & 0.961 & 0.959 \\
\cline{1-9}
\multirow[t]{3}{*}{100} & 25 & 0.927 & 0.949 & 0.898 & 0.931 & 0.964 & 0.947 & 0.965 \\
 & 50 & 0.941 & 0.944 & 0.934 & 0.945 & 0.955 & 0.956 & 0.951 \\
 & 100 & 0.959 & 0.958 & 0.958 & 0.961 & 0.956 & 0.956 & 0.955 \\
\cline{1-9}
\bottomrule
\end{tabular}
\label{tab:MC_DGP2_b}
}
\end{center}
\footnotesize
\renewcommand{\baselineskip}{11pt}
\textbf{Note:} $(\sqrt{NT})$ Scaled bias and coverage rate based on confidence intervals. From left to right, the shown estimators are conventional CCEP \citet{Pesaran2006}, and the bias corrected versions: DVS \citet{DeVos2019}, HPJ \citet{dhaene2015split}, and AN \citet{HahnKursteiner2011BiasReduction}. The label (CS) refers to the cross-sectional bootstrap as in \citet{Kapetanios2008}, while (RDn) refers to the ``naive'' implementation of the proposed recursive design bootstrap.
\end{table}
\end{landscape}
\begin{landscape}
\begin{table}[h]
\begin{center}
\caption{ARX(1): Monte Carlo results for $\widehat{\alpha}$ in DGP (3).}
\small{
\begin{tabular}{ll|cccc|ccc}
\toprule
\multicolumn{9}{c}{Bias}\\
\toprule
 &  & CCEP (CS) & DVS (CS) & HPJ (CS) & AN (CS) & CCEP (RDn) & HPJ (RDn) & AN (RDn) \\
$N$ & $T$ &  &  &  &  &  &  &  \\
\midrule
\multirow[t]{3}{*}{25} & 25 & -2.716 & 0.091 & 1.787 & -1.306 & -0.718 & 0.536 & -0.357 \\
 & 50 & -1.408 & 0.146 & 0.998 & -0.391 & -0.092 & 0.133 & 0.024 \\
 & 100 & -0.758 & 0.225 & 0.508 & 0.015 & 0.067 & 0.157 & 0.128 \\
\cline{1-9}
\multirow[t]{3}{*}{50} & 25 & -3.749 & 0.084 & 2.625 & -1.831 & -0.954 & 0.708 & -0.506 \\
 & 50 & -2.063 & 0.078 & 1.245 & -0.677 & -0.205 & 0.054 & -0.071 \\
 & 100 & -1.205 & 0.147 & 0.505 & -0.169 & -0.014 & 0.072 & 0.041 \\
\cline{1-9}
\multirow[t]{3}{*}{100} & 25 & -5.355 & 0.050 & 3.575 & -2.670 & -1.383 & 0.779 & -0.768 \\
 & 50 & -2.913 & 0.037 & 1.694 & -1.007 & -0.309 & 0.040 & -0.132 \\
 & 100 & -1.765 & 0.106 & 0.595 & -0.342 & -0.064 & 0.053 & 0.002 \\
\cline{1-9}
\bottomrule
\multicolumn{9}{c}{Coverage rates}\\
\toprule
 &  & CCEP (CS) & DVS (CS) & HPJ (CS) & AN (CS) & CCEP (RDn) & HPJ (RDn) & AN (RDn) \\
$N$ & $T$ &  &  &  &  &  &  &  \\
\midrule
\multirow[t]{3}{*}{25} & 25 & 0.302 & 0.934 & 0.710 & 0.722 & 0.915 & 0.898 & 0.927 \\
 & 50 & 0.482 & 0.940 & 0.842 & 0.877 & 0.921 & 0.965 & 0.929 \\
 & 100 & 0.679 & 0.951 & 0.939 & 0.935 & 0.917 & 0.960 & 0.936 \\
\cline{1-9}
\multirow[t]{3}{*}{50} & 25 & 0.099 & 0.944 & 0.568 & 0.541 & 0.907 & 0.862 & 0.932 \\
 & 50 & 0.200 & 0.943 & 0.744 & 0.794 & 0.930 & 0.958 & 0.936 \\
 & 100 & 0.422 & 0.940 & 0.918 & 0.902 & 0.929 & 0.952 & 0.932 \\
\cline{1-9}
\multirow[t]{3}{*}{100} & 25 & 0.013 & 0.954 & 0.440 & 0.291 & 0.875 & 0.773 & 0.930 \\
 & 50 & 0.042 & 0.942 & 0.583 & 0.646 & 0.915 & 0.936 & 0.930 \\
 & 100 & 0.148 & 0.944 & 0.875 & 0.867 & 0.930 & 0.956 & 0.942 \\
\cline{1-9}
\bottomrule
\end{tabular}
\label{tab:MC_DGP3_a}
}
\end{center}
\footnotesize
\renewcommand{\baselineskip}{11pt}
\textbf{Note:} $(\sqrt{NT})$ Scaled bias and coverage rate based on confidence intervals. From left to right, the shown estimators are conventional CCEP \citet{Pesaran2006}, and the bias corrected versions: DVS \citet{DeVos2019}, HPJ \citet{dhaene2015split}, and AN \citet{HahnKursteiner2011BiasReduction}. The label (CS) refers to the cross-sectional bootstrap as in \citet{Kapetanios2008}, while (RDn) refers to the ``naive'' implementation of the proposed recursive design bootstrap.\end{table}
\end{landscape}
\begin{landscape}
\begin{table}[h]
\begin{center}
\caption{ARX(1): Monte Carlo results for $\widehat{\beta}$ in DGP (3).}
\small{
\begin{tabular}{ll|cccc|ccc}
\toprule
\multicolumn{9}{c}{Bias}\\
\toprule
 &  & CCEP (CS) & DVS (CS) & HPJ (CS) & AN (CS) & CCEP (RDn) & HPJ (RDn) & AN (RDn) \\
$N$ & $T$ &  &  &  &  &  &  &  \\
\midrule
\multirow[t]{3}{*}{25} & 25 & 0.033 & -0.011 & 0.403 & -0.076 & -0.021 & 0.225 & -0.042 \\
 & 50 & 0.086 & -0.032 & 0.132 & -0.015 & -0.015 & -0.022 & -0.026 \\
 & 100 & 0.081 & -0.030 & 0.016 & -0.013 & -0.007 & -0.016 & -0.015 \\
\cline{1-9}
\multirow[t]{3}{*}{50} & 25 & 0.052 & -0.026 & 0.573 & -0.102 & -0.037 & 0.280 & -0.059 \\
 & 50 & 0.165 & -0.006 & 0.223 & 0.018 & 0.015 & 0.007 & 0.002 \\
 & 100 & 0.145 & -0.012 & 0.060 & 0.016 & 0.010 & 0.007 & 0.004 \\
\cline{1-9}
\multirow[t]{3}{*}{100} & 25 & 0.095 & -0.014 & 0.838 & -0.128 & -0.030 & 0.368 & -0.064 \\
 & 50 & 0.242 & -0.005 & 0.327 & 0.038 & 0.022 & 0.017 & 0.008 \\
 & 100 & 0.211 & -0.009 & 0.088 & 0.027 & 0.011 & 0.006 & 0.001 \\
\cline{1-9}
\bottomrule
\multicolumn{9}{c}{Coverage rates}\\
\toprule
 &  & CCEP (CS) & DVS (CS) & HPJ (CS) & AN (CS) & CCEP (RDn) & HPJ (RDn) & AN (RDn) \\
$N$ & $T$ &  &  &  &  &  &  &  \\
\midrule
\multirow[t]{3}{*}{25} & 25 & 0.950 & 0.948 & 0.927 & 0.942 & 0.972 & 0.934 & 0.970 \\
 & 50 & 0.968 & 0.972 & 0.961 & 0.971 & 0.968 & 0.971 & 0.963 \\
 & 100 & 0.984 & 0.986 & 0.983 & 0.985 & 0.963 & 0.960 & 0.965 \\
\cline{1-9}
\multirow[t]{3}{*}{50} & 25 & 0.941 & 0.941 & 0.898 & 0.938 & 0.969 & 0.910 & 0.963 \\
 & 50 & 0.952 & 0.962 & 0.945 & 0.960 & 0.964 & 0.963 & 0.963 \\
 & 100 & 0.964 & 0.968 & 0.966 & 0.969 & 0.961 & 0.958 & 0.956 \\
\cline{1-9}
\multirow[t]{3}{*}{100} & 25 & 0.941 & 0.941 & 0.837 & 0.926 & 0.968 & 0.900 & 0.960 \\
 & 50 & 0.939 & 0.957 & 0.936 & 0.955 & 0.967 & 0.969 & 0.963 \\
 & 100 & 0.940 & 0.955 & 0.949 & 0.956 & 0.954 & 0.954 & 0.954 \\
\cline{1-9}
\bottomrule
\end{tabular}
\label{tab:MC_DGP3_b}
}
\end{center}
\footnotesize
\renewcommand{\baselineskip}{11pt}
\textbf{Note:} $(\sqrt{NT})$ Scaled bias and coverage rate based on confidence intervals. From left to right, the shown estimators are conventional CCEP \citet{Pesaran2006}, and the bias corrected versions: DVS \citet{DeVos2019}, HPJ \citet{dhaene2015split}, and AN \citet{HahnKursteiner2011BiasReduction}. The label (CS) refers to the cross-sectional bootstrap as in \citet{Kapetanios2008}, while (RDn) refers to the ``naive'' implementation of the proposed recursive design bootstrap.\end{table}
\end{landscape}
\begin{landscape}
\begin{table}[h]
\begin{center}
\caption{ARX(1): Monte Carlo results for $\widehat{\alpha}$ in DGP (4).}
\small{
\begin{tabular}{ll|cccc|ccc}
\toprule
\multicolumn{9}{c}{Bias}\\
\toprule
 &  & CCEP (CS) & DVS (CS) & HPJ (CS) & AN (CS) & CCEP (RDn) & HPJ (RDn) & AN (RDn) \\
$N$ & $T$ &  &  &  &  &  &  &  \\
\midrule
\multirow[t]{3}{*}{25} & 25 & -2.203 & 0.157 & 0.974 & -1.024 & -0.731 & -0.044 & -0.463 \\
 & 50 & -1.195 & 0.308 & 0.668 & -0.296 & -0.210 & 0.153 & -0.130 \\
 & 100 & -0.648 & 0.356 & 0.427 & 0.052 & -0.076 & 0.104 & -0.016 \\
\cline{1-9}
\multirow[t]{3}{*}{50} & 25 & -3.077 & 0.255 & 1.417 & -1.456 & -0.952 & -0.078 & -0.624 \\
 & 50 & -1.798 & 0.317 & 0.776 & -0.574 & -0.363 & 0.118 & -0.286 \\
 & 100 & -1.091 & 0.319 & 0.354 & -0.150 & -0.196 & 0.002 & -0.157 \\
\cline{1-9}
\multirow[t]{3}{*}{100} & 25 & -4.378 & 0.318 & 1.872 & -2.136 & -1.358 & -0.220 & -0.935 \\
 & 50 & -2.545 & 0.424 & 1.018 & -0.875 & -0.461 & 0.135 & -0.394 \\
 & 100 & -1.619 & 0.374 & 0.376 & -0.330 & -0.248 & 0.002 & -0.227 \\
\cline{1-9}
\bottomrule
\multicolumn{9}{c}{Coverage rates}\\
\toprule
 &  & CCEP (CS) & DVS (CS) & HPJ (CS) & AN (CS) & CCEP (RDn) & HPJ (RDn) & AN (RDn) \\
$N$ & $T$ &  &  &  &  &  &  &  \\
\midrule
\multirow[t]{3}{*}{25} & 25 & 0.409 & 0.919 & 0.812 & 0.769 & 0.918 & 0.961 & 0.940 \\
 & 50 & 0.648 & 0.941 & 0.909 & 0.897 & 0.933 & 0.965 & 0.947 \\
 & 100 & 0.827 & 0.973 & 0.963 & 0.958 & 0.947 & 0.961 & 0.956 \\
\cline{1-9}
\multirow[t]{3}{*}{50} & 25 & 0.201 & 0.908 & 0.719 & 0.659 & 0.890 & 0.939 & 0.921 \\
 & 50 & 0.394 & 0.935 & 0.878 & 0.850 & 0.925 & 0.960 & 0.931 \\
 & 100 & 0.644 & 0.967 & 0.963 & 0.942 & 0.946 & 0.956 & 0.950 \\
\cline{1-9}
\multirow[t]{3}{*}{100} & 25 & 0.054 & 0.891 & 0.605 & 0.440 & 0.847 & 0.903 & 0.894 \\
 & 50 & 0.149 & 0.911 & 0.789 & 0.775 & 0.918 & 0.953 & 0.921 \\
 & 100 & 0.383 & 0.938 & 0.934 & 0.907 & 0.934 & 0.950 & 0.937 \\
\cline{1-9}
\bottomrule
\end{tabular}
\label{tab:MC_DGP4_a}
}
\end{center}
\footnotesize
\renewcommand{\baselineskip}{11pt}
\textbf{Note:} $(\sqrt{NT})$ Scaled bias and coverage rate based on confidence intervals. From left to right, the shown estimators are conventional CCEP \citet{Pesaran2006}, and the bias corrected versions: DVS \citet{DeVos2019}, HPJ \citet{dhaene2015split}, and AN \citet{HahnKursteiner2011BiasReduction}. The label (CS) refers to the cross-sectional bootstrap as in \citet{Kapetanios2008}, while (RDn) refers to the ``naive'' implementation of the proposed recursive design bootstrap.\end{table} 
\end{landscape}
\begin{landscape}
\begin{table}[h]
\begin{center}
\caption{ARX(1): Monte Carlo results for $\widehat{\beta}$ in DGP (4).}
\small{
\begin{tabular}{ll|cccc|ccc}
\toprule
\multicolumn{9}{c}{Bias}\\
\toprule
 &  & CCEP (CS) & DVS (CS) & HPJ (CS) & AN (CS) & CCEP (RDn) & HPJ (RDn) & AN (RDn) \\
$N$ & $T$ &  &  &  &  &  &  &  \\
\midrule
\multirow[t]{3}{*}{25} & 25 & -0.242 & -1.691 & 0.421 & -0.183 & -0.814 & -0.360 & -0.465 \\
 & 50 & -0.092 & -1.296 & 0.194 & -0.057 & -0.682 & -0.144 & -0.287 \\
 & 100 & -0.017 & -0.916 & 0.069 & -0.001 & -0.553 & -0.129 & -0.205 \\
\cline{1-9}
\multirow[t]{3}{*}{50} & 25 & -0.388 & -2.471 & 0.583 & -0.313 & -1.216 & -0.532 & -0.696 \\
 & 50 & -0.137 & -1.850 & 0.263 & -0.093 & -0.948 & -0.156 & -0.369 \\
 & 100 & -0.044 & -1.319 & 0.094 & -0.024 & -0.733 & -0.090 & -0.220 \\
\cline{1-9}
\multirow[t]{3}{*}{100} & 25 & -0.521 & -3.470 & 0.829 & -0.421 & -1.700 & -0.748 & -0.955 \\
 & 50 & -0.176 & -2.610 & 0.389 & -0.111 & -1.323 & -0.162 & -0.472 \\
 & 100 & -0.051 & -1.866 & 0.141 & -0.018 & -0.991 & -0.055 & -0.234 \\
\cline{1-9}
\bottomrule
\multicolumn{9}{c}{Coverage rates}\\
\toprule
 &  & CCEP (CS) & DVS (CS) & HPJ (CS) & AN (CS) & CCEP (RDn) & HPJ (RDn) & AN (RDn) \\
$N$ & $T$ &  &  &  &  &  &  &  \\
\midrule
\multirow[t]{3}{*}{25} & 25 & 0.951 & 0.676 & 0.933 & 0.951 & 0.861 & 0.935 & 0.940 \\
 & 50 & 0.977 & 0.839 & 0.969 & 0.977 & 0.871 & 0.959 & 0.956 \\
 & 100 & 0.992 & 0.946 & 0.991 & 0.992 & 0.906 & 0.967 & 0.965 \\
\cline{1-9}
\multirow[t]{3}{*}{50} & 25 & 0.923 & 0.347 & 0.911 & 0.929 & 0.687 & 0.893 & 0.885 \\
 & 50 & 0.959 & 0.564 & 0.954 & 0.963 & 0.754 & 0.945 & 0.931 \\
 & 100 & 0.980 & 0.818 & 0.977 & 0.982 & 0.823 & 0.956 & 0.949 \\
\cline{1-9}
\multirow[t]{3}{*}{100} & 25 & 0.901 & 0.076 & 0.856 & 0.911 & 0.429 & 0.824 & 0.806 \\
 & 50 & 0.944 & 0.187 & 0.927 & 0.945 & 0.555 & 0.934 & 0.907 \\
 & 100 & 0.965 & 0.479 & 0.959 & 0.964 & 0.699 & 0.954 & 0.944 \\
\cline{1-9}
\bottomrule
\end{tabular}
\label{tab:MC_DGP4_b}
}
\end{center}
\footnotesize
\renewcommand{\baselineskip}{11pt}
\textbf{Note:} $(\sqrt{NT})$ Scaled bias and coverage rate based on confidence intervals. From left to right, the shown estimators are conventional CCEP \citet{Pesaran2006}, and the bias corrected versions: DVS \citet{DeVos2019}, HPJ \citet{dhaene2015split}, and AN \citet{HahnKursteiner2011BiasReduction}. The label (CS) refers to the cross-sectional bootstrap as in \citet{Kapetanios2008}, while (RDn) refers to the ``naive'' implementation of the proposed recursive design bootstrap.\end{table}
\end{landscape}
\begin{landscape}
\begin{table}[h]
\begin{center}
\caption{ARX(1): Monte Carlo results for $\widehat{\alpha}$ in DGP (5).}
\small{
\begin{tabular}{ll|cccc|ccc}
\toprule
\multicolumn{9}{c}{Bias}\\
\toprule
 &  & CCEP (CS) & DVS (CS) & HPJ (CS) & AN (CS) & CCEP (RDn) & HPJ (RDn) & AN (RDn) \\
$N$ & $T$ &  &  &  &  &  &  &  \\
\midrule
\multirow[t]{3}{*}{25} & 25 & -2.635 & 0.189 & 1.096 & -1.191 & -0.882 & -0.120 & -0.503 \\
 & 50 & -1.506 & 0.265 & 0.667 & -0.417 & -0.328 & 0.081 & -0.187 \\
 & 100 & -0.860 & 0.326 & 0.406 & -0.017 & -0.135 & 0.074 & -0.046 \\
\cline{1-9}
\multirow[t]{3}{*}{50} & 25 & -3.744 & 0.280 & 1.626 & -1.746 & -1.182 & -0.179 & -0.714 \\
 & 50 & -2.214 & 0.309 & 0.839 & -0.716 & -0.455 & 0.080 & -0.317 \\
 & 100 & -1.374 & 0.303 & 0.375 & -0.228 & -0.238 & 0.018 & -0.165 \\
\cline{1-9}
\multirow[t]{3}{*}{100} & 25 & -5.250 & 0.421 & 2.224 & -2.475 & -1.596 & -0.329 & -0.984 \\
 & 50 & -3.185 & 0.397 & 1.132 & -1.088 & -0.618 & 0.103 & -0.455 \\
 & 100 & -2.051 & 0.329 & 0.384 & -0.458 & -0.336 & -0.010 & -0.266 \\
\cline{1-9}
\bottomrule
\multicolumn{9}{c}{Coverage rates}\\
\toprule
 &  & CCEP (CS) & DVS (CS) & HPJ (CS) & AN (CS) & CCEP (RDn) & HPJ (RDn) & AN (RDn) \\
$N$ & $T$ &  &  &  &  &  &  &  \\
\midrule
\multirow[t]{3}{*}{25} & 25 & 0.297 & 0.920 & 0.796 & 0.741 & 0.906 & 0.955 & 0.936 \\
 & 50 & 0.529 & 0.951 & 0.910 & 0.893 & 0.928 & 0.964 & 0.948 \\
 & 100 & 0.770 & 0.977 & 0.970 & 0.952 & 0.948 & 0.964 & 0.958 \\
\cline{1-9}
\multirow[t]{3}{*}{50} & 25 & 0.086 & 0.923 & 0.708 & 0.585 & 0.890 & 0.925 & 0.931 \\
 & 50 & 0.241 & 0.936 & 0.860 & 0.820 & 0.929 & 0.960 & 0.930 \\
 & 100 & 0.528 & 0.967 & 0.957 & 0.934 & 0.942 & 0.955 & 0.950 \\
\cline{1-9}
\multirow[t]{3}{*}{100} & 25 & 0.009 & 0.889 & 0.572 & 0.354 & 0.848 & 0.871 & 0.912 \\
 & 50 & 0.042 & 0.912 & 0.757 & 0.700 & 0.912 & 0.958 & 0.925 \\
 & 100 & 0.193 & 0.945 & 0.934 & 0.883 & 0.935 & 0.952 & 0.938 \\
\cline{1-9}
\bottomrule
\end{tabular}
\label{tab:MC_DGP5_a}
}
\end{center}
\footnotesize
\renewcommand{\baselineskip}{11pt}
\textbf{Note:} $(\sqrt{NT})$ Scaled bias and coverage rate based on confidence intervals. From left to right, the shown estimators are conventional CCEP \citet{Pesaran2006}, and the bias corrected versions: DVS \citet{DeVos2019}, HPJ \citet{dhaene2015split}, and AN \citet{HahnKursteiner2011BiasReduction}. The label (CS) refers to the cross-sectional bootstrap as in \citet{Kapetanios2008}, while (RDn) refers to the ``naive'' implementation of the proposed recursive design bootstrap.\end{table}
\end{landscape}
\begin{landscape}
\begin{table}[h]
\begin{center}
\caption{ARX(1): Monte Carlo results for $\widehat{\beta}$ in DGP (5).}
\small{
\begin{tabular}{ll|cccc|ccc}
\toprule
\multicolumn{9}{c}{Bias}\\
\toprule
 &  & CCEP (CS) & DVS (CS) & HPJ (CS) & AN (CS) & CCEP (RDn) & HPJ (RDn) & AN (RDn) \\
$N$ & $T$ &  &  &  &  &  &  &  \\
\midrule
\multirow[t]{3}{*}{25} & 25 & 0.010 & -1.761 & 0.657 & -0.260 & -0.760 & -0.213 & -0.571 \\
 & 50 & 0.155 & -1.290 & 0.228 & -0.124 & -0.573 & -0.034 & -0.321 \\
 & 100 & 0.063 & -1.007 & -0.060 & -0.175 & -0.494 & -0.104 & -0.255 \\
\cline{1-9}
\multirow[t]{3}{*}{50} & 25 & 0.146 & -2.412 & 1.027 & -0.252 & -1.006 & -0.315 & -0.717 \\
 & 50 & 0.290 & -1.787 & 0.353 & -0.114 & -0.782 & -0.020 & -0.393 \\
 & 100 & 0.238 & -1.294 & 0.044 & -0.099 & -0.585 & -0.015 & -0.211 \\
\cline{1-9}
\multirow[t]{3}{*}{100} & 25 & 0.292 & -3.351 & 1.514 & -0.278 & -1.398 & -0.413 & -0.965 \\
 & 50 & 0.490 & -2.474 & 0.573 & -0.077 & -1.085 & 0.011 & -0.495 \\
 & 100 & 0.417 & -1.767 & 0.130 & -0.056 & -0.804 & 0.023 & -0.238 \\
\cline{1-9}
\bottomrule
\multicolumn{9}{c}{Coverage rates}\\
\toprule
 &  & CCEP (CS) & DVS (CS) & HPJ (CS) & AN (CS) & CCEP (RDn) & HPJ (RDn) & AN (RDn) \\
$N$ & $T$ &  &  &  &  &  &  &  \\
\midrule
\multirow[t]{3}{*}{25} & 25 & 0.938 & 0.714 & 0.892 & 0.937 & 0.870 & 0.944 & 0.932 \\
 & 50 & 0.966 & 0.856 & 0.956 & 0.969 & 0.882 & 0.959 & 0.947 \\
 & 100 & 0.988 & 0.963 & 0.988 & 0.992 & 0.920 & 0.969 & 0.964 \\
\cline{1-9}
\multirow[t]{3}{*}{50} & 25 & 0.923 & 0.453 & 0.831 & 0.927 & 0.750 & 0.907 & 0.889 \\
 & 50 & 0.942 & 0.641 & 0.934 & 0.959 & 0.793 & 0.950 & 0.930 \\
 & 100 & 0.962 & 0.855 & 0.969 & 0.974 & 0.860 & 0.952 & 0.947 \\
\cline{1-9}
\multirow[t]{3}{*}{100} & 25 & 0.911 & 0.144 & 0.734 & 0.930 & 0.565 & 0.868 & 0.818 \\
 & 50 & 0.902 & 0.292 & 0.891 & 0.948 & 0.653 & 0.943 & 0.908 \\
 & 100 & 0.928 & 0.572 & 0.953 & 0.962 & 0.756 & 0.949 & 0.943 \\
\cline{1-9}
\bottomrule
\end{tabular}
\label{tab:MC_DGP5_b}
}
\end{center}
\footnotesize
\renewcommand{\baselineskip}{11pt}
\textbf{Note:} $(\sqrt{NT})$ Scaled bias and coverage rate based on confidence intervals. From left to right, the shown estimators are conventional CCEP \citet{Pesaran2006}, and the bias corrected versions: DVS \citet{DeVos2019}, HPJ \citet{dhaene2015split}, and AN \citet{HahnKursteiner2011BiasReduction}. The label (CS) refers to the cross-sectional bootstrap as in \citet{Kapetanios2008}, while (RDn) refers to the ``naive'' implementation of the proposed recursive design bootstrap.\end{table}
\end{landscape}

\clearpage
\subsection{Additional simulation exercises}\label{3}
    \subsubsection{Empirical Monte Carlo}\label{3.1}
    We consider another Monte Carlo design specifically tailored to the empirical illustration, with $T=20$ fixed and $N$ varying. Examining this setting in the Monte Carlo before turning to the empirical application provides a useful benchmark. To approximate the empirical setting, we set $\alpha_0=0.2$, $\theta_0=0$ and $\phi_0 \in \{0,0.6\}$. A detailed overview of the results can be found in Tables \ref{tab:MC_ED_a}, \ref{tab:MC_ED_b}, \ref{tab:MC_ED2_a}, and \ref{tab:MC_ED2_b}.

    The CCEP estimator is severely biased downward in both designs. Both the AN and HPJ methods considerably reduce the bias of the CCEP estimator. The DVS method achieves superior bias correction compared to both bias-correction methods.

    The RDn bootstrap applied to the CCEP estimator reduces bias substantially. In line with the previous Monte Carlo results, the bias can be further reduced with bias-correction tools. Across the bias-correction tools considered with RDn bootstrap, HPJ achieves the lowest bias (on average) in both designs. 

    The CCEP estimator with CS bootstrap obtains coverage far below nominal level for $\widehat{\alpha}$. Given that $\widehat{\beta}$ is strictly exogenous, the coverage is close to the nominal level. When combining the CS bootstrap with AN or HPJ, coverage decreases as $(N/T)$ increases. This is in line with previous Monte Carlo results. The DVS method is below the nominal level when $N=20$, but as $N$ increases coverage improves to the nominal level. 

    The coverage of the RDn bootstrap with CCEP is almost at the nominal level for all designs considered. As $N$ increases, we observe a decrease in the coverage. When the RDn bootstrap is supplemented with HPJ, we observe that the coverage is slightly conservative. On the other hand, the AN method has coverage rates close to the nominal level for all designs considered.
   
    Taken together, these results reinforce the preference for the RD bootstrap over the conventional CS bootstrap. Moreover, both the bias and coverage rates suggest that the RD bootstrap is a viable alternative to DVS in this setting.

\begin{landscape}
\begin{table}[h]
\begin{center}
\caption{ARX(1): Monte Carlo results for $\widehat{\alpha}$ in Empirical Design (1).}
\small{
\begin{tabular}{ll|cccc|ccc}
\toprule
\multicolumn{9}{c}{Bias}\\
\toprule
 &  & CCEP (CS) & DVS (CS) & HPJ (CS) & AN (CS) & CCEP (RDn) & HPJ (RDn) & AN (RDn) \\
$N$ & $T$ &  &  &  &  &  &  &  \\
\midrule
20 & 20 & -1.200 & 0.352 & 0.718 & -0.265 & -0.233 & 0.258 & 0.067 \\
\cline{1-9}
40 & 20 & -2.038 & 0.200 & 0.647 & -0.707 & -0.548 & 0.049 & -0.152 \\
\cline{1-9}
60 & 20 & -2.598 & 0.170 & 0.695 & -0.977 & -0.699 & -0.011 & -0.241 \\
\cline{1-9}
80 & 20 & -3.110 & 0.101 & 0.741 & -1.246 & -0.871 & -0.058 & -0.352 \\
\cline{1-9}
100 & 20 & -3.522 & 0.094 & 0.781 & -1.431 & -0.974 & -0.091 & -0.395 \\
\cline{1-9}
\bottomrule
\multicolumn{9}{c}{Coverage rates}\\
\toprule
 &  & CCEP (CS) & DVS (CS) & HPJ (CS) & AN (CS) & CCEP (RDn) & HPJ (RDn) & AN (RDn) \\
$N$ & $T$ &  &  &  &  &  &  &  \\
\midrule
20 & 20 & 0.638 & 0.903 & 0.887 & 0.862 & 0.951 & 0.985 & 0.956 \\
\cline{1-9}
40 & 20 & 0.401 & 0.932 & 0.874 & 0.818 & 0.953 & 0.980 & 0.963 \\
\cline{1-9}
60 & 20 & 0.243 & 0.937 & 0.862 & 0.766 & 0.949 & 0.985 & 0.958 \\
\cline{1-9}
80 & 20 & 0.135 & 0.943 & 0.843 & 0.695 & 0.943 & 0.982 & 0.963 \\
\cline{1-9}
100 & 20 & 0.082 & 0.948 & 0.828 & 0.645 & 0.936 & 0.984 & 0.954 \\
\cline{1-9}
\bottomrule
\end{tabular}
\label{tab:MC_ED_a}
}
\end{center}
\footnotesize
\renewcommand{\baselineskip}{11pt}
\textbf{Note:} $(\sqrt{NT})$ Scaled bias and coverage rate based on confidence intervals. From left to right, the shown estimators are conventional CCEP \citet{Pesaran2006}, and the bias corrected versions: DVS \citet{DeVos2019}, HPJ \citet{dhaene2015split}, and AN \citet{HahnKursteiner2011BiasReduction}. The label (CS) refers to the cross-sectional bootstrap as in \citet{Kapetanios2008}, while (RDn) refers to the ``naive'' implementation of the proposed recursive design bootstrap.\end{table}
\end{landscape}
\begin{landscape}
\begin{table}[h]
\begin{center}
\caption{ARX(1): Monte Carlo results for $\widehat{\beta}$ in Empirical Design (1).}
\small{
\begin{tabular}{ll|cccc|ccc}
\toprule
\multicolumn{9}{c}{Bias}\\
\toprule
 &  & CCEP (CS) & DVS (CS) & HPJ (CS) & AN (CS) & CCEP (RDn) & HPJ (RDn) & AN (RDn) \\
$N$ & $T$ &  &  &  &  &  &  &  \\
\midrule
20 & 20 & -0.119 & 0.018 & 0.447 & -0.070 & -0.058 & 0.081 & -0.025 \\
\cline{1-9}
40 & 20 & -0.225 & 0.002 & 0.688 & -0.130 & -0.101 & 0.147 & -0.045 \\
\cline{1-9}
60 & 20 & -0.315 & -0.016 & 0.799 & -0.202 & -0.142 & 0.107 & -0.089 \\
\cline{1-9}
80 & 20 & -0.358 & 0.000 & 0.966 & -0.231 & -0.138 & 0.147 & -0.087 \\
\cline{1-9}
100 & 20 & -0.393 & 0.015 & 1.112 & -0.239 & -0.131 & 0.188 & -0.069 \\
\cline{1-9}
\bottomrule
\multicolumn{9}{c}{Coverage rates}\\
\toprule
 &  & CCEP (CS) & DVS (CS) & HPJ (CS) & AN (CS) & CCEP (RDn) & HPJ (RDn) & AN (RDn) \\
$N$ & $T$ &  &  &  &  &  &  &  \\
\midrule
20 & 20 & 0.928 & 0.936 & 0.922 & 0.929 & 0.971 & 0.980 & 0.971 \\
\cline{1-9}
40 & 20 & 0.947 & 0.956 & 0.921 & 0.952 & 0.977 & 0.985 & 0.974 \\
\cline{1-9}
60 & 20 & 0.930 & 0.942 & 0.897 & 0.935 & 0.967 & 0.976 & 0.965 \\
\cline{1-9}
80 & 20 & 0.924 & 0.941 & 0.879 & 0.932 & 0.966 & 0.976 & 0.966 \\
\cline{1-9}
100 & 20 & 0.926 & 0.945 & 0.862 & 0.936 & 0.968 & 0.978 & 0.969 \\
\cline{1-9}
\bottomrule
\end{tabular}
\label{tab:MC_ED_b}
}
\end{center}
\footnotesize
\renewcommand{\baselineskip}{11pt}
\textbf{Note:} $(\sqrt{NT})$ Scaled bias and coverage rate based on confidence intervals. From left to right, the shown estimators are conventional CCEP \citet{Pesaran2006}, and the bias corrected versions: DVS \citet{DeVos2019}, HPJ \citet{dhaene2015split}, and AN \citet{HahnKursteiner2011BiasReduction}. The label (CS) refers to the cross-sectional bootstrap as in \citet{Kapetanios2008}, while (RDn) refers to the ``naive'' implementation of the proposed recursive design bootstrap.\end{table}
\end{landscape}
\begin{landscape}
\begin{table}[h]
\begin{center}
\caption{ARX(1): Monte Carlo results for $\widehat{\alpha}$ in Empirical Design (2).}
\small{
\begin{tabular}{ll|cccc|ccc}
\toprule
\multicolumn{9}{c}{Bias}\\
\toprule
 &  & CCEP (CS) & DVS (CS) & HPJ (CS) & AN (CS) & CCEP (RDn) & HPJ (RDn) & AN (RDn) \\
$N$ & $T$ &  &  &  &  &  &  &  \\
\midrule
20 & 20 & -1.862 & 0.215 & 0.731 & -0.653 & -0.573 & 0.002 & -0.149 \\
\cline{1-9}
40 & 20 & -2.855 & 0.123 & 0.895 & -1.148 & -0.878 & -0.102 & -0.326 \\
\cline{1-9}
60 & 20 & -3.643 & 0.049 & 0.991 & -1.542 & -1.142 & -0.204 & -0.477 \\
\cline{1-9}
80 & 20 & -4.201 & 0.079 & 1.129 & -1.777 & -1.261 & -0.240 & -0.511 \\
\cline{1-9}
100 & 20 & -4.754 & 0.042 & 1.254 & -2.044 & -1.426 & -0.301 & -0.594 \\
\cline{1-9}
\bottomrule
\multicolumn{9}{c}{Coverage rates}\\
\toprule
 &  & CCEP (CS) & DVS (CS) & HPJ (CS) & AN (CS) & CCEP (RDn) & HPJ (RDn) & AN (RDn) \\
$N$ & $T$ &  &  &  &  &  &  &  \\
\midrule
20 & 20 & 0.469 & 0.916 & 0.887 & 0.812 & 0.936 & 0.980 & 0.947 \\
\cline{1-9}
40 & 20 & 0.185 & 0.938 & 0.856 & 0.724 & 0.934 & 0.982 & 0.957 \\
\cline{1-9}
60 & 20 & 0.062 & 0.944 & 0.828 & 0.610 & 0.913 & 0.980 & 0.949 \\
\cline{1-9}
80 & 20 & 0.024 & 0.949 & 0.796 & 0.542 & 0.907 & 0.975 & 0.950 \\
\cline{1-9}
100 & 20 & 0.007 & 0.947 & 0.766 & 0.449 & 0.885 & 0.968 & 0.945 \\
\cline{1-9}
\bottomrule
\end{tabular}
\label{tab:MC_ED2_a}
}
\end{center}
\footnotesize
\renewcommand{\baselineskip}{11pt}
\textbf{Note:} $(\sqrt{NT})$ Scaled bias and coverage rate based on confidence intervals. From left to right, the shown estimators are conventional CCEP \citet{Pesaran2006}, and the bias corrected versions: DVS \citet{DeVos2019}, HPJ \citet{dhaene2015split}, and AN \citet{HahnKursteiner2011BiasReduction}. The label (CS) refers to the cross-sectional bootstrap as in \citet{Kapetanios2008}, while (RDn) refers to the ``naive'' implementation of the proposed recursive design bootstrap.
\end{table}
\end{landscape}
\begin{landscape}
\begin{table}[h]
\begin{center}
\caption{ARX(1): Monte Carlo results for $\widehat{\beta}$ in Empirical Design (2).}
\small{
\begin{tabular}{ll|cccc|ccc}
\toprule
\multicolumn{9}{c}{Bias}\\
\toprule
 &  & CCEP (CS) & DVS (CS) & HPJ (CS) & AN (CS) & CCEP (RDn) & HPJ (RDn) & AN (RDn) \\
$N$ & $T$ &  &  &  &  &  &  &  \\
\midrule
20 & 20 & 0.752 & -0.100 & 0.892 & 0.164 & 0.212 & 0.212 & 0.023 \\
\cline{1-9}
40 & 20 & 1.147 & -0.061 & 1.466 & 0.314 & 0.323 & 0.337 & 0.072 \\
\cline{1-9}
60 & 20 & 1.475 & -0.021 & 1.848 & 0.434 & 0.440 & 0.386 & 0.124 \\
\cline{1-9}
80 & 20 & 1.679 & -0.054 & 2.220 & 0.495 & 0.466 & 0.465 & 0.123 \\
\cline{1-9}
100 & 20 & 1.949 & 0.017 & 2.541 & 0.606 & 0.575 & 0.532 & 0.181 \\
\cline{1-9}
\bottomrule
\multicolumn{9}{c}{Coverage rates}\\
\toprule
 &  & CCEP (CS) & DVS (CS) & HPJ (CS) & AN (CS) & CCEP (RDn) & HPJ (RDn) & AN (RDn) \\
$N$ & $T$ &  &  &  &  &  &  &  \\
\midrule
20 & 20 & 0.905 & 0.938 & 0.890 & 0.933 & 0.964 & 0.981 & 0.968 \\
\cline{1-9}
40 & 20 & 0.862 & 0.946 & 0.857 & 0.935 & 0.968 & 0.974 & 0.970 \\
\cline{1-9}
60 & 20 & 0.810 & 0.939 & 0.820 & 0.931 & 0.967 & 0.975 & 0.966 \\
\cline{1-9}
80 & 20 & 0.759 & 0.946 & 0.788 & 0.923 & 0.970 & 0.973 & 0.969 \\
\cline{1-9}
100 & 20 & 0.709 & 0.943 & 0.742 & 0.928 & 0.969 & 0.973 & 0.971 \\
\cline{1-9}
\bottomrule
\end{tabular}
\label{tab:MC_ED2_b}
}
\end{center}
\footnotesize
\renewcommand{\baselineskip}{11pt}
\textbf{Note:} $(\sqrt{NT})$ Scaled bias and coverage rate based on confidence intervals. From left to right, the shown estimators are conventional CCEP \citet{Pesaran2006}, and the bias corrected versions: DVS \citet{DeVos2019}, HPJ \citet{dhaene2015split}, and AN \citet{HahnKursteiner2011BiasReduction}. The label (CS) refers to the cross-sectional bootstrap as in \citet{Kapetanios2008}, while (RDn) refers to the ``naive'' implementation of the proposed recursive design bootstrap. 
\end{table}
\end{landscape}

\clearpage
\subsubsection{Higher-order dynamics}\label{3.2}
In this exercise, we demonstrate the flexibility of the recursive design bootstrap method by allowing for an AR$(2)$ process:
\begin{align*}
    y_{i,t}=\sqrt{1-\alpha_{0,1}^2-\alpha_{0,2}^2}c_i+\alpha_{0,1} y_{i,t-1}+\alpha_{0,2} y_{i,t-2}+\gamma_if_t+\sqrt{1-\alpha_{0,1}^2-\alpha_{0,2}^2}\varepsilon_{i,t}, \quad \varepsilon_{i,t} \sim \mathcal{N}(0,1),
\end{align*}
where all notation is defined as in Section \ref{2.2}. We set $\alpha_{0,1} = 0.8$ to match the AR$(1)$-baseline and take $\alpha_{0,2} = 0$. In the factor proxy of cross-sectional averages, we must now also include $\bar{y}_{t-2}$ to span the factor space appropriately. Since the bias-correction method of \citet{DeVos2019} does not readily extend to AR$(p)$-processes, we do not implement this method here.

In Tables \ref{tab:MC_HOD_a} and \ref{tab:MC_HOD_b}, we report bias and coverage rates. We compare the estimation results to the relevant AR(1)-baseline shown in Table \ref{tab:MC_DGP1}. 

Compared to the baseline, bias increases slightly for all estimators considered. This result is most apparent for the CCEP estimator, and less pronounced with the bias-corrected alternatives. The RDn or RDs bootstrap both reduce the bias, but it remains larger compared to the baseline.

The increased bias results in a deterioration in the coverage rates, especially for CCEP and HPJ. As in the baseline, we observe that the coverage rates deteriorate proportionally to the magnitude of the bias. This deterioration is largest for the CS bootstrap, smaller with the ``naive'' construction of factors, and smallest with the ``sophisticated'' method. 

Similar to the AR$(1)$-setting, in the AR$(2)$-setting, supplementing the RD bootstrap with HPJ attains coverage rates that are close to nominal level. One noteworthy finding is that the AN method with RDn or RDs bootstrap attains coverage rates closer to the nominal level in the AR$(2)$-setting compared to HPJ when $T=25$. For larger values of $T$, both methods achieve comparable coverage rates. 

Intuitively, this arises because there are now two bias terms that differ in size. Since $\alpha_{0,1}$ is substantially larger than $\alpha_{0,2}$, the bias term of the former is larger. In turn, HPJ will tend to underestimate the bias of $\alpha_{0,1}$ and overestimate the bias of $\alpha_{0,2}$. This is most apparent when considering the bias in Tables \ref{tab:MC_HOD_a} and \ref{tab:MC_HOD_b}. Fortunately, AN does not suffer from the same problem, and the finite sample improvements from pivotal confidence intervals illustrated in the AR(1)-baseline largely carry over to the AR$(2)$-setting.

\begin{landscape}
\begin{table}[h]
\begin{center}
\caption{AR(2): Monte Carlo results for $\widehat{\alpha}_1$ in the Higher-order dynamics design.}
\small{
\begin{tabular}{ll|ccc|ccc|ccc}
\toprule
\multicolumn{11}{c}{Bias}\\
\toprule
 &  & CCEP (CS) & HPJ (CS) & AN (CS) & CCEP (RDn) & HPJ (RDn) & AN (RDn) & CCEP (RDs) & HPJ (RDs) & AN (RDs) \\
$N$ & $T$ &  &  &  &  &  &  &  &  &  \\
\midrule
\multirow[t]{3}{*}{25} & 25 & -3.744 & 1.806 & -1.761 & -1.398 & 0.239 & -0.761 & -0.841 & 0.345 & -0.452 \\
 & 50 & -2.140 & 0.758 & -0.580 & -0.335 & -0.021 & -0.116 & -0.187 & -0.050 & -0.055 \\
 & 100 & -1.400 & 0.232 & -0.256 & -0.150 & -0.048 & -0.077 & -0.105 & -0.042 & -0.058 \\
\cline{1-11}
\multirow[t]{3}{*}{50} & 25 & -5.293 & 2.487 & -2.491 & -1.966 & 0.233 & -1.052 & -1.178 & 0.372 & -0.624 \\
 & 50 & -3.068 & 0.973 & -0.873 & -0.493 & -0.113 & -0.199 & -0.298 & -0.160 & -0.120 \\
 & 100 & -1.948 & 0.357 & -0.338 & -0.159 & -0.021 & -0.066 & -0.104 & -0.027 & -0.047 \\
\cline{1-11}
\multirow[t]{3}{*}{100} & 25 & -7.481 & 3.656 & -3.528 & -2.733 & 0.365 & -1.469 & -1.636 & 0.529 & -0.869 \\
 & 50 & -4.307 & 1.404 & -1.214 & -0.660 & -0.141 & -0.253 & -0.387 & -0.212 & -0.145 \\
 & 100 & -2.727 & 0.527 & -0.455 & -0.186 & -0.003 & -0.054 & -0.114 & -0.019 & -0.034 \\
\cline{1-11}
\bottomrule
\multicolumn{11}{c}{Coverage rates}\\
\toprule
 &  & CCEP (CS) & HPJ (CS) & AN (CS) & CCEP (RDn) & HPJ (RDn) & AN (RDn) & CCEP (RDs) & HPJ (RDs) & AN (RDs) \\
$N$ & $T$ &  &  &  &  &  &  &  &  &  \\
\midrule
\multirow[t]{3}{*}{25} & 25 & 0.191 & 0.712 & 0.728 & 0.717 & 0.857 & 0.892 & 0.857 & 0.847 & 0.912 \\
 & 50 & 0.508 & 0.882 & 0.899 & 0.916 & 0.939 & 0.938 & 0.931 & 0.938 & 0.934 \\
 & 100 & 0.702 & 0.928 & 0.918 & 0.932 & 0.939 & 0.934 & 0.937 & 0.936 & 0.933 \\
\cline{1-11}
\multirow[t]{3}{*}{50} & 25 & 0.028 & 0.606 & 0.539 & 0.573 & 0.798 & 0.866 & 0.784 & 0.814 & 0.904 \\
 & 50 & 0.219 & 0.842 & 0.864 & 0.901 & 0.934 & 0.940 & 0.927 & 0.935 & 0.938 \\
 & 100 & 0.533 & 0.933 & 0.930 & 0.941 & 0.944 & 0.941 & 0.946 & 0.939 & 0.943 \\
\cline{1-11}
\multirow[t]{3}{*}{100} & 25 & 0.000 & 0.422 & 0.284 & 0.394 & 0.712 & 0.819 & 0.683 & 0.734 & 0.896 \\
 & 50 & 0.036 & 0.750 & 0.796 & 0.877 & 0.940 & 0.936 & 0.922 & 0.942 & 0.943 \\
 & 100 & 0.261 & 0.912 & 0.920 & 0.937 & 0.938 & 0.946 & 0.946 & 0.940 & 0.944 \\
\cline{1-11}
\bottomrule
\end{tabular}
\label{tab:MC_HOD_a}
}
\end{center}
\footnotesize
\renewcommand{\baselineskip}{11pt}
\textbf{Note:} $(\sqrt{NT})$ Scaled bias and coverage rate based on confidence intervals. From left to right, the shown estimators are conventional CCEP \citet{Pesaran2006}, and the bias corrected versions: HPJ \citet{dhaene2015split}, and AN \citet{HahnKursteiner2011BiasReduction}. The label (CS) refers to the cross-sectional bootstrap as in \citet{Kapetanios2008}, while (RDn) refers to the ``naive'' implementation of the proposed recursive design bootstrap. (RDs) refers to the ``sophisticated'' implementation.
\end{table}
\end{landscape}
\begin{landscape}
\begin{table}[h]
\begin{center}
\caption{AR(2): Monte Carlo results for $\widehat{\alpha}_2$ in the Higher-order dynamics design.}
\small{
\begin{tabular}{ll|ccc|ccc|ccc}
\toprule
\multicolumn{11}{c}{Bias}\\
\toprule
 &  & CCEP (CS) & HPJ (CS) & AN (CS) & CCEP (RDn) & HPJ (RDn) & AN (RDn) & CCEP (RDs) & HPJ (RDs) & AN (RDs) \\
$N$ & $T$ &  &  &  &  &  &  &  &  &  \\
\midrule
\multirow[t]{3}{*}{25} & 25 & -0.963 & 0.093 & -0.718 & -0.011 & 0.275 & -0.200 & 0.357 & 0.186 & -0.049 \\
 & 50 & -0.672 & -0.001 & -0.541 & 0.016 & -0.023 & -0.139 & 0.153 & -0.018 & -0.084 \\
 & 100 & -0.405 & 0.035 & -0.233 & 0.062 & 0.046 & 0.007 & 0.105 & 0.039 & 0.013 \\
\cline{1-11}
\multirow[t]{3}{*}{50} & 25 & -1.326 & 0.122 & -0.960 & 0.020 & 0.411 & -0.230 & 0.529 & 0.330 & -0.030 \\
 & 50 & -0.837 & 0.074 & -0.660 & 0.091 & 0.025 & -0.108 & 0.286 & 0.041 & -0.036 \\
 & 100 & -0.585 & 0.015 & -0.346 & 0.048 & 0.004 & -0.033 & 0.113 & 0.003 & -0.016 \\
\cline{1-11}
\multirow[t]{3}{*}{100} & 25 & -1.751 & 0.479 & -1.239 & 0.122 & 0.698 & -0.246 & 0.851 & 0.562 & 0.049 \\
 & 50 & -1.169 & 0.120 & -0.927 & 0.118 & 0.016 & -0.166 & 0.387 & 0.036 & -0.066 \\
 & 100 & -0.801 & 0.030 & -0.472 & 0.069 & 0.009 & -0.045 & 0.161 & 0.006 & -0.019 \\
\cline{1-11}
\bottomrule
\multicolumn{11}{c}{Coverage rates}\\
\toprule
 &  & CCEP (CS) & HPJ (CS) & AN (CS) & CCEP (RDn) & HPJ (RDn) & AN (RDn) & CCEP (RDs) & HPJ (RDs) & AN (RDs) \\
$N$ & $T$ &  &  &  &  &  &  &  &  &  \\
\midrule
\multirow[t]{3}{*}{25} & 25 & 0.811 & 0.825 & 0.879 & 0.892 & 0.896 & 0.928 & 0.894 & 0.915 & 0.918 \\
 & 50 & 0.881 & 0.896 & 0.899 & 0.918 & 0.942 & 0.928 & 0.922 & 0.938 & 0.924 \\
 & 100 & 0.922 & 0.921 & 0.929 & 0.930 & 0.943 & 0.933 & 0.937 & 0.939 & 0.932 \\
\cline{1-11}
\multirow[t]{3}{*}{50} & 25 & 0.698 & 0.724 & 0.838 & 0.865 & 0.861 & 0.924 & 0.867 & 0.892 & 0.914 \\
 & 50 & 0.837 & 0.868 & 0.885 & 0.921 & 0.945 & 0.933 & 0.922 & 0.940 & 0.930 \\
 & 100 & 0.896 & 0.930 & 0.926 & 0.931 & 0.944 & 0.939 & 0.934 & 0.942 & 0.940 \\
\cline{1-11}
\multirow[t]{3}{*}{100} & 25 & 0.588 & 0.597 & 0.770 & 0.834 & 0.787 & 0.908 & 0.810 & 0.829 & 0.903 \\
 & 50 & 0.752 & 0.820 & 0.835 & 0.919 & 0.937 & 0.935 & 0.918 & 0.943 & 0.933 \\
 & 100 & 0.865 & 0.929 & 0.918 & 0.938 & 0.947 & 0.946 & 0.935 & 0.946 & 0.942 \\
\cline{1-11}
\bottomrule
\end{tabular}
\label{tab:MC_HOD_b}
}
\end{center}
\footnotesize
\renewcommand{\baselineskip}{11pt}
\textbf{Note:} $(\sqrt{NT})$ Scaled bias and coverage rate based on confidence intervals. From left to right, the shown estimators are conventional CCEP \citet{Pesaran2006}, and the bias corrected versions: HPJ \citet{dhaene2015split},  and AN \citet{HahnKursteiner2011BiasReduction}. The label (CS) refers to the cross-sectional bootstrap as in \citet{Kapetanios2008}, while (RDn) refers to the ``naive'' implementation of the proposed recursive design bootstrap. (RDs) refers to the ``sophisticated'' implementation.
\end{table}
\end{landscape}

\clearpage
    \subsubsection{Iterative Bootstrap}\label{3.3}
In this subsection, we investigate the properties of an iterative RD bootstrap procedure studied in \citet{heller_jochmans_2026_iterated_bootstrap}, applied to our factor-augmented dynamic model. We use the benchmark AR$(1)$ model from Section \ref{2.2}:
\begin{align*}
    y_{i,t}=\sqrt{1-\alpha_0^2}c_i+\alpha_{0} y_{i,t-1}+\gamma_if_t+\sqrt{1-\alpha_0^2}\varepsilon_{i,t}, \quad \varepsilon_{i,t} \sim \mathcal{N}(0,1).
\end{align*}
In the iterative bootstrap, the second stage bootstrap is used to calibrate the nominal level of the first stage bootstrap critical values. We use $B=199$ bootstrap draws in both the first and the second stages of the bootstrap procedure. The results are summarized in Table \ref{tab:MC_IB_a}.

Several observations are worth noting when comparing Table \ref{tab:MC_IB_a} (iterative) to Table \ref{tab:MC_DGP1} (non-iterative). The RDs method attains marginally higher in coverage rates compared to the RDn method. Especially in the small $T$ designs, there is still a clear advantage to the ``sophisticated'' method in terms of coverage. We note that in the large $T$ designs, the iterative methods attain coverage that is slightly below the nominal level. This is in contrast to the non-iterated methods, which do attain coverage at the nominal level when $T$ is large.

When using RDn, the coverage rates of the iterated bootstrap are comparable to the coverage rates achieved with AN in Table \ref{tab:MC_DGP1}. Comparing the coverage rates of the iterated bootstrap to HPJ in Table \ref{tab:MC_DGP1}, the jackknife-correction provides higher coverage when $T$ is small, and coverage at the nominal level when $T \geq 50$. The results of this simulation exercise therefore indicate that the iterated bootstrap from \citet{heller_jochmans_2026_iterated_bootstrap}, in terms of coverage rates, is a competitive alternative to the bias-corrected recursive design bootstrap. Our proposed bootstrap method, however, achieves coverage closer to the nominal level across all designs considered. 

\begin{landscape}
\begin{table}[h]
\begin{center}
\caption{AR(1): Monte Carlo results for $\widehat{\alpha}$ with the Iterative Bootstrap in DGP(1).}
\small{
\begin{tabular}{ll|cc}
\toprule
\multicolumn{4}{c}{Bias}\\
\toprule
 &  & CCEP (RDn) & CCEP (RDs) \\
$N$ & $T$ &  &  \\
\midrule
\multirow[t]{3}{*}{25} & 25 & -0.957 & -0.508 \\
 & 50 & -0.281 & -0.108 \\
 & 100 & -0.082 & -0.019 \\
\cline{1-4}
\multirow[t]{3}{*}{50} & 25 & -1.345 & -0.721 \\
 & 50 & -0.346 & -0.108 \\
 & 100 & -0.113 & -0.028 \\
\cline{1-4}
\multirow[t]{3}{*}{100} & 25 & -1.773 & -0.915 \\
 & 50 & -0.455 & -0.126 \\
 & 100 & -0.135 & -0.022 \\
\cline{1-4}
\bottomrule
\multicolumn{4}{c}{Coverage rates}\\
\toprule
 &  & CCEP (RDn) & CCEP (RDs) \\
$N$ & $T$ &  &  \\
\midrule
\multirow[t]{3}{*}{25} & 25 & 0.888 & 0.924 \\
 & 50 & 0.934 & 0.938 \\
 & 100 & 0.938 & 0.937 \\
\cline{1-4}
\multirow[t]{3}{*}{50} & 25 & 0.834 & 0.907 \\
 & 50 & 0.928 & 0.941 \\
 & 100 & 0.930 & 0.932 \\
\cline{1-4}
\multirow[t]{3}{*}{100} & 25 & 0.723 & 0.882 \\
 & 50 & 0.927 & 0.940 \\
 & 100 & 0.935 & 0.937 \\
\cline{1-4}
\bottomrule
\end{tabular}
\label{tab:MC_IB_a}
}
\end{center}
\footnotesize
\renewcommand{\baselineskip}{11pt}
\textbf{Note:} $(\sqrt{NT})$ Scaled bias and coverage rate based on confidence intervals. From left to right, the shown estimator is conventional CCEP \citet{Pesaran2006}. The label (RDn) refers to the ``naive'' implementation of the proposed recursive design bootstrap. (RDs) refers to the ``sophisticated'' implementation.
\end{table}
\end{landscape}

\subsubsection{Time-series Heteroskedasticity}
We now consider time-series heteroskedasticity, adapting the benchmark AR$(1)$ model from Section \ref{2.2}.
\begin{align*}
    y_{i,t}&=\sqrt{1-\alpha_0^2}c_i+\alpha_{0} y_{i,t-1}+\gamma_if_t+\sqrt{1-\alpha_0^2}\varepsilon_{i,t}\frac{(1+f_t)}{\sqrt{2}}, \quad \varepsilon_{i,t} \sim \mathcal{N}(0,1), \\
    f_{t} &= a_0 f_{t-1} + \sqrt{1-a_0^2}e_{t}, \quad e_t \sim \mathcal{N}(0,1).
\end{align*}
The term $\frac{(1+f_t)}{\sqrt{2}}$ induces the time-series heteroskedasticity. We divide by $\sqrt{2}$ here to normalize the unconditional variance of the $\varepsilon_{i,t}$-term to $1$. 

In this exercise, we investigate whether the considered estimators are sensitive to misspecified time-series heteroskedasticity. We primarily compare the results of Table \ref{tab:MC_H_a} with the baseline results in Table \ref{tab:MC_DGP1}. 

First, for the conventional CCEP estimator, bias increases compared to the baseline in all designs. For the AN method, bias also increases compared to the baseline. On the other hand, with HPJ we observe that the bias decreases. The bias in DVS is substantially increased. This is somewhat expected, as in \citet{DeVos2019} the analytical correction of DVS is derived under homoskedasticity.

As in the previous Monte Carlo results, the RDn bootstrap enables additional bias-correction of the point estimate. In general, this does considerably reduce the bias; however, the remaining bias is larger compared to the baseline results. It remains true that RDn reduces the bias to a lesser extent compared to RDs.

In terms of coverage, the CCEP estimator with CS bootstrap achieves coverage rates that are far below nominal level. With bias-correction tools, coverage does improve, but remains below the nominal level. This result is comparable to the baseline. Whereas the DVS method obtains coverage rates at the nominal level in the baseline results, in the current setting coverage rates fall below the nominal level for all designs considered. 

We next consider the RDn bootstrap. The CCEP estimator attains coverage rates that are slightly lower compared to the baseline. The coverage rates improve as $T$ increases, and are at the nominal level whenever $T=100$. For smaller values of $T$, coverage can be improved by using either HPJ or AN to bias-correct the estimate. The differences in coverage compared to the baseline remain minor when $T \geq 50$. HPJ attains coverage at the nominal level in this setting, whereas AN has coverage slightly below the nominal level. The ``sophisticated'' construction of the factors improves coverage for the CCEP estimator. In contrast to the baseline, the RDs method only marginally improves the coverage rates over the RDn method. 

\begin{landscape}
\begin{table}[h]
\begin{center}
\caption{AR(1): Monte Carlo results for $\widehat{\alpha}$ with time-series heteroskedasticity.}
\small{
\begin{tabular}{ll|cccc|ccc|ccc}
\toprule
\multicolumn{12}{c}{Bias}\\
\toprule
 &  & CCEP (CS) & DVS (CS) & HPJ (CS) & AN (CS) & CCEP (RDn) & HPJ (RDn) & AN (RDn) & CCEP (RDs) & HPJ (RDs) & AN (RDs) \\
$N$ & $T$ &  &  &  &  &  &  &  &  &  &  \\
\midrule
\multirow[t]{3}{*}{25} & 25 & -4.311 & -0.393 & 0.115 & -2.070 & -1.166 & -0.207 & -0.691 & -0.722 & -0.138 & -0.484 \\
 & 50 & -2.814 & -0.372 & 0.448 & -1.024 & -0.345 & -0.094 & -0.229 & -0.163 & -0.121 & -0.169 \\
 & 100 & -1.872 & -0.245 & 0.211 & -0.459 & -0.098 & -0.019 & -0.071 & -0.035 & -0.025 & -0.053 \\
\cline{1-12}
\multirow[t]{3}{*}{50} & 25 & -6.077 & -0.536 & 0.134 & -2.894 & -1.622 & -0.258 & -0.930 & -1.004 & -0.164 & -0.639 \\
 & 50 & -3.889 & -0.426 & 0.759 & -1.384 & -0.411 & -0.097 & -0.263 & -0.172 & -0.153 & -0.179 \\
 & 100 & -2.614 & -0.316 & 0.272 & -0.636 & -0.124 & -0.033 & -0.085 & -0.034 & -0.045 & -0.064 \\
\cline{1-12}
\multirow[t]{3}{*}{100} & 25 & -8.368 & -0.564 & 0.452 & -3.922 & -2.101 & -0.294 & -1.175 & -1.251 & -0.190 & -0.774 \\
 & 50 & -5.486 & -0.616 & 0.966 & -1.928 & -0.561 & -0.165 & -0.332 & -0.215 & -0.236 & -0.214 \\
 & 100 & -3.643 & -0.392 & 0.457 & -0.861 & -0.121 & -0.003 & -0.076 & 0.003 & -0.021 & -0.046 \\
\cline{1-12}
\bottomrule
\multicolumn{12}{c}{Coverage rates}\\
\toprule
 &  & CCEP (CS) & DVS (CS) & HPJ (CS) & AN (CS) & CCEP (RDn) & HPJ (RDn) & AN (RDn) & CCEP (RDs) & HPJ (RDs) & AN (RDs) \\
$N$ & $T$ &  &  &  &  &  &  &  &  &  &  \\
\midrule
\multirow[t]{3}{*}{25} & 25 & 0.100 & 0.745 & 0.656 & 0.681 & 0.760 & 0.874 & 0.870 & 0.864 & 0.877 & 0.855 \\
 & 50 & 0.212 & 0.882 & 0.786 & 0.820 & 0.935 & 0.943 & 0.898 & 0.948 & 0.932 & 0.887 \\
 & 100 & 0.435 & 0.909 & 0.881 & 0.902 & 0.952 & 0.944 & 0.915 & 0.958 & 0.938 & 0.917 \\
\cline{1-12}
\multirow[t]{3}{*}{50} & 25 & 0.011 & 0.717 & 0.536 & 0.474 & 0.640 & 0.813 & 0.864 & 0.791 & 0.822 & 0.846 \\
 & 50 & 0.042 & 0.851 & 0.692 & 0.720 & 0.915 & 0.942 & 0.905 & 0.944 & 0.939 & 0.892 \\
 & 100 & 0.143 & 0.898 & 0.864 & 0.882 & 0.950 & 0.954 & 0.925 & 0.959 & 0.948 & 0.921 \\
\cline{1-12}
\multirow[t]{3}{*}{100} & 25 & 0.000 & 0.665 & 0.421 & 0.262 & 0.534 & 0.752 & 0.851 & 0.730 & 0.779 & 0.831 \\
 & 50 & 0.001 & 0.790 & 0.538 & 0.543 & 0.886 & 0.921 & 0.912 & 0.938 & 0.918 & 0.898 \\
 & 100 & 0.016 & 0.865 & 0.777 & 0.820 & 0.955 & 0.953 & 0.931 & 0.962 & 0.949 & 0.926 \\
\cline{1-12}
\bottomrule
\end{tabular}
\label{tab:MC_H_a}
}
\end{center}
\footnotesize
\renewcommand{\baselineskip}{11pt}
\textbf{Note:} $(\sqrt{NT})$ Scaled bias and coverage rate based on confidence intervals. From left to right, the shown estimators are conventional CCEP \citet{Pesaran2006}, and the bias corrected versions: DVS \citet{DeVos2019}, HPJ \citet{dhaene2015split}, and AN \citet{HahnKursteiner2011BiasReduction}. The label (CS) refers to the cross-sectional bootstrap as in \citet{Kapetanios2008}, while (RDn) refers to the ``naive'' implementation of the proposed recursive design bootstrap. (RDs) refers to the ``sophisticated'' implementation.
\end{table}
\end{landscape}

\clearpage
\subsection{Proofs}\label{4}
\subsubsection{Assumptions for AR(1) and ARX(1) Cases}\label{4.1}
\begin{assumption}
\label{ass::1}(a) The error term $\varepsilon_{i,t}$ is i.i.d. over $i,t$ with $E[\varepsilon_{i,t}]=0$, $E[\varepsilon_{i,t}]=\sigma^2$ and $E[|\varepsilon_{i,t}|^{8}]<\Delta$; (b) The error terms $(\+\nu_t,t=1,\ldots,T)$ are covariance stationary for all $i=1,\ldots,N$; (c) $E[\+\nu_{i,t}]
=\*0_{k\times 1}$, $E[\+\nu_{i,t}\+\nu_{i,t}']=\+\Sigma_{\+\nu}(0)$, $E\left[\left\|\+\nu_{i,t} \right\|^{8}\right]<\Delta$; (d) The sequence $E\left[\+\nu_{i,h}\+\nu'_{i,0}\right]=\+\Sigma_{\+\nu}(h)$ is absolutely summable.
\end{assumption}

\begin{assumption}
\label{ass::2}(a) The factors $(\+\f_t,t=1,\ldots,T)$ are covariance stationary; (b) $E[\+\f_t\+\f_t']=\+\Sigma_{\+\f}(0)\in \mathbb{R}^{R\times R}$ is positive definite, and $E[\left\|\+\f_t \right\|^{8} ]<\Delta$; (c) The sequence $E[\+\f_{h}\+\f_{0}']=\+\Sigma_{\+\f}(h)$ is absolutely summable.
\end{assumption}

\begin{assumption}
\label{ass::3}(a) The loadings $\+\Gamma_{i}\in \mathbb{R}^{R\times K}$; (b) $\mathrm{rk}(\overline{\+\Gamma})=R=K$ as $N\to \infty$; (c) $\frac{1}{N}\sum_{i=1}^N\left\|\+\Gamma_i \right\|^{8}<\Delta$ for all $N$, including $N\to\infty$.
\end{assumption}

\begin{assumption}
\label{ass::4}Sequences $\+\f_t$, $\+\nu_{i,s}$ and $\varepsilon_{j,r}$ are independent for all $i,j,r,t$ and $s$.
\end{assumption}

\begin{assumption}
\label{ass::5}(a) The process $( \+\z_{i,t},t=1,\ldots,T)$ is initialized at an infinite past;  (b) $|\alpha_0|<1$, and the spectral radius of $\+\A_0$ is bounded by 1; (c) $\+\z_{i,0}$ is available for all $i=1,\ldots, N$.
\end{assumption}

\begin{assumption}
\label{ass::6}$NT^{-1}\to \kappa\in (0,\infty)$ as $N,T\to\infty$ jointly.
\end{assumption}
\noindent Assumptions \ref{ass::1} and \ref{ass::2} are standard in the literature of dynamic panel data models. Independence of the error terms is not necessary and is primarily used to simplify the proofs. For example, it can be a martingale difference that is dependent over time, and where dependence is regulated through the summability conditions of its cumulants (see \citealp{gonccalves2015bootstrap}). Unconditional cross-sectional heteroskedasticity is allowed as long as $\max_i \sigma^2_i <\infty$, $\lim_{N\to \infty}\frac{1}{N}\sum_{i=1}^N\sigma_i^2=\sigma^2<\infty$ and $\lim_{N\to \infty}\frac{1}{N}\sum_{i=1}^N\sigma^4_i=\tau<\infty$. In most cases, it is enough to have $4+\delta$ ($\delta>0$) moments that are uniformly bounded, but setting $\delta=4$ (eighth moment) is sufficient to verify the conditions for bootstrap central limit theorem. Assumption \ref{ass::3} implies the loadings are non-random parameters with an informative average as $N\to \infty$, similarly to \cite{Karabiyik2017}. However, they can be random and follow, for example, additive structure, so that $\+\Gamma_i=\+\Gamma+\+\eta_i$, where $E[\+\eta_i]=\*0_{R\times K}$ (i.i.d.), similarly to \cite{Pesaran2006}. Assumption \ref{ass::4} is also standard, while Assumption \ref{ass::5} allows us to conveniently write $y_{i,0}$ (initial value) as a linear process in terms of past shocks and factor realizations, and so we we can recursively re-write $y_{i,t}$ in this fashion for every $i,t$. It imposes stability on $\{\*z_{i,t}\}$, while Assumption \ref{ass::6} declares our asymptotic setting which allows us to explore an asymptotic bias. \\

\noindent When analyzing the results for the AR(1) model (next section), we will specify results assuming that all Assumptions \ref{ass::1}-\ref{ass::6} are satisfied. this simply means that that the corresponding parts associated with $\+\x_{i,t}$ are omitted (i.e. $\+\beta_0=\*0_{k\times 1}$). 
\subsubsection{Discussion of Weights Choice}
There are different possible choices of bootstrap weight $\omega_{i,t}$. \cite{Goncalves2014} or \cite{gonccalves2015bootstrap} specify $\omega_{i,t}$ as $i.i.d.(0,1)$ for each $i$ and $t$. Another possibility found in the literature is to let $\omega_{i,t}=w_i\nu_t$, where $w_i$ is i.i.d. over $i$ and $\nu_t$ is i.i.d. over $t$ and they are mutually independent (see e.g. \citealp{menzel2021bootstrap}, or \citealp{Juodis2020}). Because now the \textit{multiplicative} $\omega_{i,t}$ is not jointly i.i.d., it can capture two-way clustering in the errors. Since, in both cases, $E[\omega_{i,t}]=0$, $E[\omega_{i,t}^2]=1$ and $E[\omega_{i,t}\omega_{j,s}]=0$ when $i\neq j$ even if $t=s$ (or, vice versa), both weighting schemes are \textit{typically} asymptotically equivalent when the distribution of the bootstrap statistic is driven by the second moments. The numerator of the CCEP estimator is an example of this. \\ 
\indent However, in our recursive bootstrap setting, the equivalence breaks down. The multiplicative $\omega_{i,t}$ generates cross sectional dependence under the bootstrap measure, which prohibits the use of standard bootstrap CLT results (see Theorem A.2 in \cite{gonccalves2015bootstrap}). In particular, a typical component analyzed in the upcoming proofs that generates the asymptotic bootstrap distribution is
\begin{align}
    c_{N,T}^*=\frac{1}{\sqrt{NT}}\sum_{i=1}^N\sum_{t=2}^Ta_{i,t-1}^*b_{i,t}^*=\frac{1}{\sqrt{NT}}\sum_{i=1}^N\sum_{t=2}^T\omega_{i,t-1}\omega_{i,t}a_{i,t-1}b_{i,t}.
\end{align}
for some $a_{i,t}$ and $b_{i,t}$ with second moments and $a_{i,t-1}$ that potentially includes lags of $b_{i,t}$. Clearly, under non-multiplicative weights, $|c_{N,T}^*|=O_{P^*}(1)$, which coincides with $O_P(1)$ result in original sample under our assumptions. However, under multiplicative weights, we obtain 
\begin{align}
    c_{N,T}^*=\frac{1}{\sqrt{NT}}\sum_{i=1}^N\sum_{t=2}^Tw_{i}^2\nu_{t-1}\nu_{t}a_{i,t-1}b_{i,t}, 
\end{align}
where the summands are dependent over $i$ (under bootstrap measure) in this case. Note that $w_{i}^2=1$ (with probability 1) under Rademacher weight, which eliminates randomness over $i$ altogether. These theoretical considerations determine our prescription of $\omega_{i,t}\sim i.i.d.(0,1)$ in the main text. For our proof strategy, we analyze our interim results (Lemma 1 and Lemma 2, and Lemma X1 and Lemma X2) under multiplicative weights to show that they hold under a higher degree of dependence. Clearly, they hold under non-multiplicative weights that reduce dependence, thus, they are invariant to the choice of between these two common conventions. Our CLT results (Lemma 3 and Lemma X3) will be based on non-multiplicative weights only. We do not claim that CLT does not apply in the case of multiplicative $\omega_{i,t}$, but recursive bootstrap under cross-section dependence is beyond the scope of this paper.    
\subsubsection{AR(1) Case}\label{4.2}
\paragraph{The Setup}\label{4.2.1}
We focus on a panel AR(1) model with an interactive effect ($R=1$): 
\begin{align}
    y_{i,t}=\alpha_0y_{i,t-1}+\gamma_if_t+\varepsilon_{i,t},
\end{align}
where we remove fixed-effect without loss of generality. We estimate $f_t$ using $\overline{y}_t-\alpha_0\overline{y}_{t-1}=\overline{\gamma}f_t+\overline{\varepsilon}_t$. Let $\+\alpha_0=[1,\alpha_0]'$, thus $f_t=\underbrace{(\overline{y}_t-\alpha_0\overline{y}_{t-1})}_{\widehat{f}_{\+\alpha_0,t}}\overline{\gamma}^{-1}-\overline{\varepsilon}_t\overline{\gamma}^{-1}$. Further, we can write $y_{i,t-1}$ in the reduced form 
\begin{align}
    y_{i,t-1}&=\gamma_i\sum_{j=0}^\infty (\alpha_0)^jf_{t-j-1} + \underbrace{\sum_{j=0}^\infty(\alpha_0)^j\varepsilon_{i,t-j-1}}_{u_{i,t-1}}\notag\\
    &=\gamma_i\overline{\gamma}^{-1}\sum_{j=0}^\infty(\alpha_0)^j\widehat{f}_{\+\alpha_0,t-j-1} - \gamma_i\overline{\gamma}^{-1}\sum_{j=0}^\infty(\alpha_0)^j\overline{\varepsilon}_{t-j-1} + u_{i,t-1}\notag\\
    &=\gamma_i\overline{\gamma}^{-1}\sum_{j=0}^\infty(\alpha_0)^j(\overline{y}_{t-j-1}-\alpha_0\overline{y}_{t-j-2})- \overline{u}_{t-1}\overline{\gamma}^{-1}\gamma_i+u_{i,t-1}\notag\\
    &=\overline{y}_{t-1}\overline{\gamma}^{-1}\gamma_i - \overline{u}_{t-1}\overline{\gamma}^{-1}\gamma_i+u_{i,t-1}
\end{align}
by using the telescoping sum argument. Letting $\widehat{\F}=[\overline{\*y}, \overline{\*y}_{-1}]$ in stacked notation, the CCEP estimator of $\alpha_0$ gives
\begin{align}
    \sqrt{NT}(\widehat{\alpha}_{CCEP}-\alpha_0)&=\left(\frac{1}{NT}\sum_{i=1}^N\*y_{i,-1}'\*M_{\widehat{\*F}}\*y_{i,-1}\right)^{-1}\frac{1}{\sqrt{NT}}\sum_{i=1}^N\*y_{i,-1}'\*M_{\widehat{\*F}}(\*f\gamma_i+\+\varepsilon_i)\notag\\
    &=\left(\frac{1}{NT}\sum_{i=1}^N\*y_{i,-1}'\*M_{\widehat{\*F}}\*y_{i,-1}\right)^{-1}\frac{1}{\sqrt{NT}}\sum_{i=1}^N(\*u_{i,-1}-\overline{\*u}_{-1}\overline{\gamma}^{-1}\gamma_i)'\*M_{\widehat{\*F}}(\+\varepsilon_i-\overline{\+\varepsilon}\overline{\gamma}^{-1}\gamma_i),
\end{align}
where we project out $\overline{\*y}_{-1}\overline{\gamma}^{-1}\gamma_i$. We will reconstruct this setup in bootstrap realm.
\paragraph {Bootstrap DGP: Naive Method}\label{4.2.2}
\noindent In this section, the factor space is re-produced by using the \textit{naive} method. We call the reconstruction method this way, because we simply use two averages $\overline{\*y}$ and $\overline{\*y}_{-1}$ instead of the true factors in the bootstrap DGP. This bypasses the theoretical prescription to use $\overline{\*y}-\alpha_0\overline{\*y}_{-1}$ (where we replace $\alpha_0$ with $\widehat{\alpha}_{CCEP}$). We call the latter the \textit{sophisticated} proxy: $\widehat{\*f}_{\widehat{\+\alpha}}=\overline{\*y}-\widehat{\alpha}_{CCEP}\overline{\*y}_{-1}$, where we let $\widehat{\+\alpha}=[1,-\widehat{\alpha}_{CCEP}]'$. In the naive method, we have $\widehat{\*F}=[\overline{\*y}, \overline{\*y}_{-1}]$. The connection between the two is: 
\begin{align}
    \widehat{\*F}\widehat{\+\alpha}=\overline{\*y}-\widehat{\alpha}_{CCEP}\overline{\*y}_{-1}=\widehat{\*f}_{\widehat{\+\alpha}},
\end{align}
where we indicate that the sophisticated factor proxy is a function of $\widehat{\+\alpha}$. The naive method induces an additional factor in the bootstrap world. To illustrate this point, without loss of generality, we omit the fixed-effect. We generate the bootstrap dataset with 
\begin{align}
    y_{i,t}^*=\widehat{\alpha}_{CCEP}y^*_{i,t-1}+\underbrace{[\overline{\*y}, \overline{\*y}_{-1}]}_{\widehat{\*F}}\widehat{\+\gamma}_i+\varepsilon^*_{i,t},
\end{align}
where $\varepsilon_{i,t}^*=\omega_{i,t}\widehat{\varepsilon}_{i,t}$, such that $\widehat{\varepsilon}_{i,t}=y_{i,t}-\widehat{\alpha}_{CCEP}y_{i,t-1}-\widehat{\+\gamma}_i'\widehat{\*f}_t $. We use weights $\omega_{i,t}$  which are i.i.d. $(0,1)$, as in \cite{gonccalves2015bootstrap} (for instance, Rademacher weights).  Note that here
\begin{align}
    &\widehat{\+\gamma}_i=(\widehat{\*F}'\widehat{\*F})^{-1}\widehat{\*F}'(\*y_i-\widehat{\alpha}_{CCEP}\*y_{i,-1}),\\
    &\overline{\widehat{\+\gamma}}=(\widehat{\*F}'\widehat{\*F})^{-1}\widehat{\*F}'(\overline{\*y}-\widehat{\alpha}_{CCEP}\overline{\*y}_{-1})=(\widehat{\*F}'\widehat{\*F})^{-1}\widehat{\*F}'\widehat{\*F}\widehat{\+\alpha}=\widehat{\+\alpha}.
\end{align}
Clearly, this implies that 
\begin{align}\label{naive_overline_gamma}
    \widehat{\*F}\overline{\widehat{\gamma}}=[\overline{\*y}, \overline{\*y}_{-1}]\widehat{\+\alpha}=\overline{\*y}-\widehat{\alpha}_{CCEP}\overline{\*y}_{-1}=\widehat{\f}_{\widehat{\+\alpha}},
\end{align}
which coincides with the sophisticated factor proxy. Next, we introduce a rotation matrix $\*R=\begin{bmatrix}1 & 0\\
-\widehat{\alpha}_{CCEP} & 1\end{bmatrix}$, which leads to:  
\begin{align}
    \*y_i^*&=\*y^*_{i,-1}\widehat{\alpha}_{CCEP}+[ \overline{\*y}, \overline{\*y}_{-1}]\widehat{\+\gamma}_i+\+\varepsilon_i^*\notag\\
    &= \*y^*_{i,-1}\widehat{\alpha}_{CCEP} +[ \overline{\*y}, \overline{\*y}_{-1}]\*R(\*R^{-1}\widehat{\+\gamma}_i)+\+\varepsilon_i^*\notag\\
    &= \*y^*_{i,-1}\widehat{\alpha}_{CCEP} + [ \overline{\*y}, \overline{\*y}_{-1}]\*R\widetilde{\+\gamma}_i+\+\varepsilon_i^*\notag\\
    &= \*y^*_{i,-1}\widehat{\alpha}_{CCEP} +\widetilde{\gamma}_{1,i}\underbrace{(\overline{\*y}-\widehat{\alpha}_{CCEP}\overline{\*y}_{-1})}_{\widehat{\*f}_{\widehat{\+\alpha}}} + \widetilde{\gamma}_{2,i}\overline{\*y}_{-1}+\+\varepsilon_i^*,
\end{align} 
This clearly demonstrates that naive reconstruction method utilizes the sophisticated factor proxy but induces and additional interactive effect in the bootstrap DGP in $\widetilde{\gamma}_{2,i}\overline{\*y}_{-1}$. Directly from (\ref{naive_overline_gamma}) and calculation of $\overline{\*y}^*$, we know that the average of $\widetilde{\gamma}_{1,i}$ is 1 and the average of $\widetilde{\gamma}_{2,i}$ is zero. By using this, we can solve for the $\widehat{\*f}_{\widehat{\+\alpha}}$ (the true factor in the bootstrap DGP) in terms of the bootstrap primitives: 
\begin{align}
    \widehat{\*f}_{\widehat{\+\alpha}}=\overline{\*y}^*-\overline{\*y}^*_{-1}\widehat{\alpha}_{CCEP}-\overline{\+\varepsilon}^*
\end{align}
 Thus
\begin{align}\label{y_i*}
    \*y_i^*=\*y_{i,-1}^*\widehat{\alpha}_{CCEP}+(\overline{\*y}^*-\overline{\*y}^*_{-1}\widehat{\alpha}_{CCEP})\widetilde{\gamma}_{1,i}-\overline{\+\varepsilon}^*\widetilde{\gamma}_{1,i}+\overline{\*y}_{-1}\widetilde{\gamma}_{2,i}+\+\varepsilon_i^{*}.
\end{align}
Before inserting this into the bootstrap CCEP estimator, we use the choice of $\*y_{i,0}^*=\*y_{i,0}$. Because of this choice, we know that 
\begin{align}
&\overline{y}^*_1=\widehat{\alpha}_{CCEP}\overline{y}_0+\widehat{\+\alpha}'[\overline{y}_1,\overline{y}_0]'+\overline{\varepsilon}_1^*=\widehat{\alpha}_{CCEP}\overline{y}_0+[1, -\widehat{\alpha}_{CCEP}][\overline{y}_1,\overline{y}_0]'+ \overline{\varepsilon}_1^*=\overline{y}_1+\overline{\varepsilon}_1^*\notag\\
&\overline{y}_{2}^*=\widehat{\alpha}_{CCEP}\overline{y}_1^*+\widehat{\+\alpha}'[\overline{y}_2,\overline{y}_1]'+\overline{\varepsilon}_2^*=\widehat{\alpha}_{CCEP}\overline{y}_1^*+[1, -\widehat{\alpha}_{CCEP}][\overline{y}_2,\overline{y}_1]'+ \overline{\varepsilon}_2^*\notag\\
&=\overline{y}_2+\widehat{\alpha}_{CCEP}\overline{y}_1-\widehat{\alpha}_{CCEP}\overline{y}_1+\widehat{\alpha}_{CCEP}\overline{\varepsilon}_1^*+\overline{\varepsilon}_2^*=\overline{y}_2+\widehat{\alpha}_{CCEP}\overline{\varepsilon}_1^*+\overline{\varepsilon}_2^*\notag\\
&\vdots \notag\\
&\overline{y}_t^*=\overline{y}_t+\sum_{j=0}^{t-1}\widehat{\alpha}^j
_{CCEP}\overline{\varepsilon}^*_{t-j}=\overline{y}_t+\overline{u}_t^*
\end{align}
and thus it holds for any $t$, and so $\overline{\*y}_{-1}=\overline{\*y}^*_{-1}-\overline{\*u}_{-1}^*$. Note that different choices are possible, but this allows to simplify algebra. Combining this with (\ref{y_i*}), we obtain
\begin{align}
    \sqrt{NT}(\widehat{\alpha}^*_{CCEP}&-\widehat{\alpha}_{CCEP})=\left( \frac{1}{NT}\sum_{i=1}^N\*y_{i,-1}^{*\prime}\*M_{\widehat{\*F}^*}\*y_{i,-1}^*\right)^{-1}\notag\\
    &\times \left(\frac{1}{\sqrt{NT}}\sum_{i=1}^N\*y_{i,-1}^{*\prime}\*M_{\widehat{\*F}^*}\overline{\*y}_{-1}\widetilde{\gamma}_{2,i}-\frac{1}{\sqrt{NT}}\sum_{i=1}^N\*y_{i,-1}^{*\prime}\*M_{\widehat{\*F}^*}\overline{\+\varepsilon}^*\widetilde{\gamma}_{1,i}+\frac{1}{\sqrt{NT}}\sum_{i=1}^N\*y_{i,-1}^{*\prime}\*M_{\widehat{\*F}^*}\+\varepsilon_i^*\right)\notag\\
    &= \left( \frac{1}{NT}\sum_{i=1}^N\*y_{i,-1}^{*\prime}\*M_{\widehat{\*F}^*}\*y_{i,-1}^*\right)^{-1}\notag\\
    &\times \left(\frac{1}{\sqrt{NT}}\sum_{i=1}^N\*y_{i,-1}^{*\prime}\*M_{\widehat{\*F}^*}\+\varepsilon_i^*-\frac{1}{\sqrt{NT}}\sum_{i=1}^N\*y_{i,-1}^{*\prime}\*M_{\widehat{\*F}^*}\overline{\*u}^*_{-1}\widetilde{\gamma}_{2,i} -\frac{1}{\sqrt{NT}}\sum_{i=1}^N\*y_{i,-1}^{*\prime}\*M_{\widehat{\*F}^*}\overline{\+\varepsilon}^*\widetilde{\gamma}_{1,i}\right)\notag\\
    &=D^{-1}\left(I+II +III \right)
\end{align}
 In what follows, we need to re-write $\*y_{i,-1}^*$. In scalar notation, and using the previous choice of the initial value, we recursively obtain 
\begin{align}
    y_{i,t}^*&=\widehat{\alpha}_{CCEP}^ty_{i,0} +\widehat{\+\gamma}_i'\sum_{j=0}^{t-1}(\widehat{\alpha}_{CCEP})^j[\overline{y}_{t-j},\overline{y}_{t-j-1}]'+\sum_{j=0}^{t-1}\widehat{\alpha}_{CCEP}^j\varepsilon_{i,t-j}^*\notag\\
    &= \widetilde{\gamma}_{1,i}\sum_{j=0}^{t-1}(\widehat{\alpha}_{CCEP})^j(\overline{y}_{t-j}-\widehat{\alpha}_{CCEP}\overline{y}_{t-j-1}) +\widetilde{\gamma}_{2,i}\underbrace{\sum_{j=0}^{t-1}(\widehat{\alpha}_{CCEP})^j\overline{y}_{t-j-1}}_{b_t}+\widehat{\alpha}_{CCEP}^ty_{i,0}+\underbrace{\sum_{j=0}^{t-1}(\widehat{\alpha}_{CCEP})^j\varepsilon_{i,t-j}^*}_{u_{i,t}^*}\notag\\
    &= \widetilde{\gamma}_{1,i}\overline{y}_t+\widehat{\alpha}_{CCEP}^t(y_{i,0}-\widetilde{\gamma}_{1,i}\overline{y}_0)+\widetilde{\gamma}_{2,i}b_t+u_{i,t}^*,
\end{align}
where we used the fact that $\sum_{j=0}^{t-1}(\widehat{\alpha}_{CCEP})^j(\overline{y}_{t-j}-\widehat{\alpha}_{CCEP}\overline{y}_{t-j-1})=\overline{y}_t-\widehat{\alpha}_{CCEP}^t\overline{y}_0$, which comes from the property of telescoping sum. Clearly, $\frac{1}{N}\sum_{i=1}^N \widehat{\alpha}_{CCEP}^t(y_{i,0}-\widetilde{\gamma}_{1,i}\overline{y}_0)=0$ due to $\overline{\widetilde{\gamma}}_1=1$. In stacked notation, we can write $\*y_{i}^*$ and its lag $\*y_{i,-1}^*$ as 
\begin{align}
    &\*y_i^*=\overline{\*y}\widetilde{\gamma}_{1,i}+\widehat{\+\A}_{CCEP} \+\iota_T(y_{i,0}-\overline{y}_0\widetilde{\gamma}_{1,i})+\*b\widetilde{\gamma}_{2,i}+\*u_i^*,\\
    & \*y_{i,-1}^*=\overline{\*y}_{-1}\widetilde{\gamma}_{1,i}+\widehat{\+\A}_{CCEP,-1} \+\iota_T(y_{i,0}-\overline{y}_0\widetilde{\gamma}_{1,i})+\*b_{-1}\widetilde{\gamma}_{2,i}+\*u_{i,-1}^*
\end{align}
where $\widehat{\*A}_{CCEP}=\mathrm{diag}(\widehat{\alpha}_{CCEP}^T, \widehat{\alpha}_{CCEP}^{T-1}, \ldots, \widehat{\alpha}_{CCEP}^2 )$ and $\widehat{\*A}_{CCEP,-1}=\mathrm{diag}(\widehat{\alpha}_{CCEP}^{T-1}, \widehat{\alpha}_{CCEP}^{T-2}, \ldots, \widehat{\alpha}_{CCEP})$, i.e. $\widehat{\*A}_{CCEP}=\widehat{\alpha}_{CCEP}\widehat{\*A}_{CCEP,-1}$. 
\paragraph  {Useful Interim Lemmas}\label{4.2.3}

To demonstrate the order of many upcoming terms, it is useful to prove the following result on residuals and estimated loadings.\\

\noindent \textbf{Lemma A}. \textit{Under Assumptions \ref{ass::1} - \ref{ass::6}, we have as $(N,T)\to \infty$
\begin{enumerate}[(a)]
    \item $\frac{1}{NT}\sum_{i=1}^N\sum_{t=2}^T|\widehat{\varepsilon}_{i,t}|^p=O_P(1)$,
    \item $\frac{1}{N}\sum_{i=1}^N\left\|\widehat{\+\gamma}_i\right\|^p=O_P(1)$.
\end{enumerate}}
\bigskip 

\noindent \textbf{Proof.} (a) We work out the precise expression of the residuals. Notice that using
\begin{equation*}
\*y_{i,-1}=\gamma_{i}\* g_{-1}+\* u_{i,-1},
\end{equation*}
where $g_{t-1}=\sum_{j=0}^\infty(\alpha_0)^jf_{t-j-1}$ and $u_{i,t-1}=\sum_{j=0}^\infty(\alpha_0)^j\varepsilon_{i,t-j-1}$, we can expand
\begin{align}\label{residual_expnasion}
    &\widehat{\+\varepsilon}_i=\*M_{\widehat{\*F}}(\*y_i-\*y_{i,-1}\widehat{\alpha}_{CCEP})\notag\\
    &=\*M_{\widehat{\*F}}(\*f\gamma_i+\*y_{i,-1}\alpha_0+\+\varepsilon_i-\*y_{i,-1}\widehat{\alpha}_{CCEP})\notag\\
    &= \*M_{\widehat{\*F}}(-\overline{\+\varepsilon}\overline{\gamma}^{-1}\gamma_i+[\overline{\*y}-\overline{\*y}_{-1}\alpha_0]\overline{\gamma}^{-1}\gamma_i+(\alpha_0-\widehat{\alpha}_{CCEP})\*g_{-1}\gamma_i+(\alpha_0-\widehat{\alpha}_{CCEP})\*u_{i,-1}+\+\varepsilon_i)\notag\\
    &= \*M_{\widehat{\*F}}(-\overline{\+\varepsilon}\overline{\gamma}^{-1}\gamma_i + \overline{\*y} \overline{\gamma}^{-1}\gamma_i - (\*g_{-1}\overline{\gamma}+\overline{\*u}_{-1})\alpha_0\overline{\gamma}^{-1}\gamma_i +\*g_{-1}\+\gamma_i\alpha_0 - \*g_{-1}\+\gamma_i\widehat{\alpha}_{CCEP}  +(\alpha_0-\widehat{\alpha}_{CCEP})\*u_{i,-1}+\+\varepsilon_i)\notag\\
    &= \*M_{\widehat{\*F}}(-\overline{\+\varepsilon}\overline{\gamma}^{-1}\gamma_i + \overline{\*y}\overline{\gamma}^{-1}\gamma_i - \alpha_0 \overline{\*u}_{-1}\overline{\gamma}^{-1}\gamma_i - \*g_{-1}\+\gamma_i\widehat{\alpha}_{CCEP}+(\alpha_0-\widehat{\alpha}_{CCEP})\*u_{i,-1}+\+\varepsilon_i)\notag\\
    &=\*M_{\widehat{\*F}}(-\overline{\+\varepsilon}\overline{\gamma}^{-1}\gamma_i +[\overline{\*y}-\widehat{\alpha}_{CCEP}\overline{\*y}_{-1}]\overline{\gamma}^{-1}\gamma_i +\widehat{\alpha}_{CCEP}(\*g_{-1}\overline{\gamma}+\overline{\*u}_{-1})\overline{\gamma}^{-1}\gamma_i- \alpha_0 \overline{\*u}_{-1}\overline{\gamma}^{-1}\gamma_i - \*g_{-1}\+\gamma_i\widehat{\alpha}_{CCEP}\notag \\
    &+(\alpha_0-\widehat{\alpha}_{CCEP})\*u_{i,-1}+\+\varepsilon_i )\notag\\
    &=\*M_{\widehat{\*F}}(-\overline{\+\varepsilon}\overline{\gamma}^{-1}\gamma_i +[\overline{\*y}-\widehat{\alpha}_{CCEP}\overline{\*y}_{-1}]\overline{\gamma}^{-1}\gamma_i - (\alpha_0 -\widehat{\alpha}_{CCEP})\overline{\*u}_{-1}\overline{\gamma}^{-1}\gamma_i+(\alpha_0-\widehat{\alpha}_{CCEP})\*u_{i,-1}+\+\varepsilon_i )\notag\\
    &= \*M_{\widehat{\*F}}(-\overline{\+\varepsilon}\overline{\gamma}^{-1}\gamma_i - (\alpha_0 -\widehat{\alpha}_{CCEP})\overline{\*u}_{-1}\overline{\gamma}^{-1}\gamma_i+(\alpha_0-\widehat{\alpha}_{CCEP})\*u_{i,-1}+\+\varepsilon_i ).
\end{align}
By using this, we can define $\+\varepsilon_i^*=\*V_i\widehat{\+\varepsilon}_i$ under multiplicative weights, where  $\V_{i}=w_{i} \mathrm{diag}(\nu_{T},\nu_{T-1},\ldots,\nu_{1})$. Here $\omega_{i}$ are cross-sectional bootstrap weight, while $\nu_{T},\nu_{T-1},\ldots,\nu_{1}$ are time-series bootstrap weights. \\

\noindent We can use (\ref{residual_expnasion}) to obtain the actual residual for each $i,t$:
\begin{align}\label{resid_expansion}
    \widehat{\varepsilon}_{i,t}&=-\left(\overline{\varepsilon}_t -\widehat{\*f}_t'\left(T^{-1}\widehat{\*F}'\widehat{\*F} \right)^{-1}\frac{1}{T}\sum_{s=2}^T\widehat{\*f}_s\overline{\varepsilon}_s \right)\overline{\gamma}^{-1}\gamma_i \notag\\
    &- (NT)^{-1/2}\left(\overline{u}_{t-1}- \widehat{\*f}_t'\left(T^{-1}\widehat{\*F}'\widehat{\*F} \right)^{-1}\frac{1}{T}\sum_{s=2}^T\widehat{\*f}_s\overline{u}_{s-1}\right)\sqrt{NT}(\alpha_0 - \widehat{\alpha}_{CCEP})\overline{\gamma}^{-1}\gamma_i\notag\\
    &+(NT)^{-1/2}\left(u_{i,t-1}- \widehat{\*f}_t'\left(T^{-1}\widehat{\*F}'\widehat{\*F} \right)^{-1}\frac{1}{T}\sum_{s=2}^T\widehat{\*f}_su_{i,s-1}\right)\sqrt{NT}(\alpha_0 - \widehat{\alpha}_{CCEP})\notag\\
    &+ \left(\varepsilon_{i,t}- \widehat{\*f}_t'\left(T^{-1}\widehat{\*F}'\widehat{\*F} \right)^{-1}\frac{1}{T}\sum_{s=2}^T\widehat{\*f}_s\varepsilon_{i,s}\right)=-a_{i,t}-b_{i,t}+c_{i,t}+d_{i,t}. 
\end{align}
Then we use that for any $p\geq 2$ we have $(\sum_{i=1}^nx_i)^p\leq n^{p-1}\sum_{i=1}^nx_i^p$. For our purposes, we will set $p=4+\delta$ for $\delta\in (0, 2]$. Hence, 
\begin{align}\label{residual_moments_start}
    \frac{1}{NT}\sum_{i=1}^N\sum_{t=2}^T|\widehat{\varepsilon}_{i,t}|^p\leq M_0\frac{1}{NT}\sum_{i=1}^N\sum_{t=2}^T(|a_{i,t}|^p+|b_{i,t}|^p+|c_{i,t}|^p+|d_{i,t}|^p)
\end{align}
 for a constant $M_0=4^{3+\delta}$, and by applying the same inequality again in connection to $\frac{1}{N}\sum_{i=1}^N|\gamma_i|^p=O(1)$, we have 
\begin{align}
    \frac{1}{NT}\sum_{i=1}^N\sum_{t=2}^T|a_{i,t}|^p&\leq M_1|\overline{\gamma}^{-1}|^p\frac{1}{N}\sum_{i=1}^N|\gamma_i|^p\frac{1}{T}\sum_{t=2}^T|\overline{\varepsilon}_t|^p\notag\\
    &+M_1|\overline{\gamma}^{-1}|^p\frac{1}{N}\sum_{i=1}^N|\gamma_i|^p\frac{1}{T}\sum_{t=2}^T\left\|\widehat{\*f}_t \right\|^p \left\| \frac{1}{T}\sum_{s=2}^T\widehat{\*f}_s\overline{\varepsilon}_s\right\|^p\left\|\left(T^{-1}\widehat{\*F}'\widehat{\*F} \right)^{-1} \right\|^p=o_P(1).
\end{align}
Here, it is useful to write $\widehat{\*F}=\*Q\overline{\*C}+\overline{\*V}$, and it comes from the discussion on (2.4) and (2.5) in \cite{juodis2021robustness}. Here, $\*q_t=[f_t, \alpha_0(L)f_{t-1}]'=[f_t, g_{t-1}]'$, $\overline{\*v}_t=[\alpha_0(\alpha_0(L)\overline{\varepsilon}_{t-1})+\overline{\varepsilon}_t, \alpha_0(L)\overline{\varepsilon}_{t-1}]'=[\alpha_0 \overline{u}_{t-1}+\overline{\varepsilon}_t, \overline{u}_{t-1}]'$ and $\overline{\*C}=\frac{1}{N}\sum_{i=1}^N\*C_i$, where
\begin{align}
    \*C_i=\begin{bmatrix}
        \gamma_i & 0\\
        \alpha_0 \gamma_i & \gamma_i
    \end{bmatrix},
\end{align}
with rank of 2 as in (2.8) of \cite{juodis2021robustness}. This means that the initial model can be written as a model with 2 factors. Note that 
\begin{align}
    T^{-1}\*Q'\*Q\to _p \+\Sigma_\*q=\begin{bmatrix} \+\Sigma_\*f & \+\Sigma_{\*f\*g}\\
   \+\Sigma_{\*f\*g}' & \+\Sigma_{\*g}
    \end{bmatrix},
\end{align}
which is positive definite under our assumptions, since $\*g_t$ is a covariance stationary process. Then 
\begin{align}
    \frac{1}{T}\sum_{t=2}^T|\overline{\varepsilon}_t|^p=\frac{1}{T}\sum_{t=2}^T\left|\frac{1}{N}\sum_{i=1}^N\varepsilon_{i,t} \right|^p=O_P(N^{-p/2}),
\end{align}
because the moments are bounded uniformly in $t$: $\max_tE\left[\right|\frac{1}{N}\sum_{i=1}^N\varepsilon_{i,t}\left|^2  \right]=\frac{\sigma^2}{N}=O(N^{-1})$. Next, 
\begin{align}
\frac{1}{T}\sum_{t=2}^T\left\|\widehat{\*f}_t \right\|^p=\frac{1}{T}\sum_{t=2}^T\left\|\overline{\*C}'\*q_t+\overline{\*v}_t \right\|^p&\leq 2^{p-1}\left\|\overline{\*C}\right\|^p\frac{1}{T}\sum_{t=2}^T\left\|\*q_t\right\|^p+ 2^{p-1}\frac{1}{T}\sum_{t=2}^T\left\|\overline{\*v}_t \right\|^p\notag\\
&\leq 4^{p-1}\left\|\overline{\*C}\right\|^p\frac{1}{T}\sum_{t=2}^T|f_t|^p+ 4^{p-1}\left\|\overline{\*C}\right\|^p\frac{1}{T}\sum_{t=2}^T|g_{t-1}|^p\notag\\
&+ 4^{p-1}\left\|\overline{\*C}\right\|^p\frac{1}{T}\sum_{t=2}^T\left|\alpha_0\frac{1}{N}\sum_{i=1}^Nu_{i,t-1} +\frac{1}{N}\sum_{i=1}^N\varepsilon_{i,t}\right|^p\notag\\
&+ 4^{p-1}\left\|\overline{\*C}\right\|^p\frac{1}{T}\sum_{t=2}^T\left|\frac{1}{N}\sum_{i=1}^Nu_{i,t-1}\right|^p\notag\\
&=O_P(1),
\end{align}
which holds if $E[|f_{t}|^p]$ is bounded for $p=4+\delta$, which is the case. In relation to linear process $g_{t-1}$, we bound it in the following way. Notice that for a positive constant $K\to \infty$ as $(N,T)\to \infty$, we have 
\begin{align}\label{O_P(1)_def}
    \mathbb{P}\left( \frac{1}{T}\sum_{t=2}^T|g_{t-1}|^p >K\right)\leq K^{-1}\frac{1}{T}\sum_{t=2}^TE[|g_{t-1}|^p]\leq K^{-1}\frac{M}{(1-\alpha_0)^p}<\epsilon
\end{align}
for all $\epsilon>0$ for $K$ sufficiently large, which implies $O_P(1)$. The latter result is derived from Minkowski's inequality: 
\begin{align}\label{minkowski1}
    E[|g_{t-1}|^p]^{1/p}=E\left[\left|\sum_{j=0}^\infty (\alpha_0)^jf_{t-j-1}\right|^p\right]^{1/p}\leq \sum_{j=0}^\infty (\alpha^{0})^jE[|f_{t-j-1}|^p]^{1/p}\leq \frac{M^{1/p}}{1-\alpha_0},
\end{align}
which implies that $E[|g_{t-1}|^p]\leq \frac{M}{(1-\alpha_0)^p}$ for $p=4+\delta$. Clearly,
\begin{align}
    \left|\frac{1}{N}\sum_{i=1}^Nu_{i,t-1} \right|^p=O_P(N^{-p/2})
\end{align}
because it is a CA of cross-sectionally independent linear process. Next, 
\begin{align}
    \left\| \frac{1}{T}\sum_{s=2}^T\widehat{\*f}_s\overline{\varepsilon}_s\right\|^p&\leq \left\|\overline{\*C} \right\|^p2^{p-1}\left\|\frac{1}{T}\sum_{s=2}^T\*q_s\overline{\varepsilon}_s \right\|^p+2^{p-1}\left\|\frac{1}{T}\sum_{s=2}^T\overline{\*v}_s\overline{\varepsilon}_s \right\|^p=(o_P(1))^p=o_P(1),
\end{align}
because 
\begin{align}
    \mathbb{P}\left(\left\| \frac{1}{T}\sum_{s=2}^T\*q_s\overline{\varepsilon}_s \right\|>\epsilon \right)\leq \frac{1}{\epsilon^2}\frac{1}{NT^2}\sum_{s=2}^TE[\left\| \*q_s\right\|^2]E[N\overline{\varepsilon}_s^2]=\sigma^2\frac{1}{NT^2}\sum_{s=2}^TE[\left\| \*q_s\right\|^2]=O((NT)^{-1})
\end{align}
For the second term, we notice that 
\begin{align}
  \left\|\frac{1}{T}\sum_{s=2}^T\overline{\*v}_s\overline{\varepsilon}_s \right\|^p=\left\|\frac{1}{T}\sum_{s=2}^T[a\overline{u}_{s-1}\overline{\varepsilon}_s+\overline{\varepsilon}^2_s, \overline{u}_{s-1}\overline{\varepsilon}_s]' \right\|^p&\leq \left( \left|\frac{1}{NT}\sum_{s=2}^TN(\overline{\varepsilon}_s^2+\alpha_0\overline{u}_{s-1}\overline{\varepsilon}_s) \right| + \left| \frac{1}{NT}\sum_{s=2}^TN\overline{u}_{s-1}\overline{\varepsilon}_s\right|\right)^p\notag\\
  &\leq 3^{p-1}\left|\frac{1}{NT}\sum_{s=2}^TN\overline{\varepsilon}_s ^2\right|^p+3^{p-1}(|\alpha_0|^p+1)\left| \frac{1}{NT}\sum_{s=2}^TN\overline{u}_{s-1}\overline{\varepsilon}_s\right|^p \notag\\
  &=(o_P(1))^p=o_P(1).
\end{align}
The result comes because $\varepsilon_{i,s}$ is independent from $u_{i,s-1}$ and hence $\{u_{i,t-1}\varepsilon_{i,t} \}$ is and MDS:
\begin{align}\label{MDS_term}
\mathbb{P}\left( \left| \frac{1}{NT}\sum_{s=2}^TN\overline{u}_{s-1}\overline{\varepsilon}_s\right|>\epsilon\right)&\leq \frac{1}{\epsilon^2}\frac{1}{N^2T^2}\sum_{s=2}^TE\left[\sqrt{N}\overline{u}_{s-1}\right]^2E\left[\sqrt{N}\overline{\varepsilon}_s\right]^2\notag\\
&=\sigma^2\frac{1}{\epsilon^2}\frac{1}{N^2T^2}\sum_{s=2}^T\frac{1}{N}\sum_{i=1}^NE[u_{i,s-1}^2] \notag\\
&=\sigma^2\frac{1}{\epsilon^2}\frac{1}{N^2T^2}\sum_{s=2}^T\frac{1}{N}\sum_{i=1}^N\left(\sum_{j=0}^\infty\sum_{k=0}^\infty(\alpha_0)^{j+k}E[\varepsilon_{i,t-j-1}\varepsilon_{i,t-k-1}]\right)\notag\\
&=\sigma^4\frac{1}{\epsilon^2}\frac{1}{N^2T^2}\sum_{s=2}^T\sum_{j=0}^\infty(\alpha_0)^{2j}=O(N^{-2}T^{-1})
\end{align}
Overall, this is sufficient to demonstrate that
$ \frac{1}{NT}\sum_{t=2}^T\sum_{i=1}^Na_{i,t}^p=o_P(1)$. 
The next component behaves similarly:
\begin{align}
    \frac{1}{NT}&\sum_{i=1}^N\sum_{t=2}^T|b_{i,t}|^p\leq M_2 (NT)^{-p/2}\left\| \sqrt{NT}(\alpha_0 - \widehat{\alpha}_{CCEP})\right\|^p|\overline{\gamma}^{-1}|^p\frac{1}{N}\sum_{i=1}^N|\gamma_i|^p \frac{1}{T}\sum_{t=2}^T|\overline{u}_{t-1}|^p\notag\\
    &+ M_2(NT)^{-p/2}\left\| \sqrt{NT}(\alpha_0 - \widehat{\alpha}_{CCEP})\right\|^p|\overline{\gamma}^{-1}|^p\frac{1}{N}\sum_{i=1}^N|\gamma_i|^p\frac{1}{T}\sum_{t=2}^T\left\|\widehat{\*f}_t \right\|^p \left\| \frac{1}{T}\sum_{s=2}^T\widehat{\*f}_s\overline{u}_{s-1}\right\|^p\left\|\left(T^{-1}\widehat{\*F}'\widehat{\*F} \right)^{-1} \right\|^p\notag\\
    &=o_P(1),
\end{align}
where similarly to the case of $a_{i,t}^p$, we used 
\begin{align}\label{double_sum_component}
     \left\| \frac{1}{T}\sum_{s=2}^T\widehat{\*f}_s\overline{u}_{s-1}\right\|^p&=\left( \left\|\overline{\*C}\right\|\left\|\frac{1}{T}\sum_{
     s=2}^T\*q_s\overline{u}_{s-1}\right\|+\left\|\frac{1}{T}\sum_{s=2}^T\overline{\*v}_s\overline{u}_{s-1} \right\|\right)^p\notag\\
     &\leq 2^{p-1}\left\| \overline{\*C}\right\|^p\left\|\frac{1}{T}\sum_{
     s=2}^T\*q_s\overline{u}_{s-1}\right\|^p + 2^{p-1}\left\| \frac{1}{T}\sum_{s=2}^T\overline{\*v}_s\overline{u}_{s-1}\right\|^p\notag\\
     &= 2^{p-1}\left\| \overline{\*C}\right\|^p\left\|\frac{1}{T}\sum_{
     s=2}^T\*q_s\overline{u}_{s-1}\right\|^p +4^{p-1}\left|\frac{1}{NT}\sum_{s=2}^TN(\overline{u}_{s-1}\overline{\varepsilon}_s+\alpha_0 \overline{u}_{s-1}^2) \right|^p \notag\\
     &+4^{p-1}\left|\frac{1}{NT}\sum_{s=2}^TN\overline{u}_{s-1}^2 \right|^p \notag\\
     &=(o_P(1))^p=o_P(1).
\end{align}
Here, it is sufficient to check the first component:
\begin{align}
    \mathbb{P}\left(\left\|\frac{1}{T}\sum_{
     s=2}^T\*q_s\overline{u}_{s-1} \right\|>\epsilon\right)&\leq \frac{1}{\epsilon^2}E\left[\left\|\frac{1}{T}\sum_{s=2}^T\*q_s\overline{u}_{s-1}\right\|^2 \right]\notag\\
    &=\frac{1}{\epsilon^2}E\left[\left\|\sum_{p=0}^\infty (\alpha_0)^p\frac{1}{T}\sum_{s=2}^T\*q_s\overline{\varepsilon}_{s-p-1} \right\|^2\right]\notag\\
    &=\frac{1}{\epsilon^2}\sum_{p=0}^\infty\sum_{q=0}^\infty(\alpha_0)^{p+q}\frac{1}{T^2}\sum_{s=2}^T\sum_{r=2}^TE[\*q_s'\*q_r]E\left[\overline{\varepsilon}_{s-p-1}\overline{\varepsilon}_{r-q-1} \right]\notag\\
    &= \frac{1}{N\epsilon^2}\sum_{p=0}^\infty\sum_{q=0}^\infty(\alpha_0)^{p+q}\frac{1}{T^2}\sum_{s=2}^T\sum_{r=2}^TE[\*q_s'\*q_r]\left(\frac{1}{N}\sum_{i=1}^N\sum_{j=1}^NE[\varepsilon_{i,s-p-1}\varepsilon_{j,r-q-1}] \right)\notag\\
    &=\frac{1}{N\epsilon^2}\sum_{p=0}^\infty\sum_{q=0}^\infty(\alpha_0)^{p+q}\frac{1}{T^2}\sum_{s=2}^T\sum_{r=2}^TE[\*q_s'\*q_r]\left(\frac{1}{N}\sum_{i=1}^NE[\varepsilon_{i,s-p-1}\varepsilon_{i,r-q-1}] \right)\notag\\
    &=\sigma^2\frac{1}{N\epsilon^2}\sum_{p=0}^\infty\sum_{q=0}^\infty(\alpha_0)^{p+q}\frac{1}{T^2}\sum_{s=2}^T\sum_{r=2}^TE[\*q_s'\*q_r]\times \mathbb{1}\{s-p=r-q \}\notag\\
    &=\sigma^2\frac{1}{NT\epsilon^2}\sum_{p=0}^\infty\sum_{q=0}^\infty(\alpha_0)^{p+q}\frac{1}{T}\sum_{s=2}^TE[\*q_s'\*q_{s-p+q}]=O((NT)^{-1})
\end{align}
for $E[\|\*q_t\|^2]<C$. The next term gives
\begin{align}
    \frac{1}{NT}&\sum_{i=1}^N\sum_{t=2}^T|c_{i,t}|^p\leq M_3 (NT)^{-p/2}\left\| \sqrt{NT}(\alpha_0 - \widehat{\alpha}_{CCEP})\right\|^p\frac{1}{NT}\sum_{i=1}^N\sum_{t=2}^T|u_{i,t-1}|^p\notag\\
    &+  M_3 (NT)^{-p/2}\left\| \sqrt{NT}(\alpha_0 - \widehat{\alpha}_{CCEP})\right\|^p\frac{1}{T}\sum_{t=2}^T\left\| \widehat{\*f}_t\right\|^p \frac{1}{N}\sum_{i=1}^N\left\| \frac{1}{T}\sum_{s=2}^T\widehat{\*f}_su_{i,s-1}\right\|^p\left\|\left(T^{-1}\widehat{\*F}'\widehat{\*F} \right)^{-1} \right\|^p\notag\\
    &=o_P(1),
\end{align}
where we utilized 
\begin{align}\label{hatf_u-1}
    \frac{1}{N}\sum_{i=1}^N\left\| \frac{1}{T}\sum_{s=2}^T\widehat{\*f}_su_{i,s-1}\right\|^p&\leq 2^{p-1}\left\| \overline{\*C}\right\|^p\frac{1}{N}\sum_{i=1}^N\left\| \frac{1}{T}\sum_{s=2}^T\*q_su_{i,s-1}\right\|^p +2^{p-1}\frac{1}{N}\sum_{i=1}^N\left\| \frac{1}{T}\sum_{s=2}^T\overline{\*v}_su_{i,s-1}\right\|^p\notag\\
    &\leq 2^{p-1}\left\| \overline{\*C}\right\|^p\frac{1}{N}\sum_{i=1}^N\left\| \frac{1}{T}\sum_{s=2}^T\*q_su_{i,s-1}\right\|^p + 4^{p-1}\frac{1}{N}\sum_{i=1}^N\left|\frac{1}{T}\sum_{s=2}^T(\overline{\varepsilon}_su_{i,s-1}+\alpha_0\overline{u}_{s-1}u_{i,s-1}) \right|^p\notag\\
    &+4^{p-1}\frac{1}{N}\sum_{i=1}^N\left|\frac{1}{T}\sum_{s=2}^T\overline{u}_{s-1}u_{i,s-1} \right|^p=o_P(1),
\end{align}
where we need to check that the moments of the inner sum are uniformly bounded in $i$. Similarly, to (\ref{double_sum_component}), we get 
\begin{align}
    \max_iE\left[\left\|\frac{1}{\sqrt{T}}\sum_{s=2}^T\*q_su_{i,s-1}\right\|^2 \right]
&=\sigma^2\sum_{p=0}^\infty\sum_{q=0}^\infty(\alpha_0)^{p+q}\frac{1}{T}\sum_{s=2}^TE[\*q_s'\*q_{s-p+q}]=O(1)
\end{align}
for $E[\|\*q_t\|^2]<C$ and so $\frac{1}{N}\sum_{i=1}^N\left\| \frac{1}{T}\sum_{s=2}^T\*q_su_{i,s-1}\right\|^p=O_P(T^{-p/2})$. The second term can be shown to be negligible, as well, because the summands are MDS, similarly to (\ref{MDS_term}). Lastly, $\frac{1}{T}\sum_{s=2}^T\overline{u}_{s-1}u_{i,s-1}=\frac{1}{NT}\sum_{s=2}^Tu_{i,s-1}^2+\frac{1}{NT}\sum_{s=2}^T\sum_{j\neq i}^Nu_{j,s-1}u_{i,s-1}$, so that 
\begin{align}
    \max_i \frac{1}{T}\sum_{s=2}^TE[u_{i,s-1}^2]=\frac{\sigma^2}{T}\sum_{s=2}^T\sum_{j=0}^\infty(\alpha_0)^{2j}=O(1),
\end{align}
and
\begin{align}
    \max_i E\left[\left|\frac{1}{\sqrt{NT}}\sum_{s=2}^T\sum_{j\neq i}^Nu_{j,s-1}u_{i,s-1} \right|^2\right]&=\max_i\frac{1}{NT}\sum_{s=2}^T\sum_{r=2}^T\sum_{j\neq i}^N\sum_{k\neq i}^NE[u_{j,s-1}u_{k,r-1}u_{i,s-1}u_{i,r-1}]\notag\\
    &=\max_i\frac{1}{NT}\sum_{j\neq i}^N\sum_{s=2}^T\sum_{r=2}^TE[u_{j,s-1}u_{j,r-1}]E[u_{i,s-1}u_{i,r-1}]\notag\\
    &\leq \max_i\frac{1}{NT}\sum_{j\neq i}^N\sum_{s=2}^T\sum_{r=2}^T|E[u_{j,s-1}u_{j,r-1}]||E[u_{i,s-1}u_{i,r-1}]|\notag\\
    &\leq \max_{r,s}|E[u_{i,s-1}u_{i,r-1}]|\frac{1}{NT}\sum_{j\neq i}^N\sum_{s=2}^T\sum_{r=2}^T|E[u_{j,s-1}u_{j,r-1}]|=O(1),
\end{align}
Because the expectation does not depend on $i$, the covariances of the linear process are summable, and individual covariances are bounded. This implies that $\frac{1}{N}\sum_{i=1}^N\left|\frac{1}{T}\sum_{s=2}^T\overline{u}_{s-1}u_{i,s-1} \right|^p=O_P(N^{-p})+O_P((NT)^{-p/2})$.
Finally, 
\begin{align}
    \frac{1}{NT}\sum_{i=1}^N\sum_{t=2}^T|d_{i,t}|^p\leq M_4\underbrace{\frac{1}{NT}\sum_{i=1}^N\sum_{t=2}^T|\varepsilon_{i,t}|^p}_{O_P(1)}+M_4\frac{1}{T}\sum_{t=2}^T\left\| \widehat{\*f}_t\right\|^p \frac{1}{N}\sum_{i=1}^N\left\| \frac{1}{T}\sum_{s=2}^T\widehat{\*f}_s\varepsilon_{i,s}\right\|^p\left\|\left(T^{-1}\widehat{\*F}'\widehat{\*F} \right)^{-1} \right\|^p=O_P(1) 
\end{align}
By exactly the same arguments as the previous terms. \\

\noindent (b) By using (\ref{residual_expnasion}) in the formula for loadings, we obtain
\begin{align}\label{hat_gamma}
    \widehat{\+\gamma}_i&=(\widehat{\*F}'\widehat{\F})^{-1}\widehat{\*F'}(-\overline{\+\varepsilon}\overline{\gamma}^{-1}\gamma_i +[\overline{\*y}-\widehat{\alpha}_{CCEP}\overline{\*y}_{-1}]\overline{\gamma}^{-1}\gamma_i - (\alpha_0 -\widehat{\alpha}_{CCEP})\overline{\*u}_{-1}\overline{\gamma}^{-1}\gamma_i+(\alpha_0-\widehat{\alpha}_{CCEP})\*u_{i,-1}+\+\varepsilon_i )\notag\\
    &= (\widehat{\*F}'\widehat{\F})^{-1}\widehat{\*F'}(-\overline{\+\varepsilon}\overline{\gamma}^{-1}\gamma_i +\widehat{\*F}'[1,-\widehat{\alpha}_{CCEP}]'\overline{\gamma}^{-1}\gamma_i - (\alpha_0 -\widehat{\alpha}_{CCEP})\overline{\*u}_{-1}\overline{\gamma}^{-1}\gamma_i+(\alpha_0-\widehat{\alpha}_{CCEP})\*u_{i,-1}+\+\varepsilon_i )\notag\\
    &= [1,-\widehat{\alpha}_{CCEP}]'\overline{\gamma}^{-1}\gamma_i + (T^{-1}\widehat{\*F}'\widehat{\F})^{-1}T^{-1}\widehat{\*F}'(-\overline{\+\varepsilon}\overline{\gamma}^{-1}\gamma_i  - (\alpha_0 -\widehat{\alpha}_{CCEP})\overline{\*u}_{-1}\overline{\gamma}^{-1}\gamma_i+(\alpha_0-\widehat{\alpha}_{CCEP})\*u_{i,-1}+\+\varepsilon_i )\notag\\
    &= [1,-\alpha_0]'\overline{\gamma}^{-1}\gamma_i + [0, (\alpha_0-\widehat{\alpha}_{CCEP})]'\overline{\gamma}^{-1}\gamma_i\notag\\
    &+(T^{-1}\widehat{\*F}'\widehat{\F})^{-1}T^{-1}\widehat{\*F}'(-\overline{\+\varepsilon}\overline{\gamma}^{-1}\gamma_i  - (\alpha_0 -\widehat{\alpha}_{CCEP})\overline{\*u}_{-1}\overline{\gamma}^{-1}\gamma_i+(\alpha_0-\widehat{\alpha}_{CCEP})\*u_{i,-1}+\+\varepsilon_i ),
\end{align}
which means that 
\begin{align}
    \frac{1}{N}\sum_{i=1}^N\left\|\widehat{\+\gamma}_i \right\|^p&\leq\ C_1|[1,-\alpha_0]'\overline{\gamma}^{-1} |^p\frac{1}{N}\sum_{i=1}^N|\gamma_i |^p\notag\\
    &+C_2(NT)^{-p/2}\left\|\sqrt{NT}(\alpha_0 -\widehat{\alpha}_{CCEP}) \right\|^p|\overline{\gamma}^{-1}|^p\frac{1}{N}\sum_{i=1}^N|\gamma_i|^p \notag\\
    &+C_3\left\|(T^{-1}\widehat{\*F}'\widehat{\F})^{-1}\right\|^p\left\| T^{-1}\widehat{\*F}'\overline{\+\varepsilon}\overline{\gamma}^{-1} \right\|^p\frac{1}{N}\sum_{i=1}^N| \gamma_i|^p \notag\\
    &+C_4(NT)^{-p/2}\left\|(T^{-1}\widehat{\*F}'\widehat{\F})^{-1}\right\|^p\left\|\sqrt{NT}(\alpha_0 -\widehat{\alpha}_{CCEP}) \right\|^p\left\|T^{-1}\widehat{\*F}'\overline{\*u}_{-1} \overline{\gamma}^{-1}\right\|^p\frac{1}{N}\sum_{i=1}^N| \gamma_i|^p\notag\\
    &+C_5(NT)^{-p/2}\left\|(T^{-1}\widehat{\*F}'\widehat{\F})^{-1}\right\|^p\left\|\sqrt{NT}(\alpha_0 -\widehat{\alpha}_{CCEP}) \right\|^p\frac{1}{N}\sum_{i=1}^N\left\|T^{-1}\widehat{\*F}'\*u_{i,-1} \right\|^p\notag\\
    &+C_4\left\|(T^{-1}\widehat{\*F}'\widehat{\F})^{-1}\right\|^p\frac{1}{N}\sum_{i=1}^N\left\| T^{-1}\widehat{\*F}'\+\varepsilon_i\right\|^{p}=O_P(1).
\end{align}
which is given by the exact same argument as part a).
\paragraph {Expansions and Rates of the Numerator }\label{4.2.4}
\noindent 
\textbf{Lemma 1.} \textit{Under Assumptions \ref{ass::1} - \ref{ass::6}, and let the weights be either multiplicative or non-multiplicative. Then as $(N,T)\to \infty$ with $NT^{-1}\to \kappa\in (0, \infty)$, we have
\begin{enumerate}[a) ]
    \item $I=\frac{1}{\sqrt{NT}}\sum_{t=2}^T\sum_{i=1}^Nu^*_{i,t-1}\varepsilon_{i,t}^*-\frac{1}{\sqrt{NT}}\sum_{i=1}^N\*u_{i,-1}^{*\prime}\*P_{\widehat{\*F}^*}\+\varepsilon_i^*+o_{P^*}(1)$. 
    \item $|II|=o_{P^*}(1)$. 
    \item $|III|=o_{P^*}(1)$.
\end{enumerate}
}
\noindent \textbf{Proof.} \textbf{a)} We start our analysis with the numerator. In particular, 
\begin{align}
    I=\frac{1}{\sqrt{NT}}\sum_{i=1}^N\*y_{i,-1}^{*\prime}\*M_{\widehat{\*F}^*}\+\varepsilon_i^*&= \frac{1}{\sqrt{NT}}\sum_{i=1}^N\widetilde{\gamma}_{1,i}\overline{\*y}_{-1}'\*M_{\widehat{\*F}^*}\+\varepsilon_i^*\notag\\
    &+\frac{1}{\sqrt{NT}}\sum_{i=1}^N(y_{i,0}-\overline{y}_0\widetilde{\gamma}_{1,i})\+\iota_T'\widehat{\+\A}_{CCEP,-1}'\*M_{\widehat{\*F}^*}\+\varepsilon_i^*\notag\\
    &+\frac{1}{\sqrt{NT}}\sum_{i=1}^N\widetilde{\gamma}_{2,i}\*b_{-1}'\*M_{\widehat{\*F}^*}\+\varepsilon_i^*\notag\\
    &+ \frac{1}{\sqrt{NT}}\sum_{i=1}^N\*u_{i,-1}^{*\prime}\*M_{\widehat{\*F}^*}\+\varepsilon_i^*\notag\\
    &=I_a+I_b+I_c+I_d.
\end{align}
Starting with $I_a$, we can use the fact that  $\overline{\*y}_{-1}=\overline{\*y}^*_{-1}-\overline{\*u}^*_{-1}$, which means that 
\begin{align}
    I_a=\frac{1}{\sqrt{NT}}\sum_{i=1}^N\widetilde{\gamma}_{1,i}\overline{\*y}_{-1}'\*M_{\widehat{\*F}^*}\+\varepsilon_i^*&=-\frac{1}{\sqrt{NT}}\sum_{i=1}^N\widetilde{\gamma}_{1,i}\overline{\*u}^{*\prime}_{-1}\*M_{\widehat{\*F}^*}\+\varepsilon_i^*\notag\\
    &=\frac{1}{\sqrt{NT}}\sum_{i=1}^N\widetilde{\gamma}_{1,i}\overline{\*u}^{*\prime}_{-1}\*P_{\widehat{\*F}^*}\+\varepsilon_i^*-\frac{1}{\sqrt{NT}}\sum_{i=1}^N\widetilde{\gamma}_{1,i}\overline{\*u}^{*\prime}_{-1}\+\varepsilon_i^*=I_{a,1} - I_{a,2},
\end{align}
where 
\begin{align}\label{I_a2}
    I_{a,2}=\frac{1}{\sqrt{NT}}\sum_{i=1}^N\widetilde{\gamma}_{1,i}\left(\sum_{t=2}^{T}\overline{u}^*_{t-1}\varepsilon_{i,t}^* \right)=\frac{1}{N}\sum_{i=1}^N\widetilde{\gamma}_{1,i}\left(\frac{1}{\sqrt{T}}\sum_{t=2}^{T}\sqrt{N}\overline{u}^*_{t-1}\varepsilon_{i,t}^*\right)=O_{P^*}(N^{-1/2}).
\end{align}
This follows from $\overline{u}_{t-1}^*\varepsilon_{i,t}^*=\frac{1}{N}\sum_{j=1}^N u_{j,t-1}^*\varepsilon_{i,t}^*=N^{-1}u_{i,t-1}^*\varepsilon^*_{i,t}+\frac{1}{N}\sum_{j\neq i}^Nu_{j,t-1}^*\varepsilon_{i,t}^*$.
 In fact, with respect to bootstrap measure, $\{ u_{i,t-1}^*\varepsilon^*_{i,t}\}$ is an MDS in both cases, which means that $\left| \frac{1}{\sqrt{T}}\sum_{t=2}^{T}u^*_{i,t-1}\varepsilon_{i,t}^*\right|=O_{P^*}(1)$. We can further split $I_{a,2}$ in the following way: 
\begin{align}\label{I_a2_exp}
    I_{a,2}=\frac{1}{\sqrt{N}}\sum_{i=1}^N\widetilde{\gamma}_{1,i}\frac{1}{N}\sum_{j=1}^N\left(\frac{1}{\sqrt{T}}\sum_{t=2}^{T}u^*_{j,t-1}\varepsilon_{i,t}^* \right)&=\frac{1}{N\sqrt{N}}\sum_{i=1}^N\widetilde{\gamma}_{1,i}\left(\frac{1}{\sqrt{T}}\sum_{t=2}^{T}u^*_{i,t-1}\varepsilon_{i,t}^* \right)\notag\\
    &+\frac{1}{\sqrt{N}}\sum_{i=1}^N\widetilde{\gamma}_{1,i}\frac{1}{N}\sum_{j\neq i}^N\left(\frac{1}{\sqrt{T}}\sum_{t=2}^{T}u^*_{j,t-1}\varepsilon_{i,t}^* \right)\notag\\
    &=I_{a,2,1}+I_{a,2,2}.
\end{align}
To analyze the components above, let us define
\begin{align}
    u_{i,t}^{**}=\begin{cases}
        w_i^{-1}u_{i,t}^*=\sum_{l=0}^{t-1}\widehat{\alpha}_{CCEP}^l\nu_{t-l}\widehat{\varepsilon}_{i,t-l} \hspace{1mm} \text{under multiplicative weights},\\
        \sum_{l=0}^{t-1}\widehat{\alpha}_{CCEP}^l\omega_{i,t-l}\widehat{\varepsilon}_{i,t-l}=u^{*}_{i,t}\hspace{1mm} \text{under non-multiplicative weights}.
    \end{cases} 
\end{align}
 This implies that 
 \begin{align}
     E^*\left[u_{i,t-1}^*\varepsilon^*_{i,t}u_{j,t-1}^*\varepsilon^*_{j,t} \right]=\begin{cases}
         \underbrace{E^*[w_i^2w_j^2\nu_t^2]}_{=1}\widehat{\varepsilon}_{i,t}\widehat{\varepsilon}_{j,t}E^*[u_{i,t-1}^{**}u_{j,t-1}^{**}] \hspace{1mm} \text{for $i\neq j$} \hspace{1mm} \text{under multiplicative weights},\\
         0 \hspace{1mm}\text{for $i\neq j$} \hspace{1mm} \text{under non-multiplicative weights},
     \end{cases}
 \end{align}
 which immediately reveals that multiplicative weights induce cross-section dependence in such cases. We proceed with $u^{**}_{i,t}$ under multiplicative weights to demonstrate stronger results, which means that the rates of terms we will analyze below will not get worse under non-multiplicative weights. Let $\widehat{b}_{i,t-1}\equiv E^*[(u^{**}_{i,t-1})^{2}]=\sum_{l=0}^{t-2}\widehat{\alpha}_{CCEP}^{2l}(\widehat{\varepsilon}_{i,t-l-1})^{2}$, which is an important quantity in the upcoming steps. Notice how
\begin{align}
    \frac{1}{NT}\sum_{t=2}^T\sum_{i=1}^N\widehat{b}_{i,t-1}^2&=\frac{1}{NT}\sum_{t=2}^T\sum_{i=1}^N\sum_{l=0}^{t-2}\sum_{k=0}^{t-2}\widehat{\alpha}_{CCEP}^{2(l+k)}\widehat{\varepsilon}^2_{i,t-l-1}\widehat{\varepsilon}_{i,t-k-1}^2\notag\\
    &=\sum_{l=0}^{T-2}\sum_{k=0}^{T-2}\widehat{\alpha}_{CCEP}^{2(l+k)}\frac{1}{N}\sum_{i=1}^N\frac{1}{T}\sum_{t=\max\{l,k \}+2}^T\widehat{\varepsilon}^2_{i,t-l-1}\widehat{\varepsilon}_{i,t-k-1}^2\notag\\
    &\leq \sum_{l=0}^{T-2}\sum_{k=0}^{T-2}\widehat{\alpha}_{CCEP}^{2(l+k)}\frac{1}{N}\sum_{i=1}^N\left(\frac{1}{T}\sum_{t=\max\{l,k \}+2}^T\widehat{\varepsilon}^4_{i,t-l-1} \right)^{1/2}\left( \frac{1}{T}\sum_{t=\max\{l,k \}+2}^T\widehat{\varepsilon}^4_{i,t-k-1} \right)^{1/2}\notag\\
    &\leq \sum_{l=0}^{T-2}\sum_{k=0}^{T-2}\widehat{\alpha}_{CCEP}^{2(l+k)}\frac{1}{NT}\sum_{i=1}^N\sum_{t=1}^T\widehat{\varepsilon}^4_{i,t}\notag\\
    &=\left(\frac{1-\widehat{\alpha}_{CCEP}^{2(T-1)}}{1-\widehat{\alpha}_{CCEP}^{2}} \right)^2\frac{1}{NT}\sum_{i=1}^N\sum_{t=1}^T\widehat{\varepsilon}^4_{i,t}\notag\\
    &=O_P(1),
\end{align}
due to the consistency of $\widehat{\alpha}_{CCEP}$. Here we used the fact that necessarily $\frac{1}{T}\sum_{t=\max\{l,k \}+2}^T\widehat{\varepsilon}^4_{i,t-k-1}\leq \frac{1}{T}\sum_{t=1}^T\widehat{\varepsilon}^4_{i,t}$.
Then, using $E^*[w_i^2w_j^2\nu_t^2]=E^*[w_i^2w_j^2]E^*[\nu_t^2]=E^*[w_i^2w_j^2]$, the rate of $I_{a,2,1}$ is given by: 
\begin{align}
    \mathbb{P}^*(|I_{a,2,1}|>\epsilon)&\leq \frac{1}{\epsilon^2}E^*\left[\left|\frac{1}{N\sqrt{N}}\sum_{i=1}^N\widetilde{\gamma}_{1,i}\left(\frac{1}{\sqrt{T}}\sum_{t=2}^{T}u^*_{i,t-1}\varepsilon_{i,t}^* \right) \right|^2\right]\notag\\
    &=\frac{1}{\epsilon^2}\frac{1}{N}\frac{1}{N^2}\sum_{i=1}^N\sum_{j=1}^N\widetilde{\gamma}_{1,i}\widetilde{\gamma}_{1,j}\frac{1}{T}\sum_{t=2}^TE^*[u^*_{i,t-1}\varepsilon_{i,t}^*u_{j,t-1}^*\varepsilon_{j,t}^*]\notag\\
    &= \frac{1}{\epsilon^2}\frac{1}{N}\frac{1}{N^2}\sum_{i=1}^N\sum_{j=1}^N\widetilde{\gamma}_{1,i}\widetilde{\gamma}_{1,j}\frac{1}{T}\sum_{t=2}^TE^*[w_i^2w_j^2\nu_t^2]\widehat{\varepsilon}_{i,t}\widehat{\varepsilon}_{j,t}E^*[u_{i,t-1}^{**}u_{j,t-1}^{**}]\notag\\
    &=\frac{1}{\epsilon^2}\frac{1}{N}\frac{1}{N^2}\sum_{i=1}^N\sum_{j=1}^N\widetilde{\gamma}_{1,i}\widetilde{\gamma}_{1,j}E^*[w_i^2w_j^2]\frac{1}{T}\sum_{t=2}^T\widehat{\varepsilon}_{i,t}\widehat{\varepsilon}_{j,t}\notag\\
    &\times E^*\left[\left(\sum_{l=0}^{t-2}\widehat{\alpha}_{CCEP}^l\nu_{t-l-1}\widehat{\varepsilon}_{i,t-l-1} \right) \left(\sum_{k=0}^{t-2}\widehat{\alpha}_{CCEP}^k\nu _{t-k-1}\widehat{\varepsilon}_{j,t-k-1} \right) \right]\notag\\
    &=\frac{1}{\epsilon^2}\frac{1}{N}\frac{1}{N^2}\sum_{i=1}^N\sum_{j=1}^N\widetilde{\gamma}_{1,i}\widetilde{\gamma}_{1,j}E^*[w_i^2w_j^2]\frac{1}{T}\sum_{t=2}^T\widehat{\varepsilon}_{i,t}\widehat{\varepsilon}_{j,t}\sum_{l=0}^{t-2}\widehat{\alpha}_{CCEP}^{2l}\widehat{\varepsilon}_{i,t-l-1}\widehat{\varepsilon}_{j,t-l-1}\notag\\
    &\leq \frac{E^*[w_i^4]}{\epsilon^2}\frac{1}{N}\frac{1}{N^2}\sum_{i=1}^N\sum_{j=1}^N|\widetilde{\gamma}_{1,i}||\widetilde{\gamma}_{1,j}|\frac{1}{T}\sum_{t=2}^T|\widehat{\varepsilon}_{i,t}||\widehat{\varepsilon}_{j,t}|\left(\sum_{l=0}^{t-2}\widehat{\alpha}_{CCEP}^{2l}\widehat{\varepsilon}_{i,t-l-1}^2\right)^{1/2}\left(\sum_{l=0}^{t-2}\widehat{\alpha}_{CCEP}^{2l}\widehat{\varepsilon}_{j,t-l-1}^2\right)^{1/2}\notag\\
    &=\frac{E^*[w_i^4]}{\epsilon^2}\frac{1}{N} \frac{1}{T}\sum_{t=2}^T\left(\frac{1}{N}\sum_{i=1}^N|\widetilde{\gamma}_{1,i}||\widehat{\varepsilon}_{i,t}|\widehat{b}_{i,t-1}^{1/2} \right)\left(\frac{1}{N}\sum_{j=1}^N|\widetilde{\gamma}_{1,j}||\widehat{\varepsilon}_{j,t}|\widehat{b}_{j,t-1}^{1/2} \right)\notag\\
    &\leq \frac{E^*[w_i^4]}{\epsilon^2}\frac{1}{N}\frac{1}{T}\sum_{t=2}^T\left(\frac{1}{N}\sum_{i=1}^N\widetilde{\gamma}_{1,i}^2 \right)^{1/2}\left(\frac{1}{N}\sum_{i=1}^N\widehat{\varepsilon}_{i,t}^2b_{i,t-1}\right)^{1/2}\left(\frac{1}{N}\sum_{j=1}^N\widetilde{\gamma}_{1,j}^2 \right)^{1/2}\left(\frac{1}{N}\sum_{j=1}^N\widehat{\varepsilon}_{j,t}^2b_{j,t-1}\right)^{1/2}\notag\\
    &\leq \frac{E^*[w_i^4]}{\epsilon^2}\frac{1}{N}\frac{1}{N}\sum_{i=1}^N\widetilde{\gamma}_{1,i}^2\left(\frac{1}{T}\sum_{t=1}^T\frac{1}{N}\sum_{i=1}^N\widehat{\varepsilon}_{i,t}^2\widehat{b}_{i,t-1} \right)^{1/2} \left(\frac{1}{T}\sum_{t=1}^T\frac{1}{N}\sum_{j=1}^N\widehat{\varepsilon}_{j,t}^2\widehat{b}_{j,t-1} \right)^{1/2}\notag\\
    &= \frac{E^*[w_i^4]}{\epsilon^2}\frac{1}{N}\left(\frac{1}{N}\sum_{i=1}^N\widetilde{\gamma}_{1,i}^2\right)\left(\frac{1}{T}\sum_{t=2}^T\frac{1}{N}\sum_{i=1}^N\widehat{\varepsilon}_{i,t}^2b_{i,t-1} \right)\notag\\
    &\leq \frac{E^*[w_i^4]}{\epsilon^2}\frac{1}{N}\left(\frac{1}{N}\sum_{i=1}^N\widetilde{\gamma}_{1,i}^2\right) \frac{1}{T}\sum_{t=2}^T\left(\frac{1}{N}\sum_{i=1}^N\widehat{\varepsilon}_{i,t}^4 \right) ^{1/2}\left( \frac{1}{N}\sum_{i=1}^N\widehat{b}_{i,t-1}^2\right)^{1/2}\notag \\
    &\leq \frac{1}{\epsilon^2}\frac{1}{N}\left(\frac{1}{N}\sum_{i=1}^N\widetilde{\gamma}_{1,i}^2\right)\left(\frac{1}{NT}\sum_{t=2}^T\sum_{i=1}^N\widehat{\varepsilon}_{i,t}^4 \right)^{1/2} \left(\frac{1}{NT}\sum_{t=2}^T\sum_{i=1}^N\widehat{b}_{i,t-1}^2 \right)^{1/2}\notag\\
    &=O_P(N^{-1})
\end{align}

As for $I_{a,2,2}$, a similar bound can be derived by noticing that the expectation with respect to bootstrap weights is non-zero when $i=k$ and $\ell=j$, or when $k=j$ and $i=\ell$. Then 
\begin{align}
    &\mathbb{P}^*\left(\left|I_{a,2,2} \right| > \epsilon\right)\leq \frac{1}{N^{3}\epsilon^{2}}\frac{1}{T}\sum_{t=2}^{T}\sum_{i=1}^{N}\sum_{j\neq i}^{N}\sum_{k=1}^{N}\sum_{\ell\neq k}^{N} \widetilde{\gamma}_{1,i}\widetilde{\gamma}_{1,k}E^*\left[u^*_{j,t-1}\varepsilon_{i,t}^*u^*_{\ell,t-1}\varepsilon_{k,t}^*\right]\notag\\
&= \frac{E^*[w_i^4]}{N^{3}\epsilon^{2}}\frac{1}{T}\sum_{t=2}^{T}\sum_{i=1}^{N}\sum_{j\neq i}^{N}\widetilde{\gamma}_{1,i}^{2}\widehat{\varepsilon}_{i,t}^{2}\widehat{b}_{j,t-1}+\frac{1}{N^{3}\epsilon^{2}}\frac{1}{T}\sum_{t=2}^{T}\sum_{i=1}^{N}\sum_{j\neq i}^{N}\widetilde{\gamma}_{1,i}\widetilde{\gamma}_{1,j}E^*\left[u^{*}_{i,t-1}\varepsilon_{i,t}^{*}u^{*}_{j,t-1}\varepsilon_{j,t}^{*}\right]\notag\\
&\leq \frac{E^*[w_i^4]}{N^{3}\epsilon^{2}}\frac{1}{T}\sum_{t=2}^{T}\sum_{i=1}^{N}\sum_{j\neq i}^{N}\widetilde{\gamma}_{1,i}^{2}\widehat{\varepsilon}_{i,t}^{2}\widehat{b}_{j,t-1}+\underbrace{\frac{1}{N^{3}\epsilon^{2}}\frac{1}{T}\sum_{t=2}^{T}\sum_{i=1}^{N}\sum_{j\neq i}^{N}|\widetilde{\gamma}_{1,i}||\widetilde{\gamma}_{1,j}|\left|E^*\left[u^{*}_{i,t-1}\varepsilon_{i,t}^{*}u^{*}_{j,t-1}\varepsilon_{j,t}^{*}\right]\right|}_{\text{order of}\hspace{1mm}I_{a,2,1}}\notag\\
&\leq \frac{E^*[w_i^4]}{N^{3}\epsilon^{2}}\frac{1}{T}\sum_{t=2}^{T}\sum_{i=1}^{N}\sum_{j=1}^{N}\widetilde{\gamma}_{1,i}^{2}\widehat{\varepsilon}_{i,t}^{2}\widehat{b}_{j,t-1}+\frac{E^*[w_i^4]}{N\varepsilon^2}\left(\frac{1}{N}\sum_{i=1}^N\widetilde{\gamma}_{1,i}^2\right)\left(\frac{1}{NT}\sum_{t=2}^T\sum_{i=1}^N\widehat{\varepsilon}_{i,t}^4 \right)^{1/2} \left(\frac{1}{NT}\sum_{t=2}^T\sum_{i=1}^N\widehat{b}_{i,t-1}^2 \right)^{1/2}\notag\\
&\leq \frac{E^*[w_i^4]}{N\epsilon^2}\frac{1}{T}\sum_{t=2}^T\left(\frac{1}{N}\sum_{i=1}^N\widetilde{\gamma}_{1,i}^4 \right)^{1/2}\left( \frac{1}{N}\sum_{i=1}^N\widehat{\varepsilon}_{i,t}^4\right)^{1/2}\left(\frac{1}{N}\sum_{j=1}^N\widehat{b}_{j,t-1} \right) + O_P(N^{-1})\notag\\
&= \frac{E^*[w_i^4]}{N\epsilon^2}\left(\frac{1}{N}\sum_{i=1}^N\widetilde{\gamma}_{1,i}^4 \right)^{1/2}\frac{1}{T}\sum_{t=2}^T\left( \frac{1}{N}\sum_{i=1}^N\widehat{\varepsilon}_{i,t}^4\right)^{1/2}\left(\frac{1}{N}\sum_{j=1}^N\widehat{b}_{j,t-1} \right) + O_P(N^{-1})\notag\\
&\leq \frac{E^*[w_i^4]}{N\epsilon^2}\left(\frac{1}{N}\sum_{i=1}^N\widetilde{\gamma}_{1,i}^4 \right)^{1/2}\left(\frac{1}{NT}\sum_{t=2}^T\sum_{i=1}^N\widehat{\varepsilon}_{i,t}^4\right)^{1/2}\left(\frac{1}{T}\sum_{t=2}^T\left( \frac{1}{N}\sum_{j=1}^N\widehat{b}_{j,t-1}\right)^2 \right)^{1/2}+O_P(N^{-1})\notag\\
&\leq \frac{E^*[w_i^4]}{N\epsilon^2}\left(\frac{1}{N}\sum_{i=1}^N\widetilde{\gamma}_{1,i}^4 \right)^{1/2}\left(\frac{1}{NT}\sum_{t=2}^T\sum_{i=1}^N\widehat{\varepsilon}_{i,t}^4\right)^{1/2}\left(\frac{1}{NT}\sum_{t=2}^T\sum_{j=1}^N\widehat{b}_{j,t-1}^2\right)^{1/2}+O_P(N^{-1})\notag\\
&=O_P(N^{-1})
\end{align}

\noindent Here, we used the fact that $\sum_{i=1}^N\sum_{j\neq i}^Na_ib_ic_j=\sum_{i=1}^Na_ib_i\sum_{j=1}^N(c_j-c_i)=\sum_{i=1}^N\sum_{j=1}^Na_ib_ic_j-\sum_{i=1}^Na_ib_ic_i\leq \sum_{i=1}^N\sum_{j=1}^Na_ib_ic_j$ for non-negative $a_i$, $b_i$ and $c_i$. Indeed, $\widehat{b}_{j,t-1}$ is non-negative and also $\frac{1}{N}\sum_{j=1}^N\widehat{b}_{j,t-1}\leq \left(\frac{1}{N}\sum_{j=1}^N\widehat{b}_{j,t-1}^2 \right)^{1/2}$.    Next, we have go to 
\begin{align}
    \left|I_{a,1}\right|&= \left|\frac{1}{\sqrt{NT}}\sum_{i=1}^N\overline{\*u}^{*\prime}_{-1}\widehat{\*F}^*(T^{-1}\widehat{\*F}^{*\prime}\widehat{\*F}^*)^{-1}T^{-1}\widehat{\*F}^{*\prime}\+\varepsilon_i^*\widetilde{\gamma}_{1,i}\right|\notag\\
&\leq\left\|T^{-1/2}\overline{\*u}_{-1}^{*\prime}\widehat{\*F}^{*} \right\|\left\|\frac{1}{\sqrt{N}}\sum_{i=1}^NT^{-1} \widehat{\*F}^{*\prime}\+\varepsilon^{* }_i\widetilde{\gamma}_{i,1}\right\|    \left\| (T^{-1}\widehat{\*F}^{*\prime}\widehat{\*F}^*)^{-1}\right\|\notag\\
&= \left\|\frac{1}{N\sqrt{T}}\sum_{i=1}^N\widehat{\*F}^{*\prime} \*u_{i,-1}^{*}\right\|\left\|\frac{1}{\sqrt{N}}\sum_{i=1}^NT^{-1} \widehat{\*F}^{*\prime}\+\varepsilon^{* }_i\widetilde{\gamma}_{i,1}\right\|    \left\| (T^{-1}\widehat{\*F}^{*\prime}\widehat{\*F}^*)^{-1}\right\|\notag\\
&= \sqrt{\frac{T}{N}}\left\|\frac{1}{\sqrt{N}}\sum_{i=1}^NT^{-1}\widehat{\*F}^{*\prime} \*u_{i,-1}^{*}\right\|\left\|\frac{1}{\sqrt{N}}\sum_{i=1}^NT^{-1} \widehat{\*F}^{*\prime}\+\varepsilon^{* }_i\widetilde{\gamma}_{i,1}\right\|    \underbrace{\left\| (T^{-1}\widehat{\*F}^{*\prime}\widehat{\*F}^*)^{-1}\right\|}_{O_{P^*}(1)}\notag\\
&=O_{P^*}(C_{N,T}^{-1}),
\end{align}
where $C_{N,T}=\min \{ \sqrt{N}, \sqrt{T}\}$. A closer look at the second component will help us determine this overall order. Note that 
\begin{align}\label{hatF*}
    \widehat{\*F}^*=[\overline{\*y}^*, \overline{\*y}_{-1}^*]= [\overline{\*y}+\overline{\*u}^*,\overline{\*y}_{-1}+\overline{\*u}^*_{-1} ]=\widehat{\*F}+ [\overline{\*u}^*,\overline{\*u}^*_{-1}]=\widehat{\*F}+\overline{\*V}^*,
\end{align}
By using this, we obtain the following:  
\begin{align}\label{F_eps_gamma}
\left\|\frac{1}{\sqrt{N}}\sum_{i=1}^NT^{-1} \widehat{\*F}^{*\prime}\+\varepsilon^{* }_i\widetilde{\gamma}_{i,1}\right\|
&\leq \left\|\frac{1}{\sqrt{N}T}\sum_{i=1}^N\sum_{t=2}^{T}\widehat{\*f}_t\varepsilon_{i,t}^*\widetilde{\gamma}_{i,1}\right\|+ \left\|\frac{1}{\sqrt{N}}\sum_{i=1}^N\begin{bmatrix}
    T^{-1}\overline{\*u}^{*\prime}\+\varepsilon_i^*\widetilde{\gamma}_{i,1}\\
    T^{-1}\overline{\*u}^{*\prime}_{-1}\+\varepsilon_i^*\widetilde{\gamma}_{i,1}
\end{bmatrix} \right\|
 \notag\\
&\leq\left\|\frac{1}{\sqrt{N}T}\sum_{i=1}^N\sum_{t=2}^{T}\widehat{\*f}_t\varepsilon_{i,t}^*\widetilde{\gamma}_{i,1}\right\|+\left|\frac{1}{\sqrt{N}}\sum_{i=1}^N
T^{-1}\overline{\*u}^{*\prime}\+\varepsilon_i^*\widetilde{\gamma}_{i,1}
\right|\notag\\
&+\left|\frac{1}{\sqrt{N}}\sum_{i=1}^NT^{-1}\overline{\*u}^{*\prime}_{-1}\+\varepsilon_i^*\widetilde{\gamma}_{i,1}\right|=O_{P^*}(C_{N,T}^{-1}),
\end{align}
because
\begin{align}
    E^*\left[ \left\|\frac{1}{\sqrt{N}T}\sum_{i=1}^N\sum_{t=2}^{T}\widehat{\*f}_t\varepsilon_{i,t}^*\widetilde{\gamma}_{i,1}\right\|^2\right]&=\frac{1}{NT^2}\sum_{i=1}^N\sum_{t=2}^T\left\|\widehat{\*f}_t\right\|^2\widehat{\varepsilon}_{i,t}^2\widetilde{\gamma}_{i,1}^2\notag\\
    &=\frac{1}{T^2}\sum_{t=2}^T\left\|\widehat{\*f}_t\right\|^2\left(\frac{1}{N}\sum_{i=1}^N \widehat{\varepsilon}_{i,t}^2\widetilde{\gamma}_{i,1}^2\right)\notag\\
    &\leq \left( \frac{1}{T^2}\sum_{t=2}^T\left\|\widehat{\*f}_t\right\|^4\right)^{1/2}\left( \frac{1}{T^2}\sum_{t=2}^T\left( \frac{1}{N}\sum_{i=1}^N \widehat{\varepsilon}_{i,t}^2\widetilde{\gamma}_{i,1}^2\right)^2\right)^{1/2}\notag\\
    &\leq \frac{1}{T}\left( \frac{1}{T}\sum_{t=2}^T\left\|\widehat{\*f}_t\right\|^4\right)^{1/2}\left( \frac{1}{T}\sum_{t=2}^T\left(\frac{1}{N}\sum_{i=1}^N\widehat{\varepsilon}_{i,t}^4 \right) \left(\frac{1}{N}\sum_{i=1}^N \widetilde{\gamma}_{i,1}^4\right)\right)^{1/2}\notag\\
    &= \frac{1}{T} \left(\frac{1}{N}\sum_{i=1}^N \widetilde{\gamma}_{i,1}^4 \right)^{1/2} \left(  \frac{1}{T}\sum_{t=2}^T\left\|\widehat{\*f}_t\right\|^4\right)^{1/2} \left(\frac{1}{NT}\sum_{i=1}^N\sum_{t=2}^T\widehat{\varepsilon}_{i,t}^4 \right)^{1/2}\notag\\
    &=O_P(T^{-1})
\end{align}
and therefore $\left\|\frac{1}{\sqrt{N}T}\sum_{i=1}^N\sum_{t=2}^{T}\widehat{\*f}_t\varepsilon_{i,t}^*\widetilde{\gamma}_{i,1}\right\| =O_{P^*}(T^{-1/2})$ in probability. Also, 
\begin{align}
    \left|\frac{1}{\sqrt{N}}\sum_{i=1}^NT^{-1}\overline{\*u}^{*\prime}_{-1}\+\varepsilon_i^*\widetilde{\gamma}_{i,1}\right|=O_{P^*}(N^{-1/2}T^{-1/2})
\end{align}
 in probability, because it is the same term as (\ref{I_a2}), but scaled by $T^{-1/2}$. The remaining term requires more work as it is not an MDS with respect to the bootstrap measure. We get 
\begin{align}
      \left|\frac{1}{\sqrt{N}}\sum_{i=1}^N
T^{-1}\overline{\*u}^{*\prime}\+\varepsilon_i^*\widetilde{\gamma}_{i,1}\right|&\leq \frac{1}{T}\sum_{t=2}^T\left|\frac{1}{N}\sum_{i=1}^N\widetilde{\gamma}_{i,1}\varepsilon_{i,t}^*\right|\left|\frac{1}{\sqrt{N}}\sum_{j=1}^N u^*_{j,t}\right|\notag\\
&\leq \frac{1}{\sqrt{N}}\left( \frac{1}{T}\sum_{t=2}^T\left(\frac{1}{\sqrt{N}}\sum_{i=1}^N\widetilde{\gamma}_{i,1}\varepsilon_{i,t}^* \right)^2 \right)^{1/2} \left(\frac{1}{T}\sum_{t=2}^T\left(\frac{1}{\sqrt{N}}\sum_{j=1}^N u^*_{j,t} \right)^2 \right)^{1/2}\notag\\
&=N^{-1/2}a^{1/2}b^{1/2}=O_{P^*}(N^{-1/2}),
\end{align}
because both $a$ and $b$ are $O_{P^*}(1)$ in probability: 
\begin{align}
    \mathbb{P}^*(a>M)\leq \frac{1}{M}\frac{1}{T}\sum_{t=2}^T\frac{1}{N}\sum_{i=1}^N\widetilde{\gamma}_{i,1}^2E^*[\varepsilon_{i,t}^{*2}]&\leq \frac{1}{M}\left( \frac{1}{N}\sum_{i=1}^N\widetilde{\gamma}_{i,1}^4\right)^{1/2}\frac{1}{T}\sum_{t=2}^T\left(\frac{1}{N}\sum_{i=1}^N\widehat{\varepsilon}_{i,t}^4\right)^{1/2}\notag\\
    &\leq \frac{1}{M}\left( \frac{1}{N}\sum_{i=1}^N\widetilde{\gamma}_{i,1}^4\right)^{1/2}\left(\frac{1}{NT}\sum_{t=2}^T\sum_{i=1}^N\widehat{\varepsilon}_{i,t}^4 \right)^{1/2}\notag\\
    &\leq \epsilon,
\end{align}
as $M\to \infty$ for any $\epsilon>0$ conditionally on sample, implying $O_{P^*}(1)$, and a similar bound holds for $b$:
\begin{align}\label{u^2_bound}
    \mathbb{P}^*(b>M)\leq \frac{1}{M}\frac{1}{NT}\sum_{t=2}^T\sum_{j=1}^NE^*[u_{j,t}^{*2}]=\frac{1}{M}\frac{1}{NT}\sum_{t=2}^T\sum_{j=1}^N\widehat{b}_{j,t}\leq \epsilon
\end{align}
as $M\to \infty$ for any $\epsilon>0$ conditionally on the sample. This implies that 
\begin{align}\label{rate_a1}
    \left\|\frac{1}{\sqrt{N}}\sum_{i=1}^NT^{-1} \widehat{\*F}^{*\prime}\+\varepsilon^{* }_i\widetilde{\gamma}_{i,1}\right\|=O_{P^*}(C_{N,T}^{-1}).
\end{align}
Next, we have 
\begin{align}
    \left\|\frac{1}{\sqrt{N}}\sum_{i=1}^NT^{-1}\widehat{\*F}^{*\prime}\*u_{i,-1}^{*}\right\|&\leq \left\|\frac{1}{\sqrt{N}}\sum_{i=1}^NT^{-1}\widehat{\*F}'\*u_{i,-1}^* \right\| + \left\|\frac{1}{\sqrt{N}}\sum_{i=1}^NT^{-1}\left[\overline{\*u}^{*}, \overline{\*u}^{*}_{-1}\right]'\*u_{i,-1}^*\right\|\notag\\
    &=a+b,
\end{align}
which is sufficient to be $O_{P^*}(1)$ for our purposes. Here
 \begin{align}
    a=\left\|\frac{1}{\sqrt{N}T}\sum_{i=1}^N\sum_{t=2}^T\widehat{\*f}_tu_{i,t-1}^* \right\|&\leq \frac{1}{T}\sum_{t=2}^T\left\|\widehat{\*f}_t \right\|\left|\frac{1}{\sqrt{N}} \sum_{i=1}^Nu_{i,t-1}^*\right|\notag\\
     &\leq \left( \frac{1}{T}\sum_{t=2}^T\left\|\widehat{\*f}_t \right\|^2 \right)^{1/2}\left(\frac{1}{T}\sum_{t=2}^T\left( \frac{1}{\sqrt{N}} \sum_{i=1}^Nu_{i,t-1}^*\right)^{2} \right)^{1/2}\notag\\
     &=c^{1/2}d^{1/2}=O_{P^*}(1),
 \end{align}
 because 
 \begin{align}
     \mathbb{P}^*(d>M)\leq \frac{1}{M}E^*\left[ \frac{1}{T}\sum_{t=2}^T\left( \frac{1}{\sqrt{N}} \sum_{i=1}^Nu_{i,t-1}^*\right)^{2} \right]=\frac{1}{M}\frac{1}{NT}\sum_{t=2}^T\sum_{i=1}^N\widehat{b}_{i,t-1}\leq \epsilon
 \end{align}
as $M\to \infty$ for any $\epsilon>0$, conditionally on the sample. Now, onto $b$:
\begin{align}
    b\leq \left|\frac{1}{\sqrt{N}}\sum_{i=1}^NT^{-1}\overline{\*u}^{*\prime}\*u_{i,-1}^* \right|+\left| \frac{1}{\sqrt{N}}\sum_{i=1}^NT^{-1}\overline{\*u}_{-1}^{*\prime}\*u_{i,-1}^*\right|=e+f,
\end{align}
so that 
\begin{align*}
    e&=\frac{1}{\sqrt{N}}\left|\frac{1}{T}\sum_{t=2}^T\frac{1}{\sqrt{N}}\sum_{j=1}^Nu_{j,t}^*\frac{1}{\sqrt{N}}\sum_{i=1}^Nu_{i,t-1}^* \right|\leq \frac{1}{\sqrt{N}}\frac{1}{T}\sum_{t=2}^T\left|\frac{1}{\sqrt{N}}\sum_{j=1}^Nu_{j,t}^* \right| \left|\frac{1}{\sqrt{N}}\sum_{i=1}^Nu_{i,t-1}^* \right|\notag\\
    &\leq \frac{1}{\sqrt{N}}\left( \frac{1}{T}\sum_{t=2}^T\left( \frac{1}{\sqrt{N}}\sum_{j=1}^Nu_{j,t}^*\right)^2\right)^{1/2}\left( \frac{1}{T}\sum_{t=2}^T\left(\frac{1}{\sqrt{N}}\sum_{i=1}^Nu_{i,t-1}^* \right)^2 \right)^{1/2}\notag\\
    &=O_{p^{*}}(N^{-1/2})
\end{align*}
and 
\begin{align}
    f&=\frac{1}{\sqrt{N}}\left|\frac{1}{T}\sum_{t=2}^T\frac{1}{\sqrt{N}}\sum_{j=1}^Nu_{j,t-1}^*\frac{1}{\sqrt{N}}\sum_{i=1}^Nu_{i,t-1}^* \right|\leq \frac{1}{\sqrt{N}}\frac{1}{T}\sum_{t=2}^T\left|\frac{1}{\sqrt{N}}\sum_{j=1}^Nu_{j,t-1}^* \right| \left|\frac{1}{\sqrt{N}}\sum_{i=1}^Nu_{i,t-1}^* \right|\notag\\
    &\leq \frac{1}{\sqrt{N}}\left( \frac{1}{T}\sum_{t=2}^T\left( \frac{1}{\sqrt{N}}\sum_{j=1}^Nu_{j,t-1}^*\right)^2\right)^{1/2}\left( \frac{1}{T}\sum_{t=2}^T\left(\frac{1}{\sqrt{N}}\sum_{i=1}^Nu_{i,t-1}^* \right)^2 \right)^{1/2}\notag\\
    &=O_{p^{*}}(N^{-1/2}),
\end{align}
because the components under the square root in both $e$ and $f$ are $O_{P^*}(1)$ by the exact same argument as in (\ref{u^2_bound}). Therefore, 
\begin{align}\label{f*u_-1*}
    \left\|\frac{1}{\sqrt{N}}\sum_{i=1}^NT^{-1}\widehat{\*F}^{*\prime}\*u_{i,-1}^{*}\right\|=O_{P^*}(1),
\end{align}
which is sufficient to show that 
under $TN^{-1}\to \kappa^{-1}\in (0,\infty)$, we have that 
\begin{align}
    |I_a|\leq |I_{a,1}|+ |I_{a,2}|=O_{P^*}(N^{-1/2})+O_{p^{*}}(T^{-1/2}).
\end{align}
in probability.
\\

\noindent Next, we have 
\begin{align}
    I_b&= \frac{1}{\sqrt{NT}}\sum_{i=1}^N(y_{i,0}-\overline{y}_0\widetilde{\gamma}_{1,i})\+\iota_T'\widehat{\+\A}_{CCEP,-1}'\+\varepsilon_i^*-\frac{1}{\sqrt{NT}}\sum_{i=1}^N(y_{i,0}-\overline{y}_0\widetilde{\gamma}_{1,i})\+\iota_T'\widehat{\+\A}_{CCEP,-1}'\*P_{\widehat{\*F}^*}\+\varepsilon_i^*\notag\\
    &=I_{b,1}-I_{b,2},
\end{align}
where, by using $\max_{i,j}\left(x_{i,j}^2-\frac{1}{n_1n_2}\sum_{i=1}^{n_1}\sum_{j=1}^{n_2}x_{i,j}^2\right)\leq 2\left( \sum_{i=1}^{n_1}\sum_{j=1}^{n_2}x_{i,j}^p\right)^{2/p}$ for $p>1$:
\begin{align}\label{alpha_epsilon_rate}
\mathbb{P}^*\left( |I_{b,1}|>\epsilon  \right)
    &\leq \frac{1}{\epsilon^2}\frac{1}{NT}\sum_{i=1}^N(y_{i,0}-\overline{y}_0\widetilde{\gamma}_{1,i})^2\sum_{t=2}^T\widehat{\alpha}_{CCEP}^{2(t-1)}E^*[\varepsilon_{i,t}^{*2}]\notag\\
    &=\frac{1}{\epsilon^2}\frac{1}{NT}\sum_{i=1}^N(y_{i,0}-\overline{y}_0\widetilde{\gamma}_{1,i})^2\sum_{t=2}^T\widehat{\alpha}_{CCEP}^{2(t-1)}\widehat{\varepsilon}_{i,t}^{2}\notag\\
    &=\frac{1}{\epsilon^2}\frac{1}{NT}\sum_{i=1}^N(y_{i,0}-\overline{y}_0\widetilde{\gamma}_{1,i})^2\sum_{t=2}^T\widehat{\alpha}_{CCEP}^{2(t-1)}\left(\frac{1}{NT}\sum_{i=1}^N\sum_{t=1}^T \widehat{\varepsilon}_{i,t}^{2}\right)\notag\\
    &+\frac{1}{\epsilon^2}\frac{1}{NT}\sum_{i=1}^N(y_{i,0}-\overline{y}_0\widetilde{\gamma}_{1,i})^2\sum_{t=2}^T\widehat{\alpha}_{CCEP}^{2(t-1)}\left(\widehat{\varepsilon}_{i,t}^{2}-\frac{1}{NT}\sum_{i=1}^N\sum_{t=1}^T \widehat{\varepsilon}_{i,t}^{2}\right)\notag\\
    &\leq \max_{i,t}\left(\widehat{\varepsilon}_{i,t}^{2}-\frac{1}{NT}\sum_{i=1}^N\sum_{t=1}^T \widehat{\varepsilon}_{i,t}^{2}\right)\frac{1}{\epsilon^2}\frac{1}{NT}\sum_{i=1}^N(y_{i,0}-\overline{y}_0\widetilde{\gamma}_{1,i})^2\sum_{t=2}^T\widehat{\alpha}_{CCEP}^{2(t-1)} + O_P(T^{-1})\notag\\
    &\leq 2 T^{-1}(NT)^{2/(4+\delta)}\left(\frac{1}{NT}\sum_{i=1}^N\sum_{t=1}^T|\widehat{\varepsilon}_{i,t}|^{4+\delta} \right)^{2/(4+\delta)}\underbrace{\frac{1}{\epsilon^2}\frac{1}{N}\sum_{i=1}^N(y_{i,0}-\overline{y}_0\widetilde{\gamma}_{1,i})^2\sum_{t=2}^T\widehat{\alpha}_{CCEP}^{2(t-1)}}_{O_P(1)}+O_P(T^{-1})\notag\\
    &=O_P(T^{-1}(NT)^{1/(2+\delta/2)}) \notag\\
    &\propto O_P(T^{-1}(T^2)^{1/(2+\delta/2)})=O_P(T^{-1}T^{2/(2+\delta/2)})=O_P(T^{-\delta/(4+\delta)})
\end{align}
for $\delta>0$ under $TN^{-1}=O(1)$, provided consistency of $\widehat{\alpha}_{CCEP}$ and convergence of the infinite sum. Note that the rate at which the component vanishes improves in $\delta$. Note that 
\begin{align}
    \frac{1}{N}\sum_{i=1}^N(y_{i,0}-\overline{y}_0\widetilde{\gamma}_{1,i})^2\leq 2\frac{1}{N}\sum_{i=1}^Ny_{i,0}^2+2\overline{y}_0\frac{1}{N}\sum_{i=1}^N\widetilde{\gamma}_{1,i}^2=O_P(1),
\end{align}
as long as $\frac{1}{N}\sum_{i=1}^Ny_{i,0}^p=O_P(1)$ for $p\geq 2$. We can verify this by noticing that 
\begin{align}\label{initial_value_bound}
    \frac{1}{N}\sum_{i=1}^N|y_{i,0}|^p\leq 4^{p-1} \underbrace{\left(\frac{1}{N}\sum_{i=1}^N|\gamma_i|^p\right)}_{O_P(1)}|g_0|^p+4^{p-1}\frac{1}{N}\sum_{i=1}^N|u_{i,0}|^p=O_P(1),
    \end{align}
    because both $g_0$ and $u_{i,0}$ are linear processes and we can form a moment bound in terms of (\ref{minkowski1}) for both of them. Next, 
\begin{align}
    &|I_{b,2}|=\left| \frac{1}{\sqrt{NT}}\sum_{i=1}^N(y_{i,0}-\overline{y}_0\widetilde{\gamma}_{1,i})\+\iota_T'\widehat{\+\A}_{CCEP,-1}'\*P_{\widehat{\*F}^*}\+\varepsilon_i^*\right|\notag\\
    &\leq \left\|\frac{1}{\sqrt{N}}\sum_{i=1}^NT^{-1}\widehat{\*F}^{*\prime}\+\varepsilon_i^*(y_{i,0}-\overline{y}_0\widetilde{\gamma}_{1,i}) \right\|\left\|T^{-1/2}\+\iota_T'\widehat{\+\A}_{CCEP,-1}'\widehat{\*F}^* \right\|\left\|\left( T^{-1}\widehat{\*F}^{*\prime}\widehat{\*F}^*\right)^{-1} \right\|=abc,
\end{align}
where 
\begin{align}
    a\leq \left\|\frac{1}{\sqrt{N}}\sum_{i=1}^NT^{-1}\widehat{\*F}^{*\prime}\+\varepsilon_i^*y_{i,0} \right\|+|\overline{y}_0|\left\| \frac{1}{\sqrt{N}}\sum_{i=1}^NT^{-1}\widehat{\*F}^{*\prime}\+\varepsilon_i^*\widetilde{\gamma}_{i,1}\right\|=O_{p*}(C_{N,T}^{-1}),
\end{align}
in probability, because the second term is exactly (\ref{F_eps_gamma}), whereas the first admits the bound with the same order as long as $\frac{1}{N}\sum_{i=1}^Ny_{i,0}^p=O_P(1)$ for $p\geq 4$, which we can show by the same steps as in (\ref{initial_value_bound}). Hence, it is sufficient to show that $b$ and $c$ are bounded, where $c$ clearly is: we have two averages that proxy two factors in this setup. Then, by using $\max_{i}|x_i|\leq \left(\sum_{i=1}^nx_i^2 \right)^{1/2}$
    \begin{align}\label{accumulated_f_star}
        b=\left\|T^{-1/2}\sum_{t=2}^T\widehat{\alpha}_{CCEP}^{t-1}\widehat{\*f}_t^* \right\|&\leq T^{-1/2}\sum_{t=2}^T\left| \widehat{\alpha}_{CCEP}^{t-1}\right|\left\| \widehat{\*f}_t^*\right\|\notag\\
        &\leq T^{-1/2}\sum_{t=2}^T\left| \widehat{\alpha}_{CCEP}^{t-1}\right|\left\| \widehat{\*f}_t\right\|+T^{-1/2}\sum_{t=2}^T\left| \widehat{\alpha}_{CCEP}^{t-1}\right| |\overline{u}_t^*|\notag\\
        &+T^{-1/2}\sum_{t=2}^T\left| \widehat{\alpha}_{CCEP}^{t-1}\right| |\overline{u}_{t-1}^*|\notag\\
        &\leq \underbrace{\left( \frac{1}{T}\sum_{t=2}^T\left\| \widehat{\*f}_t\right\|^2\right)^{1/2}}_{O_P(1)}\sum_{t=2}^T\left|\widehat{\alpha}_{CCEP}\right| ^{t-1}+ N^{-1/2}\left( \frac{1}{T}\sum_{t=2}^TN\overline{u}_t^{*2}\right)^{1/2}\sum_{t=2}^T \left|\widehat{\alpha}_{CCEP}\right| ^{t-1}\notag\\
        &+N^{-1/2}\left( \frac{1}{T}\sum_{t=2}^TN\overline{u}_{t-1}^{*2}\right)^{1/2}\sum_{t=2}^T \left|\widehat{\alpha}_{CCEP}\right| ^{t-1}=O_P(1),
    \end{align}
  due to consistency and convergence of the infinite sum, and where clearly, 
  \begin{align}\label{u_Op1}
      \mathbb{P}^*\left(\frac{1}{T}\sum_{t=2}^TN\overline{u}_t^{*2}>M \right)\leq \frac{1}{M}\frac{1}{T}\sum_{t=2}^TE^*[N\overline{u}_t^{*2}]=\frac{1}{M}\frac{1}{NT}\sum_{t=2}^T\sum_{i=1}^N\widehat{b}_{i,t}\leq \epsilon,
  \end{align}
 as $M\to \infty$ for any $\epsilon>0$ conditionally on the sample, implying $O_{P^*}(1)$, and the same holds for the third term. Therefore, 
 \begin{align}
     |I_b|\leq |I_{b,1}|+ |I_{b,2}|=O_{P^*}(T^{-\delta/2(4+\delta)}) + O_{P^*}(C_{N,T}^{-1})
 \end{align}
 in probability. \\

 \noindent We move on to $I_c$: 
 \begin{align}
   I_c=  \frac{1}{\sqrt{NT}}\sum_{i=1}^N\widetilde{\gamma}_{2,i}\*b_{-1}'\*M_{\widehat{\*F}^*}\+\varepsilon_i^*&= \frac{1}{\sqrt{NT}}\sum_{i=1}^N\widetilde{\gamma}_{2,i}\*b_{-1}'\+\varepsilon_{i}^*-\frac{1}{\sqrt{NT}}\sum_{i=1}^N\widetilde{\gamma}_{2,i}\*b_{-1}'\*P_{\widehat{\*F}^*}\+\varepsilon_i^*\notag\\
   &=I_{c,1}-I_{c,2}.
 \end{align}
We now use Holder's inequality for non-negative $x_i$ and $y_i$: $\sum_{i=1}^Nx_iy_i\leq \left(\sum_{i=1}^nx_i^p \right)^{1/p}\left(\sum_{i=1}^ny_i^q \right)^{1/q}\leq\left(\sum_{i=1}^nx_i^p \right)^{1/p} \sum_{i=1}^ny_i $ to obtain
\begin{align}
    E^*\left[ I_{c,1}^2\right]&=\sum_{i=1}^{N}(\widetilde{\gamma}_{2,i})^{2}\frac{1}{NT}\sum_{t=2}^{T}(b_{t-1}\widehat{\varepsilon}_{i,t})^{2}\notag\\
    &\leq \left(\sum_{i=1}^{N}(\widetilde{\gamma}_{2,i})^{4+2\delta}\right)^{1/(2+\delta)}\frac{1}{N}\sum_{i=1}^{N}\frac{1}{T}\sum_{t=2}^{T}(b_{t-1}\widehat{\varepsilon}_{i,t})^{2}\notag\\
    &\leq \left(\sum_{i=1}^{N}(\widetilde{\gamma}_{2,i})^{4+2\delta}\right)^{1/(2+\delta)}\left(\frac{1}{T}\sum_{t=2}^{T}(b_{t-1})^{4}\right)^{1/2}\left(\frac{1}{NT}\sum_{i=1}^{N}\sum_{t=2}^{T}\widehat{\varepsilon}_{i,t}^{4}\right)^{1/2}\notag\\
    &=N^{1/(2+\delta)}\left(\frac{1}{N}\sum_{i=1}^{N}(\widetilde{\gamma}_{2,i})^{4+2\delta}\right)^{1/(2+\delta)}\left(\frac{1}{T}\sum_{t=2}^{T}(b_{t-1})^{4}\right)^{1/2}\left(\frac{1}{NT}\sum_{i=1}^{N}\sum_{t=2}^{T}\widehat{\varepsilon}_{i,t}^{4}\right)^{1/2},
\end{align}
  where at this stage we let $\delta\in (0,2]$ to preserve the overall highest 8th moment. Here, the structure of $\widetilde{\gamma}_{2,i}$ is the key. Note that it is the second coordinate of $\widetilde{\+\gamma}_i=\*R^{-1}\widehat{\+\gamma}_i$, where $\*R^{-1}=\begin{bmatrix} 1& 0\\ \widehat{\alpha}_{CCEP} & 1
      \end{bmatrix}$, which means that $\widetilde{\gamma}_{2,i}=[\widehat{\alpha}_{CCEP}, 1]\widehat{\+\gamma}_i$. Crucially, the leading terms in (\ref{hat_gamma}) are nullified. In particular,
    \begin{align}\label{gamma_tilde_2}
        \widetilde{\gamma}_{2,i}&=\begin{bmatrix}
            \widehat{\alpha}_{CCEP}\\
            1
        \end{bmatrix}'[1,-\widehat{\alpha}_{CCEP}]'\overline{\gamma}^{-1}\gamma_i \notag\\
        &+\begin{bmatrix}
            \widehat{\alpha}_{CCEP}\\
            1
        \end{bmatrix}'(T^{-1}\widehat{\*F}'\widehat{\F})^{-1}T^{-1}\widehat{\*F}'(-\overline{\+\varepsilon}\overline{\gamma}^{-1}\gamma_i  - (\alpha -\widehat{\alpha}_{CCEP})\overline{\*u}_{-1}\overline{\gamma}^{-1}\gamma_i+(\alpha_0-\widehat{\alpha}_{CCEP})\*u_{i,-1}+\+\varepsilon_i )\notag\\
        &=\begin{bmatrix}
            \widehat{\alpha}_{CCEP}\\
            1
        \end{bmatrix}'(T^{-1}\widehat{\*F}'\widehat{\F})^{-1}T^{-1}\widehat{\*F}'(-\overline{\+\varepsilon}\overline{\gamma}^{-1}\gamma_i  - (\alpha -\widehat{\alpha}_{CCEP})\overline{\*u}_{-1}\overline{\gamma}^{-1}\gamma_i+(\alpha_0-\widehat{\alpha}_{CCEP})\*u_{i,-1}+\+\varepsilon_i ).
    \end{align}
    Because $\left\|[\widehat{\alpha}_{CCEP}, 1]' \right\|^{4+2\delta}=O_P(1)$ and $\left\|(T^{-1}\widehat{\*F}'\widehat{\F})^{-1} \right\|^{4+2\delta}=O_P(1)$, we have that 
    \begin{align}
        \left\| T^{-1}\widehat{\*F}'\overline{\+\varepsilon}\overline{\gamma}^{-1}\gamma_i\right\|^{4+2\delta}&\leq 2^{3+2\delta}\left\|\overline{\gamma}^{-1}\gamma_i \right\|^{4+2\delta}\left\| \overline{\*C}\right\|^{4+2\delta}\left\| T^{-1}\*Q'\overline{\+\varepsilon}\right\|^{4+2\delta}\notag\\
        &+2^{3+2\delta}\left\|\overline{\gamma}^{-1}\gamma_i \right\|^{4+2\delta}\left\| T^{-1}\overline{\*V}'\overline{\+\varepsilon}\right\|^{4+2\delta}\notag\\
        &=O_P(N^{-(4+2\delta)})+O_P((NT)^{-(2+\delta)}),
    \end{align}
 and 
 \begin{align}
     \left\| T^{-1}\widehat{\*F}'\overline{\*u}_{-1}(\widehat{\alpha}_{CCEP}-\alpha_0)\overline{\gamma}^{-1}\gamma_i\right\|^{4+2\delta}&\leq 2^{3+2\delta}\left\| \widehat{\alpha}_{CCEP}-\alpha_0\right\|^{4+2\delta}\left\|\overline{\gamma}^{-1}\gamma_i \right\|^{4+2\delta}\left\| \overline{\*C}\right\|^{4+2\delta}\left\| T^{-1}\*Q'\overline{\*u}_{-1}\right\|^{4+2\delta}\notag\\
        &+2^{3+2\delta}\left\| \widehat{\alpha}_{CCEP}-\alpha_0\right\|^{4+2\delta}\left\|\overline{\gamma}^{-1}\gamma_i \right\|^{4+2\delta}\left\| T^{-1}\overline{\*V}'\overline{\*u}_{-1}\right\|^{4+2\delta}\notag\\
        &=O_P((NT)^{-(2+\delta)})\times (O_P(N^{-(4+2\delta)})+O_P((NT)^{-(2+\delta)})),
 \end{align}
 and 
 \begin{align}
     \left\| T^{-1}\widehat{\*F}'\*u_{i,-1}(\widehat{\alpha}_{CCEP}-\alpha_0)\right\|^{4+2\delta}&\leq 2^{3+2\delta}\left\| \widehat{\alpha}_{CCEP}-\alpha_0\right\|^{4+2\delta}\left\| \overline{\*C}\right\|^{4+2\delta}\left\|T^{-1}\*Q'\*u_{i,-1} \right\|^{4+2\delta}\notag\\
     &+2^{3+2\delta}\left\| \widehat{\alpha}_{CCEP}-\alpha_0\right\|^{4+2\delta}\left\|T^{-1}\overline{\*V}'\*u_{i,-1} \right\|^{4+2\delta}\notag\\
     &=O_P((NT)^{-(2+\delta)})\times (O_P(T^{-(2+\delta)})+O_P(N^{-(4+2\delta)}).
 \end{align}
 Ultimately, 
 \begin{align}
     \left\| T^{-1}\widehat{\*F}'\+\varepsilon_i\right\|^{4+2\delta}&\leq 2^{3+2\delta}\left\|\overline{\*C}\right\|^{4+2\delta}\left\|T^{-1}\*Q'\+\varepsilon_{i} \right\|^{4+2\delta}+2^{3+2\delta}\left\|T^{-1}\overline{\*V}'\+\varepsilon_i \right\|^{4+2\delta}\notag\\
     &=O_P(T^{-(2+\delta)}),
 \end{align}
 which is the dominating order. This implies that under $TN^{-1}=O(1)$ we have
 \begin{align}
     N^{1/(2+\delta)}\left(\frac{1}{N}\sum_{i=1}^{N}(\widetilde{\gamma}_{2,i})^{4+2\delta}\right)^{1/(2+\delta)}&=N^{1/(2+\delta)}O_P(T^{-1})\propto O_P(N^{1/(2+\delta)}N^{-(2+\delta)/(2+\delta)})\notag\\
     &=O_P(N^{-(1+\delta)/(2+\delta)})=o_P(1),
 \end{align}
 where again the rate improves in $\delta$. Therefore, we are left to demonstrate that the second component in the product is $O_P(1)$. Note how: 
 \begin{align}\label{b4_rate}
     \frac{1}{T}&\sum_{t=2}^Tb_{t-1}^4=\frac{1}{T}\sum_{t=2}^T\left( \sum_{j=0}^{t-2}(\widehat{\alpha}_{CCEP})^j\overline{y}_{t-j-2}\right)^4\notag\\
     &=\frac{1}{T}\sum_{t=2}^T\sum_{j=0}^{t-2}\sum_{k=0}^{t-2}\sum_{l=0}^{t-2}\sum_{p=0}^{t-2}\widehat{\alpha}_{CCEP}^{j+l+k+p}\overline{y}_{t-j-2}\overline{y}_{t-l-2}\overline{y}_{t-k-2}\overline{y}_{t-p-2}\notag\\
     &\leq\sum_{j=0}^{T-2}\sum_{k=0}^{T-2}\sum_{l=0}^{T-2}\sum_{p=0}^{T-2}|\widehat{\alpha}_{CCEP}|^{j+l+k+p}\frac{1}{T}\sum_{t=\max \{j,l,k,p \}+2}^T|\overline{y}_{t-j-2}||\overline{y}_{t-l-2}||\overline{y}_{t-k-2}||\overline{y}_{t-p-2}|\notag\\
     &\leq \sum_{j=0}^{T-2}\sum_{k=0}^{T-2}\sum_{l=0}^{T-2}\sum_{p=0}^{T-2}|\widehat{\alpha}_{CCEP}|^{j+l+k+p}\left( \frac{1}{T}\sum_{t=\max \{j,l,k,p \}+2}^T\overline{y}_{t-j-2}^2\overline{y}_{t-l-2}^2\right)^{1/2}\left(  \frac{1}{T}\sum_{t=\max \{j,l,k,p \}+2}^T\overline{y}_{t-k-2}^2\overline{y}_{t-p-2}^2\right)^{1/2}\notag\\
     &\leq \sum_{j=0}^{T-2}\sum_{k=0}^{T-2}\sum_{l=0}^{T-2}\sum_{p=0}^{T-2}|\widehat{\alpha}_{CCEP}|^{j+l+k+p} \frac{1}{T}\sum_{t=2}^T\overline{y}_{t-1}^4\notag\\
     &\leq \left(\frac{1-|\widehat{\alpha}_{CCEP}|^{T-1}}{1-|\widehat{\alpha}_{CCEP}|} \right)^4\left(8\overline{\gamma}^4\frac{1}{T}\sum_{t=2}^Tg_{t-1}^4 + 8\frac{1}{T}\sum_{t=2}^T\overline{u}_{t-1}^4  \right)\notag\\
     &\leq C_1\frac{1}{T}\sum_{t=2}^Tg_{t-1}^4 + C_2\frac{1}{T}\sum_{t=2}^T\left(\frac{1}{N}\sum_{i=1}^Nu_{i,t-1} \right)^4=O_P(1)
 \end{align}
 directly by (\ref{O_P(1)_def}) - (\ref{minkowski1}) and consistency of $\widehat{\alpha}_{CCEP}$, where we moved from the second to the third inequality by another iteration of CS inequality. This implies that overall
 \begin{align}
     |I_{c,1}|=O_{P^*}(N^{-(1+\delta)/(4+2\delta)})
 \end{align}
 in probability.\\
 

\noindent For the second component, we will use the fact that $|\sqrt{T}\widetilde{\gamma}_{2,i}|=O_P(1)$, because from (\ref{gamma_tilde_2}), we see that the dominant term is of the order $O_P(T^{-1/2})$. Therefore,
\begin{align} |I_{c,2}|&\leq\left\|T^{-1}\*b_{-1}'\widehat{\*F}^*\right\| \left\|\left(T^{-1}{(\widehat{\*F}^*)'}\widehat{\*F}^*\right)^{-1}\right\|\left\|\frac{1}{\sqrt{NT}}\sum_{i=1}^N\widetilde{\gamma}_{2,i}{(\widehat{\*F}^*)'}\+\varepsilon_i^*\right\|\notag\\
&=\left\|\frac{1}{T}\sum_{t=2}^Tb_{t-1}\widehat{\*f}_t^*\right\|\left\|\left(T^{-1}{(\widehat{\*F}^*)'}\widehat{\*F}^*\right)^{-1}\right\|\left\|\frac{1}{\sqrt{NT}}\sum_{i=1}^N\widetilde{\gamma}_{2,i}{(\widehat{\*F}^*)'}\+\varepsilon_i^*\right\|\notag\\
&=\left\|\left(T^{-1}{(\widehat{\*F}^*)'}\widehat{\*F}^*\right)^{-1}\right\|\left(\frac{1}{T}\sum_{t=2}^Tb_{t-1}^2 \right)^{1/2}\left(\frac{1}{T}\sum_{t=2}^T\left\|\widehat{\*f}_t^* \right\|^2 \right)^{1/2}\left\|\frac{1}{\sqrt{N}T}\sum_{i=1}^N(\sqrt{T}\widetilde{\gamma}_{2,i}){(\widehat{\*F}^*)'}\+\varepsilon_i^*\right\|\notag\\
&=O_{P^*}(C_{N,T}^{-1})
\end{align}
in probability, because the last component drives the order and in this form it exactly coincides with (\ref{rate_a1}). Thus, overall,
\begin{align}
    |I_c|\leq |I_{c,1}|+|I_{c,2}|=O_{P^*}(N^{-(1+\delta)/(4+2\delta)})+O_{P^*}(C_{N,T}^{-1})
\end{align}
in probability. Lastly,
\begin{align}\label{I_d}
    I_d=\frac{1}{\sqrt{NT}}\sum_{i=1}^N\*u_{i,-1}^{*\prime}\*M_{\widehat{\*F}^*}\+\varepsilon_i^*&=\frac{1}{\sqrt{NT}}\sum_{i=1}^N\*u_{i,-1}^{*\prime}\+\varepsilon_i^*-\frac{1}{\sqrt{NT}}\sum_{i=1}^N\*u_{i,-1}^{*\prime}\*P_{\widehat{\*F}^*}\+\varepsilon_i^*\notag\\
    &=I_{d,1}-I_{d,2},
\end{align}
where $I_{d,1}$ generates the asymptotic distribution, whereas $I_{d,2}$ gives the Nickel-type bias. Regarding the former,
\begin{align}
    I_{d,1}=\frac{1}{\sqrt{NT}}\sum_{i=1}^N\sum_{t=2}^Tu_{i,t-1}^*\varepsilon_{i,t}^*&=\frac{1}{\sqrt{NT}}\sum_{i=1}^N\sum_{t=2}^T\sum_{j=0}^{t-2}\widehat{\alpha}_{CCEP}^{j}\varepsilon_{i,t-j-1}^*\varepsilon_{i,t}^*\notag\\
    &=\sum_{j=0}^{T-2}\widehat{\alpha}_{CCEP}^j\frac{1}{\sqrt{NT}}\sum_{i=1}^N\sum_{t=j+2}^T\varepsilon_{i,t-j-1}^*\varepsilon_{i,t}^*\notag\\
    &=\sum_{j=0}^{T-2}(\alpha_0)^j\frac{1}{\sqrt{NT}}\sum_{i=1}^N\sum_{t=j+2}^T\varepsilon_{i,t-j-1}^*\varepsilon_{i,t}^*\notag\\
    &+\sum_{j=0}^{T-2}(\widehat{\alpha}_{CCEP}^j-(\alpha_0)^j)\frac{1}{\sqrt{NT}}\sum_{i=1}^N\sum_{t=j+2}^T\varepsilon_{i,t-j-1}^*\varepsilon_{i,t}^*\notag\\
    &=I_{d,1,1}+I_{d,1,2},
\end{align}
which is a decomposition analogous to the one in \cite{gonccalves2004bootstrapping}. We will use it in the further analysis of (\ref{I_d}
)\\

\noindent \textbf{b)} Moving forwards, we have
\begin{align}
    II=\frac{1}{\sqrt{NT}}\sum_{i=1}^N\*y_{i,-1}^{*\prime}\*M_{\widehat{\*F}^*}\overline{\*u}_{-1}^*\widetilde{\gamma}_{2,i}&= \frac{1}{\sqrt{NT}}\sum_{i=1}^N\widetilde{\gamma}_{1,i}\overline{\*y}_{-1}'\*M_{\widehat{\*F}^*}\overline{\*u}_{-1}^*\widetilde{\gamma}_{2,i}\notag\\
    &+\frac{1}{\sqrt{NT}}\sum_{i=1}^N(y_{i,0}-\overline{y}_0\widetilde{\gamma}_{1,i})\+\iota_T'\widehat{\+\A}_{CCEP,-1}'\*M_{\widehat{\*F}^*}\overline{\*u}_{-1}^*\widetilde{\gamma}_{2,i}\notag\\
    &+\frac{1}{\sqrt{NT}}\sum_{i=1}^N\*b_{-1}'\*M_{\widehat{\*F}^*}\overline{\*u}_{-1}^*\widetilde{\gamma}_{2,i}^2\notag\\
    &+ \frac{1}{\sqrt{NT}}\sum_{i=1}^N\*u_{i,-1}^{*\prime}\*M_{\widehat{\*F}^*}\overline{\*u}_{-1}^*\widetilde{\gamma}_{2,i}\notag\\
    &=II_a+II_b+II_c+II_d.
\end{align}
Again, by using $\overline{\*y}_{-1}=\overline{\*y}_{-1}^*-\overline{\*u}_{-1}^*$, we have 
\begin{align}
    |II_a|=\left|\frac{1}{\sqrt{NT}}\sum_{i=1}^N\widetilde{\gamma}_{1,i}\overline{\*u}_{-1}^{*\prime}\*M_{\widehat{\*F}^*}\overline{\*u}_{-1}^*\widetilde{\gamma}_{2,i} \right|&\leq \sqrt{\frac{T}{N}} \left\|NT^{-1} \overline{\*u}_{-1}^{*\prime}\*M_{\widehat{\*F}^*}\overline{\*u}_{-1}^*\right\|\frac{1}{N}\sum_{i=1}^N|\widetilde{\gamma}_{1,i}||\widetilde{\gamma}_{2,i}|\notag\\
    &\leq\sqrt{\frac{T}{N}} \left\|NT^{-1} \overline{\*u}_{-1}^{*\prime}\*M_{\widehat{\*F}^*}\overline{\*u}_{-1}^*\right\| \left( \frac{1}{N}\sum_{i=1}^N\widetilde{\gamma}_{1,i}^2\right)^{1/2} \left(\frac{1}{N}\sum_{i=1}^N\widetilde{\gamma}_{2,i}^2 \right)^{1/2}\notag\\
    &=O_{P^*}(T^{-1/2}), 
\end{align}
under $TN^{-1}=O(1)$ as it is the order of the dominant term in $\widetilde{\gamma}_{2,i}$. Note that 
\begin{align}
    \left\|NT^{-1} \overline{\*u}_{-1}^{*\prime}\*M_{\widehat{\*F}^*}\overline{\*u}_{-1}^* \right\|&\leq \left\|\frac{1}{T}\sum_{t=2}^TN\overline{u}_{t-1}^{*2}\right\|+\left\|(T^{-1}\widehat{\*F}^{*\prime}\widehat{\*F}^{*})^{-1} \right\|\left\|\frac{1}{\sqrt{N}}\sum_{i=1}^N\widehat{\*F}^{*\prime}\*u_{i,-1}^* \right\|^2\notag\\
    &=O_{P^*}(1)
\end{align}
by (\ref{u_Op1}) and (\ref{f*u_-1*}). Next, 
\begin{align}
    II_b=\frac{1}{\sqrt{NT}}\sum_{i=1}^N(y_{i,0}-\overline{y}_0\widetilde{\gamma}_{1,i})\+\iota_T'\widehat{\+\A}_{CCEP,-1}'\overline{\*u}_{-1}^*\widetilde{\gamma}_{2,i}&-\frac{1}{\sqrt{NT}}\sum_{i=1}^N(y_{i,0}-\overline{y}_0\widetilde{\gamma}_{1,i})\+\iota_T'\widehat{\+\A}_{CCEP,-1}'\*P_{\widehat{\*F}^*}\overline{\*u}_{-1}^*\widetilde{\gamma}_{2,i}\notag\\
    &=II_{b,1}-II_{b,2},
\end{align}
where by using $\max_i |x_i|\leq \left(\sum_{i=1}^nx_i^2\right)^{1/2}$, we obtain 
\begin{align}
    |II_{b,1}|&\leq \left|\frac{1}{N}\sum_{i=1}^N(y_{i,0}-\overline{y}_0\widetilde{\gamma}_{1,i})\widetilde{\gamma}_{2,i}\right|\left\|T^{-1/2} \+\iota_T'\widehat{\+\A}_{CCEP,-1}'\sqrt{N}\overline{\*u}_{-1}^*\right\|\notag\\
    &\leq \left(\frac{1}{N}\sum_{i=1}^N(y_{i,0}-\overline{y}_0\widetilde{\gamma}_{1,i})^2\right)^{1/2}\left( \frac{1}{N}\sum_{i=1}^N\widetilde{\gamma}_{2,i}^2\right)^{1/2}\left\|T^{-1/2} \+\iota_T'\widehat{\+\A}_{CCEP,-1}'\sqrt{N}\overline{\*u}_{-1}^*\right\| \notag\\
    &= \left(\frac{1}{N}\sum_{i=1}^N(y_{i,0}-\overline{y}_0\widetilde{\gamma}_{1,i})^2\right)^{1/2}\left( \frac{1}{N}\sum_{i=1}^N\widetilde{\gamma}_{2,i}^2\right)^{1/2}\left\|T^{-1/2} \sum_{t=2}^T\widehat{\alpha}_{CCEP}^{t-1}\sqrt{N}\overline{u}_{t-1}^*\right\|\notag\\
    &\leq \frac{1}{\sqrt{T}}\left(\frac{1}{N}\sum_{i=1}^N(y_{i,0}-\overline{y}_0\widetilde{\gamma}_{1,i})^2\right)^{1/2}\left( \frac{1}{N}\sum_{i=1}^N(\sqrt{T}\widetilde{\gamma}_{2,i})^2\right)^{1/2} \left(\frac{1}{T}\sum_{t=2}^TN\overline{u}_{t-1}^{*2} \right)^{1/2}\sum_{t=2}^T|\widehat{\alpha}_{CCEP}|^{t-1}\notag\\
    &=O_{P^*}(T^{-1/2})
\end{align}
in probability. Next, similarly we obtain 
\begin{align}
    |II_{b,2}|&\leq  \left(\frac{1}{N}\sum_{i=1}^N(y_{i,0}-\overline{y}_0\widetilde{\gamma}_{1,i})^2\right)^{1/2}\left( \frac{1}{N}\sum_{i=1}^N\widetilde{\gamma}_{2,i}^2\right)^{1/2}\left\|T^{-1/2}\+\iota_T'\widehat{\+\A}_{CCEP,-1}'\*P_{\widehat{\*F}^*}\sqrt{N}\overline{\*u}_{-1}^* \right\|\notag\\
    &\leq \left(\frac{1}{N}\sum_{i=1}^N(y_{i,0}-\overline{y}_0\widetilde{\gamma}_{1,i})^2\right)^{1/2}\left( \frac{1}{N}\sum_{i=1}^N\widetilde{\gamma}_{2,i}^2\right)^{1/2}\left\|(T^{-1}\widehat{\*F}^{*\prime}\widehat{\*F}^*)^{-1} \right\|\left\|T^{-1/2}\sum_{t=2}^T\widehat{\alpha}_{CCEP}^{t-1}\widehat{\*f}_t^* \right\|\notag\\
    &\times \left( \frac{1}{T}\sum_{t=2}^T\left\| \widehat{\*f}_t^{*}\right\|^2 \right)^{1/2}\left(  \frac{1}{T}\sum_{t=2}^T N\overline{u}_{t-1}^{*2}\right)^{1/2}\notag\\
    &=O_{P^*}(T^{-1/2})
\end{align}
in probability by (\ref{accumulated_f_star}), while 
\begin{align}
    |II_{c}| &\leq \left|\frac{1}{N}\sum_{i=1}^N\sqrt{T}\widetilde{\gamma}_{2,i}^2\right|\left\|\sqrt{N}T^{-1} \*b_{-1}'\*M_{\widehat{\*F}^*}\overline{\*u}_{-1}^*\right\|=O_{P^*}(T^{-1/2}),
\end{align}
because 
\begin{align}
    \left\|\sqrt{N}T^{-1} \*b_{-1}'\*M_{\widehat{\*F}^*}\overline{\*u}_{-1}^*\right\|&\leq \left\|\frac{1}{T} \sum_{t=2}^Tb_{t-1}\sqrt{N}\overline{u}_{t-1}^* \right\|+\left\| (T^{-1}\widehat{\*F}^{*\prime}\widehat{\*F}^*)^{-1} \right\|\left\| \frac{1}{T}\sum_{t=2}^Tb_{t-1}\widehat{\*f}_t^{*}\right\|\left\| \frac{1}{T}\sum_{t=2}^T\widehat{\*f}_t^*\sqrt{N}\overline{u}_{t-1}^*\right\|\notag\\
    &\leq \left(\frac{1}{T}\sum_{t=2}^Tb_{t-1}^2 \right)^{1/2}\left(\frac{1}{T}\sum_{t=2}^TN\overline{u}_{t-1}^{*2} \right)^{1/2}\notag\\
    &+\left\| (T^{-1}\widehat{\*F}^{*\prime}\widehat{\*F}^*)^{-1} \right\|\left(\frac{1}{T}\sum_{t=2}^Tb_{t-1}^2 \right)^{1/2}\frac{1}{T}\sum_{t=2}^T\left\|\widehat{\*f}_{t}^*\right\|^2 \left(\frac{1}{T}\sum_{t=2}^TN\overline{u}_{t-1}^{*2} \right)^{1/2}\notag\\
    &=O_{P^*}(1).
\end{align}
Eventually, 

\begin{align}
   II_d= \frac{1}{\sqrt{NT}}\sum_{i=1}^N\*u_{i,-1}^{*\prime}\overline{\*u}_{-1}^*\widetilde{\gamma}_{2,i}-\frac{1}{\sqrt{NT}}\sum_{i=1}^N\*u_{i,-1}^{*\prime}\*P_{\widehat{\*F}^*}\overline{\*u}_{-1}^*\widetilde{\gamma}_{2,i}=II_{d,1}-II_{d,2},
\end{align}
where 
\begin{align}\label{II_d1}
    |II_{d,1}|&\leq \frac{1}{\sqrt{N}T}\sum_{t=2}^T\left|\frac{1}{\sqrt{N}}\sum_{i=1}^N u_{i,t-1}^*\sqrt{T}\widetilde{\gamma}_{2,i}\right|\left|\frac{1}{\sqrt{N}}\sum_{j=1}^N u_{j,t}^*\right|\notag\\
    &\leq \frac{1}{\sqrt{N}}\left( \frac{1}{T}\sum_{t=2}^T\left( \frac{1}{\sqrt{N}}\sum_{i=1}^N u_{i,t-1}^*\sqrt{T}\widetilde{\gamma}_{2,i}\right)^2\right)^{1/2}\left(\frac{1}{T}\sum_{t=2}^T\left( \frac{1}{\sqrt{N}}\sum_{i=1}^N u_{j,t}^*\right)^2 \right)^{1/2}\notag\\
    &=O_{P^*}(N^{-1/2}),
\end{align}
because 
\begin{align}\label{var_verification}
    E^*\left[  \frac{1}{T}\sum_{t=2}^T\left( \frac{1}{\sqrt{N}}\sum_{i=1}^N u_{i,t-1}^*\sqrt{T}\widetilde{\gamma}_{2,i}\right)^2\right]&=\frac{1}{T}\sum_{t=2}^T\frac{1}{N}\sum_{i=1}^NT\widetilde{\gamma}_{2,1}^2\widehat{b}_{i,t-1}\notag\\
    &\leq \left( \frac{1}{N}\sum_{i=1}^N(\sqrt{T}\widetilde{\gamma}_{2,1})^4\right)^{1/2}\frac{1}{T}\sum_{t=2}^T\left( \frac{1}{N}\sum_{i=1}^N\widehat{b}_{i,t-1}^2\right)^{1/2}\notag\\
    &\leq  \left( \frac{1}{N}\sum_{i=1}^N(\sqrt{T}\widetilde{\gamma}_{2,1})^4\right)^{1/2}\left(\frac{1}{NT}\sum_{t=2}^T\sum_{i=1}^N\widehat{b}_{i,t-1}^2 \right)^{1/2}\notag\\
    &=O_P(1). 
\end{align}
Next, 
\begin{align}
    |II_{d,2}|\leq \left\|\frac{1}{NT}\sum_{i=1}^N\sqrt{T}\widetilde{\gamma}_{2,i}\*u_{i,-1}^{*\prime}\widehat{\*F}^*\right\| \left\|(T^{-1}\widehat{\*F}^{*\prime}\widehat{\*F}^*)^{-1} \right\|\left\|\sqrt{N}T^{-1}\widehat{\*F}^{*\prime} \overline{\*u}_{-1}^*\right\|=O_{P^*}(N^{-1/2}),
\end{align}
because 
\begin{align}\label{II_d2}
    \left\|\frac{1}{NT}\sum_{i=1}^N\sqrt{T}\widetilde{\gamma}_{2,i}\*u_{i,-1}^{*\prime}\widehat{\*F}^*\right\|&=\left\|\frac{1}{NT}\sum_{t=2}^T\widehat{\*f}_t^*\sum_{i=1}^Nu_{i,t-1}^*\sqrt{T}\widetilde{\gamma}_{2,i} \right\|\notag\\
    &\leq N^{-1/2}\frac{1}{T}\sum_{t=2}^T\left\|\widehat{\*f}_t^*\right\|\left| \frac{1}{\sqrt{N}}\sum_{i=1}^Nu_{i,t-1}^*\sqrt{T}\widetilde{\gamma}_{2,i}\right|\notag\\
    &\leq N^{-1/2}\left(\frac{1}{T}\sum_{t=2}^T\left\|\widehat{\*f}_t^*\right\|^2 \right)^{1/2}\left(\frac{1}{T}\sum_{t=2}^T\left(\frac{1}{\sqrt{N}}\sum_{i=1}^Nu_{i,t-1}^*\sqrt{T}\widetilde{\gamma}_{2,i} \right)^2 \right)^{1/2}\notag\\
    &=O_{p^{*}}(N^{-1/2})
\end{align}
due to (\ref{var_verification}) and the fact that the third component coincides with (\ref{f*u_-1*}), so it is $O_{P^*}(1)$. Therefore, overall,
\begin{align}
    |II|=O_{P^*}(C_{N,T}^{-1}). 
\end{align}
\noindent \textbf{c)} Ultimately, we have
\begin{align}
    III=\frac{1}{\sqrt{NT}}\sum_{i=1}^N\*y_{i,-1}^{*\prime}\*M_{\widehat{\*F}^*}\overline{\+\varepsilon}^*\widetilde{\gamma}_{1,i}&= \frac{1}{\sqrt{NT}}\sum_{i=1}^N\widetilde{\gamma}_{1,i}\overline{\*y}_{-1}'\*M_{\widehat{\*F}^*}\overline{\+\varepsilon}^*\widetilde{\gamma}_{1,i}\notag\\
    &+\frac{1}{\sqrt{NT}}\sum_{i=1}^N(y_{i,0}-\overline{y}_0\widetilde{\gamma}_{1,i})\+\iota_T'\widehat{\+\A}_{CCEP,-1}'\*M_{\widehat{\*F}^*}\overline{\+\varepsilon}^*\widetilde{\gamma}_{1,i}\notag\\
    &+\frac{1}{\sqrt{NT}}\sum_{i=1}^N\widetilde{\gamma}_{2,i}\*b_{-1}'\*M_{\widehat{\*F}^*}\overline{\+\varepsilon}^*\widetilde{\gamma}_{1,i}\notag\\
    &+ \frac{1}{\sqrt{NT}}\sum_{i=1}^N\*u_{i,-1}^{*\prime}\*M_{\widehat{\*F}^*}\overline{\+\varepsilon}^*\widetilde{\gamma}_{1,i}\notag\\
    &=III_a+III_b+III_c+III_d.
\end{align}
Here,
\begin{align}
    III_a=\frac{1}{\sqrt{NT}}\sum_{i=1}^N\widetilde{\gamma}_{1,i}^2\overline{\*u}_{-1}^{*\prime}\*P_{\widehat{\*F}^*}\overline{\+\varepsilon}^* - \frac{1}{\sqrt{NT}}\sum_{i=1}^N\widetilde{\gamma}_{1,i}^2\overline{\*u}_{-1}^{*\prime}\overline{\+\varepsilon}^*=III_{a,1}-III_{a,2},
\end{align}
where 
\begin{align}
    |III_{a,2}|&\leq \frac{1}{N}\sum_{i=1}^N\widetilde{\gamma}_{1,i}^2 \left|\sqrt{N}T^{-1/2}\overline{\*u}_{-1}^{*\prime}\overline{\+\varepsilon}^*\right|\notag\\
    &=\frac{1}{N}\sum_{i=1}^N\widetilde{\gamma}_{1,i}^2 \left|\frac{1}{\sqrt{N}N}\sum_{i=1}^N\sum_{l=1}^N\left(\frac{1}{\sqrt{T}}\sum_{t=2}^Tu_{j,t-1}^*\varepsilon_{l,t}^* \right) \right|=O_{P^*}(N^{-1/2}),
    \end{align}
    because the second component follows the structure of (\ref{I_a2_exp}). Also,
    \begin{align}
        |III_{a,1}|&\leq \frac{1}{N}\sum_{i=1}^N\widetilde{\gamma}_{1,i}^2\left|\sqrt{N}T^{-1/2} \overline{\*u}_{-1}^{*\prime}\*P_{\widehat{\*F}^*}\overline{\+\varepsilon}^*\right|\notag\\
        &\leq \frac{1}{N}\sum_{i=1}^N\widetilde{\gamma}_{1,i}^2 \left\|(T^{-1}\widehat{\*F}^{*\prime}\widehat{\*F}^*)^{-1} \right\|\sqrt{\frac{T}{N}}\left\|\frac{1}{\sqrt{N}}\sum_{i=1}^NT^{-1} \widehat{\*F}^{*\prime}\*u_{i,-1}^*\right\|\left\| \frac{1}{\sqrt{N}}\sum_{i=1}^NT^{-1} \widehat{\*F}^{*\prime}\+\varepsilon_{i}^*\right\|\notag\\
        &=O_{P^*}(C_{N,T}^{-1})
    \end{align}
    under $TN^{-1}=O(1)$, because the order is driven by the last component which coincides with (\ref{F_eps_gamma}), and the second-to-last component corresponds to (\ref{f*u_-1*}). Next, 
    \begin{align}
        III_{b}&=\frac{1}{\sqrt{NT}}\sum_{i=1}^N(y_{i,0}-\overline{y}_0\widetilde{\gamma}_{1,i})\+\iota_T'\widehat{\+\A}_{CCEP,-1}'\overline{\+\varepsilon}^*\widetilde{\gamma}_{1,i}-\frac{1}{\sqrt{NT}}\sum_{i=1}^N(y_{i,0}-\overline{y}_0\widetilde{\gamma}_{1,i})\+\iota_T'\widehat{\+\A}_{CCEP,-1}'\*P_{\widehat{\*F}^*}\overline{\+\varepsilon}^*\widetilde{\gamma}_{1,i}\notag\\
        &=III_{b,1}-III_{b,2},
    \end{align}
    where 
   \begin{align}
       |III_{b,1}|\leq \underbrace{\left|\frac{1}{N} \sum_{i=1}^N(y_{i,0}-\overline{y}_0\widetilde{\gamma}_{1,i})\widetilde{\gamma}_{1,i}\right|}_{O_P(1)}\left| \frac{1}{\sqrt{T}}\sum_{t=2}^T\widehat{\alpha}_{CCEP}^{t-1}\sqrt{N}\overline{\varepsilon}_t^*\right|=ab=O_P(T^{-\delta/2(4+\delta)}),
   \end{align}
because $b$ has this rate: 
\begin{align}\label{b_term_slow_rate}
    \mathbb{P}^*\left(b>\epsilon\right)&\leq \frac{1}{\epsilon^2}\frac{1}{T}\sum_{t=2}^T\widehat{\alpha}_{CCEP}^{2(t-1)}\frac{1}{N}\sum_{i=1}^N\widehat{\varepsilon}_{i,t}^2\notag\\
    &=\frac{1}{\epsilon^2}\frac{1}{T}\sum_{t=2}^T\widehat{\alpha}_{CCEP}^{2(t-1)}\frac{1}{N}\sum_{i=1}^N\left(\widehat{\varepsilon}_{i,t}^2-\frac{1}{NT}\sum_{t=1}^T\sum_{i=1}^N\widehat{\varepsilon}_{i,t}^2\right)\notag\\
    &+\frac{1}{\epsilon^2}\frac{1}{T}\sum_{t=2}^T\widehat{\alpha}_{CCEP}^{2(t-1)}\left(\frac{1}{NT}\sum_{t=1}^T\sum_{i=1}^N\widehat{\varepsilon}_{i,t}^2\right)\notag\\
    &\leq \max_{i,t}\left(\widehat{\varepsilon}_{i,t}^2-\frac{1}{NT}\sum_{t=1}^T\sum_{i=1}^N\widehat{\varepsilon}_{i,t}^2\right)\frac{1}{\epsilon^2}\frac{1}{T}\sum_{t=2}^T\widehat{\alpha}_{CCEP}^{2(t-1)} + O_P(T^{-1})\notag\\
    &\leq 2 T^{-1}(NT)^{2/(4+\delta)}\left(\frac{1}{NT}\sum_{i=1}^N\sum_{t=1}^T|\widehat{\varepsilon}_{i,t}|^{4+\delta} \right)^{2/(4+\delta)}\frac{1}{\epsilon^2}\frac{1}{T}\sum_{t=2}^T\widehat{\alpha}_{CCEP}^{2(t-1)} + O_P(T^{-1})\notag\\
    &=O_P(T^{-\delta/(4+\delta)})
\end{align}
under $TN^{-1}=O(1)$. The next term can be handled similarly:
\begin{align}
    |III_{b,2}|&\leq \left|\frac{1}{N} \sum_{i=1}^N(y_{i,0}-\overline{y}_0\widetilde{\gamma}_{1,i})\widetilde{\gamma}_{1,i} \right|\left|\sqrt{N}T^{-1/2}\+\iota_T'\widehat{\+\A}_{CCEP,-1}'\*P_{\widehat{\*F}^*}\overline{\+\varepsilon}^* \right|\notag\\
    &\leq \left|\frac{1}{N} \sum_{i=1}^N(y_{i,0}-\overline{y}_0\widetilde{\gamma}_{1,i})\widetilde{\gamma}_{1,i} \right|\left\|(T^{-1}\widehat{\*F}^{*\prime}\widehat{\*F}^*)^{-1} \right\|\underbrace{\left\|\frac{1}{\sqrt{T}}\sum_{t=2}^T\widehat{\alpha}_{CCEP}^{t-1}\widehat{\*f}_t^* \right\|}_{O_{P^*}(1)}\left\|\frac{1}{\sqrt{N}}\sum_{i=1}^NT^{-1}\widehat{\*F}^{*\prime}\+\varepsilon_i^* \right\|\notag\\
    &=O_{p^{*}}(C_{N,T}^{-1}),
\end{align}
which is the order of the last component, which dictates the behavior of $III_{b,2}$. Thus, overall
\begin{align}
    |III_b|\leq |III_{b,1}| + |III_{b,2}|=O_{p^{*}}(C_{N,T}^{-1})+O_{p^{*}}(T^{-\delta/2(4+\delta)}).
\end{align}
Moving on 
\begin{align}
    III_c= \frac{1}{\sqrt{NT}}\sum_{i=1}^N\widetilde{\gamma}_{2,i}\*b_{-1}'\overline{\+\varepsilon}^*\widetilde{\gamma}_{1,i}-\frac{1}{\sqrt{NT}}\sum_{i=1}^N\widetilde{\gamma}_{2,i}\*b_{-1}'\*P_{\widehat{\*F}^*}\overline{\+\varepsilon}^*\widetilde{\gamma}_{1,i}=III_{c,1}-III_{c,2},
\end{align}
where 
\begin{align}
    |III_{c,1}|&\leq \left|\frac{1}{N}\sum_{i=1}^N\widetilde{\gamma}_{2,i}\widetilde{\gamma}_{1,i} \right|\left|\sqrt{N}T^{-1/2}\*b_{-1}'\overline{\+\varepsilon}^* \right|\notag\\
    &\leq \left( \frac{1}{N}\sum_{i=1}^N\widetilde{\gamma}_{2,i}^2\right)^{1/2}\left(\frac{1}{N}\sum_{i=1}^N\widetilde{\gamma}_{1,i}^2 \right)^{1/2}\left|\frac{1}{\sqrt{T}}\sum_{t=2}^Tb_{t-1} \sqrt{N}\overline{\varepsilon}_t^*\right|=ab=O_{P^*}(T^{-1/2}),
\end{align}
because $b$ is $O_{P^*}(1)$: 
\begin{align}
    E^*[b^2]=\frac{1}{T}\sum_{t=2}^Tb_{t-1}^2\frac{1}{N}\sum_{i=1}^N\widehat{\varepsilon}_{i,t}^2&\leq \left( \frac{1}{T}\sum_{t=2}^Tb_{t-1}^4\right)^{1/2}\left( \frac{1}{T}\sum_{t=2}^T\left( \frac{1}{N}\sum_{i=1}^N\widehat{\varepsilon}_{i,t}^2\right)^2\right)^{1/2}\notag\\
    &\leq \left( \frac{1}{T}\sum_{t=2}^Tb_{t-1}^4\right)^{1/2} \left(\frac{1}{NT}\sum_{t=2}^T\sum_{i=1}^N\widehat{\varepsilon}_{i,t}^4\right)^{1/2}\notag\\
    &=O_P(1).
\end{align}
Next, 
\begin{align}
    |III_{c,2}|&\leq \left|\frac{1}{N}\sum_{i=1}^N\sqrt{T}\widetilde{\gamma}_{2,1}\widetilde{\gamma}_{1,i} \right|\left|\sqrt{N} T^{-1}\*b_{-1}'\*P_{\widehat{\*F}^*}\overline{\+\varepsilon}^*\right|\notag\\
    &\leq \left( \frac{1}{N}\sum_{i=1}^NT\widetilde{\gamma}_{2,i}^2\right)^{1/2}\left(\frac{1}{N}\sum_{i=1}^N\widetilde{\gamma}_{1,i}^2 \right)^{1/2}\left\|\frac{1}{T}\sum_{t=2}^Tb_{t-1}\widehat{\*f}_t^* \right\|\left\|(T^{-1}\widehat{\*F}^{*\prime}\widehat{\*F}^*)^{-1} \right\|\left\|\frac{1}{\sqrt{N}}\sum_{i=1}^NT^{-1} \widehat{\*F}^{*\prime}\+\varepsilon_i^*\right\|\notag\\
    &\leq \left( \frac{1}{N}\sum_{i=1}^NT\widetilde{\gamma}_{2,i}^2\right)^{1/2}\left(\frac{1}{N}\sum_{i=1}^N\widetilde{\gamma}_{1,i}^2 \right)^{1/2}\left( \frac{1}{T}\sum_{t=2}^Tb_{t-1}^2\right)^{1/2}\left(  \frac{1}{T}\sum_{t=2}^T\left\| \widehat{\*f}^*_t\right\|^2\right)^{1/2}\notag\\
    &\times \left\|(T^{-1}\widehat{\*F}^{*\prime}\widehat{\*F}^*)^{-1} \right\|\left\|\frac{1}{\sqrt{N}}\sum_{i=1}^NT^{-1} \widehat{\*F}^{*\prime}\+\varepsilon_i^*\right\|=O_{P^*}(C_{N,T}^{-1}),
\end{align}
based on (\ref{F_eps_gamma}). Lastly, 
\begin{align}
    III_{d}=\frac{1}{\sqrt{NT}}\sum_{i=1}^N\*u_{i,-1}^{*\prime}\overline{\+\varepsilon}^*\widetilde{\gamma}_{1,i}-\frac{1}{\sqrt{NT}}\sum_{i=1}^N\*u_{i,-1}^{*\prime}\*P_{\widehat{\*F}^*}\overline{\+\varepsilon}^*\widetilde{\gamma}_{1,i}=III_{d,1}-III_{d,2},
\end{align}
where 
\begin{align}
    III_{d,1}=\frac{1}{N\sqrt{N}}\sum_{i=1}^N\widetilde{\gamma}_{1,i}\sum_{j=1}^N\left( \frac{1}{\sqrt{T}}\sum_{t=2}^Tu_{i,t-1}^*\varepsilon_{j,t}^*\right)=O_{p^{*}}(N^{-1/2})
\end{align}
as it has exactly the same structure as (\ref{I_a2_exp}). Then
\begin{align}
    |III_{d,2}|&\leq \sqrt{\frac{T}{N}}\underbrace{\left\|\frac{1}{\sqrt{N}}\sum_{i=1}^NT^{-1}\widehat{\*F}^{*\prime}\*u_{i,-1}^*\widetilde{\gamma}_{1,i}\ \right\|}_{O_{P^*}(1)}\left\|(T^{-1}\widehat{\*F}^{*\prime}\widehat{\*F}^*)^{-1} \right\|\left\| \frac{1}{\sqrt{N}}\sum_{i=1}^NT^{-1}\widehat{\*F}^{*\prime}\+\varepsilon^*_i\right\|\notag\\
    &=O_{P^*}(C_{N,T}^{-1}),
\end{align}
with the order driven by the last term. 
\paragraph{Expansion and Rates of the Denominator }\label{4.2.5}
We now establish the asymptotic behavior of the denominator of the bootstrap CCEP estimator. In particular, 
\begin{align}
    D=
   & \frac{1}{NT}\sum_{i=1}^N\*y_{i,-1}^{*\prime}\*M_{\widehat{\*F}^*}\*y_{i,-1}^{*}\notag\\
   &=\frac{1}{NT}\sum_{i=1}^N\*u_{i,-1}^{*\prime}\*M_{\widehat{\*F}^*}\*u_{i,-1}^{*}\notag\\
   &+ \frac{1}{NT}\sum_{i=1}^N\widetilde{\gamma}_{1,i}^2\overline{\*y}_{-1}'\*M_{\widehat{\*F}^*}\overline{\*y}_{-1}\notag\\
    &+\frac{1}{NT}\sum_{i=1}^N(y_{i,0}-\overline{y}_0\widetilde{\gamma}_{1,i})^2\+\iota_T'\widehat{\+\A}_{CCEP,-1}'\*M_{\widehat{\*F}^*}\widehat{\+\A}_{CCEP,-1}\+\iota_T\notag\\
    &+\frac{1}{NT}\sum_{i=1}^N\widetilde{\gamma}_{2,i}^2\*b_{-1}'\*M_{\widehat{\*F}^*}\*b_{-1}\notag\\
    &= D_1 + D_2 + D_3 + D_4 +D_{5-16},
    \end{align}
    where $D_{5-16}$ represents the cross-terms. \\

    \noindent \textbf{Lemma 2.} \textit{Under Assumptions \ref{ass::1} - \ref{ass::6}, and let the weights be either multiplicative or non-multiplicative. Then as $(N,T)\to \infty$ and $NT^{-1}\to \kappa\in (0, \infty)$, we have 
    \begin{enumerate}[a)]
        \item $D_1=\frac{1}{NT}\sum_{t=2}^T\sum_{i=1}^Nu_{i,t-1}^{*2}+o_{P^*}(1)$.
        \item $|D_{2-16}|=o_{P^*}(1)$. 
    \end{enumerate}
    }
\bigskip 

    \noindent \textbf{Proof.} \textbf{a)} We decompose $D_1$ as 
    \begin{align}
        D_1=\frac{1}{NT}\sum_{i=1}^N\*u_{i,-1}^{*\prime}\*u_{i,-1}^* - \frac{1}{N}\sum_{i=1}^NT^{-1}\*u_{i,-1}^{*\prime}\widehat{\*F}^*(T^{-1}\widehat{\*F}^{*\prime}\widehat{\*F})^{-1}T^{-1}\widehat{\*F}^{*\prime}\*u_{i,-1}^*=D_{1,1}-D_{1,2},
    \end{align}
    where 
    \begin{align}
        |D_{1,2}|\leq \left\|(T^{-1}\widehat{\*F}^{*\prime}\widehat{\*F}^*)^{-1} \right\|&\frac{1}{N}\sum_{i=1}^N\left\|T^{-1}\widehat{\*F}^{*\prime}\*u_{i,-1}^* \right\|^2\notag\\
        &\leq  2\left\|(T^{-1}\widehat{\*F}^{*\prime}\widehat{\*F}^*)^{-1} \right\|\frac{1}{N}\sum_{i=1}^N\left\|T^{-1}\widehat{\*F}'\*u_{i,-1}^* \right\|^2\notag\\
        &+2\left\|(T^{-1}\widehat{\*F}^{*\prime}\widehat{\*F}^*)^{-1} \right\|\frac{1}{N}\sum_{i=1}^N\left\|\begin{bmatrix} T^{-1}\overline{\*u}^{*\prime}\*u_{i,-1}^*\\
        T^{-1}\overline{\*u}_{-1}^{*\prime}\*u_{i,-1}^*
        \end{bmatrix} \right\|^2\notag\\
        &\leq T^{-1}2\left\|(T^{-1}\widehat{\*F}^{*\prime}\widehat{\*F}^*)^{-1}\right\|\frac{1}{N}\sum_{i=1}^N\left\|\underbrace{\sum_{j=0}^{T-2}\widehat{\alpha}_{CCEP}^j\frac{1}{\sqrt{T}}\sum_{t=j+2}^T\varepsilon_{i,t-j-1}\widehat{\*f}_t}_{D_{i,1,2,1}}\right\|^2\notag\\
        &+4\left\|(T^{-1}\widehat{\*F}^{*\prime}\widehat{\*F}^*)^{-1}\right\|\frac{1}{N}\sum_{i=1}^N\left|\underbrace{T^{-1}\overline{\*u}^{*\prime}\*u_{i,-1}^*}_{D_{i,1,2,2}}\right|^2\notag\\
        &+4\left\|(T^{-1}\widehat{\*F}^{*\prime}\widehat{\*F}^*)^{-1}\right\|\frac{1}{N}\sum_{i=1}^N\left|\underbrace{T^{-1}\overline{\*u}_{-1}^{*\prime}\*u_{i,-1}^* }_{D_{i,1,2,3}}\right|^2=o_{P^*}(1).
    \end{align}
    Firstly, this can be seen from $D_{i,1,2,1}$, which can be analyzed along the lines of Lemma A.5 in \cite{gonccalves2004bootstrapping}. By using their strategy, let us fix some finite $m$, so that for fixed $m$ and $i$, we have
    \begin{align}\label{D_121m}
        D_{i,1,2,1,m}=\sum_{j=0}^{m-2}\widehat{\alpha}_{CCEP}^j\frac{1}{\sqrt{T}}\sum_{t=j+2}^T\varepsilon^*_{i,t-j-1}\widehat{\*f}_t&=\sum_{j=0}^{m-2}(\alpha_0)^j\frac{1}{\sqrt{T}}\sum_{t=j+2}^T\varepsilon^*_{i,t-j-1}\widehat{\*f}_t\notag\\
        &-\sum_{j=0}^{m-2}(\widehat{\alpha}_{CCEP}^j-(\alpha_0)^j)\frac{1}{\sqrt{T}}\sum_{t=j+2}^T\varepsilon^*_{i,t-j-1}\widehat{\*f}_t\notag\\
        &=\sum_{j=0}^{m-2}(\alpha_0)^j\frac{1}{\sqrt{T}}\sum_{t=j+2}^T\varepsilon^*_{i,t-j-1}\widehat{\*f}_t + o_{P^*}(1) \notag\\
        &=O_{P^*}(1)
    \end{align}
    due to consistency of $\widehat{\alpha}_{CCEP}$ for for fixed $m$. Clearly, $\frac{1}{\sqrt{T}}\sum_{t=j+2}^T\varepsilon^*_{i,t-j-1}\widehat{\*f}_t=O_{P^*}(1)$ as $\widehat{\*f}_t$ is constant and the residuals are i.i.d. for each $i$ in bootstrap realm. Therefore, we need to check how $D_{i,1,2,1,m}$ approximates $D_{i,1,2,1}$. Following the proof of Lemma A.5. in \cite{gonccalves2004bootstrapping}, we take $\+\lambda\in \mathbb{R}^2$, so that $\+\lambda'\+\lambda=1$, and so by $|xy|\leq (x^2+y^2)/2$ (Young's inequality), we obtain 
    \begin{align}
        E^*\Big[ \+\lambda'(D_{i,1,2,1}-D_{i,1,2,1,m})^2&\Big]=\sum_{j=m}^{T-2}\sum_{r=m}^{T-2}\widehat{\alpha}_{CCEP}^{j+r}E^*\left[\frac{1}{T}\sum_{t=j+2}^T\sum_{s=r+2}^T\varepsilon_{i,t-j-1}^*\varepsilon_{i,s-r-1}^*\+\lambda'\widehat{\*f}_{t}\widehat{\*f}_{s}'\+\lambda \right]\notag\\
        &=\sum_{j=m}^{T-2}\sum_{r=m}^{T-2}\widehat{\alpha}^{j+r}_{CCEP}\frac{1}{T}\sum_{t=\max(j,r)+2}^T\widehat{\varepsilon}_{i,t-j-1}^2\+\lambda'\widehat{\*f}_{t}\widehat{\*f}_{t+r-j}'\+\lambda \notag\\
        &\leq \sum_{j=m}^{T-2}\sum_{r=m}^{T-2}|\widehat{\alpha}_{CCEP}|^{j+r}\frac{1}{2}\frac{1}{T}\sum_{t=\max(j,r)+2}^T\widehat{\varepsilon}_{i,t-j-1}^2(|\+\lambda'\widehat{\*f}_{t}|^2+|\widehat{\*f}_{t+r-j}'\+\lambda|^2)\notag\\
        &\leq  \frac{1}{2}\sum_{j=m}^{T-2}\sum_{r=m}^{T-2}|\widehat{\alpha}_{CCEP}|^{j+r}\left( \frac{1}{T}\sum_{t=\max(j,r)+2}^T\widehat{\varepsilon}_{i,t-j-1}^4\right)^{1/2}\left(\frac{1}{T}\sum_{t=\max(j,r)+2}^T |\+\lambda'\widehat{\*f}_{t}|^4\right)^{1/2}  \notag\\
        & +   \frac{1}{2}\sum_{j=m}^{T-2}\sum_{r=m}^{T-2}|\widehat{\alpha}_{CCEP}|^{j+r}\left( \frac{1}{T}\sum_{t=\max(j,r)+2}^T\widehat{\varepsilon}_{i,t-j-1}^4\right)^{1/2}\left(\frac{1}{T}\sum_{t=\max(j,r)+2}^T |\widehat{\*f}_{t+r-j}'\+\lambda|^4\right)^{1/2}\notag\\
        &\leq  \left\| \+\lambda\right\|^2\left(\sum_{j=m}^{T-2}|\widehat{\alpha}_{CCEP} |^j\right)^2 \left(\frac{1}{T}\sum_{t=1}^T\widehat{\varepsilon}_{i,t}^4 \right)^{1/2} \left(\frac{1}{T}\sum_{t=1}^T\left\|\widehat{\*f}_{t} \right\|^4 \right)^{1/2}\notag\\
        &=o_P(1)
    \end{align}
    because clearly $\lim_{m\to \infty}\limsup_{(N,T)\to \infty}E^*\left[ \+\lambda'(D_{i,1,2,1}-D_{i,1,2,1,m})^2\right]=o_P(1)$, for each $i$ due to consistency of $\widehat{\alpha}_{CCEP}$ and vanishing tail of an infinite sum. We created the last inequality by summing the maximum number of non-negative elements. Next, 
    \begin{align}\label{D122}
        |D_{i,1,2,2}|\leq \frac{1}{\sqrt{N}}\frac{1}{T}\sum_{t=2}^T|\sqrt{N}\overline{u}^*_{t}u_{i,t-1}^*|&\leq \frac{1}{\sqrt{N}}\left( \frac{1}{T}\sum_{t=2}^T\left(\frac{1}{\sqrt{N}}\sum_{j=1}^Nu_{j,t}^* \right)^2\right)^{1/2} \left(\frac{1}{T}\sum_{t=2}^Tu_{i,t-1}^{*2}\right)^{1/2}\notag\\
        &=O_{P^*}(N^{-1/2}),
    \end{align}
    by the results (\ref{u^2_bound}) but for a fixed $i$. Similarly, 
    \begin{align}\label{D123}
        |D_{i,1,2,3}|\leq \frac{1}{\sqrt{N}}\frac{1}{T}\sum_{t=2}^T|\sqrt{N}\overline{u}^*_{t-1}u_{i,t-1}^*|&\leq \frac{1}{\sqrt{N}}\left( \frac{1}{T}\sum_{t=2}^T\left(\frac{1}{\sqrt{N}}\sum_{j=1}^Nu_{j,t-1}^* \right)^2\right)^{1/2} \left(\frac{1}{T}\sum_{t=2}^Tu_{i,t-1}^{*2}\right)^{1/2}\notag\\
        &=O_{P^*}(N^{-1/2})
    \end{align}
    by the same argument. Therefore, by taking averages over $i$ of $o_{P^*}(1)$ terms whose rate does not depend on $i$, we have that 
    \begin{align}
         D_1=\frac{1}{NT}\sum_{i=1}^N\*u_{i,-1}^{*\prime}\*u_{i,-1}^* + o_{P^*}(1)=\frac{1}{NT}\sum_{t=2}^T\sum_{i=1}^Nu_{i,t-1}^{*2}+o_{P^*}(1) 
    \end{align}
    as $(N,T)\to \infty$. \\

    \noindent \textbf{b)} In what follows, we work out the quadratic terms. By using $\overline{\*y}=\overline{\*y}^*-\overline{\*u}^*$, we get that 
    \begin{align}
        |D_2|&\leq \frac{1}{N}\left(\frac{1}{N} \sum_{i=1}^N\widetilde{\gamma}_{1,i}^2\right)\left|NT^{-1} \overline{\*u}_{-1}^{*\prime}\*M_{\widehat{\*F}^*}\overline{\*u}_{-1}^{*}\right|\notag\\
        &\leq \frac{1}{N}\left(\frac{1}{N} \sum_{i=1}^N\widetilde{\gamma}_{1,i}^2\right) \underbrace{\left(\frac{1}{T}\sum_{t=2}^TN\overline{u}_{t-1}^{*2} + \left\| \frac{1}{\sqrt{N}T}\sum_{i=1}^N\widehat{\*F}^{*\prime}\*u_{i,-1}^*\right\|^2\left\|(T^{-1}\widehat{\*F}^{*\prime}\widehat{\*F}^*)^{-1} \right\| \right)}_{O_{P^*}(1)}\notag\\
        &=O_{P^*}(N^{-1})
    \end{align}
  due to (\ref{f*u_-1*}) and because 
       \begin{align}
        E^*\left[\frac{1}{T}\sum_{t=2}^TN\overline{u}_{t-1}^{*2} \right]=\frac{1}{T}\sum_{t=2}^T\sum_{i=1}^NE^*[u_{i,t-1}^{**2}] =\frac{1}{NT}\sum_{t=2}^T\sum_{i=1}^N\widehat{b}_{i,t-1}=O_P(1).
    \end{align}
  Next, 
    \begin{align}
        |D_3|&\leq \left(\frac{1}{N}\sum_{i=1}^N(y_{i,0}-\overline{y}_0\widetilde{\gamma}_{1,i})^2\right)\left(T^{-1}\left|\+\iota_T'\widehat{\*A}'_{CCEP,-1}\widehat{\*A}_{CCEP,-1}\+\iota_T\right| +T^{-1} \left|\+\iota_T'\widehat{\*A}'_{CCEP,-1}\*P_{\widehat{\*F}^*}\widehat{\*A}_{CCEP,-1}\+\iota_T\right|\right)\notag\\
        &\leq  \left(\frac{1}{N}\sum_{i=1}^N(y_{i,0}-\overline{y}_0\widetilde{\gamma}_{1,i})^2\right)\left(\frac{1}{T}\sum_{t=2}^T\widehat{\alpha}_{CCEP}^{2(t-1)} +\frac{1}{T}\left\|\frac{1}{\sqrt{T}}\sum_{t=2}^T\widehat{\alpha}_{CCEP}^{t-1}\widehat{\*f}_t^*\right\|^2\left\|(T^{-1}\widehat{\*F}^{*\prime}\widehat{\*F}^*)^{-1} \right\| \right)\notag\\
        &=O_{P^*}(T^{-1}),
    \end{align}
    due to consistency of $\widehat{\alpha}_{CCEP}$, and 
    \begin{align}
        |D_4|&\leq  \frac{1}{T}\frac{1}{N}\sum_{i=1}^N(\sqrt{T}\widetilde{\gamma}_{2,i})^2\left(\frac{1}{T}\sum_{t=2}^Tb_{t-1}^2\right) + \frac{1}{T}\frac{1}{N}\sum_{i=1}^N(\sqrt{T}\widetilde{\gamma}_{2,i})^2\left\|\frac{1}{T}\sum_{t=2}^Tb_t\widehat{\*f}_t^
        *\right\|^2\left\| (T^{-1}\widehat{\*F}^{*\prime}\widehat{\*F}^*)^{-1}\right\|\notag\\
        &\leq \frac{1}{T}\frac{1}{N}\sum_{i=1}^N(\sqrt{T}\widetilde{\gamma}_{2,i})^2\left(\frac{1}{T}\sum_{t=2}^Tb_{t-1}^2\right) + \frac{1}{T}\frac{1}{N}\sum_{i=1}^N(\sqrt{T}\widetilde{\gamma}_{2,i})^2 \left(\frac{1}{T}\sum_{t=2}^Tb_{t-1}^2\right) \left( \frac{1}{T}\sum_{t=2}^T\left\|\widehat{\*f}^*_t \right\|^2\right)\left\| (T^{-1}\widehat{\*F}^{*\prime}\widehat{\*F}^*)^{-1}\right\|\notag\\
        &=O_{P^*}(T^{-1}).
    \end{align}
    Next, we examine the cross-terms which behave similarly. For example, 
    \begin{align}
        |D_5|&= \left|\frac{1}{NT}\sum_{i=1}^N\*u_{i,-1}^{*\prime}\*M_{\widehat{\*F}^*}\overline{\*y}_{-1}\widetilde{\gamma}_{1,i} \right|=\left|\frac{1}{NT}\sum_{i=1}^N\widetilde{\gamma}_{1,i}\*u_{i,-1}^{*\prime}\*M_{\widehat{\*F}^*}\overline{\*u}^*_{-1} \right|\notag\\
        &\leq \left|\frac{1}{NT}\sum_{i=1}^N\widetilde{\gamma}_{1,i}\*u_{i,-1}^{*\prime}\overline{\*u}_{-1}^*\right|+\left\|\frac{1}{NT}\sum_{i=1}^N\widetilde{\gamma}_{1,i}\*u_{i,-1}^{*\prime}\widehat{\*F}^* \right\|\left\|\frac{1}{NT}\sum_{j=1}^N\widehat{\*F}^{*\prime}\*u_{j,-1}^* \right\|\left\|(T^{-1}\widehat{\*F}^{*\prime}\widehat{\*F}^*)^{-1} \right\|\notag\\
        &=O_{P^*}(N^{-1/2})
    \end{align}
    by (\ref{II_d1}) and (\ref{II_d2}). Next, 
    \begin{align}
        |D_6|&=\left|\frac{1}{NT}\sum_{i=1}^N(y_{i,0}-\overline{y}_0\widetilde{\gamma}_{1,i})\*u_{i,-1}^{*\prime}\*M_{\widehat{\*F}^*}\widehat{\*A}_{CCEP,-1}\+\iota_T \right|\notag\\
        &\leq \left|\frac{1}{NT}\sum_{i=1}^N(y_{i,0}-\overline{y}_0\widetilde{\gamma}_{1,i})\*u_{i,-1}^{*\prime}\widehat{\*A}_{CCEP,-1}\+\iota_T \right| + \left| \frac{1}{NT}\sum_{i=1}^N(y_{i,0}-\overline{y}_0\widetilde{\gamma}_{1,i})\*u_{i,-1}^{*\prime}\*P_{\widehat{\*F}^*}\widehat{\*A}_{CCEP,-1}\+\iota_T\right|\notag\\
        &=|D_{6,1}|+ |D_{6,2}|=O_{P^*}((NT)^{-1/4}), 
    \end{align}
    which comes from the fact that
    \begin{align}
         \mathbb{P}^*\left(|D_{6,1}|>\epsilon \right)&\leq \frac{1}{\epsilon^2}\frac{1}{N^2T^2}\sum_{i=1}^N(y_{i,0}-\overline{y}_0\widetilde{\gamma}_{1,i})^2\sum_{t=2}^T\sum_{s=2}^T\widehat{\alpha}_{CCEP}^{t+s-2}E^*\left[u_{i,t-1}^{**}u_{i,s-1}^{**} \right]\notag\\
         &\leq  \frac{1}{\epsilon^2}\frac{1}{N^2T^2}\sum_{i=1}^N(y_{i,0}-\overline{y}_0\widetilde{\gamma}_{1,i})^2\sum_{t=2}^T\sum_{s=2}^T|\widehat{\alpha}_{CCEP}|^{t+s-2}E^*\left[|u_{i,t-1}^{**}||u_{i,s-1}^{**}| \right]\notag\\
         &\leq  \frac{1}{\epsilon^2}\frac{1}{N^2T^2}\sum_{i=1}^N(y_{i,0}-\overline{y}_0\widetilde{\gamma}_{1,i})^2\sum_{t=2}^T\sum_{s=2}^T|\widehat{\alpha}_{CCEP}|^{t+s-2}E^*[u_{i,t-1}^{**2}]^{1/2}E^*[u_{i,s-1}^{**2}]^{1/2}\notag\\
         &=\frac{1}{\epsilon^2}\frac{1}{N^2}\sum_{i=1}^N(y_{i,0}-\overline{y}_0\widetilde{\gamma}_{1,i})^2\left( \frac{1}{T}\sum_{t=2}^T|\widehat{\alpha}_{CCEP}|^{t-1}\widehat{b}_{i,t-1}^{1/2}\right)\left(\frac{1}{T}\sum_{s=2}^T|\widehat{\alpha}_{CCEP}|^{s-1}\widehat{b}_{i,s-1}^{1/2}\right)\notag\\
         &=\frac{1}{\epsilon^2}\frac{1}{N^2}\sum_{i=1}^N(y_{i,0}-\overline{y}_0\widetilde{\gamma}_{1,i})^2\left( \frac{1}{T}\sum_{t=2}^T|\widehat{\alpha}_{CCEP}|^{t-1}\widehat{b}_{i,t-1}^{1/2}\right)^2\notag\\
         &\leq \frac{1}{\epsilon^2}\frac{1}{N^2}\sum_{i=1}^N(y_{i,0}-\overline{y}_0\widetilde{\gamma}_{1,i})^2 \frac{1}{T}\sum_{t=2}^T\widehat{\alpha}_{CCEP}^{2(t-1)}\widehat{b}_{i,t-1}\notag\\
         &\leq \frac{1}{\sqrt{NT}}\left(\frac{1}{NT}\sum_{t=2}^T\sum_{i=1}^N\widehat{b}_{i,t-1}^2 \right)^{1/2} \frac{1}{\epsilon^2}\frac{1}{N}\sum_{i=1}^N(y_{i,0}-\overline{y}_0\widetilde{\gamma}_{1,i})^2\sum_{t=2}^T\widehat{\alpha}_{CCEP}^{2(t-1)}\notag\\
         &=O_P((NT)^{-1/2}),
    \end{align}
    where we used the fact that $\max_{i,t}\widehat{b}_{i,t-1}\leq \left(\sum_{t=2}^T\sum_{i=1}^N\widehat{b}_{i,t-1}^2 \right)^{1/2}$ for a positive $\widehat{b}_{i,t-1}$. This implies that $|D_{6,1}|=O_{p^{*}}((NT)^{-1/4})$, which is sufficient for our purposes. Moving on,
    \begin{align}
        |D_{6,2}|&\leq \frac{1}{\sqrt{NT}}\left\|\frac{1}{\sqrt{N}T}\sum_{i=1}^N(y_{i,0}-\overline{y}_0\widetilde{\gamma}_{1,i})\*u_{i,-1}^{*\prime}\widehat{\*F}^* \right\|\underbrace{\left\|\frac{1}{\sqrt{T}}\sum_{t=2}^T\widehat{\alpha}_{CCEP}^{t-1}\widehat{\*f}_t^* \right\|}_{O_{P^*}(1)}\left\|(T^{-1}\widehat{\*F}^{*\prime}\widehat{\*F}^*)^{-1} \right\|\notag\\
        &=O_{P^*}((NT)^{-1/2})
    \end{align}
    by the same structure of (\ref{II_d2}). Next, 
    \begin{align}
        |D_7|&=\left|\frac{1}{NT}\sum_{i=1}^N\widetilde{\gamma}_{2,i}\*u_{i,-1}^{*\prime}\*M_{\widehat{\*F}^*}\*b_{-1} \right|\leq \left|\frac{1}{NT^{3/2}}\sum_{i=1}^N(\sqrt{T}\widetilde{\gamma}_{2,i})\*u_{i,-1}^{*\prime}\*b_{-1} \right| + \left|\frac{1}{NT^{3/2}}\sum_{i=1}^N(\sqrt{T}\widetilde{\gamma}_{2,i})\*u_{i,-1}^{*\prime}\*P_{\widehat{\*F}^*}\*b_{-1} \right|\notag\\
        &\leq \left|\frac{1}{N}\sum_{i=1}^N(\sqrt{T}\widetilde{\gamma}_{2,i})\frac{1}{T^{3/2}}\sum_{t=2}^Tu_{i,t-1}^*b_{t-1} \right|\notag\\
        &+  \frac{1}{\sqrt{NT}}\left\|\frac{1}{\sqrt{N}T}\sum_{i=1}^N(\sqrt{T}\widetilde{\gamma}_{2,i})\*u_{i,-1}^{*\prime}\widehat{\*F}^*  \right\| \left\| \frac{1}{T}\sum_{t=2}^T\widehat{\*f}^*_tb_{t-1}\right\|\left\|(T^{-1}\widehat{\*F}^{*\prime}\widehat{\*F}^*)^{-1} \right\|\notag\\
        &\leq \frac{1}{\sqrt{T}}\frac{1}{N}\sum_{i=1}^N|\sqrt{T}\widetilde{\gamma}_{2,i}|\left(\frac{1}{T}\sum_{t=2}^Tu_{i,t-1}^{*2} \right)^{1/2}\left(\frac{1}{T}\sum_{t=2}^Tb_{t-1}^2 \right)^{1/2}\notag\\
        &+\frac{1}{\sqrt{NT}}\left\|\frac{1}{\sqrt{N}T}\sum_{i=1}^N(\sqrt{T}\widetilde{\gamma}_{2,i})\*u_{i,-1}^{*\prime}\widehat{\*F}^*  \right\| \left(\frac{1}{T}\sum_{t=2}^T\left\|\widehat{\*f}^*_t\right\|^2\right)^{1/2}\left(\frac{1}{T}\sum_{t=2}^Tb_{t-1}^2\right)^{1/2}\left\|(T^{-1}\widehat{\*F}^{*\prime}\widehat{\*F}^*)^{-1} \right\|\notag\\
        &\leq \frac{1}{\sqrt{T}}\left(\frac{1}{N}\sum_{i=1}^N(\sqrt{T}\widetilde{\gamma}_{2,i})^2 \right)^{1/2}\left(\frac{1}{NT}\sum_{i=1}^N\sum_{t=2}^Tu_{i,t-1}^{*2} \right)^{1/2}\left(\frac{1}{T}\sum_{t=2}^Tb_{t-1}^2 \right)^{1/2}\notag\\
        &+\frac{1}{\sqrt{NT}}\left\|\frac{1}{\sqrt{N}T}\sum_{i=1}^N(\sqrt{T}\widetilde{\gamma}_{2,i})\*u_{i,-1}^{*\prime}\widehat{\*F}^*  \right\| \left(\frac{1}{T}\sum_{t=2}^T\left\|\widehat{\*f}^*_t\right\|^2\right)^{1/2}\left(\frac{1}{T}\sum_{t=2}^Tb_{t-1}^2\right)^{1/2}\left\|(T^{-1}\widehat{\*F}^{*\prime}\widehat{\*F}^*)^{-1} \right\|\notag\\
        &=O_{P^*}(T^{-1/2})
    \end{align}
    because of (\ref{II_d2}) and the fact that 
    \begin{align}
        E^*\left[\frac{1}{NT}\sum_{i=1}^N\sum_{t=2}^Tu_{i,t-1}^{*2} \right]=\frac{1}{NT}\sum_{i=1}^N\sum_{t=2}^TE^*[u_{i,t-1}^{**2}]=\frac{1}{NT}\sum_{i=1}^N\sum_{t=2}^T\widehat{b}_{i,t-1}=O_P(1). 
    \end{align}
    We move on to $D_8$, where 
    \begin{align}
        |D_8|&= \left|\frac{1}{NT}\sum_{i=1}^N(y_{i,0}-\overline{y}_0\widetilde{\gamma}_{1,i})\widetilde{\gamma}_{1,i}\overline{\*y}_{-1}'\M_{\widehat{\*F}^*}\widehat{\*A}_{CCEP,-1}\+\iota_T \right| \notag\\
        &\leq \left| \frac{1}{N}\sum_{i=1}^N(y_{i,0}-\overline{y}_0\widetilde{\gamma}_{1,i})\widetilde{\gamma}_{1,i}\right|\left|T^{-1}\overline{\*u}_{-1}^*\*M_{\widehat{\*F}^*}\widehat{\*A}_{CCEP,-1}\+\iota_T \right|\notag\\
        &\leq  \left| \frac{1}{N}\sum_{i=1}^N(y_{i,0}-\overline{y}_0\widetilde{\gamma}_{1,i})\widetilde{\gamma}_{1,i}\right|\left(\left|\frac{1}{T}\sum_{t=2}^T\widehat{\alpha}_{CCEP}^{t-1}\overline{u}_{t-1}^*\right| + \left\|\frac{1}{NT}\sum_{j=1}^N\widehat{\*F}^{*\prime}\*u_{j,-1}^* \right\|\left\|T^{-1}\sum_{t=2}^T\widehat{\alpha}_{CCEP}^{t-1}\widehat{\*f}_t^* \right\|\left\|(T^{-1}\widehat{\*F}^{*\prime}\widehat{\*F}^*)^{-1} \right\| \right) \notag\\
        &\leq \frac{1}{\sqrt{NT}}\left( \frac{1}{N}\sum_{i=1}^N(y_{i,0}-\overline{y}_0\widetilde{\gamma}_{1,i})^2\right)^{1/2}\left(\frac{1}{N}\sum_{i=1}^N\widetilde{\gamma}_{1,i}^2 \right)^{1/2}\left(\frac{1}{T}\sum_{t=2}^TN\overline{u}_{t-1}^{*2} \right)^{1/2}\left(\sum_{t=2}^T\widehat{\alpha}_{CCEP}^{2(t-1)} \right)^{1/2} \notag\\
        &+ O_{P^*}((NT)^{-1/2})\notag\\
        &=O_{P^*}((NT)^{-1/2})
    \end{align}
    due to (\ref{f*u_-1*}) and again because of 
    \begin{align}
        E^*\left[\frac{1}{T}\sum_{t=2}^TN\overline{u}_{t-1}^{*2} \right] =\frac{1}{NT}\sum_{t=2}^T\sum_{i=1}^N\widehat{b}_{i,t-1}=O_P(1).
    \end{align}
    In what follows, 
    \begin{align}
        |D_9|&= \left|\frac{1}{NT}\sum_{i=1}^N\widetilde{\gamma}_{1,i}\widetilde{\gamma}_{2,i}\overline{\*y}_{-1}'\*M_{\widehat{\*F}^*}\*b_{-1} \right|= \left|\frac{1}{NT^{3/2}}\sum_{i=1}^N\widetilde{\gamma}_{1,i}(\sqrt{T}\widetilde{\gamma}_{2,i})\overline{\*u}_{-1}^{*\prime}\*M_{\widehat{\*F}^*}\*b_{-1} \right|\notag\\
        &\leq \left|\frac{1}{N}\sum_{i=1}^N\widetilde{\gamma}_{1,i}(\sqrt{T}\widetilde{\gamma}_{2,i}) \right|\left(\frac{1}{\sqrt{T}}\left|\frac{1}{T}\sum_{t=2}^T\overline{u}^*_{t-1}b_{t-1} \right| +\left\|\frac{1}{NT^{3/2}}\sum_{j=1}^N\*u_{j,-1}^{*\prime}\widehat{\*F}^* \right\|\underbrace{\left\|\frac{1}{T}\sum_{t=2}^Tb_{t-1}\widehat{\*f}_t^* \right\|}_{O_{P^*}(1)}\left\|(T^{-1}\widehat{\*F}^{*\prime}\widehat{\*F}^*)^{-1} \right\| \right)\notag\\
        &\leq \frac{1}{\sqrt{NT}}\left(\frac{1}{N}\sum_{i=1}^N\widetilde{\gamma}_{1,i}^2 \right)^{1/2}\left(\frac{1}{N}\sum_{i=1}^N(\sqrt{T}\widetilde{\gamma}_{2,i})^2 \right)^{1/2}\left( \frac{1}{T}\sum_{t=2}^TN\overline{u}_{t-1}^{*2}\right)^{1/2}\left( \frac{1}{T}\sum_{t=2}^Tb_{t-1}^2\right)^{1/2} \notag\\
        &+ O_{P^*}((NT)^{-1/2})\notag\\
        &=O_{P^*}((NT)^{-1/2}).
    \end{align}
    Next we have 
    \begin{align}
        &|D_{10}|= \left|\frac{1}{NT}\sum_{i=1}^N(y_{i,0}-\overline{y}_0\widetilde{\gamma}_{1,i})\widetilde{\gamma}_{2,i}\*b_{-1}'\*M_{\widehat{\*F}^*}\widehat{\*A}_{CCEP,-1}\+\iota_T \right|\notag\\
        &\leq \left|\frac{1}{N}\sum_{i=1}^N (y_{i,0}-\overline{y}_0\widetilde{\gamma}_{1,i})(\sqrt{T}\widetilde{\gamma}_{2,i})\right| \left|T^{-3/2}\*b_{-1}'\*M_{\widehat{\*F}^*}\widehat{\*A}_{CCEP,-1}\+\iota_T \right|\notag\\
        &\leq \frac{1}{\sqrt{T}} \left|\frac{1}{N}\sum_{i=1}^N (y_{i,0}-\overline{y}_0\widetilde{\gamma}_{1,i})(\sqrt{T}\widetilde{\gamma}_{2,i})\right| \left(\underbrace{\left|\frac{1}{T}\sum_{t=2}^T\widehat{\alpha}_{CCEP}^{t-1}b_{t-1}\right|}_{O_{P^*}(T^{-1/2})} +\left\|\frac{1}{T}\sum_{t=2}^Tb_{t-1}\widehat{\*f}_t^*\right\| \underbrace{\left\|\frac{1}{T}\sum_{t=2}^T\widehat{\alpha}_{CCEP}^{t-1}\widehat{\*f}_t^* \right\|}_{O_{P^*}(T^{-1/2})}\left\|(T^{-1}\widehat{\*F}^{*\prime}\widehat{\*F}^*)^{-1} \right\|\right)\notag\\
        &=O_{P^*}(T^{-1}).
    \end{align}
    The rest of cross-terms in the denominator have the same orders by symmetry. 
    \paragraph  {Bootstrap Central Limit Theorem and Bias}\label{4.2.6}
    We have established that 
    \begin{align}
        I=&\frac{1}{\sqrt{NT}}\sum_{t=2}^T\sum_{i=1}^Nu_{i,t-1}^*\varepsilon^*_{i,t}-\frac{1}{\sqrt{NT}}\sum_{i=1}^N\*u_{i,-1}^{*\prime}\*P_{\widehat{\*F}^*}\+\varepsilon_{i}^* + o_{P^*}(1)\notag\\
        &=I_{d,1}-I_{d,2} + o_{P^*}(1).
    \end{align}
    \noindent \textbf{Lemma 3.} \textit{Under Assumptions \ref{ass::1} - \ref{ass::6}, and let the weights be non-multiplicative. Then as $(N,T)\to \infty$ with $NT^{-1}\to \kappa\in (0, \infty)$, we have
    \begin{enumerate}[a)]
        \item $I_{d,1}\to_{d^*}\mathcal{N}(0, \sigma^4\varphi)$, where $\varphi=\frac{1}{1-(\alpha_0)^2}$,
        \item $I_{d,2}\to_{p^*} \sigma^2\sqrt{\kappa}\sum_{h=1}^{\infty}(\alpha_0)^{h-1}\mathrm{tr}\left(\+\Sigma_\*q(h)\+\Sigma_\*q^{-1}\right)$
    \end{enumerate}
    }
    \bigskip 
    \noindent \textbf{Proof.} \textbf{a)} We can re-write $I_{d,1}$ and its counterpart for some fixed $m$ as 
    \begin{align}\label{I_d1}
        &I_{d,1}= \sum_{j=0}^{T-2}(\alpha_0)^j\frac{1}{\sqrt{NT}}\sum_{i=1}^N\sum_{t=j+2}^T\varepsilon_{i,t-j-1}^*\varepsilon_{i,t}^*+\sum_{j=0}^{T-2}(\widehat{\alpha}_{CCEP}^j-(\alpha_0)^j)\frac{1}{\sqrt{NT}}\sum_{i=1}^N\sum_{t=j+2}^T\varepsilon_{i,t-j-1}^*\varepsilon_{i,t}^*\notag\\
        & I_{d,1,m}= \sum_{j=0}^{m-2}(\alpha_0)^j\frac{1}{\sqrt{NT}}\sum_{i=1}^N\sum_{t=j+2}^T\varepsilon_{i,t-j-1}^*\varepsilon_{i,t}^*+\sum_{j=0}^{m-2}(\widehat{\alpha}_{CCEP}^j-(\alpha_0)^j)\frac{1}{\sqrt{NT}}\sum_{i=1}^N\sum_{t=j+2}^T\varepsilon_{i,t-j-1}^*\varepsilon_{i,t}^*\notag\\
        &=I_{d,1,m,1}+ I_{d,1,m,2}
    \end{align}
    where we follow the steps similr to the ones implemented in (\ref{D_121m}). The component $\frac{1}{\sqrt{NT}}\sum_{i=1}^N\sum_{t=j+2}^T\varepsilon_{i,t-j-1}^*\varepsilon_{i,t}^*$ is $O_{P^*}(1)$ and it readily satisfies the CLT by Lemma B.2 in \cite{gonccalves2015bootstrap} under our assumptions, and under $\delta=4$, so that we utilize the eighth moment. In particular, 
    \begin{align}
        \frac{1}{\sqrt{NT}}\sum_{i=1}^N\sum_{t=j+2}^T\varepsilon_{i,t-j-1}^*\varepsilon_{i,t}^*\to_{d^*}\mathcal{N}\left(0, E[\varepsilon_{i,t}^2\varepsilon_{i,t-j-1}^2]\right)\overset{d}{=}\mathcal{N}(0, \sigma^4) 
    \end{align}
    as $(N,T)\to \infty$ under full homoskedasticity and independence over time. Therefore, for a fixed $m$ and due to consistency of $\widehat{\alpha}_{CCEP}$, we have that $|I_{d,1,m,2}|=o_{P^*}(1)$. Then for fixed $m$, we have 
    \begin{align}
        I_{d,1,m,1}\to_{d^*}\mathcal{N}\left(0, \sigma^4\sum_{j=0}^{m-2}a^{2j}\right)
    \end{align}
    as $(N,T)\to \infty$. Therefore, we need to check how $I_{d,1,m}$ approximates $I_{d,1}$ as $m\to \infty$ and $(N,T)\to \infty$, similarly to (\ref{D_121m}). By cross-section independence induced by non-multiplicative weights, we get 
    \begin{align}
        \mathbb{P}^*\left(|I_{d,1,m}- I_{d,1}|>\epsilon \right)&\leq \frac{1}{\epsilon^2}\sum_{j=m}^{T-2}\sum_{r=m}^{T-2}\widehat{\alpha}_{CCEP}^{j+r}E^*\left[\frac{1}{NT}\sum_{i=1}^N\sum_{k=1}^N\sum_{t=j+2}^T\sum_{s=r+2}^T\varepsilon^*_{i,t-j-1}\varepsilon^*_{i,t}\varepsilon^*_{j,s-r-1}\varepsilon^*_{k,s} \right]\notag\\
        &=\frac{1}{\epsilon^2}\sum_{j=m}^{T-2}\widehat{\alpha}_{CCEP}^{2j}\frac{1}{NT}\sum_{i=1}^N\sum_{t=j+2}^T\widehat{\varepsilon}_{i,t-j-1}^2\widehat{\varepsilon}_{i,t}^2\notag\\
        &\leq \frac{1}{\epsilon^2}\sum_{j=m}^{T-2}\widehat{\alpha}_{CCEP}^{2j} \frac{1}{N}\sum_{i=1}^N\left(\frac{1}{T}\sum_{t=j+2}^T\widehat{\varepsilon}^4_{i,t-j-1} \right)^{1/2}\left(\frac{1}{T}\sum_{t=j+2}^T\widehat{\varepsilon}^4_{i,t} \right)^{1/2}\notag\\
        &\leq \left(\frac{1}{NT}\sum_{i=1}^N\sum_{t=1}^T\widehat{\varepsilon}^4_{i,t} \right)\sum_{j=m}^{T-2}\widehat{\alpha}_{CCEP}^{2j}=o_P(1),
    \end{align}
    as $m\to \infty$ with $(N,T)\to \infty$ due to consistency of $\widehat{\alpha}_{CCEP}$ and vanishing tail of an infinite sum. Therefore,
    \begin{align}
        I_{d,1}\to_{d^*}\mathcal{N}\left(0, \sigma^4 \sum_{j=0}^{\infty}(\alpha_0)^{2j}\right)\overset{d}{=}\mathcal{N}\left( 0, \sigma^4\phi\right),
    \end{align}
    where $\phi=\frac{1}{1-(\alpha_0)^2}$. \\

    \noindent \textbf{b)} Here, we analyze $I_{d,2}$, where 
\begin{align}
    I_{d,2}&=\frac{1}{\sqrt{NT}}\sum_{i=1}^N\*u_{i,-1}^{*\prime}\*P_{\widehat{\*F}^*}\+\varepsilon_{i}^* = \frac{1}{\sqrt{NT}}\sum_{i=1}^N\*u_{i,-1}^{*\prime}\*P_{\widehat{\*F}}\+\varepsilon_{i}^*+\frac{1}{\sqrt{NT}}\sum_{i=1}^N\*u_{i,-1}^{*\prime}(\*P_{\widehat{\*F}^*}-\*P_{\widehat{\*F}})\+\varepsilon_{i}^*\notag\\
    &=\frac{1}{\sqrt{NT}}\sum_{i=1}^N\*u_{i,-1}^{*\prime}\*P_{\*Q}\+\varepsilon_{i}^* +\frac{1}{\sqrt{NT}}\sum_{i=1}^N\*u_{i,-1}^{*\prime}(\*P_{\widehat{\*F}}-\*P_{\*Q})\+\varepsilon_{i}^*+\frac{1}{\sqrt{NT}}\sum_{i=1}^N\*u_{i,-1}^{*\prime}(\*P_{\widehat{\*F}^*}-\*P_{\widehat{\*F}})\+\varepsilon_{i}^*\notag\\
    &= I_{d,2,1}+I_{d,2,2}+I_{d,2,3}.
\end{align}
Here, 
    \begin{align}
        I_{d,2,1}&=\sqrt{\frac{N}{T}}\frac{1}{N}\sum_{i=1}^NT^{-1}\*u_{i,-1}^{*\prime}\*Q(T^{-1}\*Q'\*Q)^{-1}\*Q'\+\varepsilon_i^*=\sqrt{\frac{N}{T}}\frac{1}{N}\sum_{i=1}^N\frac{1}{T}\sum_{t=2}^T\sum_{s=2}^Tu_{i,t-1}^*\varepsilon^*_{i,s}\*q_t(T^{-1}\*Q'\*Q)^{-1}\*q_s\notag\\
        &=\sqrt{\frac{N}{T}}\frac{1}{N}\sum_{i=1}^N\frac{1}{T}\sum_{t=2}^T\sum_{s=2}^Tu_{i,t-1}^*\varepsilon^*_{i,s}\mathrm{tr}\left(\*q_s\*q_t'(T^{-1}\*Q'\*Q)^{-1} \right).
    \end{align}
    We further inspect the expected value of $I_{d,2,1}$ conditionally on the data. Note that for any $s<t$, the expectation is non zero. For example,  
    \begin{align}
        &E^*[u_{i,t-1}^*\varepsilon_{i,t-1}^*]=E^*\left[\sum_{l=0}^{t-2}\widehat{\alpha}_{CCEP}^l\omega_{i,t-l-1}\omega_{i,t-1}\widehat{\varepsilon}_{i,t-l-1}\widehat{\varepsilon}_{i,t-1} \right]=\widehat{\varepsilon}_{i,t-1}^2,\notag\\
        &E^*[u_{i,t-1}^*\varepsilon_{i,t-2}^*]=E^*\left[\sum_{l=0}^{t-2}\widehat{\alpha}_{CCEP}^l\omega_{i,t-l-1}\omega_{i,t-2}\widehat{\varepsilon}_{i,t-l-1}\widehat{\varepsilon}_{i,t-2} \right]=\widehat{\alpha}_{CCEP}\widehat{\varepsilon}_{i,t-2}^2\notag\\
        &\vdots \\
        &E^*[u_{i,t-1}^*\varepsilon_{i,2}^*]=E^*\left[\sum_{l=0}^{t-2}\widehat{\alpha}_{CCEP}^l\omega_{i,t-l-1}\omega_{i,2}\widehat{\varepsilon}_{i,t-l-1}\widehat{\varepsilon}_{i,2} \right]=\widehat{\alpha}_{CCEP}^{t-3}\widehat{\varepsilon}_{i,2}^2,
    \end{align}
    and so we can represent the total expectation with 
    \begin{align}
        E^*[I_{d,2,1}]&=\sqrt{\frac{N}{T}}\frac{1}{N}\sum_{i=1}^N \frac{1}{T}\sum_{t=2}^{T}\sum_{s=2}^{t-1}\widehat{\alpha}_{CCEP}^{t-s-1}\widehat{\varepsilon}_{i,s}^2\mathrm{tr}\left(\*q_s\*q_t'(T^{-1}\*Q'\*Q)^{-1}\right)\notag\\
        &=\sqrt{\frac{N}{T}}\frac{1}{N}\sum_{i=1}^N \frac{1}{T}\sum_{t=2}^{T}\sum_{s=2}^{t-1}\widehat{\alpha}_{CCEP}^{t-s-1}\widehat{\varepsilon}_{i,s}^2\mathrm{tr}\left(\*q_s\*q_t'\+\Sigma_{\*q}^{-1}\right) + o_P(1),
    \end{align}
    where we further introduce $E[\*q_s\*q_t']=\+\Sigma_\*q(t-s)$ for $s<t$, which is an absolutely summable covariance matrix. Let $t-s=h$, which means that $h=1,\ldots, T-2$. Then, as $(N,T)\to \infty$, by following the steps in A.8 in \cite{juodis2021robustness}, we obtain
    \begin{align}
         E^*[I_{d,2,1}]&=\sqrt{\frac{N}T}\sum_{h=1}^{T-2}\widehat{\alpha}_{CCEP}^{h-1}\left(\frac{1}{N}\sum_{i=1}^N\frac{1}{T}\sum_{s=2}^{T-h}\widehat{\varepsilon}_{i,s}^2\right)\mathrm{tr}\left(\+\Sigma_\*q(h)\+\Sigma_\*q^{-1}\right)+o_P(1)\notag\\
         &\to_p\sigma^2\sqrt{\kappa}\sum_{h=1}^{\infty}(\alpha_0)^{h-1}\mathrm{tr}\left(\+\Sigma_\*q(h)\+\Sigma_{\*q}^{-1}\right)
    \end{align}
    since $NT^{-1}\to \kappa\in (0, \infty)$. We next use the fact that 
    \begin{align}\label{P_F-P_Q}
\*P_{\widehat{\*F}}-\*P_{\*Q}= \*P_{\widehat{\*F}}-\*P_{\*Q\overline{\*C}}&=\overline{\*V}(\widehat{\*F}'\widehat{\*F})^{-1}\overline{\*V}'+ \overline{\*V}(\widehat{\*F}'\widehat{\*F})^{-1} \overline{\*C}'\*Q' + \*Q\overline{\*C}(\widehat{\*F}'\widehat{\*F})^{-1}\overline{\*V}'\notag\\
&+ \*Q\overline{\*C}\left( (\widehat{\*F}'\widehat{\*F})^{-1} - (\overline{\*C}'\*Q'\*Q\overline{\*C})^{-1}\right)\overline{\*C}'\*Q',
    \end{align}
    and by inserting (\ref{P_F-P_Q}) into $I_{d,2,2}$, we obtain $I_{d,2,2}=I_{d,2,2,1}+I_{d,2,2,2}+I_{d,2,2,3}+I_{d,2,2,4}$. Here, 
    \begin{align}
        |I_{d,2,2,1}|&\leq \frac{1}{N}\sum_{i=1}^N \left\|T^{-1}\*u_{i,-1}^{*\prime}\overline{\*V} \right\|\left\| (T^{-1}\widehat{\*F}'\widehat{\*F})^{-1}\right\|\left\|\sqrt{N}T^{-1/2}\overline{\*V}'\+\varepsilon_i^*\right\|\notag\\
        &=\left\| (T^{-1}\widehat{\*F}'\widehat{\*F})^{-1}\right\|\frac{1}{N}\sum_{i=1}^N\left\|\frac{1}{T}\sum_{t=2}^T\overline{\*v}_tu_{i,t-1}^* \right\|\left\|\frac{1}{\sqrt{T}}\sum_{s=2}^T(\sqrt{N}\overline{\*v}_s)\varepsilon_{i,s}^* \right\|\notag\\
        &\leq \left\| (T^{-1}\widehat{\*F}'\widehat{\*F})^{-1}\right\|\left( \frac{1}{T}\sum_{t=2}^T\left\|\overline{\*v}_t \right\|^2\right)^{1/2}\frac{1}{N}\sum_{i=1}^N\left(\frac{1}{T}\sum_{t=2}^Tu_{i,t-1}^{*2} \right)^{1/2}\left\|\frac{1}{\sqrt{T}}\sum_{s=2}^T(\sqrt{N}\overline{\*v}_s)\varepsilon_{i,s}^* \right\|\notag\\
        &=O_{P^*}(N^{-1/2})
    \end{align}
    because 
    \begin{align}
        E^*\left[ \left\|\frac{1}{\sqrt{T}}\sum_{s=2}^T(\sqrt{N}\overline{\*v}_s)\varepsilon_{i,s}^* \right\|^2\right]=\frac{1}{T}\sum_{s=2}^TN\overline{\*v}_s'\overline{\*v}_s\widehat{\varepsilon}_{i,s}^2\leq \left(\frac{1}{T}\sum_{t=2}^T\widehat{\varepsilon}_{i,s}^4 \right)^{1/2} \left(\frac{1}{T}\sum_{s=2}^T\left\| \sqrt{N}\overline{\*v}_s\right\|^4 \right)^{1/2}=O_{P^*}(1)
    \end{align}
    for every $i$, and therefore we average $O_P(1)$ variables. Next, 
    \begin{align}
        |I_{d,2,2,2}|\leq \left\| (T^{-1}\widehat{\*F}'\widehat{\*F})^{-1}\right\|\frac{1}{N}\sum_{i=1}^N\left\|\underbrace{\frac{1}{\sqrt{T}}\sum_{t=2}^T(\sqrt{N}\overline{\*v}_t)u_{i,t-1}^*}_{:=\*z_i^*} \right\|\left\|\overline{\*C}\right\|\underbrace{\left\|\frac{1}{T}\sum_{s=2}^T\*q_s\varepsilon_{i,t}^* \right\|}_{O_{P^*}(T^{-1/2})}=O_{P^*}(T^{-1/2})
    \end{align}
    because for some fixed $m$, we have that for every $i$
\begin{align}
    \*z^*_{i,m} = \sum_{j=0}^{m-2}\widehat{\alpha}_{CCEP}^j\frac{1}{\sqrt{T}}\sum_{t=j+2}^T\varepsilon_{i,t-j-1}^*(\sqrt{N}\overline{\*v}_t)=O_{P^*}(1)
\end{align}
by the structure of (\ref{D_121m}). Therefore, we again need to check how well $\*z_{i,m}^*$ approximates $\*z_{i}^*$ as $m\to \infty$ and $(N,T)\to \infty$. By following the exact same chain of inequalities in (\ref{D_121m}), for some unit-length $\+\lambda$, we have that 
\begin{align}
    E^*\left[\+\lambda'(\*z_{i}^*-\*z_{i,m}^*)^2 \right] \leq  \left\| \+\lambda\right\|^2\left(\sum_{j=m}^{T-2}|\widehat{\alpha}_{CCEP} |^{j}\right)^2\left(\frac{1}{T}\sum_{t=1}^T\widehat{\varepsilon}_{i,t}^4 \right)^{1/2} \left(\frac{1}{T}\sum_{t=1}^T\left\|(\sqrt{N}\overline{\*v}_t) \right\|^4 \right)^{1/2}=o_P(1)
\end{align}
as $m\to \infty$ with $(N,T)\to \infty$ due to consistency of $\widehat{\alpha}_{CCEP}$ and vanishing tail of an infinite sum. Next, 
\begin{align}
    |I_{d,{2,2,3}}|\leq\left\| (T^{-1}\widehat{\*F}'\widehat{\*F})^{-1}\right\|\left\|\overline{\*C} \right\| \frac{1}{N}\sum_{i=1}^N\left\|\frac{1}{\sqrt{T}}\sum_{t=2}^T\*q_t u_{i,t-1}^* \right\|\left\|\frac{1}{T}\sum_{s=2}^T(\sqrt{N}\overline{\*v}_s) \varepsilon_{i,s}^*\right\|=O_{P^*}(T^{-1/2})
\end{align}
by the same argument used in $I_{d,2,2,2}$. Lastly, 
\begin{align}
    |I_{d,2,2,4}|&\leq \sqrt{\frac{N}{T}}\left\|\overline{\*C} \right\|^2 \frac{1}{N}\sum_{i=1}^N\left\|\frac{1}{\sqrt{T}}\sum_{t=2}^T\*q_t u_{i,t-1}^* \right\|\left\|\frac{1}{\sqrt{T}}\sum_{s=2}^T\*q_s\varepsilon_{i,s}^*\right\|\notag\\
    &\times \left\|  (T^{-1}\widehat{\*F}'\widehat{\*F})^{-1} - (\overline{\*C}'T^{-1}\*Q'\*Q\overline{\*C})^{-1}\right\|\notag\\
    &=o_{P^*}(1),
\end{align}
under $NT^{-1}=O(1)$, which is sufficient for our purposes. Overall, this implies that 
\begin{align}
|I_{d,2,2}|=o_{P^*}(1).
\end{align}
We move on to $I_{d,2,3}$, where we need to use the decomposition based on the expression of $\widehat{\*F}^*$ in (\ref{hatF*}):
\begin{align}
\*P_{\widehat{\*F}^*}-\*P_{\widehat{\*F}}&=\overline{\*V}^*(\widehat{\*F}^{*\prime}\widehat{\*F}^*)^{-1}\overline{\*V}^{*\prime}+\overline{\*V}^*(\widehat{\*F}^{*\prime}\widehat{\*F}^*)^{-1}\widehat{\*F}'+\widehat{\*F}(\widehat{\*F}^{*\prime}\widehat{\*F}^*)^{-1}\overline{\*V}^{*\prime}\notag\\
&+\widehat{\*F}\left((\widehat{\*F}^{*\prime}\widehat{\*F}^*)^{-1}-(\widehat{\*F}'\widehat{\*F})^{-1} \right)\widehat{\*F}',
\end{align}
which produces $I_{d,2,3}=I_{d,2,3,1}+I_{d,2,3,2}+I_{d,2,3,3}+I_{d,2,3,4}$, when inserted into $I_{d,2,3}$. Before engaging in the term-by-term analysis, it is very useful to obtain a refined rate for $\frac{1}{T}\sum_{t=2}^Tu_{i,t-1}^*\sqrt{N}\overline{\*v}_{t}^*$, which is a component that will determine the asymptotic behavior of the upcoming terms. In particular, it is sufficient to look at its second coordinate as it retains the most dependence conditionally on the data. Particularly, 
\begin{align}
    \frac{1}{T}\sum_{t=2}^Tu_{i,t-1}^*\sqrt{N}\overline{u}_{t-1}^*&=\frac{1}{T\sqrt{N}}\sum_{t=2}^T\sum_{j=1}^Nu_{i,t-1}^*u_{j,t-1}^*=\frac{1}{\sqrt{N}}\frac{1}{T}\sum_{t=2}^Tu_{i,t-1}^{*2} + \frac{1}{\sqrt{N}T}\sum_{t=2}^T\sum_{j\neq i}^Nu_{i,t-1}^*u_{j,t-1}^*\notag\\
    &=a+b=b + O_{P^*}(N^{-1/2}),
\end{align}
since the term $a$ is clearly negligible for every $i$. Now, we need to show that $b$ is negligible as well. For this, we will use the fact that $|\widehat{\alpha}_{CCEP}-\alpha_0|=O_P((NT)^{-1/2})$. Now, let $u_{i,t-1}^*=u_{i,t-1}^*(\widehat{\alpha})$, because it is a function of $\widehat{\alpha}_{CCEP}$. Then, we can proceed with the Taylor approximation around the true $\alpha_0$:
\begin{align}\label{u_star_taylor_approx}
    u_{i,t-1}^*(\widehat{\alpha})&= u_{i,t-1}^*(\alpha_0)+(\widehat{\alpha}_{CCEP}-\alpha_0)\sum_{l=1}^{t-2}l(\alpha_0)^{l-1}\varepsilon_{i,t-l-1}^*+o_{P^*}((NT)^{-1/2})=u_{i,t-1}^*(\alpha_0)+R^*_{i,t-1}\notag\\
    &+o_{P^*}((NT)^{-1/2}),
\end{align}
where $|R^*_{i,t-1}|=O_{P^*}((NT)^{-1/2})$, because the sum over $l$ is finite due to exponential decay ($|\alpha_0|<1$) dominating the polynomial expansion. In particular, 
\begin{align}
    &E^*\left[\left| (\widehat{\alpha}_{CCEP}-\alpha_0)\sum_{l=1}^{t-2}l(\alpha_0)^{l-1}\varepsilon_{i,t-l-1}^* \right|^2\right]= (\sqrt{NT}(\widehat{\alpha}_{CCEP}-\alpha_0))^2\frac{1}{NT}\sum_{l=1}^{t-2}l^2(\alpha_0)^{2(l-1)}\widehat{\varepsilon}_{i,t-l-1}^2\notag\\
    &= (\sqrt{NT}(\widehat{\alpha}_{CCEP}-\alpha_0))^2\left(\frac{1}{NT}\sum_{l=1}^{t-2}l^2(\alpha_0)^{2(l-1)} \varepsilon_{i,t-l-1}^2
    +C_{N,T}^{-1}\frac{1}{NT}\sum_{l=1}^{t-2}l^2(\alpha_0)^{2(l-1)}C_{N,T}(\widehat{\varepsilon}_{i,t-l-1}^2 - \varepsilon_{i,t-l-1}^2)\right)\notag\\
    &=O_P((NT)^{-1}).
\end{align}
Here we used $|\widehat{\varepsilon}_{i,t}^2-\varepsilon_{i,t}^2|=O_P(N^{-1/2})+O_P(T^{-1/2})=O_P(C_{N,T}^{-1})$ for each $i,t$ based on (\ref{resid_expansion})\footnote{Based on this expansion, we can write $\widehat{\varepsilon}_{i,t}^2=\varepsilon_{i,t}^{2}+2\varepsilon_{i,t}b_{i,t}+b_{i,t}^2$, where $|b_{i,t}|=O_P(C_{N,T}^{-1})$. Thus $|\widehat{\varepsilon}_{i,t}^2-\varepsilon_{i,t}^{2}|=O_P(C_{N,T}^{-1})$, as well. In particular, the leading terms in $b_{i,t}$ are $\overline{\varepsilon}_t$ and $\widehat{\*f}_t'\left(T^{-1}\widehat{\*F}'\widehat{\*F} \right)^{-1}\frac{1}{T}\sum_{s=2}^T\widehat{\*f}_s\varepsilon_{i,s}$.}, and the fact that $\sum_{l=1}^{t-2}l^2(\alpha_0)^{2(l-1)}\leq \sum_{l=1}^{\infty}l^2(\alpha_0)^{2(l-1)}<\infty$, leading to the rate of the expression in parentheses equaling this\footnote{Using $\widehat{\varepsilon}_{i,t}^2=\varepsilon_{i,t}^{2}+2\varepsilon_{i,t}b_{i,t}+b_{i,t}^2$, where $|b_{i,t}|=O_P(C_{N,T}^{-1})$, we have $C_{N,T}^{-1}\sum_{l=1}^{t-2}l^2(\alpha_0)^{2(l-1)}C_{N,T}(\widehat{\varepsilon}_{i,t-l-1}^2 - \varepsilon_{i,t-l-1}^2)=2C_{N,T}^{-1}\sum_{l=1}^{t-2}l^2(\alpha_0)^{2(l-1)}C_{N,T}\varepsilon_{i,t-l-1}b_{i,t-l-1}+C_{N,T}^{-2}\sum_{l=1}^{t-2}l^2(\alpha_0)^{2(l-1)}C_{N,T}^2b_{i,t-l-1}^2=1+2$, where $2=C_{N,T}^{-2}\sum_{l=1}^{t-2}l^2(\alpha_0)^{2(l-1)}C_{N,T}^2b_{i,t-l-1}^2\leq \left(C_{N,T}^{-2}\sum_{l=1}^Tl^4(\alpha_0)^{4(l-1)}\right)^{1/2}\left((C_{N,T}^{-2}T)\frac{1}{T}\sum_{l=1}^TC_{N,T}^4b_{i,l}^4\right)^{1/2}=O_P(C_{N,T}^{-1})$ under $TN^{-1}=O(1)$. As of 1, from (\ref{resid_expansion}) we know the leading terms that generate $O_P(C_{N,T}^{-1})$: $\overline{\varepsilon}_t$ and $\widehat{\*f}_t'\left(T^{-1}\widehat{\*F}'\widehat{\*F} \right)^{-1}\frac{1}{T}\sum_{s=2}^T\widehat{\*f}_s\varepsilon_{i,s}$. Thus, $1$ is driven by $E\left[\left|N^{-1/2}\sum_{l=1}^{t-2}l^2(\alpha_0)^{2(l-1)}\varepsilon_{i,t-l-1}\sqrt{N}\overline{\varepsilon}_{t-l-1} \right| \right]\leq BN^{-1/2}\sum_{l=1}^{t-2}l^2(\alpha_0)^{2(l-1)}=O(N^{-1/2})$, where $B\geq E[\varepsilon_{i,t-l-1}^2]^{1/2}E[\sqrt{N}\overline{\varepsilon}_{t-l-1}^2]^{1/2}$ due to uniformly bounded moments up to 8. Also, $\left|\sum_{l=1}^{t-2}l^2(\alpha_0)^{2(l-1)}\varepsilon_{i,t-l-1}\widehat{\*f}_{t-l-1}'\left(T^{-1}\widehat{\*F}'\widehat{\*F} \right)^{-1}\frac{1}{T}\sum_{s=2}^T\widehat{\*f}_s\varepsilon_{i,s} \right| \leq O_P(1) \frac{1}{\sqrt{T}}\sum_{l=1}^{T}l^2(\alpha_0)^{2(l-1)}|\varepsilon_{i,l}|\left\|\widehat{\*f}_l \right\|=O_P(T^{-1/2})$ since we can uniformly bound the expectation of the second component of the product.}. Therefore, we have that 
\begin{align}
    b= \frac{1}{\sqrt{N}T}\sum_{t=2}^T\sum_{j\neq i}^Nu_{i,t-1}^*u_{j,t-1}^*&=\underbrace{\frac{1}{\sqrt{N}T}\sum_{t=2}^T\sum_{j\neq i}^Nu_{i,t-1}^*(\alpha_0)u_{j,t-1}^*(\alpha_0)}_{b(\alpha_0)}\notag\\
    &+ \frac{1}{NT^{3/2}}\sum_{t=2}^T\sum_{j\neq i}^Nu_{i,t-1}^*(\sqrt{NT}R^*_{j,t-1})\notag\\
    &+\frac{1}{NT^{3/2}}\sum_{t=2}^T\sum_{j\neq i}^Nu_{j,t-1}^*(\sqrt{NT}R^*_{i,t-1})\notag\\
    &+\frac{1}{N^{3/2}T^{2}}\sum_{t=2}^T\sum_{j\neq i}^N(NTR^*_{i,t-1}R^*_{j,t-1})+o_{P^*}(1)\notag\\
    &=b(\alpha_0)+o_{P^*}(1),
\end{align}
where it is clear that the remaining terms are negligible due to cross-section independence, and therefore the ``biggest'' variance will expand as $N^{-2}T^{-3}(NT^2)=o(1)$. Let $\widehat{g}_{i,t-1}=\sum_{l=1}^{t-2}l^2(\alpha_0)^{2(l-1)}\widehat{\varepsilon}_{i,t-l-1}^2$, Then, for example, 
\begin{align}\label{taylor_remainder_var}
    &E^*\left[\left| \frac{1}{NT^{3/2}}\sum_{t=2}^T\sum_{j\neq i}^Nu_{i,t-1}^*(\sqrt{NT}R^*_{j,t-1}) \right|^2 \right]=\frac{1}{N^2T^3}\sum_{j\neq i}^N\sum_{t=2}^T\sum_{s=2}^TE^*[u_{i,t-1}^*u_{i,s-1}^*]E^*[NTR^*_{j,t-1}R^*_{j,s-1}]\notag\\
    &\leq \frac{1}{N^2T^{3}}\sum_{j\neq i}^N\sum_{t=2}^T\sum_{s=2}^T E^*[|u_{i,t-1}^*|^2]^{1/2}E^*[|u_{i,s-1}^*|^2]^{1/2}E^*[NT|R^*_{j,t-1}|^2]^{1/2}E^*[NT|R^*_{j,s-1}|^2]^{1/2}\notag\\
    &=\frac{1}{NT} \left|\sqrt{NT}(\widehat{\alpha}_{CCEP}-\alpha_0)\right|^2 \frac{1}{N}\sum_{j\neq i}^N \left(\frac{1}{T}\sum_{t=2}^T\widehat{b}_{i,t-1}^{1/2}E^*\left[\left|\sum_{l=1}^{t-2}l(\alpha_0)^{l-1}\varepsilon_{j,t-l-1}^* \right|^2 \right]^{1/2} \right)^2\notag\\
    &=O_P((NT)^{-1})\frac{1}{N}\sum_{j\neq i}^N\left(\frac{1}{T}\sum_{t=2}^T\widehat{b}_{i,t-1}^{1/2}\widehat{g}_{j,t-1} ^{1/2}\right)^2\notag\\
    &\leq O_P((NT)^{-1}) \left(\frac{1}{T}\sum_{t=2}^T\widehat{b}_{i,t-1}\right)\frac{1}{NT}\sum_{j\neq i}^{N}\sum_{t=2}^T\widehat{g}_{j,t-1}\notag\\
    &=O_P((NT)^{-1}) \left(\frac{1}{T}\sum_{t=2}^T\widehat{b}_{i,t-1}\right) \sum_{l=1}^{T-2}l^2(\alpha_0)^{2(l-1)}\frac{1}{N}\sum_{j\neq i}^N\frac{1}{T}\sum_{t=l+2}^T\widehat{\varepsilon}_{j,t-l-1}^2\notag\\
    &\leq 
    O_P((NT)^{-1}) \left(\frac{1}{T}\sum_{t=2}^T\widehat{b}_{i,t-1}\right) \left(\frac{1}{NT}\sum_{i=1}^N\sum_{t=2}^T\widehat{\varepsilon}^2_{i,t} \right) \sum_{l=1}^{\infty}l^2(\alpha_0)^{2(l-1)}=O_P((NT)^{-1}).
\end{align}
Now, due to cross-section independence, we have 
\begin{align}\label{E*(b0^2)}
E^*\left[b(\alpha_0)^2 \right]&=\frac{1}{NT^2}\sum_{j\neq i}^N\sum_{t=2}^T\sum_{s=2}^TE^*[u_{i,t-1}^*(\alpha_0)u_{i,s-1}^*(\alpha_0)]E^*[u_{j,t-1}^*(\alpha_0)u_{j,s-1}^*(\alpha_0)]\notag\\
&=\frac{1}{NT^2}\sum_{j\neq i}^N \sum_{t=2}^T\sum_{s=2}^T\underbrace{\left(\sum_{l=0}^{t-2}\sum_{k=0}^{s-2}(\alpha_0)^{l+k}E^*[\varepsilon_{i,t-l-1}^*\varepsilon_{i,s-k-1}^*] \right) \left(\sum_{m=0}^{t-2}\sum_{r=0}^{s-2}(\alpha_0)^{m+r}E^*[\varepsilon_{j,t-m-1}^*\varepsilon_{j,s-r-1}^*] \right)}_{\textit{(non-zero when $\underbrace{t-l=s-k}_{k=s-t+l}$ and $\underbrace{t-m=s-r}_{r=s-t+m}$)}}\notag\\
&= \frac{1}{NT^2}\sum_{j\neq i}^N\sum_{t=2}^T\sum_{s=2}^T\left(\sum_l(\alpha_0)^{2l+s-t}\widehat{\varepsilon}_{i,t-l-1}^2 \right) \left(\sum_m(\alpha_0)^{2m+s-t}\widehat{\varepsilon}_{j,t-m-1}^2 \right) \notag\\
&=\frac{1}{NT^2}\sum_{j\neq i}^N\sum_{t=2}^T\sum_{s=2}^T\left(\sum_l(\alpha_0)^{2l+s-t}\varepsilon_{i,t-l-1}^2 \right) \left(\sum_m(\alpha_0)^{2m+s-t}\varepsilon_{j,t-m-1}^2 \right)+O_P(C_{N,T}^{-1}),
\end{align}
because again $|\widehat{\varepsilon}_{i,t}^2-\varepsilon_{i,t}^2|=O_P(C_{N,T}^{-1})$ for each $i,t$ based on (\ref{resid_expansion})\footnote{Note that we can safely implement the replacement, because $\left|C_{N,T}^{-1}[\sum_{l} (\alpha_0)^{2l+s-t}C_{N,T}(\widehat{\varepsilon}^2_{i,t-l-1}-\varepsilon^2_{i,t-l-1})\right|\leq (\alpha^{0})^{s-t}C_{N,T}^{-1}\sum_{l=0}^\infty(\alpha_0)^{2l}C_{N,T}|\widehat{\varepsilon}^2_{i,t-l-1}-\varepsilon^2_{i,t-l-1}|=O(C_{N,T}^{-1})$ for each $i,s,t$ by employing equivalent decompositions and tracking leading terms similarly to Footnote 14.}. Therefore, the unconditional expectation is:
\begin{align}
    E\left[\left| E^*\left[b(\alpha_0)^2\right]\right|\right]&\leq\sigma^4\left(\frac{N-1}{N}\right)\frac{1}{T^2}\sum_{t=2}^T\sum_{s=2}^T|\alpha_0|^{2|s-t|}\left(\sum_{m=0}^\infty |\alpha_0|^{2m}\right)^2+O(C_{N,T}^{-1})\notag\\
    &=\left(\frac{N-1}{N} \right) \frac{\sigma^4}{(1-(\alpha_0)^2)^2}\frac{1}{T^2}\sum_{t=2}^T\sum_{s=2}^T|\alpha_0|^{2|s-t|}+O(C_{N,T}^{-1})\notag\\
    &=\left(\frac{N-1}{N} \right) \frac{\sigma^4}{(1-(\alpha_0)^2)^2}\frac{1}{T^2}\sum_{h}(T-|h|)|\alpha_0|^{2|h|}=O(C_{N,T}^{-1}),
\end{align}
because the sum over $h$ expands with $T$. Therefore, we overall have that 
\begin{align}\label{u_tbaru_t}
    \left| \frac{1}{T}\sum_{t=2}^Tu_{i,t-1}^*\sqrt{N}\overline{u}_{t-1}^*\right|=O_{P^*}(N^{-1/4}) + O_{P^*}(T^{-1/4})=O_{P^*}(C_{N,T}^{-1/2}).
\end{align}
for every $i$.\\

\noindent Next, we can use these results in $I_{d,2,3,1}$. We can bound it with 
\begin{align}
    |I_{d,2,3,1}|&\leq \sqrt{\frac{T}{N}}\left\|(T^{-1}\widehat{\*F}^{*\prime}\widehat{\*F}^*)^{-1}\right\|\frac{1}{N}\sum_{i=1}^N\left\|T^{-1}\*u_{i,-1}^{*\prime }\sqrt{N}\overline{\*V}^* \right\|\left\|\sqrt{N}T^{-1}\overline{\*V}^{*\prime}\+\varepsilon_{i}^* \right\|\notag\\
    &\leq \left\|(T^{-1}\widehat{\*F}^{*\prime}\widehat{\*F}^*)^{-1}\right\| \sqrt{\frac{T}{N}}\frac{1}{N}\sum_{i=1}^N \left\|\begin{bmatrix} T^{-1}\sum_{t=2}^Tu_{i,t-1}^*\sqrt{N}\overline{u}_{t}^*\\
   T^{-1}\sum_{t=2}^Tu_{i,t-1}^*\sqrt{N}\overline{u}_{t-1}^*
    \end{bmatrix} \right\|\left\|\begin{bmatrix} 
    T^{-1}\sum_{s=2}^T\sqrt{N}\overline{u}_{s}^*\varepsilon^*_{i,s}\\
    T^{-1}\sum_{s=2}^T\sqrt{N}\overline{u}_{s-1}^*\varepsilon^*_{i,s}
    \end{bmatrix} \right\| \notag\\
    &=O_{P^*}(T^{-3/4})+O_{P^*}(N^{-1/4}T^{-1/2}),
\end{align}
 because $\left| T^{-1}\sum_{s=2}^T\sqrt{N}\overline{u}_{s-1}^*\varepsilon^*_{i,s}\right|=O_{P^*}(T^{-1/2})$ at least as it is an MDS for each $i$ with respect to the bootstrap measure. However, the order is driven by the second coordinate: 
\begin{align}
    \left|\frac{1}{T}\sum_{s=2}^T\sqrt{N}\overline{u}_{s}^*\varepsilon^*_{i,s}\right|&=\left|\frac{1}{\sqrt{N}T}\sum_{s=2}^T\sum_{j=1}^Nu_{j,s}^*\varepsilon_{i,s}^* \right|=\frac{1}{\sqrt{N}}\left|\frac{1}{T}\sum_{s=2}^Tu_{i,s}^*\varepsilon_{i,s}^* \right|+\left|\frac{1}{\sqrt{N}T}\sum_{t=2}^T\sum_{j\neq i}^Nu_{j,s}^*\varepsilon^*_{i,s} \right|\notag\\
    &\leq \frac{1}{\sqrt{N}}\left( \frac{1}{T}\sum_{s=2}^Tu_{i,s}^{*2}\right)^{1/2}\left(\frac{1}{T}\sum_{s=2}^T\varepsilon_{i,s}^{*2} \right)^{1/2} + \left|\frac{1}{\sqrt{N}T}\sum_{s=2}^T\sum_{j\neq i}^Nu_{j,s}^*\varepsilon^*_{i,s} \right|\notag\\
    &=O_{P^*}(N^{-1/2}) +O_{P^*}(T^{-1/2})
\end{align}
 for each $i$, because the second component is $O_{P^*}(T^{-1/2})$ as its summands are i.i.d. under bootstrap measure for $j\neq i$ . We now move on to $I_{d,2,3,2}$:
\begin{align}
    |I_{d,2,3,2}|\leq \left\|(T^{-1}\widehat{\*F}^{*\prime}\widehat{\*F}^*)^{-1}\right\|\frac{1}{N}\sum_{i=1}^N\left\|\frac{1}{T}\sum_{t=2}^Tu_{i,t-1}^*\sqrt{N}\overline{\*v}_{t}^* \right\|\left\|\frac{1}{\sqrt{T}}\sum_{s=2}^T\widehat{\*f}_t\varepsilon_{i,s}^* \right\|=O_{P^*}(C_{N,T}^{-1/2}),
\end{align}
which comes directly from (\ref{u_tbaru_t}) and the fact that 
\begin{align}
    E^*\left[\left\|\frac{1}{\sqrt{T}}\sum_{s=2}^T\widehat{\*f}_s\varepsilon_{i,s}^*\right\|^2 \right]=\frac{1}{T}\sum_{s=2}^T\widehat{\*f}_s'\widehat{\*f}_s\widehat{\varepsilon}_{i,s}^2\leq \left(\frac{1}{T}\sum_{s=2}^T\left\|\widehat{\*f}_s \right\|^4 \right)^{1/2} \left(\frac{1}{T}\sum_{s=2}^T\widehat{\varepsilon}_{i,s}^4 \right)^{1/2}=O_P(1). 
\end{align}
Next, we have 
\begin{align}
    |I_{d,2,3,3}|&\leq \left\|(T^{-1}\widehat{\*F}^{*\prime}\widehat{\*F}^*)^{-1} \right\|\frac{1}{N}\sum_{i=1}^N\underbrace{\left\| \frac{1}{\sqrt{T}}\sum_{t=2}^Tu_{i,t-1}^*\widehat{\*f}_t\right\|}_{O_{P^*}(1)}\left\|\frac{1}{T}\sum_{s=1}^T\sqrt{N}\overline{\*v}_s^*\varepsilon_{i,s}^* \right\|\notag\\
    &= O_{P^*}(N^{-1/2})+  O_{P^*}(T^{-1/2})
\end{align}
by the same argument as (\ref{u_tbaru_t}). Lastly, 
\begin{align}
    |I_{d,2,3,4}|&\leq \sqrt{\frac{N}{T}}\frac{1}{N}\sum_{i=1}^N\left\| \frac{1}{\sqrt{T}}\sum_{t=2}^Tu_{i,t-1}^*\widehat{\*f}_t\right\|\left\|\frac{1}{\sqrt{T}}\sum_{s=2}^T\widehat{\*f}_t\varepsilon^*_{i,s} \right\| \left\| (T^{-1}\widehat{\*F}^{*\prime}\widehat{\*F}^*)^{-1} - (T^{-1}\widehat{\*F}'\widehat{\*F})^{-1}\right\|\notag\\
    &=o_{P^*}(1),
\end{align}
which is sufficient for our purposes under $NT^{-1}=O(1)$. 
\paragraph  {Bootstrap Distribution of CCEP: AR(1)}\label{4.2.7}
Here, we combine the results from Lemma 1 - 3. \\

\noindent \textbf{Theorem 3.1} \textit{Let conditions of Lemma 2 and 3 hold, then as $(N,T)\to \infty$ with $NT^{-1}\to \kappa\in (0, \infty)$, we have } 
\begin{align}
    \sqrt{NT}(\widehat{\alpha}_{CCEP}^*-\widehat{\alpha}_{CCEP})\to_{d^*}\mathcal{N}\left(0, \phi^{-1} \right) - \sqrt{\kappa}\phi^{-1}b,
\end{align}
\textit{where $b=\sum_{h=1}^{\infty}(\alpha_0)^{h-1}\mathrm{tr}\left(\+\Sigma_\*q(h)\+\Sigma_\*q^{-1}\right)$}, and so 
\begin{align*}
    \sup_{x\in \mathbb{R}}\left|\mathbb{P}^*\left( \sqrt{NT}(\widehat{\alpha}_{CCEP}^*-\widehat{\alpha}_{CCEP})\leq x \right)-\mathbb{P}\left(\sqrt{NT}(\widehat{\alpha}_{CCEP}-\alpha_0)\leq x \right) \right|\to_p0.
\end{align*}
\\
\noindent \textbf{Proof.} By Lemma 2, we have that 
\begin{align}
    \frac{1}{NT}\sum_{i=1}^N\*y^{*\prime}_{i,-1}\*M_{\widehat{\*F}^*}\*y_{i,-1}^*=\frac{1}{NT}\sum_{i=1}^N\sum_{t=2}^Tu_{i,t-1}^{*2}+o_{P^*}(1) \to_{p^*}\frac{\sigma^2}{1-(\alpha_0)^2}=\sigma^2\phi, 
\end{align}
where the convergence is the results of Lemma B.3 (a) in \cite{gonccalves2015bootstrap}. Then by Lemma 1 in connection to 2 and 3 and Continuous Mapping Theorem, we have that 
\begin{align}
     \sqrt{NT}(\widehat{\alpha}_{CCEP}^*-\widehat{\alpha}_{CCEP})&=\left(\sigma^2\phi \right)^{-1}\left( I_{d,1}-I_{d,2,1}\right)+o_{P^*}(1)\notag\\
     &\to_{d^*} \left(\sigma^2\phi \right)^{-1}\mathcal{N}\left(0, \sigma^4\phi \right)-\left(\sigma^2\phi \right)^{-1}\sigma^2\sqrt{\kappa}\sum_{h=1}^{\infty}(\alpha_0)^{h-1}\mathrm{tr}\left(\+\Sigma_{\*q}(h)\+\Sigma_\*q^{-1}\right)\notag\\
     &\overset{d}{=} \mathcal{N}\left(0, \phi^{-1} \right) - \sqrt{\kappa}\phi^{-1}b
\end{align}
as intended. The final distribution approximation statement follows from Polya's Theorem and the argument analogous to the one in the proof of Corollary 3.1 in \cite{Goncalves2014}. 
\paragraph {Bootstrap DGP: Sophisticated Method}\label{4.2.8}
When we reconstruct the DGP in bootstrap using the sophisticated method, we get 
\begin{align}
    y_{i,t}^*=\widehat{\alpha}_{CCEP}y_{i,t-1}^*+  \widehat{\gamma}_{\widehat{\+\alpha},i}\widehat{f}_{\widehat{\+\alpha},t} + \varepsilon_{i,t}^*,
\end{align}
where clearly now the estimated loading is a function of $\widehat{\+\alpha}$, as well. Note that 
\begin{align}
\widehat{\gamma}_{\widehat{\+\alpha},i}=\left( \widehat{\*f}'_{\widehat{\+\alpha}}\widehat{\*f}_{\widehat{\+\alpha}}\right)^{-1}\widehat{\*f}'_{\widehat{\+\alpha}}(\*y_i-\widehat{\alpha}_{CCEP}\*y_{i,-1}),
\end{align}
which implies that the average of the estimated loading is given by
\begin{align}\label{gamma_delta_avg}
    \overline{\widehat{\gamma}}_{\widehat{\+\alpha}}=\left( \widehat{\*f}'_{\widehat{\+\alpha}}\widehat{\*f}_{\widehat{\+\alpha}}\right)^{-1}\widehat{\*f}'_{\widehat{\+\alpha}}(\overline{\*y}-\widehat{\alpha}_{CCEP}\overline{\*y}_{-1})=\left( \widehat{\*f}'_{\widehat{\+\alpha}}\widehat{\*f}_{\widehat{\+\alpha}}\right)^{-1}\widehat{\*f}'_{\widehat{\+\alpha}}\widehat{\*f}_{\widehat{\+\alpha}}=1.
\end{align}
Therefore, similarly to the naive reconstruction method, we can solve for $\widehat{\*f}_{\widehat{\+\alpha}}$ (in stacked notation):
\begin{align}
    \widehat{\*f}_{\widehat{\+\alpha}}=\overline{\*y}^*-\overline{\*y}^*_{-1}\widehat{\alpha}_{CCEP}-\overline{\+\varepsilon}^*,
\end{align}
and so 
\begin{align}
    \*y^*_i=\*y_{-1}^*\widehat{\alpha}_{CCEP}+(\overline{\*y}^*-\overline{\*y}^*_{-1}\widehat{\alpha}_{CCEP})\widehat{\gamma}_{\widehat{\+\alpha},i}-\overline{\+\varepsilon}^*\widehat{\gamma}_{\widehat{\+\alpha},i}+\+\varepsilon_i^*,
\end{align}
which is the bootstrap DGP similar to the one under the naive method, but here $\widetilde{\gamma}_{2,i}=0$ immediately. This leads to 
\begin{align}\label{sophisticated_CCEP}
     \sqrt{NT}(\widehat{\alpha}^*_{CCEP}-\widehat{\alpha}_{CCEP})=\left( \frac{1}{NT}\sum_{i=1}^N\*y_{i,-1}^{*\prime}\*M_{\widehat{\*F}^*}\*y_{i,-1}^*\right)^{-1} \left(\frac{1}{\sqrt{NT}}\sum_{i=1}^N\*y_{i,-1}^{*\prime}\*M_{\widehat{\*F}^*}\+\varepsilon_i^*-\frac{1}{\sqrt{NT}}\sum_{i=1}^N\*y_{i,-1}^{*\prime}\*M_{\widehat{\*F}^*}\overline{\+\varepsilon}^*\widehat{\gamma}_{\widehat{\+\alpha},i}\right).
\end{align}
Also, due to (\ref{gamma_delta_avg}), we still have that $\overline{\*y}^*=\overline{\*y}+\overline{\*u}^*$ if we use $y_{i,0}^*=y_{i,0}$. Therefore, we can once again expand $y_{i,t}^*$:
\begin{align}
   y_{i,t}^*&= \widehat{\gamma}_{\widehat{\+\alpha},i}\sum_{j=0}^{t-1}(\widehat{\alpha}_{CCEP})^j(\overline{y}_{t-j}-\widehat{\alpha}_{CCEP}\overline{y}_{t-j-1}) +\widehat{\alpha}_{CCEP}^ty_{i,0}+\underbrace{\sum_{j=0}^{t-1}(\widehat{\alpha}_{CCEP})^j\varepsilon_{i,t-j}^*}_{u_{i,t}^*}\notag\\
    &= \widehat{\gamma}_{\widehat{\+\alpha},i}\overline{y}_t+\widehat{\alpha}_{CCEP}^t(y_{i,0}-\widehat{\gamma}_{\widehat{\+\alpha},i}\overline{y}_0)+u_{i,t}^*
\end{align}
due to the telescoping sum argument again. Hence, we insert 
\begin{align}
    \*y_{i,-1}^*&=\overline{\*y}_{-1}\widehat{\gamma}_{\widehat{\+\alpha},i}+\widehat{\+\A}_{CCEP,-1} \+\iota_T(y_{i,0}-\overline{y}_0\widehat{\gamma}_{\widehat{\+\alpha},i})+\*u_{i,-1}^*\notag\\
    &=\overline{\*y}^*_{-1}\widehat{\gamma}_{\widehat{\+\alpha},i}-\overline{\*u}^*_{-1}\widehat{\gamma}_{\widehat{\+\alpha},i}+\widehat{\+\A}_{CCEP,-1} \+\iota_T(y_{i,0}-\overline{y}_0\widehat{\gamma}_{\widehat{\+\alpha},i})+\*u_{i,-1}^*
\end{align}
into (\ref{sophisticated_CCEP}), where the first component will be projected out. Therefore, the analysis will be exactly the same as the one leading to Theorem 3.1 under the naive reconstruction method, as long as $\frac{1}{N}\sum_{i=1}^N\left|\widehat{\gamma}_{\widehat{\+\alpha},i}\right|^{4+\delta}=O_P(1)$ for $\delta \in (0, 4]$. We can demonstrate this by writing 
\begin{align}
    \widehat{\gamma}_{\widehat{\+\alpha},i}&=\left( \widehat{\*f}'_{\widehat{\+\alpha}}\widehat{\*f}_{\widehat{\+\alpha}}\right)^{-1}\widehat{\*f}'_{\widehat{\+\alpha}}(\*y_i-\widehat{\alpha}_{CCEP}\*y_{i,-1})\notag\\
    &=\left( \widehat{\*f}'_{\widehat{\+\alpha}}\widehat{\*f}_{\widehat{\+\alpha}}\right)^{-1}\widehat{\*f}'_{\widehat{\+\alpha}} (-\overline{\+\varepsilon}\overline{\gamma}^{-1}\gamma_i +[\overline{\*y}-\widehat{\alpha}_{CCEP}\overline{\*y}_{-1}]\overline{\gamma}^{-1}\gamma_i - (\alpha_0 -\widehat{\alpha}_{CCEP})\overline{\*u}_{-1}\overline{\gamma}^{-1}\gamma_i+(\alpha_0-\widehat{\alpha}_{CCEP})\*u_{i,-1}+\+\varepsilon_i )\notag\\
    &= \left( \widehat{\*f}'_{\widehat{\+\alpha}}\widehat{\*f}_{\widehat{\+\alpha}}\right)^{-1}\widehat{\*f}'_{\widehat{\+\alpha}} (-\overline{\+\varepsilon}\overline{\gamma}^{-1}\gamma_i +\widehat{\*f}_{\widehat{\+\alpha}}\overline{\gamma}^{-1}\gamma_i - (\alpha_0 -\widehat{\alpha}_{CCEP})\overline{\*u}_{-1}\overline{\gamma}^{-1}\gamma_i+(\alpha_0-\widehat{\alpha}_{CCEP})\*u_{i,-1}+\+\varepsilon_i )\notag\\
    &= \overline{\gamma}^{-1}\gamma_i +\left(T^{-1} \widehat{\*f}'_{\widehat{\+\alpha}}\widehat{\*f}_{\widehat{\+\alpha}}\right)^{-1}T^{-1}\widehat{\*f}'_{\widehat{\+\alpha}}(-\overline{\+\varepsilon}\overline{\gamma}^{-1}\gamma_i - (\alpha_0 -\widehat{\alpha}_{CCEP})\overline{\*u}_{-1}\overline{\gamma}^{-1}\gamma_i+(\alpha_0-\widehat{\alpha}_{CCEP})\*u_{i,-1}+\+\varepsilon_i )\notag\\
    &= \overline{\gamma}^{-1}\gamma_i +\left(T^{-1} \widehat{\*f}'_{\widehat{\+\alpha}}\widehat{\*f}_{\widehat{\+\alpha}}\right)^{-1}T^{-1}\widehat{\*f}'_{\+\alpha_0}(-\overline{\+\varepsilon}\overline{\gamma}^{-1}\gamma_i - (\alpha_0 -\widehat{\alpha}_{CCEP})\overline{\*u}_{-1}\overline{\gamma}^{-1}\gamma_i+(\alpha_0-\widehat{\alpha}_{CCEP})\*u_{i,-1}+\+\varepsilon_i )\notag\\
    &+\left(T^{-1} \widehat{\*f}'_{\widehat{\+\alpha}}\widehat{\*f}_{\widehat{\+\alpha}}\right)^{-1}T^{-1}(\widehat{\*f}_{\widehat{\+\alpha}}-\widehat{\*f}_{\+\alpha_0})'(-\overline{\+\varepsilon}\overline{\gamma}^{-1}\gamma_i - (\alpha_0 -\widehat{\alpha}_{CCEP})\overline{\*u}_{-1}\overline{\gamma}^{-1}\gamma_i+(\alpha_0-\widehat{\alpha}_{CCEP})\*u_{i,-1}+\+\varepsilon_i )\notag\\
    &= \overline{\gamma}^{-1}\gamma_i + a+b.
\end{align}
Firstly, note that 
\begin{align}
    T^{-1} \widehat{\*f}'_{\widehat{\+\alpha}}\widehat{\*f}_{\widehat{\+\alpha}}=\widehat{\+\alpha}'T^{-1}\widehat{\*F}'\widehat{\*F}\widehat{\+\alpha}=\widehat{\+\alpha}'\*C'\+\Sigma_\*q\*C\widehat{\+\alpha}+o_P(1)\to_p d>0,
\end{align}
because $\+\Sigma_\*q$ is positive definite and so is $\*C'\+\Sigma_\*q\*C$ since $\*C$ is $2\times 2$ and has a rank of 2 (full rank). Moreover, $\left\| \widehat{\+\alpha}-\+\alpha_0\right\|=|\widehat{\alpha}_{CCEP}-\alpha_0|=O_P((NT)^{-1/2})$. Next, $T^{-1/2}\left\|\widehat{\*f}_{\widehat{\+\alpha}}-\widehat{\*f}_{\+\alpha_0}\right\|\leq\left|\widehat{\alpha}_{CCEP}-\alpha_0\right|T^{-1/2}\left\|\overline{\*y}_{-1}\right\|=O_P((NT)^{-1/2})$, since 
\begin{align}
    T^{-1/2}\left\|\overline{\*y}_{-1}\right\|=\sqrt{\mathrm{tr}\left(T^{-1}\overline{\*y}_{-1}'\overline{\*y}_{-1}\right)}\to_p \sqrt{\mathrm{tr}([\*C'\+\Sigma_\*q\*C]_{2,2})}
\end{align}
 (trace of the lower-right block) by definition of Frobenius norm. Given that $T^{-1/2}\left\| \+\varepsilon_i\right\|=O_P(1)$ for every $i$, we have that $|b|=o_P(1)$. Similarly, we have that $\left\|T^{-1}\widehat{\*f}_{\+\alpha}'\+\varepsilon_i \right\|=o_P(1)$ (by the similar argument to the one in (\ref{hatf_u-1})), and so $|a|=o_P(1)$. To summarize, we can use $|\widehat{\gamma}_{\widehat{\+\alpha},i}|^{4+\delta}\leq C(|\overline{\gamma}^{-1}\gamma_i|^{4+\delta}+|a|^{4+\delta}+|b|^{4+\delta})$ for a positive constant $C$. Then have that 
\begin{align}
    \frac{1}{N}\sum_{i=1}^N\left|\widehat{\gamma}_{\widehat{\+\alpha},i}\right|^{4+\delta} \leq C |\overline{\gamma}^{-1}|^{{4+\delta}}\frac{1}{N}\sum_{i=1}^N|\gamma_i|^{4+\delta} + o_P(1)=O_P(1).
\end{align}
Because of this, the analysis of (\ref{sophisticated_CCEP}) is analogous to the one conducted in Lemmas 1-3. This brings Corollary 1. \\

\noindent \textbf{Corollary 1.} \textit{Let the conditions of Lemma 2 and 3 hold and let the reconstruction method be sophisticated, then as $(N,T)\to \infty$ with $NT^{-1}\to \kappa\in (0, \infty)$, we have }
\begin{align}
    \sqrt{NT}(\widehat{\alpha}_{CCEP}^*-\widehat{\alpha}_{CCEP})\to_{d^*}\mathcal{N}\left(0, \phi^{-1} \right) - \sqrt{\kappa}\phi^{-1}b,
\end{align}
\textit{where $b=\sum_{h=1}^{\infty}(\alpha_0)^{h-1}\mathrm{tr}\left(\+\Sigma_\*q(h)\+\Sigma_\*q^{-1}\right)$}.
\paragraph{Comment on Deeper Dynamics: AR($p$)}\label{4.2.9}
We can extend our model for a finite $p$ to 
\begin{align}
    y_{i,t}=\sum_{j=1}^p\alpha_{0,j}y_{i,t-j}+\gamma_if_t+\varepsilon_{i,t}.
\end{align}
Let $\theta_0(L)=1-\sum_{j=1}^p\alpha_{0,j}L^j$ be a lag polynomial such that $\theta(z)=0$ has all its roots outside of the unit circle. Then we obtain $f_t$ using 
\begin{align}
    f_t=\overline{\gamma}^{-1}\theta_0(L)\overline{y}_t-\overline{\gamma}^{-1}\overline{\varepsilon}_t.
\end{align}
Note that $y_{i,t-j}$ for each $j=1,\ldots,p$ can be written as 
\begin{align}
    y_{i,t-j}=\gamma_i\theta_0(L)^{-1}f_{t-j}+\theta_0(L)^{-1}\varepsilon_{i,t-1}&=\gamma_i\theta_0(L)^{-1}(\overline{\gamma}^{-1}\theta_0(L)\overline{y}_{t-j}-\overline{\gamma}^{-1}\overline{\varepsilon}_{t-j})+\theta_0(L)^{-1}\varepsilon_{i,t-1}\notag\\
    &=\overline{y}_{t-j}\overline{\gamma}^{-1}\gamma_i-\theta_0(L)^{-1}\overline{\varepsilon}_{t-j}\overline{\gamma}^{-1}\gamma_i+\theta_0(L)^{-1}\varepsilon_{i,t-1}\notag\\
    &=\overline{y}_{t-j}\overline{\gamma}^{-1}\gamma_i-\overline{u}_{t-j}\overline{\gamma}^{-1}\gamma_i+u_{i,t-j},
\end{align}
where now $u_{i,t-j}=\sum_{l=0}^\infty\theta_{0,l}\varepsilon_{i,t-j-l}$, and $\sum_{l=0}^\infty |\theta_{0,l}|<\infty$. Let $\*Y_{i,-}=[\*y_{i,-1}, \ldots, \*y_{i,-p}]\in \mathbb{R}^{(T-p-1)\times p}$, $\widehat{\*F}=[\overline{\*y}, \overline{\*y}_{-1}, \ldots, \overline{\*y}_{-p}]\in \mathbb{R}^{(T-p-1)\times (p+1)}$ and $\+\alpha_0=[\alpha_{0,1}, \ldots, \alpha_{0,p}]'$. Then, letting $\widehat{\+\alpha}_{CCEP}$ be its estimator, we have 
\begin{align}\label{CCEP_ARp}
    \sqrt{NT}(\widehat{\+\alpha}_{CCEP}-\+\alpha_0)&=\left(\frac{1}{NT}\sum_{i=1}^N\*Y_{i,-}'\*M_{\widehat{\*F}}\*Y_{i,-} \right)^{-1}\frac{1}{\sqrt{NT}}\sum_{i=1}^N\*Y_{i,-}'\*M_{\widehat{\*F}}(\*f\gamma_i+\+\varepsilon_i)\notag\\
    &=\left(\frac{1}{NT}\sum_{i=1}^N\*Y_{i,-}'\*M_{\widehat{\*F}}\*Y_{i,-} \right)^{-1}\frac{1}{\sqrt{NT}}\sum_{i=1}^N(\*U_{i,-}-\overline{\*U}\overline{\gamma}^{-1}\gamma_i)'\*M_{\widehat{\*F}}(\+\varepsilon_i-\overline{\+\varepsilon}\overline{\gamma}^{-1}\gamma_i),
\end{align}
where $\*U_{i,-}=[\*u_{i,-1}, \ldots, \*u_{i,-p}]$. The asymptotic normality and the ``Nickell bias'' can be derived by using steps analogous to those in \cite{juodis2021robustness}.  \\

\noindent We need to replicate (\ref{CCEP_ARp}) in the bootstrap realm. Under the naive implementation of recursive bootstrap, we have 
\begin{align}
    &\widehat{\+\gamma}_i=(\widehat{\*F}'\widehat{\*F})^{-1}\widehat{\*F}'\left(\*y_i-\sum_{j=1}^p\*y_{i,-j}\widehat{\alpha}_{j,CCEP}\right),\\
    &\widehat{\+\varepsilon}_i=\*M_{\widehat{\*F}}\left(\*y_i-\sum_{j=1}^p\*y_{i,-j}\widehat{\alpha}_{j,CCEP}\right),
\end{align}
where also 
\begin{align}\label{bar_gamma_ARp}
    \overline{\widehat{\+\gamma}}=(\widehat{\*F}'\widehat{\*F})^{-1}\widehat{\*F}'\left(\overline{\*y}_i-\sum_{j=1}^p\overline{\*y}_{i,-j}\widehat{\alpha}_{j,CCEP}\right)=(\widehat{\*F}'\widehat{\*F})^{-1}\widehat{\*F}'\widehat{\*F}[1,-\widehat{\alpha}_{1,CCEP},\ldots,-\widehat{\alpha}_{p,CCEP}]'=\widehat{\+\alpha}_p,
\end{align}
where we keep a similar notation to AR(1) case, but stress dependence on $p$ lags. Therefore, the connection between a naive an sophisticated implementation method is also similar: $\widehat{\*f}_{\widehat{\alpha}_p}=\widehat{\*F}\widehat{\+\alpha}_p$. Therefore, the naive bootstrap DGP is given by
\begin{align}
    \*y_i^*=\sum_{j=1}^p\widehat{\alpha}_{j,CCEP}\*y^*_{i,-j}+\underbrace{[\overline{\*y}, \overline{\*y}_{-1},\ldots,\overline{\*y}_{-p}]}_{\widehat{\*F}}\widehat{\+\gamma}_i+\+\varepsilon_i^*.
\end{align}
We further introduce a $(p+1)\times (p+1)$ rotation matrix
\begin{align}
    \*R=\begin{bmatrix}
        1 & \*0_{1\times p}\\
        -\widehat{\+\alpha}_{CCEP} & \*I_{p\times p}
    \end{bmatrix},
\end{align}
so that 
\begin{align}\label{y*_ARp}
\*y_i^*&=\sum_{j=1}^p\widehat{\alpha}_{j,CCEP}\*y^*_{i,-j}+[\overline{\*y}, \overline{\*y}_{-1},\ldots,\overline{\*y}_{-p}]\*R\*R^{-1}\widehat{\+\gamma}_i+\+\varepsilon_i^*\notag\\
&=\sum_{j=1}^p\widehat{\alpha}_{j,CCEP}\*y^*_{i,-j}+\left[\overline{\*y}-\sum_{j=1}^p\widehat{\alpha}_{j,CCEP}\overline{\*y}_{-j}, \overline{\*y}_{-1},\ldots, \overline{\*y}_{-p} \right]\*R^{-1}\widehat{\+\gamma}_i+\+\varepsilon_i^*\notag\\
&=\sum_{j=1}^p\widehat{\alpha}_{j,CCEP}\*y^*_{i,-j}+\left[\widehat{\*f}_{\widehat{\+\alpha}_p},\underbrace{\overline{\*y}_{-1},\ldots, \overline{\*y}_{-p}}_{\overline{\*Y}_{-}}\right]\left[\widetilde{\gamma}_{1,i}, \widetilde{\+\gamma}_{2,i}'\right]'+\+\varepsilon_i^*\notag\\
    &=\sum_{j=1}^p\widehat{\alpha}_{j,CCEP}\*y^*_{i,-j}+\widehat{\*f}_{\widehat{\+\alpha}_p}\widetilde{\gamma}_{1,i}+\overline{\*Y}_{-}\widetilde{\+\gamma}_{2,i}+\+\varepsilon_i^*,
\end{align}
where we again know from (\ref{bar_gamma_ARp}) that the average of $\widetilde{\gamma}_{1,i}$ is 1 and the average of $\widetilde{\+\gamma}_{2,i}$ is $\*0_{p\times 1}$. Using this and switching back to the scalar notation, we can demonstrate that $\overline{\*y}_t^*=\overline{y}_t+\sum_{j=0}^{t-1}\widehat{\theta}_j\overline{\varepsilon}_{t-j}^*$, where $\widehat{\theta}_0=1$ and for $j=1,...,t-1$ they admit the AR$(p)$ coefficient recursion. In particular,
\begin{align}\label{boot_avg_ARp}
    &\overline{y}_{1}^*=\sum_{j=0}^{p-1}\widehat{\alpha}_{j+1,CCEP}\overline{y}_{-j}+[\overline{y}_1,\overline{y}_0,\ldots, \overline{y}_{-p+1}]\widehat{\+\alpha}_p+\overline{\varepsilon}_1^*=\overline{y}_1+\overline{\varepsilon}_1^*\notag\\
    &\overline{y}^*_2=\sum_{j=0}^{p-1}\widehat{\alpha}_{j+1,CCEP}\overline{y}_{1-j}^*+[\overline{y}_2,\overline{y}_1,\ldots, \overline{y}_{-p+2}]\widehat{\+\alpha}_p+\overline{\varepsilon}_2^*\notag\\
    &=\widehat{\alpha}_{1,CCEP}\overline{y}_1^*+\sum_{j=1}^{p-1}\widehat{\alpha}_{j+1,CCEP}\overline{y}_{1-j}^*+[\overline{y}_2,\overline{y}_1,\ldots, \overline{y}_{-p+2}]\widehat{\+\alpha}_p+\overline{\varepsilon}_2^*\notag\\
&=\widehat{\alpha}_{1,CCEP}\overline{y}_1+\widehat{\alpha}_{1,CCEP}\overline{\varepsilon}_1^*+\sum_{j=1}^{p-1}\widehat{\alpha}_{j+1,CCEP}\overline{y}_{1-j}^*+[\overline{y}_2,\overline{y}_1,\ldots, \overline{y}_{-p+2}]\widehat{\+\alpha}_p+\overline{\varepsilon}_2^*=\overline{y}_2+\widehat{\alpha}_{1,CCEP}\overline{\varepsilon}_1^*+\overline{\varepsilon}_2^*\notag\\
&\overline{y}^*_3=\sum_{j=0}^{p-1}\widehat{\alpha}_{j+1,CCEP}\overline{y}_{2-j}^*+[\overline{y}_3,\overline{y}_2,\ldots, \overline{y}_{-p+3}]\widehat{\+\alpha}_p+\overline{\varepsilon}_3^*\notag\\
&=\widehat{\alpha}_{1,CCEP}\overline{y}^*_{2}+\widehat{\alpha}_{2,CCEP}\overline{y}_1^*+\sum_{j=2}^{p-1}\widehat{\alpha}_{j+1,CCEP}\overline{y}_{2-j}^*+[\overline{y}_3,\overline{y}_2,\ldots, \overline{y}_{-p+3}]\widehat{\+\alpha}_p+\overline{\varepsilon}_3^*\notag\\
&=\overline{y}_3+(\widehat{\alpha}_{1,CCEP}^2+\widehat{\alpha}_{2,CCEP})\overline{\varepsilon}_{1}^*+\widehat{\alpha}_{1,CCEP}\overline{\varepsilon}_{2}^*+\overline{\varepsilon}_3^*\notag\\
&\vdots\notag\\ 
&\overline{y}^*_t=\overline{y}_t+\sum_{j=0}^{t-1}\widehat{\theta}_j\overline{\varepsilon}_{t-j}^*=\overline{y}_t+\overline{u}_{t}^*. 
\end{align}
 Here, $\widehat{\theta}_j=\widehat{\alpha}_{1,CCEP}\widehat{\theta}_{j-1}+\widehat{\alpha}_{2,CCEP}\widehat{\theta}_{j-2}+\ldots+\widehat{\alpha}_{p,CCEP}\widehat{\theta}_{j-p}$. For example,  $\widehat{\theta}_3=\widehat{\alpha}_{1,CCEP}^3+2\widehat{\alpha}_{2,CCEP}\widehat{\alpha}_{1,CCEP}+\widehat{\alpha}_{3,CCEP}$, and so on. Thus, using (\ref{y*_ARp}) and denoting $\*Y^*_{i,-}=[\*y_{i,-1}^*,\ldots, \*y_{i,-p}^*]$, we obtain 
\begin{align}
    \widehat{\*f}_{\widehat{\+\alpha}_p}=\overline{\*y}^*-\overline{\*Y}^*_{-}\widehat{\+\alpha}_{CCEP}-\overline{\+\varepsilon}_i^*,
\end{align}
leading to 
\begin{align}
    \*y_i^*=\*Y^*_{i,-}\widehat{\+\alpha}_{CCEP}+(\overline{\*y}^*-\overline{\*Y}^*_{-}\widehat{\+\alpha}_{CCEP}-\overline{\+\varepsilon}_i^*)\widetilde{\gamma}_{1,i}+\overline{\*Y}_{-}\widetilde{\gamma}_{2,i}+\+\varepsilon_i^*,
\end{align}
and thus 
\begin{align}\label{CCEP*_ARp}
    \sqrt{NT}(\widehat{\+\alpha}^*_{CCEP}-\widehat{\+\alpha}_{CCEP})=\left(\frac{1}{NT}\sum_{i=1}^N\*Y_{i,-}^{*\prime}\*M_{\widehat{\*F}^*}\*Y^*_{i,-} \right)^{-1}\frac{1}{\sqrt{NT}}\sum_{i=1}^N\*Y_{i,-}^{*\prime}\*M_{\widehat{\*F}^*}(\+\varepsilon_{i}^*-\overline{\+\varepsilon}^*\widetilde{\gamma}_{1,i}-\overline{\*U}_{-}^*\widetilde{\+\gamma}_{2,i}),
\end{align}
where $\overline{\*U}^*_{-}=[\overline{\*u}_{-1}^*,\ldots, \overline{\*u}_{-p}^*]$, and we used the relationship in (\ref{boot_avg_ARp}). To complete, note that by switching to the scalar notation, we can recursively write
\begin{align}\label{y*_recursive_ARp}
    y_{i,t}^*&=\widehat{\+\gamma}_i'\sum_{j=0}^{t-1}\widehat{\theta}_j[\overline{y}_{t-j}, \overline{y}_{t-j-1}, \ldots, \overline{y}_{t-j-p}] + \sum_{j=0}^{t-1}\widehat{\theta}_j\varepsilon^*_{i,t-j}+\widehat{c}_{0,i,t}\notag\\
    &=\widetilde{\gamma}_{1,i}\sum_{j=0}^{t-1}\widehat{\theta}_j\left(\overline{y}_{t-j}-\sum_{l=1}^p\widehat{\alpha}_{l,CCEP}\overline{y}_{t-l-j} \right)+\widetilde{\+\gamma}_{2,i}'\underbrace{\sum_{j=0}^{t-1}\widehat{\theta}_j[ \overline{y}_{t-j-1}, \ldots, \overline{y}_{t-j-p}]}_{\*b_{t}}+  u_{i,t}^*+\widehat{c}_{0,i,t}\notag\\
    &=\widetilde{\gamma}_{1,i}\overline{y}_t+\widetilde{\+\gamma}_{2,i}'\*b_{t}+u_{i,t}^*+\widehat{r}_{0,i,t},
\end{align}
where $\widehat{r}_{0,i,t}$ summarizes the impact of initial values, which originates from backward substitutions similar to the AR(1) case. In the first term, we used the telescoping sum, which holds in the AR$(p)$ case as well, because the $\widehat{\theta}_j$ itself follows a recursion of AR$(p)$. In (\ref{CCEP*_ARp}), we use (\ref{y*_recursive_ARp}) stacked over time and $p$ lags, and in connection to (\ref{boot_avg_ARp}), so that the bootstrap average will be projected out:
\begin{align}
    \*Y_{i,-}=\overline{\*Y}^*_{-}\widetilde{\gamma}_{1,i}-\overline{\*U}^*_{-}\widetilde{\gamma}_{1,i}+\*B_{-}\widetilde{\+\gamma}_{2,i}+\*U_{i,-}^*+\widehat{\*R}_{0,-},
\end{align}
where $\*B_{-}=[\*B_{-1}, \ldots, \*B_{-p}]$ and $\*B_j$ is a stack of $\*b_{t-j}$ over time for $j=1,\ldots,p$. $\widehat{\*R}_{0,-}$ is defined analogously.  Hence, we can see a clear extension of AR$(1)$ representation to the one of AR$(p)$. The complete proof follows analogous steps taking into account 1) the summability of $\widehat{\theta}_j$, which is guaranteed under consistency of the CCEP and the stability of the AR$(p)$ polynomial, and 2) the fact that $\widetilde{\+\gamma}_{2,i}$ is negligible. This can be seen from the following manipulation. Let $\*G_{-}=[\*g_{-1},\ldots, \*g_{-p}]$, where $g_{t-j}=\theta_0(L)^{-1}f_{t-j}$ for $j=1,\ldots,p$, so that 
\begin{align}
    \widehat{\+\gamma}_i&=(\widehat{\*F}'\widehat{\*F})^{-1}\widehat{\*F}'(\*y_i-\*Y_{i,-}\widehat{\+\alpha}_{CCEP})\notag\\
    &=(\widehat{\*F}'\widehat{\*F})^{-1}\widehat{\*F}'(\*f\gamma_i+\+\varepsilon_i-\*Y_{i,-}(\widehat{\+\alpha}_{CCEP}-\+\alpha_0))\notag\\
    &=(\widehat{\*F}'\widehat{\*F})^{-1}\widehat{\*F}'(\widehat{\*F}[1, -\+\alpha_0']'\overline{\gamma}^{-1}\gamma_i-\overline{\+\varepsilon}\overline{\gamma}^{-1}\gamma_i-\*G_{-}(\widehat{\+\alpha}_{CCEP}-\+\alpha_0)\gamma_i-\*U_{i,-}(\widehat{\+\alpha}_{CCEP}-\+\alpha_0)+\+\varepsilon_i)\notag\\
    &=[1, -\+\alpha_0']'\overline{\gamma}^{-1}\gamma_i+(T^{-1}\widehat{\*F}'\widehat{\*F})^{-1}T^{-1}\widehat{\*F}'(\overline{\+\varepsilon}\overline{\gamma}^{-1}\gamma_i-\*G_{-}(\widehat{\+\alpha}_{CCEP}-\+\alpha_0)\gamma_i-\*U_{i,-}(\widehat{\+\alpha}_{CCEP}-\+\alpha_0)+\+\varepsilon_i)\notag\\
    &=[1, -\widehat{\+\alpha}_{CCEP}']'\overline{\gamma}^{-1}\gamma_i+\underbrace{[0,(\widehat{\+\alpha}_{CCEP}-\+\alpha_0)']\overline{\gamma}^{-1}\gamma_i}_{O_P((NT)^{-1/2})}\notag\\
    &+\underbrace{(T^{-1}\widehat{\*F}'\widehat{\*F})^{-1}T^{-1}\widehat{\*F}'(\overline{\+\varepsilon}\overline{\gamma}^{-1}\gamma_i-\*G_{-}(\widehat{\+\alpha}_{CCEP}-\+\alpha_0)\gamma_i-\*U_{i,-}(\widehat{\+\alpha}_{CCEP}-\+\alpha_0)+\+\varepsilon_i)}_{O_P(T^{-1/2})}.
\end{align}
Therefore, using $\*R^{-1}=\begin{bmatrix}
    1 & \*0_{1\times p}\\
    \widehat{\+\alpha}_{CCEP} & \*I_{p\times p}
\end{bmatrix}$, we have 
\begin{align}
    \widetilde{\+\gamma}_{2,i}=[\widehat{\+\alpha}_{CCEP}, \*I_{p\times p}][1,-\widehat{\+\alpha}_{CCEP}']'\overline{\gamma}^{-1}\gamma_i+O_P(T^{-1/2})=O_P(T^{-1/2}).
\end{align}
The latter represented an extension of the naive method to AR$(p)$. As for the sophisticated version, we again notice that $\widehat{\*F}[1,-\widehat{\+\alpha}_{CCEP}']'=\widehat{\*F}\widehat{\+\alpha}_{p}=\widehat{\*f}_{\widehat{\+\alpha}_p}$, where now $\widehat{\+\alpha}_{p}$ is $(p+1)\times 1$. Then, similarly to AR$(1)$ case, we have 
    \begin{align}
        \widehat{\gamma}_{\widehat{\+\alpha}_{p},i}=\left( \widehat{\*f}_{\widehat{\+\alpha}_p}'\widehat{\*f}_{\widehat{\+\alpha}_p}\right)^{-1}\widehat{\*f}_{\widehat{\+\alpha}_p}'(\*y_i-\*Y_{i,-}\widehat{\+\alpha}_{CCEP})
    \end{align}
    and \begin{align}
        \overline{\widehat{\gamma}}_{\widehat{\+\alpha}_{p}}=\left( \widehat{\*f}_{\widehat{\+\alpha}_p}'\widehat{\*f}_{\widehat{\+\alpha}_p}\right)^{-1}\widehat{\*f}_{\widehat{\+\alpha}_p}'(\overline{\*y}-\overline{\*Y}_{-}\widehat{\+\alpha}_{CCEP})=\left( \widehat{\*f}_{\widehat{\+\alpha}_p}'\widehat{\*f}_{\widehat{\+\alpha}_p}\right)^{-1}\widehat{\*f}_{\widehat{\+\alpha}_p}'\widehat{\*f}_{\widehat{\+\alpha}_p}=1.
    \end{align}
   Hence, our bootstrap DGP is 
   \begin{align}
       y_{i,t}^*=\sum_{j=1}^p\widehat{\alpha}_{CCEP,j}y^*_{i,t-j}+\widehat{\gamma}_{\widehat{\+\alpha}_{p},i}\widehat{f}_{\widehat{\alpha}_{p},t}+\varepsilon_{i,t}^*,
   \end{align}
   which, in stacked notation, implies 
   \begin{align}
       \widehat{\*f}_{\widehat{\+\alpha}_p}=\overline{\*y}^*-\overline{\*Y}^*_{-}\widehat{\+\alpha}_{CCEP}-\overline{\+\varepsilon}^*,
   \end{align}
   and therefore, 
   \begin{align}
        \sqrt{NT}(\widehat{\+\alpha}^*_{CCEP}-\widehat{\+\alpha}_{CCEP})=\left(\frac{1}{NT}\sum_{i=1}^N\*Y_{i,-}^{*\prime}\*M_{\widehat{\*F}^*}\*Y^*_{i,-} \right)^{-1}\frac{1}{\sqrt{NT}}\sum_{i=1}^N\*Y_{i,-}^{*\prime}\*M_{\widehat{\*F}^*}(\+\varepsilon_{i}^*-\overline{\+\varepsilon}^*\widetilde{\gamma}_{\widehat{\+\alpha}_p,i}).
   \end{align}
   The rest of the analysis proceeds similarly to the naive approach. Since $\overline{\widehat{\gamma}}_{\widehat{\+\alpha}_p}=1$, we still have $\overline{y}_t^*=\overline{y}_t+\overline{u}_t^*$, where the definition of $\overline{u}^*_t$ is the same. Hence, $\overline{\*Y}^*_{-}=\overline{\*Y}_{-}+\overline{\*U}^*_{-}$, as well. Thus, we repeat the steps in (\ref{y*_recursive_ARp}), which leads to 
   \begin{align}
       \*Y^*_{i,-}=\overline{\*Y}^*_{-}\widehat{\gamma}_{\widehat{\+\alpha}_p,i}-\overline{\*U}^*_{-}\widehat{\gamma}_{\widehat{\+\alpha}_p,i}+\*U_{i,-}^*+\widehat{\*R}_{0,-},
   \end{align}
   where $\widehat{\*R}_{0,-}$ again summarizes the initial values, and $\overline{\*Y}^*_{-}\widehat{\gamma}_{\widehat{\+\alpha}_p,i}$ will be projected out. To finalize, we notice that 
   \begin{align}
       \widehat{\gamma}_{\widehat{\+\alpha}_{p},i}&=\left( \widehat{\*f}_{\widehat{\+\alpha}_p}'\widehat{\*f}_{\widehat{\+\alpha}_p}\right)^{-1}\widehat{\*f}_{\widehat{\+\alpha}_p}'(\*y_i-\*Y_{i,-}\widehat{\+\alpha}_{CCEP})\notag\\
       &=\left( \widehat{\*f}_{\widehat{\+\alpha}_p}'\widehat{\*f}_{\widehat{\+\alpha}_p}\right)^{-1}\widehat{\*f}_{\widehat{\+\alpha}_p}'(\*f\gamma_i+\+\varepsilon_i-\*Y_{i,-}(\widehat{\+\alpha}_{CCEP}-\+\alpha_0))\notag\\
       &=\left( \widehat{\*f}_{\widehat{\+\alpha}_p}'\widehat{\*f}_{\widehat{\+\alpha}_p}\right)^{-1}\widehat{\*f}_{\widehat{\+\alpha}_p}'([\overline{\*y}-\overline{\*Y}_{-}\+\alpha_0]\overline{\gamma}^{-1}\gamma_i-\overline{\+\varepsilon}\overline{\gamma}^{-1}\gamma_i-\*G_{-}(\widehat{\+\alpha}_{CCEP}-\+\alpha_0)\gamma_i-\*U_{i,-}(\widehat{\+\alpha}_{CCEP}-\+\alpha_0)+\+\varepsilon_i)\notag\\
       &= \left( \widehat{\*f}_{\widehat{\+\alpha}_p}'\widehat{\*f}_{\widehat{\+\alpha}_p}\right)^{-1}\widehat{\*f}_{\widehat{\+\alpha}_p}'([\overline{\*y}-\overline{\*Y}_{-}\widehat{\+\alpha}_{CCEP}]\overline{\gamma}^{-1}\gamma_i+\overline{\*Y}_{-}(\widehat{\+\alpha}_{CCEP}-\+\alpha_0)\overline{\gamma}^{-1}\gamma_i-\overline{\+\varepsilon}\overline{\gamma}^{-1}\gamma_i\notag\\
       &-\*G_{-}(\widehat{\+\alpha}_{CCEP}-\+\alpha_0)\gamma_i-\*U_{i,-}(\widehat{\+\alpha}_{CCEP}-\+\alpha_0)+\+\varepsilon_i)\notag\\
       &=\overline{\gamma}^{-1}\gamma_i+\left( \widehat{\*f}_{\widehat{\+\alpha}_p}'\widehat{\*f}_{\widehat{\+\alpha}_p}\right)^{-1}\widehat{\*f}_{\widehat{\+\alpha}_p}'(-\overline{\+\varepsilon}\overline{\gamma}^{-1}\gamma_i-\overline{\*U}_{-}(\widehat{\+\alpha}_{CCEP}-\+\alpha_0)\overline{\gamma}^{-1}\gamma_i-\*U_{i,-}(\widehat{\+\alpha}_{CCEP}-\+\alpha_0)+\+\varepsilon_i)\notag\\
       &=\overline{\gamma}^{-1}\gamma_i+\left(T^{-1} \widehat{\*f}_{\widehat{\+\alpha}_p}'\widehat{\*f}_{\widehat{\+\alpha}_p}\right)^{-1}T^{-1}\widehat{\*f}_{\+\alpha_{0,p}}'(-\overline{\+\varepsilon}\overline{\gamma}^{-1}\gamma_i-\overline{\*U}_{-}(\widehat{\+\alpha}_{CCEP}-\+\alpha_0)\overline{\gamma}^{-1}\gamma_i-\*U_{i,-}(\widehat{\+\alpha}_{CCEP}-\+\alpha_0)+\+\varepsilon_i)\notag\\
       &+\left( T^{-1}\widehat{\*f}_{\widehat{\+\alpha}_p}'\widehat{\*f}_{\widehat{\+\alpha}_p}\right)^{-1}T^{-1}(\widehat{\*f}_{\widehat{\+\alpha}_{p}}-\widehat{\*f}_{\+\alpha_{0,p}})'(-\overline{\+\varepsilon}\overline{\gamma}^{-1}\gamma_i-\overline{\*U}_{-}(\widehat{\+\alpha}_{CCEP}-\+\alpha_0)\overline{\gamma}^{-1}\gamma_i-\*U_{i,-}(\widehat{\+\alpha}_{CCEP}-\+\alpha_0)+\+\varepsilon_i)\notag\\
       &=\overline{\gamma}^{-1}\gamma_i+a+b,
   \end{align}
   where $T^{-1/2}\left\|\widehat{\*f}_{\widehat{\+\alpha}_{p}}-\widehat{\*f}_{\+\alpha_{0,p}} \right\|\leq \left\|\widehat{\+\alpha}_{CCEP}-\+\alpha_0 \right\|T^{-1/2}\left\|\overline{\*Y}_{-}\right\|=O_P((NT)^{-1/2})$ for a finite $p$. Together with $T^{-1/2}\left\|\+\varepsilon_i\right\|=O_P(1)$ and $T^{-1}\widehat{\*f}_{\widehat{\+\alpha}_p}'\widehat{\*f}_{\widehat{\+\alpha}_p}=T^{-1}\widehat{\+\alpha}_p'\widehat{\*F}'\widehat{\*F}\widehat{\+\alpha}_p$ being strictly positive in the limit, we have $|b|=o_P(1)$. Since $\left\|T^{-1}\widehat{\*f}_{\+\alpha_0}'\+\varepsilon_i \right\|=o_P(1)$ for each $i$, we know that $|a|=o_P(1)$. Ultimately, 
   \begin{align}
       \frac{1}{N}\sum_{i=1}^N|\widehat{\gamma}_{\widehat{\+\alpha}_p,i}|^{4+\delta}\leq C|\overline{\gamma}^{-1}|^{4+\delta}\frac{1}{N}\sum_{i=1}^N|\gamma_i|^{4+\delta}+o_P(1)=O_P(1)
   \end{align}
   as needed. 
\subsubsection{ARX(1) Case}\label{4.3}
\paragraph {The Setup} \label{4.3.1}
We update our pure AR(1) DGP by including $k$ explanatory variables $\*x_{i,t}$
\begin{align}
    &y_{i,t}=\alpha_0y_{i,t-1}+\+\beta_0^{\prime}\*x_{i,t}+\+\gamma_{y,i}'\*f_t+\varepsilon_{i,t}=\+\delta_0^{\prime}\*b_{i,t}+\+\gamma_{y,i}'\*f_t+\varepsilon_{i,t},\\
    &\*x_{i,t}=\+\theta_0y_{i,t-1}+\+\Gamma_{x,i}'\*f_t+\+\nu_{i,t},
\end{align}
where $\+\delta_0=[\alpha_0, \+\beta_0^{\prime}]'$, $\*b_{i,t}=[y_{i,t-1}, \*x_{i,t}']'$ and $\+\theta_0=[\theta_{0,1},\ldots, \theta_{0,k}]'$ which implies that $x_{i,t}$ is weakly exogenous unless $\+\theta_0=\*0_{k\times 1}$. Here, $\*f_t\in \mathbb{R}^R$, and let $k+1=K$. We consider the situation where $R=K$, because otherwise we would not be able to consistently estimate the loadings.\\

  \noindent Let $\*z_{i,t}=[y_{i,t}, \*x_{i,t}']'\in \mathbb{R}^K$, then we can write it as 
\begin{align}\label{z_VAR}
    \*z_{i,t}=\*A_0^{\prime}\*z_{i,t-1}+\+\Gamma_i'\*f_t+\*e_{i,t}, 
\end{align}
where 
\begin{align}
    &\*A_0=\begin{bmatrix} \alpha_0+\+\beta_0^{\prime}\+\theta_0 & \+\theta_0^{\prime}\\
   \*0_{k\times 1}  & \*0_{k\times k}\end{bmatrix},\\
    &\+\Gamma_i=[\+\Gamma_{x,i}\+\beta_0+\+\gamma_{y,i}, \+\Gamma_{x,i}],\\
    &\*e_{i,t}=[\+\nu_{i,t}'\+\beta_0+\varepsilon_{i,t}, \+\nu_{i,t}']'.
\end{align}
From (\ref{z_VAR}), we have that 
\begin{align}
    \overline{\*z}_t=\*A_0^{\prime}\overline{\*z}_{t-1}+\overline{\+\Gamma}'\*f_t+\overline{\*e}_t,
\end{align}
and so provided that $\mathrm{rk}(\overline{\+\Gamma})=R$, we have
\begin{align}
    \*f_t=(\overline{\+\Gamma}')^{-1}(\overline{\*z}_t-\*A_0^{\prime}\overline{\*z}_{t-1}-\overline{\*e}_t),
\end{align}
which means that we need $\overline{\*z}_t$ and its lag to proxy $\*f_t$ ($R$ factors). Hence, the \textit{sophisticated} factor proxy for the bootstrap is now given by 
\begin{align}\label{f_hat_A_0}
\widehat{\*f}_{\mathbb{A}_0,t}=\overline{\*z}_t-\*A_0^{\prime}\overline{\*z}_{t-1}=\overline{\+\Gamma}'\*f_t+\overline{\*e}_t,
\end{align}
where $\*A_0$ should be replaced by an appropriate $\widehat{\*A}$, \textit{if possible}. Let $\widehat{\*f}_t=[\overline{\*z}_t,\overline{\*z}_{t-1}]'\in \mathbb{R}^{2K}$, which represents the \textit{naive} proxy. By letting $\mathbb{A}_0=[\*I_{K\times K}, -\*A_0^{\prime}]'$ and $\widehat{\mathbb{A}}=[\*I_{K\times K}, -\widehat{\*A}^{\prime}]'$ its estimator, the connection between the two is 
\begin{align}
\widehat{\*f}_{\widehat{\mathbb{A}},t}=\widehat{\mathbb{A}}^{\prime}\widehat{\*f}_t \quad (\widehat{\*F}_{\widehat{\mathbb{A}}}=\widehat{\*F}\widehat{\mathbb{A}}).
\end{align}
To provide an explicit representation of the naive proxy $\widehat{\*f}_t$ and proceed with the analysis, we update our assumptions list.\\

\noindent With Assumption \ref{ass::6}, we can write
\begin{align}\label{z_it_X}
    \*z_{i,t}&=\sum_{l=0}^\infty(\*A_0^{\prime})^l \+\Gamma_i'\*f_{t-l} +\sum_{l=0}^\infty(\*A^{0})^l \*e_{i,t-l}\notag\\
    &=\sum_{l=0}^\infty (\*f_{t-l}'\otimes (\*A_0^{\prime})^l)\mathrm{vec}(\+\Gamma_i')+\sum_{l=0}^\infty(\*A_0^{\prime})^l \*e_{i,t-l} \notag\\
     &=\sum_{l=0}^\infty \*I_{K\times K}(\*f_{t-l}'\otimes (\*A_0^{\prime})^l)\mathrm{vec}(\+\Gamma_i')+\sum_{l=0}^\infty(\*A_0^{\prime})^l \*e_{i,t-l} \notag\\
    &= (\mathrm{\mathrm{vec}(\+\Gamma_i')\otimes \*I_{K\times K}})'\underbrace{\sum_{l=0}^\infty \mathrm{vec}(\*f_{t-l}'\otimes (\*A_0^{\prime})^l) }_{\*g_{t}}+\underbrace{\sum_{l=0}^\infty(\*A_0^{\prime})^l \*e_{i,t-l}}_{\*u_{i,t}} \notag\\
   &= \+\Lambda_i'\*g_{t}+\*u_{i,t},
\end{align}
 where we applied $\mathrm{vec}(\*A\*B\*C)=(\*C'\otimes \*A)\mathrm{vec}(\*B)$ on a vector two times to factor out $\+\Lambda_i=\mathrm{vec}(\+\Gamma_i')\otimes \*I_{K\times K}$. Effectively, (\ref{z_it_X}) is a re-structuring of the same term in \cite{Chudik2015}, where we separate loadings form the factors and collect them outside from the infinite sum. Let $\*d_{i,t}=[\*z_{i,t}', \*z_{i,t-1}']'\in \mathbb{R}^{2K}$, where we lagged (\ref{z_it_X}), which gives
\begin{align}
    \*d_{i,t}=\begin{bmatrix}\+\Gamma_i'& \*A_0\+\Lambda_i' \\
    \*0_{K\times K} & \+\Lambda_i'\end{bmatrix} \begin{bmatrix} \*f_t \\ \*g_{t-1} \end{bmatrix} + \begin{bmatrix} \*A_0\*u_{i,t-1}+\*e_{i,t}\\ \*u_{i,t-1}
    \end{bmatrix},
\end{align}
which is a similar representation of the factor estimate to the one in AR(1) case, where 2 averages effectively proxied 2 factors under $R=1$.  Here, due to the presence of $\*x_{i,t}$, $2K$ averages proxy $2K^2R=2K^3$ factors under $R=K$ ($\mathrm{rk}\left(\begin{bmatrix}\+\Gamma_i'& \*A_0\+\Lambda_i' \\
    \*0_{K\times K} & \+\Lambda_i'\end{bmatrix} \right)=2K$). However, it is more convenient to represent $\*d_{i,t}$ in the spirit of \cite{DeVos2019}. Let $\*C_i=\*I_{2 \times 2}\otimes \+\Lambda_i$, then 
    \begin{align}\label{d_it}
        \*d_{i,t}=(\*I_{2 \times 2}\otimes \+\Lambda _i)'\begin{bmatrix} \*g_t \\ \*g_{t-1} \end{bmatrix} + \begin{bmatrix} \underbrace{\*A_0'\*u_{i,t-1}+\*e_{i,t}}_{\*u_{i,t}}\\ \*u_{i,t-1}
    \end{bmatrix}=\*C_i'\*q_t+\*v_{i,t}
    \end{align}
    and so 
    \begin{align}\label{hatF_total_X}
        \widehat{\*f}_t=\overline{\*d}_t=\overline{\*C}'\*q_t+\overline{\*v}_t.
    \end{align}
    Therefore, as $(N,T)\to \infty$,
    \begin{align}
        T^{-1}\widehat{\*F}'\widehat{\*F}=\frac{1}{T}\sum_{t=2}^T\widehat{\*f}_t\widehat{\*f}_t'\to_p\*C'\+\Sigma_\*q\*C,
    \end{align}
   which is positive definite as a whole.\footnote{Note that $\widehat{\*f}_t$ is just a restructuring of the same vector (only under stochastic loadings) as in \cite{Chudik2015}, whose second sample moment converges to a positive definite matrix under analogous assumptions.}  Then by letting $\*S_{\*b}=\begin{bmatrix}
0 & \*0_{1\times k} & 1 & \*0_{1\times k} \\
\*0_{k\times 1} & \*I_{k\times k} & \*0_{k\times 1} & \*0_{k\times k}
\end{bmatrix}'$ so that $\*S_\*b'\*d_{i,t}=\*b_{i,t}$, we have
\begin{align}
    \widehat{\+\delta}_{CCEP}&=\left(\frac{1}{NT}\sum_{i=1}^N\*B_{i}'\*M_{\widehat{\*F}}\*B_{i} \right)^{-1}\frac{1}{NT}\sum_{i=1}^N\*B_{i}'\*M_{\widehat{\*F}}\*y_i\notag\\
    &=\left(\frac{1}{NT}\sum_{i=1}^N\*S_\*b'\*D_{i}'\*M_{\widehat{\*F}}\*D_{i}\*S_\*b \right)^{-1}\frac{1}{NT}\sum_{i=1}^N\*S_\*b'\*D_{i}'\*M_{\widehat{\*F}}\*y_i
\end{align}
where we know that $\left\|\widehat{\+\delta}_{CCEP}-\+\delta_0 \right\|=O_P((NT)^{-1/2})$ as $(N,T)\to \infty$ with $NT^{-1}\to \kappa\in (0,\infty)$ (see \citealp{Juodis2022CCER}). 
    \paragraph {Fixed Regressors and Weak Exogeneity}\label{4.3.2}
If $\+\theta_0\neq \*0_{k\times 1}$, $\*x_{i,t}$ is weakly exogenous. Importantly, we cannot estimate $\+\theta_0$ and $\widehat{\*A}$ cannot be constructed. Therefore, must keep the regressors fixed in the bootstrap world, which means that we will not be able to reproduce (\ref{z_VAR}) in the bootstrap world exactly. At the same time, we will avoid making strong assumptions about the structure of $\*x_{i,t}$. Therefore, we let $\*x_{i,t}^*=\*x_{i,t}$. Then
\begin{align}
    \*y_{i}^*&=\widehat{\alpha}_{CCEP}\*y_{i,-1}^*+\*X_i\widehat{\+\beta}_{CCEP}+\widehat{\*F}\widehat{\+\gamma}_{y,i}+\+\varepsilon_{i}^*,
\end{align}
Here, using the definition of $\*B_i$ and (\ref{d_it}), we obtain the following: 
\begin{align}\label{estimated_gamma_with_X}
    \widehat{\+\gamma}_{y,i}&=(\widehat{\*F}'\widehat{\*F})^{-1}\widehat{\*F}'(\*y_i-\*y_{i,-1}\widehat{\alpha}_{CCEP}-\*X_i\widehat{\+\beta}_{CCEP})\notag\\
    &=(\widehat{\*F}'\widehat{\*F})^{-1}\widehat{\*F}'(\*y_{i,-1}\alpha_0+\*X_i\+\beta_0+\*F\+\gamma_{y,i}-\*y_{i,-1}\widehat{\alpha}_{CCEP}-\*X_i\widehat{\+\beta}_{CCEP}+\+\varepsilon_i)\notag\\
    &=(\widehat{\*F}'\widehat{\*F})^{-1}\widehat{\*F}'(-\*B_i(\widehat{\+\delta}_{CCEP}-\+\delta_0)+\*F\+\gamma_{y,i}+\+\varepsilon_i )\notag\\
    &=(\widehat{\*F}'\widehat{\*F})^{-1}\widehat{\*F}'(-\*B_i(\widehat{\+\delta}_{CCEP}-\+\delta_0)+ (\overline{\*Z}-\overline{\*Z}_{-1}\*A_0)\overline{\+\Gamma}^{-1}\+\gamma_{y,i}-\overline{\*E}\overline{\+\Gamma}^{-1}\+\gamma_{y,i}+\+\varepsilon_i)\notag\\
    &=\mathbb{A}_0\overline{\+\Gamma}^{-1}\+\gamma_{y,i}+(\widehat{\*F}'\widehat{\*F})^{-1}\widehat{\*F}'(-\*D_i\*S_{\*b}(\widehat{\+\delta}_{CCEP}-\+\delta_0)-\overline{\*E}\overline{\+\Gamma}^{-1}\+\gamma_{y,i}+\+\varepsilon_i)\\
    &=\mathbb{A}_0\overline{\+\Gamma}^{-1}\+\gamma_{y,i}+ (\widehat{\*F}'\widehat{\*F})^{-1}\widehat{\*F}'(-\*Q\*C_i\*S_\*b(\widehat{\+\delta}_{CCEP}-\+\delta_0)-\*V_i\*S_\*b(\widehat{\+\delta}_{CCEP}-\+\delta_0)-\overline{\*E}\overline{\+\Gamma}^{-1}\+\gamma_{y,i}+\+\varepsilon_i)
\end{align}
and so 
\begin{align}\label{resid_with_X}
\widehat{\+\varepsilon}_i=\*M_{\widehat{\*F}}(-\*Q\*C_i\*S_\*b(\widehat{\+\delta}_{CCEP}-\+\delta_0)-\*V_i\*S_\*b(\widehat{\+\delta}_{CCEP}-\+\delta_0)-\overline{\*E}\overline{\+\Gamma}^{-1}\+\gamma_{y,i}+\+\varepsilon_i).
\end{align}
Further, notice that 
\begin{align}
    \overline{\widehat{\+\gamma}}_{y}=(\widehat{\*F}'\widehat{\*F})^{-1}\widehat{\*F}'(\overline{\*y}-\overline{\*y}_{-1}\widehat{\alpha}_{CCEP}-\overline{\*X}\widehat{\+\beta}_{CCEP})&=(\widehat{\*F}'\widehat{\*F})^{-1}\widehat{\*F}'(\overline{\*y}-\overline{\*y}_{-1}\widehat{\alpha}_{CCEP}-\overline{\*X}\widehat{\+\beta}_{CCEP}+\overline{\*X}_{-1}\*0_{k\times 1})\notag\\
    &=(\widehat{\*F}'\widehat{\*F})^{-1}\widehat{\*F}'\widehat{\*F}[1, -\widehat{\+\beta}_{CCEP}', -\widehat{\alpha}_{CCEP}, \*0_{1\times k}]'\notag\\
    &=[1, -\widehat{\+\beta}_{CCEP}', -\widehat{\alpha}_{CCEP}, \*0_{1\times k}]'.
\end{align}
This immediately implies that 
\begin{align}
    \overline{\*y}^*&=\widehat{\alpha}_{CCEP}\overline{\*y}_{-1}^*+\overline{\*X}\widehat{\+\beta}_{CCEP}+\widehat{\*F}\overline{\widehat{\+\gamma}}_{y}+\overline{\+\varepsilon}^*\notag\\
    &=\widehat{\alpha}_{CCEP}\overline{\*y}_{-1}^*+\overline{\*X}\widehat{\+\beta}_{CCEP}+(\overline{\*y}-\widehat{\alpha}_{CCEP}\overline{\*y}_{-1}-\overline{\*X}\widehat{\+\beta}_{CCEP}) +\overline{\+\varepsilon}^*\notag\\
    &= \widehat{\alpha}_{CCEP}\overline{\*y}_{-1}^*+\underbrace{(\overline{\*y}-\widehat{\alpha}_{CCEP}\overline{\*y}_{-1})}_{\widehat{\*f}_{\widehat{\+\alpha}}}+\overline{\+\varepsilon}^*,
\end{align}
where we uncover the definition of $\widehat{\*f}_{\widehat{\+\alpha}}$ which comes from the pure AR(1) case, but clearly now it does not approximate the full factor space of $R$ factors. Note that we again have the following useful expression
\begin{align}\label{hat_F_a_withX}
    \widehat{\*f}_{\widehat{\+\alpha}}=\overline{\*y}^*-\widehat{\alpha}_{CCEP}\overline{\*y}_{-1}^*-\overline{\+\varepsilon}^*.
\end{align}
Switching to vector notation and again using the availability of $y_{i,0}$, we obtain 
\begin{align}
\overline{y}_1^*=\widehat{\alpha}_{CCEP}\overline{y}_0+\widehat{\+\beta}_{CCEP}'\overline{\*x}_{1}+\overline{\widehat{\gamma}}_{y}'[\overline{\*z}_{1}', \overline{\*z}_0']'+\overline{\varepsilon}^*_1&=\widehat{\alpha}_{CCEP}\overline{y}_0+\widehat{\+\beta}_{CCEP}'\overline{\*x}_{1}+\overline{\widehat{\gamma}}_{y}'[\overline{y}_1, \overline{\*x}_1',\overline{y}_0, \overline{\*x}_0' ]'+\overline{\varepsilon}^*_1\notag\\
    &=\widehat{\alpha}_{CCEP}\overline{y}_0+\widehat{\+\beta}_{CCEP}'\overline{\*x}_{1}-\widehat{\alpha}_{CCEP}\overline{y}_0-\widehat{\+\beta}_{CCEP}'\overline{\*x}_{1}+\overline{y}_1+\overline{\varepsilon}^*_1\notag\\
    &=\overline{y}_1+\overline{\varepsilon}^*_1,
\end{align}
and 
\begin{align}
\overline{y}_2^*=\widehat{\alpha}_{CCEP}\overline{y}_1^*+\widehat{\+\beta}_{CCEP}'\overline{\*x}_{2}+\overline{\widehat{\gamma}}_{y}'[\overline{\*z}_{2}', \overline{\*z}_1']'+\overline{\varepsilon}^*_2&=\widehat{\alpha}_{CCEP}\overline{y}_1^*+\widehat{\+\beta}_{CCEP}'\overline{\*x}_{2}-\widehat{\alpha}_{CCEP}\overline{y}_{1}-\widehat{\+\beta}_{CCEP}'\overline{\*x}_{2}+\overline{y}_2+\overline{\varepsilon}^*_2\notag\\
&=\widehat{\alpha}_{CCEP}\overline{y}_1^*-\widehat{\alpha}_{CCEP}\overline{y}_{1}+\overline{y}_2+\overline{\varepsilon}^*_2\notag\\
&=\overline{y}_2+\widehat{\alpha}_{CCEP}\overline{\varepsilon}_1+\overline{\varepsilon}^*_2,
\end{align}
which implies that again 
\begin{align}\label{y_property_with_X}
\overline{y}_{t}^*=\overline{y}_t+\sum_{l=0}^{t-1}\widehat{\alpha}_{CCEP}^l\overline{\varepsilon}_{t-l}^*=\overline{y}_t+\overline{u}_t^*,  
\end{align}
and therefore we can use $\overline{\*y}=\overline{\*y}^*-\overline{\*u}^*$. In what follows, we use the fact that $\widehat{\*F}^*=[\overline{\*Z}^*, \overline{\*Z}^*_{-1}]=[\overline{\*y}^*, \overline{\*X}, \overline{\*y}^*_{-1},\overline{\*X}_{-1}]$, which means that $\overline{\*X}, \overline{\*X}_{-1}$ are \textit{observed factors} in the bootstrap world, which will be projected out of $\*y^*_i$ equation. Let $\*S_{\gamma_1,\gamma_{K+1}}=\begin{bmatrix}
    \*e'_{1}\\
    \*e_{K+1}'
\end{bmatrix}$ be a selector matrix with $\*e_j$ having 1 in $j$-th position for $j\in \{1, 2K \}$, such that $\*S_{\gamma_1,\gamma_{K+1}}\widehat{\+\gamma}_{y,i}=[\widehat{\gamma}_{y,i,1}, \widehat{\gamma}_{y,i,K+1}]'$, so that we select only the relevant estimated loadings. Then using the usual $\*R=\begin{bmatrix}
    1 & 0\\
    -\widehat{\alpha}_{CCEP} & 1
\end{bmatrix}$ we can write the bootstrap CCEP estimator as 
\begin{align}\label{CCEP_X_Endog}
    \sqrt{NT}(\widehat{\+\delta}^*_{CCEP}-\widehat{\+\delta}_{CCEP})&=\left(\frac{1}{NT}\sum_{i=1}^N\*B_i^{*\prime}\*M_{\widehat{\*F}^*}\*B_i^* \right)^{-1}\frac{1}{\sqrt{NT}}\sum_{i=1}^N\*B_i^{*\prime}\*M_{\widehat{\*F}^*}(\widehat{\*F}\widehat{\+\gamma}_{y,i}+\+\varepsilon_i^*)\notag\\
    &=\left(\frac{1}{NT}\sum_{i=1}^N\*B_i^{*\prime}\*M_{\widehat{\*F}^*}\*B_i^* \right)^{-1}\frac{1}{\sqrt{NT}}\sum_{i=1}^N\*B_i^{*\prime}\*M_{\widehat{\*F}^*}([\overline{\*y}, \overline{\*y}_{-1}](\*S_{\gamma_1, \gamma_{K+1}}\widehat{\+\gamma}_{y,i})+\+\varepsilon_i^*)\notag\\
    &=\left(\frac{1}{NT}\sum_{i=1}^N\*B_i^{*\prime}\*M_{\widehat{\*F}^*}\*B_i^* \right)^{-1}\frac{1}{\sqrt{NT}}\sum_{i=1}^N\*B_i^{*\prime}\*M_{\widehat{\*F}^*}([\overline{\*y}, \overline{\*y}_{-1}]\*R\*R^{-1}(\*S_{\gamma_1, \gamma_{K+1}}\widehat{\+\gamma}_{y,i})+\+\varepsilon_i^*)\notag\\
    &= \left(\frac{1}{NT}\sum_{i=1}^N\*B_i^{*\prime}\*M_{\widehat{\*F}^*}\*B_i^* \right)^{-1}\frac{1}{\sqrt{NT}}\sum_{i=1}^N\*B_i^{*\prime}\*M_{\widehat{\*F}^*}(\underbrace{[\overline{\*y}-\overline{\*y}_{-1}\widehat{\alpha}_{CCEP}]}_{\widehat{\*f}_{\widehat{\+\alpha}}}\widetilde{\gamma}_{y,i,1}+\overline{\*y}_{-1}\widetilde{\gamma}_{y,i,2}+\+\varepsilon_i^*)\notag\\
    &=\left(\frac{1}{NT}\sum_{i=1}^N\*B_i^{*\prime}\*M_{\widehat{\*F}^*}\*B_i^* \right)^{-1}\frac{1}{\sqrt{NT}}\sum_{i=1}^N\*B_i^{*\prime}\*M_{\widehat{\*F}^*}(\+\varepsilon_i^*-\overline{\+\varepsilon}^*\widetilde{\gamma}_{y,i,1}-\overline{\*u}_{-1}^*\widetilde{\gamma}_{y,i,2})
\end{align}
where we substituted $\overline{\*y}_{-1}=\overline{\*y}_{-1}^*-\overline{\*u}^*_{-1}$ and (\ref{hat_F_a_withX}). We further investigate $\widetilde{\gamma}_{y,i,1}$ and $\widetilde{\gamma}_{y,i,2}$. In particular, using the fact that $\*R^{-1}=\begin{bmatrix}
    1 & 0\\
    \widehat{\alpha}_{CCEP} & 1
\end{bmatrix}$, we obtain
\begin{align}\label{gamma_y_2_X}
    [\widetilde{\gamma}_{y,i,1},\widetilde{\gamma}_{y,i,2}]'&=\begin{bmatrix}
    1 & 0\\
    \widehat{\alpha}_{CCEP} & 1
\end{bmatrix}\*S_{\gamma_1, \gamma_{K+1}}\widehat{\+\gamma}_{y,i}\notag\\
    &=\begin{bmatrix}
    1 & 0\\
    \widehat{\alpha}_{CCEP} & 1
\end{bmatrix}\*S_{\gamma_1, \gamma_{K+1}}\mathbb{A}_0\overline{\+\Gamma}^{-1}\+\gamma_{y,i}\notag\\
    &+ \begin{bmatrix}
    1 & 0\\
    \widehat{\alpha}_{CCEP} & 1
\end{bmatrix}\*S_{\gamma_1, \gamma_{K+1}}(\widehat{\*F}'\widehat{\*F})^{-1}\widehat{\*F}'(-\*Q\*C_i\*S_\*b(\widehat{\+\delta}_{CCEP}-\+\delta_0)-\*V_i\*S_\*b(\widehat{\+\delta}_{CCEP}-\+\delta_0)-\overline{\*E}\overline{\+\Gamma}^{-1}\+\gamma_{y,i}+\+\varepsilon_i)\notag\\
    &= \begin{bmatrix}
    1 & 0\\
    \widehat{\alpha}_{CCEP} & 1
\end{bmatrix}\*S_{\gamma_1, \gamma_{K+1}}\begin{bmatrix}
        \*I_{K\times K}\\
        -\*A_0
    \end{bmatrix}\overline{\+\Gamma}^{-1}\+\gamma_{y,i} + O_P(T^{-1/2})\notag\\
    &=\begin{bmatrix}
    1 & 0\\
    \widehat{\alpha}_{CCEP} & 1
\end{bmatrix} \begin{bmatrix}
        1 & \*0_{1\times k}\\
        -\alpha_0-\+\beta_0^{\prime}\+\theta_0 & -\+\theta_0^{\prime}
    \end{bmatrix}\overline{\+\Gamma}^{-1}\+\gamma_{y,i}+ O_P(T^{-1/2})\notag\\
    &=\begin{bmatrix}
       1 & \*0_{1\times k} \\
        (\widehat{\alpha}_{CCEP}-\alpha_0)-\+\beta_0^{\prime}\+\theta_0 & -\+\theta_0^{\prime}
    \end{bmatrix}\overline{\+\Gamma}^{-1}\+\gamma_{y,i}+ O_P(T^{-1/2}),
\end{align}
where $\widetilde{\gamma}_{y,1,i}$ asymptotically picks the first element of $\overline{\+\Gamma}^{-1}\+\gamma_{y,i}$, while $\widetilde{\gamma}_{y,1,2}$ is negligible only if $\+\theta_0=\*0_{k\times 1}$. The intuition of this result is as follows. The rotation matrix, either $\*R$ or $\mathbb{R}$ (used in VAR case), separates the CAs that serve as the true factors in the bootstrap world into two groups. The first group corresponds to a linear combination of CAs (effectively, the \textit{sophisticated method}) whose dimension matches the true number of factors in the original sample ($R=K$ in our case). Another one contains CAs that serve as true factors in bootstrap, but act as excess factors relative to the original sample. Because the latter should not exist under the original sample, their loadings should asymptotically vanish in the bootstrap world if the scheme is consistent. However, for this to happen, $\*R$ ($\mathbb{R}$) must be a function of $\widehat{\alpha}_{CCEP}$ ($\widehat{\*A}$) - the true coefficients in the bootstrap world. Although given automatically in a pure AR(1) case, it is only possible in a strictly exogenous case under VAR(1). In the latter case, if we choose to re-create $\*X_i$ in the bootstrap world under strong assumptions, then the needed behavior of the rotation is also given automatically. If we treat $\*X_i$ as fixed, where $\overline{\*X}$ and $\overline{\*X}_{-1}$ are observed factors in the bootstrap world, we still need strict exogeneity ($\+\theta_0=\*0_{k\times 1}$). The reason is because the initial $\widehat{\+\gamma}_{y,i}$ is still obtained using $\overline{\*X}$ and $\overline{\*X}_{-1}$ where the effect of $\+\theta_0$ leaks in, even though they are projected out of the equation $\*y_i^*$, as seen from the result (\ref{gamma_y_2_X}). \\

\noindent We need to obtain a workable representation of the regressors $\*B_i^*$. It is convenient to write it in terms of a single system. Note that by using (\ref{z_VAR}) we can express 
\begin{align}
    \*X_i=\*y_{i,-1} \+\theta_0^{\prime}+\*F\+\Gamma_{x,i}+\+\nu_i= \*y_{i,-1} \+\theta_0^{\prime}+\widehat{\*F}_{\mathbb{A}_0}\overline{\+\Gamma}^{-1}\+\Gamma_{x,i}-\overline{\*E}\overline{\+\Gamma}^{-1}\+\Gamma_{x,i}+\+\nu_i.
\end{align}
Next, we can write 
\begin{align}
    \*y_i^*&=\widehat{\alpha}_{CCEP}\*y_{i,-1}^*+\*X_i\widehat{\+\beta}_{CCEP}+\widehat{\*F}\widehat{\gamma}_{y,i}+\+\varepsilon_i^*\notag\\
    &=\widehat{\alpha}_{CCEP}\*y_{i,-1}^*+\*X_i\widehat{\+\beta}_{CCEP}+\widehat{\*F}\mathbb{R}_0(\mathbb{R}_0)^{-1}\widehat{\gamma}_{y,i}+\+\varepsilon_i^*\notag\\
    &=\widehat{\alpha}_{CCEP}\*y_{i,-1}^*+\*X_i\widehat{\+\beta}_{CCEP}+\widehat{\*F}_{\mathbb{A}_0}\widetilde{\+\gamma}_{0,y,i,1}+\overline{\*Z}_{-1}\widetilde{\+\gamma}_{0,y,i,2}+\+\varepsilon_i^*,
    \end{align}
where we used $\mathbb{R}_0=\begin{bmatrix}
    \*I_{K\times K} & \*0_{K\times K}\\ -\*A_0 & \*I_{K\times K}, 
\end{bmatrix}$. This implies that 
\begin{align}
\*y^*_i&=\*y_{i,-1}^*\widehat{\alpha}_{CCEP}+\widehat{\*F}_{\mathbb{A}_0}(\overline{\+\Gamma}^{-1}\+\Gamma_{x,i}\widehat{\+\beta}_{CCEP}+\widetilde{\+\gamma}_{0,y,i,1})+(\*y_{i,-1}\+\theta_0^{\prime})\widehat{\+\beta}_{CCEP}-\overline{\*E}\overline{\+\Gamma}^{-1}\+\Gamma_{x,i}\widehat{\+\beta}_{CCEP}+\notag\\
&+\overline{\*Z}_{-1}\widetilde{\+\gamma}_{0,y,i,2}+\+\nu_i\widehat{\+\beta}_{CCEP}+\+\varepsilon_i^*,
\end{align}
which means that letting $\*Z_i^*=[\*y_i^*,\*X_i]$ be the bootstrap VAR(1) under the fixed $\*X_i$ design, we obtain
\begin{align}\label{Z_Boot_Endog_X}
   \*Z_{i}^*=\*Z_{i,-1}^*\widehat{\*A}+\widehat{\*F}_{\mathbb{A}_0}\+\Pi_{0,1,i}+\*y_{i,-1}\+\theta_0^{\prime}\+\Pi_2+\overline{\*Z}_{-1}\+\Pi_{0,3,i}+\+\zeta_i^*,
\end{align}
where $\+\Pi_{0,1,i}=[\overline{\+\Gamma}^{-1}\+\Gamma_{x,i}\widehat{\+\beta}_{CCEP}+\widetilde{\+\gamma}_{0,y,i,1}, \overline{\+\Gamma}^{-1}\+\Gamma_{x,i}]\in \mathbb{R}^{K\times K}$, $\+\Pi_2=[\widehat{\+\beta}_{CCEP}, \*I_{k\times k}]\in \mathbb{R}^{k\times K}$, $\+\Pi_{0,3,i}=[\widetilde{\+\gamma}_{0,y,i,2},\*0_{K\times k} ]\\\in \mathbb{R}^{K\times K}$ and $\+\zeta_i^*=[\overline{\*E}\overline{\+\Gamma}^{-1}\+\Gamma_{x,i}\widehat{\+\beta}_{CCEP}+\+\nu_i\widehat{\+\beta}_{CCEP}+\+\varepsilon_i^*, \overline{\*E}\overline{\+\Gamma}^{-1}\+\Gamma_{x,i}+\+\nu_i]\in \mathbb{R}^{T\times K}$. The matrix $\widehat{\*A}$ is given by 
\begin{align}
    \widehat{\*A}=\begin{bmatrix}
        \widehat{\alpha}_{CCEP} & \*0_{1\times k}\\
        \*0_{k\times 1} & \*0_{k\times k}
    \end{bmatrix}.
\end{align}
Under our assumptions, $\mathrm{rk}(\overline{\+\Pi}_{0,1})=K$, and so we can solve for the true factor in the bootstrap world:
\begin{align}\label{true_F-Boot_endog}
    \widehat{\*F}_{\mathbb{A}_0}=(\overline{\*Z}^*-\overline{\*Z}_{-1}^*\widehat{\*A}-\overline{\*y}_{-1}\+\theta_0^{\prime}\+\Pi_2-\overline{\*Z}_{-1}\overline{\+\Pi}_{0,3}-\overline{\+\zeta}^*)(\overline{\+\Pi}^{0}_1)^{-1}.
\end{align}
To further prepare for the analysis of the bootstrap CCEP estimator, we switch to the vector notation and notice that 
\begin{align}
    \*z_{i,t}^*&=\widehat{\*A}^t\*z_{i,0}+\sum_{j=0}^{t-1}\widehat{\*A}^{j}\+\Pi_{0,1,i}^{\prime}\widehat{\*f}_{\mathbb{A}_0,t-j}+\sum_{j=0}^{t-1}\widehat{\*A}^j\+\Pi_2'\+\theta_0y_{i,t-j-1}+\sum_{j=0}^{t-1}\widehat{\*A}^j\+\Pi_{0,3,i}^{\prime}\overline{\*z}_{t-j-1}+\underbrace{\sum_{j=0}^{t-1}\widehat{\*A}^{j}\+\zeta_{i,t-j}^*}_{\+\xi^*_{i,t}}\notag\\
    &= \widehat{\*A}^t\*z_{i,0}+(\mathrm{vec}(\+\Pi_{0,1,i}^{\prime})\otimes \*I_{K\times K})'\sum_{j=0}^{t-1}\mathrm{vec}(\widehat{\*f}_{\mathbb{A}_0,t-j}'\otimes \widehat{\*A}^j)+ (\mathrm{vec}(\+\Pi_2'\+\theta_0)\otimes \*I_{K\times K})'\sum_{j=0}^{t-1}\mathrm{vec}(y_{i,t-j-1}\otimes \widehat{\*A}^j)\notag\\
    &+(\mathrm{vec}(\+\Pi_{0,3,i}^{\prime})\otimes \*I_{K\times K})'\sum_{j=0}^{t-1}\mathrm{vec}(\overline{\*z}_{t-j-1}'\otimes \widehat{\*A}^j)+\+\xi_{i,t}^*\notag\\
    &=\widehat{\*A}^t\*z_{i,0} +\widetilde{\+\Pi}_{0,1,i}^{\prime}\widehat{\*g}_{0,t}+\widetilde{\+\Pi}_2'\mathcal{\+\eta}_{i,t} + \widetilde{\+\Pi}_{0,3,i}^{\prime}\*b_t+\+\xi_{i,t}^*,
\end{align}
where $\widetilde{\+\Pi}_{0,1,i}=\mathrm{vec}(\+\Pi_{0,1,i}^{\prime})\otimes \*I_{K\times K}$, $\widetilde{\+\Pi}_2=\mathrm{vec}(\+\Pi_2'\+\theta_0)\otimes \*I_{K\times K}$, $\widetilde{\+\Pi}_{0,3,i}=\mathrm{vec}(\+\Pi_{0,3,i}^{\prime})\otimes \*I_{K\times K}$, and the definitions of the factors are obvious. Note that the expression is structurally very similar to the one in (\ref{z_star_X_exog}), which appears when (\ref{z_VAR}) is re-constructed in bootstrap exactly. In fact, $\*b_t$ is the same term, and $\widehat{\*g}_{0,t}$ only stresses the fact that we use $\widehat{\*f}_{\mathbb{A}_0}$ and not $\widehat{\*f}_{\widehat{\mathbb{A}}}$. We still have $\*z_{i,0}$ as we assume availability of the initial values that we utilize in the bootstrap, as well. The fundamental difference arises due to the presence of the third term, which exactly generates weak exogeneity in our regressors, unless $\+\theta_0=\*0_{k\times 1}$. Thus, in line with (\ref{d_it_stat_exog}), we obtain 
\begin{align}
    \*d_{i,t}^*=\begin{bmatrix}
        \widehat{\*A}\\ \*I_{K\times K}
    \end{bmatrix}\widehat{\*A}^{t-1}\*z_{i,0}+\mathcal{\+P}_{0,1,i}^{\prime}\widehat{\*q}_{0,t}+ \mathcal{\+P}_{2}'\+\kappa_{i,t}+\mathcal{\+P}_{0,3,i}^{\prime}\*h_t+\*u_{i,t}^*,
\end{align}
where we define $\mathcal{\+P}_{0,1,i}=\*I_{2 \times 2}\otimes \widetilde{\+\Pi}_{0,1,i}$, and $\mathcal{\+P}_{2}$ and $\mathcal{\+P}_{0,3,i}$ are defined analogously. Also, $\widehat{\*q}_{0,t}=[\widehat{\*g}^{\prime}_{0,t},\widehat{\*g}^{\prime}_{0,t-1}]' $, $\+\kappa_{i,t}=[\+\eta_{i,t}', \+\eta_{i,t-1}']'$, $\*h_t=[\*b_t', \*b_{t-1}']'$ and $\*u_{i,t}^*=[\+\xi^{*\prime}_{i,t}, \+\xi^{*\prime}_{i,t-1}]'$. This leads to 
\begin{align}
    \*B_{i}^*=[\*y_{i,-1}^*, \*X_i]=\*D_i^*\*S_\*b= (\widehat{\*A}_{\*z_0,i}[ \widehat{\*A}, \*I_{K\times K}]+\widehat{\*Q}_0\mathcal{\+P}_{0,1,i}+\+\kappa_i\mathcal{\+P}_2+\*H\mathcal{\+P}_{0,3,i}+\*U_i^*)\*S_\*b,
\end{align}
which can be substituted into (\ref{CCEP_X_Endog}). 
\paragraph {Useful Interim Lemmas}\label{4.3.3}
\noindent Before we start, we state and prove two useful lemmas that will allow us to derive the limits of the asymptotic bias terms in the bootstrap realm. Lemma XA reveals that unlike $\widehat{\gamma}_{y,i,2}$, which is contaminated by $\+\theta_0$, $\widetilde{\+\gamma}_{0,y,i,2}$ is actually negligible due to utilization of the full $\*A_0$ in the rotation matrix. Lemma XB shows that the bootstrap loadings for the $K=R$ factors are consistent for the true loadings. 
\\

\noindent \textbf{Lemma XA.} \textit{Under Assumptions \ref{ass::1} - \ref{ass::6}, we have as $(N,T)\to \infty$
\begin{enumerate}[(a)]
    \item $\widetilde{\+\gamma}_{0,y,i,1}=\overline{\+\Gamma}^{-1}\+\gamma_{yi}+O_P(T^{-1/2})$.
    \item $\widetilde{\+\gamma}_{0,y,i,2}=O_P(T^{-1/2})$
\end{enumerate}}
 \noindent \textbf{Proof.} (a) Using the fact that $(\mathbb{R}_0)^{-1}=\begin{bmatrix}
     \*I_{K\times K} & \*0_{K\times K}\\
     \*A_0 & \*I_{K\times K}
 \end{bmatrix}$, we have that 
\begin{align}
    \widetilde{\+\gamma}_{0,y,i,1}&=[\*I_{K\times K}, \*0_{K\times K}]\mathbb{A}_0\overline{\+\Gamma}^{-1}\+\gamma_{y,i}\notag\\
    &+ [\*I_{K\times K}, \*0_{K\times K}](\widehat{\*F}'\widehat{\*F})^{-1}\widehat{\*F}'(-\*Q\*C_i\*S_\*b(\widehat{\+\delta}_{CCEP}-\+\delta_0)-\*V_i\*S_\*b(\widehat{\+\delta}_{CCEP}-\+\delta_0)-\overline{\*E}\overline{\+\Gamma}^{-1}\+\gamma_{y,i}+\+\varepsilon_i)\notag\\
    &= \overline{\+\Gamma}^{-1}\+\gamma_{y,i}+[\*I_{K\times K}, \*0_{K\times K}](\widehat{\*F}'\widehat{\*F})^{-1}\widehat{\*F}'(-\*Q\*C_i\*S_\*b(\widehat{\+\delta}_{CCEP}-\+\delta_0)-\*V_i\*S_\*b(\widehat{\+\delta}_{CCEP}-\+\delta_0)-\overline{\*E}\overline{\+\Gamma}^{-1}\+\gamma_{y,i}+\+\varepsilon_i)\notag\\
    &=\overline{\+\Gamma}^{-1}\+\gamma_{y,i} + O_P(T^{-1/2}),
\end{align}
 since the leading term in the second component is 
 \begin{align}
     \left\| [\*I_{K\times K}, \*0_{K\times K}](\widehat{\*F}'\widehat{\*F})^{-1}\widehat{\*F}'\+\varepsilon_i\right\|\leq  \left\| [\*I_{K\times K}, \*0_{K\times K}]\right\| \left\|(T^{-1}\widehat{\*F}'\widehat{\*F})^{-1} \right\|\left\|T^{-1}\widehat{\*F}'\+\varepsilon_i \right\|=O_P(T^{-1/2})
 \end{align}
 for each $i$ from $\left\|T^{-1}\*Q'\+\varepsilon_i \right\|=O_P(T^{-1/2})$. \\

 \noindent (b) Using the same definition of $(\mathbb{R}_0)^{-1}$, we have that 
\begin{align}
    &\widetilde{\+\gamma}_{0,y,i,2}=\underbrace{\begin{bmatrix}
        \*A_0, \*I_{K\times K}
    \end{bmatrix} \mathbb{A}_0\overline{\+\Gamma}^{-1}\+\gamma_{y,i}}_{\*0_{K\times 1}}\notag\\
    &+ \begin{bmatrix}
        \*A_0, \*I_{K\times K}
    \end{bmatrix} (T^{-1}\widehat{\*F}'\widehat{\*F})^{-1}T^{-1}\widehat{\*F}'(-\*Q\*C_i\*S_\*b(\widehat{\+\delta}_{CCEP}-\+\delta_0)-\*V_i\*S_\*b(\widehat{\+\delta}_{CCEP}-\+\delta_0)-\overline{\*E}\overline{\+\Gamma}^{-1}\+\gamma_{y,i}+\+\varepsilon_i)\notag\\
    &=O_P(T^{-1/2}),
\end{align}
where the second term follows directly from (a).\\

\noindent \textbf{Lemma XB.} \textit{Under Assumptions \ref{ass::1} -\ref{ass::6}, we have that
\begin{enumerate}[(a)]
    \item $\widehat{\*Q}_0=\*Q_{\widehat{\alpha}}(\*I_{2 \times 2}\otimes \+\Pi_{\overline{\+\Gamma}})+O_P(\sqrt{T}N^{-1/2})$, where $\*Q_{\widehat{\alpha}}=[\*G_{\widehat{\alpha}},\*G_{\widehat{\alpha},-1} ]$, with $\*g_{\widehat{\alpha},t}=\sum_{j=0}^{t-1}\mathrm{vec}(\*f_{t-j}'\otimes \widehat{\*A}^j)$ and $\+\Pi_{\overline{\+\Gamma}}=\overline{\+\Gamma}\otimes \*I_{K^2 \times K^2}$. 
    \item $\+\Pi_{\overline{\+\Gamma}}\mathcal{\+P}_{0,1,i}=\*C_i+O_P(T^{-1/2})$.
    \item $\mathcal{\+P}_{0,1,i}^{\prime}T^{-1}\widehat{\*Q}_0^{\prime}\*Q=\*C_i'T^{-1}\*Q'_{\widehat{\alpha}}\*Q+O_P(C_{N,T}^{-1})$. 
\end{enumerate}}

\noindent \textbf{Proof.} (a) Note that $\widehat{\*Q}_0=[\widehat{\*G}_0, \widehat{\*G}_{0,-1}]$, where we can evaluate a single $ \widehat{\*g}_{0,t}$. We can write
 \begin{align}
     \widehat{\*g}_{0,t}&=\sum_{j=0}^{t-1}\mathrm{vec}(\widehat{\*f}'_{\mathbb{A}_0,t-j}\otimes \widehat{\*A}^j)=\sum_{j=0}^{t-1}\mathrm{vec}((\*f_{t-j}'\overline{\+\Gamma}+\overline{\*e}_{t-j}')\otimes \widehat{\*A}^j)\notag\\
     &=\sum_{j=0}^{t-1}\mathrm{vec}(\*f_{t-j}'\overline{\+\Gamma}\otimes \widehat{\*A}^j)+\underbrace{N^{-1/2}\sum_{j=0}^{t-1}\mathrm{vec}((\sqrt{N}\overline{\*e}_{t-j}')\otimes \widehat{\*A}^j)}_{\widehat{\*g}_{\overline{\*e},t}}\notag\\
     &=\sum_{j=0}^{t-1}\mathrm{vec}((\*f_{t-j}'\otimes \widehat{\*A}^j)(\overline{\+\Gamma}\otimes \*I_{K\times K}))+ \widehat{\*g}_{\overline{\*e},t}\notag\\
     &=((\overline{\+\Gamma}\otimes \*I_{K\times K})' \otimes \*I_{K\times K})\sum_{j=0}^{t-1}\mathrm{vec}(\*f_{t-j}'\otimes \widehat{\*A}^j)+\widehat{\*g}_{\overline{\*e},t}\notag\\
     &=\+\Pi_{\overline{\+\Gamma}}'\*g_{\widehat{\alpha},t}+ \widehat{\*g}_{\overline{\*e},t},
 \end{align}
 where we let and $\+\Pi_{\overline{\+\Gamma}}=(\overline{\+\Gamma}\otimes \*I_{K^2 \times K^2})$ and $\*g_{\widehat{\alpha},t}=\sum_{j=0}^{t-1}\mathrm{vec}(\*f_{t-j}'\otimes \widehat{\*A}^j)$ to stress the dependence \textit{only} on $\widehat{\alpha}_{CCEP}$ due to the design of $\widehat{\*A}$. Note that crucially for each $t$ we have $\left\| \widehat{\*g}_{\overline{\*e},t}\right\|=O_{P^*}(N^{-1/2})$ because of consistency of $\widehat{\alpha}_{CCEP}$. Using this, we have that 
 \begin{align}
    \widehat{\*Q}_0= [\widehat{\*G}_0, \widehat{\*G}_{0,-1}]&=\left[\*G_{\widehat{\alpha}}\+\Pi_{\overline{\+\Gamma}}, \*G_{\widehat{\alpha},-1}\+\Pi_{\overline{\+\Gamma}} \right] + \left[\widehat{\*G}_{\overline{\*e}}, \widehat{\*G}_{\overline{\*e},-1} \right]\notag\\
    &=\*Q_{\widehat{\alpha}}(\*I_{2 \times 2}\otimes \+\Pi_{\overline{\+\Gamma}})+\widehat{\*Q}_{\overline{\*e}},
 \end{align}
 which can be inserted into the derived bias/distribution terms by accounting for $\left\|\widehat{\*Q}_{\overline{\*e}} \right\|=O_P(\sqrt{T}N^{-1/2})$.\\

 \noindent (b) By using Kronecker and vectorization properties repeatedly, we obtain
 \begin{align}
     (\*I_{2 \times 2}\otimes \+\Pi_{\overline{\+\Gamma}} )\mathcal{\+P}_{0,1,i}&=(\*I_{2 \times 2}\otimes \+\Pi_{\overline{\+\Gamma}} )(\*I_{2 \times 2}\otimes \widetilde{\+\Pi}_{0,1,i})=\*I_{2 \times 2}\otimes \+\Pi_{\overline{\+\Gamma}}(\mathrm{vec}(\+\Pi_{0,1,i}^{\prime})\otimes \*I_{K\times K}) \notag\\
     &=\*I_{2 \times 2}\otimes \+\Pi_{\overline{\+\Gamma}}((\+\Pi_{0,1,i}\otimes \*I_{K\times K})\mathrm{vec}(\*I_{K\times K})\otimes \*I_{K\times K})\notag\\
     &=\*I_{2 \times 2}\otimes (\overline{\+\Gamma}\otimes \*I_{K^2 \times K^2})((\+\Pi_{0,1,i}\otimes \*I_{K\times K})\mathrm{vec}(\*I_{K\times K})\otimes \*I_{K\times K})\notag\\
     &= \*I_{2 \times 2}\otimes ((\overline{\+\Gamma}\otimes \*I_{K\times K})\otimes \*I_{K\times K})((\+\Pi_{0,1,i}\otimes \*I_{K\times K})\mathrm{vec}(\*I_{K\times K})\otimes \*I_{K\times K})\notag\\
     &=\*I_{2 \times 2}\otimes ((\overline{\+\Gamma}\+\Pi_{0,1,i}\otimes \*I_{K\times K})\mathrm{vec}(\*I_{K\times K})\otimes \*I_{K\times K})\notag\\
     &= \*I_{2 \times 2}\otimes ((\widetilde{\+\Gamma}_{0,i}\otimes \*I_{K\times K})\mathrm{vec}(\*I_{K\times K}) \otimes \*I_{K\times K})\notag\\
     &=\*I_{2 \times 2}\otimes (\mathrm{vec}(\widetilde{\+\Gamma}_{0,i}^{\prime})\otimes \*I_{K\times K})=\widetilde{\*C}_{0,i},
 \end{align}
 where $\widetilde{\+\Gamma}_{0,i}= \overline{\+\Gamma}\+\Pi_{0,1,i}=\overline{\+\Gamma}[\overline{\+\Gamma}^{-1}\+\Gamma_{x,i}\widehat{\+\beta}_{CCEP}+\widetilde{\+\gamma}_{0,y,i,1}, \overline{\+\Gamma}^{-1}\+\Gamma_{x,i}]=[\+\Gamma_{x,i}\widehat{\+\beta}_{CCEP}+\overline{\+\Gamma}\widetilde{\+\gamma}_{0,y,i,1}, \+\Gamma_{x,i}]$. Notice that using Lemma X1 (a), we obtain 
 \begin{align}
     \overline{\+\Gamma}\widetilde{\+\gamma}_{0,y,i,1}=\overline{\+\Gamma}[\*I_{K\times K}, \*0_{K\times K}]\mathbb{A}_0\overline{\+\Gamma}^{-1}\+\gamma_{y,i}+ O_P(T^{-1/2})=\+\gamma_{y,i}+ O_P(T^{-1/2}),
 \end{align}
 and, therefore, by the consistency of the CCEP, we have that $\widetilde{\*C}_{0,i}=\*C_i+O_P(T^{-1/2})$.\\

 \noindent (c)  By merging parts (a) and (b), we obtain 
 \begin{align}
     \mathcal{\+P}_{0,1,i}^{\prime}T^{-1}\widehat{\*Q}_0^{\prime}\*Q&=\*C_i'T^{-1}\*Q'_{\widehat{\alpha}}\*Q+N^{-1/2}\*C_i'T^{-1}(\sqrt{N}\widehat{\*Q}_{\overline{\*e}})'\*Q+O_P(T^{-1/2})\notag\\
     &=\*C_i'T^{-1}\*Q'_{\widehat{\alpha}}\*Q+O_P(C_{N,T}^{-1}),
 \end{align}
which will be used to simplify the asymptotic expressions. \\

\noindent \textbf{Lemma XC}. \textit{Under Assumptions \ref{ass::1} - \ref{ass::6}, we have for $p=4+\delta$ as $(N,T)\to \infty$
\begin{enumerate}[(a)]
    \item $\frac{1}{NT}\sum_{i=1}^N\sum_{t=2}^T|\widehat{\varepsilon}_{i,t}|^p=O_P(1)$,
    \item $\frac{1}{N}\sum_{i=1}^N\left\|\widehat{\+\gamma}_i\right\|^p=O_P(1)$.
\end{enumerate}}
\bigskip 

\noindent \textbf{Proof.} The proofs of (a) and (b) are analogous to Lemma A, where we now use the expansions in (\ref{resid_with_X}) and (\ref{estimated_gamma_with_X}), respectively. 
\paragraph{Expansion and Rates of the Numerator}\label{4.3.4}
\noindent We decompose the numerator into 
\begin{align}
    \frac{1}{\sqrt{NT}}\sum_{i=1}^N\*B_i^{*\prime}\*M_{\widehat{\*F}^*}(\+\varepsilon_i^*-\overline{\+\varepsilon}^*\widetilde{\gamma}_{y,i,1}-\overline{\*u}_{-1}^*\widetilde{\gamma}_{y,i,2})=\*I  -\mathbf{II}-\mathbf{III}
\end{align}
and will derive the limits in turn.
\bigskip 

\noindent \textbf{Lemma X1.} \textit{Under Assumptions \ref{ass::1} - \ref{ass::6}, with either multiplicative or non-multiplicative weights, as $(N,T)\to \infty$ with $NT^{-1}\to \kappa>0$, we have}
\begin{enumerate}[(a)]
    \item \begin{align*}
       \mathbf{II}&=\*S_\*b'\+\Sigma_{\gamma_{y,1}\*C}'\frac{1}{\sqrt{N}}\sum_{i=1}^NT^{-1/2}\*Q'_{\widehat{\alpha}}\+\varepsilon_i^*+\*S_\*b'\mathcal{\+P}_{0,2}^{\prime}\frac{1}{\sqrt{N}}\sum_{i=1}^NT^{-1/2}\overline{\+\kappa}_{\widetilde{\gamma}_{y,1}}'\+\varepsilon_i^*\notag\\
     &-\*S_\*b'\+\Theta_{\gamma_{y,1},\alpha_0}'\*C(\*C'\+\Sigma_\*q\*C)^{-1}\*C'\frac{1}{\sqrt{N}}\sum_{i=1}^NT^{-1/2}\*Q'\+\varepsilon_i^*\notag\\
      &-\kappa^{-1/2}\*S_\*b'\+\Sigma_{\gamma_{y,1}\*C}'\+\Sigma_{\*q_{\alpha_0}}\*C(\*C'\+\Sigma_\*q\*C)^{-1}[\sigma^2, \*0_{1\times (2K-1)}]'\notag\\
        &-\kappa^{-1/2}\*S_\*b'\mathcal{\+P}_{0,2}^{\prime}\+\Sigma_{\gamma_{y,1}\*q\+\kappa}'\*C(\*C'\+\Sigma_\*q\*C)^{-1}[\sigma^2, \*0_{1\times (2K-1)}]'\notag\\
        &+o_{P^*}(1),
    \end{align*}
    where $\overline{\+\kappa}_{\widetilde{\gamma}_{y,1}}=\frac{1}{N}\sum_{i=1}^N\widetilde{\gamma}_{y,i,1}\+\kappa_i$, $\+\Sigma_{\*q_{\alpha_0}}=\plim_{(N,T)\to \infty}T^{-1}\*Q_{\widehat{\alpha}}'\*Q$, $\mathcal{\+P}_{0,2}=\plim_{(N,T)\to \infty}\mathcal{\+P}_2$,   $\+\Theta_{\gamma_{y,1},\alpha_0}=\+\Sigma_{\*q_{\alpha_0}}'\+\Sigma_{\gamma_{y,1}\*C}+\+\Sigma_{\gamma_{y,1}\*q\+\kappa}\mathcal{\+P}_{0,2}$, $\+\Sigma_{\gamma_{y,1}\*C}=\plim_{(N,T)\to \infty}\frac{1}{N}\sum_{i=1}^N\widetilde{\gamma}_{y,i,1}\*C_i$ and $\+\Sigma_{\gamma_{y,1}\*q\+\kappa}=\plim_{(N,T)\to \infty}\frac{1}{N}\sum_{i=1}^N\widetilde{\gamma}_{y,i,1}T^{-1}\*Q'\+\kappa_i$. 
    \item  \begin{align*}
     \mathbf{III}&=\*S_\*b'\+\Sigma_{\gamma_{y,2}\*C}'\frac{1}{\sqrt{N}}\sum_{i=1}^NT^{-1/2}\*Q'_{\widehat{\alpha}}\*u_{i,-1}^*+\*S_\*b'\mathcal{\+P}_{0,2}^{\prime}\frac{1}{\sqrt{N}}\sum_{i=1}^NT^{-1/2}\overline{\+\kappa}_{\widetilde{\gamma}_{y,2}}'\*u_{i,-1}^*\notag\\
     &-\*S_\*b'\+\Theta_{\gamma_{y,2},\alpha_0}'\*C(\*C'\+\Sigma_\*q\*C)^{-1}\*C'\frac{1}{\sqrt{N}}\sum_{i=1}^NT^{-1/2}\*Q'\*u_{i,-1}^*\notag\\
       &-\kappa^{-1/2}\frac{\sigma^2}{1-(\alpha_0)^2}\*S_\*b'\+\Sigma_{\gamma_{y,2}\*C}'\+\Sigma_{\*q_{\alpha_0}}\*C(\*C'\+\Sigma_\*q\*C)^{-1}[
        \alpha_0 , \*0_{1\times k}, 1, \*0_{1\times k}   ]'\notag\\
        &-\kappa^{-1/2}\frac{\sigma^2}{1-(\alpha_0)^2}\*S_\*b'\mathcal{\+P}_{0,2}^{\prime}\+\Sigma_{\gamma_{y,2}\*q\+\kappa}'\*C(\*C'\+\Sigma_\*q\*C)^{-1} [
        \alpha_0 , \*0_{1\times k}, 1, \*0_{1\times k}   ]'\notag\\
        &+ \kappa^{-1/2}\frac{1}{1-(\alpha_0)^2}\left[\gamma_{y,2}, \*0_{1\times k}\right]'+ o_{P^*}(1).
 \end{align*}
   where $\overline{\+\kappa}_{\widetilde{\gamma}_{y,2}}=\frac{1}{N}\sum_{i=1}^N\widetilde{\gamma}_{y,i,2}\+\kappa_i$, and  $\+\Theta_{\gamma_{y,2},\alpha_0}=\+\Sigma_{\*q_{\alpha_0}}'\+\Sigma_{\gamma_{y,2}\*C}+\+\Sigma_{\gamma_{y,2}\*q\+\kappa}\mathcal{\+P}_{0,2}$, and $ \gamma_{y,2}=\plim_{N\to \infty}\frac{1}{N}\sum_{i=1}^N\widetilde{\gamma}_{y,i,2}$,  $\+\Sigma_{\gamma_{y,2}\*C}=\plim_{(N,T)\to \infty}\frac{1}{N}\sum_{i=1}^N\widetilde{\gamma}_{y,i,2}\*C_i$, \\and $\+\Sigma_{\gamma_{y,2}\*q\+\kappa}=\plim_{(N,T)\to \infty}\frac{1}{N}\sum_{i=1}^N\widetilde{\gamma}_{y,i,2}T^{-1}\*Q'\+\kappa_i$. 
    \item \begin{align*}
          \*I&=\*S_\*b'\frac{1}{\sqrt{NT}}\sum_{i=1}^N\*C_i'\*Q_{\widehat{\alpha}}^{\prime}\+\varepsilon_i^* + \*S_\*b'\mathcal{\+P}_{0,2}^{\prime}\frac{1}{\sqrt{NT}}\sum_{i=1}^N\+\kappa_i'\+\varepsilon_i^*+\frac{1}{\sqrt{NT}}\sum_{i=1}^N\mathcal{\+V}_i^{*\prime}\+\varepsilon_i^*\notag\\
     &-\+\Psi_{\alpha_0}'\frac{1}{\sqrt{N}}\sum_{i=1}^N\mathrm{vec}\left(T^{-1/2}\+\varepsilon_i^{*\prime}\*Q \otimes \*C_i' \right)\notag\\
     &-\+\Psi_{\mathcal{\+P}_{0,2}}'\frac{1}{\sqrt{N}}\sum_{i=1}^N\mathrm{vec}\left(T^{-1/2}\+\varepsilon_i^{*\prime}\*Q \otimes \+\Sigma_{\*q\+\kappa,i}' \right)\notag\\
        &- \sqrt{\kappa}\sum_{h=1}^\infty\+\Sigma_{\varepsilon\mathcal{\+v}}(-h)'\mathrm{tr}\left(\+\Sigma_\*q(h)\*C(\*C'\+\Sigma_\*q\*C)^{-1}\*C'\right)\notag\\
        &-\kappa^{-1/2}\*S_\*b'\*C'\+\Sigma_{\*q_{\alpha_0}}\*C(\*C'\+\Sigma_\*q\*C)^{-1}[\sigma^2, \*0_{1\times (2K-1)}]'\notag\\
        &-\kappa^{-1/2}\*S_\*b'\mathcal{\+P}_{0,2}^{\prime}\+\Sigma_{\*q\+\kappa}'\*C(\*C'\+\Sigma_\*q\*C)^{-1}[\sigma^2, \*0_{1\times (2K-1)} ]'   + o_{P^*}(1),
    \end{align*}
    where $\+\Sigma_{\*q\+\kappa}=\plim_{(N,T)\to \infty}\frac{1}{N}\sum_{i=1}^NT^{-1}\*Q'\+\kappa_i$, $\+\Sigma_{\*q\+\kappa,i}=\plim_{T\to \infty}T^{-1}\*Q'\+\kappa_i$, $\mathcal{\+V}^*_{i}=[\mathcal{\+v}^*_{i,2},\ldots, \mathcal{\+v}_{i,T}^*]'$ with $\mathcal{\+v}^*_{i,t}=[\widehat{\+\beta}_{CCEP}'\+\nu_{i,t}^{-}+u_{i,t-1}^*, \+\nu_{i,t}']'$ and $\+\Psi_{\alpha_0}=\mathrm{vec}(\+\Sigma_{\*q_{\alpha_0}}\*C(\*C'\+\Sigma_\*q\*C)^{-1}\*C')\otimes \*S_\*b$ and $\+\Psi_{\mathcal{\+P}_{0,2}}= \mathrm{vec}(\*C(\*C'\+\Sigma_\*q\*C)^{-1}\*C')\otimes \mathcal{\+P}_{0,2}\*S_\*b $. Also, $\+\nu_{i,t}^-=\sum_{j=0}^{t-2}\widehat{\alpha}_{CCEP}^j\+\nu_{i,t-j-1}$, and  $\+\Sigma_{\+\varepsilon\mathcal{\+v}}(-h)'=[\sigma^2(\alpha^{0})^{h-1}, \*0_{1\times k}]'$ in the notation of \cite{juodis2021robustness}.
\end{enumerate}
\bigskip 

\noindent \textbf{Proof.} (a) We start the numerator analysis from $\mathbf{II}$:
\begin{align}
\mathbf{II}=\frac{1}{\sqrt{NT}}\sum_{i=1}^N\*B_i^{*\prime}\*M_{\widehat{\*F}^*}\overline{\+\varepsilon}^*\widetilde{\gamma}_{y,i,1}&=\frac{1}{\sqrt{NT}}\sum_{i=1}^N\*B_i^{*\prime}\overline{\+\varepsilon}^*\widetilde{\gamma}_{y,i,1}-\frac{1}{\sqrt{NT}}\sum_{i=1}^N\*B_i^{*\prime}\*P_{\widehat{\*F}^*}\overline{\+\varepsilon}^*\widetilde{\gamma}_{y,i,1}\notag\\
&=\mathbf{II}_1-\mathbf{II}_2,
\end{align}
where 
\begin{align}
    \mathbf{II}_1&=\frac{1}{\sqrt{NT}}\sum_{i=1}^N\*S_\*b'[\widehat{\*A}, \*I_{K\times K}]'\widehat{\*A}_{\*z_0,i}'\overline{\+\varepsilon}^*\widetilde{\gamma}_{y,i,1}+\frac{1}{\sqrt{NT}}\sum_{i=1}^N\*S_\*b'\mathcal{\+P}_{0,1,i}^{\prime}\widehat{\*Q}_0^{\prime}\overline{\+\varepsilon}^*\widetilde{\gamma}_{y,i,1}\notag\\
    &+\frac{1}{\sqrt{NT}}\sum_{i=1}^N\*S_\*b'\mathcal{\+P}_{2}'\+\kappa_i'\overline{\+\varepsilon}^*\widetilde{\gamma}_{y,i,1}+\frac{1}{\sqrt{NT}}\sum_{i=1}^N\*S_\*b'\mathcal{\+P}_{0,3,i}^{\prime}\*H'\overline{\+\varepsilon}^*\widetilde{\gamma}_{y,i,1}\notag\\
&+\frac{1}{\sqrt{NT}}\sum_{i=1}^N\*S_\*b'\*U_i^{*\prime}\overline{\+\varepsilon}^*\widetilde{\gamma}_{y,i,1}\notag\\
&=\mathbf{II}_{1,a}+\mathbf{II}_{1,b}+\mathbf{II}_{1,c}+\mathbf{II}_{1,d}+\mathbf{II}_{1,e}
\end{align}
Notice that by using the structure of $\widehat{\*A}$ we get
\begin{align}
    \mathbf{II}_{1,a}&=\frac{1}{\sqrt{NT}}\sum_{i=1}^N\*S_\*b'[\widehat{\*A}, \*I_{K\times K}]'\sum_{t=2}^T\widehat{\*A}^{t-1}\*z_{i,0}\overline{\varepsilon}^*_t\widetilde{\gamma}_{y,i,1}\notag\\
    &=\*S_\*b'[\widehat{\*A}, \*I_{K\times K}]'\frac{1}{\sqrt{NT}}\sum_{i=1}^N\sum_{t=2}^T\widehat{\*A}^{t-1}\*z_{i,0}\overline{\varepsilon}^*_t\widetilde{\gamma}_{y,i,1}\notag\\
    &=\*S_\*b'[\widehat{\*A}, \*I_{K\times K}]'\begin{bmatrix}
        \frac{1}{\sqrt{NT}}\sum_{i=1}^N\sum_{t=2}^T\widehat{\alpha}_{CCEP}^{t-1}y_{i,0}\overline{\varepsilon}_t^*\widetilde{\gamma}_{y,i,1}\\
        \*0_{k\times 1}
    \end{bmatrix}\notag\\
    &= \*S_\*b'[\widehat{\*A}, \*I_{K\times K}]'\begin{bmatrix}
        \frac{1}{N}\sum_{i=1}^Ny_{i,0}\widetilde{\gamma}_{y,i,1}\frac{1}{\sqrt{T}}\sum_{t=2}^T\widehat{\alpha}_{CCEP}^{t-1}\sqrt{N}\overline{\varepsilon}_t^*\\
        \*0_{k\times 1}
    \end{bmatrix},
\end{align}
which means that 
\begin{align}\label{II_1a_X}
    \left\| \mathbf{II}_{1,a}\right\|&\leq \left\|\*S_\*b'[\widehat{\*A}, \*I_{K\times K}]' \right\|\left| \frac{1}{N}\sum_{i=1}^Ny_{i,0}\widetilde{\gamma}_{y,i,1} \right|\left|\frac{1}{\sqrt{T}}\sum_{t=2}^T\widehat{\alpha}_{CCEP}^{t-1}\sqrt{N}\overline{\varepsilon}_t^* \right|\notag\\
    &\leq \left\|\*S_\*b'[\widehat{\*A}, \*I_{K\times K}]' \right\|\left( \frac{1}{N}\sum_{i=1}^Ny_{i,0}^2\right)^{1/2}\left(\frac{1}{N}\sum_{i=1}^N\widetilde{\gamma}_{y,i,1}^2\right)^{1/2}\left|\frac{1}{\sqrt{NT}}\sum_{i=1}^N\sum_{t=2}^T\widehat{\alpha}_{CCEP}^{t-1}\varepsilon_{i,t}^* \right|\notag\\
    &= O_{P^*}(T^{-\delta/2(4+\delta)})
\end{align}
under $TN^{-1}\to c>0$, which stems directly from (\ref{b_term_slow_rate}). Therefore, the effect of the initial value disappears asymptotically, similarly to the pure AR(1) case. Next, 
\begin{align}
    \mathbf{II}_{1,b}=\frac{1}{\sqrt{NT}}\sum_{i=1}^N\*S_\*b'\mathcal{\+P}_{0,1,i}^{\prime}\widehat{\*Q}_0^{\prime}\overline{\+\varepsilon}^*\widetilde{\gamma}_{y,i,1}=\left(\frac{1}{N}\sum_{i=1}^N\widetilde{\gamma}_{y,i,1}\mathcal{\+P}_{0,1,i}\*S_\*b\right)'\sqrt{N} T^{-1/2}\widehat{\*Q}_0^{\prime}\overline{\+\varepsilon}^*=O_{P^*}(1),
\end{align}
which is similar to the distribution generating term $\*A^{\*F}_{NT,1}$ on p. 39 in the Supplement of \cite{DeVos2019}. The key difference is that the effect of $\overline{\*X}$ and $\overline{\*X}_{-1}$ is projected outs as they are the observed factors. The other  term survives only if $\+\theta_0\neq \*0_{k\times 1}$: 
\begin{align}
    \mathbf{II}_{1,c}=\*S_\*b'\mathcal{\+P}_2'\left(\frac{1}{N}\sum_{i=1}^N\widetilde{\gamma}_{y,i,1}\+\kappa_i \right)'\sqrt{N}T^{-1/2}\overline{\+\varepsilon}^*=O_{P^*}(1),
\end{align}
while
\begin{align}
    \left\|\mathbf{II}_{1,d}\right\|&\leq\left\|\left(\frac{1}{N}\sum_{i=1}^N\widetilde{\gamma}_{y,i,1}\mathcal{\+P}_{0,3,i}\*S_\*b\right)\right\|\left\|\sqrt{N} T^{-1/2}\*H'\overline{\+\varepsilon}^*\right\|\notag\\
    &= \frac{1}{\sqrt{T}}\left\|\left(\frac{1}{N}\sum_{i=1}^N\widetilde{\gamma}_{y,i,1}(\sqrt{T}\mathcal{\+P}_{0,3,i})\*S_\*b\right)\right\|\left\|\sqrt{N} T^{-1/2}\*H'\overline{\+\varepsilon}^*\right\|=O_{P^*}(T^{-1/2})
\end{align}
Note that if $\+\theta_0=\*0_{k\times 1}$, then $\mathbf{II}_{1,c}=\*0_{K\times 1}$ exactly by the definition of $\mathcal{\+P}_2$. The last term will require more work. Note that by definition of $\*S_\*b$, $\*U_i^*$ and $\widehat{\*A}$, we have that $t$-th row of $\*U_i^*$ gives 
\begin{align}\label{S_bU_def}
    \*S_\*b'\*u_{i,t}^*&=\*S_\*b'\begin{bmatrix} \sum_{j=0}^{t-1}\widehat{\*A}^{j}\+\zeta_{i,t-j}^*\\
    \sum_{j=0}^{t-2}\widehat{\*A}^{j}\+\zeta_{i,t-1-j}^*
    \end{bmatrix}\notag\\
    &=\
    \begin{bmatrix}
\widehat{\+\beta}_{CCEP}'\+\Gamma_{x,i}'(\overline{\+\Gamma}^{-1})'\sum_{j=0}^{t-2}\widehat{\alpha}_{CCEP}^j\overline{\*e}_{t-1-j}+\widehat{\+\beta}_{CCEP}'\sum_{j=0}^{t-2}\widehat{\alpha}_{CCEP}^j\+\nu_{i,t-1-j}+u_{i,t-1}^*\\
\+\Gamma_{x,i}'(\overline{\+\Gamma}^{-1})'\overline{\*e}_t+\+\nu_{i,t}
    \end{bmatrix},
\end{align}
where, again $u_{i,t-1}^*=\sum_{j=0}^{t-2}\widehat{\alpha}^j_{CCEP}\varepsilon_{i,t-1-j}^*$. Also, for further reference let $\overline{\*e}_t^-=\sum_{j=0}^{t-1}\widehat{\alpha}_{CCEP}^j\overline{\*e}_{t-j}$ and $\+\nu_{i,t}^-=\sum_{j=0}^{t-1}\widehat{\alpha}^j_{CCEP}\+\nu_{i,t-j}$. Therefore,
    $\mathbf{II}_{1,e}=\begin{bmatrix}
        i+ii+iii\\
        \mathbf{iv}+\*v
    \end{bmatrix}$, where 
    \begin{align}
        |iii|=\left|\frac{1}{\sqrt{NT}}\sum_{i=1}^N\sum_{t=2}^Tu_{i,t-1}^*\overline{\varepsilon}_t^*\widetilde{\gamma}_{y,i,1} \right|=\left|\frac{1}{\sqrt{N}N}\sum_{i=1}^N\widetilde{\gamma}_{y,i,1}\sum_{j=1}^N\frac{1}{\sqrt{T}}\sum_{t=2}^Tu_{i,t-1}^*\varepsilon_{j,t}^*\right|=O_{P^*}(N^{-1/2}),
    \end{align}
which stems directly from (\ref{I_a2_exp}). Then
\begin{align}
    |i|&=\left|\widehat{\+\beta}_{CCEP}'\frac{1}{\sqrt{NT}}\sum_{i=1}^N\+\Gamma_{x,i}'(\overline{\+\Gamma}^{-1})'\sum_{t=2}^T\overline{\*e}_{t-1}^-\overline{\varepsilon}_{t}^*\widetilde{\gamma}_{y,i,1}  \right|\notag\\
    &\leq N^{-1/2}\left\|\widehat{\+\beta}_{CCEP} \right\|\left\| \overline{\+\Gamma}^{-1}\right\|\frac{1}{N}\sum_{i=1}^N\left\| \+\Gamma_{x,i}\right\||\widetilde{\gamma}_{y,i,1}| \left\|\frac{1}{\sqrt{T}}\sum_{t=2}^TN\overline{\*e}_{t-1}^-\overline{\varepsilon}_{t}^*\right\|\notag\\
    &\leq N^{-1/2}\left\|\widehat{\+\beta}_{CCEP} \right\|\left\| \overline{\+\Gamma}^{-1}\right\|\left(\frac{1}{N}\sum_{i=1}^N\left\| \+\Gamma_{x,i}\right\|^2\right)^{1/2}\left(\frac{1}{N}\sum_{i=1}^N|\widetilde{\gamma}_{y,i,1}|^2\right)^{1/2} \left\|\frac{1}{\sqrt{T}}\sum_{t=2}^TN\overline{\*e}_{t-1}^-\overline{\varepsilon}_{t}^*\right\|\notag\\
    &=O_{p^{*}}(N^{-1/2})
\end{align}
because 
\begin{align}
    E^*\left[\left\|\frac{1}{\sqrt{T}}\sum_{t=2}^TN\overline{\*e}_{t-1}^-\overline{\varepsilon}_{t}^*\right\|^2 \right]&=\frac{1}{T}\sum_{t=2}^T(N\overline{\*e}_{t-1}^{-\prime}\overline{\*e}_{t-1}^-)E^*[N\overline{\varepsilon}_t^{*2} ]\notag\\
    &=\frac{1}{T}\sum_{t=2}^T(N\overline{\*e}_{t-1}^{-\prime}\overline{\*e}_{t-1}^-)\left(\frac{1}{N}\sum_{i=1}^N\widehat{\varepsilon}_{i,t}^2 \right)\notag\\
    &\leq \left(\frac{1}{T} \sum_{t=2}^T(N\overline{\*e}_{t-1}^{-\prime}\overline{\*e}_{t-1}^-)^2\right)^{1/2} \left(\frac{1}{T} \sum_{t=2}^T \left(\frac{1}{N}\sum_{i=1}^N\widehat{\varepsilon}_{i,t}^2 \right)^2\right)^{1/2}\notag\\
    &\leq \left(\frac{1}{T} \sum_{t=2}^T(N\overline{\*e}_{t-1}^{-\prime}\overline{\*e}_{t-1}^-)^2\right)^{1/2} \left(\frac{1}{NT} \sum_{t=2}^T\sum_{i=1}^N\widehat{\varepsilon}_{i,t}^4 \right)^{1/2}=O_P(1)
\end{align}
due to 
\begin{align}\label{Ne'e_X}
    \frac{1}{T} \sum_{t=2}^T(N\overline{\*e}_{t-1}^{-\prime}\overline{\*e}_{t-1}^-)^2&=\frac{1}{T}\sum_{t=2}^TN^2\sum_{j=0}^{t-2}\sum_{k=0}^{t-2}\sum_{l=0}^{t-2}\sum_{r=0}^{t-2}\widehat{\alpha}_{CCEP}^{j+k+l+r}\overline{\*e}_{t-1-j}'\overline{\*e}_{t-1-k}\overline{\*e}_{t-1-l}'\overline{\*e}_{t-1-r}\notag\\
    &=\sum_{j=0}^{T-2}\sum_{k=0}^{T-2}\sum_{l=0}^{T-2}\sum_{r=0}^{T-2}\widehat{\alpha}_{CCEP}^{j+k+l+r} \frac{1}{T}\sum_{t=\max \{j,k,l,r \}+2}^TN^2\overline{\*e}_{t-1-j}'\overline{\*e}_{t-1-k}\overline{\*e}_{t-1-l}'\overline{\*e}_{t-1-r}\notag\\
    &\leq \sum_{j=0}^{T-2}\sum_{k=0}^{T-2}\sum_{l=0}^{T-2}\sum_{r=0}^{T-2}|\widehat{\alpha}_{CCEP}|^{j+k+l+r}\frac{1}{T}\sum_{t=2}^T\left\| \sqrt{N}\overline{\*e}_t\right\|^4\notag\\
    &=\left(\frac{1-|\widehat{\alpha}_{CCEP}|^T}{1-|\widehat{\alpha}_{CCEP}|} \right)^4\frac{1}{T}\sum_{t=2}^T\left\| \sqrt{N}\overline{\*e}_t\right\|^4=O_P(1)
\end{align}
under our assumptions, which follows the structure of (\ref{b4_rate}). Clearly, $\mathbf{iv}$ behaves very similarly, 
\begin{align}
    \left\| \mathbf{iv}\right\|&=\left\|\frac{1}{\sqrt{NT}}\sum_{i=1}^N\sum_{t=2}^T\+\Gamma_{x,i}'(\overline{\+\Gamma}^{-1})'\overline{\*e}_t\overline{\varepsilon}_t^*\widetilde{\gamma}_{i,i,1}\right\|\leq N^{-1/2}\left\|\overline{\+\Gamma}^{-1} \right\| \frac{1}{N}\sum_{i=1}^N\left\|\+\Gamma_{x,i}\widetilde{\gamma}_{i,i,1}\right\| \left\|\frac{1}{\sqrt{T}}\sum_{t=2}^TN \overline{\*e}_t\overline{\varepsilon}_t^*\right\|\notag\\
    &\leq N^{-1/2}\left\|\overline{\+\Gamma}^{-1} \right\| \left(\frac{1}{N}\sum_{i=1}^N\left\|\+\Gamma_{x,i}\right\|^2\right)^{1/2}\left(\frac{1}{N}\sum_{i=1}^N|\widetilde{\gamma}_{i,i,1}|^2 \right)^{1/2} \left\|\frac{1}{\sqrt{T}}\sum_{t=2}^TN \overline{\*e}_t\overline{\varepsilon}_t^*\right\|\notag\\
    &=O_{P^*}(N^{-1/2}). 
\end{align}
We move on to the remaining two terms. In particular, 
\begin{align}
    \left\|\*v \right\|=\left\|\frac{1}{\sqrt{NT}}\sum_{i=1}^N\sum_{t=2}^T\+\nu_{i,t}\overline{\varepsilon}_t^*\widetilde{\gamma}_{y,i,1} \right\|=O_P(N^{-1/2})+O_P(T^{-1/4}),
\end{align}
which is a rate sufficient for our purposes. Specifically,
\begin{align}\label{v_term_with_x}
    E^*[\left\| \*v\right\|^2]= E^*\left[\left\|\frac{1}{\sqrt{NT}}\sum_{i=1}^N\sum_{t=2}^T\+\nu_{i,t}\overline{\varepsilon}_t^*\widetilde{\gamma}_{y,i,1} \right\|^2 \right]&=\frac{1}{N^2}\sum_{i=1}^N\sum_{j=1}^N\frac{1}{T}\sum_{t=2}^T\widetilde{\gamma}_{y,i,1}\widetilde{\gamma}_{y,j,1}\+\nu_{i,t}'\+\nu_{j,t}E^*[N\overline{\varepsilon}_t^{*2}]\notag\\
    &=\frac{1}{N^2}\sum_{i=1}^N\sum_{j=1}^N\frac{1}{T}\sum_{t=2}^T\widetilde{\gamma}_{y,i,1}\widetilde{\gamma}_{y,j,1}\+\nu_{i,t}'\+\nu_{j,t}\left(\frac{1}{N}\sum_{k=1}^N\widehat{\varepsilon}_{k,t}^2 \right)\notag\\
    &=\frac{1}{N^2}\sum_{i=1}^N\sum_{j=1}^N\frac{1}{T}\sum_{t=2}^T\widetilde{\gamma}_{y,i,1}\widetilde{\gamma}_{y,j,1}\+\nu_{i,t}'\+\nu_{j,t}\left(\frac{1}{N}\sum_{k=1}^N\varepsilon_{k,t}^2 \right)+O_P(C_{N,T}^{-1})\notag\\
    &= \frac{1}{N^2}\sum_{i=1}^N\sum_{j=1}^N\frac{1}{T}\sum_{t=2}^T\gamma_{y,i,1}\gamma_{y,j,1}\+\nu_{i,t}'\+\nu_{j,t}\left(\frac{1}{N}\sum_{k=1}^N\varepsilon_{k,t}^2 \right)+O_P(N^{-1/2})\notag\\
    &+O_P(T^{-1/2})\notag\\
    &=\sigma^2\frac{1}{N^2}\sum_{i=1}^N\gamma_{y,i,1}^2\frac{1}{T}\sum_{t=2}^T\+\nu_{i,t}'\+\nu_{i,t}\notag\\
    &+\sigma^2\frac{1}{N^2}\sum_{i=1}^N\sum_{j\neq i}^N\frac{1}{T}\sum_{t=2}^T\gamma_{y,i,1}\gamma_{y,j,1}\+\nu_{i,t}'\+\nu_{j,t}+O_P(C_{N,T}^{-1})\notag\\
    &=O_P(C_{N,T}^{-1}),
\end{align}
where we firstly used $|\widehat{\varepsilon}_{i,t}^2-\varepsilon_{i,t}^2|=O_P(C_{N,T}^{-1})$ for each $i$ and $t$ based on (\ref{resid_with_X}). Then we safely replaced $\widetilde{\+\gamma}_{y,i,1}$ and $\widetilde{\+\gamma}_{y,j,1}$ with their (rotated) probability limits using (\ref{estimated_gamma_with_X}): $|\widetilde{\gamma}_{y,i,1}-\gamma_{y,i,1}|=O_P(T^{-1/2})$, because the double sum is bounded. Lastly $\left| \frac{1}{N}\sum_{i=1}^N(\varepsilon_{i,t}^2-\sigma^2)\right|=O_P(N^{-1/2})$. The second term is vanishing due to cross-section independence. Finally, we explore term $ii$, and we start with an approximation analogous to the one in (\ref{u_star_taylor_approx}):
\begin{align}\label{nu_taylor_expansion}
    \+\nu_{i,t-1}^-=\+\nu_{i,t-1}^-(\widehat{\alpha})&=\+\nu_{i,t-1}^-(\alpha_0)+\sqrt{NT}(\widehat{\alpha}_{CCEP}-\alpha_0)\frac{1}{\sqrt{NT}}\sum_{j=1}^{t-2}j(\alpha)^{j-1}\+\nu_{i,t-j-1}+o_P((NT)^{-1/2})\notag\\
    &=\+\nu_{i,t-1}^-(\alpha_0)+O_P((NT)^{-1/2})
\end{align}
for each $i$ and $t$, since 
\begin{align}
    E\left[\left\|\frac{1}{\sqrt{NT}}\sum_{j=1}^{t-2}j(\alpha)^{j-1}\+\nu_{i,t-j-1}\right\|^2 \right]&=\frac{1}{NT}\sum_{l=1}^{t-2}j^2(\alpha_0)^{2(j-1)}\mathrm{tr}\left(E[\+\nu_{i,t-j-1}\+\nu_{i,t-j-1}'] \right)\notag\\
    &\leq (NT)^{-1}\mathrm{tr}(\+\Omega_{\+\nu}) \sum_{j=1}^{\infty}j^2(\alpha_0)^{2(j-1)}\notag\\
    &=O((NT)^{-1}).
\end{align}
Then
\begin{align}
    E^*[|ii|^2]&=E^*\left[\left|\widehat{\+\beta}_{CCEP}'\frac{1}{\sqrt{NT}}\sum_{i=1}^N\sum_{t=2}^T\+\nu_{i,t-1}^-\overline{\varepsilon}_{t}^*\widetilde{\gamma}_{y,i,1}\right|^2\right]\notag\\
    &\leq \left\|\widehat{\+\beta}_{CCEP} \right\|^2 E^*\left[\left\|\frac{1}{\sqrt{NT}}\sum_{i=1}^N\sum_{t=2}^T\+\nu_{i,t-1}^-\overline{\varepsilon}_{t}^*\widetilde{\gamma}_{y,i,1}\right\|^2\right]\notag\\
    &=\left\|\widehat{\+\beta}_{CCEP} \right\|^2 \frac{1}{N^2}\sum_{i=1}^N\sum_{j=1}^N\frac{1}{T}\sum_{t=2}^T\widetilde{\gamma}_{y,i,1}\widetilde{\gamma}_{y,j,1}\+\nu_{i,t-1}^{-\prime}\+\nu_{j,t-1}^{-}E^*[N\overline{\varepsilon}^{*2}_t]\notag\\
    &=\left\|\widehat{\+\beta}_{CCEP} \right\|^2 \frac{1}{N^2}\sum_{i=1}^N\sum_{j=1}^N\frac{1}{T}\sum_{t=2}^T\widetilde{\gamma}_{y,i,1}\widetilde{\gamma}_{y,j,1}\+\nu_{i,t-1}^{-\prime}\+\nu_{j,t-1}^{-}\left(\frac{1}{N}\sum_{k=1}^N\varepsilon_{k,t}^2\right)+O_P(C_{N,T}^{-1})\notag\\
    &=\left\|\widehat{\+\beta}_{CCEP} \right\|^2 \frac{1}{N^2}\sum_{i=1}^N\sum_{j=1}^N\frac{1}{T}\sum_{t=2}^T\widetilde{\gamma}_{y,i,1}\widetilde{\gamma}_{y,j,1}\+\nu_{i,t-1}^{-\prime}(\alpha_0)\+\nu_{j,t-1}^{-}(\alpha_0)\left(\frac{1}{N}\sum_{k=1}^N\varepsilon_{k,t}^2\right)+O_P(C_{N,T}^{-1})\notag\\
    &=\left\|\widehat{\+\beta}_{CCEP} \right\|^2 \frac{1}{N^2}\sum_{i=1}^N\sum_{j=1}^N\frac{1}{T}\sum_{t=2}^T\gamma_{y,i,1}\gamma_{y,j,1}\+\nu_{i,t-1}^{-\prime}(\alpha_0)\+\nu_{j,t-1}^{-}(\alpha_0)\left(\frac{1}{N}\sum_{k=1}^N\varepsilon_{k,t}^2\right)+O_P(C_{N,T}^{-1})\notag\\
    &=\left\|\widehat{\+\beta}_{CCEP} \right\|^2\sigma^2 \frac{1}{N^2}\sum_{i=1}^N\sum_{j=1}^N\frac{1}{T}\sum_{t=2}^T\gamma_{y,i,1}\gamma_{y,j,1}\+\nu_{i,t-1}^{-\prime}(\alpha_0)\+\nu_{j,t-1}^{-}(\alpha_0)+O_P(C_{N,T}^{-1})\notag\\
    &=O_P(C_{N,T}^{-1})
\end{align}
by the same arguments as in (\ref{v_term_with_x}) and due to cross-section independence of the remaining summands. In summary,
\begin{align}
    \mathbf{II}_1&=\left(\frac{1}{N}\sum_{i=1}^N\widetilde{\gamma}_{y,i,1}\mathcal{\+P}_{0,1,i}\*S_\*b\right)'\sqrt{N} T^{-1/2}\widehat{\*Q}_0^{\prime}\overline{\+\varepsilon}^*\notag+\*S_\*b'\mathcal{\+P}_2'\left(\frac{1}{N}\sum_{i=1}^N\widetilde{\gamma}_{y,i,1}\+\kappa_i \right)'\sqrt{N}T^{-1/2}\overline{\+\varepsilon}^*
     + o_{P^*}(1)\notag\\
     &= \left(\frac{1}{N}\sum_{i=1}^N\widetilde{\gamma}_{y,i,1}\*C_i\*S_\*b\right)'\sqrt{N} T^{-1/2}\*Q_{\widehat{\alpha}}'\overline{\+\varepsilon}^*\notag+\*S_\*b'\mathcal{\+P}_2'\left(\frac{1}{N}\sum_{i=1}^N\widetilde{\gamma}_{y,i,1}\+\kappa_i \right)'\sqrt{N}T^{-1/2}\overline{\+\varepsilon}^*
     + o_{P^*}(1)\notag\\
     &=\*S_\*b'\+\Sigma_{\gamma_{y,1}\*C}'\sqrt{N} T^{-1/2}\*Q_{\widehat{\alpha}}'\overline{\+\varepsilon}^*\notag+\*S_\*b'\mathcal{\+P}_2'\left(\frac{1}{N}\sum_{i=1}^N\widetilde{\gamma}_{y,i,1}\+\kappa_i \right)'\sqrt{N}T^{-1/2}\overline{\+\varepsilon}^*
     + o_{P^*}(1)
\end{align}
where we substituted the results from Lemma XB, and let $\+\Sigma_{\gamma_{y,1}\*C}=\plim_{(N,T)\to \infty}\frac{1}{N}\sum_{i=1}^N\widetilde{\gamma}_{y,i,1}\*C_i$. Specifically,
\begin{align}
    \left(\frac{1}{N}\sum_{i=1}^N\widetilde{\gamma}_{y,i,1}\mathcal{\+P}_{0,1,i}\*S_\*b\right)'\sqrt{N} T^{-1/2}\widehat{\*Q}_0^{\prime}\overline{\+\varepsilon}^*&=\frac{1}{N}\sum_{i=1}^N\widetilde{\gamma}_{y,i,1}\*S_\*b'\mathcal{\+P}_{0,1,i}^{\prime} T^{-1/2}\widehat{\*Q}_0^{\prime}(\sqrt{N}\overline{\+\varepsilon}^*)\notag\\
    &=\frac{1}{N}\sum_{i=1}^N\widetilde{\gamma}_{y,i,1}\*S_\*b'\mathcal{\+P}_{0,1,i}^{\prime}(\*I_{2 \times 2}\otimes \+\Pi_{\overline{\+\Gamma}})' T^{-1/2}\*Q_{\widehat{\alpha}}'(\sqrt{N}\overline{\+\varepsilon}^*)\notag\\
    &+\underbrace{N^{-1/2}\frac{1}{N}\sum_{i=1}^N\widetilde{\gamma}_{y,i,1}\*S_\*b'\mathcal{\+P}_{0,1,i}^{\prime} (NT^{-1/2}\widehat{\*Q}_{\overline{\*e}}'\overline{\+\varepsilon}^*)}_{O_{P^*}(N^{-1/2})}\notag\\
    &= \frac{1}{N}\sum_{i=1}^N\widetilde{\gamma}_{y,i,1}\*S_\*b'\*C_i' T^{-1/2}\*Q_{\widehat{\alpha}}'(\sqrt{N}\overline{\+\varepsilon}^*) + O_{P^*}(C_{N,T}^{-1})
\end{align}
since $\left\|NT^{-1/2}\widehat{\*Q}_{\overline{\*e}}'\overline{\+\varepsilon}^* \right\|=\left\| \frac{1}{\sqrt{NT}}\sum_{i=1}^N\sum_{t=2}^T\widehat{\*q}_{\overline{\*e},t}\varepsilon_{i,t}^*\right\|=O_{P^*}(1)$, which is the substitution that we will use repeatedly going forward to simplify the expressions.
\\

\noindent We will move on to $\mathbf{II}_2$. Then 
\begin{align}
  \mathbf{II}_2&=  \frac{1}{\sqrt{NT}}\sum_{i=1}^N\*B_i^{*\prime}\*P_{\widehat{\*F}^*}\overline{\+\varepsilon}^*\widetilde{\gamma}_{y,i,1}\notag\\
  &= \frac{1}{\sqrt{NT}}\sum_{i=1}^N\*B_i^{*\prime}\*P_{\*Q\overline{\*C}}\overline{\+\varepsilon}^*\widetilde{\gamma}_{y,i,1}+\frac{1}{\sqrt{NT}}\sum_{i=1}^N\*B_i^{*\prime}(\*P_{\widehat{\*F}}-\*P_{\*Q\overline{\*C}})\overline{\+\varepsilon}^*\widetilde{\gamma}_{y,i,1}+\frac{1}{\sqrt{NT}}\sum_{i=1}^N\*B_i^{*\prime}(\*P_{\widehat{\*F}^*}-\*P_{\widehat{\*F}})\overline{\+\varepsilon}^*\widetilde{\gamma}_{y,i,1}\notag\\
  &=\mathbf{II}_{2,1}+\mathbf{II}_{2,2}+\mathbf{II}_{2,3},
\end{align}
where $\*P_{\*Q\overline{\*C}}$ is based on (\ref{hatF_total_X}). We will start with 
\begin{align}
    \mathbf{II}_{2,1}&=\frac{1}{\sqrt{NT}}\sum_{i=1}^N\*S_\*b'[\widehat{\*A}, \*I_{K\times K}]'\widehat{\*A}_{\*z_0,i}'\*P_{\*Q\overline{\*C}}\overline{\+\varepsilon}^*\widetilde{\gamma}_{y,i,1}+\frac{1}{\sqrt{NT}}\sum_{i=1}^N\*S_\*b'\mathcal{\+P}_{0,1,i}^{\prime}\widehat{\*Q}_0^{\prime}\*P_{\*Q\overline{\*C}}\overline{\+\varepsilon}^*\widetilde{\gamma}_{y,i,1}\notag\\
    &+\frac{1}{\sqrt{NT}}\sum_{i=1}^N\*S_\*b'\mathcal{\+P}_{2}'\+\kappa_i'\*P_{\*Q\overline{\*C}}\overline{\+\varepsilon}^*\widetilde{\gamma}_{y,i,1}+\frac{1}{\sqrt{NT}}\sum_{i=1}^N\*S_\*b'\mathcal{\+P}_{0,3,i}^{\prime}\*H'\*P_{\*Q\overline{\*C}}\overline{\+\varepsilon}^*\widetilde{\gamma}_{y,i,1}\notag\\
    &+\frac{1}{\sqrt{NT}}\sum_{i=1}^N\*S_\*b'\*U_i^{*\prime}\*P_{\*Q\overline{\*C}}\overline{\+\varepsilon}^*\widetilde{\gamma}_{y,i,1}\notag\\
    &=\mathbf{II}_{2,1,a}+\mathbf{II}_{2,1,b}+\mathbf{II}_{2,1,c}+\mathbf{II}_{2,1,d}+\mathbf{II}_{2,1,e},
\end{align}
where 
\begin{align}
    \left\|\mathbf{II}_{2,1,a} \right\| &\leq \left\|\*S_\*b'[\widehat{\*A}, \*I_{K\times K}]' \right\|\left\|\frac{1}{N}\sum_{i=1}^N\widetilde{\gamma}_{y,i,1} T^{-1}\widehat{\*A}_{\*z_0,i}'\*Q\overline{\*C}\right\|\left\| \left( T^{-1}\overline{\*C}'\*Q'\*Q\overline{\*C}\right)^{-1}\right\|\left\| \frac{1}{\sqrt{T}}\sum_{t=2}^T\overline{\*C}'\*q_t\sqrt{N}\overline{\varepsilon}_t^*\right\|\notag\\
    &\leq \left\|\overline{\*C} \right\|^2\left\|\*S_\*b'[\widehat{\*A}, \*I_{K\times K}]' \right\|  \frac{1}{N}\sum_{i=1}^N|\widetilde{\+\gamma}_{y,i,1}|\left\|\*z_{i,0}\right\| \frac{1}{T}\sum_{t=2}^T\left\|\widehat{\*A}^{t-1} \right\|\left\|\*q_t \right\| \notag\\
    &\times \left\| \left( T^{-1}\overline{\*C}'\*Q'\*Q\overline{\*C}\right)^{-1}\right\|\left\| \frac{1}{\sqrt{T}}\sum_{t=2}^T\*q_t\sqrt{N}\overline{\varepsilon}_t^*\right\|\notag\\
    &\leq O_P(1)\times \left(\frac{1}{N}\sum_{i=1}^N|\widetilde{\gamma}_{y,i,1} |^2 \right)^{1/2}\left( \frac{1}{N}\sum_{i=1}^N\left\|\*z_{i,0}\right\|^2\right)^{1/2}
    \left(\frac{1}{T}\sum_{t=2}^T\left\|\*q_t \right\|^2 \right)^{1/2}\notag\\
    &\times \frac{1}{\sqrt{T}}\sum_{t=2}^T|\widehat{\alpha}_{CCEP}|^{t-1} \left\| \frac{1}{\sqrt{T}}\sum_{t=2}^T\*q_t\sqrt{N}\overline{\varepsilon}_t^*\right\|=O_{P^*}(T^{-1/2})
\end{align}
due to consistency of $\widehat{\alpha}_{CCEP}$, and so the effect of the initial value vanishes asymptotically as expected. We can write the next term as
\begin{align}
    \mathbf{II}_{2,1,b}=\left(\frac{1}{N}\sum_{i=1}^N\widetilde{\gamma}_{y,i,1}\mathcal{\+P}_{0,1,i}\*S_\*b\right)'T^{-1}\widehat{\*Q}_0^{\prime}\*Q\overline{\*C}\left( T^{-1}\overline{\*C}'\*Q'\*Q\overline{\*C}\right)^{-1}\sqrt{N}T^{-1/2}\overline{\*C}'\*Q'\overline{\+\varepsilon}^*=O_{P^*}(1),
\end{align}
which corresponds to $\*A^\*F_{1,2}$ in \cite{DeVos2019} with the difference of $\overline{\*X}$, $\overline{\*X}_{-1}$ being projected out. The next term again survives if $\+\theta_0\neq \*0_{k\times 1}$: 
\begin{align}
    \mathbf{II}_{2,1,c}=\+S_\*b'\mathcal{\+P}_2'\left(\frac{1}{N}\sum_{i=1}^N\widetilde{\gamma}_{y,i,1}T^{-1}\overline{\*C}'\*Q'\+\kappa_i \right)'\left( T^{-1}\overline{\*C}'\*Q'\*Q\overline{\*C}\right)^{-1}\sqrt{N}T^{-1/2}\overline{\*C}'\*Q'\overline{\+\varepsilon}^*=O_{P^*}(1),
\end{align}
which is exactly zero under $\+\theta_0= \*0_{k\times 1}$. Then
\begin{align}
\left\|\mathbf{II}_{2,1,d}\right\|&\leq \frac{1}{\sqrt{T}}\left\|\left(\frac{1}{N}\sum_{i=1}^N\widetilde{\gamma}_{y,i,1}(\sqrt{T}\mathcal{\+P}_{0,3,i})\*S_\*b\right)\right\|\left\| \overline{\*C}\right\|^2\left\|T^{-1}\*H'\*Q\right\|\left\|\left( T^{-1}\overline{\*C}'\*Q'\*Q\overline{\*C}\right)^{-1}\right\|\left\|\sqrt{N}T^{-1/2}\*Q'\overline{\+\varepsilon}^*\right\|\notag\\
&=O_{P^*}(T^{-1/2}),
\end{align}
which since $\left\|\mathcal{\+P}_{0,3,i} \right\|=O_P(T^{-1/2})$. Eventually,
\begin{align}\label{II_21e}
    \left\|\mathbf{II}_{2,1,e}\right\|&\leq \left\|\overline{\*C} \right\|^2\left\| \frac{1}{N}\sum_{i=1}^N \widetilde{\gamma}_{y,i,1}T^{-1}\*S_\*b'\*U_i^{*\prime}\*Q\right\|\left\| \left( T^{-1}\overline{\*C}'\*Q'\*Q\overline{\*C}\right)^{-1}\right\|\left\| \sqrt{N}T^{-1/2}\*Q'\overline{\+\varepsilon}^*\right\|\notag\\
    &\leq O_P(1) \times \left( \frac{1}{N}\sum_{i=1}^N\widetilde{\gamma}_{y,i,1}^2\right)^{1/2}\left( \frac{1}{N}\sum_{i=1}^N\left\| T^{-1}\*S_\*b'\*U_i^{*\prime}\*Q\right\|^2\right)^{1/2}\left\| \sqrt{N}T^{-1/2}\*Q'\overline{\+\varepsilon}^*\right\|\notag\\    &=O_{P^*}(T^{-1/2}),
\end{align}

because for each $i$
\begin{align}\label{U^*Q}
    \left\| T^{-1}\*S_\*b'\*U_i^{*\prime}\*Q\right\|=O_{P^*}(T^{-1/2}).
\end{align}
 This rate is driven by the two components in $\left\|T^{-1} \*S_\*b'\*U_i^{*\prime}\*Q\right\|$ due to a fast rate of consistency of $\widehat{\alpha}_{CCEP}$. In particular, the leading terms are $\left\| T^{-1}\sum_{t=2}^T\+\nu_{i,t}\*q_t'\right\|=O_P(T^{-1/2})$ (unconditionally) for each $i$. The second leading term for each $i$ is $\frac{1}{T}\sum_{t=2}^T\*q_tu_{i,t-1}^*=:\*b_i$, so it is sufficient to show that it is bounded when scaled by $\sqrt{T}$. Note that for some fixed $m$, we can re-write the latter as:
\begin{align}
   \sqrt{T}\*b_{i,m}= \sum_{j=0}^{m-2}\widehat{\alpha}_{CCEP}^j\frac{1}{\sqrt{T}}\sum_{t=j+2}^T\varepsilon^*_{i,t-j-1}\*q_t&=\sum_{j=0}^{m-2}(\alpha_0)^j\frac{1}{\sqrt{T}}\sum_{t=j+2}^T\varepsilon^*_{i,t-j-1}\*q_t\notag\\
        &-\sum_{j=0}^{m-2}(\widehat{\alpha}_{CCEP}^j-(\alpha_0)^j)\frac{1}{\sqrt{T}}\sum_{t=j+2}^T\varepsilon^*_{i,t-j-1}\*q_t\notag\\
        &=\sum_{j=0}^{m-2}(\alpha_0)^j\frac{1}{\sqrt{T}}\sum_{t=j+2}^T\varepsilon^*_{i,t-j-1}\*q_t + o_{P^*}(1) \notag\\
        &=O_{P^*}(1)
\end{align}
Following the proof of Lemma A.5. in \cite{gonccalves2004bootstrapping}, we take $\+\lambda\in \mathbb{R}^{2K}$, so that $\+\lambda'\+\lambda=1$, and so the approximation of $\sqrt{T}\*b_i$ by $\sqrt{T}\*b_{i,m}$ gives
    \begin{align}\label{b_m-b}
        E^*\left[ (\+\lambda'\sqrt{T}(\*b-\*b_m))^2\right]&=\sum_{j=m}^{T-2}\sum_{r=m}^{T-2}\widehat{\alpha}_{CCEP}^{j+r}E^*\left[\frac{1}{T}\sum_{t=j+2}^T\sum_{s=r+2}^T\varepsilon_{i,t-j-1}^*\varepsilon_{i,s-r-1}^*\+\lambda'\*q_{t}\*q_{s}'\+\lambda \right]\notag\\
        &=\sum_{j=m}^{T-2}\sum_{r=m}^{T-2}\widehat{\alpha}^{j+r}_{CCEP}\frac{1}{T}\sum_{t=\max(j,r)+2}^T\widehat{\varepsilon}_{i,t-j-1}^2\+\lambda'\*q_{t}\*q_{t+r-j}'\+\lambda \notag\\
        &\leq \sum_{j=m}^{T-2}\sum_{r=m}^{T-2}|\widehat{\alpha}_{CCEP}|^{j+r}\frac{1}{2}\frac{1}{T}\sum_{t=\max(j,r)+2}^T\widehat{\varepsilon}_{i,t-j-1}^2(|\+\lambda'\*q_{t}|^2+|\*q_{t+r-j}'\+\lambda|^2)\notag\\
        &\leq  \frac{1}{2}\sum_{j=m}^{T-2}\sum_{r=m}^{T-2}|\widehat{\alpha}_{CCEP}|^{j+r}\left( \frac{1}{T}\sum_{t=\max(j,r)+2}^T\widehat{\varepsilon}_{i,t-j-1}^4\right)^{1/2}\left(\frac{1}{T}\sum_{t=\max(j,r)+2}^T |\+\lambda'\*q_{t}|^4\right)^{1/2}  \notag\\
        & +   \frac{1}{2}\sum_{j=m}^{T-2}\sum_{r=m}^{T-2}|\widehat{\alpha}_{CCEP}|^{j+r}\left( \frac{1}{T}\sum_{t=\max(j,r)+2}^T\widehat{\varepsilon}_{i,t-j-1}^4\right)^{1/2}\left(\frac{1}{T}\sum_{t=\max(j,r)+2}^T |\*q_{t+r-j}'\+\lambda|^4\right)^{1/2}\notag\\
        &\leq  \left\| \+\lambda\right\|^2\left(\sum_{j=m}^{T-2}|\widehat{\alpha}_{CCEP} |^j\right)^2 \left(\frac{1}{T}\sum_{t=1}^T\widehat{\varepsilon}_{i,t}^4 \right)^{1/2} \left(\frac{1}{T}\sum_{t=1}^T\left\|\*q_{t} \right\|^4 \right)^{1/2}\notag\\
        &=o_P(1)
        \end{align}
    as needed, and where we use $|\widehat{\alpha}_{CCEP}|^j=|\widehat{\alpha}_{CCEP}^j|$. In summary, we have 
\begin{align}
    \mathbf{II}_{2,1}&=\left(\frac{1}{N}\sum_{i=1}^N\widetilde{\gamma}_{y,i,1}\mathcal{\+P}_{0,1,i}\*S_\*b\right)'T^{-1}\widehat{\*Q}_0^{\prime}\*Q\overline{\*C}\left( T^{-1}\overline{\*C}'\*Q'\*Q\overline{\*C}\right)^{-1}\sqrt{N}T^{-1/2}\overline{\*C}'\*Q'\overline{\+\varepsilon}^*\notag\\
    &+\+S_\*b'\mathcal{\+P}_2'\left(\frac{1}{N}\sum_{i=1}^N\widetilde{\gamma}_{y,i,1}T^{-1}\overline{\*C}'\*Q'\+\kappa_i \right)'\left( T^{-1}\overline{\*C}'\*Q'\*Q\overline{\*C}\right)^{-1}\sqrt{N}T^{-1/2}\overline{\*C}'\*Q'\overline{\+\varepsilon}^* + o_{P^*}(1)\notag\\
    &=\*S_\*b'\+\Sigma_{\gamma_{y,1}\*C}'T^{-1}\*Q'_{\widehat{\alpha}}\*Q\*C(\*C'\+\Sigma_\*q\*C)^{-1}\sqrt{N}T^{-1/2}\*C'\*Q'\overline{\+\varepsilon}^*\notag\\
    &+\+S_\*b'\mathcal{\+P}_2'\+\Sigma_{\gamma_{y,1}\*q\+\kappa}'\*C(\*C'\+\Sigma_\*q\*C)^{-1}\sqrt{N}T^{-1/2}\*C'\*Q'\overline{\+\varepsilon}^* + o_{P^*}(1),
\end{align}
where we inserted the results from Lemma XB and let $\+\Sigma_{\gamma_{y,1}\*q\+\kappa}=\plim_{(N,T)\to \infty}\frac{1}{N}\sum_{i=1}^N\widetilde{\gamma}_{y,i,1}T^{-1}\*Q'\+\kappa_i$. We proceed with $\mathbf{II}_{2,2}$, where the expansion of $\*P_{\widehat{\*F}}-\*P_{\*Q\overline{\*C}}$ is essential:
\begin{align}\label{P_F-P_QC}
    \*P_{\widehat{\*F}}-\*P_{\*Q\overline{\*C}}&=\overline{\*V}(\widehat{\*F}'\widehat{\*F})^{-1}\overline{\*V}'+ \overline{\*V}(\widehat{\*F}'\widehat{\*F})^{-1} \overline{\*C}'\*Q' + \*Q\overline{\*C}(\widehat{\*F}'\widehat{\*F})^{-1}\overline{\*V}'\notag\\
    &+ \*Q\overline{\*C}\left( (\widehat{\*F}'\widehat{\*F})^{-1} - (\overline{\*C}'\*Q'\*Q\overline{\*C})^{-1}\right)\overline{\*C}'\*Q'.
\end{align}
Therefore, 
\begin{align}
    \mathbf{II}_{2,2}&=\frac{1}{\sqrt{NT}}\sum_{i=1}^N\*B_i^{*\prime}(\*P_{\widehat{\*F}}-\*P_{\*Q\overline{\*C}})\overline{\+\varepsilon}^*\widetilde{\gamma}_{y,i,1}=\frac{1}{\sqrt{NT}}\sum_{i=1}^N\*S_\*b'[\widehat{\*A}, \*I_{K\times K}]'\widehat{\*A}_{\*z_0,i}'( \*P_{\widehat{\*F}}-\*P_{\*Q\overline{\*C}})\overline{\+\varepsilon}^*\widetilde{\gamma}_{y,i,1}\notag\\
    &+\frac{1}{\sqrt{NT}}\sum_{i=1}^N\*S_\*b'\mathcal{\+P}_{0,1,i}^{\prime}\widehat{\*Q}_0^{\prime}( \*P_{\widehat{\*F}}-\*P_{\*Q\overline{\*C}})\overline{\+\varepsilon}^*\widetilde{\gamma}_{y,i,1}+\frac{1}{\sqrt{NT}}\sum_{i=1}^N\*S_\*b'\mathcal{\+P}_{2}'\+\kappa_i'( \*P_{\widehat{\*F}}-\*P_{\*Q\overline{\*C}})\overline{\+\varepsilon}^*\widetilde{\gamma}_{y,i,1}\notag\\
    &+\frac{1}{\sqrt{NT}}\sum_{i=1}^N\*S_\*b'\mathcal{\+P}_{0,3,i}^{\prime}\*H'( \*P_{\widehat{\*F}}-\*P_{\*Q\overline{\*C}})\overline{\+\varepsilon}^*\widetilde{\gamma}_{y,i,1}+\frac{1}{\sqrt{NT}}\sum_{i=1}^N\*S_\*b'\*U_i^{*\prime}( \*P_{\widehat{\*F}}-\*P_{\*Q\overline{\*C}})\overline{\+\varepsilon}^*\widetilde{\gamma}_{y,i,1}\notag\\
    &= \mathbf{II}_{2,2,1}+ \mathbf{II}_{2,2,2}+ \mathbf{II}_{2,2,3}+ \mathbf{II}_{2,2,4}+ \mathbf{II}_{2,2,5}.
\end{align}
We start form $\mathbf{II}_{2,2,1}=\mathbf{II}_{2,2,1,a}+\mathbf{II}_{2,2,1,b}+\mathbf{II}_{2,2,1,c}+\mathbf{II}_{2,2,1,d}$ with the definitions of the components following (\ref{P_F-P_QC}). In $\mathbf{II}_{2,2,1,a}$, we will use the fact that for any stack of stationary vectors $\*B$ with the second moment we have 
\begin{align}
    \left\|\frac{1}{N}\sum_{i=1}^NT^{-1}\widehat{\*A}'_{\*z_0,i}\*B\right\|\leq \frac{1}{N}\sum_{i=1}^N\left\|\*z_{i,0} \right\|\left(\frac{1}{T}\sum_{t=2}^T\left\|\*b_t\right\|^2 \right)^{1/2}\frac{1}{\sqrt{T}}\sum_{t=2}^T|\widehat{\alpha}_{CCEP}|^{t-1}=O_P(T^{-1/2}).
\end{align}
Then 
\begin{align}
    \left\|\mathbf{II}_{2,2,1,a} \right\|&=\left\|\frac{1}{\sqrt{NT}}\sum_{i=1}^N\*S_\*b'[\widehat{\*A}, \*I_{K\times K}]'T^{-1}\widehat{\*A}_{\*z_0,i}'\overline{\*V}(T^{-1}\widehat{\*F}'\widehat{\*F})^{-1}\overline{\*V}'\overline{\+\varepsilon}^*\widetilde{\gamma}_{y,i,1} \right\|\notag\\
    &\leq \frac{1}{N\sqrt{T}}\left\|\*S_\*b'[\widehat{\*A}, \*I_{K\times K}]' \right\|\left\| (T^{-1}\widehat{\*F}'\widehat{\*F})^{-1}\right\|\left(\frac{1}{N}\sum_{i=1}^N| \widetilde{\gamma}_{y,i,1}|^2\right)^{1/2}\left(\frac{1}{N}\sum_{i=1}^N\left\|\*z_{i,0} \right\|^2 \right)^{1/2}\notag\\
    &\times \left(\frac{1}{T}\sum_{t=2}^T\left\|\sqrt{N}\overline{\*v}_t \right\|^2 \right)^{1/2}\sum_{t=2}^T|\widehat{\alpha}_{CCEP} |^{t-1}\underbrace{\left\|\frac{1}{\sqrt{T}}\sum_{t=2}^TN\overline{\*v}_t\overline{\varepsilon}_{t}^* \right\|}_{O_{P^*}(1)}=O_{P^*}(N^{-1}T^{-1/2}),
\end{align}
because $\overline{\*v}_t$ is a \textit{constant} in the bootstrap world, and so $N\overline{\*v}_t\overline{\+\varepsilon}_t^*$ is mean-zero and uncorrelated process under bootstrap measure. Also,
\begin{align}
     \left\|\mathbf{II}_{2,2,1,b} \right\|&=\left\|\frac{1}{\sqrt{NT}}\sum_{i=1}^N\*S_\*b'[\widehat{\*A}, \*I_{K\times K}]'T^{-1}\widehat{\*A}_{\*z_0,i}'\*Q\overline{\*C}(T^{-1}\widehat{\*F}'\widehat{\*F})^{-1}\overline{\*V}'\overline{\+\varepsilon}^*\widetilde{\gamma}_{y,i,1} \right\|\notag\\
     &\leq  \frac{1}{\sqrt{NT}}\left\|\*S_\*b'[\widehat{\*A}, \*I_{K\times K}]' \right\|\left\|\overline{\*C} \right\|\left\| (T^{-1}\widehat{\*F}'\widehat{\*F})^{-1}\right\|\left(\frac{1}{N}\sum_{i=1}^N| \widetilde{\gamma}_{y,i,1}|^2\right)^{1/2}\left(\frac{1}{N}\sum_{i=1}^N\left\|\*z_{i,0} \right\|^2 \right)^{1/2}\notag\\
    &\times \left(\frac{1}{T}\sum_{t=2}^T\left\|\*q_t \right\|^2 \right)^{1/2}\sum_{t=2}^T|\widehat{\alpha}_{CCEP} |^{t-1}\underbrace{\left\|\frac{1}{\sqrt{T}}\sum_{t=2}^TN\overline{\*v}_t\overline{\varepsilon}_{t}^* \right\|}_{O_{P^*}(1)}=O_{P^*}((NT)^{-1/2}),
\end{align}
\begin{align}
     \left\|\mathbf{II}_{2,2,1,c} \right\|&=\left\|\frac{1}{\sqrt{NT}}\sum_{i=1}^N\*S_\*b'[\widehat{\*A}, \*I_{K\times K}]'T^{-1}\widehat{\*A}_{\*z_0,i}'\overline{\*V}(T^{-1}\widehat{\*F}'\widehat{\*F})^{-1}\overline{\C}'\*Q'\overline{\+\varepsilon}^*\widetilde{\gamma}_{y,i,1} \right\|\notag\\
    &\leq \frac{1}{\sqrt{NT}}\left\|\*S_\*b'[\widehat{\*A}, \*I_{K\times K}]' \right\|\left\|\overline{\*C}\right\|\left\| (T^{-1}\widehat{\*F}'\widehat{\*F})^{-1}\right\|\left(\frac{1}{N}\sum_{i=1}^N| \widetilde{\gamma}_{y,i,1}|^2\right)^{1/2}\left(\frac{1}{N}\sum_{i=1}^N\left\|\*z_{i,0} \right\|^2 \right)^{1/2}\notag\\
    &\times \left(\frac{1}{N}\sum_{t=2}^T\left\|\sqrt{N}\overline{\*v}_t \right\|^2 \right)^{1/2}\sum_{t=2}^T|\widehat{\alpha}_{CCEP} |^{t-1}\underbrace{\left\|\frac{1}{\sqrt{T}}\sum_{t=2}^T\sqrt{N}\*q_t\overline{\varepsilon}_{t}^* \right\|}_{O_{P^*}(1)}=O_{P^*}((NT)^{-1/2}),
\end{align}
and finally, 
\begin{align}
    \left\|\mathbf{II}_{2,2,1,d} \right\|&=\left\|\frac{1}{\sqrt{NT}}\sum_{i=1}^N\*S_\*b'[\widehat{\*A}, \*I_{K\times K}]'T^{-1}\widehat{\*A}_{\*z_0,i}'\*Q\overline{\*C}\left( (T^{-1}\widehat{\*F}'\widehat{\*F})^{-1} - (T^{-1}\overline{\*C}'\*Q'\*Q\overline{\*C})^{-1}\right)\overline{\C}'\*Q'\overline{\+\varepsilon}^*\widetilde{\gamma}_{y,i,1} \right\|\notag\\
    &\leq  \frac{1}{\sqrt{T}}\left\|\*S_\*b'[\widehat{\*A}, \*I_{K\times K}]' \right\|\left\|\overline{\*C} \right\|^2\left(\frac{1}{N}\sum_{i=1}^N| \widetilde{\gamma}_{y,i,1}|^2\right)^{1/2}\left(\frac{1}{N}\sum_{i=1}^N\left\|\*z_{i,0} \right\|^2 \right)^{1/2} \left(\frac{1}{T}\sum_{t=2}^T\left\|\*q_t \right\|^2 \right)^{1/2}\notag\\
&\times\sum_{t=2}^T|\widehat{\alpha}_{CCEP} |^{t-1}\underbrace{\left\|\frac{1}{\sqrt{T}}\sum_{t=2}^T\sqrt{N}\*q_t\overline{\varepsilon}_{t}^* \right\|}_{O_{P^*}(1)}\left\| (T^{-1}\widehat{\*F}'\widehat{\*F})^{-1} - (T^{-1}\overline{\*C}'\*Q'\*Q\overline{\*C})^{-1}\right\|\notag\\
&=O_{P^*}(N^{-1}T^{-1/2})+O_{P^*}(N^{-1/2}T^{-1}),
\end{align}
which means that overall 
\begin{align}
    \left\|\mathbf{II}_{2,2,1} \right\|=O_{P^*}((NT)^{-1/2}),
\end{align}
and the effect of the initial value is eliminated since we can verify that $\left\| (T^{-1}\widehat{\*F}'\widehat{\*F})^{-1} - (T^{-1}\overline{\*C}'\*Q'\*Q\overline{\*C})^{-1}\right\|\\=O_P(N^{-1})+O_P((NT)^{-1/2})$ under $K=R$. we now move on to $\mathbf{II}_{2,2,2}=\mathbf{II}_{2,2,2,a}+\mathbf{II}_{2,2,2,b}+\mathbf{II}_{2,2,2,c}+\mathbf{II}_{2,2,2,d}$, where 
\begin{align}
    \left\| \mathbf{II}_{2,2,2,a}\right\|&=\left\| \frac{1}{\sqrt{NT}}\sum_{i=1}^N\*S_\*b'\mathcal{\+P}_{0,1,i}^{\prime}T^{-1}\widehat{\*Q}_0^{\prime}\overline{\*V}(T^{-1}\widehat{\*F}'\widehat{\*F})^{-1}\overline{\*V}'\overline{\+\varepsilon}^*\widetilde{\gamma}_{y,i,1}\right\|\notag\\
    &\leq N^{-1}\left\|\*S_\*b \right\|\left( \frac{1}{N}\sum_{i=1}^N\left\|\mathcal{\+P}_{0,1,i} \right\|^2\right)^{1/2}\left( \frac{1}{N}\sum_{i=1}^N|\widetilde{\gamma}_{y,i,1}|^2\right)^{1/2}\left\|(T^{-1}\widehat{\*F}'\widehat{\*F})^{-1} \right\|\notag\\
    &\times \left\|\sqrt{N}T^{-1}\widehat{\*Q}_0^{\prime}\overline{\*V} \right\|\left\|\frac{1}{\sqrt{T}}\sum_{t=2}^TN\overline{\*v}_t\overline{\varepsilon}_{t}^* \right\| =O_{P^*}(N^{-1}),
\end{align}
which is the rate that is sufficient for our purposes, but it can be improved since we can demonstrate that $\left\|\sqrt{N}T^{-1}\widehat{\*Q}_0^{\prime}\overline{\*V} \right\|=o_P(1)$ by using (\ref{f_hat_A_0}) in the definition of $\widehat{\*Q}_0$. Next, 
\begin{align}
     \left\| \mathbf{II}_{2,2,2,b}\right\|&=\left\| \frac{1}{\sqrt{NT}}\sum_{i=1}^N\*S_\*b'\mathcal{\+P}_{0,1,i}^{\prime}T^{-1}\widehat{\*Q}_0^{\prime}\*Q\overline{\*C}(T^{-1}\widehat{\*F}'\widehat{\*F})^{-1}\overline{\*V}'\overline{\+\varepsilon}^*\widetilde{\gamma}_{y,i,1}\right\|\notag\\
    &\leq N^{-1/2}\left\|\*S_\*b \right\|\left\| \overline{\*C}\right\|\left( \frac{1}{N}\sum_{i=1}^N\left\|\mathcal{\+P}_{0,1,i} \right\|^2\right)^{1/2}\left( \frac{1}{N}\sum_{i=1}^N|\widetilde{\gamma}_{y,i,1}|^2\right)^{1/2}\left\|(T^{-1}\widehat{\*F}'\widehat{\*F})^{-1} \right\|\notag\\
    &\times \left\|T^{-1}\widehat{\*Q}_0^{\prime}\*Q\right\|\left\|\frac{1}{\sqrt{T}}\sum_{t=2}^TN\overline{\*v}_t\overline{\varepsilon}_{t}^* \right\|=O_{P^*}(N^{-1/2}),
\end{align}
\begin{align}
     \left\| \mathbf{II}_{2,2,2,c}\right\|&=\left\| \frac{1}{\sqrt{NT}}\sum_{i=1}^N\*S_\*b'\mathcal{\+P}_{0,1,i}^{\prime}T^{-1}\widehat{\*Q}_0^{\prime}\overline{\*V}(T^{-1}\widehat{\*F}'\widehat{\*F})^{-1}\overline{\*C}'\*Q'\overline{\+\varepsilon}^*\widetilde{\gamma}_{y,i,1}\right\|\notag\\
    &\leq N^{-1/2}\left\|\*S_\*b \right\|\left\| \overline{\*C}\right\|\left( \frac{1}{N}\sum_{i=1}^N\left\|\mathcal{\+P}_{0,1,i} \right\|^2\right)^{1/2}\left( \frac{1}{N}\sum_{i=1}^N|\widetilde{\gamma}_{y,i,1}|^2\right)^{1/2}\left\|(T^{-1}\widehat{\*F}'\widehat{\*F})^{-1} \right\|\notag\\
    &\times \left\|\sqrt{N}T^{-1}\widehat{\*Q}_0^{\prime}\overline{\*V}\right\|\left\|\frac{1}{\sqrt{T}}\sum_{t=2}^T\sqrt{N}\*q_t\overline{\varepsilon}_{t}^* \right\|=O_{P^*}(N^{-1/2}),
\end{align}
and 
\begin{align}
     \left\| \mathbf{II}_{2,2,2,d}\right\|&=\left\|\frac{1}{\sqrt{NT}}\sum_{i=1}^N\*S_\*b'\mathcal{\+P}_{0,1,i}^{\prime}T^{-1}\widehat{\*Q}_0^{\prime}\*Q\overline{\*C}((T^{-1}\widehat{\*F}'\widehat{\*F})^{-1} - (T^{-1}\overline{\*C}'\*Q'\*Q\overline{\*C})^{-1})\overline{\*C}'\*Q'\overline{\+\varepsilon}^*\widetilde{\gamma}_{y,i,1}\right\|\notag\\
     &\leq \left\|\*S_\*b \right\|\left\| \overline{\*C}\right\|^2\left( \frac{1}{N}\sum_{i=1}^N\left\|\mathcal{\+P}_{0,1,i} \right\|^2\right)^{1/2}\left( \frac{1}{N}\sum_{i=1}^N|\widetilde{\gamma}_{y,i,1}|^2\right)^{1/2}\left\|(T^{-1}\widehat{\*F}'\widehat{\*F})^{-1} \right\|\notag\\
    &\times \left\|T^{-1}\widehat{\*Q}_0^{\prime}\*Q\right\|\left\|\frac{1}{\sqrt{T}}\sum_{t=2}^T\sqrt{N}\*q_t\overline{\varepsilon}_{t}^* \right\|\left\| (T^{-1}\widehat{\*F}'\widehat{\*F})^{-1} - (T^{-1}\overline{\*C}'\*Q'\*Q\overline{\*C})^{-1}\right\|\notag\\
    &=O_{P^*}(N^{-1})+O_{P^*}((NT)^{-1/2}),
\end{align}
which implies that 
\begin{align}
     \left\| \mathbf{II}_{2,2,2}\right\|=O_{P^*}(N^{-1/2}).
\end{align}
The next term exhibits a similar asymptotic behavior. We have $\mathbf{II}_{2,2,3}=\mathbf{II}_{2,2,3,a}+\mathbf{II}_{2,2,3,b}+\mathbf{II}_{2,2,3,c}+\mathbf{II}_{2,2,3,d}$, where 
\begin{align}
    \left\| \mathbf{II}_{2,2,3,a}\right\|=&\left\| \frac{1}{\sqrt{NT}}\sum_{i=1}^N\*S_\*b'\mathcal{\+P}_{2}'T^{-1}\+\kappa_i'\overline{\*V}(T^{-1}\widehat{\*F}'\widehat{\*F})^{-1}\overline{\*V}'\overline{\+\varepsilon}^*\widetilde{\gamma}_{y,i,1}\right\|\notag\\
    &\leq N^{-1}\left\|\*S_\*b \right\|\left\| \mathcal{\+P}_2\right\|\left\| (T^{-1}\widehat{\*F}'\widehat{\*F})^{-1}\right\|\left(\frac{1}{N}\sum_{i=1}^N\left\| \sqrt{N}T^{-1}\+\kappa_i'\overline{\*V}\right\|^2\right)^{1/2}\left(\frac{1}{N}\sum_{i=1}^N|\widetilde{\gamma}_{y,i,1}|^2 \right)^{1/2}\notag\\
    &\times \left\|\frac{1}{\sqrt{T}}\sum_{t=2}^TN\overline{\*v}_t\overline{\varepsilon}_{t}^* \right\|=O_{P^*}(N^{-1}),
\end{align}
\begin{align}
    \left\| \mathbf{II}_{2,2,3,b}\right\|=&\left\| \frac{1}{\sqrt{NT}}\sum_{i=1}^N\*S_\*b'\mathcal{\+P}_{2}'T^{-1}\+\kappa_i'\*Q\overline{\*C}(T^{-1}\widehat{\*F}'\widehat{\*F})^{-1}\overline{\*V}'\overline{\+\varepsilon}^*\widetilde{\gamma}_{y,i,1}\right\|\notag\\
    &\leq N^{-1/2}\left\|\*S_\*b \right\|\left\| \mathcal{\+P}_2\right\|\left\| \overline{\*C}\right\|\left\| (T^{-1}\widehat{\*F}'\widehat{\*F})^{-1}\right\|\left(\frac{1}{N}\sum_{i=1}^N\left\| T^{-1}\+\kappa_i'\*Q\right\|^2\right)^{1/2}\left(\frac{1}{N}\sum_{i=1}^N|\widetilde{\gamma}_{y,i,1}|^2 \right)^{1/2}\notag\\
    &\times \left\|\frac{1}{\sqrt{T}}\sum_{t=2}^TN\overline{\*v}_t\overline{\varepsilon}_{t}^* \right\|=O_{P^*}(N^{-1/2}),
\end{align}
\begin{align}
    \left\| \mathbf{II}_{2,2,3,c}\right\|&=\left\|  \frac{1}{\sqrt{NT}}\sum_{i=1}^N\*S_\*b'\mathcal{\+P}_{2}'T^{-1}\+\kappa_i'\overline{\*V}(T^{-1}\widehat{\*F}'\widehat{\*F})^{-1}\overline{\*C}'\*Q'\overline{\+\varepsilon}^*\widetilde{\gamma}_{y,i,1}\right\| \notag\\
    &\leq \frac{1}{\sqrt{N}}\left\|\*S_\*b \right\|\left\| \mathcal{\+P}_2\right\|\left\| \overline{\*C}\right\|\left\| (T^{-1}\widehat{\*F}'\widehat{\*F})^{-1}\right\|\left(\frac{1}{N}\sum_{i=1}^N\left\| \sqrt{N}T^{-1}\+\kappa_i'\overline{\*V}\right\|^2\right)^{1/2}\left(\frac{1}{N}\sum_{i=1}^N|\widetilde{\gamma}_{y,i,1}|^2 \right)^{1/2}\notag\\
    &\times \left\|\frac{1}{\sqrt{T}}\sum_{t=2}^T\sqrt{N}\*q_t\overline{\varepsilon}_{t}^* \right\|=O_{P^*}(N^{-1/2}),
\end{align}
and finally 
\begin{align}
   \left\| \mathbf{II}_{2,2,3,d}\right\|&=\left\|\frac{1}{\sqrt{NT}}\sum_{i=1}^N\*S_\*b'\mathcal{\+P}_{2}'T^{-1}\+\kappa_i'\*Q\overline{\*C}((T^{-1}\widehat{\*F}'\widehat{\*F})^{-1} - (T^{-1}\overline{\*C}'\*Q'\*Q\overline{\*C})^{-1})\overline{\*C}'\*Q'\overline{\+\varepsilon}^*\widetilde{\gamma}_{y,i,1}\right\|\notag\\
   &\leq \frac{1}{\sqrt{N}}\left\|\*S_\*b \right\|\left\| \mathcal{\+P}_2\right\|\left\| \overline{\*C}\right\|^2\left\| (T^{-1}\widehat{\*F}'\widehat{\*F})^{-1}\right\|\left(\frac{1}{N}\sum_{i=1}^N\left\| T^{-1}\+\kappa_i'\*Q\right\|^2\right)^{1/2}\left(\frac{1}{N}\sum_{i=1}^N|\widetilde{\gamma}_{y,i,1}|^2 \right)^{1/2}\notag\\
    &\times \left\|\frac{1}{\sqrt{T}}\sum_{t=2}^T\sqrt{N}\*q_t\overline{\varepsilon}_{t}^* \right\|\left\| (T^{-1}\widehat{\*F}'\widehat{\*F})^{-1} - (T^{-1}\overline{\*C}'\*Q'\*Q\overline{\*C})^{-1}\right\|\notag\\
    &=O_{P^*}(N^{-1})+O_{P^*}((NT)^{-1/2}),
\end{align}
giving 
    \begin{align}
     \left\| \mathbf{II}_{2,2,3}\right\|=O_{P^*}(N^{-1/2}).
\end{align}
Note that $\mathbf{II}_{2,2,4}$ asymptotically behaves exactly the same as $\mathbf{II}_{2,2,2}$, except that it is additionally scaled by $\frac{1}{\sqrt{T}}$ because $\left(\frac{1}{N}\sum_{i=1}^N\left\|\mathcal{\+P}_{0,3,i}\right\|^2\right)^{1/2}=O_P(T^{-1/2})$. Therefore, 
\begin{align}
   \left\|\mathbf{II}_{2,2,4}\right\|= \left\|\frac{1}{\sqrt{NT}}\sum_{i=1}^N\*S_\*b'\mathcal{\+P}_{0,3,i}^{\prime}\*H'( \*P_{\widehat{\*F}}-\*P_{\*Q\overline{\*C}})\overline{\+\varepsilon}^*\widetilde{\gamma}_{y,i,1}\right\|=O_{P^*}((NT)^{-1/2}).
\end{align}
Ultimately, we have $\mathbf{II}_{2,2,5}=\mathbf{II}_{2,2,5,a}+\mathbf{II}_{2,2,5,b}+\mathbf{II}_{2,2,5,c}+\mathbf{II}_{2,2,5,d}$, where 
\begin{align}
    \left\| \mathbf{II}_{2,2,5,a}\right\|&=\left\|\frac{1}{\sqrt{NT}}\sum_{i=1}^N\*S_\*b'T^{-1}\*U_i^{*\prime}\overline{\*V}(T^{-1}\widehat{\*F}'\widehat{\*F})^{-1}\overline{\*V}'\overline{\+\varepsilon}^*\widetilde{\gamma}_{y,i,1}\right\|\notag\\
    &\leq N^{-1} \left\| (T^{-1}\widehat{\*F}'\widehat{\*F})^{-1}\right\|\left(\frac{1}{N}\sum_{i=1}^N\left\|\sqrt{N}T^{-1}\*S_\*b'\*U_i^{*\prime}\overline{\*V} \right\|^2 \right)^{1/2}\left( \frac{1}{N}\sum_{i=1}^N|\widetilde{\gamma}_{y,i,1}|^2\right)^{1/2}\notag\\
    &\times \left\|\frac{1}{\sqrt{T}}\sum_{t=2}^TN\overline{\*v}_t\overline{\varepsilon}_{t}^* \right\|=O_{P^*}(N^{-1}T^{-1/2}),
\end{align}
Next, 
\begin{align}
    \left\| \mathbf{II}_{2,2,5,b}\right\|&=\left\|\frac{1}{\sqrt{NT}}\sum_{i=1}^N\*S_\*b'T^{-1}\*U_i^{*\prime}\*Q\overline{\*C}(T^{-1}\widehat{\*F}'\widehat{\*F})^{-1}\overline{\*V}'\overline{\+\varepsilon}^*\widetilde{\gamma}_{y,i,1}\right\|\notag\\
    &\leq N^{-1/2}\left\| \overline{\*C}\right\| \left\| (T^{-1}\widehat{\*F}'\widehat{\*F})^{-1}\right\|\left(\frac{1}{N}\sum_{i=1}^N\left\|T^{-1}\*S_\*b'\*U_i^{*\prime}\*Q \right\|^2 \right)^{1/2}\left( \frac{1}{N}\sum_{i=1}^N|\widetilde{\gamma}_{y,i,1}|^2\right)^{1/2}\notag\\
    &\times \left\|\frac{1}{\sqrt{T}}\sum_{t=2}^TN\overline{\*v}_t\overline{\varepsilon}_{t}^* \right\|=O_{P^*}((NT)^{-1/2}),
\end{align}
\begin{align}
     \left\| \mathbf{II}_{2,2,5,c}\right\|&=\left\|\frac{1}{\sqrt{NT}}\sum_{i=1}^N\*S_\*b'T^{-1}\*U_i^{*\prime}\overline{\*V}(T^{-1}\widehat{\*F}'\widehat{\*F})^{-1}\overline{\*C}'\*Q'\overline{\+\varepsilon}^*\widetilde{\gamma}_{y,i,1}\right\|\notag\\
    &\leq N^{-1/2}\left\| \overline{\*C}\right\| \left\| (T^{-1}\widehat{\*F}'\widehat{\*F})^{-1}\right\|\left(\frac{1}{N}\sum_{i=1}^N\left\|\sqrt{N}T^{-1}\*S_\*b'\*U_i^{*\prime}\overline{\*V}\right\|^2 \right)^{1/2}\left( \frac{1}{N}\sum_{i=1}^N|\widetilde{\gamma}_{y,i,1}|^2\right)^{1/2}\notag\\
    &\times \left\|\frac{1}{\sqrt{T}}\sum_{t=2}^T\sqrt{N}\*q_t\overline{\varepsilon}_{t}^* \right\|=O_{P^*}((NT)^{-1/2}),
\end{align}
where $\left\|\sqrt{N}T^{-1}\*S_\*b'\*U_i^{*\prime}\overline{\*V} \right\|=O_{P^*}(T^{-1/2})$ by an argument similar (\ref{b_m-b}). Finally, 
\begin{align}
     \left\| \mathbf{II}_{2,2,5,d}\right\|&=\left\|\frac{1}{\sqrt{NT}}\sum_{i=1}^N\*S_\*b'T^{-1}\*U_i^{*\prime}\*Q\overline{\*C}(T^{-1}\widehat{\*F}'\widehat{\*F})^{-1}\overline{\*C}'\*Q'\overline{\+\varepsilon}^*\widetilde{\gamma}_{y,i,1}\right\|\notag\\
    &\leq\left\| \overline{\*C}\right\|^2\left(\frac{1}{N}\sum_{i=1}^N\left\|T^{-1}\*S_\*b'\*U_i^{*\prime}\*Q\right\|^2 \right)^{1/2}\left( \frac{1}{N}\sum_{i=1}^N|\widetilde{\gamma}_{y,i,1}|^2\right)^{1/2}\notag\\
    &\times \left\|\frac{1}{\sqrt{T}}\sum_{t=2}^T\sqrt{N}\*q_t\overline{\varepsilon}_{t}^* \right\|\left\| (T^{-1}\widehat{\*F}'\widehat{\*F})^{-1} - (T^{-1}\overline{\*C}'\*Q'\*Q\overline{\*C})^{-1}\right\|\notag\\
    &=O_{P^*}(N^{-1}T^{-1/2})+O_{P^*}(N^{-1/2}T^{-1}).
\end{align}
 This implies that 
\begin{align}
     \left\| \mathbf{II}_{2,2,5}\right\|=O_{P^*}((NT)^{-1/2}),
\end{align}
and overall
\begin{align}
     \left\| \mathbf{II}_{2,2}\right\|=O_{P^*}(N^{-1/2}).
\end{align}
We move on to $\mathbf{II}_{2,3}$, where we will use (\ref{y_property_with_X}), which means that 
\begin{align}
    \widehat{\*F}^*=[\overline{\*y}^*,\overline{\*X}, \overline{\*y}_{-1}^*, \overline{\*X}_{-1}]&=[\overline{\*y},\overline{\*X}, \overline{\*y}_{-1}, \overline{\*X}_{-1}]+[\overline{\*u}^*, \*0_{(T-1)\times k}, \overline{\*u}^*_{-1}, \*0_{(T-1)\times k}]\notag\\
    &=\widehat{\*F}+\overline{\*W}^*,
\end{align}
and so 
\begin{align}\label{Pf*-Pfhat}
   \*P_{\widehat{\*F}^*}-\*P_{\widehat{\*F}}&=\overline{\*W}^*(\widehat{\*F}^{*\prime}\widehat{\*F}^*)^{-1}\overline{\*W}^{*\prime}+\overline{\*W}^*(\widehat{\*F}^{*\prime}\widehat{\*F}^*)^{-1}\widehat{\*F}'+\widehat{\*F}(\widehat{\*F}^{*\prime}\widehat{\*F}^*)^{-1}\overline{\*W}^{*\prime}\notag\\
&+\widehat{\*F}\left((\widehat{\*F}^{*\prime}\widehat{\*F}^*)^{-1}-(\widehat{\*F}'\widehat{\*F})^{-1} \right)\widehat{\*F}',
\end{align}
which we will use in 
\begin{align}
    \mathbf{II}_{2,3}&=\frac{1}{\sqrt{NT}}\sum_{i=1}^N\*B_i^{*\prime}( \*P_{\widehat{\*F}^*}-\*P_{\widehat{\*F}})\overline{\+\varepsilon}^*\widetilde{\gamma}_{y,i,1}=\frac{1}{\sqrt{NT}}\sum_{i=1}^N\*S_\*b'[\widehat{\*A}, \*I_{K\times K}]'\widehat{\*A}_{\*z_0,i}'( \*P_{\widehat{\*F}^*}-\*P_{\widehat{\*F}})\overline{\+\varepsilon}^*\widetilde{\gamma}_{y,i,1}\notag\\
    &+\frac{1}{\sqrt{NT}}\sum_{i=1}^N\*S_\*b'\mathcal{\+P}_{0,1,i}^{\prime}\widehat{\*Q}_0^{\prime}(  \*P_{\widehat{\*F}^*}-\*P_{\widehat{\*F}})\overline{\+\varepsilon}^*\widetilde{\gamma}_{y,i,1}+\frac{1}{\sqrt{NT}}\sum_{i=1}^N\*S_\*b'\mathcal{\+P}_{2}'\+\kappa_i'(  \*P_{\widehat{\*F}^*}-\*P_{\widehat{\*F}})\overline{\+\varepsilon}^*\widetilde{\gamma}_{y,i,1}\notag\\
    &+\frac{1}{\sqrt{NT}}\sum_{i=1}^N\*S_\*b'\mathcal{\+P}_{0,3,i}^{\prime}\*H'(  \*P_{\widehat{\*F}^*}-\*P_{\widehat{\*F}})\overline{\+\varepsilon}^*\widetilde{\gamma}_{y,i,1}+\frac{1}{\sqrt{NT}}\sum_{i=1}^N\*S_\*b'\*U_i^{*\prime}(  \*P_{\widehat{\*F}^*}-\*P_{\widehat{\*F}})\overline{\+\varepsilon}^*\widetilde{\gamma}_{y,i,1}\notag\\
    &= \mathbf{II}_{2,3,1}+ \mathbf{II}_{2,3,2}+ \mathbf{II}_{2,3,3}+ \mathbf{II}_{2,3,4}+ \mathbf{II}_{2,3,5}.
\end{align}
We will start with $\mathbf{II}_{2,3,1}=\mathbf{II}_{2,3,1,a}+\mathbf{II}_{2,3,1,b}+\mathbf{II}_{2,3,1,c}+\mathbf{II}_{2,3,1,d}$, where by using similar arguments we learn that the effect of the initial value is asymptotically negligible. In particular, 
\begin{align}
    \left\|\mathbf{II}_{2,3,1,a} \right\|&=\left\| \frac{1}{\sqrt{NT}}\sum_{i=1}^N\*S_\*b'[\widehat{\*A}, \*I_{K\times K}]'T^{-1}\widehat{\*A}_{\*z_0,i}'\overline{\*W}^*(T^{-1}\widehat{\*F}^{*\prime}\widehat{\*F}^*)^{-1}\overline{\*W}^{*\prime}\overline{\+\varepsilon}^*\widetilde{\gamma}_{y,i,1}\right\|\notag\\
    &\leq N^{-1}\left\|\*S_\*b'[\widehat{\*A}, \*I_{K\times K}]' \right\|\left\| (T^{-1}\widehat{\*F}^{*\prime}\widehat{\*F}^*)^{-1}\right\|\left(\frac{1}{N}\sum_{i=1}^N| \widetilde{\gamma}_{y,i,1}|^2\right)^{1/2}\left(\frac{1}{N}\sum_{i=1}^N\left\|\*z_{i,0} \right\|^2 \right)^{1/2}\notag\\
    &\times \left(\frac{1}{T}\sum_{t=2}^T\left\|\sqrt{N}\overline{\*w}^*_t \right\|^2 \right)^{1/2}\sum_{t=2}^T|\widehat{\alpha}_{CCEP} |^{t-1}\underbrace{\left\|NT^{-1}\overline{\*W}^{*\prime}\overline{\+\varepsilon}^* \right\|}_{O_{P^*}(1)}=O_{P^*}(N^{-1}),
\end{align}
\begin{align}
    \left\|\mathbf{II}_{2,3,1,b} \right\|&= \left\| \frac{1}{\sqrt{NT}}\sum_{i=1}^N\*S_\*b'[\widehat{\*A}, \*I_{K\times K}]'T^{-1}\widehat{\*A}_{\*z_0,i}'\widehat{\*F}(T^{-1}\widehat{\*F}^{*\prime}\widehat{\*F}^*)^{-1}\overline{\*W}^{*\prime}\overline{\+\varepsilon}^*\widetilde{\gamma}_{y,i,1}\right\|\notag\\
     &\leq N^{-1/2}\left\|\*S_\*b'[\widehat{\*A}, \*I_{K\times K}]' \right\|\left\| (T^{-1}\widehat{\*F}^{*\prime}\widehat{\*F}^*)^{-1}\right\|\left(\frac{1}{N}\sum_{i=1}^N| \widetilde{\gamma}_{y,i,1}|^2\right)^{1/2}\left(\frac{1}{N}\sum_{i=1}^N\left\|\*z_{i,0} \right\|^2 \right)^{1/2}\notag\\
    &\times \left(\frac{1}{T}\sum_{t=2}^T\left\|\widehat{\*f}_t \right\|^2 \right)^{1/2}\sum_{t=2}^T|\widehat{\alpha}_{CCEP} |^{t-1}\underbrace{\left\|NT^{-1}\overline{\*W}^{*\prime}\overline{\+\varepsilon}^* \right\|}_{O_{P^*}(1)}=O_{P^*}(N^{-1/2}),
\end{align}
\begin{align}
     \left\|\mathbf{II}_{2,3,1,c} \right\|&=\left\| \frac{1}{\sqrt{NT}}\sum_{i=1}^N\*S_\*b'[\widehat{\*A}, \*I_{K\times K}]'T^{-1}\widehat{\*A}_{\*z_0,i}'\overline{\*W}^*(T^{-1}\widehat{\*F}^{*\prime}\widehat{\*F}^*)^{-1}\widehat{\*F}'\overline{\+\varepsilon}^*\widetilde{\gamma}_{y,i,1}\right\|\notag\\
    &\leq (NT)^{-1/2}\left\|\*S_\*b'[\widehat{\*A}, \*I_{K\times K}]' \right\|\left\| (T^{-1}\widehat{\*F}^{*\prime}\widehat{\*F}^*)^{-1}\right\|\left(\frac{1}{N}\sum_{i=1}^N| \widetilde{\gamma}_{y,i,1}|^2\right)^{1/2}\left(\frac{1}{N}\sum_{i=1}^N\left\|\*z_{i,0} \right\|^2 \right)^{1/2}\notag\\
    &\times \left(\frac{1}{T}\sum_{t=2}^T\left\|\sqrt{N}\overline{\*w}^*_t \right\|^2 \right)^{1/2}\sum_{t=2}^T|\widehat{\alpha}_{CCEP} |^{t-1}\underbrace{\left\|\sqrt{N}T^{-1/2}\widehat{\*F}'\overline{\+\varepsilon}^* \right\|}_{O_{P^*}(1)}=O_{P^*}((NT)^{-1/2}),
\end{align}
because $\left\|\sqrt{N}T^{-1/2}\widehat{\*F}'\overline{\+\varepsilon}^*\right\|\leq \left\| \overline{\*C}\right\|\left\|\sqrt{N}T^{-1/2}\*Q'\overline{\+\varepsilon}^* \right\|+\left\|\sqrt{N}T^{-1/2}\overline{\*V}'\overline{\+\varepsilon}^*  \right\|=O_{P^*}(1)+O_{P^*}(N^{-1/2})$. Lastly, 
\begin{align}
    \left\| \mathbf{II}_{2,3,1,d}\right\|&=\left\| \frac{1}{\sqrt{NT}}\sum_{i=1}^N\*S_\*b'[\widehat{\*A}, \*I_{K\times K}]'T^{-1}\widehat{\*A}_{\*z_0,i}'\widehat{\*F}((T^{-1}\widehat{\*F}^{*\prime}\widehat{\*F}^*)^{-1}-(T^{-1}\widehat{\*F}'\widehat{\*F})^{-1})\widehat{\*F}'\overline{\+\varepsilon}^*\widetilde{\gamma}_{y,i,1}\right\|\notag\\
    &\leq T^{-1/2}\left\|\*S_\*b'[\widehat{\*A}, \*I_{K\times K}]' \right\|\left(\frac{1}{N}\sum_{i=1}^N| \widetilde{\gamma}_{y,i,1}|^2\right)^{1/2}\left(\frac{1}{N}\sum_{i=1}^N\left\|\*z_{i,0} \right\|^2 \right)^{1/2}\notag\\
    &\times \left(\frac{1}{T}\sum_{t=2}^T\left\|\widehat{\*f}_t \right\|^2 \right)^{1/2}\sum_{t=2}^T|\widehat{\alpha}_{CCEP} |^{t-1}\underbrace{\left\|\sqrt{N}T^{-1/2}\widehat{\*F}'\overline{\+\varepsilon}^* \right\|}_{O_{P^*}(1)} \left\| (T^{-1}\widehat{\*F}^{*\prime}\widehat{\*F}^*)^{-1}-(T^{-1}\widehat{\*F}'\widehat{\*F})^{-1}\right\|\notag\\
    &= O_{P^*}(N^{-1}T^{-1/2})+O_{P^*}(N^{-1/2}T^{-1})
\end{align}
due to consistency of $\widehat{\alpha}_{CCEP}$, and because 
\begin{align}
    \left\| (T^{-1}\widehat{\*F}^{*\prime}\widehat{\*F}^*)^{-1}-(T^{-1}\widehat{\*F}'\widehat{\*F})^{-1}\right\|&\leq \left\|(T^{-1}\widehat{\*F}'\widehat{\*F})^{-1} \right\| \left\| (T^{-1}\widehat{\*F}^{*\prime}\widehat{\*F}^*)^{-1}\right\|\left\| T^{-1}\widehat{\*F}^{*\prime}\widehat{\*F}^*-T^{-1}\widehat{\*F}'\widehat{\*F} \right\|\notag\\
    &= \left\|(T^{-1}\widehat{\*F}'\widehat{\*F})^{-1} \right\| \left\| (T^{-1}\widehat{\*F}^{*\prime}\widehat{\*F}^*)^{-1}\right\|\left\|T^{-1}\widehat{\*F}'\overline{\*W}^*+T^{-1}\overline{\*W}^{*\prime}\widehat{\*F} +T^{-1}\overline{\*W}^{*\prime}\overline{\*W}^{*}\right\|\notag\\
    &= O_{P^*}(N^{-1})+O_{P^*}((NT)^{-1/2}),
\end{align}
where we can verify that 
\begin{align}\label{FW*}
    \left\|\sqrt{N}T^{-1/2}\widehat{\*F}'\overline{\*W}^* \right\|&\leq \left\| \overline{\*C}\right\| \left\|\begin{bmatrix}
        \frac{1}{\sqrt{T}}\sum_{t=2}^T\*q_t\sqrt{N}\overline{u}^*_t & \*0_{2K\times k} & \frac{1}{\sqrt{T}}\sum_{t=2}^T\*q_t\sqrt{N}\overline{u}_{t-1}^* & \*0_{2K\times k}
    \end{bmatrix} \right\|\notag\\
    & +\left\|\begin{bmatrix}
        \frac{1}{\sqrt{T}}\sum_{t=2}^T\overline{\*v}_t\sqrt{N}\overline{u}^*_t & \*0_{2K\times k} & \frac{1}{\sqrt{T}}\sum_{t=2}^T\overline{\*v}_t\sqrt{N}\overline{u}_{t-1}^* & \*0_{2K\times k}
    \end{bmatrix} \right\|\notag\\
    &\leq \left\| \overline{\*C}\right\| \left\|\frac{1}{\sqrt{T}}\sum_{t=2}^T\*q_t\sqrt{N}\overline{u}^*_t \right\| + \left\| \overline{\*C}\right\| \left\|\frac{1}{\sqrt{T}}\sum_{t=2}^T\*q_t\sqrt{N}\overline{u}^*_{t-1} \right\|\notag\\
    &+\left\| \frac{1}{\sqrt{T}}\sum_{t=2}^T\overline{\*v}_t\sqrt{N}\overline{u}^*_t \right\| + \left\|  \frac{1}{\sqrt{T}}\sum_{t=2}^T\overline{\*v}_t\sqrt{N}\overline{u}^*_{t-1}\right\|=O_{P^*}(1),
\end{align}
 The first two components follow (\ref{b_m-b}). The last two components are $O_{P^*}(N^{-1/2})$. Denote the last component $\*
 d$, so its scaled counterpart for a fixed $m$ is 
 \begin{align}
    \sqrt{N}\*d_m&=\sum_{j=0}^{m-2}\widehat{\alpha}_{CCEP}^j\frac{1}{\sqrt{NT}}\sum_{i=1}^N\sum_{t=j+2}^T\varepsilon_{i,t-j-1}^*\sqrt{N}\overline{\*v}_t \notag\\
    &=\sum_{j=0}^{m-2}(\alpha_0)^j\underbrace{\frac{1}{\sqrt{NT}}\sum_{i=1}^N\sum_{t=j+2}^T\varepsilon_{i,t-j-1}^*\sqrt{N}\overline{\*v}_t}_{\sqrt{N}\*d_{j,m}=O_{P^*}(1)} + o_{P^*}(1) =O_{P^*}(1)
 \end{align}
  Then by again applying $|xy|\leq (x^2+y^2)/2$ we get 
 \begin{align}\label{d_m-d}
     E^*[(\+\lambda'\sqrt{N}(\*d_m&-\*d))^2 ]=\sum_{j=m}^{T-2}\sum_{r=m}^{T-2}\widehat{\alpha}_{CCEP}^{j+r}E^*\left[\frac{1}{NT}\sum_{i=1}^N\sum_{j=1}^N\sum_{t=j+2}^T\sum_{s=r+2}^T\varepsilon_{i,t-j-1}^*\varepsilon_{i,s-r-1}^*\+\lambda'N\overline{\*v}_{t}\overline{\*v}_{s}'\+\lambda \right]\notag\\
     &=\sum_{j=m}^{T-2}\sum_{r=m}^{T-2}\widehat{\alpha}_{CCEP}^{j+r}\frac{1}{N}\sum_{i=1}^N\frac{1}{T}\sum_{t=\max(j,r)+2}^T\widehat{\varepsilon}_{i,t-j-1}^2\+\lambda'N\overline{\*v}_{t}\overline{\*v}_{t+r-j}'\+\lambda\notag\\
     &\leq  \sum_{j=m}^{T-2}\sum_{r=m}^{T-2}|\widehat{\alpha}_{CCEP}|^{j+r}\frac{1}{2}\frac{1}{N}\sum_{i=1}^N\frac{1}{T}\sum_{t=\max(j,r)+2}^T\widehat{\varepsilon}_{i,t-j-1}^2(|\+\lambda'\sqrt{N}\overline{\*v}_{t}|^2+|\sqrt{N}\overline{\*v}_{t+r-j}'\+\lambda|^2)\notag\\
        &\leq  \frac{1}{2}\sum_{j=m}^{T-2}\sum_{r=m}^{T-2}|\widehat{\alpha}_{CCEP}|^{j+r}\frac{1}{N}\sum_{i=1}^N\left( \frac{1}{T}\sum_{t=\max(j,r)+2}^T\widehat{\varepsilon}_{i,t-j-1}^4\right)^{1/2}\left(\frac{1}{T}\sum_{t=\max(j,r)+2}^T |\+\lambda'\sqrt{N}\overline{\*v}|^4\right)^{1/2}  \notag\\
        & +   \frac{1}{2}\sum_{j=m}^{T-2}\sum_{r=m}^{T-2}|\widehat{\alpha}_{CCEP}|^{j+r}\frac{1}{N}\sum_{i=1}^N\left( \frac{1}{T}\sum_{t=\max(j,r)+2}^T\widehat{\varepsilon}_{i,t-j-1}^4\right)^{1/2}\left(\frac{1}{T}\sum_{t=\max(j,r)+2}^T |\sqrt{N}\overline{\*v}_{t+r-j}'\+\lambda|^4\right)^{1/2}\notag\\
        &\leq  \left\| \+\lambda\right\|^2\left(\sum_{j=m}^{T-2}|\widehat{\alpha}_{CCEP} |^j\right)^2 \left(\frac{1}{NT}\sum_{i=1}^N\sum_{t=1}^T\widehat{\varepsilon}_{i,t}^4 \right)^{1/2} \left(\frac{1}{T}\sum_{t=1}^T\left\|\sqrt{N}\overline{\*v}_{t} \right\|^4 \right)^{1/2}\notag\\
        &=o_P(1),
 \end{align}
and the same holds for the second-to-last term. Overall,
\begin{align}
    \left\|\mathbf{II}_{2,3,1} \right\|=O_{P^*}(N^{-1/2}).
\end{align}

\noindent We now go to $\mathbf{II}_{2,3,2}=\mathbf{II}_{2,3,2,a}+\mathbf{II}_{2,3,2,b}+\mathbf{II}_{2,3,2,c}+\mathbf{II}_{2,3,2,d}$, where 
\begin{align}
    \left\|\mathbf{II}_{2,3,2,a} \right\|&=\left\| \frac{1}{\sqrt{NT}}\sum_{i=1}^N\*S_\*b'\mathcal{\+P}_{0,1,i}^{\prime}T^{-1}\widehat{\*Q}_0^{\prime}\overline{\*W}^*(T^{-1}\widehat{\*F}^{*\prime}\widehat{\*F}^*)^{-1}\overline{\*W}^{*\prime}\overline{\+\varepsilon}^*\widetilde{\gamma}_{y,i,1}\right\|\notag\\
    &\leq N^{-1/2}\sqrt{\frac{T}{N}}\left\|\*S_\*b \right\|\left( \frac{1}{N}\sum_{i=1}^N\left\|\mathcal{\+P}_{0,1,i} \right\|^2\right)^{1/2}\left( \frac{1}{N}\sum_{i=1}^N|\widetilde{\gamma}_{y,i,1}|^2\right)^{1/2}\left\|(T^{-1}\widehat{\*F}^{*\prime}\widehat{\*F}^*)^{-1} \right\|\notag\\
    &\times \left\|\sqrt{N}T^{-1}\widehat{\*Q}_0^{\prime}\overline{\*W}^* \right\|\left\|NT^{-1}\overline{\*W}^{*\prime}\overline{\+\varepsilon}^*\right\| =O_{P^*}(N^{-1/2})
\end{align}
under $NT^{-1}\to \kappa>0$, which is sufficient for our purposes. The next term generates an asymptotic bias: 
\begin{align}\label{II_232b}
    \mathbf{II}_{2,3,2,b}&=\frac{1}{\sqrt{NT}}\sum_{i=1}^N\*S_\*b'\mathcal{\+P}_{0,1,i}^{\prime}T^{-1}\widehat{\*Q}_0^{\prime}\widehat{\*F}(T^{-1}\widehat{\*F}^{*\prime}\widehat{\*F}^*)^{-1}\overline{\*W}^{*\prime}\overline{\+\varepsilon}^*\widetilde{\gamma}_{y,i,1}\notag\\
    &=\sqrt{\frac{T}{N}}\left(\frac{1}{N}\sum_{i=1}^N\widetilde{\gamma}_{y,i,1}\mathcal{\+P}_{0,1,i}\*S_\*b \right)'T^{-1}\widehat{\*Q}_0^{\prime}\*Q\overline{\*C}(T^{-1}\widehat{\*F}^{*\prime}\widehat{\*F}^*)^{-1}(NT^{-1}\overline{\*W}^{*\prime}\overline{\+\varepsilon}^*)\notag\\
    &+\frac{1}{\sqrt{N}}\sqrt{\frac{T}{N}}\left(\frac{1}{N}\sum_{i=1}^N\widetilde{\gamma}_{y,i,1}\mathcal{\+P}_{0,1,i}\*S_\*b \right)'(\sqrt{N}T^{-1}\widehat{\*Q}_0^{\prime}\overline{\*V})(T^{-1}\widehat{\*F}^{*\prime}\widehat{\*F}^*)^{-1}(NT^{-1}\overline{\*W}^{*\prime}\overline{\+\varepsilon}^*)\notag\\
    &=\sqrt{\frac{T}{N}}\left(\frac{1}{N}\sum_{i=1}^N\widetilde{\gamma}_{y,i,1}\mathcal{\+P}_{0,1,i}\*S_\*b \right)'T^{-1}\widehat{\*Q}_0^{\prime}\*Q\overline{\*C}(T^{-1}\widehat{\*F}^{*\prime}\widehat{\*F}^*)^{-1}(NT^{-1}\overline{\*W}^{*\prime}\overline{\+\varepsilon}^*) + O_{P^*}(N^{-1/2})\notag\\
    &=\sqrt{\frac{T}{N}}\left(\frac{1}{N}\sum_{i=1}^N\widetilde{\gamma}_{y,i,1}\mathcal{\+P}_{0,1,i}\*S_\*b \right)'T^{-1}\widehat{\*Q}_0^{\prime}\*Q\overline{\*C}(T^{-1}\widehat{\*F}^{*\prime}\widehat{\*F}^*)^{-1}\begin{bmatrix}
        NT^{-1}\overline{\*u}^{*\prime}\overline{\+\varepsilon}^* & \*0_{1\times (2K-1)}    \end{bmatrix}'\notag\\
        &+O_{P^*}(C_{N,T}^{-1}),
\end{align}
because we notice that 
\begin{align}
    |NT^{-1}\overline{\*u}^{*\prime}_{-1}\overline{\+\varepsilon}^*|=\sqrt{\frac{N}{T}}\underbrace{\left|\frac{1}{N}\sum_{i=1}^N\frac{1}{\sqrt{T}}\sum_{t=2}^{T}\sqrt{N}\overline{u}^*_{t-1}\varepsilon_{i,t}^* \right|}_{O_{P^*}(N^{-1/2})}=O_{P^*}(T^{-1/2}),
\end{align}
which we already demonstrated in (\ref{I_a2}). Next, 
\begin{align}
    \left\|  \mathbf{II}_{2,3,2,c}\right\|&=\left\| \frac{1}{\sqrt{NT}}\sum_{i=1}^N\*S_\*b'\mathcal{\+P}_{0,1,i}^{\prime}T^{-1}\widehat{\*Q}_0^{\prime}\overline{\*W}^*(T^{-1}\widehat{\*F}^{*\prime}\widehat{\*F}^*)^{-1}\widehat{\*F}^{\prime}\overline{\+\varepsilon}^*\widetilde{\gamma}_{y,i,1}\right\|\notag\\
    &\leq  N^{-1/2}
    \left\|\*S_\*b \right\|\left( \frac{1}{N}\sum_{i=1}^N\left\|\mathcal{\+P}_{0,1,i} \right\|^2\right)^{1/2}\left( \frac{1}{N}\sum_{i=1}^N|\widetilde{\gamma}_{y,i,1}|^2\right)^{1/2}\left\|(T^{-1}\widehat{\*F}^{*\prime}\widehat{\*F}^*)^{-1} \right\|\notag\\
    &\times \left\|\sqrt{N}T^{-1}\widehat{\*Q}_0^{\prime}\overline{\*W}^* \right\|\left\|\sqrt{N}T^{-1/2}\widehat{\*F}'\overline{\+\varepsilon}^*\right\| =O_{P^*}(N^{-1/2}),
\end{align}
and finally 
\begin{align}
     \left\|  \mathbf{II}_{2,3,2,d}\right\|&=\left\| \frac{1}{\sqrt{NT}}\sum_{i=1}^N\*S_\*b'\mathcal{\+P}_{0,1,i}^{\prime}T^{-1}\widehat{\*Q}_0^{\prime}\widehat{\*F}((T^{-1}\widehat{\*F}^{*\prime}\widehat{\*F}^*)^{-1}-(T^{-1}\widehat{\*F}'\widehat{\*F})^{-1})\widehat{\*F}^{\prime}\overline{\+\varepsilon}^*\widetilde{\gamma}_{y,i,1}\right\|\notag\\
     &\leq \left\|\*S_\*b \right\|\left( \frac{1}{N}\sum_{i=1}^N\left\|\mathcal{\+P}_{0,1,i} \right\|^2\right)^{1/2}\left( \frac{1}{N}\sum_{i=1}^N|\widetilde{\gamma}_{y,i,1}|^2\right)^{1/2}\notag\\
    &\times \left\|T^{-1}\widehat{\*Q}_0^{\prime}\widehat{\*F} \right\|\left\|\sqrt{N}T^{-1/2}\widehat{\*F}'\overline{\+\varepsilon}^*\right\| \left\| (T^{-1}\widehat{\*F}^{*\prime}\widehat{\*F}^*)^{-1}-(T^{-1}\widehat{\*F}'\widehat{\*F})^{-1}\right\|\notag\\
    &= O_{P^*}(N^{-1})+O_{P^*}((NT)^{-1/2}),
\end{align}
and thus 
\begin{align}
    \left\|\mathbf{II}_{2,3,2} \right\|&=\sqrt{\frac{T}{N}}\left(\frac{1}{N}\sum_{i=1}^N\widetilde{\gamma}_{y,i,1}\mathcal{\+P}_{0,1,i}\*S_\*b \right)'T^{-1}\widehat{\*Q}_0^{\prime}\*Q\overline{\*C}(T^{-1}\widehat{\*F}^{*\prime}\widehat{\*F}^*)^{-1}\begin{bmatrix}
        NT^{-1}\overline{\*u}^{*\prime}\overline{\+\varepsilon}^* & \*0_{1\times (2K-1)}    \end{bmatrix}'\notag\\
        &+O_{P^*}(C_{N,T}^{-1}).
\end{align}
The upcoming term, which is driven by the weak exogeneity parameter $\+\theta_0$, is $\mathbf{II}_{2,3,3}=\mathbf{II}_{2,3,3,a}+\mathbf{II}_{2,3,3,b}+\mathbf{II}_{2,3,3,c}+\mathbf{II}_{2,3,3,d}$ asymptotically behaves similarly to $\mathbf{II}_{2,3,2}$. Specifically,
\begin{align}
    \left\|\mathbf{II}_{2,3,3,a} \right\|&=\left\| \frac{1}{\sqrt{NT}}\sum_{i=1}^N\*S_\*b'\mathcal{\+P}_{2}'T^{-1}\+\kappa_i'\overline{\*W}^*(T^{-1}\widehat{\*F}^{*\prime}\widehat{\*F}^*)^{-1}\overline{\*W}^{*\prime}\overline{\+\varepsilon}^*\widetilde{\gamma}_{y,i,1}\right\|\notag\\
    &\leq N^{-1/2}\sqrt{\frac{T}{N}}\left\|\*S_\*b \right\|\left\|\mathcal{\+P}_2 \right\|\left(\frac{1}{N}\sum_{i=1}^N |\widetilde{\gamma}_{y,i,1}|^2\right)^{1/2}\left(\frac{1}{N}\sum_{i=1}^N\left\|\sqrt{N}T^{-1}\+\kappa_i'\overline{\*W}^* \right\|^2 \right)^{1/2}\notag\\
    &\times \left\| (T^{-1}\widehat{\*F}^{*\prime}\widehat{\*F}^*)^{-1}\right\|\left\|NT^{-1}\overline{\*W}^{*\prime}\overline{\+\varepsilon}^* \right\|=O_{P^*}(N^{-1/2}),
\end{align}
where the upcoming component produces asymptotic bias:
\begin{align}
    \mathbf{II}_{2,3,3,b}&=\frac{1}{\sqrt{NT}}\sum_{i=1}^N\*S_\*b'\mathcal{\+P}_{2}'T^{-1}\+\kappa_i'\widehat{\*F}(T^{-1}\widehat{\*F}^{*\prime}\widehat{\*F}^*)^{-1}\overline{\*W}^{*\prime}\overline{\+\varepsilon}^*\widetilde{\gamma}_{y,i,1}\notag\\
    &=\sqrt{\frac{T}{N}}\left(\frac{1}{N}\sum_{i=1}^N\widetilde{\gamma}_{y,i,1}\overline{\*C}'T^{-1}\*Q'\+\kappa_i\mathcal{\+P}_2\*S_{\*b} \right)'(T^{-1}\widehat{\*F}^{*\prime}\widehat{\*F}^*)^{-1}(NT^{-1}\overline{\*W}^{*\prime}\overline{\+\varepsilon}^*)\notag\\
    &+N^{-1/2}\sqrt{\frac{T}{N}}\left(\frac{1}{N}\sum_{i=1}^N\widetilde{\gamma}_{y,i,1}\sqrt{N}T^{-1}\overline{\*V}'\+\kappa_i\mathcal{\+P}_2\*S_{\*b} \right)'(T^{-1}\widehat{\*F}^{*\prime}\widehat{\*F}^*)^{-1}(NT^{-1}\overline{\*W}^{*\prime}\overline{\+\varepsilon}^*)\notag\\
    &= \sqrt{\frac{T}{N}}\left(\frac{1}{N}\sum_{i=1}^N\widetilde{\gamma}_{y,i,1}\overline{\*C}'T^{-1}\*Q'\+\kappa_i\mathcal{\+P}_2\*S_{\*b} \right)'(T^{-1}\widehat{\*F}^{*\prime}\widehat{\*F}^*)^{-1}\begin{bmatrix}
        NT^{-1}\overline{\*u}^{*\prime}\overline{\+\varepsilon}^* & \*0_{1\times (2K-1)}    \end{bmatrix}'\notag\\
        &+O_{P^*}(C_{N,T}^{-1}),
\end{align}
by the same argument as in (\ref{II_232b}). Next, 
\begin{align}
    \left\| \mathbf{II}_{2,3,3,c}\right\|&=\left\|\frac{1}{\sqrt{NT}}\sum_{i=1}^N\*S_\*b'\mathcal{\+P}_{2}'T^{-1}\+\kappa_i'\overline{\*W}^*(T^{-1}\widehat{\*F}^{*\prime}\widehat{\*F}^*)^{-1}\widehat{\*F}'\overline{\+\varepsilon}^*\widetilde{\gamma}_{y,i,1} \right\|\notag\\
    &\leq N^{-1/2}\left\| \*S_\*b \right\|\left\| \mathcal{\+P}_{2}\right\|\left(\frac{1}{N}\sum_{i=1}^N |\widetilde{\gamma}_{y,i,1}|^2\right)^{1/2}\left(\frac{1}{N}\sum_{i=1}^N\left\|\sqrt{N}T^{-1}\+\kappa_i'\overline{\*W}^* \right\|^2 \right)^{1/2}\notag\\
    &\times \left\| (T^{-1}\widehat{\*F}^{*\prime}\widehat{\*F}^*)^{-1}\right\|\left\|\sqrt{N}T^{-1/2}\widehat{\*F}' \overline{\+\varepsilon}^* \right\|=O_{P^*}(N^{-1/2}),
\end{align}
and 
\begin{align}
    \left\|  \mathbf{II}_{2,3,3,d}\right\|&=\left\|\frac{1}{\sqrt{NT}}\sum_{i=1}^N\*S_\*b'\mathcal{\+P}_{2}'T^{-1}\+\kappa_i'\widehat{\*F}((T^{-1}\widehat{\*F}^{*\prime}\widehat{\*F}^*)^{-1}-(T^{-1}\widehat{\*F}'\widehat{\*F})^{-1})\widehat{\*F}'\overline{\+\varepsilon}^*\widetilde{\gamma}_{y,i,1} \right\|\notag\\
    &\leq \left\| \*S_\*b \right\|\left\| \mathcal{\+P}_{2}\right\|\left(\frac{1}{N}\sum_{i=1}^N |\widetilde{\gamma}_{y,i,1}|^2\right)^{1/2}\left(\frac{1}{N}\sum_{i=1}^N\left\|T^{-1}\+\kappa_i'\widehat{\*F}\right\|^2 \right)^{1/2}\left\|\sqrt{N}T^{-1/2}\widehat{\*F}' \overline{\+\varepsilon}^* \right\|\notag\\
    &\times  \left\| (T^{-1}\widehat{\*F}^{*\prime}\widehat{\*F}^*)^{-1}-(T^{-1}\widehat{\*F}'\widehat{\*F})^{-1}\right\|= O_{P^*}(N^{-1})+O_{P^*}((NT)^{-1/2}),
\end{align}
and so in total we have that 
\begin{align}
    \mathbf{III}_{2,3,3}= \sqrt{\frac{T}{N}}\left(\frac{1}{N}\sum_{i=1}^N\widetilde{\gamma}_{y,i,1}\overline{\*C}'T^{-1}\*Q'\+\kappa_i\mathcal{\+P}_2\*S_\*b \right)'(T^{-1}\widehat{\*F}^{*\prime}\widehat{\*F}^*)^{-1}\begin{bmatrix}
        NT^{-1}\overline{\*u}^{*\prime}\overline{\+\varepsilon}^* & \*0_{1\times (2K-1)}    \end{bmatrix}'+O_{P^*}(C_{N,T}^{-1}),
\end{align}
which disappears if $\+\theta_0=\*0_{k\times1}$. The term $\mathbf{II}_{2,3,4}=\mathbf{II}_{2,3,4,a}+\mathbf{II}_{2,3,4,b}+\mathbf{II}_{2,3,4,c}+\mathbf{II}_{2,3,4,d}$ exhibits the same asymptotic behavior as $\mathbf{II}_{2,3,2}$, however, instead of producing a bias term, it will be negligible due to $\left\| \mathcal{\+P}_{0,3,i}\right\|=O_P(T^{-1/2})$. In fact, its leading term is $\mathbf{II}_{2,3,4,b}$, which implies that 
\begin{align}
    \left\|\mathbf{II}_{2,3,4}\right\|&= \left\|\frac{1}{\sqrt{NT}}\sum_{i=1}^N\*S_\*b'\mathcal{\+P}_{0,3,i}^{\prime}\*H'( \*P_{\widehat{\*F}^*}-\*P_{\widehat{\*F}})\overline{\+\varepsilon}^*\widetilde{\gamma}_{y,i,1}\right\|\notag\\
    &\leq \left\| \mathbf{II}_{2,3,4,b}\right\|+O_{P^*}(T^{-1/2}C_{N,T}^{-1})\notag\\
    &\leq N^{-1/2}\left\|\*S_\*b \right\|\left(\frac{1}{N}\sum_{i=1}^N|\widetilde{\gamma}_{y,i,1}|^2 \right)^{1/2}\left( \frac{1}{N}\sum_{i=1}^N\left\|\sqrt{T}\mathcal{\+P}_{3,i} \right\|^2\right)^{1/2}\left\| T^{-1}\*H'\widehat{\*F} \right\|\left\| (T^{-1}\widehat{\*F}^{*\prime}\widehat{\*F}^*)^{-1}\right\|\notag\\
    &\times \left\|NT^{-1}\overline{\*W}^{*\prime}\overline{\+\varepsilon}^* \right\| + O_{P^*}(T^{-1/2}C_{N,T}^{-1})=O_{P^*}(N^{-1/2}). 
\end{align}
The last term $\mathbf{II}_{2,3,5}=\mathbf{II}_{2,3,5,a}+\mathbf{II}_{2,3,5,b}+\mathbf{II}_{2,3,5,c}+\mathbf{II}_{2,3,5,d}$ is also negligible: 
\begin{align}
    \left\|\mathbf{II}_{2,3,5,a} \right\|&=\left\| \frac{1}{\sqrt{NT}}\sum_{i=1}^N\*S_\*b'T^{-1}\*U_i^{*\prime}\overline{\*W}^*(T^{-1}\widehat{\*F}^{*\prime}\widehat{\*F}^*)^{-1}\overline{\*W}^{*\prime}\overline{\+\varepsilon}^*\widetilde{\gamma}_{y,i,1}\right\|\notag\\
    &\leq N^{-1/2}\sqrt{\frac{T}{N}}\left(\frac{1}{N}\sum_{i=1}^N|\widetilde{\gamma}_{y,i,1}|^2 \right)^{1/2}\underbrace{\left(\frac{1}{N}\sum_{i=1}^N\left\|\sqrt{N}T^{-1}\*S_\*b'\*U_i^{*\prime}\overline{\*W}^* \right\|^2 \right)^{1/2}}_{O_{P^*}(C_{N,T}^{-1/2})}\left\|(T^{-1}\widehat{\*F}^{*\prime}\widehat{\*F}^*)^{-1} \right\|\notag\\
    &\times \left\|NT^{-1} \overline{\*W}^{*\prime}\overline{\+\varepsilon}^*\right\|=O_{P^*}( N^{-3/4})+O_{p^*} (N^{-1/2}T^{-1/4}),
\end{align}
where the rate is fastened by (\ref{u_tbaru_t}), which is the leading element in $\sqrt{N}T^{-1}\*S_\*b'\*U_i^{*\prime}\overline{\*W}^*$. Next, \begin{align}
    \left\| \mathbf{II}_{2,3,5,b}\right\|&=\left\| \frac{1}{\sqrt{NT}}\sum_{i=1}^N\*S_\*b'T^{-1}\*U_i^{*\prime}\widehat{\*F}(T^{-1}\widehat{\*F}^{*\prime}\widehat{\*F}^*)^{-1}\overline{\*W}^{*\prime}\overline{\+\varepsilon}^*\widetilde{\gamma}_{y,i,1}\right\|\notag\\
    &= \sqrt{\frac{T}{N}}\left( \frac{1}{N}\sum_{i=1}^N|\widetilde{\gamma}_{y,i,1}|^2\right)^{1/2}\left(\frac{1}{N}\sum_{i=1}^N\left\|T^{-1}\*S_\*b'\*U_i^{*\prime}\widehat{\*F}\ \right\|^2 \right)^{1/2}\left\|(T^{-1}\widehat{\*F}^{*\prime}\widehat{\*F}^*)^{-1} \right\|\notag\\
    &\times \left\|NT^{-1} \overline{\*W}^{*\prime}\overline{\+\varepsilon}^*\right\|=O_{P^*}(T^{-1/2})
\end{align}
under $TN^{-1}=O(1)$ following the rate in (\ref{U^*Q}), which is sufficient for our purposes. Then 
\begin{align}
    \left\| \mathbf{II}_{2,3,5,c}\right\|&=\left\| \frac{1}{\sqrt{NT}}\sum_{i=1}^N\*S_\*b'T^{-1}\*U_i^{*\prime}\overline{\*W}^*(T^{-1}\widehat{\*F}^{*\prime}\widehat{\*F}^*)^{-1}\widehat{\*F}'\overline{\+\varepsilon}^*\widetilde{\gamma}_{y,i,1}\right\|\notag\\
    &\leq N^{-1/2}\left(\frac{1}{N}\sum_{i=1}^N|\widetilde{\gamma}_{y,i,1}|^2 \right)^{1/2}\left(\frac{1}{N}\sum_{i=1}^N\left\|\sqrt{N}T^{-1}\*S_\*b'\*U_i^{*\prime}\overline{\*W}^* \right\|^2 \right)^{1/2}\left\| (T^{-1}\widehat{\*F}^{*\prime}\widehat{\*F}^*)^{-1}\right\|\notag\\
    &\times \left\|\sqrt{N}T^{-1/2}\widehat{\*F}'\overline{\+\varepsilon}^* \right\|=O_{P^*}( N^{-3/4})+O_{p^*} (N^{-1/2}T^{-1/4}),
\end{align}
and, ultimately, 
\begin{align}
      \left\| \mathbf{II}_{2,3,5,d}\right\|&=\left\| \frac{1}{\sqrt{NT}}\sum_{i=1}^N\*S_\*b'T^{-1}\*U_i^{*\prime}\widehat{\*F}((T^{-1}\widehat{\*F}^{*\prime}\widehat{\*F}^*)^{-1}-(T^{-1}\widehat{\*F}'\widehat{\*F})^{-1})\widehat{\*F}'\overline{\+\varepsilon}^*\widetilde{\gamma}_{y,i,1}\right\|\notag\\
      &\leq \left(\frac{1}{N}\sum_{i=1}^N|\widetilde{\gamma}_{y,i,1}|^2 \right)^{1/2}\left(\frac{1}{N}\sum_{i=1}^N\left\|T^{-1}\*S_\*b'\*U_i^{*\prime}\widehat{\*F}\right\|^2 \right)^{1/2}\left\|\sqrt{N}T^{-1/2} \widehat{\*F}'\overline{\+\varepsilon}^*\right\|\notag\\
      &\times \left\| (T^{-1}\widehat{\*F}^{*\prime}\widehat{\*F}^*)^{-1}-(T^{-1}\widehat{\*F}'\widehat{\*F})^{-1}\right\|=O_{P^*}(T^{-1/2}N^{-1})+O_{P^*}(N^{-1/2}T^{-1}),
\end{align}
which overall gives that 
\begin{align}
    \left\|  \mathbf{II}_{2,3,5}\right\|=O_{P^*}(T^{-1/2}). 
\end{align}
In summary, we know that 
\begin{align}
    \mathbf{II}_{2,3}&= \sqrt{\frac{T}{N}}\left(\frac{1}{N}\sum_{i=1}^N\widetilde{\gamma}_{y,i,1}\mathcal{\+P}_{0,1,i}\*S_\*b \right)'T^{-1}\widehat{\*Q}_0^{\prime}\*Q\overline{\*C}(T^{-1}\widehat{\*F}^{*\prime}\widehat{\*F}^*)^{-1}\begin{bmatrix}
        NT^{-1}\overline{\*u}^{*\prime}\overline{\+\varepsilon}^* & \*0_{1\times (2K-1)}    \end{bmatrix}'\notag\\
    &+\sqrt{\frac{T}{N}}\left(\frac{1}{N}\sum_{i=1}^N\widetilde{\gamma}_{y,i,1}\overline{\*C}'T^{-1}\*Q'\+\kappa_i\mathcal{\+P}_2\*S_{\*b} \right)'(T^{-1}\widehat{\*F}^{*\prime}\widehat{\*F}^*)^{-1}\begin{bmatrix}
NT^{-1}\overline{\*u}^{*\prime}\overline{\+\varepsilon}^* & \*0_{1\times (2K-1)}    \end{bmatrix}'+o_{P^*}(1)\notag\\
&=\kappa^{-1/2}\*S_\*b'\+\Sigma_{\gamma_{y,1}\*C}'T^{-1}\*Q'_{\widehat{\alpha}}\*Q\*C(\*C'\+\Sigma_\*q\*C)^{-1}[\sigma^2, \*0_{1\times (2K-1)}]'\notag\\
&+\kappa^{-1/2}\*S_\*b'\mathcal{\+P}_2'\+\Sigma_{\gamma_{y,1}\*q\+\kappa}'\*C(\*C'\+\Sigma_\*q\*C)^{-1}[\sigma^2, \*0_{1\times (2K-1)}]'+o_{P^*}(1), 
\end{align}
where we substituted Lemma A2, and noticed that $\frac{1}{T}\sum_{t=2}^TN\overline{u}_t^{*}\overline{\varepsilon}_t^*=\frac{1}{T}\sum_{t=2}^TN\overline{\varepsilon}_t^{*2}+\frac{1}{T}\sum_{t=2}^TN\left(\sum_{j=1}^{t-1}\widehat{\alpha}_{CCEP}^j\overline{\varepsilon}_{t-j}^* \right)\overline{\varepsilon}_t^*=NT^{-1}\overline{\+\varepsilon}^{*\prime}\overline{\+\varepsilon}^*+o_{P^*}(1)=\sigma^2+o_{P^*}(1)$. \\

\noindent (b) We proceed to a new term of the numerator $\mathbf{III}$. In particular, 
\begin{align}
    \mathbf{III}=\frac{1}{\sqrt{NT}}\sum_{i=1}^N\*B_i^{*\prime}\*M_{\widehat{\*F}^*}\overline{\*u}_{-1}^*\widetilde{\gamma}_{y,i,2}&=\frac{1}{\sqrt{NT}}\sum_{i=1}^N\*B_i^{*\prime}\overline{\*u}_{-1}^*\widetilde{\gamma}_{y,i,2}-\frac{1}{\sqrt{NT}}\sum_{i=1}^N\*B_i^{*\prime}\*P_{\widehat{\*F}^*}\overline{\*u}_{-1}^*\widetilde{\gamma}_{y,i,2}\notag\\
    &=\mathbf{III}_1 - \mathbf{III}_2,
\end{align}
whose analysis will be similar to the one in $\mathbf{II}$, but with modifications because $\overline{\+\varepsilon}^*\widetilde{\gamma}_{y,i,1}$ is replaced by $\overline{\*u}^*_{-1}\widetilde{\gamma}_{y,i,2}$, and we will illustrate the differences. Note that 
\begin{align}
    \mathbf{III}_1&=\frac{1}{\sqrt{NT}}\sum_{i=1}^N\*S_\*b'[\widehat{\*A}, \*I_{K\times K}]'\widehat{\*A}_{\*z_0,i}'\overline{\*u}^*_{-1}\widetilde{\gamma}_{y,i,2}+\frac{1}{\sqrt{NT}}\sum_{i=1}^N\*S_\*b'\mathcal{\+P}_{0,1,i}^{\prime}\widehat{\*Q}_0^{\prime}\overline{\*u}^*_{-1}\widetilde{\gamma}_{y,i,2}\notag\\
    &+\frac{1}{\sqrt{NT}}\sum_{i=1}^N\*S_\*b'\mathcal{\+P}_{2}'\+\kappa_i'\overline{\*u}^*_{-1}\widetilde{\gamma}_{y,i,2}+\frac{1}{\sqrt{NT}}\sum_{i=1}^N\*S_\*b'\mathcal{\+P}_{0,3,i}^{\prime}\*H'\overline{\*u}^*_{-1}\widetilde{\gamma}_{y,i,2}\notag\\
&+\frac{1}{\sqrt{NT}}\sum_{i=1}^N\*S_\*b'\*U_i^{*\prime}\overline{\*u}^*_{-1}\widetilde{\gamma}_{y,i,2}\notag\\
&=\mathbf{III}_{1,a}+\mathbf{III}_{1,b}+\mathbf{III}_{1,c}+\mathbf{III}_{1,d}+\mathbf{III}_{1,e},
\end{align}
where the asymptotic behavior of $\mathbf{III}_{1,b}$, $\mathbf{III}_{1,c}$ and $\mathbf{III}_{1,d}$ is the same demonstrated by their counterparts in $\mathbf{II}_1$. Specifically, 
\begin{align}
    \mathbf{III}_{1,b}=\frac{1}{\sqrt{NT}}\sum_{i=1}^N\*S_\*b'\mathcal{\+P}_{0,1,i}^{\prime}\widehat{\*Q}_0^{\prime}\overline{\*u}^*_{-1}\widetilde{\gamma}_{y,i,2}=\left(\frac{1}{N}\sum_{i=1}^N\widetilde{\gamma}_{y,i,2}\mathcal{\+P}_{0,1,i}\*S_\*b\right)'\sqrt{N} T^{-1/2}\widehat{\*Q}_0^{\prime}\overline{\*u}^*_{-1}=O_{P^*}(1),
\end{align}
which is again similar to the distribution generating term $\*A^{\*F}_{NT,1}$ on p. 39 in the Supplement of \cite{DeVos2019}, but when $\overline{\*X}$ and $\overline{\*X}_{-1}$ are projected out. Also,  
\begin{align}
    \mathbf{III}_{1,c}=\*S_\*b'\mathcal{\+P}_2'\left(\frac{1}{N}\sum_{i=1}^N\widetilde{\gamma}_{y,i,2}\+\kappa_i \right)'\sqrt{N}T^{-1/2}\overline{\*u}^*_{-1}=O_{P^*}(1),
\end{align}
which vanishes if $\+\theta_0=\*0_{k\times 1}$, and 
\begin{align}
\left\|\mathbf{III}_{1,d}\right\|&\leq\left\|\left(\frac{1}{N}\sum_{i=1}^N\widetilde{\gamma}_{y,i,2}\mathcal{\+P}_{0,3,i}\*S_\*b\right)\right\|\left\|\sqrt{N} T^{-1/2}\*H'\overline{\*u}^*_{-1}\right\|\notag\\
    &= \frac{1}{\sqrt{T}}\left\|\left(\frac{1}{N}\sum_{i=1}^N\widetilde{\gamma}_{y,i,2}(\sqrt{T}\mathcal{\+P}_{0,3,i})\*S_\*b\right)\right\|\left\|\sqrt{N} T^{-1/2}\*H'\overline{\*u}^*_{-1}\right\|=O_{P^*}(T^{-1/2}),
    \end{align}
    as before, since the last component is bounded by an argument identical to the one in (\ref{d_m-d}). We elaborate on the two remaining terms. Following (\ref{II_1a_X}), we know that 
    \begin{align}
         \left\| \mathbf{III}_{1,a}\right\|&\leq \left\|\*S_\*b'[\widehat{\*A}, \*I_{K\times K}]' \right\|\left| \frac{1}{N}\sum_{i=1}^Ny_{i,0}\widetilde{\gamma}_{y,i,2} \right|\left|\frac{1}{\sqrt{T}}\sum_{t=2}^T\widehat{\alpha}_{CCEP}^{t-1}\sqrt{N}\overline{u}_{t-1}^* \right|\notag\\
    &\leq \left\|\*S_\*b'[\widehat{\*A}, \*I_{K\times K}]' \right\|\left( \frac{1}{N}\sum_{i=1}^Ny_{i,0}^2\right)^{1/2}\left(\frac{1}{N}\sum_{i=1}^N\widetilde{\gamma}_{y,i,2}^2\right)^{1/2}\left|\frac{1}{\sqrt{NT}}\sum_{i=1}^N\sum_{t=2}^T\widehat{\alpha}_{CCEP}^{t-1}u_{i,t-1}^* \right|\notag\\
    &=O_{P^*}(T^{-1/2}),
    \end{align}
    which comes from re-writing the last term for a fixed $m$ as 
    \begin{align}
        h_{m}=\sum_{j=0}^{m-2}\widehat{\alpha}_{CCEP}^j\frac{1}{\sqrt{NT}}\sum_{i=1}^N\sum_{t=j+2}^T\varepsilon^*_{i,t-j-1}\widehat{\alpha}_{CCEP}^{t-1}=\sum_{j=0}^{m-2}(\alpha_0)^j\underbrace{\frac{1}{\sqrt{NT}}\sum_{i=1}^N\sum_{t=j+2}^T\varepsilon^*_{i,t-j-1}\widehat{\alpha}_{CCEP}^{t-1}}_{h_{j,m}=O_{P^*}(T^{-1/2})}+o_{P^*}(1),
    \end{align}
    so we need to check whether the term remains negligible when $m\to \infty$, as in the case of all similar terms. Therefore, 
    \begin{align}
        E^*[|h_m-h|^2]&=\sum_{j=m}^{T-2}\sum_{r=m}^{T-2}\widehat{\alpha}_{CCEP}^{j+r}E^*\left[\frac{1}{NT}\sum_{i=1}^N\sum_{j=1}^N\sum_{t=j+2}^T\sum_{s=r+2}^T\varepsilon_{i,t-j-1}^*\varepsilon_{i,s-r-1}^*\widehat{\alpha}_{CCEP}^{t+s-2} \right]\notag\\
        &=\sum_{j=m}^{T-2}\sum_{r=m}^{T-2}\widehat{\alpha}_{CCEP}^{j+r}\frac{1}{N}\sum_{i=1}^N\frac{1}{T}\sum_{t=\max(j,r)+2}^T\widehat{\varepsilon}_{i,t-j-1}^2\widehat{\alpha}_{CCEP}^{t-1}\widehat{\alpha}_{CCEP}^{t+r-j-1}\notag\\
        &\leq \frac{1}{2}\sum_{j=m}^{T-2}\sum_{r=m}^{T-2}\widehat{\alpha}_{CCEP}^{j+r}\frac{1}{N}\sum_{i=1}^N\left(\frac{1}{T}\sum_{t=\max(j,r)+2}^T\widehat{\varepsilon}_{i,t-j-1}^4 \right)^{1/2}\left(\frac{1}{T}\sum_{t=\max(j,r)+2}^T\widehat{\alpha}_{CCEP}^{2(t-1)} \right)^{1/2}\notag\\
        &+\frac{1}{2}\sum_{j=m}^{T-2}\sum_{r=m}^{T-2}\widehat{\alpha}_{CCEP}^{j+r}\frac{1}{N}\sum_{i=1}^N\left(\frac{1}{T}\sum_{t=\max(j,r)+2}^T\widehat{\varepsilon}_{i,t-j-1}^4 \right)^{1/2}\left(\frac{1}{T}\sum_{t=\max(j,r)+2}^T\widehat{\alpha}_{CCEP}^{2(t+r-j-1)} \right)^{1/2}\notag\\
        &\leq \left(\sum_{j=m}^{T-2}|\widehat{\alpha}_{CCEP}|^{j}\right)^2\left(\frac{1}{NT}\sum_{i=1}^N\sum_{t=1}^T\widehat{\varepsilon}_{i,t}^4 \right)^{1/2}\left( \frac{1}{T}\sum_{t=1}^T|\widehat{\alpha}_{CCEP}|^{2t}\right)^{1/2}=o_P(T^{-1/2}),
    \end{align}
    as expected. Lastly, for $\mathbf{III}_{1,e}$, we can use the same definition of $\*S_\*b'\*u_{i,t}^*$ in (\ref{S_bU_def}), and obtain 
    \begin{align}
        \mathbf{III}_{1,e}=\begin{bmatrix}
       iii\\
       \*0_{k\times 1 }\end{bmatrix}+O_{P^*}(C_{N,T}^{-1}),
    \end{align}
    where only one component survives. In particular, $|i|$ and $\left\| \mathbf{iv}\right\|$ are both $O_{P^*}(N^{-1/2})$ because the difference in terms is the presence of $\overline{\*u}_{-1}^*$. Hence,  using the analysis in $(\ref{d_m-d})$, we know that $\left\|\frac{1}{\sqrt{T}}\sum_{t=2}^TN\overline{\*e}_{t-1}^-\overline{u}_{t-1}^*\right\|=O_{P^*}(1)$ and $\left\|\frac{1}{\sqrt{T}}\sum_{t=2}^TN\overline{\*e}_{t}\overline{u}_{t-1}^* \right\|=O_{P^*}(1)$, which is sufficient for the order of the upper bounds of both terms. The remaining 3 terms require more work. For the negligible ones, we focus on $ii$, because it is more complex and the simpler $\*v$ will follow. Its key part is $\*g=\frac{1}{\sqrt{NT}}\sum_{i=1}^N\sum_{t=2}^T\+\nu_{i,t-1}^-\overline{u}_{t-1}^*\widetilde{\gamma}_{y,i,2}$ and for a finite $m$ it becomes 
    \begin{align}
    \*g_m=\sum_{j=0}^{m-2}\widehat{\alpha}_{CCEP}^j\underbrace{\frac{1}{N\sqrt{T}}\sum_{i=1}^N\sum_{t=j+2}^T\widetilde{\gamma}_{y,i,2}\sqrt{N}\overline{\varepsilon}_{t-j-1}^*\+\nu_{i,t-1}^-}_{\*g_{j,m}=O_{P^*}(N^{-1/2})},     
    \end{align}
    because 
    \begin{align}
        E^*[\left\|\*g_{j,m} \right\|^2]&=\frac{1}{N^2}\sum_{i=1}^N\sum_{l=1}^N\widetilde{\gamma}_{y,i,2}\widetilde{\gamma}_{y,l,2}\frac{1}{T}\sum_{t=j+2}^T\left(\frac{1}{N}\sum_{k=1}^N\widehat{\varepsilon}^2_{k,t} \right)\+\nu_{i,t-1}^{-\prime}\+\nu_{l,t-1}^{-}\notag\\
        &=\sigma^2\frac{1}{N^2}\sum_{i=1}^N\sum_{l=1}^N\gamma_{y,i,2}\gamma_{y,l,2}\frac{1}{T}\sum_{t=j+2}^T\+\nu_{i,t-1}^{-\prime}\+\nu_{l,t-1}^{-}+O_P(C_{N,T}^{-1})\notag\\
        &=\sigma^2\frac{1}{N^2}\sum_{i=1}^N\sum_{l=1}^N\gamma_{y,i,2}\gamma_{y,l,2}\frac{1}{T}\sum_{t=j+2}^T\+\nu_{i,t-1}(\alpha_0)^{-\prime}\+\nu_{l,t-1}(\alpha_0)^{-}+O_P(C_{N,T}^{-1})\notag\\
        &=O_P(C_{N,T}^{-1})
    \end{align}
     due to cross-section independence and $|\widetilde{\gamma}_{y,i,2}-\gamma_{y,i,2}|=O_P(T^{-1/2})$, and so $\*g_m$ have the same rate for a fixed $m$. Thus, $\*g$ is negligible due to 
    \begin{align}
        E^*[(\+\lambda'&(\*g_m-\*g))^2]\sum_{j=m}^{T-2}\sum_{r=m}^{T-2}\widehat{\alpha}_{CCEP}^{j+r}\frac{1}{N^2}\sum_{i=1}^N\sum_{l=1}^N\widetilde{\gamma}_{y,i,2}\widetilde{\gamma}_{y,l,2}\frac{1}{T}\sum_{t=\max(j,r)+2}^TE^*[N\overline{\varepsilon}_{t-j-1}^{*2}]\+\nu_{i,t-1}^{-\prime}\+\nu_{l,t+r-j-1}^-\notag\\
        &\leq\sum_{j=m}^{T-2}\sum_{r=m}^{T-2}|\widehat{\alpha}_{CCEP}|^{j+r} \frac{1}{2}\frac{1}{N^2}\sum_{i=1}^N\sum_{l=1}^N|\widetilde{\gamma}_{y,i,2}||\widetilde{\gamma}_{y,l,2}|\frac{1}{T}\sum_{t=\max(j,r)+2}^TE^*[N\overline{\varepsilon}_{t-j-1}^{*2}]\left(\left\|\+\nu^-_{i,t-1}\right\|^2+ \left\|\+\nu_{l,t+r-j-1}^-\right\|^2\right)\notag\\
        &\leq \frac{1}{2}\sum_{j=m}^{T-2}\sum_{r=m}^{T-2}|\widehat{\alpha}_{CCEP}|^{j+r} \left(\frac{1}{T}\sum_{t=1}^TE^*[N\overline{\varepsilon}_{t}^{*2}]^2 \right)^{1/2}\frac{1}{N^2}\sum_{i=1}^N\sum_{l=1}^N|\widetilde{\gamma}_{y,i,2}||\widetilde{\gamma}_{y,l,2}|\left(\frac{1}{T} \sum_{t=1}^T\left\|\+\nu_{i,t}^{-} \right\|^4\right)^{1/2}\notag\\
        &+\frac{1}{2}\sum_{j=m}^{T-2}\sum_{r=m}^{T-2}|\widehat{\alpha}_{CCEP}|^{j+r} \left(\frac{1}{T}\sum_{t=1}^TE^*[N\overline{\varepsilon}_{t}^{*2}]^2 \right)^{1/2}\frac{1}{N^2}\sum_{i=1}^N\sum_{l=1}^N|\widetilde{\gamma}_{y,i,2}||\widetilde{\gamma}_{y,l,2}|\left(\frac{1}{T} \sum_{t=1}^T\left\|\+\nu_{l,t}^{-} \right\|^4\right)^{1/2}\notag\\
        &= \left(\sum_{j=m}^{T-2}|\widehat{\alpha}_{CCEP}|^j \right)^2 \left(\frac{1}{T}\sum_{t=1}^T\left( \frac{1}{N}\sum_{k=1}^N\widehat{\varepsilon}_{k,t}^2\right)^2  \right)^{1/2}\frac{1}{N^2}\sum_{i=1}^N\sum_{l=1}^N|\widetilde{\gamma}_{y,i,2}||\widetilde{\gamma}_{y,l,2}|\left(\frac{1}{T} \sum_{t=1}^T\left\|\+\nu_{i,t}^{-} \right\|^4\right)^{1/2}\notag\\
        &\leq \left(\sum_{j=m}^{T-2}|\widehat{\alpha}_{CCEP}|^j \right)^2 \left(\frac{1}{NT}\sum_{k=1}^N\sum_{t=1}^T\widehat{\varepsilon}^4_{k,t}  \right)^{1/2}\frac{1}{N^2}\sum_{i=1}^N\sum_{l=1}^N|\widetilde{\gamma}_{y,i,2}||\widetilde{\gamma}_{y,l,2}|\left(\frac{1}{T} \sum_{t=1}^T\left\|\+\nu_{i,t}^{-} \right\|^4\right)^{1/2}\notag\\
        &=o_{P^*}(1)
    \end{align}
    as intended, where $\frac{1}{T} \sum_{t=1}^T\left|\+\nu_{i,t}^{-\prime} \+\nu_{j,t}^-\right|^2=O_P(1)$ for each $i,j$ by an argument akin to the one in (\ref{Ne'e_X}). The component $\*v$ behaves asymptotically in a similar way. Lastly, 
    \begin{align}
        iii&=\frac{1}{\sqrt{NT}}\sum_{i=1}^N\sum_{t=2}^Tu_{i,t-1}^*\overline{u}_{t-1}^*\widetilde{\gamma}_{y,i,2}=\underbrace{\frac{1}{\sqrt{NT}}\frac{1}{N}\sum_{i=1}^N\widetilde{\gamma}_{y,i,2}\sum_{t=2}^Tu_{i,t-1}^{*2}}_{iii_1} + \underbrace{\frac{1}{\sqrt{NT}}\frac{1}{N}\sum_{i=1}^N\widetilde{\gamma}_{y,i,2}\sum_{j\neq i}^N\sum_{t=2}^Tu_{i,t-1}^*u_{j,t-1}^*}_{iii_2}\notag\\
        &=\sqrt{\frac{T}{N}}\frac{1}{N}\sum_{i=1}^N\widetilde{\gamma}_{y,i,2}\frac{1}{T}\sum_{t=2}^Tu_{i,t-1}^{*2} + o_{p^{*}}(1),
    \end{align}
    because 
    \begin{align}
        E^*[|iii_2|^2]&= \frac{T}{N}\frac{1}{N^2}\sum_{i=1}^N\widetilde{\gamma}_{y,i,2}^2\sum_{j\neq i}^N \frac{1}{T^2}\sum_{t=2}^T\sum_{s=2}^TE^*[u_{i,t}^*u_{i,s}^*]E^*[u_{j,t}^*u_{j,s}^*]\notag\\
        &+\frac{T}{N}\frac{1}{N^2}\sum_{i=1}^N\sum_{j\neq i}^N\widetilde{\gamma}_{y,i,2}\widetilde{\gamma}_{y,j,2} \frac{1}{T^2}\sum_{t=2}^T\sum_{s=2}^TE^*[u_{i,t}^*u_{i,s}^*]E^*[u_{j,t}^*u_{j,s}^*]=O_P(C_{N,T}^{-1}),
    \end{align}
   under $TN^{-1}=O(1)$, which follows from the exact same analysis as in (\ref{E*(b0^2)}).  In summary, 
    \begin{align}
        \mathbf{III}_{1}&=\sqrt{\frac{T}{N}}\begin{bmatrix}
            \frac{1}{N}\sum_{i=1}^N\widetilde{\gamma}_{y,i,2}\frac{1}{T}\sum_{t=2}^Tu_{i,t-1}^{*2} & \*0_{1\times k}
        \end{bmatrix}'+\left(\frac{1}{N}\sum_{i=1}^N\widetilde{\gamma}_{y,i,2}\mathcal{\+P}_{0,1,i}\*S_\*b\right)'\sqrt{N} T^{-1/2}\widehat{\*Q}_0^{\prime}\overline{\*u}^*_{-1}\notag\\
        &+\*S_\*b'\mathcal{\+P}_2'\left(\frac{1}{N}\sum_{i=1}^N\widetilde{\gamma}_{y,i,2}\+\kappa_i \right)'\sqrt{N}T^{-1/2}\overline{\*u}^*_{-1} + o_{P^*}(1)\notag\\
        &=\kappa^{-1/2}\frac{1}{1-(\alpha_0)^2}\left[\gamma_{y,2}, \*0_{1\times k}\right]'+\*S_\*b'\+\Sigma_{\+\gamma_{y,2}\*C}'\sqrt{N} T^{-1/2}\*Q_{\widehat{\alpha}}^{\prime}\overline{\*u}^*_{-1} \notag\\
        &+ \*S_\*b'\mathcal{\+P}_2'\left(\frac{1}{N}\sum_{i=1}^N\widetilde{\gamma}_{y,i,2}\+\kappa_i \right)'\sqrt{N}T^{-1/2}\overline{\*u}^*_{-1} + o_{P^*}(1),
    \end{align}
    where we let $\gamma_{y,2}=\plim_{(N,T)\to \infty}\frac{1}{N}\sum_{i=1}^N\widetilde{\gamma}_{y,i,2}$ and $\+\Sigma_{\+\gamma_{y,2}\*C}=\plim_{(N,T)\to \infty}\frac{1}{N}\sum_{i=1}^N\widetilde{\gamma}_{y,i,2}\*C_i$. Further, we have 
        \begin{align}
  \mathbf{III}_2&=  \frac{1}{\sqrt{NT}}\sum_{i=1}^N\*B_i^{*\prime}\*P_{\widehat{\*F}^*}\overline{\*u}_{-1}^*\widetilde{\gamma}_{y,i,2}\notag\\
  &= \frac{1}{\sqrt{NT}}\sum_{i=1}^N\*B_i^{*\prime}\*P_{\*Q\overline{\*C}}\overline{\*u}_{-1}^*\widetilde{\gamma}_{y,i,2}+\frac{1}{\sqrt{NT}}\sum_{i=1}^N\*B_i^{*\prime}(\*P_{\widehat{\*F}}-\*P_{\*Q\overline{\*C}})\overline{\*u}_{-1}^*\widetilde{\gamma}_{y,i,2}+\frac{1}{\sqrt{NT}}\sum_{i=1}^N\*B_i^{*\prime}(\*P_{\widehat{\*F}^*}-\*P_{\widehat{\*F}})\overline{\*u}_{-1}^*\widetilde{\gamma}_{y,i,2}\notag\\
  &=\mathbf{III}_{2,1}+\mathbf{III}_{2,2}+\mathbf{III}_{2,3}.
\end{align}
The crucial point here is that all 3 components follow exactly the structure as the ones in $\mathbf{II}_2$. Because we swap $\overline{\+\varepsilon}^*\widetilde{\gamma}_{y,i,1}$ for $\overline{\*u}^*_{-1}\widetilde{\gamma}_{y,i,2}$, we need $\left\|\sqrt{N}T^{-1/2}\widehat{\*F}'\overline{\*u}^*_{-1} \right\|=O_{P^*}(1)$ instead of $\left\|\sqrt{N}T^{-1/2}\widehat{\*F}'\overline{\+\varepsilon}^* \right\|=O_{P^*}(1)$. We demonstrated this in (\ref{FW*}). Therefore, 
\begin{align}
    \mathbf{III}_{2,1}=&\left(\frac{1}{N}\sum_{i=1}^N\widetilde{\gamma}_{y,i,2}\mathcal{\+P}_{0,1,i}\*S_\*b\right)'T^{-1}\widehat{\*Q}_0^{\prime}\*Q\overline{\*C}\left( T^{-1}\overline{\*C}'\*Q'\*Q\overline{\*C}\right)^{-1}\sqrt{N}T^{-1/2}\overline{\*C}'\*Q'\overline{\*u}^*_{-1}\notag\\
    &+\+S_\*b'\mathcal{\+P}_2'\left(\frac{1}{N}\sum_{i=1}^N\widetilde{\gamma}_{y,i,2}T^{-1}\overline{\*C}'\*Q'\+\kappa_i \right)'\left( T^{-1}\overline{\*C}'\*Q'\*Q\overline{\*C}\right)^{-1}\sqrt{N}T^{-1/2}\overline{\*C}'\*Q'\overline{\*u}^*_{-1} + o_{P^*}(1)\notag\\
    &=\*S_\*b'\+\Sigma_{\gamma_{y,2}\*C}'T^{-1}\*Q'_{\widehat{\alpha}}\*Q\*C(\*C'\+\Sigma_\*q\*C)^{-1} \sqrt{N}T^{-1/2}\*C'\*Q'\overline{\*u}^*_{-1}\notag\\
    &+\*S_\*b'\mathcal{\+P}_2'\+\Sigma_{\gamma_{y,2}\*q\+\kappa}'\*C(\*C'\+\Sigma_\*q\*C)^{-1} \sqrt{N}T^{-1/2}\*C'\*Q'\overline{\*u}^*_{-1}+o_{P^*}(1),
\end{align}
where we again substituted Lemma XB and defined $\+\Sigma_{\gamma_{y,2}\*C}$ and $\+\Sigma_{\gamma_{y,2}\*q\+\kappa}$ analogously, but $\widetilde{\gamma}_{y,i,1}$ is substituted for $\widetilde{\gamma}_{y,i,2}$. Next, 
\begin{align}
    \left\|\mathbf{III}_{2,2} \right\|=O_{P^*}(N^{-1/2}),
\end{align}
and 
\begin{align}
    &\mathbf{III}_{2,3}=\sqrt{\frac{T}{N}}\left(\frac{1}{N}\sum_{i=1}^N\widetilde{\gamma}_{y,i,2}\mathcal{\+P}_{0,1,i}\*S_\*b \right)'T^{-1}\widehat{\*Q}_0^{\prime}\*Q\overline{\*C}(T^{-1}\widehat{\*F}^{*\prime}\widehat{\*F}^*)^{-1}\begin{bmatrix}
        NT^{-1}\overline{\*u}^{*\prime}\overline{\*u}^*_{-1} & \*0_{1\times k}& NT^{-1}\overline{\*u}_{-1}^{*\prime}\overline{\*u}^*_{-1}&\*0_{1\times k}    \end{bmatrix}'\notag\\
    &+\sqrt{\frac{T}{N}}\left(\frac{1}{N}\sum_{i=1}^N\widetilde{\gamma}_{y,i,2}\overline{\*C}'T^{-1}\*Q'\+\kappa_i\mathcal{\+P}_2\*S_{\*b} \right)'(T^{-1}\widehat{\*F}^{*\prime}\widehat{\*F}^*)^{-1}\begin{bmatrix}
        NT^{-1}\overline{\*u}^{*\prime}\overline{\*u}^*_{-1} & \*0_{1\times k}& NT^{-1}\overline{\*u}_{-1}^{*\prime}\overline{\*u}^*_{-1}&\*0_{1\times k}    \end{bmatrix}'\notag\\
        &+o_{P^*}(1)\notag\\
        &=\sqrt{\frac{T}{N}}\left(\frac{1}{N}\sum_{i=1}^N\widetilde{\gamma}_{y,i,2}\mathcal{\+P}_{0,1,i}\*S_\*b \right)'T^{-1}\widehat{\*Q}_0^{\prime}\*Q\overline{\*C}\left( T^{-1}\overline{\*C}'\*Q'\*Q\overline{\*C}\right)^{-1}\begin{bmatrix}
        NT^{-1}\overline{\*u}^{*\prime}\overline{\*u}^*_{-1} & \*0_{1\times k}& NT^{-1}\overline{\*u}_{-1}^{*\prime}\overline{\*u}^*_{-1}&\*0_{1\times k}    \end{bmatrix}'\notag\\
    &+\sqrt{\frac{T}{N}}\left(\frac{1}{N}\sum_{i=1}^N\widetilde{\gamma}_{y,i,2}\overline{\*C}'T^{-1}\*Q'\+\kappa_i\mathcal{\+P}_2\*S_{\*b} \right)'\left( T^{-1}\overline{\*C}'\*Q'\*Q\overline{\*C}\right)^{-1}\begin{bmatrix}
        NT^{-1}\overline{\*u}^{*\prime}\overline{\*u}^*_{-1} & \*0_{1\times k}& NT^{-1}\overline{\*u}_{-1}^{*\prime}\overline{\*u}^*_{-1}&\*0_{1\times k}    \end{bmatrix}'\notag\\
       & +o_{P^*}(1)\notag\\
        &=\kappa^{-1/2}\frac{\sigma^2}{1-(\alpha_0)^2}\*S_\*b'\+\Sigma_{\gamma_{y,2}\*C}'T^{-1}\*Q'_{\widehat{\alpha}}\*Q\*C(\*C'\+\Sigma_\*q\*C)^{-1}[
        \alpha_0 , \*0_{1\times k}, 1, \*0_{1\times k}   ]'\notag\\
        &+ \kappa^{-1/2}\frac{\sigma^2}{1-(\alpha_0)^2}\*S_\*b'\mathcal{\+P}_2'\+\Sigma_{\gamma_{y,2}\*q\+\kappa}'\*C(\*C'\+\Sigma_\*q\*C)^{-1} [
        \alpha_0 , \*0_{1\times k}, 1, \*0_{1\times k}   ]' + o_{P^*}(1),
\end{align}
because $\left\|(T^{-1}\widehat{\*F}^{*\prime}\widehat{\*F}^*)^{-1}- \left( T^{-1}\overline{\*C}'\*Q'\*Q\overline{\*C}\right)^{-1}\right\|=o_{P^*}(1)$ and $E[u_{i,t}u_{i,t-1}]=\frac{\sigma^2\alpha_0}{1-(\alpha_0)^2}$. Note that under $\+\theta_0=\*0_{k\times 1}$, both $\left\| \mathbf{III}_{2,1}\right\|=O_{P^*}(T^{-1/2})$ and $\left| \mathbf{III}_{2,3}\right\|=O_{P^*}(T^{-1/2})$ since $|\widetilde{\gamma}_{y,i,2}|=O_P(T^{-1/2})$.\\

\noindent (c) We now turn to the final term: 
\begin{align}
 \mathbf{I}=\frac{1}{\sqrt{NT}}\sum_{i=1}^N\*B_i^{*\prime}\*M_{\widehat{\*F}^*}\+\varepsilon_i^*&=\frac{1}{\sqrt{NT}}\sum_{i=1}^N\*B_i^{*\prime}\+\varepsilon_i^*-\frac{1}{\sqrt{NT}}\sum_{i=1}^N\*B_i^{*\prime}\*P_{\widehat{\*F}^*}\+\varepsilon_i^*=\mathbf{I}_1 - \mathbf{I}_2,
    \end{align}
    where 
    \begin{align}
          \mathbf{I}_1&=\frac{1}{\sqrt{NT}}\sum_{i=1}^N\*S_\*b'[\widehat{\*A}, \*I_{K\times K}]'\widehat{\*A}_{\*z_0,i}'\+\varepsilon_i^*+\frac{1}{\sqrt{NT}}\sum_{i=1}^N\*S_\*b'\mathcal{\+P}_{0,1,i}^{\prime}\widehat{\*Q}_0^{\prime}\+\varepsilon_i^*\notag\\
    &+\frac{1}{\sqrt{NT}}\sum_{i=1}^N\*S_\*b'\mathcal{\+P}_{2}'\+\kappa_i'\+\varepsilon_i^*+\frac{1}{\sqrt{NT}}\sum_{i=1}^N\*S_\*b'\mathcal{\+P}_{0,3,i}^{\prime}\*H'\+\varepsilon_i^*\notag\\
&+\frac{1}{\sqrt{NT}}\sum_{i=1}^N\*S_\*b'\*U_i^{*\prime}\+\varepsilon_i^*\notag\\
&=\mathbf{I}_{1,a}+\mathbf{I}_{1,b}+\mathbf{I}_{1,c}+\mathbf{I}_{1,d}+\mathbf{I}_{1,e},
    \end{align}
    and all the terms, except $\*I_{1,a}$ (vanishing initial value effect) and $\*I_{1,d}$, will contribute to the asymptotic distribution. Indeed, 
    \begin{align}
        \left\|\*I_{1,a} \right\| \leq \left\|\*S_\*b'[\widehat{\*A}, \*I_{K\times K}]' \right\|\left| \frac{1}{\sqrt{N}}\sum_{i=1}^Ny_{i,0} \frac{1}{\sqrt{T}}\sum_{t=2}^T\widehat{\alpha}_{CCEP}^{t-1}\varepsilon_{i,t}^* \right|=O_{P^*}(T^{-\delta/2(4+\delta)}),
    \end{align}
    which is the order of the second component in the product that originates directly from (\ref{alpha_epsilon_rate}) under $NT^{-1}\to \kappa>0$, and 
    \begin{align}
        \left\|\*I_{1,d} \right\|\leq \frac{1}{\sqrt{T}}\left\| \frac{1}{\sqrt{NT}}\sum_{i=1}^N\*S_\*b'(\sqrt{T}\mathcal{\+P}_{0,3,i}^{\prime})\*H'\+\varepsilon_i^*\right\|=O_{P^*}(T^{-1/2})
    \end{align}
    Therefore, 
    \begin{align}
        \*I_1&=\mathbf{I}_{1,b}+\mathbf{I}_{1,c}+\mathbf{I}_{1,e}+O_{P^*}(T^{-\delta/2(4+\delta)})+O_{P^*}(T^{-1/2})\notag\\
        &=\frac{1}{\sqrt{NT}}\sum_{i=1}^N\*S_\*b'\mathcal{\+P}_{0,1,i}^{\prime}\widehat{\*Q}_0^{\prime}\+\varepsilon_i^*+\frac{1}{\sqrt{NT}}\sum_{i=1}^N\*S_\*b'\mathcal{\+P}_{2}'\+\kappa_i'\+\varepsilon_i^*+\frac{1}{\sqrt{NT}}\sum_{i=1}^N\*S_\*b'\*U_i^{*\prime}\+\varepsilon_i^*+O_{P^*}(T^{-\delta/2(4+\delta)})+O_{P^*}(T^{-1/2})\notag\\
        &=\frac{1}{\sqrt{NT}}\sum_{i=1}^N\*S_\*b'\*C_i'\*Q_{\widehat{\alpha}}^{\prime}\+\varepsilon_i^* + \frac{1}{\sqrt{NT}}\sum_{i=1}^N\*S_\*b'\mathcal{\+P}_{2}'\+\kappa_i'\+\varepsilon_i^*+\frac{1}{\sqrt{NT}}\sum_{i=1}^N\mathcal{\+V}_i^{*\prime}\+\varepsilon_i^*+o_{P^*}(1)
    \end{align}
   where we, again, safely substituted Lemma XB as $\varepsilon_{i,t}^*$ is independent of other components. Also, we notice that $\mathcal{\+V}_i^*=[\mathcal{\+v}_{i,2}^*, \ldots, \mathcal{\+v}_{i,T}^*]'$, where  $\*S_\*b'\*u_{i,t}^*=\mathcal{\+v}_{i,t}^*+o_P(1)=[\widehat{\+\beta}_{CCEP}'\+\nu_{i,t-1}^{-}+u_{i,t-1}^*, \+\nu_{i,t}']'+o_P(1)$, which produces the bootstrap equivalent of one of the normality generators in \cite{DeVos2019}. Note that non-multiplicative weights are again essential for $\*I_{1,e}$ to generate normality. We continue with 
    \begin{align}
        \mathbf{I}_2&=  \frac{1}{\sqrt{NT}}\sum_{i=1}^N\*B_i^{*\prime}\*P_{\widehat{\*F}^*}\+\varepsilon_{i}^*\notag\\
  &= \frac{1}{\sqrt{NT}}\sum_{i=1}^N\*B_i^{*\prime}\*P_{\*Q\overline{\*C}}\+\varepsilon_{i}^*+\frac{1}{\sqrt{NT}}\sum_{i=1}^N\*B_i^{*\prime}(\*P_{\widehat{\*F}}-\*P_{\*Q\overline{\*C}})\+\varepsilon_{i}^*+\frac{1}{\sqrt{NT}}\sum_{i=1}^N\*B_i^{*\prime}(\*P_{\widehat{\*F}^*}-\*P_{\widehat{\*F}})\+\varepsilon_{i}^*\notag\\
  &=\mathbf{I}_{2,1}+\mathbf{I}_{2,2}+\mathbf{I}_{2,3},
    \end{align}
    where we get 
    \begin{align}
        \mathbf{I}_{2,1}&=\frac{1}{\sqrt{NT}}\sum_{i=1}^N\*S_\*b'[\widehat{\*A}, \*I_{K\times K}]'\widehat{\*A}_{\*z_0,i}'\*P_{\*Q\overline{\*C}}\+\varepsilon_i^*+\frac{1}{\sqrt{NT}}\sum_{i=1}^N\*S_\*b'\mathcal{\+P}_{0,1,i}^{\prime}\widehat{\*Q}_0^{\prime}\*P_{\*Q\overline{\*C}}\+\varepsilon_i^*\notag\\
    &+\frac{1}{\sqrt{NT}}\sum_{i=1}^N\*S_\*b'\mathcal{\+P}_{2}'\+\kappa_i'\*P_{\*Q\overline{\*C}}\+\varepsilon_i^*+\frac{1}{\sqrt{NT}}\sum_{i=1}^N\*S_\*b'\mathcal{\+P}_{0,3,i}^{\prime}\*H'\*P_{\*Q\overline{\*C}}\+\varepsilon_i^*\notag\\
    &+\frac{1}{\sqrt{NT}}\sum_{i=1}^N\*S_\*b'\*U_i^{*\prime}\*P_{\*Q\overline{\*C}}\+\varepsilon_i^*\notag\\
    &=\mathbf{I}_{2,1,a}+\mathbf{I}_{2,1,b}+\mathbf{I}_{2,1,c}+\mathbf{I}_{2,1,d}+\mathbf{I}_{2,1,e}.
    \end{align}
    Here, the effect of the initial value vanishes and so does $\*I_{2,1,d}$, while the rest of the terms contribute either to the asymptotic distribution or to the ``Nickell bias''. In particular, by using $\mathrm{vec}(\*A\*B\*C)=(\*C'\otimes \*A)\mathrm{vec}(\*B)$
    \begin{align}
        \left\|\mathbf{I}_{2,1,a} \right\|&\leq \left\| \*S_\*b'[\widehat{\*A}, \*I_{K\times K}]'\right\| \left\|\frac{1}{\sqrt{NT}}\sum_{i=1}^N\widehat{\*A}_{\*z_0,i}'\*P_{\*Q\overline{\*C}}\+\varepsilon_i^* \right\|\notag\\
        &\leq \left\| \*S_\*b'[\widehat{\*A}, \*I_{K\times K}]'\right\| \underbrace{\left\|\frac{1}{\sqrt{N}}\sum_{i=1}^N \left( T^{-1}\+\varepsilon_i^{*\prime}\*Q\overline{\*C}\otimes T^{-1/2}\widehat{\*A}_{\*z_0,i}'\*Q\overline{\*C} \right)\right\|}_{a=O_{P^*}(T^{-1/2})}\underbrace{\left\|(T^{-1}\overline{\*C}'\*Q'\*Q\overline{\*C})^{-1}\right\|}_{b=O_P(1)}\notag\\
        &=O_{P^*}(T^{-1/2}), 
    \end{align}
    since by $\mathrm{tr}(\*A'\*A)=\mathrm{tr}(\*A\*A')$
    \begin{align}
        E^*[a^2]&=\frac{1}{N}\sum_{i=1}^N \mathrm{tr}\left[T^{-2}\+\varepsilon_i^{*\prime}\*Q\overline{\*C}\overline{\*C}'\*Q'\+\varepsilon^*_i \otimes T^{-1}\widehat{\*A}_{\*z_0,i}'\*Q\overline{\*C}\overline{\*C}'\*Q'\widehat{\*A}_{\*z_0,i} \right]\notag\\
        &=\frac{1}{N}\sum_{i=1}^NE^*[T^{-2}\+\varepsilon_i^{*\prime}\*Q\overline{\*C}\overline{\*C}'\*Q'\+\varepsilon^*_i ]\mathrm{tr}\left[ T^{-1}\widehat{\*A}_{\*z_0,i}'\*Q\overline{\*C}\overline{\*C}'\*Q'\widehat{\*A}_{\*z_0,i}\right]\notag\\
        &=\frac{1}{N}\sum_{i=1}^NE^*\left[\left\| T^{-1}\+\varepsilon_i^{*\prime}\*Q\overline{\*C}\right\|^2\right]\left\| T^{-1/2}\widehat{\*A}_{\*z_0,i}'\*Q\overline{\*C}\right\|^2=O_P(T^{-1})
    \end{align}
    driven by the first component in the summand as$\left\|T^{-1/2}\widehat{\*A}_{\*z_0,i}'\*Q\right\|\leq \left\|\*z_{i,0} \right\|\left(\frac{1}{T}\sum_{t=2}^T\left\|\*q_t\right\|^2 \right)^{1/2}\sum_{t=2}^T|\widehat{\alpha}_{CCEP}|^{t-1}\\=O_P(1)$ for each $i$. Also,
    \begin{align}
        \left\| \*I_{2,1,d}\right\|&\leq \frac{1}{\sqrt{T}}\left\| \frac{1}{\sqrt{N}}\sum_{i=1}^N \left(T^{-1/2}\+\varepsilon_i^{*\prime}\*Q\overline{\*C} \otimes \*S_\*b'(\sqrt{T}\mathcal{\+P}_{0,3,i})' \right) \right\|\left\| \overline{\*C}\right\|\left\|T^{-1}\*H'\*Q \right\| \left\|(T^{-1}\overline{\*C}'\*Q'\*Q\overline{\*C})^{-1} \right\|\notag\\
        &=O_{P^*}(T^{-1/2}),
    \end{align}
    and thus by substituting Lemma XB,
    \begin{align}
         \mathbf{I}_{2,1}&=\mathbf{I}_{2,1,b}+\mathbf{I}_{2,1,c}+\mathbf{I}_{2,1,e} +O_{P^*}(T^{-1/2})\notag\\ 
         &=\frac{1}{\sqrt{NT}}\sum_{i=1}^N\*S_\*b'\mathcal{\+P}_{0,1,i}^{\prime}\widehat{\*Q}_0^{\prime}\*P_{\*Q\overline{\*C}}\+\varepsilon_i^*+\frac{1}{\sqrt{NT}}\sum_{i=1}^N\*S_\*b'\mathcal{\+P}_{2}'\+\kappa_i'\*P_{\*Q\overline{\*C}}\+\varepsilon_i^*
    +\frac{1}{\sqrt{NT}}\sum_{i=1}^N\*S_\*b'\*U_i^{*\prime}\*P_{\*Q\overline{\*C}}\+\varepsilon_i^*+O_{P^*}(T^{-1/2})\notag\\
    &=\*S_\*b' \left(\frac{1}{\sqrt{N}}\sum_{i=1}^N T^{-1/2}\*C'\*Q'\+\varepsilon_i^* \otimes \*C_i \right)'\mathrm{vec}(T^{-1}\*Q'_{\widehat{\alpha}}\*Q\*C(\*C'\+\Sigma_\*q\*C)^{-1})\notag\\
    &+\*S_\*b'\mathrm{\+P}_2'\left(\frac{1}{\sqrt{N}}\sum_{i=1}^N T^{-1/2}\*C'\*Q'\+\varepsilon_i^* \otimes \*C'\+\Sigma_{\*q\+\kappa,i} \right)'\mathrm{vec}((\*C'\+\Sigma_\*q\*C)^{-1})\notag\\
    &+\sqrt{\frac{N}{T}}\frac{1}{T}\sum_{t=2}^T\sum_{s=2}^T\frac{1}{N}\sum_{i=1}^N\mathcal{\+v}_{i,t}^*\varepsilon_{i,s}^*\mathrm{tr}\left(\*q_s\*q_t'\overline{\*C} (T^{-1}\overline{\*C}'\*Q'\*Q\overline{\*C})\overline{\*C}'\right) + o_{P^*}(1)\notag\\
    &= \left(\mathrm{vec}(T^{-1}\*Q'_{\widehat{\alpha}}\*Q\*C(\*C'\+\Sigma_\*q\*C)^{-1}\*C')\otimes \*S_\*b \right)'\frac{1}{\sqrt{N}}\sum_{i=1}^N\mathrm{vec}\left(T^{-1/2}\+\varepsilon_i^{*\prime}\*Q \otimes \*C_i' \right)\notag\\
    &+ \left( \mathrm{vec}(\*C(\*C'\+\Sigma_\*q\*C)^{-1}\*C')\otimes \mathcal{\+P}_2\*S_\*b\right)'\frac{1}{\sqrt{N}}\sum_{i=1}^N\mathrm{vec}\left( T^{-1/2}\+\varepsilon_i^{*\prime}\*Q \otimes \+\Sigma_{\*q\+\kappa,i}'\right)\notag\\
    &+\sqrt{\frac{N}{T}}\frac{1}{T}\sum_{t=2}^T\sum_{s=2}^T\frac{1}{N}\sum_{i=1}^N\mathcal{\+v}_{i,t}^*\varepsilon_{i,s}^*\mathrm{tr}\left(\*q_s\*q_t'\overline{\*C} (T^{-1}\overline{\*C}'\*Q'\*Q\overline{\*C})\overline{\*C}'\right) + o_{P^*}(1),
    \end{align}
    where $\*I_{2,1,e}$ is the ``Nickell bias'' of the same form as in \cite{juodis2021robustness}. Note how 
    \begin{align}
        E^*[\*I_{2,1,e}]&=\sqrt{\frac{N}{T}}\frac{1}{N}\sum_{i=1}^N\begin{bmatrix}
            \frac{1}{T}\sum_{t=2}^{T}\sum_{s=2}^{t-1}\widehat{\alpha}_{CCEP}^{t-s-1}\widehat{\varepsilon}_{i,s}^2\\
            \*0_{k\times 1}
        \end{bmatrix}\mathrm{tr}\left(\*q_s\*q_t'\overline{\*C} (T^{-1}\overline{\*C}'\*Q'\*Q\overline{\*C})\overline{\*C}'\right)\notag\\
        &= \sqrt{\frac{N}{T}} \sum_{h=1}^{T-2}\widehat{\alpha}_{CCEP}^{h-1}\begin{bmatrix}
            \frac{1}{N}\sum_{i=1}^N\frac{1}{T}\sum_{s=2}^{T-h}\widehat{\varepsilon}_{i,s}^2\\
            \*0_{k\times 1}
            \end{bmatrix}\mathrm{tr}\left(\*q_s\*q_t'\overline{\*C} (T^{-1}\overline{\*C}'\*Q'\*Q\overline{\*C})\overline{\*C}'\right)
    \end{align}
    Thus, 
    \begin{align}
        \*I_{2,1,e}\to_{p^*} \sqrt{\kappa}\sum_{h=1}^\infty\+\Sigma_{\varepsilon\mathcal{\+v}}(-h)'\mathrm{tr}\left(\+\Sigma_\*q(h)\*C(\*C'\+\Sigma_\*q\*C)^{-1}\*C'\right),
    \end{align}
    where $\+\Sigma_{\+\varepsilon\mathcal{\+v}}(-h)'=[\sigma^2(\alpha^{0})^{h-1}, \*0_{1\times k}]'$ in the notation of \cite{juodis2021robustness}. Importantly, when $\+\theta_0=\*0_{k\times 1}$, this ``Nickell bias'' coincides to the one in \cite{DeVos2019} under no dynamics in $\*x_{i,t}$. Moving on to $\mathbf{II}_{2,2}$, we will again use the expansion in (\ref{P_F-P_QC}), which is a constant in the bootstrap realm, so that 
    \begin{align}
        \mathbf{I}_{2,2}&= \frac{1}{\sqrt{NT}}\sum_{i=1}^N\*S_\*b'[\widehat{\*A}, \*I_{K\times K}]'\widehat{\*A}_{\*z_0,i}'( \*P_{\widehat{\*F}}-\*P_{\*Q\overline{\*C}})\+\varepsilon_i^*\notag\\
    &+\frac{1}{\sqrt{NT}}\sum_{i=1}^N\*S_\*b'\mathcal{\+P}_{0,1,i}^{\prime}\widehat{\*Q}_0^{\prime}( \*P_{\widehat{\*F}}-\*P_{\*Q\overline{\*C}})\+\varepsilon_i^*+\frac{1}{\sqrt{NT}}\sum_{i=1}^N\*S_\*b'\mathcal{\+P}_{2}'\+\kappa_i'( \*P_{\widehat{\*F}}-\*P_{\*Q\overline{\*C}})\+\varepsilon_i^*\notag\\
    &+\frac{1}{\sqrt{NT}}\sum_{i=1}^N\*S_\*b'\mathcal{\+P}_{0,3,i}^{\prime}\*H'( \*P_{\widehat{\*F}}-\*P_{\*Q\overline{\*C}})\+\varepsilon_i^*+\frac{1}{\sqrt{NT}}\sum_{i=1}^N\*S_\*b'\*U_i^{*\prime}( \*P_{\widehat{\*F}}-\*P_{\*Q\overline{\*C}})\+\varepsilon_i^*\notag\\
    &= \mathbf{I}_{2,2,1}+ \mathbf{I}_{2,2,2}+ \mathbf{I}_{2,2,3}+ \mathbf{I}_{2,2,4}+ \mathbf{I}_{2,2,5},
    \end{align}
    where the first component $\mathbf{I}_{2,2,1}=\mathbf{I}_{2,2,1,a}+\mathbf{I}_{2,2,1,b}+\mathbf{I}_{2,2,1,c}+\mathbf{I}_{2,2,1,d}$ proceeds very similarly to $\mathbf{II}_{2,2,1}$, where the key difference is the presence of $\left\|\frac{1}{\sqrt{T}}\sum_{t=2}^T\sqrt{N}\overline{\*v}_t\varepsilon^*_{i,t}\right\|$ instead of $\left\|\frac{1}{\sqrt{T}}\sum_{t=2}^TN\overline{\*v}_t\overline{\varepsilon}^*_{t}\right\|$, and $\left\|\frac{1}{\sqrt{T}}\sum_{t=2}^T\*q_t\varepsilon_{i,t}\right\|$ instead of $\left\|\frac{1}{\sqrt{T}}\sum_{t=2}^T\sqrt{N}\*q_t\overline{\varepsilon}_{t}\right\|$, and this will be used in many upcoming terms. Specifically, 
    \begin{align}
        \left\|\mathbf{I}_{2,2,1,a}\right\|&=\left\|\frac{1}{\sqrt{NT}}\sum_{i=1}^N\*S_\*b'[\widehat{\*A}, \*I_{K\times K}]'T^{-1}\widehat{\*A}_{\*z_0,i}'\overline{\*V}(T^{-1}\widehat{\*F}'\widehat{\*F})^{-1}\overline{\*V}'\+\varepsilon_i^* \right\|\notag\\
    &\leq \frac{1}{\sqrt{NT}}\left\|\*S_\*b'[\widehat{\*A}, \*I_{K\times K}]' \right\|\left\| (T^{-1}\widehat{\*F}'\widehat{\*F})^{-1}\right\|\left(\frac{1}{N}\sum_{i=1}^N\left\|\frac{1}{\sqrt{T}}\sum_{t=2}^T\sqrt{N}\overline{\*v}_t\varepsilon_{i,t}^* \right\|^2\right)^{1/2}\left(\frac{1}{N}\sum_{i=1}^N\left\|\*z_{i,0} \right\|^2 \right)^{1/2}\notag\\
    &\times \left(\frac{1}{T}\sum_{t=2}^T\left\|\sqrt{N}\overline{\*v}_t \right\|^2 \right)^{1/2}\sum_{t=2}^T|\widehat{\alpha}_{CCEP} |^{t-1}=O_{P^*}((NT)^{-1/2}),
    \end{align}
    and 
    \begin{align}
        \left\|\mathbf{I}_{2,2,1,b}\right\|&=\left\|\frac{1}{\sqrt{NT}}\sum_{i=1}^N\*S_\*b'[\widehat{\*A}, \*I_{K\times K}]'T^{-1}\widehat{\*A}_{\*z_0,i}'\*Q\overline{\*C}(T^{-1}\widehat{\*F}'\widehat{\*F})^{-1}\overline{\*V}'\+\varepsilon_i^* \right\|\notag\\
    &\leq \frac{1}{\sqrt{T}}\left\|\*S_\*b'[\widehat{\*A}, \*I_{K\times K}]' \right\|\left\| (T^{-1}\widehat{\*F}'\widehat{\*F})^{-1}\right\|\left\|\overline{\*C} \right\|\left(\frac{1}{N}\sum_{i=1}^N\left\|\frac{1}{\sqrt{T}}\sum_{t=2}^T\sqrt{N}\overline{\*v}_t\varepsilon_{i,t}^* \right\|^2\right)^{1/2}\left(\frac{1}{N}\sum_{i=1}^N\left\|\*z_{i,0} \right\|^2 \right)^{1/2}\notag\\
    &\times \left(\frac{1}{T}\sum_{t=2}^T\left\|\*q_t \right\|^2 \right)^{1/2}\sum_{t=2}^T|\widehat{\alpha}_{CCEP} |^{t-1}=O_{P^*}(T^{-1/2}),
    \end{align}
    with 
    \begin{align}
        \left\|\mathbf{I}_{2,2,1,c}\right\|&=\left\|\frac{1}{\sqrt{NT}}\sum_{i=1}^N\*S_\*b'[\widehat{\*A}, \*I_{K\times K}]'T^{-1}\widehat{\*A}_{\*z_0,i}'\overline{\*V}(T^{-1}\widehat{\*F}'\widehat{\*F})^{-1}\overline{\*C}'\*Q'\+\varepsilon_i^* \right\|\notag\\
    &\leq \frac{1}{\sqrt{T}}\left\|\*S_\*b'[\widehat{\*A}, \*I_{K\times K}]' \right\|\left\| (T^{-1}\widehat{\*F}'\widehat{\*F})^{-1}\right\|\left\|\overline{\*C} \right\|\left(\frac{1}{N}\sum_{i=1}^N\left\|\frac{1}{\sqrt{T}}\sum_{t=2}^T\*q_t\varepsilon_{i,t}^* \right\|^2\right)^{1/2}\left(\frac{1}{N}\sum_{i=1}^N\left\|\*z_{i,0} \right\|^2 \right)^{1/2}\notag\\
    &\times \left(\frac{1}{T}\sum_{t=2}^T\left\|\sqrt{N}\overline{\*v}_t\right\|^2 \right)^{1/2}\sum_{t=2}^T|\widehat{\alpha}_{CCEP} |^{t-1}=O_{P^*}(T^{-1/2}).
    \end{align}
    Finally, 
    \begin{align}
        \left\|\mathbf{I}_{2,2,1,d}\right\|&=\left\|\frac{1}{\sqrt{NT}}\sum_{i=1}^N\*S_\*b'[\widehat{\*A}, \*I_{K\times K}]'T^{-1}\widehat{\*A}_{\*z_0,i}'\overline{\*V}(T^{-1}\widehat{\*F}'\widehat{\*F})^{-1}\overline{\*C}'\*Q'\+\varepsilon_i^* \right\|\notag\\
    &\leq\left\|\*S_\*b'[\widehat{\*A}, \*I_{K\times K}]' \right\|\left\|\overline{\*C} \right\|^2\left(\frac{1}{N}\sum_{i=1}^N\left\|\frac{1}{\sqrt{T}}\sum_{t=2}^T\*q_t\varepsilon_{i,t}^* \right\|^2\right)^{1/2}\left(\frac{1}{N}\sum_{i=1}^N\left\|\*z_{i,0} \right\|^2 \right)^{1/2}\notag\\
    &\times \left(\frac{1}{T}\sum_{t=2}^T\left\|\*q_t\right\|^2 \right)^{1/2}\sum_{t=2}^T|\widehat{\alpha}_{CCEP} |^{t-1}\sqrt{N}\left\| (T^{-1}\widehat{\*F}'\widehat{\*F})^{-1}-(T^{-1}\overline{\*C}'\*Q'\*Q\overline{\*C})^{-1}\right\|\notag\\
    &=O_{P^*}(C_{N,T}^{-1}).
    \end{align}
    Therefore, 
    \begin{align}
        \left\| \mathbf{I}_{2,2,1}\right\|=O_{P^*}(C_{N,T}^{-1}).
    \end{align}
    Next, $\mathbf{I}_{2,2,2}=\mathbf{I}_{2,2,2,a}+\mathbf{I}_{2,2,2,b}+\mathbf{I}_{2,2,2,c}+\mathbf{I}_{2,2,2,d}$, where 
    \begin{align}
        \left\| \mathbf{I}_{2,2,2,a}\right\|&=\left\| \frac{1}{\sqrt{NT}}\sum_{i=1}^N\*S_\*b'\mathcal{\+P}_{0,1,i}^{\prime}T^{-1}\widehat{\*Q}_0^{\prime}\overline{\*V}(T^{-1}\widehat{\*F}'\widehat{\*F})^{-1}\overline{\*V}'\+\varepsilon_i^*\right\|\notag\\
    &\leq N^{-1/2}\left\|\*S_\*b \right\|\left( \frac{1}{N}\sum_{i=1}^N\left\|\mathcal{\+P}_{0,1,i} \right\|^2\right)^{1/2}\left( \frac{1}{N}\sum_{i=1}^N\left\|\frac{1}{\sqrt{T}}\sum_{t=2}^T\sqrt{N}\overline{\*v}_t\varepsilon_{i,t}^* \right\|^2\right)^{1/2}\left\|(T^{-1}\widehat{\*F}'\widehat{\*F})^{-1} \right\|\notag\\
    &\times \left\|\sqrt{N}T^{-1}\widehat{\*Q}_0^{\prime}\overline{\*V} \right\|=O_{P^*}(N^{-1/2}),
    \end{align}
    while 
    \begin{align}
        \left\| \mathbf{I}_{2,2,2,b}\right\|&=\left\| \frac{1}{\sqrt{NT}}\sum_{i=1}^N\*S_\*b'\mathcal{\+P}_{0,1,i}^{\prime}T^{-1}\widehat{\*Q}_0^{\prime}\*Q\overline{\*C}(T^{-1}\widehat{\*F}'\widehat{\*F})^{-1}\overline{\*V}'\+\varepsilon_i^*\right\|\notag\\
        &\leq \underbrace{\left\|\frac{1}{N}\sum_{i=1}^N(\sqrt{N}T^{-1/2}\overline{\*V}'\+\varepsilon_i^*\otimes \*S_\*b'\mathcal{\+P}_{0,1,i}^{\prime}) \right\|}_{O_{P^*}(N^{-1/2})}\left\| T^{-1}\widehat{\*Q}_0^{\prime}\*Q \right\|\left\| \overline{\*C} \right\|\left\| (T^{-1}\widehat{\*F}'\widehat{\*F})^{-1}\right\|=O_{P^*}(N^{-1/2}),
    \end{align}
    and 
    \begin{align}
         \left\| \mathbf{I}_{2,2,2,c}\right\|&=\left\| \frac{1}{\sqrt{NT}}\sum_{i=1}^N\*S_\*b'\mathcal{\+P}_{0,1,i}^{\prime}T^{-1}\widehat{\*Q}_0^{\prime}\overline{\*V}(T^{-1}\widehat{\*F}'\widehat{\*F})^{-1}\overline{\*C}'\*Q'\+\varepsilon_i^*\right\|\notag\\
         &\leq \left\|\frac{1}{N}\sum_{i=1}^N(T^{-1/2}\+\varepsilon_i^{*\prime} \*Q\otimes \*S_\*b'\mathcal{\+P}_{0,1,i}^{\prime} ) \right\| \left\|\sqrt{N}T^{-1}\widehat{\*Q}_0^{\prime}\overline{\*V} \right\| \left\|(T^{-1}\widehat{\*F}'\widehat{\*F})^{-1}\right\|\left\| \overline{\*C} \right\|  = O_{P^*}(N^{-1/2}),
    \end{align}
    and finally,
    \begin{align}
         \left\| \mathbf{I}_{2,2,2,d}\right\|&=\left\|\frac{1}{\sqrt{NT}}\sum_{i=1}^N\*S_\*b'\mathcal{\+P}_{0,1,i}^{\prime}T^{-1}\widehat{\*Q}_0^{\prime}\*Q\overline{\*C}((T^{-1}\widehat{\*F}'\widehat{\*F})^{-1} - (T^{-1}\overline{\*C}'\*Q'\*Q\overline{\*C})^{-1})\overline{\*C}'\*Q'\+\varepsilon_i^*\right\|\notag\\
         &\leq \left\|\frac{1}{\sqrt{N}}\sum_{i=1}^N\left(T^{-1/2}\+\varepsilon_i^{*\prime}\*Q\otimes \*S_\*b'\mathcal{\+P}_{0,1,i}^{\prime} \right) \right\| \left\| \overline{\*C}\right\|^2\left\|T^{-1}\widehat{\*Q}_0^{\prime}\*Q \right\|\notag\\
         &\times \left\| (T^{-1}\widehat{\*F}'\widehat{\*F})^{-1} - (T^{-1}\overline{\*C}'\*Q'\*Q\overline{\*C})^{-1}\right\|=O_{P^*}(N^{-1})+O_{P^*}((NT)^{-1/2}),
    \end{align}
    giving 
    \begin{align}
        \left\| \mathbf{I}_{2,2,2}\right\|=O_{P^*}(N^{-1/2}).
    \end{align}
    The term $\mathbf{I}_{2,2,3}=\mathbf{I}_{2,2,3,a}+\mathbf{I}_{2,2,3,b}+\mathbf{I}_{2,2,3,c}+\mathbf{I}_{2,2,3,d}$ exhibits a similar asymptotic behavior: 
    \begin{align}
        \left\| \mathbf{I}_{2,2,3,a}\right\|&=\left\| \frac{1}{\sqrt{NT}}\sum_{i=1}^N\*S_\*b'\mathcal{\+P}_{2}'T^{-1}\+\kappa_i'\overline{\*V}(T^{-1}\widehat{\*F}'\widehat{\*F})^{-1}\overline{\*V}'\+\varepsilon_i^*\right\|\notag\\
    &\leq N^{-1/2}\left\|\*S_\*b \right\|\left\| \mathcal{\+P}_2\right\|\left\| (T^{-1}\widehat{\*F}'\widehat{\*F})^{-1}\right\|\left(\frac{1}{N}\sum_{i=1}^N\left\| \sqrt{N}T^{-1}\+\kappa_i'\overline{\*V}\right\|^2\right)^{1/2}\left(\frac{1}{N}\sum_{i=1}^N\left\|\frac{1}{\sqrt{T}}\sum_{t=2}^T\sqrt{N}\overline{\*v}_t\varepsilon_{i,t}^* \right\|^2 \right)^{1/2}\notag\\
    &=O_{P^*}(N^{-1/2}),
    \end{align}
    \begin{align}
        \left\|\mathbf{I}_{2,2,3,b} \right\|&=\left\|\frac{1}{\sqrt{NT}}\sum_{i=1}^N\*S_\*b'\mathcal{\+P}_{2}'T^{-1}\+\kappa_i'\*Q\overline{\*C}(T^{-1}\widehat{\*F}'\widehat{\*F})^{-1}\overline{\*V}'\+\varepsilon_i^* \right\|\notag\\
        &\leq \left\|\*S_\*b \right\|\left\|\mathcal{\+P}_2 \right\|\left\|\frac{1}{N}\sum_{i=1}^N\left(\sqrt{N}T^{-1/2}\+\varepsilon_i^{*\prime}\overline{\*V}\otimes T^{-1}\+\kappa_i'\*Q \right) \right\|\left\| \overline{\*C}\right\|\left\| (T^{-1}\widehat{\*F}'\widehat{\*F})^{-1}\right\|=O_{P^*}(N^{-1/2}),
    \end{align}
    and 
    \begin{align}
         \left\|\mathbf{I}_{2,2,3,c} \right\|&=\left\|\frac{1}{\sqrt{NT}}\sum_{i=1}^N\*S_\*b'\mathcal{\+P}_{2}'T^{-1}\+\kappa_i'\overline{\*V}(T^{-1}\widehat{\*F}'\widehat{\*F})^{-1}\overline{\*C}'\*Q'\+\varepsilon_i^* \right\|\notag\\
         &\leq \left\|\*S_\*b \right\|\left\|\mathcal{\+P}_2 \right\|\left\|\frac{1}{N}\sum_{i=1}^N\left( T^{-1/2}\+\varepsilon_i^{*\prime}\*Q\otimes \sqrt{N}T^{-1}\+\kappa_i'\overline{\*V}\right) \right\|\left\|\overline{\*C} \right\|\left\| (T^{-1}\widehat{\*F}'\widehat{\*F})^{-1}\right\|=O_{P^*}(N^{-1/2}),
    \end{align}
   with 
   \begin{align}
       \left\|\mathbf{I}_{2,2,3,d} \right\|&=\left\|\frac{1}{\sqrt{NT}}\sum_{i=1}^N\*S_\*b'\mathcal{\+P}_{2}'T^{-1}\+\kappa_i'\*Q\overline{\*C}((T^{-1}\widehat{\*F}'\widehat{\*F})^{-1} - (T^{-1}\overline{\*C}'\*Q'\*Q\overline{\*C})^{-1})\overline{\*C}'\*Q'\+\varepsilon_i^* \right\|\notag\\
         &\leq \left\|\*S_\*b \right\|\left\|\mathcal{\+P}_2 \right\|\left\|\frac{1}{\sqrt{N}}\sum_{i=1}^N\left( T^{-1/2}\+\varepsilon_i^{*\prime}\*Q\otimes T^{-1}\+\kappa_i'\*Q\right) \right\|\left\|\overline{\*C} \right\|^2\left\|(T^{-1}\widehat{\*F}'\widehat{\*F})^{-1} - (T^{-1}\overline{\*C}'\*Q'\*Q\overline{\*C})^{-1} \right\|\notag\\
         &=O_{P^*}(N^{-1})+O_{P^*}((NT)^{-1/2}).
   \end{align}
   Therefore, 
   \begin{align}
       \left\| \mathbf{I}_{2,2,3}\right\|=O_{P^*}(N^{-1/2}).
   \end{align}
   The term $\mathbf{I}_{2,2,4}=\mathbf{I}_{2,2,4,a}+\mathbf{I}_{2,2,4,b}+\mathbf{I}_{2,2,4,c}+\mathbf{I}_{2,2,4,d}$ has the same form as $\mathbf{I}_{2,2,2}$, but it vanishes faster due to the presence of $\left\| \mathcal{\+P}_{0,3,i}\right\|=O_P(T^{-1/2})$, and so 
   \begin{align}
        \left\| \mathbf{I}_{2,2,4}\right\|=O_{P^*}((NT)^{-1/2}).
   \end{align}
   Ultimately, we work out $\mathbf{I}_{2,2,5}=\mathbf{I}_{2,2,4,a}+\mathbf{I}_{2,2,5,b}+\mathbf{I}_{2,2,5,c}+\mathbf{I}_{2,2,5,d}$: \begin{align}
    \left\| \mathbf{I}_{2,2,5,a}\right\|&=\left\|\frac{1}{\sqrt{NT}}\sum_{i=1}^N\*S_\*b'T^{-1}\*U_i^{*\prime}\overline{\*V}(T^{-1}\widehat{\*F}'\widehat{\*F})^{-1}\overline{\*V}'\+\varepsilon_i\right\|\notag\\
    &\leq N^{-1/2} \left\| (T^{-1}\widehat{\*F}'\widehat{\*F})^{-1}\right\|\left(\frac{1}{N}\sum_{i=1}^N\left\|\sqrt{N}T^{-1}\*S_\*b'\*U_i^{*\prime}\overline{\*V} \right\|^2 \right)^{1/2}\left( \frac{1}{N}\sum_{i=1}^N \left\|\frac{1}{\sqrt{T}}\sum_{t=2}^T\sqrt{N}\overline{\*v}_t\varepsilon_{i,t}^* \right\|^2\right)^{1/2}\notag\\
    &=O_{P^*}((NT)^{-1/2}),
\end{align}
where $\left\|\sqrt{N}T^{-1}\*S_\*b'\*U_i^{*\prime}\overline{\*V} \right\|=O_{P^*}(T^{-1/2})$ by an argument similar (\ref{b_m-b}) when focusing on the element $u_{i,t-1}^*$ (leading component). The other error components will vanish at least at the same rate.  Next, 
\begin{align}
    \left\| \mathbf{I}_{2,2,5,b}\right\|&=\left\|\frac{1}{\sqrt{NT}}\sum_{i=1}^N\*S_\*b'T^{-1}\*U_i^{*\prime}\*Q\overline{\*C}(T^{-1}\widehat{\*F}'\widehat{\*F})^{-1}\overline{\*V}'\+\varepsilon_i\right\|\notag\\
    &\leq\left\| \overline{\*C}\right\| \left\| (T^{-1}\widehat{\*F}'\widehat{\*F})^{-1}\right\|\left(\frac{1}{N}\sum_{i=1}^N\left\|T^{-1}\*S_\*b'\*U_i^{*\prime}\*Q \right\|^2 \right)^{1/2}\left( \frac{1}{N}\sum_{i=1}^N\left\|\frac{1}{\sqrt{T}}\sum_{t=2}^T\sqrt{N}\overline{\*v}_t\varepsilon_{i,t}^* \right\|^2\right)^{1/2}\notag\\
    &=O_{P^*}(T^{-1/2}),
\end{align}
and 
\begin{align}
     \left\| \mathbf{I}_{2,2,5,c}\right\|&=\left\|\frac{1}{\sqrt{NT}}\sum_{i=1}^N\*S_\*b'T^{-1}\*U_i^{*\prime}\overline{\*V}(T^{-1}\widehat{\*F}'\widehat{\*F})^{-1}\overline{\*C}'\*Q'\+\varepsilon_{i}^*\right\|\notag\\
    &\leq \left\| \overline{\*C}\right\| \left\| (T^{-1}\widehat{\*F}'\widehat{\*F})^{-1}\right\|\left(\frac{1}{N}\sum_{i=1}^N\left\|\sqrt{N}T^{-1}\*S_\*b'\*U_i^{*\prime}\overline{\*V}\right\|^2 \right)^{1/2}\left( \frac{1}{N}\sum_{i=1}^N\left\|\frac{1}{\sqrt{T}}\sum_{t=2}^T\*q_t\varepsilon_{i,t}^* \right\|^2\right)^{1/2}\notag\\
    &=O_{P^*}(T^{-1/2}),
\end{align}
 Finally, 
\begin{align}
     \left\| \mathbf{I}_{2,2,5,d}\right\|&=\left\|\frac{1}{\sqrt{NT}}\sum_{i=1}^N\*S_\*b'T^{-1}\*U_i^{*\prime}\*Q\overline{\*C}((T^{-1}\widehat{\*F}'\widehat{\*F})^{-1} - (T^{-1}\overline{\*C}'\*Q'\*Q\overline{\*C})^{-1})\overline{\*C}'\*Q'\+\varepsilon_{i}^*\right\|\notag\\
    &\leq\left\| \overline{\*C}\right\|^2\left(\frac{1}{N}\sum_{i=1}^N\left\|T^{-1}\*S_\*b'\*U_i^{*\prime}\*Q\right\|^2 \right)^{1/2}\left( \frac{1}{N}\sum_{i=1}^N\left\|\frac{1}{\sqrt{T}}\sum_{t=2}^T\*q_t\varepsilon_{i,t}^* \right\|^2\right)^{1/2}\notag\\
    &\times \sqrt{N}\left\| (T^{-1}\widehat{\*F}'\widehat{\*F})^{-1} - (T^{-1}\overline{\*C}'\*Q'\*Q\overline{\*C})^{-1}\right\|=O_{P^*}(C_{N,T}^{-1}),
\end{align}
and so 
\begin{align}
    \left\| \*I_{2,2,5}\right\|=O_{P^*}(C_{N,T}^{-1}). 
\end{align}
We focus on the very last term $\mathbf{I}_{2,3}$: 
\begin{align}
     \mathbf{I}_{2,3}&=\frac{1}{\sqrt{NT}}\sum_{i=1}^N\*B_i^{*\prime}( \*P_{\widehat{\*F}^*}-\*P_{\widehat{\*F}})\+\varepsilon_{i}^*=\frac{1}{\sqrt{NT}}\sum_{i=1}^N\*S_\*b'[\widehat{\*A}, \*I_{K\times K}]'\widehat{\*A}_{\*z_0,i}'( \*P_{\widehat{\*F}^*}-\*P_{\widehat{\*F}})\+\varepsilon_{i}^*\notag\\
    &+\frac{1}{\sqrt{NT}}\sum_{i=1}^N\*S_\*b'\mathcal{\+P}_{0,1,i}^{\prime}\widehat{\*Q}_0^{\prime}(  \*P_{\widehat{\*F}^*}-\*P_{\widehat{\*F}})\+\varepsilon_{i}^*+\frac{1}{\sqrt{NT}}\sum_{i=1}^N\*S_\*b'\mathcal{\+P}_{2}'\+\kappa_i'(  \*P_{\widehat{\*F}^*}-\*P_{\widehat{\*F}})\+\varepsilon_{i}^*\notag\\
    &+\frac{1}{\sqrt{NT}}\sum_{i=1}^N\*S_\*b'\mathcal{\+P}_{0,3,i}^{\prime}\*H'(  \*P_{\widehat{\*F}^*}-\*P_{\widehat{\*F}})\+\varepsilon_{i}^*+\frac{1}{\sqrt{NT}}\sum_{i=1}^N\*S_\*b'\*U_i^{*\prime}(  \*P_{\widehat{\*F}^*}-\*P_{\widehat{\*F}})\+\varepsilon_{i}^*\notag\\
    &= \mathbf{I}_{2,3,1}+ \mathbf{I}_{2,3,2}+ \mathbf{I}_{2,3,3}+ \mathbf{I}_{2,3,4}+ \mathbf{I}_{2,3,5},
\end{align}
where we again use the expansion in (\ref{Pf*-Pfhat}). Firstly, we demonstrate that the effect of the initial value in $\mathbf{I}_{2,3,1}=\mathbf{I}_{2,3,1,a}+\mathbf{I}_{2,3,1,b}+\mathbf{I}_{2,3,1,c}+\mathbf{I}_{2,3,1,d}$ vanishes 
\begin{align}
    \left\|\mathbf{I}_{2,3,1,a} \right\|&=\left\| \frac{1}{\sqrt{NT}}\sum_{i=1}^N\*S_\*b'[\widehat{\*A}, \*I_{K\times K}]'T^{-1}\widehat{\*A}_{\*z_0,i}'\overline{\*W}^*(T^{-1}\widehat{\*F}^{*\prime}\widehat{\*F}^*)^{-1}\overline{\*W}^{*\prime}\+\varepsilon_i^*\right\|\notag\\
    &\leq N^{-1/2}\left\|\*S_\*b'[\widehat{\*A}, \*I_{K\times K}]' \right\|\left\| (T^{-1}\widehat{\*F}^{*\prime}\widehat{\*F}^*)^{-1}\right\|\underbrace{\left(\frac{1}{N}\sum_{i=1}^N\left\|\sqrt{N}T^{-1}\overline{\*W}^{*\prime}\+\varepsilon^*_i \right\|^2\right)^{1/2}}_{O_{P^*}(C_{N,T}^{-1})}\left(\frac{1}{N}\sum_{i=1}^N\left\|\*z_{i,0} \right\|^2 \right)^{1/2}\notag\\
    &\times \left(\frac{1}{T}\sum_{t=2}^T\left\|\sqrt{N}\overline{\*w}^*_t \right\|^2 \right)^{1/2}\sum_{t=2}^T|\widehat{\alpha}_{CCEP} |^{t-1}=O_{P^*}( N^{-1})+O_{p^*} ((NT)^{-1/2}),
\end{align}
where the leading component in $\sqrt{N}T^{-1}\overline{\*W}^{*\prime}\+\varepsilon^*_i $ is $\sqrt{N}T^{-1}\overline{\*u}^{*\prime}\+\varepsilon^*_i=N^{-1/2}T^{-1}\*u_{i}^{*\prime}\+\varepsilon_i^*+\frac{1}{\sqrt{N}T}\sum_{j\neq i}^N\*u_{j}^{*\prime}\+\varepsilon_i^*=N^{-1/2}(T^{-1}\+\varepsilon_i^{*\prime}\+\varepsilon_i^*+\widehat{\alpha}_{CCEP}T^{-1}\*u_{i,-1}^{*\prime}\+\varepsilon_i^*)+T^{-1/2}(\frac{\widehat{\alpha}_{CCEP}}{\sqrt{NT}}\sum_{j\neq i}^N\*u_{j,-1}^{*\prime}\+\varepsilon_i^*+\frac{1}{\sqrt{NT}}\sum_{j\neq i}^N\+\varepsilon_{j}^{*\prime}\+\varepsilon_i^*)=O_P(C_{N,T}^{-1})$. Next, 
\begin{align}
    \left\| \mathbf{I}_{2,3,1,b}\right\|&=\left\|\frac{1}{\sqrt{NT}}\sum_{i=1}^N\*S_\*b'[\widehat{\*A}, \*I_{K\times K}]'T^{-1}\widehat{\*A}_{\*z_0,i}'\widehat{\*F}(T^{-1}\widehat{\*F}^{*\prime}\widehat{\*F}^*)^{-1}\overline{\*W}^{*\prime}\+\varepsilon_i^* \right\|\notag\\
    &\leq\left\|\*S_\*b'[\widehat{\*A}, \*I_{K\times K}]' \right\|\left\| (T^{-1}\widehat{\*F}^{*\prime}\widehat{\*F}^*)^{-1}\right\|\underbrace{\left(\frac{1}{N}\sum_{i=1}^N\left\|\sqrt{N}T^{-1}\overline{\*W}^{*\prime}\+\varepsilon^*_i \right\|^2\right)^{1/2}}_{O_{P^*}(C_{N,T}^{-1})}\left(\frac{1}{N}\sum_{i=1}^N\left\|\*z_{i,0} \right\|^2 \right)^{1/2}\notag\\
    &\times \left(\frac{1}{T}\sum_{t=2}^T\left\|\widehat{\*f}_t \right\|^2 \right)^{1/2}\sum_{t=2}^T|\widehat{\alpha}_{CCEP} |^{t-1}= O_{P^*}(C_{N,T}^{-1}),
\end{align}
and 
\begin{align}
     \left\| \mathbf{I}_{2,3,1,c}\right\|&=\left\|\frac{1}{\sqrt{NT}}\sum_{i=1}^N\*S_\*b'[\widehat{\*A}, \*I_{K\times K}]'T^{-1}\widehat{\*A}_{\*z_0,i}'\widehat{\*F}(T^{-1}\widehat{\*F}^{*\prime}\widehat{\*F}^*)^{-1}\overline{\*W}^{*\prime}\+\varepsilon_i^* \right\|\notag\\
    &\leq T^{-1/2}\left\|\*S_\*b'[\widehat{\*A}, \*I_{K\times K}]' \right\|\left\| (T^{-1}\widehat{\*F}^{*\prime}\widehat{\*F}^*)^{-1}\right\|\left(\frac{1}{N}\sum_{i=1}^N\left\|T^{-1/2}\widehat{\*F}'\+\varepsilon^*_i \right\|^2\right)^{1/2}\left(\frac{1}{N}\sum_{i=1}^N\left\|\*z_{i,0} \right\|^2 \right)^{1/2}\notag\\
    &\times \left(\frac{1}{T}\sum_{t=2}^T\left\|\sqrt{N}\overline{\*w}^*_t \right\|^2 \right)^{1/2}\sum_{t=2}^T|\widehat{\alpha}_{CCEP} |^{t-1}=O_{P^*}(T^{-1/2}),
\end{align}
 and finally,
 \begin{align}
      \left\| \mathbf{I}_{2,3,1,d}\right\|&=\left\|\frac{1}{\sqrt{NT}}\sum_{i=1}^N\*S_\*b'[\widehat{\*A}, \*I_{K\times K}]'T^{-1}\widehat{\*A}_{\*z_0,i}'\widehat{\*F}((T^{-1}\widehat{\*F}^{*\prime}\widehat{\*F}^*)^{-1}-(T^{-1}\widehat{\*F}'\widehat{\*F})^{-1})\widehat{\*F}'\+\varepsilon_i^* \right\|\notag\\
    &\leq \sqrt{\frac{N}{T}}\left\|\*S_\*b'[\widehat{\*A}, \*I_{K\times K}]' \right\|\left(\frac{1}{N}\sum_{i=1}^N\left\|T^{-1/2}\widehat{\*F}'\+\varepsilon^*_i \right\|^2\right)^{1/2}\left(\frac{1}{N}\sum_{i=1}^N\left\|\*z_{i,0} \right\|^2 \right)^{1/2}\notag\\
    &\times \left(\frac{1}{T}\sum_{t=2}^T\left\|\widehat{\*f}_t \right\|^2 \right)^{1/2}\sum_{t=2}^T|\widehat{\alpha}_{CCEP} |^{t-1}\left\| (T^{-1}\widehat{\*F}^{*\prime}\widehat{\*F}^*)^{-1}-(T^{-1}\widehat{\*F}'\widehat{\*F})^{-1}\right\|\notag\\
    &=O_{P^*}(N^{-1})+O_{P^*}((NT)^{-1/2}),
 \end{align}
 giving 
 \begin{align}
     \left\|\mathbf{II}_{2,3,1} \right\|=O_{P^*}(C_{N,T}^{-1}).
 \end{align}
 Next, we have $\mathbf{I}_{2,3,2}=\mathbf{I}_{2,3,2,a}+\mathbf{I}_{2,3,2,b}+\mathbf{I}_{2,3,2,c}+\mathbf{I}_{2,3,2,d}$. Before analyzing it, we need to sharpen the rates of certain components that we will insert when forming the bounds. Specifically, knowing the structure of $\overline{\*W}^*$, we can simply apply (\ref{FW*}) to show that $\left\|\sqrt{N} T^{-1}\widehat{\*Q}_0^{\prime}\overline{\*W}^*\right\|=O_{P^*}(T^{-1/2})$ and $\left\|\sqrt{N} T^{-1}\+\kappa_i^{\prime}\overline{\*W}^*\right\|=O_{P^*}(T^{-1/2})$ (for each $i$). Therefore, 
 \begin{align}
     \left\|\mathbf{I}_{2,3,2,a} \right\|&=\left\| \frac{1}{\sqrt{NT}}\sum_{i=1}^N\*S_\*b'\mathcal{\+P}_{0,1,i}^{\prime}T^{-1}\widehat{\*Q}_0^{\prime}\overline{\*W}^*(T^{-1}\widehat{\*F}^{*\prime}\widehat{\*F}^*)^{-1}\overline{\*W}^{*\prime}\+\varepsilon_i^*\right\|\notag\\
    &\leq N^{-1/2}\left\|\*S_\*b \right\|\left( \frac{1}{N}\sum_{i=1}^N\left\|\mathcal{\+P}_{0,1,i} \right\|^2\right)^{1/2}\left( \frac{1}{N}\sum_{i=1}^N\left\|\sqrt{N}T^{-1}\overline{\*W}^{*\prime}\+\varepsilon^*_i\right\|^2\right)^{1/2}\left\|(T^{-1}\widehat{\*F}^{*\prime}\widehat{\*F}^*)^{-1} \right\|\notag\\
    &\times \left\|\sqrt{N}T^{-1/2}\widehat{\*Q}_0^{\prime}\overline{\*W}^* \right\|=O_{P^*}(N^{-1})+O_{P^*}((NT)^{-1/2}).
 \end{align}
 Next, we obtain another candidate for an asymptotic bias. Note that $\left\| \sqrt{N}T^{-1/2}\overline{\*W}^{*\prime}\+\varepsilon_i^*\right\|=O_{P^*}(1)+O_{P^*}(\sqrt{T}N^{-1/2})$, thus 
 \begin{align}
     \*I_{2,3,2,b}&=\frac{1}{\sqrt{NT}}\sum_{i=1}^N\*S_\*b'\mathcal{\+P}_{0,1,i}^{\prime}T^{-1}\widehat{\*Q}_0^{\prime}\widehat{\*F}(T^{-1}\widehat{\*F}^{*\prime}\widehat{\*F}^*)^{-1}\overline{\*W}^{*\prime}\+\varepsilon_i^*\notag\\
     &=\frac{1}{N}\sum_{i=1}^N\*S_\*b'\mathcal{\+P}_{0,1,i}^{\prime}T^{-1}\widehat{\*Q}_0^{\prime}\*Q\overline{\*C}(T^{-1}\widehat{\*F}^{*\prime}\widehat{\*F}^*)^{-1}\sqrt{N}T^{-1/2}\overline{\*W}^{*\prime}\+\varepsilon_i^*\notag\\
     &+N^{-1/2}\frac{1}{N}\sum_{i=1}^N\*S_\*b'\mathcal{\+P}_{0,1,i}^{\prime}\sqrt{N}T^{-1}\widehat{\*Q}_0^{\prime}\overline{\*V}(T^{-1}\widehat{\*F}^{*\prime}\widehat{\*F}^*)^{-1}\sqrt{N}T^{-1/2}\overline{\*W}^{*\prime}\+\varepsilon_i^*\notag\\
     &=   \frac{1}{N}\sum_{i=1}^N\*S_\*b'\mathcal{\+P}_{0,1,i}^{\prime}T^{-1}\widehat{\*Q}_0^{\prime}\*Q\overline{\*C}(T^{-1}\widehat{\*F}^{*\prime}\widehat{\*F}^*)^{-1}               \begin{bmatrix}\sqrt{N}T^{-1/2}\overline{\*u}^{*\prime}\+\varepsilon_i^* &\*0_{1\times k} &\sqrt{N}T^{-1/2}\overline{\*u}_{-1}^{*\prime}\+\varepsilon_i^* & \*0_{1\times k}\end{bmatrix}'\notag\\
     &+O_{P^*}(N^{-1/2})\notag\\
     &= \frac{1}{N}\sum_{i=1}^N\*S_\*b'\mathcal{\+P}_{0,1,i}^{\prime}T^{-1}\widehat{\*Q}_0^{\prime}\*Q\overline{\*C}(T^{-1}\widehat{\*F}^{*\prime}\widehat{\*F}^*)^{-1}  \begin{bmatrix}
       \sqrt{N}T^{-1/2}\overline{\*u}^{*\prime}\+\varepsilon_i^* &\*0_{1\times k} &0 & \*0_{1\times k}\end{bmatrix}'  \notag\\
       &+\frac{1}{N}\sum_{i=1}^N\*S_\*b'\mathcal{\+P}_{0,1,i}^{\prime}T^{-1}\widehat{\*Q}_0^{\prime}\*Q\overline{\*C}(T^{-1}\widehat{\*F}^{*\prime}\widehat{\*F}^*)^{-1} \begin{bmatrix}0&\*0_{1\times k} &\sqrt{N}T^{-1/2}\overline{\*u}_{-1}^{*\prime}\+\varepsilon_i^* & \*0_{1\times k}\end{bmatrix}' \notag\\
       &+O_{P^*}(N^{-1/2})=\*I_{2,3,2,b}^1 + \*I_{2,3,2,b}^{2} +O_{P^*}(N^{-1/2}),
 \end{align}
 where using the fact that $\sqrt{N}T^{-1/2}\overline{\*u}^{*\prime}\+\varepsilon_i^* =\sqrt{\frac{T}{N}}\frac{1}{T}\*u_{i}^{*\prime}\+\varepsilon^*_i+\frac{1}{\sqrt{NT}}\sum_{j\neq i}^N\*u_{j}^{*\prime }\+\varepsilon_i^*$, we can make another split
 \begin{align}\label{I_i_232b}
     \*I_{2,3,2,b}^1&= \sqrt{\frac{T}{N}}\frac{1}{N}\sum_{i=1}^N\*S_\*b'\mathcal{\+P}_{0,1,i}^{\prime}T^{-1}\widehat{\*Q}_0^{\prime}\*Q\overline{\*C}(T^{-1}\widehat{\*F}^{*\prime}\widehat{\*F}^*)^{-1}  \begin{bmatrix}
      T^{-1}\*u_{i}^{*\prime}\+\varepsilon^*_i&\*0_{1\times (2K-1)}\end{bmatrix}' \notag\\
       &+\frac{1}{N}\sum_{i=1}^N\*S_\*b'\mathcal{\+P}_{0,1,i}^{\prime}T^{-1}\widehat{\*Q}_0^{\prime}\*Q\overline{\*C}(T^{-1}\widehat{\*F}^{*\prime}\widehat{\*F}^*)^{-1}  \begin{bmatrix}
       \frac{1}{\sqrt{NT}}\sum_{j\neq i}^N\*u_{j}^{*\prime }\+\varepsilon_i^*&\*0_{1\times (2K-1)}\end{bmatrix}' =\*I_{2,3,2,b}^{1,a}+\*I_{2,3,2,b}^{1,b}, 
 \end{align}
 where $\*I_{2,3,2,b}^{1,b}$ is negligible: 
 \begin{align}
     \left\|\*I_{2,3,2,b}^{1,b}\right\|&\leq \left\|\frac{1}{N}\sum_{i=1}^N\left( \left[ \frac{1}{\sqrt{NT}}\sum_{j\neq i}^N\*u_{j}^{*\prime }\+\varepsilon_i^*, \*0_{1\times (2K-1)} \right]\otimes \*S_\*b'\mathcal{\+P}_{0,1,i}^{\prime}\right)\right\|\left\| T^{-1}\widehat{\*Q}_0^{\prime}\*Q\right\| \left\|\overline{\*C} \right\|\left\| (T^{-1}\widehat{\*F}^{*\prime}\widehat{\*F}^*)^{-1}\right\|\notag\\
     &\leq O_{P^*}(1)\times \left\|\frac{1}{N}\sum_{i=1}^N\left( \left[ \frac{1}{\sqrt{NT}}\sum_{j\neq i}^N\+\varepsilon_{j}^{*\prime }\+\varepsilon_i^*, \*0_{1\times (2K-1)} \right]\otimes \*S_\*b'\mathcal{\+P}_{0,1,i}^{\prime}\right)\right\|\notag\\
     &+ O_{P^*}(1)\times \left\|\widehat{\alpha}_{CCEP} \right\| \left\|\frac{1}{N}\sum_{i=1}^N\left( \left[ \frac{1}{\sqrt{NT}}\sum_{j\neq i}^N\*u_{j,-1}^{*\prime }\+\varepsilon_i^*, \*0_{1\times (2K-1)} \right]\otimes \*S_\*b'\mathcal{\+P}_{0,1,i}^{\prime}\right)\right\|=O_{P^*}(N^{-1/2}),
 \end{align}
 since the second component is effectively (\ref{I_a2_exp}), while in the first we used $u_{j,t}^*=\widehat{\alpha}_{CCEP}u_{j,t-1}^{*}+\varepsilon_{j,t}^*$, and the fact that its second moment is $N^{-3}T^{-1}O_P(N^2T)=O_{P^*}(N^{-1})$ due to $\varepsilon_{i,t}^*$ i.i.d. over time in bootstrap realm.  In case of $\*I_{2,3,2,b}^{2}$, we notice that $\sqrt{N}T^{-1/2}\overline{\*u}_{-1}^{*\prime}\+\varepsilon_i^*=N^{-1/2} T^{-1/2}\*u_{i,-1}^{*\prime}\+\varepsilon_i^*+\frac{1}{\sqrt{NT}}\sum_{j\neq i}^N\*u_{j,-1}^{*\prime}\+\varepsilon_i^*=\frac{1}{\sqrt{NT}}\sum_{j\neq i}^N\*u_{j,-1}^{*\prime}\+\varepsilon_i^* +O_{P^*}(N^{-1/2})$ for each $i$, and so  and so
 \begin{align}
     \left\| \*I_{2,3,2,b}^{2}\right\|&=\left\| \frac{1}{N}\sum_{i=1}^N\*S_\*b'\mathcal{\+P}_{0,1,i}^{\prime}T^{-1}\widehat{\*Q}_0^{\prime}\*Q\overline{\*C}(T^{-1}\widehat{\*F}^{*\prime}\widehat{\*F}^*)^{-1} \begin{bmatrix}0&\*0_{1\times k} &\sqrt{N}T^{-1/2}\overline{\*u}_{-1}^{*\prime}\+\varepsilon_i^* & \*0_{1\times k}\end{bmatrix}'\right\|\notag\\
     &=O_{P^*}(N^{-1/2})
 \end{align}
  by the same argument as $\*I_{2,3,2,b}^{1,b}$. In summary,
 \begin{align}
     \*I_{2,3,2,b}&=\*I_{2,3,2,b}^{1,a}+o_{P^*}(1)=\sqrt{\frac{T}N{}}\frac{1}{N}\sum_{i=1}^N\*S_\*b'\mathcal{\+P}_{0,1,i}^{\prime}T^{-1}\widehat{\*Q}_0^{\prime}\*Q\overline{\*C}(T^{-1}\widehat{\*F}^{*\prime}\widehat{\*F}^*)^{-1}  \begin{bmatrix}
      T^{-1}\*u_{i}^{*\prime}\+\varepsilon^*_i&\*0_{1\times (2K-1)}\end{bmatrix}'+o_{P^*}(1)\notag\\
      &=\sqrt{\frac{T}N{}}\frac{1}{N}\sum_{i=1}^N\*S_\*b'\mathcal{\+P}_{0,1,i}^{\prime}T^{-1}\widehat{\*Q}_0^{\prime}\*Q\overline{\*C}(T^{-1}\overline{\*C}'\*Q'\*Q\overline{\*C})^{-1}  \begin{bmatrix}
      T^{-1}\*u_{i}^{*\prime}\+\varepsilon^*_i&\*0_{1\times (2K-1)}\end{bmatrix}'+o_{P^*}(1)
 \end{align}
 It is important to note that this bias component is absent in \cite{DeVos2019}. The reason is that they utilize the random loadings assumption to reduce its order (see $\*A^\varepsilon_{124}$ in their Supplement). Next, 
 \begin{align}
     \left\|\*I_{2,3,2,c} \right\|&=\left\| \frac{1}{\sqrt{NT}}\sum_{i=1}^N\*S_\*b'\mathcal{\+P}_{0,1,i}^{\prime}T^{-1}\widehat{\*Q}_0^{\prime}\overline{\*W}^*(T^{-1}\widehat{\*F}^{*\prime}\widehat{\*F}^*)^{-1}\widehat{\*F}'\+\varepsilon_i^*\right\|\notag\\
    &\leq T^{-1/2}\left\|\*S_\*b \right\|\left( \frac{1}{N}\sum_{i=1}^N\left\|\mathcal{\+P}_{0,1,i} \right\|^2\right)^{1/2}\left( \frac{1}{N}\sum_{i=1}^N\left\|T^{-1}\widehat{\*F}'\+\varepsilon^*_i\right\|^2\right)^{1/2}\left\|(T^{-1}\widehat{\*F}^{*\prime}\widehat{\*F}^*)^{-1} \right\|\notag\\
    &\times \left\|\sqrt{N}T^{-1/2}\widehat{\*Q}_0^{\prime}\overline{\*W}^* \right\|=O_{P^*}(T^{-1/2}),
 \end{align}
 and \begin{align}
     \left\|\*I_{2,3,2,d} \right\|&=\left\| \frac{1}{\sqrt{NT}}\sum_{i=1}^N\*S_\*b'\mathcal{\+P}_{0,1,i}^{\prime}T^{-1}\widehat{\*Q}_0^{\prime}\widehat{\*F}((T^{-1}\widehat{\*F}^{*\prime}\widehat{\*F}^*)^{-1}-(T^{-1}\widehat{\*F}'\widehat{\*F})^{-1})\widehat{\*F}^{\prime}\+\varepsilon_i^*\right\|\notag\\
     &\leq \left\|\*S_\*b \right\|\left( \frac{1}{N}\sum_{i=1}^N\left\|\mathcal{\+P}_{0,1,i} \right\|^2\right)^{1/2}\left( \frac{1}{N}\sum_{i=1}^N\left\|T^{-1/2}\widehat{\*F}'\+\varepsilon_i^*\right\| ^2\right)^{1/2}\notag\\
    &\times \left\|T^{-1}\widehat{\*Q}_0^{\prime}\widehat{\*F} \right\|\sqrt{N}\left\| (T^{-1}\widehat{\*F}^{*\prime}\widehat{\*F}^*)^{-1}-(T^{-1}\widehat{\*F}'\widehat{\*F})^{-1}\right\|=O_{P^*}(C_{N,T})^{-1},
 \end{align}
 and so 
 \begin{align}
     \*I_{2,3,2}&= \sqrt{\frac{T}N{}}\frac{1}{N}\sum_{i=1}^N\*S_\*b'\mathcal{\+P}_{0,1,i}^{\prime}T^{-1}\widehat{\*Q}_0^{\prime}\*Q\overline{\*C}(T^{-1}\overline{\*C}'\*Q'\*Q\overline{\*C})^{-1}  \begin{bmatrix}
      T^{-1}\*u_{i}^{*\prime}\+\varepsilon^*_i&\*0_{1\times (2K-1)}\end{bmatrix}'+o_{P^*}(1)\notag\\
     &= \sqrt{\frac{T}N{}}\frac{1}{N}\sum_{i=1}^N\*S_\*b'\*C_i'T^{-1}\*Q'_{\widehat{\alpha}}\*Q\overline{\*C}(T^{-1}\overline{\*C}'\*Q'\*Q\overline{\*C})^{-1}  \begin{bmatrix}
      T^{-1}\*u_{i}^{*\prime}\+\varepsilon^*_i&\*0_{1\times (2K-1)}\end{bmatrix}'+o_{P^*}(1)\notag\\
      &=\kappa^{-1/2}\*S_\*b'\*C'T^{-1}\*Q'_{\widehat{\alpha}}\*Q\*C(\*C'\+\Sigma_\*q\*C)^{-1}[\sigma^2, \*0_{1\times (2K-1)}]' + o_{P^*}(1). 
 \end{align}
 Next, $\mathbf{I}_{2,3,3}=\mathbf{I}_{2,3,3,a}+\mathbf{I}_{2,3,3,b}+\mathbf{I}_{2,3,3,c}+\mathbf{I}_{2,3,3,d}$ has nearly identical properties:
 \begin{align}
     \left\|\mathbf{I}_{2,3,3,a} \right\|&=\left\| \frac{1}{\sqrt{NT}}\sum_{i=1}^N\*S_\*b'\mathcal{\+P}_{2}'T^{-1}\+\kappa_i'\overline{\*W}^*(T^{-1}\widehat{\*F}^{*\prime}\widehat{\*F}^*)^{-1}\overline{\*W}^{*\prime}\+\varepsilon_i^*\right\|\notag\\
    &\leq N^{-1/2}\left\|\*S_\*b \right\|\left\|\mathcal{\+P}_2 \right\|\left(\frac{1}{N}\sum_{i=1}^N \left\|\sqrt{N}T^{-1}\overline{\*W}^{*\prime}\+\varepsilon_i^* \right\|^2\right)^{1/2}\left(\frac{1}{N}\sum_{i=1}^N\left\|\sqrt{N}T^{-1/2}\+\kappa_i'\overline{\*W}^* \right\|^2 \right)^{1/2}\notag\\
    &\times \left\| (T^{-1}\widehat{\*F}^{*\prime}\widehat{\*F}^*)^{-1}\right\|=O_{P^*}(N^{-1}) + O_{P^*}((NT)^{-1/2}),
 \end{align}
 and 
 \begin{align}
     \*I_{2,3,3,b}=\*S_\*b'\mathcal{\+P}_2'\sqrt{\frac{T}{N}}\frac{1}{N}\sum_{i=1}^NT^{-1}\+\kappa_i'\*Q\overline{\*C}(T^{-1}\overline{\*C}'\*Q'\*Q\overline{\*C})^{-1} \left[T^{-1}\*u_{i}^{*\prime}\+\varepsilon^*_i, \*0_{1\times (2K-1)} \right]'   + o_{P^*}(1),
 \end{align}
 which follows the exact same steps as $\*I_{2,3,2,b}$. Then
 \begin{align}
    \left\|\mathbf{I}_{2,3,3,c} \right\|&=\left\|\frac{1}{\sqrt{NT}}\sum_{i=1}^N\*S_\*b'\mathcal{\+P}_{2}'T^{-1}\+\kappa_i'\overline{\*W}^*(T^{-1}\widehat{\*F}^{*\prime}\widehat{\*F}^*)^{-1}\widehat{\*F}'\+\varepsilon_i^* \right\|\notag\\
    &\leq \left\| \*S_\*b \right\|\left\| \mathcal{\+P}_{2}\right\|\left(\frac{1}{N}\sum_{i=1}^N \left\|T^{-1/2}\widehat{\*F}' \+\varepsilon_{i}^* \right\|^2\right)^{1/2}\left(\frac{1}{N}\sum_{i=1}^N\left\|\sqrt{N}T^{-1}\+\kappa_i'\overline{\*W}^* \right\|^2 \right)^{1/2}\notag\\
    &\times \left\| (T^{-1}\widehat{\*F}^{*\prime}\widehat{\*F}^*)^{-1}\right\|=O_{P^*}(T^{-1/2})
 \end{align}
 and 
 \begin{align}
     \left\|\mathbf{I}_{2,3,3,d} \right\|&=\left\|\frac{1}{\sqrt{NT}}\sum_{i=1}^N\*S_\*b'\mathcal{\+P}_{2}'T^{-1}\+\kappa_i'\widehat{\*F}((T^{-1}\widehat{\*F}^{*\prime}\widehat{\*F}^*)^{-1}-(T^{-1}\widehat{\*F}'\widehat{\*F})^{-1})\widehat{\*F}'\+\varepsilon_i^*\right\|\notag\\
    &\leq \left\| \*S_\*b \right\|\left\| \mathcal{\+P}_{2}\right\|\left(\frac{1}{N}\sum_{i=1}^N \left\|T^{-1/2}\widehat{\*F}' \+\varepsilon_i^* \right\|^2\right)^{1/2}\left(\frac{1}{N}\sum_{i=1}^N\left\|T^{-1}\+\kappa_i'\widehat{\*F}\right\|^2 \right)^{1/2}\notag\\
    &\times  \sqrt{N}\left\| (T^{-1}\widehat{\*F}^{*\prime}\widehat{\*F}^*)^{-1}-(T^{-1}\widehat{\*F}'\widehat{\*F})^{-1}\right\|=O_{P^*}(C_{N,T}^{-1}),
 \end{align}
 in total giving 
 \begin{align}
     \mathbf{I}_{2,3,3}& = \*S_\*b'\mathcal{\+P}_2'\sqrt{\frac{T}{N}}\frac{1}{N}\sum_{i=1}^NT^{-1}\+\kappa_i'\*Q\overline{\*C}(T^{-1}\overline{\*C}'\*Q'\*Q\overline{\*C})^{-1} \left[T^{-1}\*u_{i}^{*\prime}\+\varepsilon^*_i, \*0_{1\times (2K-1)} \right]'   + o_{P^*}(1)\notag\\
     &= \kappa^{-1/2}\*S_\*b'\mathcal{\+P}_2'\+\Sigma_{\*q\+\kappa}'\*C(\*C'\+\Sigma_\*q\*C)^{-1}[\sigma^2, \*0_{1\times (2K-1)} ]'   + o_{P^*}(1),
 \end{align}
where we let $\+\Sigma_{\*q\+\kappa}=\plim_{(N,T)\to \infty}\frac{1}{N}\sum_{i=1}^NT^{-1}\*Q'\+\kappa_i$. Notice that as before $\mathbf{I}_{2,3,4}=\mathbf{I}_{2,3,4,a}+\mathbf{I}_{2,3,4,b}+\mathbf{I}_{2,3,4,c}+\mathbf{I}_{2,3,4,d}$ follows the structure of $\mathbf{I}_{2,3,2}$, but it is of a lower order due to $\left\|\mathcal{\+P}_{0,3,i} \right\|=O_P(T^{-1/2})$ for each $i$. Therefore, 
 \begin{align}
     \left\|\*I_{2,3,4} \right\|=O_{P^*}(T^{-1/2}). 
 \end{align}
 Ultimately, we go to the last term $\mathbf{I}_{2,3,5}=\mathbf{I}_{2,3,5,a}+\mathbf{I}_{2,3,5,b}+\mathbf{I}_{2,3,5,c}+\mathbf{I}_{2,3,5,d}$, where
 \begin{align}
     \left\|\mathbf{I}_{2,3,5,a} \right\|&=\left\| \frac{1}{\sqrt{NT}}\sum_{i=1}^N\*S_\*b'T^{-1}\*U_i^{*\prime}\overline{\*W}^*(T^{-1}\widehat{\*F}^{*\prime}\widehat{\*F}^*)^{-1}\overline{\*W}^{*\prime}\+\varepsilon_i\right\|\notag\\
    &\leq \sqrt{\frac{T}{N}}\left(\frac{1}{N}\sum_{i=1}^N\left\|\sqrt{N}T^{-1} \overline{\*W}^{*\prime}\+\varepsilon_i^*\right\|^2 \right)^{1/2}\underbrace{\left(\frac{1}{N}\sum_{i=1}^N\left\|\sqrt{N}T^{-1}\*S_\*b'\*U_i^{*\prime}\overline{\*W}^* \right\|^2 \right)^{1/2}}_{O_{P^*}(C_{N,T}^{-1/2})}\left\|(T^{-1}\widehat{\*F}^{*\prime}\widehat{\*F}^*)^{-1} \right\|\notag\\
    &=O_{P^*}(C_{N,T}^{-3/2}),
 \end{align}
 and 
 \begin{align}
\left\|\mathbf{I}_{2,3,5,b}\right\|&=\left\| \frac{1}{\sqrt{NT}}\sum_{i=1}^N\*S_\*b'T^{-1}\*U_i^{*\prime}\widehat{\*F}(T^{-1}\widehat{\*F}^{*\prime}\widehat{\*F}^*)^{-1}\overline{\*W}^{*\prime}\+\varepsilon_i^*\right\|\notag\\
    &=\underbrace{\left( \frac{1}{N}\sum_{i=1}^N \left\|\sqrt{N}T^{-1/2} \overline{\*W}^{*\prime}\+\varepsilon_i^*\right\|^2\right)^{1/2}}_{O_{P^*}(1)}\left(\frac{1}{N}\sum_{i=1}^N\left\|T^{-1}\*S_\*b'\*U_i^{*\prime}\widehat{\*F}\ \right\|^2 \right)^{1/2}\left\|(T^{-1}\widehat{\*F}^{*\prime}\widehat{\*F}^*)^{-1} \right\|=O_{P^*}(T^{-1/2})
 \end{align}
 under $TN^{-1}=O(1)$, which follows from the analysis in (\ref{I_i_232b}).  Also,
 \begin{align}
     \left\|\mathbf{I}_{2,3,5,c}\right\|&=\left\| \frac{1}{\sqrt{NT}}\sum_{i=1}^N\*S_\*b'T^{-1}\*U_i^{*\prime}\overline{\*W}^*(T^{-1}\widehat{\*F}^{*\prime}\widehat{\*F}^*)^{-1}\widehat{\*F}'\+\varepsilon_i^*\right\|\notag\\
    &\leq \left(\frac{1}{N}\sum_{i=1}^N\left\|T^{-1/2}\widehat{\*F}'\+\varepsilon_i^* \right\|^2 \right)^{1/2}\left(\frac{1}{N}\sum_{i=1}^N\left\|\sqrt{N}T^{-1}\*S_\*b'\*U_i^{*\prime}\overline{\*W}^* \right\|^2 \right)^{1/2}\left\| (T^{-1}\widehat{\*F}^{*\prime}\widehat{\*F}^*)^{-1}\right\|\notag\\
    &=O_{P^*}(C_{N,T}^{-1/2}),
 \end{align}
 and 
 \begin{align}
     \left\|\mathbf{I}_{2,3,5,d}\right\|&=\left\| \frac{1}{\sqrt{NT}}\sum_{i=1}^N\*S_\*b'T^{-1}\*U_i^{*\prime}\widehat{\*F}((T^{-1}\widehat{\*F}^{*\prime}\widehat{\*F}^*)^{-1}-(T^{-1}\widehat{\*F}'\widehat{\*F})^{-1})\widehat{\*F}'\+\varepsilon_i^*\right\|\notag\\
      &\leq \left(\frac{1}{N}\sum_{i=1}^N\left\|T^{-1/2} \widehat{\*F}'\+\varepsilon_i^*\right\|^2 \right)^{1/2}\left(\frac{1}{N}\sum_{i=1}^N\left\|T^{-1}\*S_\*b'\*U_i^{*\prime}\widehat{\*F}\right\|^2 \right)^{1/2}\notag\\
      &\times \sqrt{N}\left\| (T^{-1}\widehat{\*F}^{*\prime}\widehat{\*F}^*)^{-1}-(T^{-1}\widehat{\*F}'\widehat{\*F})^{-1}\right\|=O_{P^*}(T^{-1})+O_{P^*}((NT)^{-1/2}).
 \end{align}
 In summary,
 \begin{align}
     \left\|\mathbf{I}_{2,3,5}\right\|=O_{P^*}(C_{N,T}^{-1/2}),
 \end{align}
 which is sufficient for us, and it completes the proof. \\

 \noindent In the end, we will clean the expressions and merge similar terms. Starting from $\mathbf{II}$, using $\sqrt{N}T^{-1/2}\*Q'\overline{\+\varepsilon}^*=\frac{1}{\sqrt{N}}\sum_{i=1}^NT^{-1/2}\*Q'\+\varepsilon_i^*$, and defining $\frac{1}{N}\sum_{i=1}^N\widetilde{\gamma}_{y,i,1}\+\kappa_i=\overline{\+\kappa}_{\widetilde{\gamma}_{y,1}}$, and  $\+\Theta_{\gamma_{y,1},\alpha_0}=\+\Sigma_{\*q_{\alpha_0}}'\+\Sigma_{\gamma_{y,1}\*C}+\+\Sigma_{\gamma_{y,1}\*q\+\kappa}\mathcal{\+P}_{0,2}$, where $\+\Sigma_{\*q_{\alpha_0}}=\plim_{(N,T)\to \infty}T^{-1}\*Q_{\widehat{\alpha}}'\*Q$ and $\mathcal{\+P}_{0,2}=\plim_{(N,T)\to \infty}\mathcal{\+P}_2$, we have 
 \begin{align}
     \mathbf{II}&=\*S_\*b'\+\Sigma_{\gamma_{y,1}\*C}'\frac{1}{\sqrt{N}}\sum_{i=1}^NT^{-1/2}\*Q'_{\widehat{\alpha}}\+\varepsilon_i^*+\*S_\*b'\mathcal{\+P}_{0,2}^{\prime}\frac{1}{\sqrt{N}}\sum_{i=1}^NT^{-1/2}\overline{\+\kappa}_{\widetilde{\gamma}_{y,1}}'\+\varepsilon_i^*\notag\\
     &-\*S_\*b'\+\Theta_{\gamma_{y,1},\alpha_0}'\*C(\*C'\+\Sigma_\*q\*C)^{-1}\*C'\frac{1}{\sqrt{N}}\sum_{i=1}^NT^{-1/2}\*Q'\+\varepsilon_i^*\notag\\
      &-\kappa^{-1/2}\*S_\*b'\+\Sigma_{\gamma_{y,1}\*C}'\+\Sigma_{\*q_{\alpha_0}}\*C(\*C'\+\Sigma_\*q\*C)^{-1}[\sigma^2, \*0_{1\times (2K-1)}]'\notag\\
        &-\kappa^{-1/2}\*S_\*b'\mathcal{\+P}_{0,2}^{\prime}\+\Sigma_{\gamma_{y,1}\*q\+\kappa}'\*C(\*C'\+\Sigma_\*q\*C)^{-1}[\sigma^2, \*0_{1\times (2K-1)}]'\notag\\
        &+o_{P^*}(1).
 \end{align}
 In case of $\mathbf{III}$, we will use analogous definitions, where we substitute $\widetilde{\gamma}_{y,1}$ to $\widetilde{\gamma}_{y,2}$, and $\gamma_{y,1}$ to $\gamma_{y,2}$. We also use $\sqrt{N}T^{-1/2}\*Q'\overline{\*u}_{-1}^*=\frac{1}{\sqrt{N}}\sum_{i=1}^NT^{-1/2}\*Q'\*u_{i,-1}^*$, and thus 
 \begin{align}
     \mathbf{III}&=\*S_\*b'\+\Sigma_{\gamma_{y,2}\*C}'\frac{1}{\sqrt{N}}\sum_{i=1}^NT^{-1/2}\*Q'_{\widehat{\alpha}}\*u_{i,-1}^*+\*S_\*b'\mathcal{\+P}_{0,2}^{\prime}\frac{1}{\sqrt{N}}\sum_{i=1}^NT^{-1/2}\overline{\+\kappa}_{\widetilde{\gamma}_{y,2}}'\*u_{i,-1}^*\notag\\
     &-\*S_\*b'\+\Theta_{\gamma_{y,2},\alpha_0}'\*C(\*C'\+\Sigma_\*q\*C)^{-1}\*C'\frac{1}{\sqrt{N}}\sum_{i=1}^NT^{-1/2}\*Q'\*u_{i,-1}^*\notag\\
       &-\kappa^{-1/2}\frac{\sigma^2}{1-(\alpha_0)^2}\*S_\*b'\+\Sigma_{\gamma_{y,2}\*C}'\+\Sigma_{\*q_{\alpha_0}}\*C(\*C'\+\Sigma_\*q\*C)^{-1}[
        \alpha_0 , \*0_{1\times k}, 1, \*0_{1\times k}   ]'\notag\\
        &-\kappa^{-1/2}\frac{\sigma^2}{1-(\alpha_0)^2}\*S_\*b'\mathcal{\+P}_{0,2}^{\prime}\+\Sigma_{\gamma_{y,2}\*q\+\kappa}'\*C(\*C'\+\Sigma_\*q\*C)^{-1} [
        \alpha_0 , \*0_{1\times k}, 1, \*0_{1\times k}   ]'\notag\\
        &+ \kappa^{-1/2}\frac{1}{1-(\alpha_0)^2}\left[\gamma_{y,2}, \*0_{1\times k}\right]'+ o_{P^*}(1).
 \end{align}
 Lastly, letting $\+\Psi_{\alpha_0}=\mathrm{vec}(\+\Sigma_{\*q_{\alpha_0}}\*C(\*C'\+\Sigma_\*q\*C)^{-1}\*C')\otimes \*S_\*b$ and $\+\Psi_{\mathcal{\+P}_{0,2}}= \mathrm{vec}(\*C(\*C'\+\Sigma_\*q\*C)^{-1}\*C')\otimes \mathcal{\+P}_{0,2}\*S_\*b $, we get 
 \begin{align}
     \*I&=\*S_\*b'\frac{1}{\sqrt{NT}}\sum_{i=1}^N\*C_i'\*Q_{\widehat{\alpha}}^{\prime}\+\varepsilon_i^* + \*S_\*b'\mathcal{\+P}_{0,2}^{\prime}\frac{1}{\sqrt{NT}}\sum_{i=1}^N\+\kappa_i'\+\varepsilon_i^*+\frac{1}{\sqrt{NT}}\sum_{i=1}^N\mathcal{\+V}_i^{*\prime}\+\varepsilon_i^*\notag\\
     &-\+\Psi_{\alpha_0}'\frac{1}{\sqrt{N}}\sum_{i=1}^N\mathrm{vec}\left(T^{-1/2}\+\varepsilon_i^{*\prime}\*Q \otimes \*C_i' \right)\notag\\
     &-\+\Psi_{\mathcal{\+P}_{0,2}}'\frac{1}{\sqrt{N}}\sum_{i=1}^N\mathrm{vec}\left(T^{-1/2}\+\varepsilon_i^{*\prime}\*Q \otimes \+\Sigma_{\*q\+\kappa,i}' \right)\notag\\
        &- \sqrt{\kappa}\sum_{h=1}^\infty\+\Sigma_{\varepsilon\mathcal{\+v}}(-h)'\mathrm{tr}\left(\+\Sigma_\*q(h)\*C(\*C'\+\Sigma_\*q\*C)^{-1}\*C'\right)\notag\\
        &-\kappa^{-1/2}\*S_\*b'\*C'\+\Sigma_{\*q_{\alpha_0}}\*C(\*C'\+\Sigma_\*q\*C)^{-1}[\sigma^2, \*0_{1\times (2K-1)}]'\notag\\
        &-\kappa^{-1/2}\*S_\*b'\mathcal{\+P}_{0,2}^{\prime}\+\Sigma_{\*q\+\kappa}'\*C(\*C'\+\Sigma_\*q\*C)^{-1}[\sigma^2, \*0_{1\times (2K-1)} ]'   + o_{P^*}(1),
 \end{align}
 which we will use in the derivation of the asymptotic distribution. 
 \paragraph {Expansion and Rates of the Denominator}\label{4.3.5}
 \noindent \textbf{Lemma X2.} \textit{Under Assumptions \ref{ass::1} - \ref{ass::6}, with either multiplicative or non-multiplicative weights, we have as $(N,T)\to \infty$, the denominator can be expanded as 
 \begin{align*}
\+\Sigma_{NT,D}^*=\frac{1}{NT}\sum_{i=1}^N\*B_i^{*\prime}\*M_{\widehat{\*F}^*}\*B_i^*&=\frac{1}{N}\sum_{i=1}^N\*S_\*b'\*C_i'(T^{-1}\*Q_{\widehat{\alpha}}'\*M_{\*Q\overline{\*C}}\*Q_{\widehat{\alpha}})\*C_i\*S_\*b+  \frac{1}{N}\sum_{i=1}^NT^{-1}\mathcal{\+V}^{*\prime}_i\mathcal{\+V}_i^*\notag\\
      &+\frac{1}{NT}\sum_{i=1}^N\*S_\*b\mathcal{\+P}_2'\+\kappa_i'\*M_{\*Q\overline{\*C}}\+\kappa_i\mathcal{\+P}_2\*S_\*b+\frac{1}{NT}\sum_{i=1}^N\*S_\*b'\*C_i'\*Q_{\widehat{\alpha}}'\*M_{\*Q\overline{\*C}}\+\kappa_i\mathcal{\+P}_2\*S_\*b \notag\\
      &+  \left( \frac{1}{NT}\sum_{i=1}^N\*S_\*b'\*C_i'\*Q_{\widehat{\alpha}}'\*M_{\*Q\overline{\*C}}\+\kappa_i\mathcal{\+P}_2\*S_\*b \right)'+\frac{1}{NT}\sum_{i=1}^N\*S_\*b'\mathcal{\+P}_2'\+\kappa_i'\mathcal{\+V}_i^* \notag\\
      &+ \left(\frac{1}{NT}\sum_{i=1}^N\*S_\*b'\mathcal{\+P}_2'\+\kappa_i'\mathcal{\+V}_i^* \right)' + o_{P^*}(1). 
 \end{align*}
Additionally, if $\+\theta_0=\*0_{k\times 1}$, the limiting denominator is equivalent to the one in \cite{DeVos2019} under our assumptions. }
 \bigskip 

 \noindent \textbf{Proof.} Using $ \*B_{i}^*=[\*y_{i,-1}^*, \*X_i]=\*D_i^*\*S_\*b= (\widehat{\*A}_{\*z_0,i}[ \widehat{\*A}, \*I_{K\times K}]+\widehat{\*Q}_0\mathcal{\+P}_{0,1,i}+\+\kappa_i\mathcal{\+P}_2+\*H\mathcal{\+P}_{0,3,i}+\*U_i^*)\*S_\*b$, it is clear that the denominator will contain 25 terms to analyze. However, based on the results from Lemma X3, we can immediately infer that terms that involve the initial value component or $\mathcal{\+P}^{0}_{3,i}$ or both will be asymptotically negligible. This reduces the number of important components to 9 only. In particular, 
 \begin{align}
    \+\Sigma_{NT,D}^*= \frac{1}{NT}\sum_{i=1}^N\*B_i^{*\prime}\*M_{\widehat{\*F}^*}\*B_i^*&=\frac{1}{NT}\sum_{i=1}^N\*S_\*b'\mathcal{\+P}^{0\prime }_{1,i}\widehat{\*Q}_0^{\prime}\*M_{\widehat{\*F}^*}\widehat{\*Q}_0\mathcal{\+P}^{0}_{1,i}\*S_\*b + \frac{1}{NT}\sum_{i=1}^N\*S_\*b'\mathcal{\+P}^{0\prime }_{1,i}\widehat{\*Q}_0^{\prime}\*M_{\widehat{\*F}^*}\+\kappa_i\mathcal{\+P}_2\*S_\*b\notag\\
     &+\frac{1}{NT}\sum_{i=1}^N\*S_\*b'\mathcal{\+P}^{0\prime }_{1,i}\widehat{\*Q}_0^{\prime}\*M_{\widehat{\*F}^*}\*U_i^*\*S_\*b + \frac{1}{NT}\sum_{i=1}^N\*S_\*b'\mathcal{\+P}'_2\+\kappa_i'\*M_{\widehat{\*F}^*}\+\kappa_i\mathcal{\+P}_2\*S_\*b\notag\\
     &+ \frac{1}{NT}\sum_{i=1}^N\*S_\*b'\mathcal{\+P}'_2\+\kappa_i'\*M_{\widehat{\*F}^*}\*U_i^*\*S_\*b + \frac{1}{NT}\sum_{i=1}^N\*S_\*b'\*U_i^{*\prime}\*M_{\widehat{\*F}^*}\*U_{i}^*\*S_\*b\notag\\
     &+\left(\frac{1}{NT}\sum_{i=1}^N\*S_\*b'\mathcal{\+P}^{0\prime }_{1,i}\widehat{\*Q}_0^{\prime}\*M_{\widehat{\*F}^*}\+\kappa_i\mathcal{\+P}_2\*S_\*b \right)' + \left( \frac{1}{NT}\sum_{i=1}^N\*S_\*b'\mathcal{\+P}^{0\prime }_{1,i}\widehat{\*Q}_0^{\prime}\*M_{\widehat{\*F}^*}\*U_i^*\*S_\*b\right)'\notag\\
     &+\left(\frac{1}{NT}\sum_{i=1}^N\*S_\*b'\mathcal{\+P}'_2\+\kappa_i'\*M_{\widehat{\*F}^*}\*U_i^*\*S_\*b\right)' + o_{P^*}(1)\notag\\
     &=\mathbf{1+2+3+4+5+6+7+8+9}+ o_{P^*}(1). 
 \end{align}
 To proceed, using the expansions in (\ref{P_F-P_QC}) and (\ref{Pf*-Pfhat}) to approximate $\*M_{\widehat{\*F}^*}$, and borrowing many results from the derivations conducted in Lemma X3, we can conclude that $\left\|\*3 \right\|=o_{P^*}(1)$ and $\left\|\*8 \right\|=o_{P^*}(1)$. Next, we can use Lemma XB, to replace $\widehat{\*Q}^{0}\mathcal{\+P}^{0}_{1,i}$ with $\*Q_{\widehat{\alpha}}\*C_i+O_P(\sqrt{T}N^{-1/2})$ in the remaining terms where needed. Also, we can replace $\*U^*_i\*S_{\*b}$ with $\mathcal{\+V}_i^*+O_P(\sqrt{T}N^{-1/2})$. Therefore, the denominator produces
 \begin{align}
     \+\Sigma_{NT,D}^*= \frac{1}{NT}\sum_{i=1}^N\*B_i^{*\prime}\*M_{\widehat{\*F}^*}\*B_i^*&=\frac{1}{N}\sum_{i=1}^N\*S_\*b'\*C_i'(T^{-1}\*Q_{\widehat{\alpha}}'\*M_{\*Q\overline{\*C}}\*Q_{\widehat{\alpha}})\*C_i\*S_\*b+  \frac{1}{N}\sum_{i=1}^NT^{-1}\mathcal{\+V}^{*\prime}_i\mathcal{\+V}_i^*\notag\\
      &+\frac{1}{NT}\sum_{i=1}^N\*S_\*b\mathcal{\+P}_{0,2}^{\prime}\+\kappa_i'\*M_{\*Q\overline{\*C}}\+\kappa_i\mathcal{\+P}_{0,2}\*S_\*b+\frac{1}{NT}\sum_{i=1}^N\*S_\*b'\*C_i'\*Q_{\widehat{\alpha}}'\*M_{\*Q\overline{\*C}}\+\kappa_i\mathcal{\+P}_{0,2}\*S_\*b \notag\\
      &+ \left( \frac{1}{NT}\sum_{i=1}^N\*S_\*b'\*C_i'\*Q_{\widehat{\alpha}}'\*M_{\*Q\overline{\*C}}\+\kappa_i\mathcal{\+P}_{0,2}\*S_\*b \right)'+\frac{1}{NT}\sum_{i=1}^N\*S_\*b'\mathcal{\+P}_{0,2}^{\prime}\+\kappa_i'\mathcal{\+V}_i^* \notag\\
      &+ \left(\frac{1}{NT}\sum_{i=1}^N\*S_\*b'\mathcal{\+P}_{0,2}^{\prime}\+\kappa_i'\mathcal{\+V}_i^* \right)' + o_{P^*}(1),
 \end{align}
 where the last two components are not negligible, because unconditionally $\kappa_{i,t}$ approximates a linear process in terms of $y_{i,t-1}$ and $y_{i,t-2}$ for each $t$, and so it depends on $\+\nu_{i,t-1}, \+\nu_{i,t-2}, \ldots$. The latter sit in $\mathcal{\+v}_{i,t}$ in the form of $\+\nu^{-}_{i,t-1}$. Note that if $\+\theta_0=\*0_{k\times 1}$, only the first two terms remain, and they give a denominator equivalent to the one in \cite{DeVos2019} under $K=R$ and no dynamics in $\*x_{i,t}$. 
 \paragraph {Bootstrap Central Limit Theorem}\label{4.3.6}
 \noindent We collect all the mean zero terms (under the bootstrap measure) from $\*I$, $\mathbf{II}$ and $\mathbf{III}$ from the numerator expansion, and merge the similar ones:  
\begin{align}
    \*b_{NT,0}^*&=\frac{1}{\sqrt{NT}}\sum_{i=1}^N\mathcal{\+V}_i^{*\prime}\+\varepsilon_i^*+ \*S_\*b'\frac{1}{\sqrt{NT}}\sum_{i=1}^N\*C_i'\*Q_{\widehat{\alpha}}^{\prime}\+\varepsilon_i^* + \*S_\*b'\mathcal{\+P}_{0,2}^{\prime}\frac{1}{\sqrt{NT}}\sum_{i=1}^N\+\kappa_i'\+\varepsilon_i^*\notag\\
     &+\*S_\*b'\+\Theta_{\gamma_{y,1},\alpha_0}'\*C(\*C'\+\Sigma_\*q\*C)^{-1}\*C'\frac{1}{\sqrt{N}}\sum_{i=1}^NT^{-1/2}\*Q'\+\varepsilon_i^*
    -\*S_\*b'\+\Sigma_{\gamma_{y,1}\*C}'\frac{1}{\sqrt{N}}\sum_{i=1}^NT^{-1/2}\*Q'_{\widehat{\alpha}}\+\varepsilon_i^*\notag\\
    &-\*S_\*b'\mathcal{\+P}_{0,2}^{\prime}\frac{1}{\sqrt{N}}\sum_{i=1}^NT^{-1/2}\overline{\+\kappa}_{\widetilde{\gamma}_{y,1}}'\+\varepsilon_i^*-\+\Psi_{\alpha_0}'\frac{1}{\sqrt{N}}\sum_{i=1}^N\mathrm{vec}\left(T^{-1/2}\+\varepsilon_i^{*\prime}\*Q \otimes \*C_i' \right)\notag\\
    &-\+\Psi_{\mathcal{\+P}_{0,2}}'\frac{1}{\sqrt{N}}\sum_{i=1}^N\mathrm{vec}\left(T^{-1/2}\+\varepsilon_i^{*\prime}\*Q \otimes \+\Sigma_{\*q\+\kappa,i}' \right)+\*S_\*b'\+\Theta_{\gamma_{y,2},\alpha_0}'\*C(\*C'\+\Sigma_\*q\*C)^{-1}\*C'\frac{1}{\sqrt{N}}\sum_{i=1}^NT^{-1/2}\*Q'\*u_{i,-1}^*\notag\\
    &-\*S_\*b'\+\Sigma_{\gamma_{y,2}\*C}'\frac{1}{\sqrt{N}}\sum_{i=1}^NT^{-1/2}\*Q'_{\alpha}\*u_{i,-1}^*-\*S_\*b'\mathcal{\+P}_{0,2}^{\prime}\frac{1}{\sqrt{N}}\sum_{i=1}^NT^{-1/2}\overline{\+\kappa}_{\widetilde{\gamma}_{y,2}}'\*u_{i,-1}^*\notag\\
    &=\frac{1}{\sqrt{NT}}\sum_{i=1}^N\mathcal{\+V}_i^{*\prime}\+\varepsilon_i^*+\*S_\*b'\frac{1}{\sqrt{NT}}\sum_{i=1}^N(\*C_i-\+\Sigma_{\gamma_{y,1}\*C})'\*Q_{\widehat{\alpha}}^{\prime}\+\varepsilon_i^*+\*S_\*b'\mathcal{\+P}_{0,2}^{\prime}\frac{1}{\sqrt{NT}}\sum_{i=1}^N(\+\kappa_i-\overline{\+\kappa}_{\widetilde{\gamma}_{y,1}})'\+\varepsilon_i^*\notag\\
    &+\*S_\*b'\+\Theta_{\gamma_{y,1},\alpha_0}'\*C(\*C'\+\Sigma_\*q\*C)^{-1}\*C'\frac{1}{\sqrt{N}}\sum_{i=1}^NT^{-1/2}\*Q'\+\varepsilon_i^*-\+\Psi_{\alpha_0}'\frac{1}{\sqrt{N}}\sum_{i=1}^N\mathrm{vec}\left(T^{-1/2}\+\varepsilon_i^{*\prime}\*Q \otimes \*C_i' \right)\notag\\
    &-\+\Psi_{\mathcal{\+P}_{0,2}}'\frac{1}{\sqrt{N}}\sum_{i=1}^N\mathrm{vec}\left(T^{-1/2}\+\varepsilon_i^{*\prime}\*Q \otimes \+\Sigma_{\*q\+\kappa,i}' \right)+\*S_\*b'\+\Theta_{\gamma_{y,2},\alpha_0}'\*C(\*C'\+\Sigma_\*q\*C)^{-1}\*C'\frac{1}{\sqrt{N}}\sum_{i=1}^NT^{-1/2}\*Q'\*u_{i,-1}^*\notag\\
    &-\*S_\*b'\+\Sigma_{\gamma_{y,2}\*C}'\frac{1}{\sqrt{N}}\sum_{i=1}^NT^{-1/2}\*Q'_{\widehat{\alpha}}\*u_{i,-1}^*-\*S_\*b'\mathcal{\+P}_{0,2}^{\prime}\frac{1}{\sqrt{N}}\sum_{i=1}^NT^{-1/2}\overline{\+\kappa}_{\widetilde{\gamma}_{y,2}}'\*u_{i,-1}^*\notag\\
    &=\*b_{NT,0,\+\varepsilon^*}+\*b_{NT,0,\*u_{-1}^*},
\end{align}
where $\*b_{NT,0,\+\varepsilon^*}=\*b_{NT,0,\+\varepsilon^*,1}+\*b_{NT,0,\+\varepsilon^*,2}$ with $\*b_{NT,0,\+\varepsilon^*,1}=\frac{1}{\sqrt{NT}}\sum_{i=1}^N\mathcal{\+V}_i^{*\prime}\+\varepsilon_i^*$, and $\*b_{NT,0,\+\varepsilon^*,2}$ summarizes the rest of the components driven by $\+\varepsilon_i^*$. Likewise, $\*b_{NT,0,\*u_{-1}^*}$ summarizes the components driven by  $\*u_{i,-1}^*$.\\

\noindent \textbf{Lemma X3.} \textit{Under Assumptions \ref{ass::1} - \ref{ass::6} and non-multiplicative weights, we have as $(N,T)\to \infty$
\begin{align*}
    \*b_{NT,0}^*\to_{d^*}\*b_0\overset{d}={}\mathcal{N}\left(\*0_K, \+\Sigma_{\+\varepsilon}+\+\Sigma_{\*u_{-1}} \right),
\end{align*}
where $\+\Sigma_{\*u_{-1}}$ is a covariance matrix generated by $\*b_{NT,0,\*u_{-1}}$, and $\+\Sigma_{\+\varepsilon}$ is generated by $\*b_{NT,0,\+\varepsilon^*}$. 
} \\

\noindent \textbf{Proof.} Clearly, $\*b_{NT,0,\*u_{-1}^*}$ is independent of the first two components under multiplicative or non-multiplicative bootstrap weights. Also, each of the terms in $\*b_{NT,0,\*u_{-1}^*}$ can be written in the form of 
\begin{align*}
    \sum_{j=0}^{T-2}\widehat{\alpha}_{CCEP}^j\frac{1}{\sqrt{NT}}\sum_{i=1}^N\sum_{t=j+2}^T\varepsilon_{i,t-j-1}^*\*a_t,
\end{align*}
 where $\*a_t$ is a placeholder for different time series, for which we can show that $E[\|\*a_t \|^p]$ is uniformly bounded for $p=4+\delta$ ($\delta\in (0,4]$) under our assumptions. For a fixed $m$, $\sum_{j=0}^{m-2}\widehat{\alpha}_{CCEP}^j\frac{1}{\sqrt{NT}}\sum_{i=1}^N\sum_{t=j+2}^T\varepsilon_{i,t-j-1}^*\*a_t$ converges to a normal variable in distribution (conditionally on the sample) by the results in \cite{gonccalves2015bootstrap}, but the fixed-$m$ counterpart asymptotically approximates the true one by an argument similar to (\ref{d_m-d}). Since all 3 parts of $\*b_{NT,0,\*u_{-1}^*}$ have distribution generated by the same $u_{i,t-1}^*$, we have joint convergence, and thus $\*b_{NT,0,\*u_{-1}^*}$ is normal by the Cramer-Wold device. Similarly, asymptotic normality is established for $\*b_{NT,0,\+\varepsilon^*}=\frac{1}{\sqrt{NT}}\sum_{i=1}^N\mathcal{\+V}_i^{*\prime}\+\varepsilon_i^*+\*b_{NT,0,\+\varepsilon^*,2}$, where $\varepsilon_{i,t}^*$ is the common random i.i.d. component conditionally on data, by the same argument as in \cite{Juodis2022CCER} (see a discussion of Theorem 1 therein). We also recall that the first coordinate of $\frac{1}{\sqrt{NT}}\sum_{i=1}^N\mathcal{\+V}_i^{*\prime}\+\varepsilon_i^*$ includes $\frac{1}{\sqrt{NT}}\sum_{i=1}^N\sum_{t=2}^Tu_{i,t-1}^*\varepsilon_{i,t}^*$, whose normality we proved in Theorem 3.1, and it is also independent of the rest of terms. 
\paragraph {Bootstrap Distribution of CCEP: ARX(1)}\label{4.3.7}
We combine the results from Lemmas X1 - X3.\\

\noindent \textbf{Theorem 3.2} \textit{Under conditions of Lemma X1, X2 and X3 with non-multiplicative weights, we have as $(N,T)\to \infty$
\begin{align*}
    \sqrt{NT}(\widehat{\+\delta}_{CCEP}^*-\widehat{\+\delta}_{CCEP})\to_{d^*}\+\Sigma_\*D^{-1}\*b_0+\+\Sigma_\*D^{-1}\left(\sqrt{\kappa}\*b_1-\kappa^{-1/2}\*b_2-\kappa^{-1/2}\*b_3\right),
\end{align*}
where $\*b_0$ is defined in Lemma X3 and $\+\Sigma_\*D$ is the probability limit of $\+\Sigma_{NT,D}^*$. The asymptotic bias terms are 
\begin{align*}
    \*b_1&=-\sum_{h=1}^\infty\+\Sigma_{\varepsilon\mathcal{\+v}}(-h)'\mathrm{tr}\left(\+\Sigma_\*q(h)\*C(\*C'\+\Sigma_\*q\*C)^{-1}\*C'\right), 
\end{align*}
and 
\begin{align*}
    \*b_2&= -\*S_\*b'\*C'\+\Sigma_{\*q_{\alpha_0}}\*C(\*C'\+\Sigma_\*q\*C)^{-1}[\sigma^2, \*0_{1\times (2K-1)}]'\notag\\
        &-\*S_\*b'\mathcal{\+P}_{0,2}^{\prime}\+\Sigma_{\*q\+\kappa}'\*C(\*C'\+\Sigma_\*q\*C)^{-1}[\sigma^2, \*0_{1\times (2K-1)} ]'\notag\\
        &-\*S_\*b'\+\Sigma_{\gamma_{y,1}\*C}'\+\Sigma_{\*q_{\alpha_0}}\*C(\*C'\+\Sigma_\*q\*C)^{-1}[\sigma^2, \*0_{1\times (2K-1)}]'\notag\\
        &-\*S_\*b'\mathcal{\+P}_{0,2}^{\prime}\+\Sigma_{\gamma_{y,1}\*q\+\kappa}'\*C(\*C'\+\Sigma_\*q\*C)^{-1}[\sigma^2, \*0_{1\times (2K-1)}],
\end{align*}
while 
\begin{align*}
    \*b_3&=-\frac{\sigma^2}{1-(\alpha_0)^2}\*S_\*b'\+\Sigma_{\gamma_{y,2}\*C}'\+\Sigma_{\*q_{\alpha_0}}\*C(\*C'\+\Sigma_\*q\*C)^{-1}[
        \alpha_0 , \*0_{1\times k}, 1, \*0_{1\times k}   ]'\notag\\
        &-\frac{\sigma^2}{1-(\alpha_0)^2}\*S_\*b'\mathcal{\+P}_{0,2}^{\prime}\+\Sigma_{\gamma_{y,2}\*q\+\kappa}'\*C(\*C'\+\Sigma_\*q\*C)^{-1} [
        \alpha_0 , \*0_{1\times k}, 1, \*0_{1\times k}   ]'\notag\\
        &+ \frac{1}{1-(\alpha_0)^2}\left[\gamma_{y,2}, \*0_{1\times k}\right]'.
\end{align*} 
}
\\
\noindent \textbf{Proof.} Follows straight from the combination of the 3 lemmas. 

\noindent \begin{remark}
    The bias terms $\*b_1$-$\*b_3$ differ from the original sample counterparts in the following ways: 
    \begin{enumerate}
        \item Current ``Nickel'' bias $\*b_1$ is driven by $\+\Sigma_{\+\varepsilon\mathcal{\+v}}(-h)'=[\sigma^2(\alpha^{0})^{h-1}, \*0_{1\times k}]'$. In the original sample, the zero vector is replaced by the ``Nickell'' bias component that is induced by the weak exogeneity of $\*x_{i,t}$ (``Nickell'' bias of $\+\beta^0$). However, we fail to capture it due to $\*x_{i,t}$ being fixed within the bootstrap iterations. 
        \item Both $\*b_2$ and $\*b_3$ are plagued by many zeros, which occur solely due to $\overline{\*X}$ and $\overline{\*X}_{-1}$ now serving as observed factors. Moreover, $\+\kappa_i$ (accumulated lags of $\*y_{i}$ living in $\*X_i$) should not contribute to the factor estimation bias ($\*b_2$). However, since it fails to generate ``Nickell'' bias in the bootstrap realm, it starts interacting with other bias components as $(N,T)\to \infty$. 
        \item Lastly, $\*b_3=\*0_{K\times 1}$ in the original sample. This term is an artifact of the ``naive'' method to construct the bootstrap DGP (we use $K$ too many factors in the bootstrap DGP). While in the pure pure AR(1) case similar terms disappear asymptotically, they only vanish under $\+\theta_0=\*0_{k\times 1}$ in the ARX1 case (see the detailed discussion below (\ref{gamma_y_2_X})). Corollary 3.1 below confirms this claim.  
    \end{enumerate}
\end{remark}

\noindent \textbf{Corollary 3.1} \textit{Let Theorem 3.2 hold, but set $\+\theta_0=\*0_{k\times 1}$, then as $(N,T)\to \infty$}
\begin{align*}
     \sqrt{NT}(\widehat{\+\delta}_{CCEP}^*-\widehat{\+\delta}_{CCEP})\to_{d^*}\+\Sigma_\*D^{-1}\*b_0+\+\Sigma_\*D^{-1}\left(\sqrt{\kappa}\*b_1-\kappa^{-1/2}\*b_2\right),
\end{align*}
\textit{where }
\begin{align*}
    &\*b_0\overset{d}{=}\mathcal{N}\left(\*0_{K},\+\Omega_{\+\varepsilon} \right),\\
    &\*b_1=\sum_{h=1}^\infty\+\Sigma_{\varepsilon\mathcal{\+v}}(-h)'\mathrm{tr}\left(\+\Sigma_\*q(h)\*C(\*C'\+\Sigma_\*q\*C)^{-1}\*C'\right)\\
        &\*b_2=-[\sigma^2, \*0_{1\times (2K-1)}]'-\*S_\*b'\+\Sigma_{\gamma_{y,1}\*C}'\+\Sigma_\*q\*C(\*C'\+\Sigma_\*q\*C)^{-1}[\sigma^2, \*0_{1\times (2K-1)}]'\\
        &\+\Sigma_\*D=\lim_{(N,T)\to \infty}\frac{1}{N}\sum_{i=1}^N\*S_\*b'\*C_i'\+\Sigma_\*q\*C_i\*S_\*b - \lim_{(N,T)\to\infty }\frac{1}{N}\sum_{i=1}^N\*S_\*b'\*C_i'\+\Sigma_\*q\*C(\*C'\+\Sigma_\*q\*C)^{-1}\*C'\+\Sigma_\*q\*C_i\*S_\*b+\+\Sigma_{\mathcal{\+V}\mathcal{\+V}},
\end{align*}
\textit{with $\+\Omega_{\+\varepsilon}$ is a modification of $\+\Sigma_{\+\varepsilon}$ under $\+\theta_0=\*0_{k\times 1}$ and $\+\Sigma_{\mathcal{\+V}\mathcal{\+V}}=\lim_{(N,T)\to \infty}\frac{1}{NT}\sum_{i=1}^N\sum_{t=2}^TE[\mathcal{\+v}_{0,i,t}\mathcal{\+v}_{0,i,t}^{\prime}]$, where $\mathcal{\+v}_{0,i,t}=[\+\beta_{0}^{\prime}\+\nu_{0,i,t-1}^{-}+u_{i,t-1}, \+\nu_{i,t}']'$ and $\+\nu_{0,i,t}^{-}=\sum_{j=0}^\infty(\alpha^{0})^j\+\nu_{i,t-j}$. The definitions of other matrices change implicitly by setting $\+\theta_0=\*0_{k\times 1}$. Here, $\*b_1$ is the original and only ``Nickel bias'' term that must originate under exogeneous regressors}.  \\

\noindent \textbf{Proof.} Follows immediately by setting $\+\theta_0=\*0_{k\times 1}$, which makes $\mathcal{\+P}_{0,2}$ into a zero matrix and $\widetilde{\gamma}_{y,i,2}=O_P(T^{-1/2}). $ Moreover, we have $\+\Sigma_{\*q_{\alpha_0}}=\+\Sigma_{\*q}$, and thus $\*C'\+\Sigma_{\*q_{\alpha_0}}\*C(\*C'\+\Sigma_\*q\*C)^{-1}=\*I_{2K\times 2K}$. As for the distribution generator, note that under $\+\theta_0=\*0_{k\times 1}$, it collapses to $\*b_{NT,0}^*=\*b_{NT,0,\+\varepsilon^*}$, and $\*b_{NT,0,\+\varepsilon^*}$ is simplified further, which gives 
\begin{align}
    &\*b_{NT,,0}=\frac{1}{\sqrt{NT}}\sum_{i=1}^N\mathcal{\+V}_i^{*\prime}\+\varepsilon_i^*+\*S_\*b'\frac{1}{\sqrt{NT}}\sum_{i=1}^N(\*C_i-\+\Sigma_{\gamma_{y,1}\*C})'\*Q_{\widehat{\alpha}}^{\prime}\+\varepsilon_i^*\notag\\
    &+\*S_\*b'\+\Theta_{\gamma_{y,1},\alpha_0}'\*C(\*C'\+\Sigma_\*q\*C)^{-1}\*C'\frac{1}{\sqrt{N}}\sum_{i=1}^NT^{-1/2}\*Q'\+\varepsilon_i^*-\+\Psi_{\alpha_0}'\frac{1}{\sqrt{N}}\sum_{i=1}^N\mathrm{vec}\left(T^{-1/2}\+\varepsilon_i^{*\prime}\*Q \otimes \*C_i' \right)\notag\\
    &=\frac{1}{\sqrt{NT}}\sum_{i=1}^N\mathcal{\+V}_i^{*\prime}\+\varepsilon_i^*+\*S_\*b'\frac{1}{\sqrt{NT}}\sum_{i=1}^N(\*C_i-\+\Sigma_{\gamma_{y,1}\*C})'\*Q_{\widehat{\alpha}}^{\prime}\+\varepsilon_i^*\notag\\
    &+\*S_\*b'\+\Sigma_{\gamma_{y,1}\*C}'\+\Sigma_\*q\*C(\*C'\+\Sigma_\*q\*C)^{-1}\*C'\frac{1}{\sqrt{N}}\sum_{i=1}^NT^{-1/2}\*Q'\+\varepsilon_i^*-\*S_\*b'\frac{1}{\sqrt{N}}\sum_{i=1}^N\*C_i'\+\Sigma_\*q\*C(\*C'\+\Sigma_\*q\*C)^{-1}\*C')T^{-1/2}\*Q '\+\varepsilon_i^{*}\notag\\
    &=\frac{1}{\sqrt{NT}}\sum_{i=1}^N\mathcal{\+V}_i^{*\prime}\+\varepsilon_i^*+\*S_\*b'\frac{1}{\sqrt{NT}}\sum_{i=1}^N(\*C_i-\+\Sigma_{\gamma_{y,1}\*C})'\*Q_{\widehat{\alpha}}^{\prime}\+\varepsilon_i^*\notag\\
    &-\*S_\*b'\frac{1}{\sqrt{N}}\sum_{i=1}^N(\*C_i-\+\Sigma_{\gamma_{y,1}\*C})'\+\Sigma_\*q\*C(\*C'\+\Sigma_\*q\*C)^{-1}\*C'T^{-1/2}\*Q'\+\varepsilon_i^*\notag\\
    &=\frac{1}{\sqrt{NT}}\sum_{i=1}^N\mathcal{\+V}_i^{*\prime}\+\varepsilon_i^*+\*S_\*b'\frac{1}{\sqrt{N}}\sum_{i=1}^N(\*C_i-\+\Sigma_{\gamma_{y,1}\*C})'(\*I_{K^22R\times K^22R}-\+\Sigma_\*q\*C(\*C'\+\Sigma_\*q\*C)^{-1}\*C')T^{-1/2}\*Q'\+\varepsilon_i^*\notag\\
    &+o_{P^*}(1),
\end{align}
since $\*Q_{\widehat{\alpha}}=\*q_{\alpha_0}+o_P(1)=\*Q+o_P(1)$ under strict exogeneity. Note that the distribution generators and the bias terms coincide with those described in Lemma 14 and Lemma 17 in \cite{DeVos2019}. The differences occur due to 1) $R=K$ (which removes many terms in the latter study), 2) $\*X_i$ being fixed in bootstrap (which induces $2k$ zeros) and 3) the assumption of deterministic loadings (the first component in $\*b_2$ is absent in \cite{DeVos2019}). Lastly, $\*b_1$ corresponds to the true ``Nickell bias'' term that is non-zero for $\widehat{\alpha}_{CCEP}$ only. That is $\+\Sigma_{\+\varepsilon\mathcal{\+v}}(-h)'$ must be equal to $[\sigma^2(\alpha^{0})^{h-1}, \*0_{1\times k}]'$ when $\+\theta_0=\*0_{k\times 1}$.

\paragraph {Comment on Strict Exogeneity and Bootstrap Replication of VAR(1)}\label{4.3.8}
We assume that $\+\theta_0=\*0_{k\times 1}$, so that $\*x_{i,t}$ is strictly exogenous, and we let $\+\nu_{i,t}$ be independent over time. Effectively, (\ref{z_VAR}) becomes the process in \cite{DeVos2019}.  In this case we can re-create bootstrap DGP for both $y_{i,t}$ and $\*x_{i,t}$, which means that the whole VAR(1) process in (\ref{z_VAR}) can be recreated in the recursive bootstrap - this is the key to extend our results from AR(1) case. In particular, in the stacked notation:
\begin{align}
    &\*y_{i}^*=\*y_{i,-1}^*\widehat{\alpha}_{CCEP}+\*X_{i}^*\widehat{\+\beta}_{CCEP}+\widehat{\*F}\widehat{\+\gamma}_{y,i}+\+\varepsilon_{i}^*=\*B_{i}^*\widehat{\+\delta}_{CCEP}+\widehat{\*F}\widehat{\+\gamma}_{y,i}+\+\varepsilon_{i}^*\\
    &\*X_{i}^*=\widehat{\*F}\widehat{\+\Gamma}_{x,i}+\+\nu_{i}^*,\\
    &\*Z_i^*=\*Z_{i,-1}^*\widehat{\*A}+\widehat{\*F}\widehat{\+\Gamma}_i+\*E_i^*,\\
    &\widehat{\*A}=\begin{bmatrix}
        \widehat{\alpha}_{CCEP} & \*0_{1\times k}\\
        \*0_{k\times 1} & \*0_{k\times k}
    \end{bmatrix},
\end{align}
such that $\widehat{\*A}=\widehat{\*A}'$. Here, $\varepsilon_{i,t}^*=\omega_{1,i,t}\widehat{\varepsilon}_{i,t}$ and $\nu_{i,t}^*=\omega_{2,i,t}\widehat{\nu}_{i,t}$, where for $j=1,2$ we have $\omega_{j,i,t}\sim i.i.d.(0,1)$, also independent for $j\neq j'$. Next, $\*Z_i^*=[\*y_i^*,\*X_i^*]$, $\widehat{\+\Gamma}_i=[\widehat{\+\Gamma}_{x,i}\widehat{\+\beta}_{CCEP}+\widehat{\+\gamma}_{y,i}, \widehat{\+\Gamma}_{x,i}]$ and $\*E_i^*=[\+\nu_{i}^*\widehat{\+\beta}_{CCEP}+\+\varepsilon^*_i, \+\nu_i^*]$, where
\begin{align}
&\widehat{\+\varepsilon}_i=\*M_{\widehat{\*F}}(\*y_{i}-\widehat{\alpha}_{CCEP}\*y_{i,-1}-\*X_{i}\widehat{\+\beta}_{CCEP}), \\
   & \widehat{\+\gamma}_{y,i}=(\widehat{\*F}'\widehat{\*F})^{-1}\widehat{\*F}'(\*y_{i}-\widehat{\alpha}_{CCEP}\*y_{i,-1}-\*X_{i}\widehat{\+\beta}_{CCEP}),\\
   &\widehat{\+\nu}_i=\*M_{\widehat{\*F}}\*X_i, \\
   &\widehat{\+\Gamma}_{x,i}=(\widehat{\*F}'\widehat{\*F})^{-1}\widehat{\*F}'\*X_i,\\
   &\widehat{\+\Gamma}_i=(\widehat{\*F}'\widehat{\*F})^{-1}\widehat{\*F}'(\*Z_i-\*Z_{i,-1}\widehat{\*A}),
\end{align}
where we conveniently have that
\begin{align}
    &\overline{\widehat{\+\Gamma}}=(\widehat{\*F}'\widehat{\*F})^{-1}\widehat{\*F}'(\overline{\*Z}-\overline{\*Z}_{-1}\widehat{\*A})=(\widehat{\*F}'\widehat{\*F})^{-1}\widehat{\*F}'\widehat{\*F}\widehat{\mathbb{A}}=\widehat{\mathbb{A}}\\
    &\widehat{\*F}\overline{\widehat{\+\Gamma}}= \widehat{\*F}\widehat{\mathbb{A}}=\widehat{\*F}_{\widehat{\mathbb{A}}}.\label{hat_F_A}
\end{align}
Next, we introduce the rotation matrix $\mathbb{R}=\begin{bmatrix} \*I_{K\times K} & \*0_{K\times K}\\
-\widehat{\*A} & \*I_{K\times K}\end{bmatrix}\in \mathbb{R}^{2K\times 2K}$, so that
\begin{align}
    \*Z_i^*&=\*Z_{i,-1}^*\widehat{\*A}+\widehat{\*F}\widehat{\+\Gamma}_i+\*E_i^*\notag\\
    &=\*Z_{i,-1}^*\widehat{\*A}+\widehat{\*F}\mathbb{R}\mathbb{R}^{-1}\widehat{\+\Gamma}_i+\*E_i^*\notag\\
    &=\*Z_{i,-1}^*\widehat{\*A}+\underbrace{(\overline{\*Z}-\overline{\*Z}_{-1}\widehat{\*A})}_{\widehat{\*F}_{\widehat{\mathbb{A}}}}\widetilde{\+\Gamma}_{1,i}+\overline{\*Z}_{-1}\widetilde{\+\Gamma}_{2,i}+\*E_i^*\notag\\
    &=\*Z_{i,-1}^*\widehat{\*A}+\widehat{\*F}_{\widehat{\mathbb{A}}}\widetilde{\+\Gamma}_{1,i}+\overline{\*Z}_{-1}\widetilde{\+\Gamma}_{2,i}+\*E_i^*,
\end{align}
and so (\ref{hat_F_A}) implies that the average of $\widetilde{\+\Gamma}_{1,i}$ is $\*I_{K\times K}$ and that of $\widetilde{\+\Gamma}_{2,i}$ is $\*0_{K\times K}$. Thus, we can proxy the true bootstrap factors by the bootstrap CAs:
\begin{align}
    \widehat{\*F}_{\widehat{\mathbb{A}}}=\overline{\*Z}^*-\overline{\*Z}^*_{-1}\widehat{\*A}-\overline{\*E}^*.
\end{align}
Then, by using the rotation $\mathbb{R}$, we can re-write $\*y_i^*$ as
\begin{align}
    \*y_i^*=\*B_{i}^*\widehat{\+\delta}_{CCEP}+\widehat{\*F}\widehat{\+\gamma}_{y,i}+\+\varepsilon_i^*&=\*B_{i}^*\widehat{\+\delta}_{CCEP}+\widehat{\*F}_{\widehat{\mathbb{A}}}\widetilde{\+\gamma}_{1,y,i}+\overline{\*Z}_{-1}\widetilde{\+\gamma}_{2,y,i}+\+\varepsilon_i^*\notag\\
    &=\*B_{i}^*\widehat{\+\delta}_{CCEP}+(\overline{\*Z}^*-\overline{\*Z}_{-1}^*\widehat{\*A})\widetilde{\+\gamma}_{1,y,i}-\overline{\*E}^*\widetilde{\+\gamma}_{1,y,i}+\overline{\*Z}_{-1}\widetilde{\+\gamma}_{2,y,i}+\+\varepsilon_i^*.
\end{align}
To proceed, we examine the structure of how bootstrap CAs evolve over time. Switching to vector notation again, and using the available $y_{i,0}$, we have that 
¨\begin{align}
    \overline{\*z}_1^*&=\begin{bmatrix}
        \widehat{\alpha}_{CCEP} & \*0_{1\times k}\\ \*0_{k\times 1} &\*0_{k\times k}
    \end{bmatrix}\begin{bmatrix}
        \overline{y}_0\\
        \overline{\*x}_0
    \end{bmatrix} - \left( \begin{bmatrix} \overline{y}_1\\
    \overline{\*x}_1\end{bmatrix}+\begin{bmatrix}
        \widehat{\alpha}_{CCEP} & \*0_{1\times k}\\ \*0_{k\times 1} &\*0_{k\times k}
    \end{bmatrix} \begin{bmatrix}
        \overline{y}_0\\
        \overline{\*x}_0
    \end{bmatrix}\right) +\begin{bmatrix}
        \widehat{\+\beta}_{CCEP}'\overline{\+\nu}_1^*+\overline{\varepsilon}_{1}^*\\
        \overline{\+\nu}_1^*
    \end{bmatrix}\notag\\
    &=\begin{bmatrix} \overline{y}_1\\
    \overline{\*x}_1\end{bmatrix}+\begin{bmatrix}
        \widehat{\+\beta}_{CCEP}'\overline{\+\nu}_1^*+\overline{\varepsilon}_{1}^*\\
        \overline{\+\nu}_1^*
    \end{bmatrix}=\overline{\*z}_1+\overline{\*e}_1^*,
\end{align}
where, similarly to the pure AR(1) case, we uncover that for every $t$:
\begin{align}
\overline{\*z}_t^*=\overline{\*z}_t+\sum_{j=0}^{t-1}\widehat{\*A}^j\overline{\*e}_{t-j}^*=\overline{\*z}_t+\overline{\*u}_t^*,
\end{align}
which implies that $\overline{\*Z}_{-1}=\overline{\*Z}_{-1}^*-\overline{\*U}_{-1}^*$. Notice that the initial value $x_{i,0}$ is used here only innocuously. Therefore,
\begin{align}\label{CCEP_boot_x}
   \*y_i^*= \*B_{i}^*\widehat{\+\delta}_{CCEP}+(\overline{\*Z}^*-\overline{\*Z}_{-1}^*\widehat{\*A})\widetilde{\+\gamma}_{1,y,i}-\overline{\*E}^*\widetilde{\+\gamma}_{1,y,i}+(\overline{\*Z}^*_{-1}-\overline{\*U}^*_{-1})\widetilde{\+\gamma}_{2,y,i}+\+\varepsilon_i^*.
\end{align}
The bootstrap CCEP estimator then gives
\begin{align}
    \sqrt{NT}(\widehat{\+\delta}^*_{CCEP}-\widehat{\+\delta}_{CCEP})=&\left(\frac{1}{NT}\sum_{i=1}^N\*B_i^{*\prime}\*M_{\widehat{\*F}^*}\*B_i^*\right)^{-1}\notag\\
    &\left(\frac{1}{\sqrt{NT}}\sum_{i=1}^N\*B_i^{*\prime}\*M_{\widehat{\*F}^*}\+\varepsilon_i^*-\frac{1}{\sqrt{NT}}\sum_{i=1}^N\*B_i^{*\prime}\*M_{\widehat{\*F}^*}\overline{\*E}^*\widetilde{\+\gamma}_{1,y,i}-\frac{1}{\sqrt{NT}}\sum_{i=1}^N\*B_i^{*\prime}\*M_{\widehat{\*F}^*}\overline{\*U}^*_{-1}\widetilde{\+\gamma}_{2,y,i} \right).
\end{align}
In what follows, we re-write $\*B_i^*$. Recall that $\*D_i^*=[\*Z_i^*, \*Z_{i,-1}^*]$. Switching to the vector notation and letting $\*u_{i,t}^*=\sum_{j=0}^{t-1}\widehat{\*A}^j\*e_{i,t-j}^*$, we can write recursively for each $t$:
\begin{align}\label{z_star_X_exog}
    \*z_{i,t}^*&= \widehat{\*A}^t\*z_{i,0}+\sum_{j=0}^{t-1}\widehat{\*A}^j\widehat{\+\Gamma}_i'[\overline{\*z}_{t-j}', \overline{\*z}_{t-j-1}']'+\*u_{i,t}^*\notag\\
    &=\widehat{\*A}^t\*z_{i,0}+\sum_{j=0}^{t-1}\widehat{\*A}^j\widehat{\+\Gamma}_i'\mathbb{R}'(\mathbb{R}')^{-1}[\overline{\*z}_{t-j}', \overline{\*z}_{t-j-1}']'+\*u_{i,t}^*\notag\\
&=\widehat{\*A}^t\*z_{i,0}+\sum_{j=0}^{t-1}\widehat{\*A}^j\widetilde{\+\Gamma}_{1,i}'\underbrace{[\overline{\*z}_{t-j} -\widehat{\*A}\overline{\*z}_{t-j-1}]}_{\widehat{\*f}_{\widehat{\mathbb{A}},t-j}}+\sum_{j=0}^{t-1}\widehat{\*A}^j\widetilde{\+\Gamma}_{2,i}'\overline{\*z}_{t-j-1}+\*u_{i,t}^*\notag\\
    &=\widehat{\*A}^t\*z_{i,0} + (\mathrm{vec}(\widetilde{\+\Gamma}_{1,i}')\otimes \*I_{K\times K})'\sum_{j=0}^{t-1}\mathrm{vec}\left( \widehat{\*f}'_{\widehat{\mathbb{A}},t-j}\otimes \widehat{\*A}^j \right)\notag\\
    &+(\mathrm{vec}(\widetilde{\+\Gamma}_{2,i}')\otimes \*I_{K\times K})'\underbrace{\sum_{j=0}^{t-1}\mathrm{vec}\left(\overline{\*z}_{t-j-1}'\otimes \widehat{\*A}^j \right)}_{\*b_t}+\*u_{i,t}^*\notag\\
    &= \widehat{\*A}^t\*z_{i,0} + \widetilde{\+\Lambda}_{1,i}'\sum_{j=0}^{t-1}\mathrm{vec}\left(\widehat{\*f}'_{\widehat{\mathbb{A}},t-j}\otimes \widehat{\*A}^j\right) + \widetilde{\+\Lambda}_{2,i}'\*b_t+\*u_{i,t}^*\notag\\
    &=\widehat{\*A}^t\*z_{i,0} + \widetilde{\+\Lambda}_{1,i}'\widehat{\*g}_t+\widetilde{\+\Lambda}_{2,i}'\*b_t+\*u_{i,t}^*
\end{align}
where we define $\widetilde{\+\Lambda}_{j,i}=(\mathrm{vec}(\widetilde{\+\Gamma}_{j,i}')\otimes \*I_{K\times K})$ for $j=1,2$. Again, we see an extra factor $\*b_t$ appearing in the bootstrap world. This implies that we can obtain a representation of $\*d_{i,t}^*$ in the spirit of \cite{DeVos2019}, again. In particular, 
\begin{align}\label{d_it_stat_exog}
    \*d_{i,t}^*&=\begin{bmatrix}
        \widehat{\*A}^t\\
          \widehat{\*A}^{t-1}
    \end{bmatrix}\*z_{i,0}+(\*I_{2 \times 2}\otimes \widetilde{\+\Lambda}_{1,i})'\begin{bmatrix}
        \widehat{\*g}_t\\
        \widehat{\*g}_{t-1} 
    \end{bmatrix}+(\*I_{2 \times 2}\otimes \widetilde{\+\Lambda}_{2,i})'\begin{bmatrix}
        \*b_t\\
        \*b_{t-1} 
    \end{bmatrix} + \*u_{i,t}^*\notag\\
    &=\begin{bmatrix}
        \widehat{\*A}\\
        \*I_{K\times K}
    \end{bmatrix} \widehat{\*A}^{t-1}\*z_{i,0}+\mathbf{\*C}_{1,i}'\widehat{\*q}_t + \mathbf{\*C}_{2,i}'\*h_t + \*u_{i,t}^*
\end{align}
or
\begin{align}
    \*D_i^*=\widehat{\*A}_{\*z_0,i}[ \widehat{\*A}, \*I_{K\times K}]+\widehat{\*Q}\*C_{1,i}+\*H\*C_{2,i}+\*U_i^*
\end{align}
for where $\widehat{\*A}_{\*z_0,i}=\left[\widehat{\*A}\*z_{i,0}, \widehat{\A}_{CCEP}^2\*z_{i,0},\ldots,\widehat{\A}_{CCEP}^{T-1}\*z_{i,0}\right]'$. Therefore, $\*B_i^*=\*D_i^*\*S_\*b$. \\

\noindent In what follows, we can express the main quantities, such $\widehat{\+\varepsilon}_i$, $\widehat{\+\nu}_i$, $\widehat{\+\Gamma}_i$ and its rotated counterparts to align their asymptotic behavior to the corresponding components in the AR(1) case. Let us define the selector $\*e_1=[1,0,\ldots, 0]'\in \mathbb{R}^K$ that extracts the first column. Then $  \widehat{\+\varepsilon}_i=\widehat{\*E}_i\*e_1-\widehat{\+\nu}_i\widehat{\+\beta}_{CCEP}$. Notice that 
\begin{align}\label{resid_exp2}
  &\widehat{\*E}_i=\*M_{\widehat{\*F}}(\*Z_i-\*Z_{i,-1}\widehat{\*A})=\*M_{\widehat{\*F}}(\*Z_{i,-1}\*A_0+\*F\+\Gamma_i+\*E_i-\*Z_{i,-1}\widehat{\*A})\notag\\
  &= \*M_{\widehat{\*F}}(-\overline{\*E}\overline{\+\Gamma}^{-1}\+\Gamma_i +[\overline{\*Z}-\overline{\*Z}_{-1}\*A_0]\overline{\+\Gamma}^{-1}\+\Gamma_i + \*Z_{i,-1}(\*A_0-\widehat{\*A})+\*E_i)\notag\\
  &=\*M_{\widehat{\*F}}(-\overline{\*E}\overline{\+\Gamma}^{-1}\+\Gamma_i +[\overline{\*Z}-\overline{\*Z}_{-1}\*A_0]\overline{\+\Gamma}^{-1}\+\Gamma_i + \*G_{-1}\+\Lambda_i(\*A_0-\widehat{\*A})+ \*U_{i,-1}(\*A_0-\widehat{\*A})+\*E_i)\notag\\
  &=\*M_{\widehat{\*F}}(-\overline{\*E}\overline{\+\Gamma}^{-1}\+\Gamma_i +\overline{\*Z}\overline{\+\Gamma}^{-1}\+\Gamma_i -\overline{\*Z}_{-1}\*A_0\overline{\+\Gamma}^{-1}\+\Gamma_i + \*G_{-1}\+\Lambda_i(\*A_0-\widehat{\*A})+ \*U_{i,-1}(\*A_0-\widehat{\*A})+\*E_i)\notag\\
  &= \*M_{\widehat{\*F}}\Big(-\overline{\*E}\overline{\+\Gamma}^{-1}\+\Gamma_i +\overline{\*Z}\overline{\+\Gamma}^{-1}\+\Gamma_i -\overline{\*Z}_{-1}\widehat{\*A}\overline{\+\Gamma}^{-1}\+\Gamma_i + \overline{\*Z}_{-1}\widehat{\*A}\overline{\+\Gamma}^{-1}\+\Gamma_i-\overline{\*Z}_{-1}\*A_0\overline{\+\Gamma}^{-1}\+\Gamma_i+\*G_{-1}\+\Lambda_i(\*A_0-\widehat{\*A})\notag\\
  &+ \*U_{i,-1}(\*A_0-\widehat{\*A})+\*E_i\Big)\notag\\
  &=\*M_{\widehat{\*F}}(-\overline{\*E}\overline{\+\Gamma}^{-1}\+\Gamma_i+(\overline{\*Z}-\overline{\*Z}_{-1}\widehat{\*A})\overline{\+\Gamma}^{-1}\+\Gamma_i+\overline{\*Z}_{-1}(\widehat{\*A}-\*A_0)\overline{\+\Gamma}^{-1}\+\Gamma_i+  \*G_{-1}\+\Lambda_i(\*A_0-\widehat{\*A})+ \*U_{i,-1}(\*A_0-\widehat{\*A})+\*E_i)\notag\\
  &=\*M_{\widehat{\*F}} \Big(-\overline{\*E}\overline{\+\Gamma}^{-1}\+\Gamma_i+(\overline{\*Z}-\overline{\*Z}_{-1}\widehat{\*A})\overline{\+\Gamma}^{-1}\+\Gamma_i+\*G_{-1}\overline{\+\Lambda}(\widehat{\*A}-\*A_0)\overline{\+\Gamma}^{-1}\+\Gamma_i +\overline{\*U}_{-1}(\widehat{\*A}-\*A_0)\overline{\+\Gamma}^{-1}\+\Gamma_i+  \*G_{-1}\+\Lambda_i(\*A_0-\widehat{\*A})\notag\\
  &+ \*U_{i,-1}(\*A_0-\widehat{\*A})+\*E_i\Big)\notag\\
  &=\*M_{\widehat{\*F}} (-\overline{\*E}\overline{\+\Gamma}^{-1}\+\Gamma_i+\*G_{-1}\overline{\+\Lambda}(\widehat{\*A}-\*A_0)\overline{\+\Gamma}^{-1}\+\Gamma_i +\overline{\*U}_{-1}(\widehat{\*A}-\*A_0)\overline{\+\Gamma}^{-1}\+\Gamma_i+  \*G_{-1}\+\Lambda_i(\*A_0-\widehat{\*A})+ \*U_{i,-1}(\*A_0-\widehat{\*A})+\*E_i)
\end{align}
which is quite similar to (\ref{residual_expnasion}), where we retain two extra terms due to $r>1$, but they cancel out under $R=1$ and no exogenous regressors. Also, 
\begin{align}
\widehat{\+\nu}_i=\*M_{\widehat{\*F}}\*X_i=\*M_{\widehat{\*F}}(\*F\+\Gamma_{x,i}+\+\nu_i)&=\*M_{\widehat{\*F}}(-\overline{\*E}\overline{\+\Gamma}^{-1}\+\Gamma_{x,i} +[\overline{\*Z}-\overline{\*Z}_{-1}\*A_0]\overline{\+\Gamma}^{-1}\+\Gamma_{x,i}+\+\nu_i)\notag\\
    &=\*M_{\widehat{\*F}}(-\overline{\*E}\overline{\+\Gamma}^{-1}\+\Gamma_{x,i} +\+\nu_i),
\end{align}
which, given that $\left\|\widehat{\*A}-\*A_0 \right\|=|\widehat{\alpha}_{CCEP}-\alpha_0|=O_P((NT)^{-1/2})$ and $\left\|\widehat{\+\beta}_{CCEP}-\+\beta_0 \right\|=O_P((NT)^{-1/2})$, we can show by using similar steps to (\ref{residual_moments_start}) under our assumptions 
\begin{align}
    \frac{1}{NT}\sum_{i=1}^N\sum_{t=1}^T|\widehat{\varepsilon}_{i,t}|^p&\leq3^{p-1}\frac{1}{NT} \sum_{i=1}^N\sum_{t=1}^T|\*e_1'\widehat{\*e}_{i,t}|^p+3^{p-1}\left\| \+\beta_0\right\|\frac{1}{NT}\sum_{i=1}^N\sum_{t=1}^T\left\| \widehat{\+\nu}_{i,t}\right\|^p\notag\\
    &+3^{p-1}\left\| \widehat{\+\beta}_{CCEP}-\+\beta_0\right\|\frac{1}{NT}\sum_{i=1}^N\sum_{t=1}^T\left\| \widehat{\+\nu}_{i,t}\right\|^p=O_P(1)
\end{align}
for $p=4+\delta$, where $\delta \in (0, 4]$. Next, we examine the loadings. In particular, using (\ref{resid_exp2}), we obtain
\begin{align}
\widehat{\+\Gamma}_i&= \widehat{\mathbb{A}}\overline{\+\Gamma}^{-1}\+\Gamma_i \notag\\
&+(\widehat{\*F}'\widehat{\*F})^{-1}\widehat{\*F}'\Big(-\overline{\*E}\overline{\+\Gamma}^{-1}\+\Gamma_i+\*G_{-1}\overline{\+\Lambda}(\widehat{\*A}-\*A_0)\overline{\+\Gamma}^{-1}\+\Gamma_i +\overline{\*U}_{-1}(\widehat{\*A}-\*A_0)\overline{\+\Gamma}^{-1}\+\Gamma_i+  \*G_{-1}\+\Lambda_i(\*A_0-\widehat{\*A})\notag\\
&+ \*U_{i,-1}(\*A_0-\widehat{\*A})+\*E_i\Big),
\end{align}
which we can use to show that 
\begin{align}\label{hat_gamma_exp}
    \frac{1}{N}\sum_{i=1}^N\left\|\widehat{\+\Gamma}_i \right\|^p&\leq 6^{p-1}\left\| \widehat{\mathbb{A}}\overline{\+\Gamma}^{-1}\right\|^p\frac{1}{N}\sum_{i=1}^N\left\|\+\Gamma_i\right\|^p +  6^{p-1}\left\|(T^{-1}\widehat{\*F}'\widehat{\*F})^{-1} \right\|^p\notag\\
    &\times \frac{1}{N}\sum_{i=1}^N\Big(\left\|T^{-1}\widehat{\*F}'\overline{\*E} \right\|^p\left\| \overline{\+\Gamma}^{-1}\+\Gamma_i\right\|^p +\left\|T^{-1}\widehat{\*F}'\*G_{-1} \right\|^p\left\|\overline{\+\Lambda} \right\|^p\left\|\overline{\+\Gamma}^{-1}\+\Gamma_i \right\|^p\left\| \widehat{\*A}-\*A_0\right\|^p\notag\\
    &+\left\|T^{-1}\widehat{\*F}'\overline{\*U}_{-1} \right\|^p\left\|\overline{\+\Gamma}^{-1}\+\Gamma_i \right\|^p\left\| \widehat{\*A}-\*A_0\right\|^p+\left\|T^{-1}\widehat{\*F}'\*G_{-1} \right\|^p\left\|\+\Lambda_i \right\|^p\left\| \widehat{\*A}-\*A_0\right\|^p\notag\\
    &+\left\|T^{-1}\widehat{\*F}'\*U_{i,-1} \right\|^p\left\| \widehat{\*A}-\*A_0\right\|^p+\left\|T^{-1}\widehat{\*F}' \*E_i\right\|^p\Big)\notag\\
    &=6^{p-1}\left\| \widehat{\mathbb{A}}\overline{\+\Gamma}^{-1}\right\|^p\frac{1}{N}\sum_{i=1}^N\left\|\+\Gamma_i\right\|^p+o_P(1)\notag\\
    &=O_P(1)
\end{align}
and the same holds for $\frac{1}{N}\sum_{i=1}^N\left\|\widehat{\+\Gamma}_{x,i} \right\|^p$ and $\frac{1}{N}\sum_{i=1}^N\left\| \widehat{\+\gamma}_{y,i}\right\|^p$ since they are elements of (\ref{hat_gamma_exp}). In addition, we demonstrate that $\widetilde{\+\Gamma}_{2,i}$ and $\widetilde{\+\gamma}_{2,y,i}$ (the loadings of the extra bootstrap factors) are negligible, which means that we can conduct analysis similar to (\ref{gamma_tilde_2}). In particular, due to the lower triangular form of the matrix, we have that $\mathbb{R}^{-1}=\begin{bmatrix}
    \*I_{K\times K} & \*0_{K\times K}\\ \widehat{\*A} & \*I_{K\times K}, 
\end{bmatrix}$
and therefore by splitting $\widehat{\+\Gamma}_i=[\widehat{\+\Gamma}_{1,i}', \widehat{\+\Gamma}_{2,1}']'$, we know that $\widetilde{\+\Gamma}_{2,i}=\widehat{\*A}\widehat{\+\Gamma}_{1,i}+\widehat{\+\Gamma}_{2,i}$, which means that 
\begin{align}\label{Gamma_2_tilde}
    &\left\|\widetilde{\+\Gamma}_{2,i}\right\|=\left\|\begin{bmatrix}
        \widehat{\*A}& \*I_{K\times K}
    \end{bmatrix}\widehat{\+\Gamma}_i\right\|\notag\\
    &=\left\|\begin{bmatrix}
        \widehat{\*A}& \*I_{K\times K}
    \end{bmatrix}\right\|\left\|(T^{-1}\widehat{\*F}'\widehat{\*F})^{-1}\right\|\Big(\left\|T^{-1}\widehat{\*F}'\overline{\*E}\overline{\+\Gamma}^{-1}\+\Gamma_i\right\|+\left\|T^{-1}\widehat{\*F}'\*G_{-1}\overline{\+\Lambda}\right\|\left\|(\widehat{\*A}-\*A_0)\overline{\+\Gamma}^{-1}\+\Gamma_i\right\|\notag\\ &+\left\|T^{-1}\widehat{\*F}'\overline{\*U}_{-1}\right\|\left\|(\widehat{\*A}-\*A_0)\overline{\+\Gamma}^{-1}\+\Gamma_i\right\|+  \left\|T^{-1}\widehat{\*F}'\*G_{-1}\+\Lambda_i\right\|\left\|(\*A_0-\widehat{\*A})\right\| \left\|T^{-1}\widehat{\*F}'\*U_{i,-1}\right\| \left\|(\*A_0-\widehat{\*A})\right\|\notag\\
    &+\left\|T^{-1}\widehat{\*F}'\*E_i\right\|\Big)=O_P(T^{-1/2}),
\end{align}
which is the same rate as in the AR(1) case. This is so, since $\begin{bmatrix}
        \widehat{\*A}& \*I_{K\times K}
    \end{bmatrix}\widehat{\mathbb{A}}=\*0_{K\times K}$ and the rate is driven by 
    \begin{align}
        \left\| T^{-1}\widehat{\*F}'\*E_i\right\|\leq \left\| \overline{\*C}'T^{-1}\*Q'\*E_i\right\| + \left\| T^{-1}\overline{\*V}'\*E_i\right\|=O_P(T^{-1/2}),
    \end{align}
    where the first term dominates. Next, by following a similar decomposition, and using (\ref{Gamma_2_tilde}), we obtain
    \begin{align}
        \widetilde{\+\gamma}_{2,y,i}=\begin{bmatrix}
        \widehat{\*A}& \*I_{K\times K}
    \end{bmatrix}\widetilde{\+\gamma}_{y,i}&=\begin{bmatrix}
        \widehat{\*A}& \*I_{K\times K}
    \end{bmatrix}\left(\widehat{\+\Gamma}_i\*e_1-\widehat{\+\Gamma}_{x,i}\widehat{\+\beta}_{CCEP}\right)\notag\\
    &=-\begin{bmatrix}
        \widehat{\*A}& \*I_{K\times K}
    \end{bmatrix}(T^{-1}\widehat{\*F}'\widehat{\*F})^{-1}T^{-1}\widehat{\*F}'(-\overline{\*E}\overline{\+\Gamma}^{-1}\+\Gamma_{x,i} +[\overline{\*Z}-\overline{\*Z}_{-1}\*A_0]\overline{\+\Gamma}^{-1}\+\Gamma_{x,i}+\+\nu_i)\widehat{\+\beta}_{CCEP}\notag\\
    &+O_P(T^{-1/2})\notag\\
    &=\begin{bmatrix}
        \widehat{\*A}& \*I_{K\times K}
    \end{bmatrix}(T^{-1}\widehat{\*F}'\widehat{\*F})^{-1}T^{-1}\widehat{\*F}'\overline{\*E}\overline{\+\Gamma}^{-1}\+\Gamma_{x,i}\widehat{\+\beta}_{CCEP}\notag\\
    &-\begin{bmatrix}
        \widehat{\*A}& \*I_{K\times K}
    \end{bmatrix}(T^{-1}\widehat{\*F}'\widehat{\*F})^{-1}T^{-1}\widehat{\*F}'[\overline{\*Z}-\overline{\*Z}_{-1}\widehat{\*A}]\overline{\+\Gamma}^{-1}\+\Gamma_{x,i}\widehat{\+\beta}_{CCEP}\notag\\
    &-\begin{bmatrix}
        \widehat{\*A}& \*I_{K\times K}
    \end{bmatrix}(T^{-1}\widehat{\*F}'\widehat{\*F})^{-1}T^{-1}\widehat{\*F}'\overline{\*Z}_{-1}(\widehat{\*A}-\*A_0)\overline{\+\Gamma}^{-1}\+\Gamma_{x,i}\widehat{\+\beta}_{CCEP}\notag\\
    &-\begin{bmatrix}
        \widehat{\*A}& \*I_{K\times K}
    \end{bmatrix}(T^{-1}\widehat{\*F}'\widehat{\*F})^{-1}T^{-1}\widehat{\*F}'\+\nu_i\widehat{\+\beta}_{CCEP}+O_P(T^{-1/2})\notag\\
    &=-\begin{bmatrix}
        \widehat{\*A}& \*I_{K\times K}
    \end{bmatrix}(T^{-1}\widehat{\*F}'\widehat{\*F})^{-1}T^{-1}\widehat{\*F}'[\overline{\*Z}-\overline{\*Z}_{-1}\widehat{\*A}]\overline{\+\Gamma}^{-1}\+\Gamma_i\widehat{\+\beta}_{CCEP}+O_P(T^{-1/2})\notag\\
    &=-\underbrace{\begin{bmatrix}
        \widehat{\*A}& \*I_{K\times K}
    \end{bmatrix}\widehat{\mathbb{A}}}_{\*0_{K\times K}}\overline{\+\Gamma}^{-1}\+\Gamma_i\widehat{\+\beta}_{CCEP}+O_P(T^{-1/2})\notag\\
    &=O_P(T^{-1/2}),
    \end{align}
    and therefore the loading of the extra bootstrap factor in $\*y_i^*$ is negligible, as well, at the same rate as in pure AR(1) case, driven by (\ref{Gamma_2_tilde}) and 
    \begin{align}
        \left\| T^{-1}\widehat{\*F}'\+\nu_i\right\|\leq\left\| \overline{\*C}'T^{-1}\*Q'\+\nu_i\right\| + \left\| T^{-1}\overline{\*V}'\+\nu_i\right\|=O_P(T^{-1/2}), 
    \end{align}
    where the first term dominates. Given that $\widehat{\+\Gamma}_{2,i}$ and $\widetilde{\+\gamma}_{2,y,i}$ are negligible, the distribution of (\ref{CCEP_boot_x}) can be derived by replicating the steps in \cite{DeVos2019}.
    
\end{document}